\documentclass[12pt]{article}
\pdfoutput=1

\newlength{\abstractwidth}
\usepackage{epsf}
\usepackage{color}
\usepackage{graphicx}
\usepackage{hyperref}
\usepackage{dsfont}
\usepackage{amsmath}
\usepackage{amssymb}
\usepackage{latexsym}
\usepackage{mathtools}
\usepackage{tikz, pgf}
\usepackage{tkz-fct}
\usepackage{pgfplots}
\usetikzlibrary{shapes.misc}
\usetikzlibrary{shapes,snakes}
\usetikzlibrary{decorations.pathmorphing}	
\usetikzlibrary{decorations.markings}

\usetikzlibrary{arrows,shapes,positioning}
\usetikzlibrary{decorations.markings}
\usepackage[rightcaption]{sidecap}
\tikzstyle arrowstyle=[scale=1]
\tikzstyle directed=[postaction={decorate,decoration={markings,
    mark=at position .65 with {\arrow[arrowstyle]{stealth}}}}]
\tikzstyle reverse directed=[postaction={decorate,decoration={markings,
    mark=at position .65 with {\arrowreversed[arrowstyle]{stealth};}}}]
\usetikzlibrary{positioning}

\renewcommand{\thefootnote}{\fnsymbol{footnote}}
\renewcommand{\thanks}[1]{\footnote{#1}}
\newcommand{\starttext}{
\setcounter{footnote}{0}
\renewcommand{\thefootnote}{\arabic{footnote}}}

\numberwithin{equation}{section}
\numberwithin{equation}{subsection}
\newcommand{\bea}{\begin{eqnarray}}
\newcommand{\eea}{\end{eqnarray}}
\newcommand{\be}{\begin{eqnarray}}
\newcommand{\ee}{\end{eqnarray}}
\newcommand{\<}{\langle}
\renewcommand{\>}{\rangle}
\newcommand{\bma}{\begin{matrix}}
\newcommand{\ema}{\end{matrix}}

\def\cA{{\cal A}}
\def\cB{{\cal B}}
\def\cC{{\cal C}}
\def\cD{{\cal D}}
\def\cE{{\cal E}}
\def\cF{{\cal F}}
\def\cG{{\cal G}}
\def\cH{{\cal H}}
\def\cI{{\cal I}}
\def\cJ{{\cal J}}
\def\cK{{\cal K}}
\def\cL{{\cal L}}
\def\cM{{\cal M}}
\def\cN{{\cal N}}
\def\cO{{\cal O}}

\def\cQ{{\cal Q}}
\def\cR{{\cal R}}
\def\cS{{\cal S}}
\def\cT{{\cal T}}
\def\cU{{\cal U}}
\def\cV{{\cal V}}
\def\cW{{\cal W}}

\def\cY{{\cal Y}}

\def\bB{{\bf B}}

\def\bE{{\bf E}}

\def\bN{{\bf N}}

\def\ba{{\bf a}}

\def\mA{\mathfrak{A}}
\def\mB{\mathfrak{B}}
\def\mC{\mathfrak{C}}

\def\mF{\mathfrak{F}}

\def\mH{\mathfrak{H}}

\def\mJ{\mathfrak{J}}

\def\ma{\mathfrak{a}}
\def\mb{\mathfrak{b}}
\def\mc{\mathfrak{c}}

\def\me{\mathfrak{e}}
\def\mf{\mathfrak{f}}
\def\mg{\mathfrak{g}}
\def\mh{\mathfrak{h}}

\def\mt{\mathfrak{t}}
\def\muu{\mathfrak{u}}

\def\AA{{\mathbb A}}
\def\CC{{\mathbb C}}
\def\FF{{\mathbb F}}

\def\NN{{\mathbb N}}
\def\QQ{{\mathbb Q}}
\def\RR{{\mathbb R}}
\def\ZZ{{\mathbb Z}}

\def\Re{{\rm Re \,}}
\def\Im{{\rm Im \,}}
\def\tr{{\rm tr}}
\def\Tr{{\rm Tr}}
\def\det{{\rm det \,}}
\def\Det{{\rm Det}}
\def\half{{1\over 2}}
\def\thalf{\tfrac{1}{2}}
\def\p{\partial}

\def\a{\alpha}
\def\g{\gamma}
\def\d{\delta}
\def\b{\beta}
\def\k{\kappa}

\def\eps{\epsilon}
\def\f{\varphi}
\def\m{\mu}
\def\n{\nu}
\def\tet{\vartheta}
\def\ep{\varepsilon}
\def\om{\omega}

\def\pbz{\p _{\bar z}}

\def\ord{{\rm ord}}

\def\sn{{\rm sn}}
\def\cn{{\rm cn}}
\def\dn{{\rm dn}}

\def\CP{{\mathds{CP}}}

\def\ti{\tilde}

\def\mod{{\rm mod ~ }}

\def\Tt{{\sf T}}
\def\HE{{\sf E}}
\def\ab{{\boldsymbol{\a}}}
\def\betb{{\boldsymbol{\b}}}
\def\TAU{{ \tau}}

\def\ss{\left [ \begin{smallmatrix} \alpha \cr \beta \cr \end{smallmatrix} \right ] }

\def\no{\nonumber}
\def\sm{\smallskip}

\definecolor{Cyan}{cmyk}{1.,0,0,0}
\definecolor{Magenta}{cmyk}{0,1.,0,0}
\definecolor{Yellow}{cmyk}{0,0,1.,0}
\definecolor{White}{cmyk}{0,0,0,0}
\definecolor{Orange}{cmyk}{0,0.61,0.87,0}
\definecolor{RedOrange}{cmyk}{0,0.77,0.87,0}
\definecolor{Red}{cmyk}{0,1.,1.,0}
\definecolor{Purple}{cmyk}{0.45,0.86,0,0}
\definecolor{Violet}{cmyk}{0.79,0.88,0,0}
\definecolor{Blue}{cmyk}{1,0.5,0,0}
\definecolor{ProcessBlue}{cmyk}{0.96,0,0,0}
\definecolor{GreenYellow}{cmyk}{0.6,0,1.,0}
\definecolor{Black}{cmyk}{0,0,0,1}
\newtheorem{theorem}{Theorem}

\newtheorem{thm}{Theorem}[section]
\newtheorem{lem}[thm]{Lemma}
\newtheorem{prop}[thm]{Proposition}
\newtheorem{cor}[thm]{Corollary}
\newtheorem{conj}[thm]{Conjecture}

\newtheorem{deff}[thm]{Definition}

\def\m{\mu}

\begin{document}
\starttext
\setcounter{footnote}{0}

\begin{flushright}
2022 August 15   \\
2022 November 18 (v2)
\end{flushright}

\vskip 0.3in

\begin{center}

{\Large \bf Lectures on modular forms and strings}

\vskip 0.2in

{\large Eric D'Hoker${}^{(a)}$ and Justin Kaidi${}^{(b)}$\footnote{Address after 1 September 2022 :  \textit{Department of Physics, University of Washington, Seattle, WA, 98195}}} 

\vskip 0.15in

{ \sl ${}^{(a)}$ Mani L. Bhaumik Institute for Theoretical Physics}\\
{\sl  Department of Physics and Astronomy}\\
{\sl University of California, Los Angeles, CA 90095, USA}

\vskip 0.15in 

{\sl ${}^{(b)}$ Simons Center for Geometry and Physics}\\
{\sl Stony Brook University}\\
{\sl Stony Brook NY 11794, USA}

\vskip 0.15in 

{\tt \small dhoker@physics.ucla.edu, jkaidi@scgp.stonybrook.edu
}

\vskip 0.2in

\begin{abstract}
\vskip 0.1in
The goal of these lectures is to present an informal but precise introduction to a body of concepts and methods of interest in number theory and string theory revolving around modular forms and their generalizations. Modular invariance lies at the heart of  conformal field theory, string perturbation theory, Montonen-Olive duality,  Seiberg-Witten theory, and S-duality in Type IIB superstring theory. Automorphic forms with respect to higher arithmetic groups as well as mock modular forms enter in toroidal string compactifications and the counting of black hole microstates. After introducing the basic mathematical concepts including elliptic functions, modular forms, Maass forms, modular forms for congruence subgroups, vector-valued modular forms, and modular graph forms, we describe a small subset of the countless applications to problems in Mathematics and Physics, including those mentioned above.

\end{abstract}
\end{center}

\newpage

\setcounter{tocdepth}{2} 
\tableofcontents

\baselineskip=15pt
\setcounter{equation}{0}
\setcounter{footnote}{0}

\newpage

\section{Introduction}
\setcounter{equation}{0}
\label{sec:Intro}

Integers form a group under addition but not under multiplication. However, matrices with integer entries can form groups under multiplication. For example $2 \times 2$ matrices of unit determinant and integer entries $a,b,c,d \in \ZZ$, 
\bea
\left ( \begin{matrix} a & b \cr c & d \cr \end{matrix} \right ) \hskip 1in ad-bc=1
\eea
form the  group $SL(2,\ZZ)$ under multiplication, referred to as the \textit{modular group}. Functions and differential forms that are invariant  under $SL(2,\ZZ)$ are referred to as \textit{modular functions} and  \textit{modular forms}, respectively. Modular functions generalize \textit{periodic functions}  and \textit{elliptic functions}, to which they are intimately related. In turn, modular functions are special cases of {\sl automorphic functions} which are invariant under more general  \textit{arithmetic groups}.

\sm

In Mathematics, the study of elliptic integrals goes back to Euler, Legendre, and Abel, while the theory of elliptic functions was developed by Gauss, Jacobi, and Weierstrass  and was motivated in part by questions ranging from  number theory to the  solvability of algebraic equations by radicals. Riemann developed the theory of general Riemann surfaces and generalized elliptic functions  to  higher rank $\tet$-functions on surfaces of higher genus. 

\sm

The development of modular forms  dates back to Eisenstein,  Kronecker, and Hecke. Automorphic functions and forms were studied  by Fuchs,  Fricke,  Klein, and Poincar\'e.   In modern times, amongst many other developments, the Taniyama-Shimura-Weil conjecture ultimately led to the proof of Fermat's Last Theorem by Wiles  and Taylor, and to a  proof of the Modularity Theorem  by Breuil, Conrad, Diamond, and Taylor. A fundamental role was played by modular forms and quasi-modular forms in the solution to the sphere packing problem in eight dimensions by Viazovska, work for which she was awarded the Fields Medal in 2022.  

\sm

In Physics, elliptic integrals arose already in the simplest classical mechanical problems, such as the pendulum, and were studied in this context as far back as the 18th century. The solution of various boundary problems in electrostatics and fluid mechanics via elliptic functions, or more generally automorphic functions, further stimulated the mathematical development of these subjects a century ago. More recently, Riemann $\tet$-functions were found to provide the solution to a host of completely integrable mechanical systems. 

\sm

The modular group $SL(2,\ZZ)$ made its appearance in string theory in 1972 when Shapiro identified it as a symmetry of the integrand that defines the one-loop closed bosonic string amplitude. Shapiro defined the amplitude as the integral over the quotient of the Poincar\'e upper half-plane  by $SL(2,\ZZ)$ and argued that the amplitude thus obtained is free of the short-distance divergences that arise in quantum field theory.  This fundamental observation extends to the five perturbative superstring theories, and to all loop orders, provided that the group $SL(2,\ZZ)$ is replaced by the modular group $Sp(2h,\ZZ)$ for genus $h$ Riemann surfaces. The absence of UV divergences is principally responsible for advocating string theory as the uniquely viable candidate for a consistent quantum theory of gravity.

\sm

A second context in which the modular group $SL(2,\ZZ)$ arose in high-energy physics is  Yang-Mills gauge theory in four space-time dimensions. The Standard Model of Particle Physics is a Yang-Mills theory with gauge group $SU(3) \times SU(2) \times U(1)$ in which the Higgs mechanism generates the masses of quarks, leptons, and the gauge bosons $W^\pm$ and $Z$ via spontaneous symmetry breaking.  A particular class of Yang-Mills theories are those with simple or semi-simple gauge groups that are spontaneously  broken to a subgroup containing at least one $U(1)$-factor. This effective $U(1)$ gauge theory contains a photon and various electrically charged particles, as standard Maxwell theory does.  However, unlike standard Maxwell theory, it also contains magnetically charged particles that arise as  `t Hooft Polyakov magnetic monopoles.  Goddard, Nuyts, and Olive conjectured in 1977 that, under certain conditions,  such theories exhibit an \textit{electric-magnetic duality} which swaps electric particles and magnetic monopoles.  Generalizing the construction to dyons which carry both electric and magnetic charges shows that the duality is actually captured by the  duality group $SL(2,\ZZ)$, or a congruence subgroup thereof. A concrete realization of electric-magnetic duality, referred to as  Montonen-Olive duality, is provided by  Yang-Mills theories with extended supersymmetry, and led to Seiberg-Witten theory in 1994. 

\sm

A third context in which $SL(2,\ZZ)$ arises takes us back to superstring theory. The five superstring theories that admit a perturbative description in ten-dimensional Minkowski space-time are the Type~IIA and Type~IIB theories with the maximal number of 32 supersymmetries, and the Type~I and two Heterotic string theories with 16 supersymmetries. The massless   states of each one of these superstring theories are described, at tree-level,  by an associated supergravity theory, which  is an extension of the Einstein-Hilbert theory of general relativity in ten dimensions.  For example, Type~IIB supergravity contains, in addition to the space-time metric field, a complex axion-dilaton scalar field $\TAU$, anti-symmetric tensor fields of rank 2 and 4, as well as their fermionic partners.  The imaginary part of $\TAU$ is related to the inverse string coupling and must be positive on physical grounds. M\"obius transformations on the field $\TAU$, 
\bea
\TAU \to { a \TAU + b \over c \TAU + d} \hskip 1in  ad-bc=1
\eea
with real parameters $a,b,c,d$ form the group $SL(2,\RR)$ and are a symmetry of Type~IIB supergravity, when accompanied by suitable transformations on the other fields. The field $\TAU$ takes values in the coset $SL(2,\RR)/SO(2)$. However, the $SL(2,\RR)$ symmetry of tree-level supergravity has an anomaly and is broken to its discrete subgroup $SL(2,\ZZ)$,  the so-called \textit{S-duality} symmetry of  Type~IIB superstring theory. Type~IIB string backgrounds whose fields are related by an $SL(2,\ZZ)$ transformation are really identical, so that the space of inequivalent theories is given by the double coset $SL(2,\ZZ) \backslash SL(2,\RR) / SO(2)$.  The implications of this conjectured symmetry were fully appreciated only with the discovery of NS- and D-brane solutions in the 1990s. 

\sm

A fourth context where $SL(2,\ZZ)$ and higher arithmetic groups emerge is when string theory is considered on a space-time of the form $\RR^{10-d} \times T_d$ where $T_d$ is a $d$-dimensional flat torus. Such a setup is often referred to as \textit{toroidal compactification}. While the Fourier analysis of supergravity fields on a torus produces only momentum modes, the compactification of a string theory produces both momentum and winding modes. The winding modes are responsible for quintessentially string-theoretic discrete symmetries, referred to as \textit{T-dualities}, that have no counterpart in  quantum field theory. For example, Type~IIB superstring theory on a circle of radius $R$ is $T$-dual to Type~IIA superstring theory on a circle of radius $\alpha'/R$. Toroidal compactification converts some of the components of the bosonic fields, such as the metric, into scalar fields. These new scalar fields combine with the axion-dilaton field $\TAU$ of Type~IIB to live on a larger coset space which enjoys a larger arithmetic symmetry group. For increasing dimensions $d$ of the toroidal compactification, we have the following arithmetic symmetry groups $G(\ZZ)$ and corresponding coset spaces $G(\RR)/K(\RR)$ starting with ten-dimensional Type~IIB superstring  theory for $d=0$,  
\begin{align}
d & =0 &&SL(2,\ZZ) & &  SL(2,\RR)/SO(2)
\no \\
d & =1 &&SL(2,\ZZ) & & SL(2,\RR)/SO(2)
\no \\
d & =2 && SL(2,\ZZ) \!\! \times \!\! SL(3,\ZZ) && SL(2,\RR) \!\! \times \!\! SL(3,\RR)/(SO(3) \!\! \times \!\! SO(2))
\no \\
d & =3 && SL(5,\ZZ) && SL(5,\RR)/SO(5) 
\no \\
d & =4 && SO(5,5,\ZZ) && SO(5,5, \RR)/(SO(5) \!\! \times \!\! SO(5))
\no \\
d & =n-1 && E_{n,n} (\ZZ)  && E_{n,n}(\RR)/SO(2n)
\no
\end{align}
On the last line $E_{n,n}$ for $n=6,7,8$ denotes a particular real form of the corresponding complexified Lie groups $E_6,E_7, E_8$. Compactification on Calabi-Yau manifolds or orbifolds exhibit similarly quintessential string theoretic relations that go under the name of \textit{mirror symmetry}, which will not be discussed in these lectures. 

\sm

A fifth context where modular symmetry is of great importance is in two-dimensional conformal field theory. 
Cardy  linked S-duality in a conformal field theory to its unitarity properties, while Erik Verlinde  provided a powerful restriction on the operator product expansion coefficients of conformal primary fields in terms of representations of the modular group. More recently, Hecke operators were used to relate different  conformal field theories to one another. Besides describing the worldsheet theory of strings, two-dimensional conformal field theory models the critical behavior of a number of simple, experimentally realizable, systems such as the critical Ising and $n$-states Potts models. Experimental realizations of these systems tend to be on lattices of trivial topology, and one might naively expect that such real-world systems do not possess any manifest modular properties. However, it turns out that modular symmetry gives rise to important constraints even for topologically trivial systems. The use of modular symmetry to constrain the basic data of a conformal theory is a program referred to as the \textit{modular bootstrap}, and has  led to a classification of certain simple families of conformal field theories. Thanks to the gauge-gravity correspondence, these constraints on conformal field theory also give rise to constraints on theories of gravity in three-dimensional anti-de Sitter space, which indeed was one of the original motivations for the modular bootstrap program.

\sm

The goal of these lecture notes is to provide the reader with some of the mathematical tools that are of most immediate use in addressing the various directions of research in physics that are related to modular forms. We have attempted to make the notes as self-contained as possible as far as the mathematics is concerned. We shall assume familiarity with little more than complex analysis and some basic group theory and differential geometry.  These lecture notes are not a review: in no way have we attempted to present a complete survey of the literature either in Mathematics or in Physics.  The task would be insurmountable: each of the Physics directions mentioned above counts thousands of research papers, while the corresponding directions in Mathematics often date back more than 150 years. In particular, many beautiful and important directions of inquiry have not been addressed at all here. They include the modular and arithmetic properties of orbifold and Calabi-Yau compactifications, F-theory, Moonshine, three-dimensional topological field theory, topological modular forms, black-hole microstate counting, and the structure of super moduli space and related topics. Perhaps the discussion of some of these topics will be included in later, revised, versions.

\section*{Organization}

These lecture notes are organized in three parts. The first part provides an introduction to elliptic functions and modular forms, as well as to numerous generalizations such as quasi-modular forms, almost-holomorphic modular forms, non-holomorphic modular forms, mock modular forms, and quantum modular forms. Full sections are dedicated to a discussion of modular forms for congruence subgroups, vector-valued modular forms, and modular graph functions. In the second part, various  mathematical and physical applications and extensions of the material of the first part are provided. The mathematical extensions include Hecke theory, complex multiplication, and Galois theory, while the physical applications include string perturbation theory, S-duality in Type IIB superstring theory, dualities in super-Yang-Mills theories with extended supersymmetry,  and Seiberg-Witten theory. 
Finally, the third part consists of four appendices of material that is central to the material of the core sections, but may be read independently thereof. This includes an introduction to arithmetic, Riemann surfaces, line bundles on Riemann surfaces, and higher rank $\tet$-functions and holomorphic and meromorphic forms on Riemann surfaces of higher genus. Each part is preceded by a one-page organizational summary. 

\sm

Bibliographical notes are provided at the end of each section and appendix. For Mathematics references, many excellent textbooks are available on elliptic functions, $\tet$-functions, modular forms, Riemann surfaces, and Galois theory. We shall provide  references to the expositions that we have found particularly useful. For Physics references, we shall refer as much as possible to the textbooks, review papers, and lecture notes that we believe will be useful to the reader. We shall  refer to original research papers whenever they are classics, or when the material is not  readily available in textbooks, reviews, and lecture notes.

\section*{Acknowledgments} 

We are very happy to acknowledge collaborations on a variety of  related topics with Sriram Bharadwaj,  Gordon Chan, Pierre Deligne, Bill Duke, Thomas Dumitrescu, Nick Geiser, Jan Gerken, Michael Green, \"Omer G\"urdogan,  Michael Gutperle, Martijn Hidding, Axel Kleinschmidt, Zohar Komargodski, Per Kraus, Igor Krichever, Ying-Hsuan Lin, Carlos Mafra, Julio Parra-Martinez, Mario Martone, Emily Nardoni, Kantaro Ohmori, Eric Perlmutter, D.H. Phong, Boris Pioline, Oliver Schlotterer, Shu-Heng Shao, Sahand Seifnashri, Pierre Vanhove, Gabi Zafrir, and Yunqin Zheng. 

\sm

We are grateful to Oliver Schlotterer for helpful comments on the manuscript.
We wish to thank Oliver Schlotterer  and Terry Tomboulis for much appreciated encouragement during this project, and  Yi Li, John Miller, and Raphael Rouquier for helpful discussions at the earliest stages of this project. 
\sm

These notes are an outgrowth of lectures presented for the 2015 workshop ``Supermoduli" at the Simons Center for Geometry and Physics; for the 2017  UCLA graduate course on Physics and Geometry; for the ``Amplitudes 2018 Summer School" at the Center for Quantum Mathematics and Physics at UC Davis; and for the 2019 conference ``Arithmetic, Geometry, and Modular Forms in honor of Bill Duke" at the ETH in Zurich.  

\sm

The  research of ED was supported in part by the National Science Foundation under research grants PHY-16-19926 and PHY-19-14412, and by a Fellowship from the Simons Foundation during the academic year 2017-18. ED's visits to the KITP in Santa Barbara in 2017 and 2022 were supported in part by the NSF grant PHY-17-48958, and funding from the Simons Foundation, which are gratefully acknowledged.

\newpage

\part{Modular forms and their variants}

In the first part of these lecture notes, we introduce basic concepts,  properties, and applications of holomorphic and non-holomorphic modular functions and forms and their variants. 

\sm

In section~\ref{sec:elliptic} we review periodic functions, the Poisson resummation formula, and  applications to analytic continuations that will be needed in the sequel. We relate Weierstrass and Jacobi elliptic functions to Jacobi $\tet$-functions and exhibit their connection with elliptic curves, Abelian differentials, and Abelian  integrals. 

\sm

In section \ref{sec:SL2}, we introduce the modular group $SL(2,\ZZ)$, discuss its structure, and review its action on the Poincar\'e upper half-plane.  We proceed to holomorphic and meromorphic modular functions, modular forms, and cusp forms, and realize them in terms of holomorphic Eisenstein series. The valence formula is  used to provide the dimensions and the generators of the ring of modular forms of a given weight. The discriminant cusp form $\Delta$, the $j$-function, and the Dedekind $\eta$-function are introduced and related to Jacobi $\tet$-functions. 

\sm

In section \ref{sec:VMF} we introduce variants of modular forms for $SL(2,\ZZ)$,  including quasi-modular forms, almost-holomorphic modular forms,  non-holomorphic modular forms (such as non-holomorphic Eisenstein series and Maass forms), mock modular forms, and quantum modular forms.  We conclude section \ref{sec:VMF} by presenting the spectral decomposition of square-integrable modular functions into non-holomorphic Eisenstein series and cusp forms. 

\sm

In section \ref{sec:TorusQFT}, we illustrate the use of elliptic functions and modular forms in the physical context of two-dimensional conformal field theory. The construction of Green functions, correlation functions, and functional determinants  is presented  for scalar and spinor fields on a torus and related to the Kronecker limit formulas.

\sm

In section \ref{sec:Cong} the classic congruence subgroups of $SL(2,\ZZ)$ and the associated modular curves are reviewed. The elliptic points and cusps are counted and used to derive formulas for the genera of modular curves. In section~\ref{sec:Forms} we  construct modular forms for the classic congruence subgroups, and apply the results  to proving the theorems of Lagrange and Jacobi on representing positive integers by sums of four squares. In section \ref{sec:MDEsvvmfs} we further generalize to modular forms that transform under multi-dimensional representations  of $SL(2,\ZZ)$ but whose individual components are invariant under a congruence subgroup. Such vector-valued modular forms satisfy modular differential equations, a notion which is reviewed.

\sm

In section \ref{sec:MGF} we introduce modular graph functions and forms. Physically, they arise  in the low energy expansion of superstring amplitudes where they emerge in integrands on the moduli space of compact Riemann surfaces.  Mathematically, they provide non-trivial generalizations of non-holomorphic Eisenstein series and multiple zeta values, and are intimately related with elliptic polylogarithms and iterated modular integrals.

\newpage

\section{Elliptic functions}
\setcounter{equation}{0}
\label{sec:elliptic}

In this section, we shall begin with a review of periodic functions of a real variable, Poisson resummation, the unfolding trick, and a simple applications to analytic continuation of the Riemann $\zeta$-function. Next, we review simply periodic functions of a complex variable from the stand point of summation over image charges, show how  their differential equation and addition formulas arise as a consequence, and provide a simple application to evaluating the Riemann $\zeta$-function at even positive integers. Elliptic  functions are defined and their general properties regarding zeros and poles are obtained. The classic constructions of  elliptic functions, based on the Weierstrass $\wp$ function, the Jacobi elliptic functions $\sn, \cn, \dn$ and Jacobi $\tet$-functions are presented. Finally, the elliptic  function theory developed here is placed in the framework of elliptic curves, Abelian differentials, and Abelian integrals.

\subsection{Periodic functions of  a real variable}

A function $f : \RR \to \CC$ is periodic with period 1 if for all $x \in \RR$ it satisfies,
\bea
f(x+1) = f(x)
\eea
Equivalently, $f$ is a function of the circle $S^1 = \RR/\ZZ \to \CC$. The function $f$  is completely specified by the values it takes in the unit interval $[0,1[$. There are various other useful systematic methods for constructing  periodic functions, some of which we shall now review.

\sm

If a function $g : \RR \to \CC$ decays sufficiently rapidly at $ \infty$ then a periodic function may be constructed by the \textit{method of images},
\bea
\label{images}
f(x) = \sum _{n \in \ZZ} g(x+n)
\eea
The functions $e^{2 \pi i mx}$ for $m\in \ZZ$ are all periodic with period 1, and in fact form a basis for the square-integrable periodic functions $L^2 (S^1)$. Their orthogonality and completeness are manifest from the following relations (in the sense of distributions),
\bea
\int _0 ^1 dx \, e^{ 2 \pi i mx} \, e^{-2 \pi i m'x} & = & \delta _{m,m'}
\no \\
\sum _{m \in \ZZ} \, e^{ 2 \pi i mx} \, e^{-2 \pi i my} & = & \sum _{n\in \ZZ} \delta (x-y+n)
\eea
where $\delta(x)$ is the Dirac $\delta$-function normalized to $\int _\RR dy \, \delta (x-y) f(y)=f(x)$.
Any periodic function $f$ with period 1 may be decomposed into a Fourier series, 
\bea
f(x) = \sum _{m \in \ZZ} f_m \, e^{ 2 \pi i mx}
\hskip 1in 
f_m = \int _0 ^1 dx \, f(x) \, e^{-2 \pi i mx}
\eea
We note that in the mathematics literature, one often uses the notation $e(x) = e^{2 \pi i x}$.

\subsubsection{Unfolding and Poisson summation formulas}
\label{sec:unfolding}
A simple but extremely useful tool, which will be substantially generalized later on in the form of trace formulas, is the \textit{unfolding trick}. If a function $g : \RR \to \CC$ decays sufficiently rapidly at $ \infty$ to make its integral over $\RR$ converge absolutely, then the unfolding trick gives,
\bea
\sum _{n\in \ZZ}  \int _0 ^1 dx \, g(x+n) = \int _{-\infty} ^\infty dx \, g(x)
\eea
Combining the method of images and Fourier decomposition,  we construct a periodic function~$f$ from a non-periodic function function $g$ using (\ref{images}), and then calculate the Fourier coefficients $f_m$ using the unfolding trick, 
\bea
f_m = \int ^1 _0 dx \, e^{- 2 \pi i mx} \, \sum _{n \in \ZZ}  g(x+n) 
= \int ^\infty _{-\infty} dx \, g(x) \, e^{- 2 \pi i mx}
\eea
The Fourier transform of $g$ will be denoted by $\hat g$ and is given by,
\bea
\label{FTg}
\hat g (y) = \int ^\infty _{-\infty} dx \, g(x) \, e^{- 2 \pi i xy}
\eea
so that $f_m= \hat g (m)$. Therefore, the function $f$ may be expressed in two different ways,
\bea
f(x) = \sum _{n \in \ZZ} g(x+n) = \sum _{m \in \ZZ} \hat g (m) \, e^{ 2 \pi i m x}
\eea
Setting $x=0$ (or any integer) gives the Poisson summation formula.

{\thm A function $g$ and its Fourier transform $\hat g$, defined in (\ref{FTg}) obey the Poisson summation formula,
\label{Poisson}
\bea
 \sum _{n \in \ZZ} g(n) = \sum _{m \in \ZZ} \hat g (m) 
\eea
assuming both sums are absolutely convergent.}

An immediate application is to the case where $g$ is a Gaussian. It will be convenient to normalize Gaussians as follows,
\bea
g(x) = e^{- \pi t x^2} \hskip 1in 
\hat g (y) = { 1 \over \sqrt{t}} \, e^{ -  \pi  y^2/t}
\eea
The formulas are valid as long as $\Re(t) > 0$, and may be continued to $\Re(t)=0$. The corresponding Poisson summation formula then reads,
\bea
\label{poisson}
\sum _{n \in \ZZ} e^{- \pi t n^2} = { 1 \over \sqrt{t}} \sum _{m \in \ZZ} e^{ -  \pi  m^2/t}
\eea
We shall see later on that this relation admits an important generalization to Jacobi $\tet$-functions, and corresponds to a special case of a modular transformation.

\subsubsection{Application to analytic continuation}

The Riemann $\zeta$-function, known already to Euler, is defined by the following series,
\bea
\zeta (s) = \sum _{n=1}^\infty { 1 \over n^s}
\eea
which converges absolutely for $\Re(s) >1$ and thus defines a holomorphic functions of $s$ in that region.
Its arithmetic significance derives from the fact that it admits a product representation, due again to Euler,
\bea
\zeta (s) = \prod_{p ~ {\rm prime}} \left ( 1 - { 1 \over p^s} \right )^{-1}
\eea
The Riemann $\zeta$-function is just one example of the family of zeta functions that may be associated with certain classes of self-adjoint operators whose spectra are discrete, free of accumulation points, and bounded from below (we shall take them to be bounded from below by zero without loss of generality). Consider such a self-adjoint operator $H$, and its associated spectrum $\lambda _n$ of discrete real eigenvalues with $n \in \NN$ and $\lambda _1 >0$.\footnote{Throughout $\NN = \{ 1,2,3, \cdots \}$ denotes the set of positive integers.} We may associate a $\zeta$-function to the operator $H$ as follows, 
\bea
\zeta _H (s) = \sum _{n=1}^ \infty { 1 \over \lambda_n ^s} = \Tr \left ( H^{-s} \right  )
\eea
provided $\lambda _n$ grows with $n$ sufficiently fast, such as $\lambda _n \sim n^\alpha$ for $\alpha >0$ at large $n$. If the growth is $\lambda _n \sim n$ for large $n$ then the series is absolutely convergent for $\Re(s) >1$, and defines a holomorphic function in $s$ in that region. The Riemann $\zeta$-function may be associated in this way with the Hamiltonian of either the harmonic oscillator or the free non-relativistic particle in an interval with periodic boundary conditions.

\sm

It is a famous result of Riemann that the function $\zeta(s)$ may be analytically continued to the entire complex plane, has a single pole at $s=1$, and admits a functional reflexion relation. The result may be derived using the Poisson summation relation (\ref{poisson}). Consider the function $\xi (s)$ defined by the integral,
\bea
\xi (s) = \int _0 ^\infty {dt \over t} \, t^{{s \over 2} } \, \left ( \sum _{n \in \ZZ} e^{- \pi t n^2} -1 \right )
\eea
For large $t$ the integral is convergent for any $s$ in the complex plane, but  for small $t$ it is convergent only for $\Re (s) > 1$, as may be seen by using (\ref{poisson}) for $t \to 0$.  For $\Re(s) >1$, the integral defining $\xi$ may be evaluated by integrating term by term, where the contribution of the $n=0$ is cancelled by the subtraction of 1, and we find,
\bea
\label{xi}
\xi (s) = 2 \pi ^{- { s \over 2}} \Gamma \left ( { s \over 2} \right ) \, \zeta (s)
\eea
 To construct a form that may be analytically continued, we partition the integral into the intervals $[0,1]$ and $[1,\infty]$. We leave the integral over $[1, \infty]$ unchanged, but change variables $ t \to 1/t$ in the integral over $[0,1]$, transforming it into an integral over $[1,\infty]$,
\bea
\xi (s) = \int _1 ^\infty { dt \over t} \, t^{-{s \over 2} } \left ( \sum _{n \in \ZZ} e^{- \pi  n^2/t} -1 \right )
+ \int _1 ^\infty { dt \over t} \, t^{{s \over 2} } \left ( \sum _{n \in \ZZ} e^{- \pi t n^2} -1 \right )
\eea
In the first integral, we rewrite the sum with the help of the Poisson relation (\ref{poisson}),
\bea
\xi (s) = \int _1 ^\infty {dt \over t} \, t^{{1-s \over 2} } \left ( \sum _{n \in \ZZ} e^{- \pi t n^2} -{ 1 \over \sqrt{t}}  \right )
+ \int _1 ^\infty { dt \over t}  \, t^{{s\over 2} } \left ( \sum _{n \in \ZZ} e^{- \pi t n^2} -1 \right )
\eea
The pure power integrals may be performed exactly and admit an obvious analytic continuation with poles at $s=0$ and $s=1$, so that the final result is given by,  
\bea
\xi (s) = - { 2 \over s(1-s)} 
+ \int _1 ^\infty {dt \over t} \, \Big ( t^{{s\over 2} } + t^{{1-s\over 2} } \Big ) \left ( \sum _{n \in \ZZ} e^{- \pi t n^2} -1 \right )
\eea
The integral gives a function that is holomorphic throughout $\CC$, and the entire expression manifestly admits the functional relation, 
\bea
\xi (s) = \xi (1-s)
\eea
which in turn gives the functional relation of the Riemann $\zeta$-function. Immediate consequences are that the residue of $\zeta (s)$ at $s=1$ is one, while $\zeta (s)$ is holomorphic elsewhere throughout $\CC$. It is also immediate that $\zeta (-2n)=0$ for all $n \in \NN$.

\subsection{Periodic functions of a complex variable}
\label{sec:2.2a}

Before turning to elliptic functions, we look at how complex analytic periodic functions and trigonometric functions arise naturally from applying the method of images. We start with a meromorphic function $g: \CC \to \CC$ that has a single simple pole $g(z) = 1/z$ and use the method of images to construct a periodic meromorphic function,
\bea
\label{2c1}
f(z) =\sum _{n \in \ZZ} { 1 \over z+n}
\eea
The sum is not absolutely convergent, but it is \textit{conditionally convergent}, which means that its value depends on the order in which we choose to sum the  terms. In quantum field theory, we refer to this process as ``regularizing" the infinite sum.  A natural way to do this is by requiring $f(-z)=-f(z)$ and taking a symmetric limit of a finite sum with this property. Equivalently we may group terms with $n$ together under the summation, 
\bea
\label{2c2}
f(z) = \lim _{N \to \infty} \sum _{n=-N} ^N { 1 \over z+n} 
= {1 \over z} + \sum _{n=1} ^\infty \left ( { 1 \over z+n} + { 1 \over z-n} \right )
\eea
The meromorphic nature of $f$ allows us to evaluate the sum by finding a known function with identical poles with unit residue at every integer, identical asymptotic behavior at $\infty$, and identical symmetry property $f(-z)=-f(z)$. 
The result is the following formula, 
\bea
\label{2c3}
f(z) = \pi \, { \cos \pi z \over \sin \pi z} = - i \pi \, { 1+ e^{ 2 \pi i z} \over 1 - e^{2 \pi i z} }
\eea
Integrating both expressions for $f$  and determining the integration constant by matching the behavior of $z=0$ gives Euler's product formula for the $\sin$-function, 
\bea
\sin \pi z = \pi z \prod _{n=1}^ \infty \left ( 1 - { z^2 \over n^2} \right )
\eea
\sm
Linearity of the method of images allows us to treat any function $g$ that admits a partial fraction decomposition into poles in $z$ in an analogous manner.  

\sm

We can also proceed by deriving a differential equation for $f$ directly from its series representation in (\ref{2c2}). For $|z| <1$, we may expand the second representation in (\ref{2c2}) in a convergent power series in $z$,
\bea
\label{2c4}
f(z) = { 1 \over z} - 2  \sum _{m=1}^\infty z^{2m-1} \zeta (2m)
\eea
Its derivative is given by,
\bea
f'(z) =- { 1 \over z^2} - 2  \sum _{m=1}^\infty (2m-1) z^{2m-2} \zeta (2m)
\eea
The combination $f' + f^2$ is free of poles, bounded in $\CC$, and therefore constant by Liouville's theorem. The constant is computed by evaluating the series near $z=0$, and we find, 
\bea
f' + f^2 = - 6 \, \zeta (2) = - \pi^2
\eea
The differential equation may be solved by quadrature as it is equivalent to,
\bea
\label{2c8} 
\int { df \over f-i \pi } - \int { df \over f+i \pi} = - 2 \pi i \int dz
\eea
Interpreting $z(f) $ as the \textit{inverse function of $f(z)$} we see that $z(f)$ must be multiple-valued as a function of $f$, since the line integrals in $df$ are logarithms and therefore multiple-valued around the poles at $f=\pm i \pi$. Since $z(f)$ is multiple-valued with period~1, its inverse function $f(z)$ must be periodic with period~1. The integral relation of (\ref{2c8}), along with the symmetry condition $f(-z)=-f(z)$ used to fix the integration constant, reproduces (\ref{2c3}). 

\sm

One application is to the derivation of the addition formulas for trigonometric functions. Using only  the fact $f(z)$ of (\ref{2c3}) has a simple pole at $z=0$ and is periodic with period~1, we can derive the addition formulas for trigonometric functions. Indeed, for fixed $w$, the function $f(z+w)$ has simple poles at $z= -w ~ ({\rm mod} \, 1)$ with unit residue, which match those poles in $f'(z)/(f(z)+f(w))$. However, the latter also has simple poles at $z \in \ZZ$ which may be matched by the function $f(z)$. Implementing also the symmetry under swapping $z$ and~$w$, we arrive at the following identity,
\bea
\label{addtrig}
2f(z+w) - { f'(z)+f'(w) \over f(z)+f(w)} -  f(z) - f(w)=0
\eea
from which every addition theorem for elliptic functions may be deduced.

\subsubsection{Application to calculating $\zeta (s)$ at special values}

As another application, we evaluate the Riemann $\zeta$-function at even positive integers by equating the representations for $f(z)$ given in (\ref{2c3}) and (\ref{2c4}), 
\bea
\label{1b1}
- i \pi \, { 1+ e^{ 2 \pi i z} \over 1 - e^{2 \pi i z} } =
{ 1 \over z} - 2  \sum _{m=1}^\infty z^{2m-1} \zeta (2m)
\eea
It follows that  $\zeta (2m)$ is $\pi^{2m}$ times a rational number. To see this, we express the power series in $z$ of the left side  in terms of Bernoulli numbers $B_m \in \QQ$, which are  defined as follows, 
\bea
{ w \over e^w -1} =  \sum _{m=0}^\infty  {B_m \over m!} w^m
\eea 
Clearly, we have $B_0=1$, $B_1=-1/2$, and $B_{2m+1}=0$ for all $m \in \NN$. The next few non-zero values for the Bernoulli numbers are as follows,
\bea
B_2= {1 \over 6} \hskip 0.2in 
B_4= B_8= -{1 \over 30} \hskip 0.2in 
B_6= {1 \over 42} \hskip 0.2in 
B_{10}= {5 \over 66}  \hskip 0.2in 
B_{12}= -{691 \over 20730} 
\eea
An  equivalent formula for Bernoulli numbers of even index is as follows,
\bea
{ w \over 2} \times { e^w + 1  \over e^w -1} =  \sum _{m=0}^\infty  {B_{2m} \over (2m)!} w^{2m}
\eea 
Setting $w=2 \pi i z$ and identifying term-by-term in the series (\ref{1b1}) gives, 
\bea
\zeta (2m) = - (2 \pi i)^{2m} { B_{2m} \over 2 (2m)!}
\eea
which, in turn,  gives the following low order values, 
\bea
\zeta (2) = {\pi^2 \over 6} \hskip 0.2in 
\zeta (4) = {\pi^4 \over 90}  \hskip 0.2in 
\zeta (6) = {\pi^6 \over 945} \hskip 0.2in
\zeta (8) = {\pi^8 \over 9450} \hskip 0.2in
\zeta (10) = {\pi^{10} \over 93555}
\eea
By using the functional relation, we may also compute the $\zeta$-function at other values of interest in physics, such as 
\bea
\zeta (0) = - \half ,
\hskip 0.5in 
\zeta' (0) = - \half \ln (2 \pi)  ,
\hskip 0.5in 
\zeta (-1) = - { 1 \over 12}
\eea 
The last formula is key to the argument that superstrings  live in ten space-time dimensions~!

\subsection{Elliptic functions}

Periodic functions of one real variable may be generalized to functions of several real variables by considering functions that are  periodic in each real variable separately. \textit{Elliptic functions} are periodic functions in two real variables, or equivalently doubly periodic functions in a complex variable $z$, supplemented with the requirement of complex analyticity in $z$.  

\sm

\begin{figure}[htb]
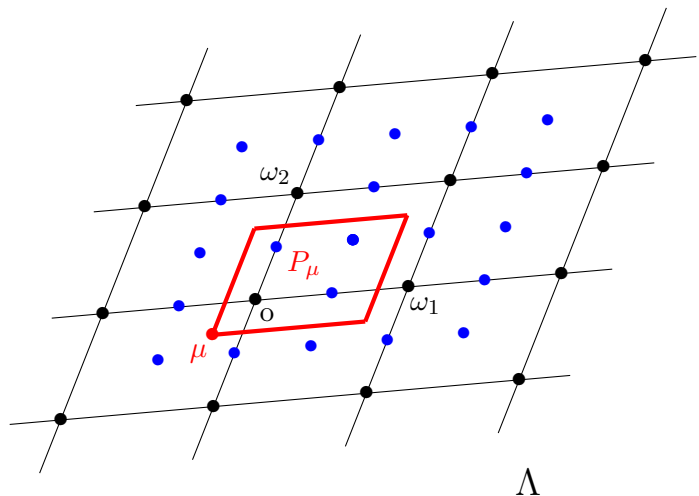

\begin{center}
\tikzpicture[scale=0.68, rotate=5]
\scope[xshift=0cm,yshift=0cm]
\draw (0,0) -- (11,0);
\draw (1,2) -- (12,2);
\draw (2,4) -- (13,4);
\draw (3,6) -- (14,6);
\draw (0.5,-1) -- (4.5,7);
\draw (3.5,-1) -- (7.5,7);
\draw (6.5,-1) -- (10.5,7);
\draw (9.5,-1) -- (13.5,7);

\draw (1,0) node{$\bullet$};
\draw (4,0) node{$\bullet$};
\draw (7,0) node{$\bullet$};
\draw (10,0) node{$\bullet$};

\draw (2,2) node{$\bullet$};
\draw (5,2) node{$\bullet$};
\draw (8,2) node{$\bullet$};
\draw (11,2) node{$\bullet$};

\draw (3,4) node{$\bullet$};
\draw (6,4) node{$\bullet$};
\draw (9,4) node{$\bullet$};
\draw (12,4) node{$\bullet$};

\draw (4,6) node{$\bullet$};
\draw (7,6) node{$\bullet$};
\draw (10,6) node{$\bullet$};
\draw (13,6) node{$\bullet$};

\draw (5.2,1.7) node{\small o};
\draw (8.3,1.6) node{\small $\om_1$};
\draw (5.6,4.4) node{\small $\om_2$};

\draw [blue]  (3,1) node{\small $\bullet$};
\draw [blue]  (4.5,1) node{\small $\bullet$};
\draw [blue]  (6,1) node{\small $\bullet$};
\draw [blue]  (7.5,1) node{\small $\bullet$};
\draw [blue]  (9,1) node{\small $\bullet$};

\draw [blue]  (3.5,2) node{\small $\bullet$};
\draw [blue] (6.5,2) node{\small $\bullet$};
\draw [blue] (9.5,2) node{\small $\bullet$};

\draw [blue]  (4,3) node{\small $\bullet$};
\draw [blue]  (5.5,3) node{\small $\bullet$};
\draw [blue]  (7,3) node{\small $\bullet$};
\draw [blue]  (8.5,3) node{\small $\bullet$};
\draw [blue]  (10,3) node{\small $\bullet$};

\draw [blue]  (4.5,4) node{\small $\bullet$};
\draw [blue] (7.5,4) node{\small $\bullet$};
\draw [blue] (10.5,4) node{\small $\bullet$};

\draw [blue]  (5,5) node{\small $\bullet$};
\draw [blue]  (6.5,5) node{\small $\bullet$};
\draw [blue]  (8,5) node{\small $\bullet$};
\draw [blue]  (9.5,5) node{\small $\bullet$};
\draw [blue]  (11,5) node{\small $\bullet$};

\draw [blue]  (7,3) node{\small $\bullet$};
\draw [blue]  (7,3) node{\small $\bullet$};
\draw [blue]  (7,3) node{\small $\bullet$};

\draw [ultra thick, red] (4.1, 1.4) -- (7.1, 1.4);
\draw [ultra thick, red] (5.1, 3.4) -- (8.1, 3.4);
\draw [ultra thick, red] (4.1, 1.4) -- (5.1, 3.4);
\draw [ultra thick, red] (7.1, 1.4) -- (8.1, 3.4);
\draw [red] (3.8,1.05) node{\small $\mu$};
\draw [red] (4.1, 1.4) node{$\bullet$};
\draw [red] (6,2.6) node{\small $P_\mu$};
\draw (10,-2) node{\large $\Lambda$};

\endscope
\endtikzpicture
\end{center}
\caption{The points of the lattice $\Lambda$ are represented in the complex plane by black dots. The lattice $\Lambda$  is generated by  the periods $\omega_1, \omega _2$ so that $\Lambda = \ZZ \om_1 \oplus \ZZ \om_2$.  The boundary $\p P_\mu$ of the fundamental parallelogram $P_\mu$ is drawn in red, where $\mu$ has been chosen so that $\p P_\mu $ contains neither periods, nor half periods  represented by blue dots. \label{2.fig:1}}
\end{figure}

Denoting the two periods in the complex variable $z$ by $\om_1, \om_2 \in \CC$,\footnote{In Bateman~\cite{Bateman2}, the periods are denoted by $2 \om$ and $2 \om '$ respectively, and are related to the periods defined here by $\om_1=2 \om$ and $\om_2 = 2 \om '$. Our notation for the periods $\om_1$ and $\om_2$   is not to be confused with the half-periods $\om_\a$ used in equation (18) in section 13.12 of \cite{Bateman2}.}  we require that the periods be linearly independent as vectors in $\RR^2$ so that $\om_2/\om_1 \not \in \RR$. The periods then generate a two-dimensional lattice $\Lambda$ defined by (see Figure \ref{2.fig:1}),
\bea
\Lambda = \ZZ \om _1 + \ZZ \om _2
\eea
The lattice $\Lambda$ is an Abelian group under addition. The quotient $\CC/\Lambda$ of the complex plane by the lattice $\Lambda$ is torus of dimension two over $\RR$, and also an Abelian group. The torus $\CC/\Lambda$ may  be represented in $\CC$ by a parallelogram $P_\mu$ anchored at a point $\mu \in \CC$ on which opposite sides are pairwise  identified with one another  (see Figure \ref{2.fig:1}), 
\bea 
P_\mu = \{ \mu + t_1 \om _1 + t_2 \om _2, ~ 0 \leq t_1, t_2 \leq 1 \}
\eea
A function of $z$ which is doubly periodic with periods $\om_1, \om_2$ is periodic under the lattice $\Lambda$, or equivalently is a well-defined function on the torus $\CC/\Lambda$. Collecting these ingredients, we obtain the following definition. 

{\deff
A meromorphic function $f: \CC \to \CC$ is elliptic with periods $\omega _1, \omega _2 \in \CC$ with $\om_2/\om_1 \not \in \RR$ if any one of the following equivalent conditions holds, 
\begin{itemize}
\itemsep=0.in
\item $f$ is periodic with periods $\omega _1, \omega_2$, namely $f(z+\om_1) = f(z+\om_2)=f(z)$ for all $z \in \CC$;
\item $f$ is invariant under translations by the lattice $\Lambda$, namely $f(z+\omega)= f(z)$ for all $\om \in \Lambda$ and for all $z \in \CC$;
\item $f$ is single-valued on the torus $\CC/\Lambda$.
\end{itemize}}
 It is often as single-valued functions on a torus that elliptic functions appear in physics. 

\sm

Sums,  products, and inverses of elliptic functions for a given lattice $\Lambda$ are elliptic functions for the same lattice $\Lambda$, and the constant function with value 1 is the identity under multiplication, so that elliptic functions for a given $\Lambda$ form a \textit{field}. 

\sm

As we shall establish shortly, elliptic functions are essentially determined by their zeros and poles, subject to certain conditions. For example, Liouville's theorem (which states that a holomorphic function on $\CC$ which is bounded on $\CC$ must be constant), implies that an elliptic function without poles must be a constant function. To obtain the relations between zeros and poles of elliptic functions more generally, we proceed by defining the residue function 
${\rm res}_f (w)$ to be the \textit{residue of $f$} at the point $w$, and  the order function ${\rm ord}_f (w)$ to be the \textit{order} of $f$ at the point $w$, namely the integer such that $f(z) (z-w)^{-{\rm ord} _f(w)} $ is a finite non-zero constant as $ z \to w$.

{\thm \label{thm:1}
The residues and orders of an arbitrary elliptic function $f$ with lattice $\Lambda$ and corresponding fundamental parallelogram $P_\mu$  satisfy the following relations, 
\bea
\label{zeropoles}
\sum _{w \in P_\mu} {\rm res} _f (w) & = & 0
\no \\
\sum _ { w \in P_\mu} {\rm ord}  _f (w) & = & 0
\no \\
  \sum _ { w \in P_\mu} w \cdot {\rm ord}  _f (w)  & = & 0 ~ ({\rm mod} \, \Lambda)
\eea}
These relations are proven  by integrating the functions $f(z)$, $f'(z)/f(z)$, and $z f'(z)/f(z)$ over the boundary $\p P_\mu$, respectively. When considering a function $f$ on a torus $\CC/\Lambda$, it will often be convenient to  choose $\mu$ such that no zeros or poles of $f$ lie on $\p P_\mu$.

\sm

In two-dimensional electro-statics, the electric potential $\Phi$ may be viewed as a function of $z$ and $\bar z$, 
and is a harmonic function away from the support of the electric charges. The electric field $\bE = - \nabla \Phi$ may be decomposed into derivatives with respect to $z$ and $\bar z$, so that $E_z = - \p_z \Phi$ is a holomorphic function away from the charges. Considering a  distribution of point-like charges (without accumulation points) on a torus $\CC/\Lambda$, the electric field is an elliptic function for the lattice $\Lambda$. 
The first condition in (\ref{zeropoles})  requires the sum of all the charges to vanish, and must hold on any compact space. The second and third conditions are specific to the torus. The conditions (\ref{zeropoles}) imply that there exists no elliptic function with just one simple pole in $P_\mu$, corresponding to the fact that the torus does not support a single point charge, but there do exist elliptic functions with two simple poles at arbitrary positions and opposite residues corresponding in electro-statics with two opposite point charges. There also exist elliptic functions with a single double pole (which may be considered as a limit of the case with two simple poles) corresponding to the electric field of an electric dipole.

\subsubsection{Construction by the method of images}

Just as periodic functions may be constructed using the method of images in (\ref{images}), an elliptic function may  be  constructed by using the method of images for the lattice $\Lambda$ with periods $\om_1, \om_2$ and $\om_2/\om_1 \not \in \RR$.  If a meromorphic  function $g:\CC \to \CC$ decays sufficiently rapidly at $\infty$ then an elliptic function $f$ may be constructed by the method of images,
\bea
\label{ell.imag}
f(z) = \sum _{\om \in \Lambda} g(z+\om) = \sum_{m,n \in \Lambda} g(z+m\om_1+n\om_2)
\eea
Any function $g(z)$ that is bounded by $|g(z)| \leq |z|^{-2-\ep}$ for $\ep>0$ satisfies this criterion. The most famous example  is given by the Weierstrass function, in terms of which all other elliptic functions may be expressed, as will be explained in the subsequent subsection.

\subsection{The Weierstrass elliptic function}

All elliptic functions for a lattice $\Lambda$ may be built up from the Weierstrass elliptic function, which has one double pole (mod $\Lambda$) and no other poles. The Weierstrass elliptic function $ \wp (z | \Lambda)$ for the lattice $\Lambda$ may be constructed with the method of images. While it may be natural to consider the sum over $\om \in \Lambda$ of translates $(z+\om)^{-2}$ of the double pole $z^{-2}$, actually  this series converges only conditionally. We shall learn how to handle such series later on. Here we shall define $\wp$ by the manifestly convergent series over the lattice $\Lambda ' = \Lambda \setminus \{ 0 \}$, 
\bea
\label{wpdef}
\wp(z| \Lambda ) = { 1 \over z^2} + \sum _{ \om \in \Lambda '} \left ( { 1 \over (z+\om)^2} - { 1 \over \om^2} \right )
\eea
The function $\wp $ is even in $z$ since the lattice summation is invariant under $\Lambda' \to - \Lambda'$.

\subsubsection{The field of elliptic functions in terms of $\wp$ and $\wp'$}

We start by describing the field of elliptic functions for a fixed  lattice $\Lambda$ in terms of $\wp(z|\Lambda)$ for  even functions in $z$. The elliptic function $\wp (z|\Lambda)- \wp (w|\Lambda)$, viewed as a  function of $z$ for fixed $w \not \in \Lambda$,  has a double pole at $z=0$ and no other poles, and therefore must have two zeros in view of the second relation in (\ref{zeropoles}). For generic $w$, the two zeros  $z=\pm w$ are distinct and must be the only two zeros. However, for the non-generic case where $ w=-w ~ ({\rm mod} \, \Lambda)$, the zero is double. There are exactly four points in $P_\mu$ with $w \equiv -w$ $ ({\rm mod} \, \Lambda)$,  namely the half periods $0$, $\om _1/2$, $\om _2/2$, and $\om _3/2 ~ ({\rm mod} \, \Lambda)$, where $\om_3=\om_1+\om_2$. At the three non-zero half periods $w$, the function  $\wp (z|\Lambda)- \wp (w|\Lambda)$ has a double zero in $z$.

\sm

The first key result is that every \textit{even} elliptic function $f$ for the lattice $\Lambda$ is a rational function of $\wp (z|\Lambda)$, given by,
\bea
\label{order}
f(z) = \prod _{w \in P_\mu} \Big ( \wp (z|\Lambda) - \wp (w|\Lambda) \Big ) ^{{\rm ord}_f(w)}
\eea
which may be established by matching poles and zeros. To represent elliptic functions which are odd under $z \to -z$, or to describe functions without definite parity, we  differentiate $\wp$, 
\bea
\wp '(z|\Lambda) = -2 \sum _{\om \in \Lambda} { 1 \over (z+\om)^3}
\eea
This series is absolutely convergent for $z \not \in \Lambda$ and we have $\wp '(-z|\Lambda) = - \wp '(z|\Lambda)$. The function $\wp '(z|\Lambda)^2$ is even and has a single pole of order 6 at $z=0$, and is therefore a polynomial of degree~3 in $\wp (z|\Lambda)$. To determine this polynomial, we proceed by analogy with the trigonometric case of subsection \ref{sec:2.2a}. We expand both $\wp $ and $\wp '$ in powers of $z$ near $z=0$ in convergent series, 
\bea
\label{2.wpex}
\wp (z|\Lambda) & = & { 1 \over z^2 } + \sum _{k=1}^\infty (k+1) G_{k+2}(\Lambda) \, z^k
\no \\
\wp ' (z|\Lambda) & = & - { 2 \over z^3 } + \sum _{k=1}^\infty k (k+1) G_{k+2} (\Lambda) \, z^{k-1}
\eea
The coefficients $G_k$ are given by the convergent series,
\bea
\label{EisenG}
G_k (\Lambda) = \sum _{\om \in \Lambda '} { 1 \over \om ^k} \hskip 1in k \geq 3
\eea
The reflection symmetry $- \Lambda = \Lambda$ implies,
\bea
G_{2k+1}(\Lambda) =0 \hskip 1in k \in \NN
\eea
so that the expansions to low order are as follows, 
\bea
\wp (z|\Lambda) & = & { 1 \over z^2} + 3 \, G_4(\Lambda)  z^2 + 5 \, G_6(\Lambda) z^4 + \cO (z^6)
\no \\
\wp '(z|\Lambda) & = & - { 2 \over z^2} + 6 \, G_4(\Lambda)  z + 20 \, G_6(\Lambda)  z^3 + \cO (z^5)
\eea
The pole in $z$ of order six, which is present in both $\wp '(z|\Lambda)^2$ and $\wp(z|\Lambda) ^3$, is cancelled in the combination $\wp '(z|\Lambda)^2-4\wp(z|\Lambda) ^3$, whose remaining pole is of order two. By matching all terms of positive or zero power in $z$ and then using Liouville's theorem, we find the relation,
\bea 
\label{wp}
\wp '(z|\Lambda)^2 = 4 \wp(z|\Lambda)^3 - g_2(\Lambda)  \wp (z|\Lambda) - g_3(\Lambda) 
\eea 
where it is conventional to introduce the notation, 
\bea
\label{g2g3}
g_2(\Lambda)  = 60 \, G_4(\Lambda) 
\hskip 1in 
g_6(\Lambda)  = 140 \, G_6(\Lambda) 
\eea
We may assemble these results in the following theorem.

{\thm
\label{thm:2}
Every elliptic function $f(z)$ for a lattice $\Lambda$ may be expressed as a rational function of the Weierstrass functions $\wp (z|\Lambda )$ and $\wp'(z|\Lambda)$ for lattice $\Lambda$, which equivalently may be reduced to a linear function in $\wp'(z)$ using the relation (\ref{wp}).}
\newline

\noindent 
To prove the theorem, we decompose $f(z)$ into its even and odd parts, $f(z)=f_+(z)+f_-(z)$ with $f_\pm (-z) = \pm f_\pm(z)$. Since $f_+(z)$ and $f_-(z)/\wp'(z|\Lambda)$ are both even elliptic functions, we use the result derived in (\ref{order})  that every even elliptic function may be expressed as a rational function of $\wp (z|\Lambda)$.

\subsubsection{The discriminant and the roots of the cubic}

The roots of the cubic polynomial in (\ref{wp}) must all produce double zeros as a function of~$z$, since the left side of (\ref{wp}) is a perfect square. But we had seen earlier that double zeros of $\wp (z|\Lambda) - \wp (w|\Lambda)$  occur if and only if $w \not \in \Lambda$ and $w \equiv -w \, ({\rm mod} \, \Lambda)$, namely at the three non-zero half-periods. Denoting the values of $\wp$ at the half periods by $e_\a$ with $\a=1,2,3$,
\bea
e_\a (\Lambda) = \wp \left ( \thalf \om _\a | \Lambda \right ) 
\eea
we obtain the following factorized form of the cubic polynomial,\footnote{Henceforth, we shall suppress the $\Lambda$-dependence of $g_2, g_3$, and $e_\a$.}
\bea
\wp '(z|\Lambda)^2 = 4 \big (\wp (z|\Lambda) - e_1 \big ) \big (\wp (z|\Lambda) - e_2 \big ) \big (\wp (z|\Lambda) - e_3 \big) 
\eea
where the three symmetric functions of $e_i$ may be identified with the coefficients $g_2$ and $g_3$ of the cubic (\ref{wp}) as follows,
\bea
\label{e}
e_1+e_2+e_3=0 
\hskip 0.7in 
e_1e_2+e_2e_3+e_3e_1= - { g_2 \over 4} 
\hskip 0.7in 
e_1e_2e_3={ g_3 \over 4}
\eea
The discriminant $\Delta$ of the cubic $4x^3-g_2x-g_3$ is defined as follows, 
\bea
\Delta = 16 (e_1-e_2)^2 (e_2-e_3)^2 (e_3-e_1)^2
\eea
In terms of the coefficients $g_2$ and $g_3$, it is given by, 
\bea
\label{Disc}
\Delta = g_2^3 - 27 g_3^2
\eea
The discriminant is non-zero if and only if all roots are distinct. When two of the roots or all three roots coincide, the discriminant vanishes and  the cubic is singular. To obtain $e_i$ in terms of $g_2$ and $g_3$, it suffices to find the roots of the cubic polynomial $4x^3-g_2x-g_3$, which are given in Lagrange's presentation of the roots by,
\bea
\label{eroots}
e_\a  =  { 1 \over 2 \sqrt{3}} \Big ( \rho^\a \delta + \rho^{2\a} g_2 \, \delta ^{-1} \Big )
\hskip 0.6in
\delta^3  =  \sqrt{27} \, g_3 + \sqrt{-\Delta} 
\eea
for $\a =1,2,3$ and $\rho= e^{2 \pi i /3}$ a cube root of unity.

\subsubsection{The Weierstrass $\zeta$-function}
\label{wpZ}

Since $\wp(z|\Lambda)$, viewed as a doubly periodic function on $\CC$, has only double poles, it is the derivative in $z$ of a single-valued meromorphic function $\zeta (z|\Lambda)$  on $\CC$, referred to as the Weierstrass $\zeta$-functions (not to be confused with the Riemann $\zeta$-function). More precisely, $\zeta (z|\Lambda)$ is defined uniquely by,
\bea
\wp(z|\Lambda) = -\zeta' (z|\Lambda) \hskip 1in \zeta (-z|\Lambda) = - \zeta (z|\Lambda)
\eea
and has simple poles in $z$ with unit residue at every point in $\Lambda$. As a result, it has a single pole in any fundamental parallelogram $P_\mu$ and therefore cannot be double periodic in the lattice $\Lambda$. Since its derivative is doubly periodic the monodromies of $\zeta(z|\Lambda)$ are constant,
\begin{align}
\label{zmon}
\zeta (z+\om_1 |\Lambda) & =  \zeta (z|\Lambda) +  \eta _1(\Lambda) & \eta _1(\Lambda) & = 2 \, \zeta(\thalf \om_1|\Lambda)
\no \\
\zeta (z+\om_2 |\Lambda) & =  \zeta (z|\Lambda) +  \eta _2(\Lambda)  & \eta _2(\Lambda)  & = 2 \, \zeta(\thalf \om_2|\Lambda)
\end{align}
The expressions for $\eta_1,\eta_2$ have been obtained by setting $z$ equal to $-\thalf \om_1, -\thalf \om_2$ respectively and using the odd parity of $\zeta(z|\Lambda)$. By integrating $z \wp(z|\Lambda)$ along $\p P_\mu$, we obtain the relation,
\bea
\label{omzet}
\eta_1 \om_2- \eta _2 \om_1 =  2 \pi i
\eea
Since the monodromies of $\zeta(z|\Lambda)$ are independent of $z$, the difference $\zeta(z-a|\Lambda) - \zeta (z-b|\Lambda)$ is an elliptic function for the lattice $\Lambda$ with simple poles at $a$ and $b$ with residues $+1$ and $-1$ respectively. One also defines the Weierstrass $\sigma$-function by $\zeta(z|\Lambda) =  \p_z \ln \sigma (z|\Lambda)$ normalized so that $\sigma (z|\Lambda) = z + \cO(z^2)$.

\subsubsection{Rescaling the periods}

\begin{figure}[tp]
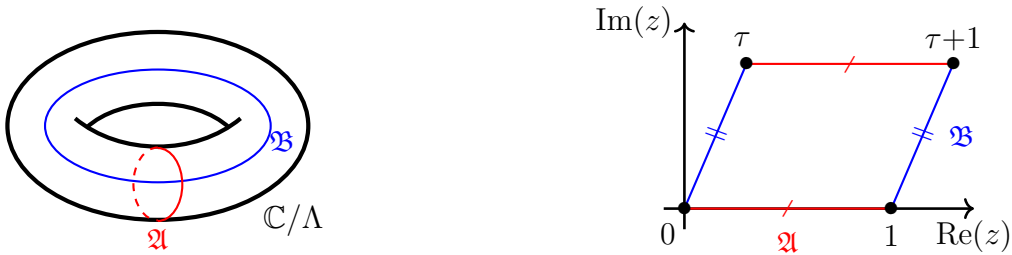

\begin{center}
\tikzpicture[scale=0.5, line width=0.35mm]
\draw[ultra thick] (0,0) ellipse  (4cm and 2.5cm);
\draw[ultra thick] (-2.2,0.2) .. controls (-1,-0.8) and (1,-0.8) .. (2.2,0.2);
\draw[ultra thick] (-1.9,-0.05) .. controls (-1,0.8) and (1,0.8) .. (1.9,-0.05);
\draw[thick, blue](0,0) ellipse  (3cm and 1.5cm);
\draw[thick, red] (0,-2.53) arc (-90:90:0.65cm and 0.97cm);
\draw[thick, red,dashed] (0,-0.575) arc (90:270:0.65cm and 0.97cm);
\draw[blue] (3.3,-0.4) node{\small $\mB$};
\draw[red] (0,-3) node{\small $\mA$};
\draw (3.6,-2.5) node{$\CC/\Lambda$};
\scope[xshift=14cm,yshift=-2.2cm,scale=1.1]
\draw[->](-0.5,0) -- (7,0) node[below]{${\rm Re}(z)$};
\draw[->](0,-0.5) -- (0,4.5) node[left]{${\rm Im}(z)$};
\draw(-0.4,-0.6)node{$0$};
\draw[thick, blue](0,0) -- (1.5,3.5);
\draw[blue] (0.75,1.75)node[rotate=-20]{$=$};
\draw(1.4,4.1)node{$\tau$};
\draw[thick, red](0,0) -- (5,0);
\draw[red] (2.5,0)node[rotate=60]{$-$};
\draw (5,-0.65)node{$1$};
\draw[thick, red](1.5,3.5) -- (6.5,3.5);
\draw[red] (4,3.5)node[rotate=60]{$-$};
\draw[thick, blue](5,0) -- (6.5,3.5);
\draw[blue] (5.75,1.75)node[rotate=-20]{$=$};
\draw(6.5,4.1)node{$\tau{+}1$};
\draw[red](2.5,-0.8)node{\small $\mA$};
\draw[blue](6.7,1.75)node{\small $\mB$};
\draw (5,0)node{$\bullet$};
\draw(0,0)node{$\bullet$};
\draw (1.5,3.5)node{$\bullet$} ;
\draw(6.5,3.5)node{$\bullet$};
\endscope
\endtikzpicture
\caption{\textit{The torus $\CC/\Lambda$  is represented in the plane by a parallelogram with complex coordinates $z,\bar z$ and opposite sides pairwise identified and is shown with a choice of canonical homology cycles $\mA$ and $\mB$ corresponding to the cycles $[0,1]$ and $[0,\tau]$ in $\CC$.}
\label{2.fig:2}}
\end{center}
\end{figure}

From the definition in (\ref{wpdef}), it is manifest that the Weierstrass functions $\zeta$, $\wp$, and $\wp'$ are homogeneous of degree $-1$, $-2$, and $-3$ respectively under simultaneous rescaling of $z$, the periods $\om_1, \om_2$, and the lattice $\Lambda$, 
\bea
\zeta(\a z| \a \Lambda) & = & \a^{-1} \zeta (z|\Lambda) 
\no \\
\wp(\a z|\a \Lambda) & = & \a^{-2} \, \wp(z|\Lambda)
\no \\
\wp'(\a z|\a \Lambda) & = & \a^{-3} \, \wp'(z|\Lambda)
\eea
As a result, the coefficients $g_2, g_3$, and $G_m$ are homogeneous of degree $4,6$, and $-m$ respectively, while the roots $e_i$ have degree $-2$ and the discriminant $\Delta$ has degree $-12$. These rescaling properties show that the overall scale of the periods is an ``easy" degree of freedom which in many, but not all, applications in mathematics and physics is immaterial. The freedom to rescale allows us to choose a canonical value for one of the periods which is usually taken to be $\om_1=1$ and $\om_2/\om_1=\tau$. In this normalization the lattice
is given by,
\bea
\Lambda = \ZZ \oplus \tau \ZZ
\eea
and the torus $\CC/\Lambda$ may be represented in $\CC$ by a parallelogram with vertices $0,1,\tau$, and $\tau+1$ and opposite sides identified pairwise, as illustrated in Figure \ref{2.fig:2}. For this normalization of the lattice the Weierstrass function will be denoted as follows by,
\bea
\wp (z) = \wp(z|\tau) = \wp(z|\Lambda)
\eea
The left most form $\wp(z)$ is reserved for when the lattice $\Lambda$ and the value of $\tau$ are fixed and clear from the context. Out of the coefficients $g_2$ and $g_3$, one may construct one combination that is invariant under scaling and, by construction, depends only on the modulus $\tau$, 
\bea
j =  1728 \, g_2^3 / \Delta
\eea
The normalization factor is chosen in such a way that $j(\tau) \approx e^{-2 \pi i \tau}$ as $\tau \to i \infty$ with unit coefficients. This so-called $j$-function will play a fundamental role in the theory of modular forms, to be studied in subsequent sections.

\subsection{Jacobi elliptic functions}

For a given lattice $\Lambda$, the elliptic function $\wp(z|\Lambda )-e_i$ associated with a root $e_i$ has a unique double pole at $z=0$ and a unique double zero at $z=\om_i/2$ (mod $\Lambda$). The latter follows from the fact that its derivative $\wp'(z|\Lambda )$ vanishes at the half periods. Hence $\sqrt{\wp(z|\Lambda)-e_i}$ is a holomorphic function which is doubly periodic, up to sign factors, in $z$ with periods $\om_1$ and $\om_2$. The Jacobi elliptic functions $\sn(u|k), \cn(u|k)$, and $\dn(u|k)$ may be defined in terms these square roots as follows,
\bea
\label{sndef}
\sn(u|k) & = & \sqrt{{ e_1-e_3 \over \wp(z|\Lambda )-e_3}}
\hskip 0.7in
\dn(u|k) = \sqrt{{ \wp(z|\Lambda )-e_2 \over \wp(z |\Lambda)-e_3}}
\no \\
\cn(u|k) & = & \sqrt{{ \wp(z|\Lambda ) - e_1 \over \wp(z| \Lambda )-e_3}}
\hskip 0.7in
k^2 = { e_2 - e_3 \over e_1 - e_3}
\eea
The variables $u$ and $z$, and the periods are related as follows,
\bea
u = \sqrt{e_1-e_3} \, z \hskip 1in K_\a=  \om_\a \, \sqrt{e_1-e_3} 
\eea
When the value of $k$ is clear from context, one uses the abbreviated notations $\sn(u)=\sn(u|k)$, $\cn(u)=\cn(u|k)$, and $\dn(u)=\dn(u|k)$.
While the functions $\sn(u)^2$, $\cn(u)^2$, $ \dn(u)^2$, and $\sn(u) \cn(u) \dn (u)$ are doubly periodic under translations of $u$ by  the lattice $\Lambda \, \sqrt{e_1-e_3} $ generated by the periods $K_1, K_2$, the functions $\sn(u)$, $ \cn(u)$, and $ \dn(u)$ are doubly periodic  only up to signs in view of the square roots in their definition. Under $u \to -u$ the function $\sn$ is odd while $\cn, \dn$ are even. Their zeros and poles are simple and may be read off from the definition.
The modulus $k^2$ plays a role for Jacobi elliptic functions analogous to the role played by the  $j$-function for Weierstrass functions, and the two may be related to one another  by expressing the roots $e_\a$ in the definition of $k^2$ in (\ref{sndef}) using (\ref{eroots}), giving,
\bea
j = 256 { (k^4-k^2+1)^3 \over k^4 (k^2-1)^2}
\eea 
 The following identities follow from the definitions of the Jacobi elliptic functions, 
\begin{align}
\label{scdn}
\sn(u)^2 + \cn(u)^2 & =  1 & \sn' (u) & =  - \cn(u) \, \dn(u)
\no \\
k^2 \sn(u)^2 + \dn(u)^2 & =  1 & \cn' (u) & =  - \sn(u) \, \dn(u)
\no \\
\dn(u)^2 - k^2 \cn(u)^2 & =  1-k^2 & \dn' (u) & =  - k^2 \sn(u) \, \cn(u)
\end{align}
where the prime denotes the derivative with respect to $u$.
The differential relations were obtained with the help of the following relation,
\bea
\label{wpscdn}
\wp'(z) = 2 (e_1-e_3)^{{3 \over 2}} { \cn (u) \, \dn (u) \over \sn (u)^3}
\eea
where the prime on the left side denotes the derivative with respect to $z$, and on the right side with respect to $u$. Upon taking the square of each differential relation and using the algebraic equations to express both sides in terms of the same function, we obtain the following differential equations, 
\bea
\label{snder}
\sn'(u)^2 & = & \left ( 1-\sn(u)^2 \right ) \left ( 1-k^2 \sn(u)^2 \right ) 
\no \\
\cn'(u)^2 & = & \left ( 1-\cn(u)^2 \right ) \left ( 1-k^2 + k^2 \cn(u)^2  \right ) 
\no \\
\dn'(u)^2 & = & \left ( 1-\dn(u)^2 \right ) \left ( k^2-1 + \dn(u)^2 \right )
\eea
 An immediate corollary for Jacobi elliptic functions following from Theorem~\ref{thm:2} is as follows.
{\cor
An arbitrary elliptic function $f(u)$ with a lattice of periods $ \Lambda \, \sqrt{e_1-e_3} $ may be expressed as a rational function of either $\sn(u)^2$, $\cn(u)^2$, or $ \dn(u)^2$ which is linear in the product $\sn(u) \, \cn(u) \, \dn (u)$. }

\subsection{Jacobi  $\vartheta$-functions}

In the Weierstrass approach to elliptic functions, the basic building block is the meromorphic $\wp$-function and its derivative $\wp'$, in terms of which every elliptic function is a rational function. Instead, the Jacobi $\tet$-function approach produces elliptic functions in terms of Jacobi $\tet$-functions, which are holomorphic, at the cost of being multiple-valued on $\CC/\Lambda$. Scaling the lattice $\Lambda$ so that $\om_1=1$ and $\om_2 = \tau$ with $\Im (\tau)>0$, the Jacobi $\tet$-function is defined by,
\bea
\tet (z |\tau) = \sum _{n \in \ZZ} e^{ i \pi \tau n^2 + 2 \pi i n z}
\eea
and is often denoted simply by $\tet (z)$ when the $\tau$-dependence is clear. The series is absolutely convergent for $\Im (\tau)>0$ and defines a holomorphic function in $z \in \CC$. The $\tet$-function satisfies a complexified version of the heat equation, 
\bea
\label{2.heat}
\p_z^2 \tet (z|\tau) = 4 \pi i \p_\tau \tet (z|\tau)
\eea
The function $\tet (z|\tau)$  is even in $z\to -z$ and transforms under shifts in the lattice $\Lambda$ by, 
\bea
\label{tetshift}
\tet (z+1 |\tau) & = & \tet (z |\tau)
\no \\
\tet (z+\tau |\tau) & = & \tet (z |\tau) \, e^{- i \pi \tau - 2 \pi i z}
\eea
Thus, $\tet(z|\tau)$ is not an elliptic function in $z$. Indeed, it could never be, as a holomorphic doubly periodic  function must be constant. We shall later see how $\tet$ can be naturally viewed as a holomorphic section of a line bundle on $\CC/\Lambda$, but for the time being we shall just work with it as a multiple-valued function.

\subsubsection{The number of zeros of $\tet$}

To find the number of zeros of $\tet(z|\tau)$ as a function of $z$, we integrate its logarithmic derivative along the closed boundary of the fundamental parallelogram $P_0$ (with vertices $0, 1, \tau, 1+\tau$), and decompose the integration as a sum of the line integrals along the four edges, 
\bea
\oint _{\p P_0} dz \,  \p_z \ln \tet (z|\tau) = \left ( \int _0 ^1 + \int _1 ^{1+\tau} + \int _{1+\tau} ^\tau + \int _\tau ^0 \right ) dz \,  \p_z \ln \tet (z|\tau)
\eea
By periodicity of $\tet (z|\tau)$ under $z\to z+1$, the contributions from the second and fourth integrals on the right side cancel
one another. The contribution of the third integral is just a translate and opposite of the first integral, so that we have,
\bea
\oint _{\p P_0}  dz \,  \p_z \ln \tet (z|\tau) = \int _0 ^1 dz \Big ( \p_z \ln \tet (z |\tau) - \p_z \ln \tet (z+\tau|\tau) \Big )
\eea
Using the second relation in (\ref{tetshift}), this integral is readily evaluated, and we find $ 2 \pi i$ which implies that $\tet (z |\tau)$ has exactly one zero in $P_0$ and, since it is holomorphic, no poles. To determine the position of this zero, it will be useful to introduce characteristicss.

\subsubsection{$\tet$-functions with characteristics}

A convenient variant of the $\tet$-function introduced above is the \textit{$\tet$-function with characteristics $\alpha, \beta \in \CC$}, defined by,
\bea
\tet \left [ \begin{smallmatrix} \alpha \cr \beta \cr \end{smallmatrix} \right ] (z |\tau)
= 
 \sum _{n \in \ZZ} e^{ i \pi \tau (n+\alpha)^2 + 2 \pi i (n +\alpha ) (z+\beta) }
\eea
It is  related to the original $\tet$-function by translation of $z$ and an exponential factor,
\bea
\label{2.tetchar}
\tet \left [ \begin{smallmatrix} \alpha \cr \beta \cr \end{smallmatrix} \right ] (z |\tau)
=
\tet (z+ \alpha \tau + \beta |\tau) \, e^{i \pi \tau \alpha ^2 + 2 \pi i \alpha (z+\beta)}
\eea
The original $\tet (z|\tau)$-function corresponds to $\alpha = \beta =0 $.
Under translations of $z$ by the lattice~$\Lambda$, and under a reflection $z\to -z$, we have, 
\bea
\label{2.theta-shift}
\tet \left [ \begin{smallmatrix}  \alpha   \cr  \beta  \cr \end{smallmatrix} \right ] ( z + 1  |\tau)
& = &
\tet \left [ \begin{smallmatrix}  \alpha   \cr  \beta  \cr \end{smallmatrix} \right ] ( z |\tau) \, e^{2 \pi i \alpha} 
\no \\
\tet \left [  \begin{smallmatrix}  \alpha   \cr  \beta  \cr \end{smallmatrix} \right ] ( z+\tau  |\tau) 
&=& 
\tet \left [  \begin{smallmatrix}  \alpha   \cr  \beta  \cr \end{smallmatrix} \right ] ( z |\tau) \, e^{- 2 \pi i \beta - i \pi \tau - 2 \pi i z } 
\no \\
\tet \left [ \begin{smallmatrix}  -\alpha   \cr  -\beta  \cr \end{smallmatrix} \right ] (- z |\tau)
& = &  \tet \left [ \begin{smallmatrix}  \alpha \cr  \beta \cr \end{smallmatrix} \right ] (z |\tau)
\eea
Under integer shift in the characteristics, the transformations are as follows,
\bea
\label{2.theta-char}
\tet \left [ \begin{smallmatrix} \alpha +1  \cr \beta \cr \end{smallmatrix} \right ] (z |\tau)
& = & \tet \left [ \begin{smallmatrix} \alpha \cr \beta \cr \end{smallmatrix} \right ] (z |\tau)
\no \\
\tet \left [ \begin{smallmatrix} \alpha   \cr \beta +1 \cr \end{smallmatrix} \right ] (z |\tau)
& = & \tet \left [ \begin{smallmatrix} \alpha \cr \beta \cr \end{smallmatrix} \right ] (z |\tau)\, e^{ 2 \pi i \alpha}
\eea
For a $\tet$-function with characteristics to have definite parity under $z \to -z$, we must have $-\alpha \equiv \alpha \, ({\rm mod} \, 1) $ and $-\beta \equiv \beta \, ({\rm mod} \, 1)$, which requires $\alpha, \beta \in \{ 0, \half \}$ and corresponds to the four half-periods. It is traditional to give these four $\tet$-functions special names,
\bea
\label{tet1234}
\tet _1 (z|\tau) = - \, \tet \left [ \begin{smallmatrix}  \half   \cr  \half  \cr \end{smallmatrix} \right ] (z |\tau)
& \hskip 1in & 
\tet _2 (z|\tau) = \tet \left [ \begin{smallmatrix}  \half   \cr  0  \cr \end{smallmatrix} \right ] (z |\tau)
\no \\
\tet _3 (z|\tau) = \tet \left [ \begin{smallmatrix}  0   \cr  0  \cr \end{smallmatrix} \right ] (z |\tau)
\hskip 0.25in
& \hskip 1in & 
\tet _4 (z|\tau) = \tet \left [ \begin{smallmatrix}  0   \cr  \half \cr \end{smallmatrix} \right ] (z |\tau)
\eea
The minus sign for $\tet_1$ is introduced for later convenience. 
From the above discussion, it is  manifest that $\tet_1$ is odd under $z\to -z$ while $\tet_2, \tet_3, \tet_4$ are even. One designates the corresponding \textit{half-integer characteristics} as \textit{odd and even characteristics} respectively.

\sm

The parity of  $\tet_1$ means  that we know the location of its zero, namely at $z=0$, and by periodicity we have zeros at every $z \in \Lambda$. Similarly, the zeros of $\tet_2, \tet_3$, and $\tet_4$ are respectively at $\half + \Lambda$, $\half +{\tau \over 2} + \Lambda$, and ${\tau \over 2} + \Lambda$. The function $\tet_1$ is particularly useful, and we record here its monodromy properties in $z$ for later use,
\bea
\label{tettrans}
\tet _1 (z+1 |\tau) & = & - \tet _1 (z|\tau)
\no \\
\tet _1 (z+\tau |\tau) & = & - \tet _1 (z|\tau) \, e^{- i \pi \tau - 2 \pi i z}
\eea

\subsubsection{The Riemann relations for Jacobi $\tet$-functions}

There are four basic Riemann relations on Jacobi $\tet$-functions, given by, 
\bea
\sum _\kappa \< \kappa |\lambda\> \, 
\prod _{i=1}^4 \tet [\kappa] (\zeta _i) 
= 2 \, \prod _{i=1}^ 4 \tet [\lambda] (\zeta _i') 
\eea
where $\kappa$ and $\lambda$ are half-integer characteristics,
\bea
\kappa = \left [ \begin{matrix} \kappa ' \cr \kappa '' \end{matrix} \right ]
\hskip 1in 
\lambda = \left [ \begin{matrix} \lambda ' \cr \lambda '' \end{matrix} \right ]
\eea
namely $\kappa', \kappa '', \lambda ', \lambda '' \in \{ 0, \half \}$. The sign factor $\< \kappa |\lambda\>$ is given by,
\bea
\< \kappa |\lambda\> = \exp \{ 4 \pi i (\kappa ' \lambda '' - \kappa '' \lambda ') \}
\eea
and the relation between $\zeta _i$ and $\zeta _i'$ is given by,
\bea
\left ( \begin{matrix} \zeta _1 ' \cr \zeta _2 ' \cr \zeta _3 ' \cr \zeta _4 ' \cr \end{matrix} \right )
= M \left ( \begin{matrix} \zeta _1  \cr \zeta _2  \cr \zeta _3  \cr \zeta _4  \cr \end{matrix} \right )
\hskip 1in 
M =
\half \left ( \begin{matrix} 
1 & 1 & 1 & 1 \cr
1 & 1 & -1 & -1 \cr
1 & -1 & 1 & -1 \cr
1 & -1 & -1 & 1 \cr \end{matrix} \right ) 
\eea
The proof of these Riemann relations is left as an exercise. An important special case is obtained by setting  $\zeta _1 =\zeta _2 = z$ and $\zeta _3=\zeta_4=w$ so that $\zeta '_1=z+w$, $\zeta _2'=z-w$, and $\zeta '_3=\zeta _4'=0$ for $i=2,3,4$. For this special case,  the Riemann relation for $\lambda '=\lambda ''=0$ becomes an addition formula,
\bea
\sum_{i=1}^4  \tet_i(z|\tau)^2 \tet_i (w|\tau)^2 
= 2 \tet _3(z+w|\tau) \tet _3(z-w|\tau) \tet _3(0|\tau)^2
\eea
and for $z=w$ becomes a duplication formula,
\bea
\tet_2(z|\tau)^4  + \tet_3(z|\tau)^4  + \tet_4(z|\tau)^4 
= 2 \tet _3(2z|\tau)  \tet _3(0|\tau)^3
\eea
Further setting $z=0$, we obtain the famous Jacobi formula, 
\bea
\tet _3(0|\tau)^4 - \tet _2(0|\tau)^4 - \tet _4(0|\tau)^4=0
\eea
whose validity guarantees the cancellation of the one-loop contribution to the cosmological constant in Type II and Heterotic string theories in flat Minkowski space-time.

\subsubsection{Constructions of elliptic functions in terms of zeros and poles}

There are various classic constructions of elliptic functions in terms of Weierstrass elliptic functions, Jacobi elliptic functions, and Jacobi $\tet$-functions, in addition to the construction given in Theorem \ref{thm:2}. The most important ones of those will be reviewed here.  The relations between these different constructions have direct physical applications, for example to Bose-Fermi equivalence in two-dimensional quantum field theory. 

\sm

$\bullet$ The first construction is convenient when the zeros and poles of an elliptic function~$f$ are specified (subject to the conditions of Theorem \ref{thm:1}) and proceeds by taking ratios of products of $\tet_1$-functions.  An elliptic function $f$ must have the same number of zeros and poles in any fundamental parallelogram $P_\mu$ by the second condition in Theorem \ref{thm:1}. We then have the following theorem.

{\thm 
\label{thm:2.5}
A meromorphic function $f(z|\tau)$ with $N$ poles $b_i$ and $N$ zeros $a_i$ in $z$ modulo $\Lambda$,  with $i=1,\cdots, N$ may be expressed in  the following general form,
\bea
\label{ftet}
f(z) =  \prod _{i=1}^N {\tet _1 (z-a_i|\tau) \over  \tet _1 (z-b_i|\tau)} 
\hskip 1in \sum _{i=1}^N (a_i-b_i) \in \Lambda
\eea
for $a_i, b_i \in \CC$, not necessarily mutually distinct.}

To prove the theorem, we observe that the poles and zeros of both sides clearly match, the right side is manifestly periodic under $z \to z+1$, while periodicity in $z \to z+\tau$ may be verified to hold using (\ref{tettrans}) if and only if the condition on $a_i,b_i$ of the theorem holds. 

\sm

$\bullet$ The second construction is convenient when the poles of an elliptic function and their residues are given  (subject to the conditions of Theorem \ref{thm:1}) and proceeds by using the Weierstrass $\zeta$-function and $\wp$-function, or equivalently by taking the logarithmic derivative of $\tet$-functions. The equivalence is made clear by the following relations, presented here for the rescaled lattice $\Lambda = \ZZ \oplus \tau \ZZ$ with periods $\om_1=1, \om_2 = \tau$,
\bea
\label{2.conv}
\wp(z|\tau) & = & - \eta_1(\tau) - \p_z^2 \ln \tet_1(z|\tau)
\no \\
\zeta (z|\tau) & = & \eta _1(\tau)  z + \p_z \ln \tet _1 (z|\tau)
\no \\
\ln \sigma (z|\tau) & = & \ln \tet_1(z|\tau) - \ln \tet_1(0|\tau) + \thalf \eta _1(\tau) z^2
\eea
The first relation follows from the fact that both sides have a double pole $1/z^2$ and no other poles and are doubly periodic. The constant $\eta_1$ may be determined in terms of $\tet$-functions by using the fact that $\tet_1(z|\tau)$ is odd in $z$ and that $\wp(z|\tau) = z^{-2} + \cO(z^2)$, and we find,
\bea
\label{2.eta1}
\eta_1(\tau) = - { 1 \over 3} \, {\tet_1'''(0|\tau) \over \tet_1'(0|\tau)}  = 2 \zeta(\thalf |\tau)
\eea
The second relation results from integrating the first and requiring $\zeta(z|\tau)$ to be odd in~$z$. Using (\ref{zmon}), (\ref{omzet}), and (\ref{tetshift}) one verifies that both sides have the same monodromy. The last equation follows from the second by integration and normalizing $\sigma(z) = z + \cO(z^3)$.
{\thm 
\label{thm:2.7}
An elliptic function $f(z|\tau)$ with $N_s$ poles $b_{i,s}$ of order $s$ and generalized residue $r_{i,s}$ for $s=1, \cdots, S $ and  $i=1,\cdots N_s$ may be represented as a sum of Weierstrass $\zeta$-functions  and its derivatives, by identifying poles and generalized residues,  
\bea
\label{2.fzeta}
f(z|\tau) = r_0 +  \sum _{s=1}^S \sum_{i=1}^{N_s} r_{i,s} \, \zeta^{(s-1)} (z- b_{i,s}|\tau)
\hskip 0.8in
\sum_{i=1}^{N_1} r_{i,1}=0
\eea
where $r_0$ and $r_{i,s}$ may depend on $\tau$ but are independent of $z$, and $\zeta'(z|\tau) = - \wp (z|\tau)$.}

\sm

Finally, expressing $\wp(z|\tau)$ as an infinite sum in (\ref{wpdef}), integrating this formula twice in $z$, and matching the integration constants, gives the product formula for the $\tet_1$-function and, by translations by half periods, for the remaining $\tet_\a$ functions,  
\bea
\label{tetprod}
\tet_1(z|\tau) & = & 2 q^{1 \over 8} \sin \pi z \prod _{n=1} ^\infty (1-q^n e^{2 \pi iz} ) (1-q^n e^{-2\pi i z}) (1-q^n)
\no \\
\tet_2(z | \tau) &=& 2 q^{1 \over 8} \cos \pi z \prod_{n=1}^\infty (1+q^n e^{2 \pi i z}) (1+q^n e^{- 2 \pi i z}) (1-q^n)
\no\\
 \tet_3(z | \tau) &=& \prod_{n=1}^\infty (1+q^{n- \half} e^{2 \pi i z}) (1+q^{n- \half} e^{- 2 \pi i z}) (1-q^n)
 \no\\
  \tet_4(z | \tau) &=& \prod_{n=1}^\infty (1-q^{n- \half} e^{2 \pi i z}) (1-q^{n- \half} e^{- 2 \pi i z}) (1-q^n)
\eea
with $q=e^{ 2 \pi i \tau}$.  We shall leave the derivation as an exercise.

\subsection{Uniformization of cubics and quartics}

Trigonometric functions may be viewed as parametrizing the points on the unit  circle, specified  by the quadratic equation  $x^2+y^2=1$, by the functions $x=\sin t$ and $y = \sin ' t = \cos t$. This process is referred to as \textit{uniformization}. 

\sm

Elliptic functions analogously parametrize all solutions to cubic and quartic equations. 
The Weierstrass function $\wp (z|\Lambda)$ uniformizes the following cubic equation,
\bea
\label{2f1}
y^2 = 4 x^3 - g_2(\Lambda)  x - g_3(\Lambda)= 4 (x-e_1) (x-e_2) (x-e_3) 
\eea
by $x=\wp (z|\Lambda) $ and $y = \wp '(z|\Lambda)$ in view of (\ref{wp}), where $g_2(\Lambda)$ and $g_3(\Lambda)$ are associated with the lattice $\Lambda$ via (\ref{g2g3}) and (\ref{EisenG}). Analogously, the Jacobi elliptic function $\sn(u|k)$ uniformizes the following quartic equation,
\bea
\label{uniJ}
y^2 = (1-x^2)(1-k^2x^2) 
\eea
by $x=\sn(u|k)$ and $y=\sn'(u|k)$ in view of the first equation in (\ref{snder}). The Jacobi elliptic functions $\cn$ and $\dn$ similarly uniformize different, but equivalent quartics.  More generally, elliptic functions may be used to uniformize an arbitrary quartic of the form,
\bea
\label{2.quart}
w^2 = (v-v_1)(v-v_2)(v-v_3)(v-v_4)= v^4 + c_3 v^3 + c_2 v^2 + c_1 v + c_0
\eea
for arbitrary points $v_1, \cdots , v_4 \in \CC$ or, equivalently, for arbitrary coefficients $c_0, \cdots, c_3 \in \CC$. 
To show this, we make use of the fact that the group $SL(2,\CC)$ acts conformally and bijectively on the compactified complex plane $\CC \cup \{ \infty \}$ by M\"obius transformations. If the action of an arbitrary $SL(2,\CC)$  transformation on the variables $v$ and $v_i$ is  given by,
\bea
\label{2.mob}
v = { ax + b \over cx +d} 
\hskip 0.8in 
v_i = { ax_i + b \over cx_i +d}
\hskip 0.8in 
\left ( \bma a & b \cr c & d \ema \right ) \in SL(2,\CC)
\eea
then $w$ and the differences $v-v_i $ obey simple transformation rules,
\bea
w =  { w_0 \, y \over (cx+d)^2} 
\hskip 1in
v-v_i = { x-x_i \over (cx+d)(cx_i+d)}
\eea
which map the general quartic (\ref{2.quart}) in $(v,w)$  into a quartic in $(x,y)$ given by,
\bea
 w_1 ^2 \, y^2 & = & (x-x_1)(x-x_2)(x-x_3)(x-x_4) 
\no \\ 
w_1^2 & = & w_0^2 (cx_1+d)(cx_2+d)(cx_3+d)(cx_4+d)
\eea
Here $w_0$ is an arbitrary constant in terms of which $w_1$ is given by the second formula above.
The choice $(x_1,x_2,x_3,x_4)=(1,-1,k,-k)$  with $w_1=1$ reproduces the quartic (\ref{uniJ}). Taking the limit  $x_4 \to \infty$  for  $(x_1,x_2,x_3)=(e_1,e_2,e_3)$ and $4 c w_0^2 \prod_{i=1}^3 (ce_i+d)=-1$ reproduces the cubic (\ref{2f1}). This completes the proof that an arbitrary quartic may be uniformized either by the Weierstrass or the Jacobi elliptic functions.

\subsection{Elliptic curves}

Since an arbitrary lattice $\Lambda$ is an Abelian subgroup of $\CC$ the lattice $\Lambda$ is a normal subgroup of $\CC$ and  the torus $\CC/\Lambda$ is an Abelian group. Since $\Lambda$ acts on $\CC$ without fixed points, the torus $\CC/ \Lambda$ is also a compact Riemann surface. Representing the torus in $\CC$ by a fundamental parallelogram $P_\mu$, the group law may be realized by addition in $\CC$  modulo the lattice $\Lambda$. The group structure requires an identity element which is just $0 \in \CC$ modulo $\Lambda$, and an additive inverse of $z$ which is just $-z$ modulo $\Lambda$. 

\sm

The procedure of uniformization, discussed in the preceding subsection, shows that elliptic functions naturally parametrize arbitrary cubics and quartics. In particular, the pair of coordinates $(x,y)=(\wp(z|\Lambda), \wp'(z|\Lambda))$ parametrizes the  cubic curve $y^2 = 4 x^3 -g_2(\Lambda)x-g_3(\Lambda)$ in terms of the points in the torus $z \in \CC/\Lambda$. The group structure of $\CC/\Lambda$ maps to an Abelian group structure on the corresponding elliptic curve, as we shall now explain in detail.

\sm

\begin{figure}[tp]
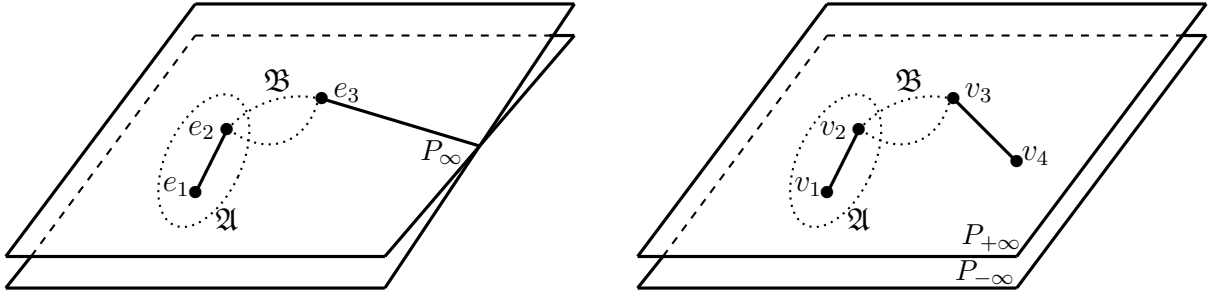

\begin{center}
\tikzpicture[scale=0.84]
\scope[xshift=10cm,yshift=0cm]
\draw[very thick] (0,0) -- (6,0);
\draw[very thick] (0,0) -- (3,4);
\draw[very thick] (6,0) -- (9,4);
\draw[very thick] (3,4) -- (9,4);
\draw[very thick] (0,-0.5) -- (6,-0.5);
\draw[very thick] (0,-0.5) -- (0.4,0);
\draw[thick, dashed] (0.4,0) -- (3,3.5);
\draw[very thick] (6,-0.5) -- (9,3.5);
\draw[thick, dashed] (3,3.5) -- (8.6,3.5);
\draw[very thick] (8.6,3.5) -- (9,3.5);
\draw (3,1) node{$\bullet$};
\draw (3.5,2) node{$\bullet$};
\draw (5,2.5) node{$\bullet$};
\draw (6,1.5) node{$\bullet$};
\draw (2.7,1.1) node{$v_1$};
\draw (3.1,2) node{$v_2$};
\draw (6.3,1.6) node{$v_4$};
\draw (5.4,2.6) node{$v_3$};
\draw (5.6,0.3) node{\small $P_{+\infty}$};
\draw (5.5,-0.27) node{\small $P_{-\infty}$};
\draw[very thick] (3,1) -- (3.5,2);
\draw[very thick] (5,2.5) -- (6,1.5);
\draw[thick, dotted,  rotate=-25] (2.2,2.7)  ellipse (17pt and 32pt);
\draw[thick, dotted,  rotate=00] (3.5,2)  arc (140:80:1.5 and 1.5);
\draw[thick, dotted,  rotate=20] (4,0.7)  arc (-140:-40:1 and 1.4);
\draw (3.5,0.6) node{$\mA$};
\draw (4.3,2.8) node{$\mB$};
\endscope
\scope[xshift=0cm,yshift=0cm]
\draw[very thick] (0,0) -- (6,0);
\draw[very thick] (0,-0.5) -- (6,-0.5);
\draw[very thick] (0,0) -- (3,4);
\draw[thick, dashed] (0.4,0) -- (3,3.5);
\draw[very thick] (6,0) -- (7.5,1.75);
\draw[very thick] (7.5,1.75) -- (9,4);
\draw[very thick] (6,-0.5) -- (7.5,1.75);
\draw[very thick] (7.5,1.75) -- (9,3.5);
\draw[very thick] (3,4) -- (9,4);
\draw[thick, dashed] (3,3.5) -- (8.6,3.5);
\draw[very thick] (0,-0.5) -- (0.4,0);
\draw[very thick] (8.6,3.5) -- (9,3.5);
\draw (3,1) node{$\bullet$};
\draw (3.5,2) node{$\bullet$};
\draw (5,2.5) node{$\bullet$};
\draw (2.7,1.1) node{$e_1$};
\draw (3.1,2) node{$e_2$};
\draw (6.9,1.65) node{\small $P_\infty$};
\draw (5.4,2.6) node{$e_3$};
\draw[very thick] (3,1) -- (3.5,2);
\draw[very thick] (5,2.5) -- (7.5,1.75);
\draw[thick, dotted,  rotate=-25] (2.2,2.7)  ellipse (17pt and 32pt);
\draw[thick, dotted,  rotate=00] (3.5,2)  arc (140:80:1.5 and 1.5);
\draw[thick, dotted,  rotate=20] (4,0.7)  arc (-140:-40:1 and 1.4);
\draw (3.5,0.6) node{$\mA$};
\draw (4.3,2.8) node{$\mB$};
\endscope

\endtikzpicture
\end{center}
\caption{\textit{The elliptic curve $\cE$ is represented in terms of a double cover of the Riemann sphere: with one branch point $P_{\infty}$ at $\infty$ in the left panel and with all four branch points finite in the right panel and two distinct points $P_{\pm \infty}$ added at $\infty$ to compactify the elliptic curve.  A choice of canonical homology cycles $\mA, \mB$ is indicated. \label{2.fig:3}}}
\end{figure}

An elliptic curve $\cE$ for given $g_2, g_3 \in \CC$ is defined as follows,
\bea
\label{calE}
\cE= \Big \{ (x,y) \in \CC \hbox{ such that } y^2 = 4 x^3-g_2x-g_3 \Big \} \cup \{ P_\infty \}
\eea
The point at infinity $P_\infty= (\infty, \infty)$  is added to obtain a compact space, just as the torus $\CC/\Lambda$ is compact.  While $y^2$ is determined in terms of $x$ by the equation, the sign of $y$ is not, which is why the second entry  $y$ is required.  For every generic value of $x$ where $y \not= 0, \infty$, there are two points in $\cE$ corresponding to $\pm y$. For the points $x=e_1,e_2,e_3,e_4=\infty$, however, we have $y=0, \infty$ and there is only a  single point in $\cE$. These points correspond to branch points of the function $y=\sqrt{4 x^3-g_2x-g_3}$ where the two copies of $\CC$ intersect (see the left panel of Figure \ref{2.fig:3}). Alternatively, an elliptic curve may be defined in terms of a quartic,
\bea
\cE= \Big \{ (v,w) \in \CC \hbox{ such that } w^2 = \prod _{i=1}^4 (v-v_i) \Big \} \cup \{ P_{+ \infty} , P_{- \infty} \}
\eea
The points $v=v_1,v_2,v_3,v_4$  correspond to the branch points of the function $w$ where the two copies of $\CC$ intersect
(see the right panel of Figure \ref{2.fig:3}). In this representation two points at infinity $P_{\pm \infty}$ are required, one on each sheet.

\subsubsection{Embedding into $\CP^2$}

Two complex-dimensional projective space $\CP^2$ is a compact space defined by an equivalence relation $\approx$ under rescaling coordinates in $(\CC^3)^*= \CC^3 \setminus \{ (0,0,0)\}$,
\bea
\CP^2 = \big \{ (x,y,z)  \in (\CC^3)^* \hbox{ such that } (\lambda x, \lambda y, \lambda z) \approx (x,y,z) \hbox { for } \lambda \in \CC^* \big \}
\eea
Actually, $\CP^2$ is a K\"ahler space isomorphic to $SU(3)/S(U(2)\times U(1))$.  Every compact (orientable) Riemann surface may be embedded in $\CP^2$. In particular, the elliptic curve $\cE$ may be embedded in $\CP^2$ by rendering its defining equation homogeneous in three variables, namely the original $x,y$ as well as the additional variable $z$, to obtain, 
\bea
\cE= \Big \{ (x,y,z) \in \CP^2 \hbox{ such that } y^2z = 4 x^3-g_2xz^2-g_3z^3 \Big \} 
\eea
Since $\cE$ is hereby defined as a closed subset of the compact space $\CP^2$, it is compact without the need to add points at infinity. The original presentation of the elliptic curve $\cE$ is recovered in the coordinate patch $z=1$, while in the other two patches it is given by,
\begin{align}
x&=1 & y^2 z & =  4 -g_2 z^2 - g_3 z^3
\no \\
y & =1 & z & =  4x^3 -g_2 xz^2 - g_3 z^3
\end{align}
The last equation shows that the point $(0,1,0) \in \CP^2$ belongs to every elliptic curve $\cE$. In fact the point $(0,1,0)$ corresponds to the point $(x,y)=P_\infty$ in (\ref{calE}). This may be seen by considering the large $x,y$ behavior in the $z=1$ patch and allowing for a suitable rescaling $(x, y, 1) \approx (xy^{-1}, 1, y^{-1}) \to (0,1,0)$ as $x,y \to \infty$ subject to $y^2=4x^3-g_2x-g_3$.

\subsection{Addition formulas and group structure of elliptic curves}

The additive group structure of the torus $\CC/\Lambda$ allows one to associate with two arbitrary points $P,Q \in \CC/\Lambda$ a third point $R \in \CC/\Lambda$ by requiring the sum of their complex coordinates $z_P, z_Q, z_R$ to vanish  $z_P+z_Q+z_R=0 ~ (\mod \Lambda)$. The addition of the points $P$ and $Q$ gives $-R$. To obtain the corresponding operation on the elliptic curve in the Weierstrass presentation, we begin by obtaining the addition formula for the Weierstrass function, which generalizes  the addition theorem of (\ref{addtrig}) for trigonometric functions,
\bea
\label{addwp}
\wp (z_P+z_Q) + \wp (z_P) + \wp (z_Q) - {1 \over 4} \left ( { \wp '(z_P) - \wp '(z_Q) \over \wp (z_P) - \wp (z_Q)} \right )^2 =0
\eea
To prove it, one begins by showing that the left side has no poles, so that it must be constant by Liouville's theorem. One then shows that the constant vanishes by evaluating the expression at special points.
Specifically, for fixed generic $z_Q$, the function $\wp (z_P+z_Q)$ is an elliptic function in $z_P$ with one double pole at $z_P=-z_Q \, ({\rm mod}\,  \Lambda)$. Thus, it is a rational function of $\wp (z_P)$ and $\wp'(z_P)$. Now $\wp (z_P) - \wp (z_Q)$ has a simple zero at $z_P=-z_Q$, so it should occur to the power $-2$, but it also has a simple zero at $z_P=z_Q$, which is cancelled by multiplying by the square of $\wp '(z_P) - \wp '(z_Q)$. The combination has a double pole in $z_P$ at $z_P=0$ which is cancelled by the addition of $\wp (z_P)$ and, by symmetry under the interchange of  $z_P$ and $z_Q$, also $\wp (z_Q)$. Thus the left side has no poles and is therefore constant. To evaluate the constant, we set $z_P=\om_1/2$ and $z_Q=\om _2/2$, use the fact that $\wp'$ vanishes at the non-zero half-periods, and then use the first formula in (\ref{e}).

\begin{figure}[tp]
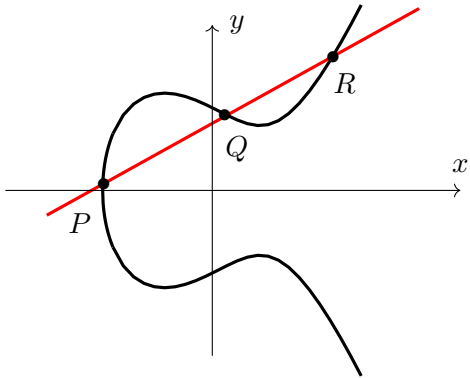

\begin{center}
\tikzpicture[scale=1.1]
\scope[xshift=10cm,yshift=0cm]
\draw [very thick, domain=-1.32472:-1.2] plot ({\x},{sqrt(\x^3-\x+1)});
\draw [very thick, domain=-1.2:1.8] plot ({\x},{sqrt(\x^3-\x+1)});
\draw [very thick, domain=-1.32472:-1.2] plot ({\x},{-sqrt(\x^3-\x+1)});
\draw [very thick, domain=-1.2:1.8] plot ({\x},{-sqrt(\x^3-\x+1)});
\draw [very thick, red] (-2,-0.3) -- (2.5,2.2);
\draw (-1.31,0.07) node{\small $\bullet$};
\draw (0.152,0.9) node{\small $\bullet$};
\draw (1.46,1.61) node{\small $\bullet$};
\draw [->] (-2.5,0) -- (3,0);
\draw [->] (0,-2) -- (0,2);
\draw (3,0.3) node{\small $x$};
\draw (0.3,2) node{\small $y$};
\draw (-1.6,-0.4) node{\small $P$};
\draw (0.3,0.5) node{\small $Q$};
\draw (1.6,1.3) node{\small $R$};
\endscope
\endtikzpicture
\end{center}
\caption{\textit{The real section of the elliptic curve $y^2 = 4 x^3 - 4 x +4$ in black,  the straight line $9y=10 x+14.72$ in red, and their three intersection points $P,Q,R$.}
 \label{2.fig:4}}
\end{figure}

\sm

On the elliptic curve $\cE$ we represent the points $P,Q,R$ by the coordinates $(x_P,y_P)$, $(x_Q,y_Q)$ and $(x_R,y_R)$ respectively and, for simplicity, we assume that the points $P$ and $Q$ are distinct.  A straight line $y= \a x + \b$ through the two distinct points $P,Q$ on $\cE$ intersects the cubic $\cE$ at exactly one further point $R$. Therefore, its coordinates $(x_R,y_R)$ satisfy, 
\bea
\label{xPQ}
y_R  = { y_P - y_Q \over x_P - x_Q} \, x_R + { x_P y_Q - x_Q y_P \over x_P -x_Q}
\hskip 1in
y_R^2  = 4 x_R^3 - g_2 x_R - g_3 
\eea
This system of equations may be solved for $x_R$ as follows,
\bea
x_R + x_P+x_Q - {1 \over 4} \left ( { y_P - y_Q \over x_P - x_Q} \right )^2 = 0
\eea
with $y_R$ given by the second equation  of (\ref{xPQ}). Parametrizing the coordinates by $x=\wp(z)$ and $y=\wp'(z)$ evaluated at $z=z_P,z_Q,z_R$, we recover the addition law (\ref{addwp}) for the Weierstrass function. The point $P_\infty$ plays the role of the unit element of the additive group~$\cE$. This may be seen by letting $x_Q, y_Q \to \infty$ which implies $(x_R,y_R)=(x_P,y_P)$.

\subsection{Meromorphic functions and Abelian differentials}

Much of the work on elliptic functions in the 19-th century was motivated  by the study of elliptic integrals, namely definite or indefinite integrals involving the square root of a cubic or quartic polynomial. In this subsection we shall  consider  meromorphic functions and Abelian differentials before addressing their integrals in the subsequent subsection. For a discussion of differentials and their properties on arbitrary Riemann surfaces, see appendix \ref{sec:RS}.

\sm

The space of meromorphic functions on the Riemann sphere $\hat \CC = \CC \cup \{ \infty \}$ is the field $\CC(x)$ of rational functions of a holomorphic coordinate $x \in \CC$.  Meromorphic functions on an elliptic curve $\cE$ correspond to elliptic functions on the torus $\CC / \Lambda$. Since elliptic functions on $\CC/\Lambda$ are rational functions of $\wp(z)=\wp(z|\Lambda)$ and $\wp'(z)=\wp'(z|\Lambda)$, the space of meromorphic functions on an elliptic curve $\cE$ is the field $\CC(x,y)$ of rational functions of $x$ and $y$.  The field $\CC(x,y)$ may be viewed as a quadratic extension of $\CC(x)$ by $y$, since $y^2 \in \CC(x)$ on an elliptic curve $\cE$. The Galois group of this extension is $\ZZ_2$, the group of involutions of $\cE$ that maps $(x,y) \to (x,-y)$ and swaps the two sheets in the elliptic representation of Figure \ref{2.fig:3}. On the torus $\CC/ \Lambda$, the $\ZZ_2$ involution simply corresponds to $z \to - z$ (mod $\Lambda$).

\subsubsection{Perspective of the torus $\CC/\Lambda$}

To investigate meromorphic differentials we first consider the perspective of the torus. On  a torus $\CC/ \Lambda$ with  local complex coordinate $z$, there exists a translation invariant $(1,0)$-form $dz$ which is unique up to a multiplicative constant. The differential $dz$ is holomorphic  and nowhere vanishing. As a result,  the vector spaces of meromorphic $(n,0)$-forms for $n \in \ZZ$ are all isomorphic to the space of meromorphic $(0,0)$-forms, i.e. elliptic functions. Thus, the most general meromorphic differential $(n,0)$-form on $\CC/\Lambda$ is given by $f(z) (dz)^n$ where $f(z)$ is an elliptic function on $\CC/ \Lambda$.  Using the general construction of elliptic functions in terms of the Weierstrass functions, Theorems \ref{thm:2} and \ref{thm:2.7} allow us to express an arbitrary meromorphic $(n,0)$-form $\varpi =  f(z) (dz)^n$ as a rational function in $\wp$ and $\wp'$  which may be chosen to be linear in $\wp'$ or, equivalently, as a sum of derivatives of the Weierstrass $\zeta$-function,
\bea
\label{omAB1}
\varpi & = &  A \big (\wp(z) \big ) (dz)^n + B \big ( \wp(z) \big ) \, \wp'(z) (dz)^n
 \\
 \label{omAB2}
\varpi & = & r_0 \, dz  +  \sum _{s=1}^S \sum_{i=1}^{N_s} r_{i,s} \, \zeta^{(s-1)} (z- b_{i,s}) \, dz
\hskip 0.8in
\sum_{i=1}^{N_1} r_{i,1}=0
\eea
where $A$ and $B$ are rational functions of their argument, while the poles $b_{i,s}$ and the coefficients $r_0, r_{i,s}$   are arbitrary aside from the vanishing of the sum of $r_{i,1}$.  We shall denote the vector space of meromorphic $(n,0)$-forms $\Omega _n= \Omega _n (\CC / \Lambda)$.

\sm

The space $d \Omega_0$ of $(1,0)$-forms obtained by taking the differential of elliptic functions is a subspace $ d \Omega _0 \subset \Omega_1$. This may be seen in the representation (\ref{omAB1}) by taking the differential of a general elliptic function $f \in \Omega_0$ and then using the  differential equation (\ref{wp}) and its derivative to recast the result again in the form (\ref{omAB1}). In the representation of (\ref{omAB2}) the differential kills the first term and  simply increases the order of the derivatives in the terms under the double sum, thereby producing a $(1,0)$-form of the type (\ref{omAB2}).

\sm

However, not every meromorphic $(1,0)$-form is the differential of an elliptic function. Constructing the space $\Omega_1 / d \Omega_0$ of  meromorphic $(1,0)$ modulo differentials of elliptic functions is a cohomology problem whose solution is fundamental in the theory of Riemann surfaces and in string theory. To obtain the generators of $\Omega_1/d \Omega_0$ we express an arbitrary $f  \in \Omega _0$ and  $\varpi  \in \Omega_1$ in terms of the representation (\ref{omAB2}),  and then take the differential of $f$, 
\begin{align}
df & =   \sum _{s=1}^{\tilde S}  \sum_{i=1}^{\tilde N_s} \tilde r_{i,s} \, \zeta^{(s)} (z- \tilde b_{i,s}) \, dz
&
\sum_{i=1}^{\tilde N_1} \tilde r_{i,1} &=0
\no \\
\varpi & =  r_0 \, dz  +  \sum _{s=1}^S \sum_{i=1}^{N_s} r_{i,s} \, \zeta^{(s-1)} (z- b_{i,s}) \, dz
&
\sum_{i=1}^{N_1} r_{i,1} &=0
\end{align}
All terms in $\varpi$ involving  $\zeta ^{(s-1)} (z-b_{i,s}) $ for $s \geq 3$ may be matched by terms in $df$ provided we choose the poles  $\tilde b_{i,s-1} = b_{i,s}$ and coefficients $\tilde r_{i,s-1}=r_{i,s}$. For $s=2$, this would also be the case if it were not for the constraint on the sum of the $\tilde r_{i,1}$, which means that a single  $\zeta ^{(1)}(z-b|\tau)$ is, manifestly,  not the differential of an elliptic function. The term $r_0 dz $ and the  terms proportional to $\zeta (z-b_{i,1})$ are not differentials of elliptic functions but, in view of the vanishing sum of residues, always occur as differences $\zeta(z-b_{i,1})- \zeta(z-b_{j,1})$.  In summary,  $ \Omega_1/d \Omega_0$ is generated by three type of $(1,0)$-forms, referred to as Abelian differentials.
\begin{enumerate}
\itemsep =0in 
\item \textit{Abelian differential of the first kind}:  holomorphic and given by a constant multiple of the coordinate differential $dz$;
\item  \textit{Abelian differentials of the second kind}: have one double pole at an arbitrary point $z_i \in \CC/\Lambda$ and no other poles and are given by a constant multiple of
\bea
\wp(z-z_i)dz = - \zeta ^{(1)} (z-z_i ) dz
\eea
\item  \textit{Abelian differentials of the third kind}: have two simple poles with residues $\pm 1$ at arbitrary points $z_i \not= z_j \in \CC/ \Lambda$ and no other poles and are given in terms of the Weierstrass $\zeta$-function by,
\bea
\big ( \zeta (z-z_i ) - \zeta (z-z_j ) +\zeta (z_i-z_j ) \big ) dz
\eea 
We note that the third term inside the parentheses is included to make the differential invariant under $\Lambda$ not only in $z$ but also in the points $z_i$ and $z_j$.
\end{enumerate}
The characterization of Abelian differentials of the second and third kind in terms of their poles does not specify them uniquely as one may add an arbitrary multiple of $dz$ to either.

\subsubsection{Perspective of the elliptic curve $\cE$}

For an elliptic curve $\cE$ with local coordinates $(x,y)$ given in terms of $g_2, g_3 \in \CC$ by, 
\bea
\label{cubic}
y^2 = 4 x^3-g_2x-g_3
\eea
the $(1,0)$ form  $dz$ on $\CC/\Lambda$ may be expressed in terms of $(x,y)=(\wp(z|\Lambda), \wp'(z|\Lambda))$ by using the relation (\ref{wp}) and we find, 
\bea
dz = { dx \over y} = { dx \over \sqrt{4 x^3 - g_2 x - g_3}}
\eea 
The differential $dx/y$ is single-valued and holomorphic on $\cE$ since  the square root at one of the branch points $e_i$  is well-defined  in a suitable local coordinate $\xi$ with $\xi^2 = x - e_i$, in which the differential is  given by
$dx / y =  c_i^{-1} d \xi+ \cO(\xi) $ where $c_i^2=(e_i-e_j)(e_i-e_k)$ for $j,k \not= i$. The coordinate differential $dx$ vanishes at the branch points $e_i$ and has a triple pole at $\infty$. 

\sm

The differential $dz$ may also be obtained for an arbitrary quartic of the form (\ref{2.quart}),
\bea
{ dv \over w} 
\hskip 1in
w^2 = (v-v_1)(v-v_2)(v-v_3)(v-v_4)
\eea
The differential $dx/y$ is mapped to $dv/w$ by the $SL(2,\CC)$ M\"obius transformation of (\ref{2.mob}) which maps $(x,y)$ to $(v,w)$,  $x_i$  to $v_i$, and the differential $dx$  to $dv =  dx / (cx+d)^2$. 

\sm

An arbitrary meromorphic differential $\om$ on the torus $\CC/\Lambda$,  expressed in terms of $\wp$ and $\wp'$  in (\ref{omAB1}),  translates to the following expression on the corresponding elliptic curve $\cE$,
\bea
\om = A(x) { dx \over y} + B(x) dx
\eea
where $A$ and $B$ are rational functions of $x$. Under the $\ZZ_2$  involution of $\cE$ which maps $(x,y) \to (x,-y)$, the first term in $\om$ is odd while the second term is even. We leave it as an exercise to show that the Abelian differentials map to the elliptic curve $\cE$ as follows,
\begin{enumerate}
\itemsep =0in 
\item \textit{Abelian differential of the first kind}: a constant multiple of $dx/y$;
\item  \textit{Abelian differentials of the second kind}: have one double pole at an arbitrary point $(x_i,y_i) \in \cE$ and no other poles and, using the addition formula of (\ref{addwp}),  are given by,
\bea
\left ( -x -x_i + { 1 \over 4}  { (y+y_i)^2 \over (x-x_i)^2 } \right ) { dx \over y}
\eea
\item  \textit{Abelian differentials of the third kind}: have two simple poles with opposite residues $\pm 1$ at arbitrary points $(x_i,y_i) \not= (x_j, y_j)$ and no other poles and are given  by,
\bea
\half \left ( { y+y_i \over x-x_i} + { y_j+y \over x_j-x} + { y_i+y_j \over x_i-x_j} \right ) {dx \over y} 
\eea
The last term inside the parentheses arises from the $\zeta(z_i-z_j)dz$ contribution in the torus representation, and is proportional to the Abelian differential of the first kind.
\end{enumerate}

\subsection{Abelian  and elliptic  integrals}

The decomposition of the space of meromorphic $(1,0)$-forms $\Omega _1$ into a sum of the space of exact differentials of elliptic functions $d \Omega_0$ plus the three kinds of basic Abelian differentials greatly simplifies the problem of  integration of $(1,0)$-forms. Integration of exact differentials $\varpi \in d \Omega_0$ is straightforward, and we shall now concentrate on the integration of the three kinds of Abelian differentials. A canonical basis of first homology 
generators of the elliptic curve $\cE$ will be denoted by $\mA, \mB$ and was shown already in Figures \ref{2.fig:2} and \ref{2.fig:3}. Parametrizing the lattice $\Lambda = \om_1 \ZZ \oplus \om_2 \ZZ$  by the periods $\om_1$ and $\om_2$ satisfying $\tau = \om_2/ \om_1$ with $\tau \not \in \RR$, the homology basis may be represented on the torus $\CC/\Lambda$ by the following simple cycles, 
\bea
\mA &: \hskip 0.2in & z \to z+ \om_1
\no \\
\mB &: \hskip 0.2in & z \to z+ \om_2
\eea
We shall study integrals of Abelian differentials along open and closed paths in $\cE$ and $\CC/ \Lambda$. We consider open paths between arbitrary points $z_1, z_2 \in \CC/\Lambda$ corresponding to the points $(x_1, y_1), (x_2, y_2) \in \cE$, related by the Weierstrass function $x_i = \wp(z_i), y_i = \wp '(z_i)$ for $i=1,2$.

\subsubsection{Abelian integrals of the first kind}

The integral of the Abelian differential $dz=dx/y$ of the first kind is given as follows,
\bea
\label{2.Abint}
 \int _{(x_1, y_1)} ^{( x_2,y_2)} { dx \over \sqrt{4 x^3 - g_2 x -g_3}} = \int _{(x_1, y_1)} ^{( x_2,y_2)} { dx \over y} =
 \int _{z_1} ^{z_2} dz = z_2-z_1
\eea
Viewed as taking values in $\CC$, the integral depends on the path of integration.  Indeed, while holomorphicity of  $dz=dx/y$ guarantees independence  under small deformations of the path of integration, the integral will depend on how many times the path circles the cycles $\mA$ and $\mB$. The integrals of $dz=dx/y$ around $\mA, \mB$ give the periods of the lattice $\Lambda$,
\bea
\oint _\mA {dx \over y} = \oint _\mA dz = \om_1
\hskip 1in
\oint _\mB {dx \over y} = \oint _\mB dz = \om_2
\eea
Thus, while the value of the  integral of (\ref{2.Abint}) depends on the path of integration when its range is $\CC$, the integral becomes single-valued as a map to $\CC / \Lambda$. 

\subsubsection{Abelian integrals of the second kind}

The integral of the Abelian differential of the second kind,
\bea
\varpi = \wp(z-z_i) dz = \left ( -x -x_i + { 1 \over 4}  { (y+y_i)^2 \over (x-x_i)^2 } \right ) { dx \over y}
\eea
 is, by definition, given by the Weierstrass $\zeta$-function in view of $\zeta'(z) = - \wp(z)$, 
\bea
\int _{z_1 }^{z_2} \varpi =  \zeta (z_1-z_i) - \zeta (z_2-z_i) 
\eea
The integral depends on the path of integration and the number of times the path winds around the cycles $\mA$ and $\mB$, which may be evaluated using (\ref{zmon}),
\bea
\oint _{\mA} \varpi = - \eta _1 
\hskip 0.8in 
\oint _{\mB} \varpi = - \eta _2
\eea
These periods do not belong to the lattice $\Lambda$, as is clear from the relation (\ref{omzet}). In fact, it is not possible to form a linear combination of $\varpi$ and the Abelian differential of the first kind $dz$ to obtain a single-valued integral taking values in $\CC/\Lambda$.  But it is possible to form combinations with either vanishing $\mA$ period or vanishing $\mB$ period. For example $\varpi + \eta_1 dz/\om_1$ has vanishing $\mA$ period while its $\mB$ period is $ 2\pi /\om_1$.

\subsubsection{Abelian integrals of the third kind}

The integral of the Abelian differential of the third kind,
\bea
\varpi & = & \Big ( \zeta(z-z_i) + \zeta(z_i-z_j) + \zeta (z_j-z) \Big ) dz 
\no \\
& = & 
\half \left ( { y+y_i \over x-x_i} + { y_i+y_j \over x_i-x_j} + { y_j+y \over x_j-x}  \right ) {dx \over y} 
\eea
is given by the logarithm of the Weierstrass $\sigma$-function in view of $\sigma (z) '/\sigma (z) = \zeta (z)$,
\bea
\int _{z_1} ^{z_2} \varpi = \ln { \sigma (z_1-z_i) \sigma (z_2-z_j) \over  \sigma (z_2-z_i) \sigma (z_1-z_j)}
\eea
Its period integrals are given by,
\bea
\oint _\mA \varpi = \om_1 \eta_1 (z_i-z_j)
\hskip 1in
\oint _\mB \varpi = \om_2 \eta_1 (z_i-z_j)
\eea
For the canonical periods $\om_1=1$ and $\om_2=\tau$, we may convert this expression in terms of Jacobi $\tet$-functions, using (\ref{2.conv}), and we obtain,
\bea
\int _{z_1} ^{z_2} \varpi = \ln { \tet_1 (z_1-z_i) \tet_1 (z_2-z_j) \over  \tet_1 (z_2-z_i) \tet_1 (z_1-z_j)}
+ \eta _1 (z_2-z_1)(z_i-z_j)
\eea
We conclude that all elliptic integrals can be evaluated in terms of Weierstrass and Jacobi $\tet$-functions.

\subsection*{$\bullet$ Bibliographical notes}

The early history of the development of elliptic functions and modular forms, with emphasis on the contributions of Eisenstein and Kronecker,  is discussed in the delightful book by Weil~\cite{Weil}. Classic treatises on Weierstrass elliptic functions, Jacobi elliptic functions, and Jacobi $\tet$-functions may be found in Whittaker and Watson~\cite{WW} as well as in the Bateman manuscript~\cite{Bateman2}, where many practical definitions and useful formulas are collected.  Mumford~\cite{Mumford1} emphasizes Jacobi $\tet$-functions, while Lang~\cite{Lang2} deals with elliptic functions as well as modular forms. An account of applications of elliptic functions to physics and technology may be found in a classic book by Oberhettinger and Magnus~\cite{Magnus}.

\newpage

\section{Modular forms for $SL(2,\ZZ)$}
\setcounter{equation}{0}
\label{sec:SL2}

In the preceding section, we studied the dependence of elliptic functions on the points in the torus $\CC/\Lambda$. In this section, we shall investigate the dependence on the lattice $\Lambda$. We begin by discussing the equivalence relations which lead to the  transformation laws of elliptic functions, Eisenstein series, and $\tet$-functions under the modular group. We shall also discuss the Poincar\'e upper half-plane and the fundamental domain of the modular group. 

\subsection{Automorphisms of lattices and the modular group}

A lattice $\Lambda$ in the complex plane $\CC$ may be generated by a pair of periods $\om_1, \om _2$ which are linearly independent over $\RR$, namely such that $\om_2/\om_1 \not \in \RR$. We shall be interested in orientable surfaces $\CC/\Lambda$, and choose a definite orientation so that the pair of periods $(\omega_1, \omega_2)$ is ordered by choosing $\Im (\om _2 / \om _1) >0$.

\sm

The lattice $\Lambda$, and thus the torus $\CC/ \Lambda$, may be generated equivalently by other choices of periods, such as for example by $\omega_1 $ and $\omega _2 + n \omega _1$ for any $n \in \ZZ$. More generally, $\Lambda$ is equivalently generated by any pair of periods $\om_1', \om_2'$ given by a  linear combination of $\om_1, \om_2$ with integer coefficients, 
\bea
\label{3a1}
\left ( \begin{matrix} \omega _2' \cr \omega _1' \cr \end{matrix}  \right ) 
= \gamma \left ( \begin{matrix} \omega _2 \cr \omega _1 \cr \end{matrix} \right ) 
\hskip 1in 
\gamma = \left ( \begin{matrix} a & b \cr c & d \cr \end{matrix} \right )
\hskip 0.6in 
a,b,c,d \in \ZZ
\eea
provided $\gamma$ has an  inverse with integer entries. This condition is equivalent to $\det \gamma =\pm 1$, or more formally $\gamma \in GL(2,\ZZ)$. On the ratio $\tau$ of periods, which is referred to as the \textit{modulus} of the lattice $\Lambda$ or the torus $\CC/\Lambda$, a transformation $\gamma \in GL(2,\ZZ)$ acts as follows, 
\bea
\tau = { \om_2 \over  \om_1}
\hskip 0.8in 
 \tau' = { \om_2' \over  \om_1'}
\hskip 1in
\gamma \tau = \tau ' = { a \tau + b \over c \tau +d}
\eea 
A transformation $\gamma \in GL(2,\ZZ)$ will preserve the orientation $\Im(\om_2/\om_1)=\Im(\tau)>0$ of the lattice  provided the new periods enjoy the same orientation $\Im (\om_2'/\om_1')= \Im (\tau')>0$. 
Using the above transformation rule, the imaginary parts are found to transform as follows,
\bea
\Im (\tau') = {\det \gamma \over |c \tau+d|^2} \, \Im (\tau) 
\eea
The orientation of the lattice $\Lambda$ will be preserved provided $\gamma$ is restricted to the $SL(2,\ZZ)$ subgroup of $GL(2,\ZZ)$ for which $\det \gamma =1$, promoting  $SL(2,\ZZ)$ to the group of orientation-preserving automorphisms of $\Lambda$. The transformation $-I \in SL(2,\ZZ)$ maps the periods into their opposites, but leaves their ratio $\tau$ invariant. Thus, the faithful action on the lattice $\Lambda$ is by $SL(2,\ZZ)$, but on $\tau$ it is by the group $ PSL(2,\ZZ) = SL(2,\ZZ) /\ZZ_2$ where $\ZZ_2= \{ \pm I \}$.

\subsubsection{Structure of the modular group}

The group $SL(2,\ZZ)$ is an infinite discrete non-Abelian group, referred to as the \textit{modular group}. It is generated by two of its elements, 
\bea
S = \left ( \begin{matrix} 0 & -1 \cr 1 & 0 \cr \end{matrix} \right )
\hskip 1in 
T = \left ( \begin{matrix} 1 & 1 \cr 0 & 1 \cr \end{matrix} \right )
 \eea
These generators satisfy three fundamental relations, 
\bea
S^2 = (ST)^3=(TS)^3=-I
\eea
Any $\gamma \in SL(2,\ZZ)$ can be written as a finite word in the letters $S,T$. By using the first of the above relations, words with two consecutive $S$ letters may be simplified and omitted when listing independent elements. Thus, any $\gamma$ has the following product decomposition,
\bea
\gamma =  \pm T^{\alpha _1} S T^{\alpha _2} S \, \cdots \, S T^{\alpha _n}  
\eea
for $\alpha _i \in \ZZ$ and some positive integer $n$.  Furthermore, any word with three consecutive combinations of $ST$ or $TS$ may also be omitted by the last two relations.

\sm

To prove the validity of this decomposition, we consider an arbitrary $\gamma \in SL(2,\ZZ)$,
\bea
\gamma = \left ( \begin{matrix} a & b \cr c & d \cr \end{matrix} \right )
\hskip 1in a,b,c,d \in \ZZ,\,\, ~ ad-bc=1
\eea
If $c=0$, then $ad=1$, and $\gamma = a T^{ab}$. If $c <0$ we multiply $\gamma$ by $-I$ to reduce this case to the case $c >0$. For $c \geq 1$ we proceed by induction on $c$. If $c=1$, then $ad-b=1$ and we have  $\gamma = T^a S T^d$. Thus, the proposition holds true for $c=0,1$. If $c \geq 2$ we shall assume that the proposition holds true for all matrices whose lower left entry is positive and less than $c$, and the proposition remains to be proven for all matrices whose left lower entry equals $c$.  Since $c \geq 2$, $c$ cannot  divide $a$, leaving the following two possibilities:

\begin{itemize}
\itemsep=0in
\item If $|a|>c$, then there exists a non-zero integer $n$ such that $a= n c + r$ with $0 < r < c$, and the matrix $\gamma '= ST^{-n} \gamma$ has lower left entry equal to $r <c$.
\item  If $|a| <c$, then the matrix $ \gamma ' = {\rm sign} (a) S \gamma$ has lower left entry $|a|<c$.
\end{itemize}
In either case,  by applying a product of $S$ and $T$ generators we have constructed from $\gamma$ a new matrix with strictly smaller value of the lower left entry. The validity of the proposition follows by induction on the value of $c$.

\subsection{The fundamental domain for $SL(2,\ZZ)$}

In this subsection, we begin with a brief review of the geometry of the Poincar\'e upper half-plane, and then discuss the construction of the fundamental domain for  $SL(2,\ZZ)$ as well as its fixed points and cusps.

\subsubsection{The Poincar\'e upper half-plane}
 
\begin{figure}[tp]
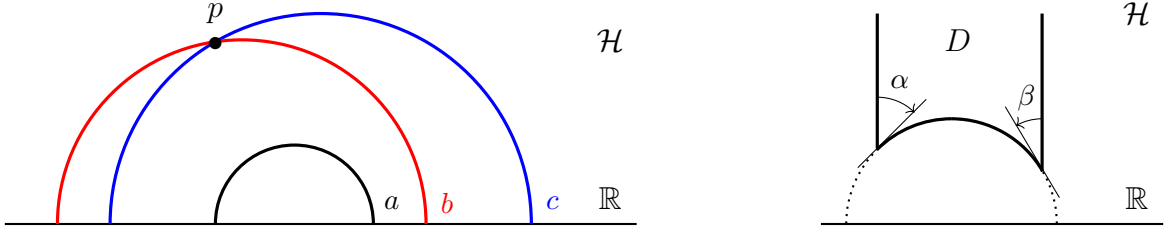

\begin{center}
\tikzpicture[scale=0.7]
\scope[xshift=0cm,yshift=0cm]
\draw [thick] (-10,0) -- (2,0);
\draw [very thick, blue] (0,0) arc (0:180:4 and 4);
\draw [very thick] (-3,0) arc (0:180:1.5 and 1.5);
\draw [very thick, red] (-2,0) arc (0:180:3.5 and 3.5);
\draw (1.5, 3.5) node{$\cH$};
\draw (1.5, 0.5) node{$\RR$};
\draw (-6, 3.43) node{$\bullet$};
\draw (-6, 4) node{$p$};
\draw (-2.65, 0.4) node{\small $a$};
\draw [red] (-1.6, 0.4) node{\small $b$};
\draw [blue] (0.4, 0.4) node{\small $c$};
\endscope
\scope[xshift=4cm,yshift=0cm]
\draw [thick] (1.5,0) -- (8,0);
\draw [very thick] (2.57,1.44) -- (2.57,4);
\draw [very thick] (5.7,1) -- (5.7,4);
\draw [very thick] (3.98,2) arc (90:30:2 and 2);
\draw [very thick] (3.98,2) arc (90:135:2 and 2);
\draw [thick, dotted] (3.98,2) arc (90:180:2 and 2);
\draw [thick, dotted] (3.98,2) arc (90:0:2 and 2);
\draw  (2.2,1.07) -- (3.5,2.37);
\draw  (6.03,0.5) -- (5,2.23);
\draw (4.1, 3.5) node{$D$};
\draw  [->] (2.57,2.41) arc (90:45:1 and 1);
\draw  [->] (5.7,2) arc (90:120:1 and 1);
\draw (5.4, 2.4) node{\small $\b$};
\draw (3, 2.7) node{\small $\a$};
\draw (7.5, 4) node{$\cH$};
\draw (7.5, 0.5) node{$\RR$};
\endscope
\endtikzpicture
\end{center}
\caption{\textit{In the left figure  the geodesics of the Poincar\'e upper half-plane $\cH$  are shown as  the semi-circles centered on $\RR$ with arbitrary radius; the geometry is non-Euclidean as several geodesics through a given point $p$, such as $b$ and $c$,  are parallel to the geodesic $a$.  In the right figure, the area enclosed by the semi-infinite triangle $D$ is ${\rm area} (D)  = \pi - \a - \b$. } 
\label{3.fig:1}}
\end{figure}

We denote the Poincar\'e upper half-plane by,
\bea
\cH = \{ \tau \in \CC, ~ \Im (\tau) >0 \}
\eea 
Under the action of $SL(2,\RR)$ by M\"obius transformations on $\tau$, the upper half-plane $\cH$ is mapped to itself and its boundary $\p \cH=\RR \cup \{ \infty \}$ is mapped to itself. The isotropy group of an arbitrary point in $\cH$ is isomorphic to $SO(2)$ and hence $\cH$ is the symmetric space,
\bea
\cH = SL(2,\RR) / SO(2)
\eea
For example, the elements which leave $\tau=i$ invariant obey $i = (ai+b)/(ci+d)$ and $ad-bc=1$ with $a,b,c,d \in \RR$ and may be parametrized by $b=-c, d=a$, and $a^2+b^2=1$ to form a group isomorphic to $SO(2)$. The Poincar\'e metric on $\cH$, defined by the line element,
\bea
ds^2 = { |d\tau |^2 \over (\Im \tau )^2}
\eea
is an $SL(2,\RR)$-invariant Riemannian metric of constant negative curvature. The Poincar\'e upper half-plane equipped with the Poincar\'e metric is a model of non-Euclidean geometry. Its geodesics (or straight lines) are the half-circles  of arbitrary radius centered on $\RR$.   Given a geodesic $a$ and a point $p$ not on this geodesic, there are an infinite number of geodesics through $p$ that do not intersect $a$ and are thus parallel to $a$, as illustrated in Figure \ref{3.fig:1}.

The area of a polygon whose sides are geodesics may be obtained by decomposing the area into  sums and differences of  semi-infinite triangles with one vertex at the cusp and two corners with opening angles $\a$ and $\b$, as shown in  Figure \ref{3.fig:1},
\bea
\label{3.area}
{\rm area}(D)= \int _D { d^2 \tau  \over (\Im \tau)^2} = \pi -\a -\b
\hskip 1in
d^2 \tau = { i \over 2} d\tau \wedge d \bar \tau
\eea
The area of a finite triangle with opening angles $\a, \b, \gamma$ is then $\pi - \a -\b -\g$, and is clearly always less than $\pi$, which is a well-known characteristic of hyperbolic geometry.

\subsubsection{The fundamental domain for $SL(2,\ZZ)$}

\begin{figure}[tp]
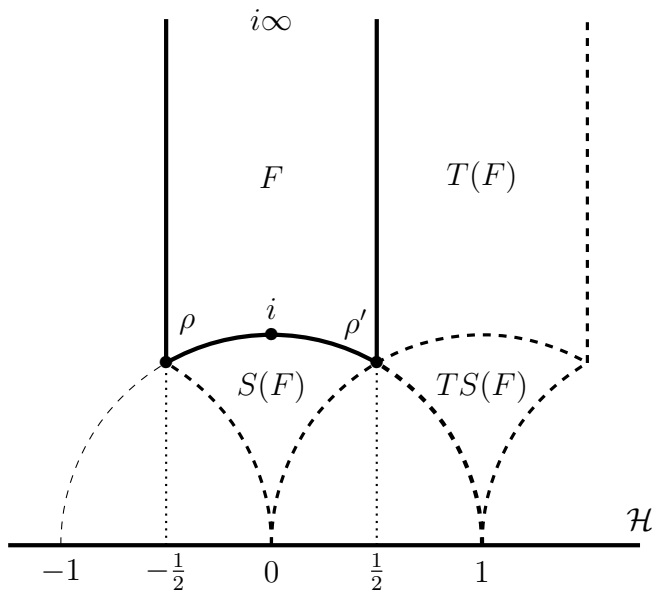

\begin{center}
\tikzpicture[scale=0.7]
\scope[xshift=-5cm,yshift=0cm]
\draw[ultra thick] (-5,0) -- (7,0);
\draw[thick, dotted] (-2,0) -- (-2,3.464);
\draw[thick, dotted] (2,0) -- (2,3.464);
\draw[ultra thick] (-2, 3.464) -- (-2, 10);
\draw[ultra thick] (2 , 3.464) -- (2, 10);
\draw [ultra thick] (-2,3.464) arc (120:60:4 and 4);
\draw [dashed] (-4,0) arc (180:120:4 and 4);
\draw [ultra thick, dashed] (4,0) arc (0:60:4 and 4);
\draw[very thick, dashed] (6 , 3.464) -- (6, 10);
\draw [very thick, dashed] (2,3.464) arc (120:60:4 and 4);
\draw [very thick, dashed] (0,0) arc (0:60:4 and 4);
\draw [very thick, dashed] (0,0) arc (180:120:4 and 4);
\draw [very thick, dashed] (4,0) arc (180:120:4 and 4);
\draw (0,-0.5) node{$0$};
\draw (-2,-0.5) node{$-\half $};
\draw (2,-0.5) node{$\half $};
\draw (-4,-0.5) node{$-1$};
\draw (4,-0.5) node{$1$};
\draw (0,7) node{$F$};
\draw (4,7) node{$T(F)$};
\draw (0,3) node{$S(F)$};
\draw (4,3) node{$TS(F)$};
\draw (7,0.5) node{$\cH$};
\draw (0,4.5) node{$i$};
\draw (0,4) node{$\bullet$};
\draw (2,3.464) node{$\bullet$};
\draw (-2,3.464) node{$\bullet$};
\draw (1.6,4.2) node{$\rho'$};
\draw (-1.6,4.2) node{$\rho$};
\draw (0,10) node{$i \infty$};
\endscope
\endtikzpicture
\end{center}
\caption{\textit{The standard fundamental domain $F$  for $SL(2,\ZZ)$ in $\cH$;  its images $T(F), S(F)$, $TS(F)$; the orbifold points $i, \rho, \rho'$ in $F$; and the cusp at $ i \infty$. }\label{3.fig:2}}
\end{figure}

Since the modular group $SL(2,\ZZ)$ is a subgroup of $SL(2,\RR)$, it also maps $\cH$ to $\cH$ and $\p \cH $ to $\p \cH$. Every oriented lattice $\Lambda$, or equivalently  oriented torus $\CC/ \Lambda$, may be specified by a point $\tau \in \cH$ up to overall rescaling of the lattice. Since the lattices corresponding to two points $\tau$ and $\tau'$ are equivalent to one another if $\tau' = \gamma \tau$ with $\gamma \in SL(2,\ZZ)$, the space of inequivalent lattices or inequivalent tori is given by the quotient,
\bea
SL(2,\ZZ) \backslash \cH = SL(2,\ZZ) \backslash SL(2,\RR) / SO(2)
\eea
Just as we represented the torus $\CC/ \Lambda$ by a fundamental parallelogram in $\CC$, we can represent the coset $SL(2,\ZZ) \backslash \cH$ by a \textit{fundamental domain} $F$ in $\cH$,  which is a connected subset of $\cH$. The fundamental domain is not unique since, for example,  if $F$ is a fundamental domain then so if $\gamma F$. The standard choice is given by the following theorem.
{\thm 
\label{thm:3.1} 
The following $F$ is a fundamental domain for $SL(2,\ZZ)$,
\bea
\label{3.fund}
F = \left \{ \tau \in \cH, ~ |\tau |  \geq 1, ~ | \Re (\tau) | \leq  \thalf \right \}
\eea
where it is understood that the boundary components at $\Re(\tau) = \pm \half $ are identified under $\tau \to \tau +1$ while the boundary components at $|\tau|=1$ for $\Re(\tau)$ positive and negative are identified with one another under $\tau \to -1/\tau$  (see Figure \ref{3.fig:2}). The area of $F$ in the Poincar\'e metric is given by ${\rm area} (F)= { \pi \over 3}$ in view of (\ref{3.area}).}
\newline

 To prove the theorem we show that for an arbitrary point $\tau \in \cH$ there exists a transformation $\gamma \in SL(2,\ZZ)$ such that $\gamma \tau \in F$. To do so, we fix $\tau$ and make use of the relation, 
\bea
\Im (\gamma \tau) = { \Im (\tau) \over |c \tau + d|^2}
\eea
to construct a $\gamma \in SL(2,\ZZ)$ with the largest possible value of $\Im(\gamma \tau)$. This is done by minimizing $|c \tau +d|$ for fixed $\tau$ as $a,b,c,d$ run over the integers satisfying $ad-bc=1$. Taking $\gamma$ to be the identity gives the bound $|c \tau +d| \leq 1$ which can be satisfied by only a finite number of integer pairs $(c,d)$. We construct $\gamma$ as a product of  translations by $T$ and inversions by $S$ using an iterative process that terminates. Starting from an arbitrary $\tau \in \cH$ which is not in $F$, apply a translation $T^\a$ for $\a \in \ZZ$ to map $\tau \to \tau'=T^\a \tau$ into the vertical strip $ \tau ' \in \{ | \Re(\tau)| \leq 1/2\}$. Under this transformation, the value of the imaginary part is unchanged  $\Im (\tau') = \Im (\tau)$.  If $\tau' \in F$, we are done. If $\tau ' \not \in F$ then it must satisfy $|\tau' |<1$. Its image under $S$ given by $\tau'' = S \tau'$  has $|\tau''|>1$ as well as $\Im (\tau'') > \Im (\tau')=\Im(\tau)$ in view of the fact that $c=1$ and $d=0$ for $S$ and that $|\tau'|<1$. If this image is in $F$ we are done, and if not, we repeat the process. Since only a finite number of pairs $(c,d)$ satisfy the bound $|c \tau +d| \leq 1$, and the iterative process strictly increases the value of $\Im(\gamma \tau)$ at each inversion $S$, the process must terminate, which completes the proof.

\subsubsection{Elliptic points}
\label{sec:3.3a}

Every point $\tau \in \cH$ is invariant under the $\ZZ_2 = \{ \pm I\}$ subgroup of $SL(2,\ZZ)$.  An elliptic point is a point in $\cH$ that is invariant under a subgroup of $SL(2,\ZZ)$ that is larger than $\ZZ_2$. We shall now show that the only elliptic points in $F$ are  $i, \rho,$ and $\rho'=\rho+1$, where $\rho^3=1$ with $\Im(\rho)>0$. Specifically, the point $i$ is invariant under the subgroup $\{ \pm I, \pm S \}$ of order four. The point $\rho$ is invariant under the subgroup $\{ \pm I, \pm (ST), \pm (ST)^2 \}$ of order six, while the point $\rho'$ is invariant under the subgroup $\{ \pm I, \pm (TS), \pm (TS)^2 \}$, also of order six. 

\sm

To prove the assertion we consider a point $\tau_\g \in \cH$ that is invariant under $\gamma \in SL(2,\ZZ)$, 
 \bea
\gamma \tau _\gamma = \tau_\gamma \hskip 1in 
\gamma = \left ( \begin{matrix} a & b \cr c & d \cr \end{matrix} \right )
\eea
and therefore satisfies the quadratic equation $c \tau_\g^2 -(a-d)\tau_\g -b=0$ subject to $\Im(\tau_\gamma) >0$. For $c=0$ there are no solutions that satisfy both conditions, while for $c \not=0$ we may assume $c>0$ without loss of generality.  The point $\tau_\g$ lies in $\cH$ provided $|a+d|<2$ in which case $\tau_\gamma$ is given by the solution with positive imaginary part,   
\bea
\tau_\g = {a-d \over 2c} + i  {\sqrt{4-(a+d)^2} \over 2c}
\eea
A transformation $\gamma \to \a \gamma \a^{-1}$ with $\a \in SL(2,\ZZ)$ maps  $\tau_\gamma \to \a \tau_\gamma$ and a suitable choice of $\a$ maps $\tau_\gamma$ to  the standard fundamental domain $F$ for $SL(2,\ZZ)$. The choice $\tau_\g \in F$ imposes the condition  $\Im(\tau_\gamma) \geq \Im (\rho)= \sqrt{3}/2$ which requires  $(a+d)^2+3c^2\leq 4$. Its solutions are  $c= 1$ and either $a+d=0$ or $a+d=\pm 1$. By a suitable translation of $\tau_\gamma$ to the fundamental domain $F$ we may set $a,d \in \{ 0,1\}$   corresponding to the following matrices,
\bea
S = \left ( \bma 0 & -1 \cr 1 & 0 \cr \ema \right )  \hskip 0.8in ST = \left ( \bma 0 & -1 \cr 1 & 1 \cr \ema \right )
\hskip 0.8in TS = \left ( \bma 1 & -1 \cr 1 & 0 \cr \ema \right )
\eea 
for which $\tau_S=i$ and $\tau_{ST} = \rho $ and $\tau_{TS} = \rho'$. Since we have $S^2= (ST)^3=-I$, their orders are 4 and 6 respectively. These are all the elliptic points of $SL(2,\ZZ)$.

\subsubsection{Cusps}
\label{sec:3.3b}

Topologically, $F$ is a sphere with one puncture at $\tau = i \infty$, referred to as the cusp of the fundamental domain of $SL(2,\ZZ)$. The cusp does not belong to $\cH$. It is invariant under the infinite Borel subgroup of translations and sign reversal,
\bea
\Gamma _\infty = \{ \pm T^n , ~ n \in \ZZ \}
\eea
Under $\gamma \in \Gamma _\infty$, the cusp in $F$ is mapped to the cusp  of the corresponding fundamental domain $\gamma F$. Under an arbitrary $\gamma \in SL(2,\ZZ)$ with $c \not=0$, the cusp is mapped to a rational number, and more precisely,
\bea
SL(2,\ZZ) \, i \infty = \QQ \cup \{ i \infty \}
\eea 
so that each rational number is a cusp of $SL(2,\ZZ)$ in the corresponding fundamental domain.

\subsection{Modular functions, modular forms, and cusp forms}

We shall now define modular functions, modular forms, and cusp forms. Let $f(\tau)$ be a meromorphic function on the upper half-plane $\cH$, which obeys the following transformation law for some integer $k \in \ZZ$, and for all $\gamma \in SL(2,\ZZ)$, 
\bea
f(\gamma \tau) = (c \tau +d)^k f(\tau) \hskip 1in \gamma = \left ( \begin{matrix} a & b \cr c & d \cr \end{matrix} \right ) 
\eea
Since $f$ is a periodic function under $\tau \to \tau+1$, it has a Fourier expansion given by,
\bea
f(\tau) = \sum _{n \in \ZZ} a_n \, q^n 
\hskip 1in q= e^{ 2 \pi i \tau}
\eea
\begin{enumerate}
\itemsep -0.02in
\item \textit{$f$ is a modular function} if $f$ is meromorphic at $i \infty$, namely has only a finite number of non-vanishing Fourier coefficients $a_n$ with negative $n$; 
\item \textit{$f$ is a modular form of weight $k$} if it is holomorphic in $\cH$ and at $i \infty$, namely its Fourier coefficients $a_n$ vanish for all negative $n$, and thus $f$ is finite at $i \infty$;
\item \textit{$f$ is a cusp form of weight $k$} if it is a modular form  of weight $k$ and vanishes at $i \infty$.
\end{enumerate}

\subsubsection{The ring of modular forms}

The space $\cM_k$ of modular forms of weight $k$ for the group $SL(2,\ZZ)$ is a vector space over~$\CC$. Since the product of two modular  forms of weight $k$ and $\ell$ is a modular form of weight $k+\ell$, the space of modular forms of arbitrary weight forms  a ring $\cM$ graded by the weight,  
\bea
\cM = \bigoplus _k \cM_k \hskip 1in 
\cM_k \cM_\ell \subset \cM_{k+\ell}
\eea
The ratio of two modular forms, in general, will not be a modular form, as dividing by a cusp form will produce an object that is not holomorphic at $\infty$. In the next few subsections, we shall show that there are no modular forms of negative weight and that $\cM$ is a polynomial ring generated by the modular forms $G_4$ and $G_6$.

\subsection{Modular forms as differential forms}
\label{3.modif}

It is useful to view modular forms as $SL(2,\ZZ)$-invariant differential forms. Using the following transformation rule for $\gamma \in SL(2,\ZZ)$ for the differential of $\tau$,
\bea
d ( \gamma \tau) = { d \tau \over (c \tau +d)^2}
\eea
we see that a modular form $f$  of weight $k$ is naturally mapped to a meromorphic differential form of 
weight $({ k \over 2},0)$, 
\bea
\mf = f(\tau) (d\tau)^{k \over 2}
\eea
which is \textit{strictly invariant under $\gamma \in SL(2,\ZZ)$}. This correspondence will be very useful in both mathematics and physics applications. The mismatch by a factor of 2 between  the modular weight $k$ of $f$ and the weight ${k \over 2}$ of the associated differential form $\mf$ is conventional.

\subsection{Eisenstein series}

Eisenstein series $G_k$ were already encountered when studying elliptic functions in (\ref{EisenG}). In this subsection, we shall show that they are modular forms of weight $k$ and obtain their Fourier series decomposition. In the next subsection we shall show that they actually generate the complete polynomial ring of all modular forms.  

\sm

We begin by considering the Eisenstein series defined in (\ref{EisenG}),\footnote{Throughout, the prime superscript on the sum instructs us to omit the term with $m=n=0$.}
\bea
\label{3c1}
G_k (\om_1, \om _2) = G_k (\Lambda) =  \sum _{\om \in \Lambda '} { 1 \over \om ^k} = \sum _{m, n \in \ZZ} ' { 1 \over (m \om _1 + n \om _2)^k}
\eea
 The sums that define $G_k$ are absolutely convergent for $k \geq 3$, and vanish for odd $k$ in view of the lattice symmetry  $\Lambda = - \Lambda$. The resulting $G_k(\om_1, \om_2)$ is invariant under $SL(2,\ZZ)$  transformations of the periods $\om_1, \om_2$ since $G_k$ depends only on the lattice~$\Lambda$ and not on the specific periods chosen to represent~$\Lambda$. The functions $G_k(\om_1, \om_2)$ are also manifestly homogeneous in the periods of degree $-k$. In summary, under an arbitrary  transformation $\gamma \in SL(2,\ZZ)$, which maps $\Lambda$ to $\Lambda$ and the periods $\om_1, \om_2$ into periods $\om_1', \om_2'$ given in terms of $\gamma$ in (\ref{3a1}),  and an arbitrary complex scaling factor $ \lambda \in \CC^*$, we have,
 \bea
\label{3c2}
G_k (\om_1', \om _2')  & = & G_k (\om_1, \om _2) 
\no \\
G_k (\lambda \om_1, \lambda \om _2)  & = & \lambda ^{-k} G_k ( \om_1, \om _2) 
\eea
Scaling allows us to define a function of $\tau = \om_2/\om_1$ on the upper half-plane $\cH$ by factoring  out a power of $\om_1$ or, equivalently, by setting $\om_1=1$,
\bea
\label{3c4}
G_k(\tau) = G_k (1, \tau) = \om _1^k \, G_k (\om_1, \om _2) 
= \sum _{m, n \in \ZZ} ' { 1 \over (m  + n \tau)^k}
\eea
Combining the lattice automorphism and the scaling transformation of  (\ref{3c2}) with the definition of $G_k$ in (\ref{3c4})  gives the modular transformation law for $G_k (\tau)$,
\bea
G_k (\gamma \tau) = \left ({ \om _1 ' \over \om _1} \right )^k G_k (\tau) = (c \tau + d )^k G_k (\tau)
\eea
This transformation may also be inferred directly from the sum in (\ref{3c4}). Thus, $G_k(\tau)$ is a modular function. To show that it is a modular form we study its Fourier decomposition.

\subsubsection{Fourier decomposition of Eisenstein series}

Since the Eisenstein series $G_k(\tau)$  is a holomorphic function for $\tau \in \cH$ as well as a periodic function of $\tau$ with period 1 it admits a Fourier expansion of the form, 
\bea
\label{3c3}
G_k (\tau) = \sum _{\nu \in \ZZ} a_\nu (k) \, q^\nu 
\hskip 1in 
q= e^{2 \pi i \tau}
\eea
To evaluate the Fourier series, we decompose the double sum of (\ref{3c4}) that defines $G_k(\tau)$ by isolating the contributions from $n=0$, 
\bea
\label{3c5}
G_k (\tau) = \sum _{m \not= 0} { 1 \over m^k } + 2 \sum _{n=1}^ \infty \, \sum _{m \in \ZZ} {1 \over (m  + n \tau)^k}
\eea
As $\tau$ approaches the cusp $\tau \to i \infty$, the double sum in (\ref{3c5}) tends to zero. 
As a result, all expansion coefficients $a_\nu(k)$ vanish  for $\nu<0$ and the first sum in (\ref{3c5}) gives 
 $a_0(k)$,
\bea
a_0(k) = 2 \zeta (k) \hskip 1in 4 \leq k \in 2\NN
\eea
Thus, $G_k(\tau)$ has a finite limit at the cusp and is a modular form of weight $k$.
To evaluate the double sum in (\ref{3c5}), we use the following formula, 
\bea
\sum _{m \in \ZZ} { 1 \over z +m} = - i \pi { 1 + e^{2 \pi i z} \over 1 - e^{2 \pi i  z}}
= - i \pi - 2 \pi i \sum _{\ell=1} ^\infty e^{2 \pi i \ell z}
\eea
which may be established by using the fact that both sides are meromorphic in $z$ with simple poles of residue 1  at all integers and are  periodic with period 1. Actually, we shall need the derivative of order $k-1$ of this formula,
\bea
\label{lif}
\sum _{m \in \ZZ} { 1 \over (z +m)^k } 
= { ( 2 \pi i )^k \over \Gamma (k)} \sum _{\ell=1} ^\infty \ell ^{k-1} \, e^{2 \pi i  \ell z}
\eea
Setting $z=n \tau$ and including the sum over $n$, we obtain the expression for the second sum in (\ref{3c5}). Putting all together gives the following Fourier series decomposition for $G_k(\tau)$, 
\bea
G_k(\tau) = 2 \zeta (k) + 2  { ( 2 \pi i )^k \over \Gamma (k)} \, \sum _{n=1}^\infty \, \sum _{\ell=1} ^\infty   \ell ^{k-1} q^{n\ell}
\eea
Changing summation variables from $(n,\ell) $ to $(N,\ell)$ with  $N=n\ell$, the sum over $\ell$ for given $N$ may be rearranged in terms of the \textit{sums of divisors} functions, defined for any $\a \in \CC$ by, 
\bea
\sigma _\alpha (N) = \sum _{\ell | N} \ell^\alpha
\eea  
The sum is over all positive divisors $\ell $ of $N$, including the divisors $1$ and $N$. In terms of sums of divisor functions, we obtain our final expression for the Fourier series of $G_k(\tau)$, 
\bea
G_k (\tau) = 2 \zeta (k) +  
 2 { ( 2 \pi i )^k \over \Gamma (k)} \, \sum _{N=1}^\infty \sigma _{k-1} (N)  q^N
\eea
Factoring out $2 \zeta(k)$, and using the relation between Bernoulli numbers and $\zeta (k)$ for $k$ even 
$(2 \pi i )^k  B_k = - 2 k! \zeta (k) $ we obtain  the \textit{normalized Eisenstein series} $\HE_k$,  defined by,
\bea
\label{Ektau}
\HE_k(\tau) =  { G_k (\tau) \over 2 \zeta (k)} = 1 + \nu_k \sum _{N=1}^\infty \sigma _{k-1} (N)  \, q^N 
\hskip 0.7in
\nu_k = - { 2 k \over B_k}
\eea
With this normalization, we manifestly have $\HE_k(\tau) \to 1$ as $\tau \to i \infty$ for any even $k \geq 4$.
The first few values for $\nu_k$ are given as follows,
\bea
\nu_4=240 \hskip 0.35in \nu_6=-504 \hskip 0.35in \nu_8= 480  \hskip 0.35in \nu_{10} = -264 \hskip 0.35in \nu_{12} = { 65520 \over 691}
\eea
In particular, the modular forms $g_2, g_3, G_4, G_6$, and the discriminant $\Delta$ of (\ref{Disc}) for the lattice $\Lambda= \ZZ \oplus \tau \ZZ$ are given in terms of the normalized Eisenstein series by, 
\bea
g_2(\tau)  =  ~ \, 60 \, G_4(\tau)  &=& { 4 \pi^4 \over 3} \HE_4(\tau)
\no \\
g_3 (\tau)  = 140 \, G_6(\tau)  &=& { 8 \pi^6 \over 27} \HE_6(\tau)
 \no \\
\Delta (\tau)  =  g_2^3 - 27 g_3^2 &=& {(2 \pi)^{12} \over 12^3} \big ( \HE_4(\tau)^3 - \HE_6(\tau)^2 \big )
\eea
Since $\HE_k(i \infty) = 1$  the discriminant vanishes at the cusp and we have $\Delta(\tau) = (2 \pi)^{12} q + \cO(q^2)$. While the functions $G_k$ are modular forms of weight $k$ for even $k \geq 4$, the discriminant $\Delta$ is  a cusp form of weight 12.

\subsubsection{Poincar\'e series representation}
\label{PoincareEk}

One construction of elliptic functions for a lattice $\Lambda$  is through the method of images, exhibited in all generality in (\ref{ell.imag}) and  shown in detail for the case of the Weierstrass function in (\ref{wpdef}). One may view the construction by the method of images as performing a sum over  the Abelian group $\Lambda$ under which every elliptic function is invariant. 

\sm

To obtain modular functions and modular forms, which are defined to be invariant or covariant under the group $SL(2,\ZZ)$, an analogous method is available through the so-called \textit{Poincar\'e series}. This method may be applied to obtain meromorphic modular forms, as will be the case here, but also non-holomorphic modular functions and forms, as will be the case in subsection \ref{sec:4.2} and section \ref{sec:MGF}. To construct a modular form by a Poincar\'e series one starts with a meromorphic function $g$ or a meromorphic differential form $\mg$ of weight $k$ and aims to sum over its images under the modular group. In many practical cases, however, the function $g$ or the form $\mg$ is invariant under a ``stability subgroup" $\Gamma _{{\rm stab}}$  of $SL(2,\ZZ)$. When this subgroup has an  infinite number of elements, as will often be the case, the sum over all images under $SL(2,\ZZ)$ would diverge. The Poincar\'e series is instead defined by summing over the images under the cosets $\Gamma_{{\rm stab}} \backslash SL(2,\ZZ)$. For the case of a meromorphic function $g$ we obtain a meromorphic modular function $f$ given by the Poincar\'e series,
\bea
f(\tau) = \sum _{\gamma\, \in\, \Gamma_{{\rm stab}} \backslash SL(2,\ZZ)} g(\gamma \tau)
\eea
The function $g$ is usually referred to as the \textit{seed of the Poincar\'e series}.  The corresponding formula defines the Poincar\'e series for a modular  form $\mf$ from a seed $\mg$. As noted originally by Poincar\'e, the seed function for a given modular function is not unique. In particular, the Poincar\'e sum of certain functions may vanish. 

\sm

By way of example, we may take the seed to be simply $\mg = (d\tau)^k$ for some positive integer~$k$. This seed is invariant under $-I \in SL(2,\ZZ)$ because $\tau$ is invariant, as well as under the following Borel subgroup of $SL(2,\ZZ)$ since $d\tau$ is invariant under shifts $\tau \to \tau + b$,
\bea
\label{Borel}
\Gamma _\infty = \left \{ {\pm} \left ( \bma 1 & b \cr 0 & 1 \ema \right ), \, b \in \ZZ \right \}
\eea 
so that  $\Gamma _{{\rm stab}} = \Gamma _\infty$. 
The remaining Poincar\'e series for even $k \geq 4$ is given by,
\bea
\mf = \sum _{\gamma \in \Gamma_\infty \backslash SL(2,\ZZ)} { (d\tau)^k \over (c \tau + d)^k}
= \half \sum_{{c,d \in \ZZ,  \atop \gcd(c,d)=1}} { (dz)^k \over (c \tau+d)^k}
\eea
The restriction $\gcd(c,d)=1$ implicitly assumes that $(c,d) \not= (0,0)$. The form $\mf$  is closely related to the Eisenstein series defined in (\ref{3c4}), as may be seen by factoring out the greatest common divisor $p=\gcd(m,n) \in \NN$ from the summation variables by parametrizing them as follows $(m,n)=(pd, pc)$ with $\gcd(c,d)=1$, 
\bea
G_k(\tau) = \sum '_{m,n \in \ZZ} { 1 \over (m + n \tau)^k}
=\zeta(k)  \sum_{{c,d \in \ZZ,  \atop \gcd(c,d)=1}} { 1 \over (c \tau+d)^k}
\eea
 so that $\mf = 2 \zeta(k) G_k(\tau) (dz)^k$ for $k>2$ even.

\subsubsection{The polynomial ring of Eisenstein series}

The Eisenstein series $G_{2k}$ is a modular form of weight $2k$ for any integer $k \geq 2$. We shall now show that the Eisenstein series $G_{2k}$ may be expressed as a  polynomial in $G_4$ and $G_6$ with rational coefficients, so that the vector space generated by the Eisenstein series $G_{2k}$ is  a  polynomial ring generated by $G_4$ and $G_6$.

\sm

To show this, we differentiate the differential equation for the Weierstrass $\wp$-function $(\wp ')^2= 4 \wp^3 - 60 G_4 \wp -140 G_6$ to obtain the following second order differential equation,
\bea
\label{3.wpder}
\wp '' = 6 \wp ^2 - 30 G_4
\eea
which is  even in $z$.  The Laurent expansion for $\wp$, obtained earlier in (\ref{2.wpex}) for an arbitrary lattice $\Lambda$, may be expressed as follows for the lattice $\Lambda = \ZZ \oplus \tau \ZZ$, 
\bea
\wp (z) = { 1 \over z^2} +\sum _{k=1}^ \infty (2k+1) G_{2k+2} \, z^{2k}
\eea
Substituting this expression and its double derivative into (\ref{3.wpder}), we see that the terms in $z^{-4}$, $z^0$, and $z^2$ automatically match. For the remaining terms, after some simplifications, we obtain the following recursion relation for $k \geq 4$, 
\bea
\label{moditer}
G_{2k}
= \sum _{\ell=1}^{k-3} {3(2\ell+1)(2k-2\ell-3) \over (k-3) (2k-1)(2k+1)} G_{2\ell+2}  G_{2k-2\ell-2}
\eea
which shows recursively that $G_{2k}$ for $k \geq 4$  is a polynomial   in $G_{2m+4}$ with $0 \leq m \leq k-4$, of homogeneous  weight $2k$. Therefore, $G_{2k}$ for $k \geq 4$ is a polynomial in $G_4$ and $G_6$.  For example, to the lowest few orders we have,
\begin{align}
\label{Glow}
G_8 & = \tfrac{ 3}{7 } \, G_4^2
&
G_{14} & = \tfrac{30}{143} \, G_4^2 \, G_6
\no \\
G_{10} & = \tfrac{ 5}{11} \, G_4 \, G_6
& 
G_{16} & = \tfrac{ 9}{221} \, G_4^4 + \tfrac{ 300}{2431} \, G_4 \, G_6^2
\no \\
G_{12} &= \tfrac{ 18}{143} \, G_4^3 + \tfrac{ 25}{143} \, G_6^2
&
G_{18} & = \tfrac{3915}{46189} \, G_4^3 \, G_6 + \tfrac{125}{4199} \, G_6^3
\end{align}
Thus, the subspace of the graded ring of all modular forms $\cM$ that is generated by the Eisenstein series $G_{2k}$ for all $k \geq 4$ is a polynomial ring generated by $G_4$ and $G_6$. While the modular forms $G_4$, $G_6$, $G_8$, $G_{10}$, and $G_{14}$ are the only Eisenstein series of weight $4,6,8, 10$, and $14$ respectively, we see that at weight 12 there are two independent modular forms, namely $G_4^3$ and $G_6^2$, whose difference involves  the discriminant $\Delta$. 

\sm

Equivalently, we may express $G_{2k}$ in terms of the normalized Eisenstein series $\HE_{2k}$, to obtain the following formula, 
\bea
\label{Emoditer}
\HE_{2k} 
= \sum _{\ell=1}^{k-3} {6 (2\ell+1)(2k-2\ell-3) \zeta(2\ell+2) \zeta (2k-2\ell-2) \over (k-3) (2k-1)(2k+1) \zeta(2k)} 
\, \HE_{2\ell+2} \, \HE_{2k-2\ell-2} 
\eea
in terms of which the relations (\ref{Glow}) become, 
\begin{align}
\HE_8 & = \HE_4^2 & \HE_{14} & = \HE_4^2\HE_6
\no \\
\HE_{10} & = \HE_4 \HE_6 & \HE_{16}& = \tfrac{1617}{3617} \, \HE_4^4 +\tfrac{2000}{3617} \, \HE_4 \HE_6^2
\no \\
\HE_{12} & = \tfrac{441}{691} \, \HE_4^3 +\tfrac{250}{691} \, \HE_6^2
& \HE_{18} & = \tfrac{38367}{43867} \, \HE_4^3 \HE_6 +\tfrac{5500}{43867} \, \HE_6^3
\end{align}
These relations, together with the Fourier series expansions of the $\HE_k$ in (\ref{Ektau}), imply an infinite number of identities between sums of divisor functions. The identities corresponding to the relations  $\HE_8 = \HE_4^2$ and $\HE_{10} = \HE_4 \HE_6$ are given as follows,
\bea
\label{3.sigma79}
\sigma _7 (N) & = & \sigma _3(N) + 120 \sum _{M=1}^{N-1} \sigma _3(N-M) \sigma _3 (M)
\no \\
11 \, \sigma _9(N) & = & 21 \sigma _5(N) - 10 \sigma _3(N) + 5040 \sum _{M=1}^{N-1} \sigma _3(N-M) \sigma _5(M)
\eea
More generally, the divisor functions $\sigma_{2k-1}$ may be expressed as polynomials in $\sigma _3$ and $\sigma _5$.

\subsection{Dimension and generators of the ring of modular forms}
 
To obtain the dimension of the space $\cM_k$ of all modular forms of weight $k$, we begin by proving the following theorem on modular functions of weight $k$. We shall denote by $\ord_f(p)$ the order of a zero (counted positively) or pole (counted negatively) of $f$ at a point $p$, and by $\ord_f(\infty )$ the exponent of the leading term in the $q$-expansion of $f(\tau)$. Meromorphicity of modular functions guarantees that $\ord_f(p)$  are integers, including for the cusp $p=i\infty$ and for the elliptic points $p= i$ and $p= \rho= e^{2 \pi i /3}$.

{\thm
\label{thm:3.1a}
 A non-zero modular function $f$ of weight $k$ satisfies the \textbf{valence formula},
 \bea
 \label{thm1a}
 \ord_f (i \infty) + \half \ord_f (i) + { 1 \over 3} \ord_f (\rho) + \sum _{p \in F\setminus \{ i \infty, i, \rho \}} \ord_f (p) = { k \over 12}
 \eea
where $F$ is the standard fundamental domain of the group $SL(2,\ZZ)$ given in (\ref{3.fund}).}
 \newline
 
\begin{figure}[tp]
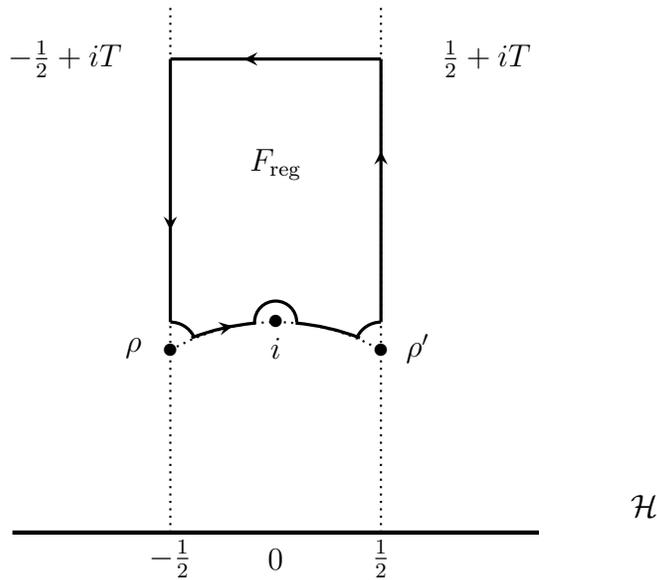

\begin{center}
\tikzpicture[scale=0.7]
\scope[xshift=-5cm,yshift=0cm]
\draw[ultra thick] (-5,0) -- (5,0);
\draw[thick, dotted] (-2,0) -- (-2,3.464);
\draw[thick, dotted] (2,0) -- (2,3.464);
\draw[thick, dotted] (-2, 3.464) -- (-2, 10);
\draw[thick, dotted] (2 , 3.464) -- (2, 10);
\draw[very thick, directed] (-2, 9) -- (-2, 4);
\draw[very thick, directed] (2 , 4) -- (2, 9);
\draw [thick, dotted] (-2,3.464) arc (120:60:4 and 4);
\draw [very thick, directed] (-1.6,3.7) arc (111:95:4.5 and 4.5);
\draw [very thick] (1.6,3.7) arc (69:85:4.5 and 4.5);
\draw [very thick, directed] (2,9) -- (-2,9);
\draw [very thick] (-2,4) arc (90:27:0.5 and 0.5);
\draw [very thick] (2,4) arc (90:153:0.5 and 0.5);
\draw [very thick] (-0.4,4) arc (180:0:0.4 and 0.4);
\draw (0,-0.5) node{$0$};
\draw (-2,-0.5) node{$-\half $};
\draw (2,-0.5) node{$\half $};
\draw (0,7) node{$F_{\rm reg}$};
\draw (-4,9) node{$-\half+iT$};
\draw (4,9) node{$\half+iT$};
\draw (7,0.5) node{$\cH$};
\draw (0,3.5) node{$i$};
\draw (0,4) node{$\bullet$};
\draw (2,3.464) node{$\bullet$};
\draw (-2,3.464) node{$\bullet$};
\draw (2.7,3.5) node{$\rho'$};
\draw (-2.7,3.5) node{$\rho$};
\endscope
\endtikzpicture
\end{center}
\caption{\textit{Integration contour of the regularized fundamental domain $F_{\rm reg}$.} \label{3.fig:3}}
\end{figure}

To prove the theorem, we count the number of zeros and poles of $f$ by integrating the logarithmic derivative $f'/f$ over a suitable contour. Doing so presents complications at the  points  $i \infty, i,$ and $\rho$, so we regularize the domain $F$ to a domain $F_{\rm reg}$ in which we cut off $F$ near those points, as shown in figure \ref{3.fig:3}. The small circular arcs near $i, \rho$, and $\rho'$ have radius $1/T $.  For simplicity, we shall assume that $f(\tau)$ has neither zeros nor poles on $\p F_{\rm reg}$, and refer to standard mathematics textbooks when this is not the case.   

\sm

On the one hand,  the residue theorem gives, 
 \bea
 \lim _{T \to \infty} { 1 \over 2 \pi i } \oint _{\p F_{\rm reg}} d \tau \, { f'(\tau) \over f(\tau)} = \sum _{p \in F_{{\rm reg}} } \ord_f(p)
 \eea
On the other hand, the integral along $\p F_{{\rm reg}}$ may be evaluated by  using the relations under modular transformations that identify the  different line segments of $\p F_{{\rm reg}}$ with one another. First, the contributions from the two vertical edges cancel one another, as their integrals are opposite. Second, we evaluate the contributions at the point $i \infty$ by using the dominant behavior at $\Im (\tau)=T \gg 1$ given by $f(\tau) \sim q^{\ord_f (i \infty )} =  e^{2 \pi i \tau \, \ord_f(i \infty)}$ to obtain the contribution $- \ord_f (i \infty)$. The minus sign arises from the orientation of the contour. Analogous contributions are collected at $i$, but are counted with a factor of $\half$ since the integration contour is only half a circle. Similarly, at $\rho$, the contribution is only ${1 \over 6}$ of a full circle, but there are identical contributions from $\rho$ and $\rho'$ which accounts for the final factor of ${1 \over 3}$. 
The above contributions account for the left side of (\ref{thm1a}). Finally, we collect the contribution along the $|\tau|=1$ segments of the contour. The contributions to the left and to the right of $i$ are related to one another by the transformation $S: \tau \to -1/\tau$, which acts on $f(\tau)$ by $f(S\tau) = \tau^k f(\tau)$, so that taking the differential of the logarithm we have, 
\bea
d(S \tau) { f'(S \tau) \over f(S\tau) }= d \tau { f'(\tau) \over f(\tau)} + k { d\tau \over \tau}
\eea
The integration arc  goes from $\tau=\rho $ to $\tau=i$ and spans ${1 \over 12}$ of a full circle, reproducing the right side of the (\ref{thm1a}) and thereby completing the proof of the theorem.

\subsubsection{Dimension and generators of $\cM_k$}

While Theorem \ref{thm:3.1a} applies to any modular function of weight $k$, whether holomorphic or not, we shall now specialize to modular forms, which are holomorphic, and use Theorem \ref{thm:3.1a} to compute the dimensions of the spaces $\cM_k$.  Modular forms $f(\tau)$ can have no poles, so that $\ord_f(p) \geq 0$, including at the points $p=i \infty, i, \rho$. 
\sm

{\thm  
\label{theorem2}
Theorem \ref{thm:3.1a} implies the following properties of $\cM_k$:
\begin{enumerate}
\itemsep=-0.03in
\item For $k$ odd,  $k<0$, and $k=2$ there are no modular forms, i.e. $\dim \cM_k=0$;
\item For $k$ even and $k \geq 0$ the space $\cM_k$ is finite-dimensional;
\item For $k=0,4,6,8,10,14$, we have $\dim \cM_k=1$ and $\cM_k = \CC G_k$;
\item The subspace $\cS_k$ of weight $k$ cusp forms is empty for $k < 12$, while $\cS_{12}= \CC \Delta$;
\item For $k \geq 16$, we have $\cS_k = \Delta \cM_{k-12}$ and $\cM_k = \CC G_k \oplus \cS_k$.
\end{enumerate}}
For low weight these results are consistent with the results of (\ref{Glow}). Moreover, the theorem implies that $\cM$ is precisely the polynomial ring freely generated by $G_4$ and $G_6$.

\sm

To prove Theorem \ref{theorem2}, we  use the valence formula (\ref{thm1a}) of Theorem \ref{thm:3.1a}. Considering the relation (\ref{thm1a}) multiplied by 12, and then mod 2, we see that there can  be no solutions for $k$ odd. 
Since modular forms have $\ord_f(p)  \geq 0$, including at the points $p=i, \rho,$ and $i \infty$, the left side of (\ref{thm1a}) is non-negative so that there can be no solutions to (\ref{thm1a}) for $k<0$. Clearly, there is also no solution when $k=2$, which concludes the proof of point 1. For $k$ even and positive, there can only be a finite number of solutions since the sum on the left of (\ref{thm1a}) is over positive or zero numbers only, proving point 2. To proceed further, it is helpful first to prove two auxiliary statements,
\bea
\label{3.G4G6}
G_4(\rho) = G_6 (i)=0
\eea
They follows from inspection of the sums, 
\bea
G_4 (\rho) & = & 
 \sum _{(m,n)}'  { 1 \over ( m \, \rho^3 + n \, \rho )^4} = { 1 \over \rho^4} \sum _{(m,n)}'  { 1 \over ( m\, \rho^2 +  n)^4}
 = \rho ^2 G_4(\rho)
 \no \\
 G_6 (i) & = & 
 \sum _{(m,n)}'  { 1 \over ( m \, i^4 + n \, i )^6} = { 1 \over i^6 } \sum _{(m,n)}'  { 1 \over ( m \, i^3 +  n)^6}
 = - G_6(i)
\eea
To recast the second sum on the first line, we use $m\rho^2+n = m(-1-\rho)+n = n-m - m \rho $ so that the sum over $(m,n)$ gives $G_4(\rho)$. To recast the second sum on the second line,  we use $m \, i^3 +n = -i m +n$, so that the sum over $(m,n)$ gives $G_6(i)$.

\sm

By inspection of (\ref{thm1a}), we must have $\ord_f (i \infty)=0$ for $k=4,6, 8, 10$ since the right side of (\ref{thm1a})  is less than 1, as well as for $k=14$ since the remaining $1/6$ cannot be accounted for by non-vanishing $\ord_f(i)$ or $\ord_f(\rho)$. Therefore, the  solutions to (\ref{thm1a}) for these values of $k$  are unique and respectively given by $(\ord_f(i), \ord_f(\rho))= (0,1), \, (1,0), \, (0,2), \, (1,1), \, (1,2)$. They are generated by $G_4$, $G_6$, $G_8\sim G_4^2$, $G_{10} \sim G_4 G_6$, $G_{14} \sim G_4^2G_6$, which proves point 3. To prove point 4, we note that a cusp form has $\ord _f(i \infty) \geq 1$, which requires $k \geq 12$, and uniqueness for $k=12$. Finally for $k\geq 16$, the solutions manifestly satisfy the recursion relation $\cS_k = \Delta \cM_{k-12}$ and $\cM_k = \CC G_k \oplus \cS_k$ by inspecting (\ref{thm1a}), proving point  5.

\subsection{Modular functions and the $j$-function}
\label{sec:jfunction}

Having classified all modular forms under $SL(2,\ZZ)$ of weight $k$, we now classify all meromorphic modular functions of weight $k$. Given an arbitrary meromorphic modular function of weight $k$, we may always divide by a modular form of weight $k$, such as $G_k$, to obtain a meromorphic modular function of weight 0. Thus, it will suffice to understand the space of modular functions of weight~0. Sums, products, and ratios of modular functions of weight~0 produce again modular functions of weight 0, so that  the corresponding space forms the field of modular functions of weight 0. We seek to obtain a complete description of this space. 

\sm

Clearly, the only holomorphic modular function of weight 0  is constant in $\tau$. 
We begin by constructing a modular function with a single pole at the cusp and no other poles. Normalizing the residue of the pole to unity, we define the famous $j$-function by,
\bea
\label{eq:jfuncdef}
j(\tau) =  {12^3 g_2^3 \over \Delta} = { 12 ^3 \, \HE_4^3 \over \HE_4^3-\HE_6^2} = { 1 \over q} + \cO(q^0)
\eea
Inspecting (\ref{thm1a}), we have $k=0$ since we are dealing with a modular function of weight zero; $\ord _\infty (j)=-1$ since $j$ has a simple  pole at $\infty$; $\ord _i(j)=0$ since $\HE_4(i)\not=0$; and $\ord _\rho(j)=3$ since $j$ has a triple zero at $\rho$ in view of the identity $G_4(\rho)=\HE_4(\rho)=0$ established in (\ref{3.G4G6}). These values indeed satisfy  (\ref{thm1a}), and we have,\footnote{The first 19 orders in the $q$-expansion of $j$ are as follows: {\small $j(\tau) = q^{-1} +744+196884\, q+21493760 \, q^2+864299970 \, q^3+20245856256 \, q^4+333202640600 \, q^5+4252023300096 \, q^6+44656994071935\, q^7+401490886656000\, q^8+3176440229784420\, q^9+22567393309593600 \, q^{10}+146211911499519294 \, q^{11}+874313719685775360 \, q^{12}+4872010111798142520 \, q^{13}+25497827389410525184 \, q^{14}+126142916465781843075 \, q^{15}+593121772421445058560 \, q^{16}+2662842413150775245160 \, q^{17} + \cO(q^{18})$.}}
\bea
j(i \infty) = \infty \hskip 1in j(i)= 1 \hskip 1in j(\rho)=0
\eea
The classification of modular functions is given as follows. 

\sm

{\thm ~ Every rational function of $j(\tau)$ is a modular function of $SL(2,\ZZ)$. Conversely, every modular function $f(\tau)$ of $SL(2,\ZZ)$ of weight 0  is a rational function of~$j$. In terms of the zeros $z_n$ and poles $p_n$ (both of which are to be counted with multiplicities, including at $i$ and $\rho$), we have the explicit formula,
\bea
f(\tau) = \prod _{n=1}^N { j(\tau) - j(z_n) \over j(\tau) - j(p_n)}
\eea}
This result completes the construction of the field of modular functions of weight 0 and, combining it with our earlier general construction of modular forms of arbitrary weight, also gives the construction for all meromorphic $SL(2,\ZZ)$-covariant functions of arbitrary weight.

\subsubsection{Modular forms with prescribed zeros}

A final comment is in order about modular forms with prescribed zeros. From (\ref{thm1a}) we see that a zero at a regular point in $F$ requires $k \geq 12$, and the addition of one regular zero requires shifting $k \to k+12$. To construct, for example, the modular forms of weight 12 with one zero at an arbitrary point $p \in F\setminus \{ i \infty, i, \rho\}$, we take a linear combination of the two modular forms available, namely $\HE_4^3$ and $\HE_6^2$, and arrange the coefficients so that a zero emerges at $p$,
\bea
f_p(\tau) = f_0 \Big ( \HE_4 (\tau)^3 \HE_6(p)^2 - \HE_4(p)^3 \HE_6(\tau)^2  \Big )
\eea
The normalization factor $f_0$ may be fixed by evaluation at any other point, including $i \infty$. One proceeds analogously for modular forms with more zeros at prescribed points.

\subsection{Modular transformations of Jacobi $\tet$-functions}

The Jacobi $\tet$-functions are subject to more delicate modular transformation laws, which we shall now derive. Recall  the definition of the basic $\tet$-function,
\bea
\tet (z|\tau) = \sum _{ n \in \ZZ} e^{ i \pi \tau n^2 + 2 \pi i n z}
\eea
Setting $z=0$ and $\tau = i t$ for $t $ real and positive, we may apply the Poisson summation formula (\ref{poisson}) to obtain the transformation rule for $S \in SL(2,\ZZ)$, 
\bea
\tet (0|-\tfrac{1}{\tau}) = (-i \tau)^{  \half}\, \tet  ( 0 | \tau   ) 
\eea
Although originally derived for $t$ real, the formula  remains valid  for $\tau \in \cH$ provided we choose the proper branch cut for the square root. Under the modular transformation $T$, however, the $\tet$-function does not transform into itself, but instead we have,  
\bea
\label{tetplusone}
\tet (z|\tau+1 ) 
 = \tet _4(z|\tau)
\eea
where $\tet_4$ is a $\tet$-function with half-integer characteristics defined in (\ref{tet1234}). The set of four $\tet$-functions with half characteristics transforms into itself, and we have,
\begin{align}
\label{tet-trans}
\tet _1(z|\tau+1) & = \ep \, \tet _1(z|\tau) 
& 
\tet_1(\tfrac{z}{\tau}|-\tfrac{1}{\tau}) & = - i (-i \tau)^\half e^{i \pi z^2/\tau} \tet_1 (z|\tau)
\no \\
\tet _2(z|\tau+1) & = \ep \, \tet _2(z|\tau) 
& 
\tet_2(\tfrac{z}{\tau}|-\tfrac{1}{\tau}) & =  (-i \tau)^\half e^{i \pi z^2/\tau} \tet_4 (z|\tau)
\no \\
\tet _3(z|\tau+1) & =  \tet _4(z|\tau) 
& 
\tet_3(\tfrac{z}{\tau}|-\tfrac{1}{\tau}) & = (-i \tau)^\half e^{i \pi z^2/\tau} \tet_3 (z|\tau)
\no \\
\tet _4(z|\tau+1) & =  \tet _3(z|\tau) 
& 
\tet_4(\tfrac{z}{\tau}|-\tfrac{1}{\tau}) & = (-i \tau)^\half e^{i \pi z^2/\tau} \tet_2 (z|\tau)
\end{align}
where $\ep=e^{2 \pi i /8}$. One may collect these transformation laws in a single expression for each transformation, valid for all half-characteristics,  
\bea
\label{4.mod2}
\tet \left [ \begin{matrix} \alpha \cr \beta \cr \end{matrix} \right ] (z |\tau+1)
& = & e^{- i \pi \alpha (1+\alpha)} \, \tet \left [ \begin{matrix} \alpha \cr \beta+ \alpha + \half  \cr \end{matrix} \right ] (z |\tau)
\no \\
\tet \left [ \begin{matrix} \alpha \cr \beta \cr \end{matrix} \right ] \left ( \tfrac{ z}{\tau}  \Big | - \tfrac{ 1}{\tau} \right )
& = & (-i \tau)^\half \, e^{2 \pi i \alpha \beta + i \pi z^2/\tau} \, 
\tet \left [ \begin{matrix} \beta \cr -\alpha \cr \end{matrix} \right ] (z |\tau)
\eea
Combining the two, one finds that under a generic element $\gamma  \in SL(2, \ZZ)$, 
the $\tet_1$ function transforms as follows, 
\bea
\label{3.tetmod}
\tet _1 (z' |  \tau') 
= \ep_\tet(\gamma)  (c\tau+d)^\half \,  e^{\pi i c z^2/(c \tau+d)} \, \tet _1 (z|\tau)
\eea
from which the transformation of the $\tet$-function with arbitrary characteristic may be deduced using (\ref{2.tetchar}), and we find,
\bea
\label{eq:Jactetmodtransf}
\tet \left [ \begin{matrix} \alpha + \half \cr \beta + \half \cr \end{matrix} \right ] \left (  z'  | \gamma \tau \right ) 
=
\ep(\a,\b\,\a',\b',\g)  (c \tau + d)^\half 
e^{\pi i c z^2/(c \tau + d)} \,\,\tet \left [ \begin{matrix}  \a' + \half \cr  \b' + \half \cr \end{matrix} \right ] (z | \tau)
\eea
where,
\bea
\ep(\a,\b\,\a',\b',\g) = \ep_\tet (\gamma) \, e^{\pi i \left \{ \a(\b+1)-\a'(\b'+1)  \right \} }
\eea
and  $z$ and the characteristics transform as follows, 
\bea
z' = { z \over c \tau +d} 
\hskip 0.7in
\left [ \bma \a' \cr \b' \ema \right ]  =\left [ \bma  a \a + c \b \cr b \a + d \b \ema \right ] 
\hskip 0.7in
\gamma = \left ( \bma a & b \cr c & d \ema \right )
\eea
Finally, $\ep_\tet(\gamma)$ and $\ep(\a,\b\,\a',\b',\g)$ are both eight roots of unity and $\ep_\tet(\gamma)$ is independent of the characteristics $\a, \b$. Its evaluation is in terms of the generalized Legendre symbol and will be given in subsection \ref{sec:Dedekindeta} on the Dedekind $\eta$ function.

\sm

Setting $z=0$ in the theta functions gives so-called \textit{theta-constants}, whose modular transformation laws are special cases of the general transformation laws presented above. None of these are modular forms under $SL(2,\ZZ)$, though we shall see later that they are modular forms under certain congruence subgroups of $SL(2,\ZZ)$. Furthermore, the derivative $\theta'_1(0|\tau)$ is closely related to the discriminant $\Delta$ and various sums of eight powers of $\tet_2,\tet_3$, and $\tet_4$ are modular forms. We now discuss both of these constructions in turn.

\subsubsection{The discriminant $\Delta$: $\tet$-constants and product formula}
\label{sec:Discriminant}

The modular transformation law for $\tet_1'(0|\tau)$ is readily obtained from (\ref{tet-trans}) and is given below along with its eighth power,
\bea
\tet _1' (0|\tau+1)  = \ep \, \tet _1' (0|\tau) 
& \hskip .8in &
\tet_1'(0|-\tfrac{1}{\tau})  =  (-i \tau)^{{ 3 \over 2}} \, \tet_1' (0|\tau)
\no \\
\tet _1' (0|\tau+1)^8  = \tet _1' (0|\tau)^8 
& &
\tet_1'(0|-\tfrac{1}{\tau})^8  =  (-i \tau)^{12} \, \tet_1' (0|\tau)^8
\eea
While $\tet'_1$ itself is not a modular form, its eight power is a modular form of weight 12, and therefore must be a linear combination of $\HE_4^3$ and $\HE_6^2$. To obtain a precise relation, we appeal to the infinite product formula for $\tet_1(z|\tau)$ given in (\ref{tetprod}), and find the following expression,
\bea
\label{3.tet1}
\tet'_1(0|\tau) = \pi \tet_2(0|\tau) \tet_3(0|\tau) \tet_4(0|\tau)  
=
2 \pi q^{1 \over 8} \prod _{n=1}^\infty (1-q^n)^3 
\eea
This expression  indeed fails to be a modular form in view of the $q^{1 \over 8}$ prefactor. However,  its eight power produces a factor of $q$ which is not only consistent with being a genuine modular form, but also shows that it is a cusp form proportional to the discriminant $\Delta$. The constant of proportionality is easily fixed by matching the term of order $q$ in the Fourier expansion of both. As a result, we find not only a relation between $\theta_1'(0|\tau)^8$ and the discriminant $\Delta$ but also a product formula for $\Delta$, 
\bea
\Delta (\tau) = (2 \pi)^4 \tet_1'(0|\tau)^8 = (2 \pi)^{12} q \prod _{n=1}^\infty (1-q^n)^{24}
\eea
The product formula for the discriminant shows that $\Delta(\tau)$ is nowhere vanishing for $\tau \in \cH$, and has a simple zero in $q$ at the cusp.

\sm

The Fourier coefficients $\tau(n)$ of $\Delta$, defined by,  
\bea
\label{3.tau}
\Delta (\tau) = (2 \pi)^{12} \sum _{n=1}^\infty \tau (n) q^n
\eea
are referred to as the Ramanujan $\tau$-function, and have special arithmetic significance. 
The lowest orders are given as follows,
\begin{align}
\tau(1) & =1 & \tau(4) & = -1472 & \tau(7) & = -16744 & \tau(10) & = -115920
\no \\
\tau(2) & = -24 & \tau(5) & = 4830 & \tau(8) & = 84480 & \tau(11) & = 534612
\no \\
\tau(3) & = 252 & \tau(6) & = -6048 & \tau(9) & = -113643 & \tau(12) & = -370944
\end{align}
The relation $1728 \Delta= (2 \pi)^{12}(E_4^3-E_6^2)$ along with $E_4^2=E_8$ allows us to express the $\tau$-function in terms of sums of divisor functions as follows,
\bea
\tau(n) & = & \sum_{m=1}^{n-1} \left ( \tfrac{200}{3} \, \sigma_3(m) \sigma _7(n-m) - 147 \, \sigma _5(m) \sigma _5(N-m) \right )
\no \\ && \quad 
+  \tfrac{5}{36} \, \sigma _3(n) + \tfrac{ 7}{12} \, \sigma _5(n) + \tfrac{5}{18} \, \sigma_ 7(n)
\eea 
Eliminating $\sigma_7$ in terms of $\sigma_3$ using the first formula in (\ref{3.sigma79}) confirms that $\tau(n)$ may be expressed solely in terms of $\sigma_3$ and $\sigma_5$.  Ramanujan conjectured that $\tau(n)$ is a \textit{multiplicative arithmetic function}, which satisfies,
\bea
\label{taumult}
\tau (mn) = \tau(m) \tau(n) \hskip 1in \gcd(m,n)=1
\eea
where $\gcd(m,n)$ stands for the greatest common divisor of $m$ and $n$. 
Ramanujan's function is, however, not \textit{completely multiplicative} (which would mean that the first equation in (\ref{taumult}) would hold for arbitrary $m,n$, not necessarily relatively prime)  and instead satisfies,
\bea
\label{taumult2}
 \tau(m) \tau(n) = \sum _{d \,|\, {\gcd} (m,n)} d^{11} \tau \left ( { mn \over d^2} \right )
\eea
Both conjectures on $\tau(n)$ were proven by Mordell in 1917 using the theory of Hecke operators applied to the cusp form $\Delta (\tau)$, and we shall provide this proof in section \ref{sec:Hecke}.

\subsubsection{Sums of eight powers}

Another way to construct a modular form using the $\tet$-constants is by sums of eighth powers. The modular transformation properties show right away that the following sums, 
\bea
T_{4k} (\tau) = \tet_2(0|\tau)^{8k} + \tet_2(0|\tau)^{8k} + \tet_2(0|\tau)^{8k} 
\eea
are modular forms of weight $4k$. For $k=1,2$, the corresponding spaces of modular forms are one-dimensional, so that we must have $T_4 = 2 \HE_4$ and $T_8 = 2  \HE_8 = 2 \HE_4^2$ by matching asymptotics at the cusp. In particular, we may express the modular $j$-function in terms of eighth powers of $\tet$-constants, 
\bea
j(\tau) = 1728 { g_2(\tau)^3 \over \Delta(\tau)} 
= 32 \, { \big (\tet _2(0|\tau)^8 + \tet _3(0|\tau)^8 + \tet _4(0|\tau)^8 \big )^3 \over 
\tet _2(0|\tau)^8 \, \tet _3(0|\tau)^8 \, \tet _4(0|\tau)^8 }
\eea
It may be verified directly from the asymptotics of the $\tet$-constants at the cusp that $j(q)$ has a simple pole at the cusp with unit residue $j(q) \sim q^{-1}$.

\subsection{The Dedekind $\eta$-function}
\label{sec:Dedekindeta}

The Dedekind $\eta$-function is defined as follows,
\bea
\eta (\tau) = q^{1 \over 24} \prod _{n=1}^\infty (1-q^n)
\eea
It is not a modular form of $SL(2,\ZZ)$, but its 24th power is a modular form of weight 12 and related to the discriminant  as follows, 
\bea
\label{eq:deletarel}
\Delta (\tau) = (2 \pi )^{12} \eta (\tau)^{24}
\eea
Under the generators $T$ and $S$ of the modular group, $\eta$ transforms as follows,
\bea
\label{3.eta1}
\eta (\tau+1) & = & e^{2\pi i /24} \, \eta (\tau)
\no \\
\eta (-\tfrac{1}{\tau}) & = & (-i \tau)^\half  \, \eta (\tau)
\eea
The first relation may be read off from the definition of $\eta$, while the second relation may be derived first for $\tau \in i \RR^+$ using Poisson resummation, and then be extended to $\tau \in \cH$.\footnote{The square root is defined via the argument function $\arg(\tau)$ for $\tau \in \CC^* $, in terms of which  $\tau = |\tau| \exp ( i \arg(\tau))$ and $-\pi \leq \arg(\tau) < \pi$ by $\tau^\half = |\tau|^\half \exp ( i \arg(\tau)/2)$ (see for example \cite{Kohler}). This definition is compatible with the fact that for $\tau = i t$ with $t \in \RR^+$ we have $\eta(it), \eta(i/t)>0$. }

\sm

The transformation of $\eta(\tau)$ under an arbitrary $\gamma \in SL(2,\ZZ)$ may be expressed as follows, 
\bea
\label{3.etamod}
\eta (\gamma \tau) = \ep_\eta (\gamma)  \, (c \tau +d) ^\half \eta (\tau)
\hskip 1in
\gamma = \left ( \bma a & b \cr c & d \ema \right )
\eea
with $\ep_\eta (\gamma)^{24}=1$, or $\ep : SL(2,\ZZ) \to \ZZ_{24}$. The function  $\ep_\eta (\gamma)$ depends on $\gamma$, but not on $\tau$. Its values may be determined from the transformation rules given in (\ref{3.eta1}). Note that the actions of the transformations $\gamma$ and $-\gamma$ coincide on $\tau$ but differ on the square root and on $\ep_\eta $ giving, in particular, the value $\ep_\eta (-I) = (-1)^\half$. In view of this property, $\ep_\eta (\gamma)$ is not a proper homomorphism of $SL(2,\ZZ) \to \ZZ_{24}$ but instead is referred to as a \textit{multiplier system}.  Special values may be deduced from the basic transformation laws (\ref{3.eta1}),
\bea
\ep_\eta (T^n) = e^{ 2 \pi i n /24} 
\hskip 0.6in
\ep_\eta (S) = e^{-2 \pi i /8}
\hskip 0.6in
 \ep_\eta (-I) =  e^{2 \pi i /4} =i
\eea
In particular, if $\gamma$ is such that either $c >0$, or $c=0$ and $d>0$, then $\ep (-\gamma) = i \ep (\gamma)$. 
The Petersson formula for $\ep_\eta (\gamma)$ for a $\gamma$ with either $c> 0$, or $c=0$ and $d>0$ is given as follows, (see \cite{Kohler})
\bea
c \hbox{ odd: } ~ & \hskip 0.2in & \ep_\eta (\gamma) 
= (d|c) \exp \left \{ \tfrac{2 \pi i }{24} \big (  (a+d)c+bd(1-c^2) -3 c \big ) \right \} 
 \\
c \hbox{ even: } & \hskip 0.2in &  \ep_\eta (\gamma) 
= {\rm sign}(d) (c|d)  \exp \left \{ \tfrac{2 \pi i }{24} \big ( (a+d)c+bd(1-c^2) +3 d +3 -3cd \big ) \right \} 
\no \eea
We recall from appendix \ref{sec:modN}, and in particular (\ref{A.jacobi}), that the Jacobi symbol $(n|N)$ for odd denominator $N$ (which is always the case here) with decomposition into distinct primes $p_i$ given by $N=\prod _i p_i ^{\a_i}$ is defined by $(n|N) = \prod_i (n|p_i)$ where $(n|p_i)$ is the Legendre symbol for an odd prime $p_i$. The definition is extended to include $n=0$ by $(0|1)=(0|-1)=1$ and negative $N$ by $(n|N)=(n| -N)$. 

\sm

Knowledge of $\ep_\eta(\gamma)$ allows us to produce a formula for the multiplier $\ep _\tet(\gamma)$  encountered in the modular transformation law of the $\tet$-functions in (\ref{eq:Jactetmodtransf}) by using the relation, 
\bea
\label{eq:tetpetarel}
\tet_1'(0|\tau) = 2 \pi \eta (\tau)^3
\eea
which is obtained by combining (\ref{3.tet1}) with the definition of the $\eta$-functions. The modular transformation of the left side may be derived from (\ref{3.tetmod}) while the modular transformation of the right side is given by (\ref{3.etamod}), and we obtain,
\bea
\ep _\tet (\gamma) = \ep_\eta (\gamma)^3
\eea
consistent with the fact that $\ep_\tet$ is an 8-th root of unity while $\ep_\eta$ is a 24-th root of unity.  

\sm

Finally, we note that the fourth power  $\ep_\eta^4(\gamma)$ no longer suffers from the subtleties of sign reversal of $\gamma$ and the square root, and is a genuine homomorphism $\ep_\eta^4 : SL(2,\ZZ) \to \ZZ_6$ with $\ep_\eta^4(-I)=1$ and $\ep_\eta^4(\gamma_1 \gamma _2) = \ep_\eta^4(\gamma_1) \ep_\eta^4(\gamma _2)$.

\newpage

\subsection*{$\bullet$ Bibliographical notes}

The books by Apostol \cite{Apostol} and Lang \cite{Lang8} and the lecture notes by Zagier \cite{Zagier123} offer classic introductions to the subject of modular forms.   The books by Shimura \cite{Shimura} and Koblitz \cite{Koblitz} provide advanced and more abstract presentations aimed at arithmetic.  A marvelously clear and detailed presentation, to which we shall refer often in these notes, is given in the fairly recent treatise by Diamond and Shurman \cite{DS}. A general perspective from the point of view of automorphic functions is given in Bateman \cite{Bateman3} and Iwaniec \cite{Iwan1}. A detailed exposition of identities involving the Dedekind $\eta$-function and the role they play in the theory of modular forms may be found in the book by K\"ohler \cite{Kohler}. Presentations with an emphasis on $L$-functions and $q$-series expansions may be found in the books by Bump \cite{Bump} and Ono \cite{Ono}, respectively.

\newpage

\section{Variants of modular forms}
\setcounter{equation}{0}
\label{sec:VMF}

In this section, we discuss a variety of objects that are closely related to modular forms, including quasi-modular forms, almost-holomorphic modular forms, non-holomorphic Eisenstein series, Maass forms, mock modular forms, and quantum modular forms. Other generalizations, including modular forms for congruence subgroups, vector-valued modular forms, and modular graph forms, will be introduced in sections \ref{sec:Forms}, \ref{sec:MDEsvvmfs}, and \ref{sec:MGF} respectively.

\subsection{Quasi-modular and almost-holomorphic modular forms}
\label{E2section}

Quasi-modular forms and almost-holomorphic modular forms are generalizations of the modular forms introduced earlier, which are obtained by relaxing certain parts of the definition of modular forms. We begin by reviewing the most famous of these objects, referred to as $\HE_2(\tau)$ and $\HE^*_2(\tau)$ respectively, before proceeding  to the definition and presentation of the general properties of quasi-modular forms and almost-holomorphic modular forms.

\subsubsection{$\HE_2$ and $\HE_2^*$}

We introduced the holomorphic Eisenstein series $G_k$ in (\ref{EisenG}) and (\ref{3c1}), or  equivalently $\HE_k$ in (\ref{Ektau}), which are given for even $k \geq 4$ in terms of absolutely convergent sums over a lattice $\Lambda = \ZZ \oplus \tau \ZZ$. For $k=2$, the corresponding  sum,
\bea
 \sum _{\om \in \Lambda '} { 1 \over \om^2} = \sum _{{m,n \in \ZZ \atop (m,n) \not= (0,0)}} { 1 \over (m + n \tau)^2}
\eea 
fails to be absolutely convergent. Instead, it is conditionally convergent and its value will depend on how the infinite sums are arranged. Physicists would  say that the sums need to be  ``regularized". There may be one or several choices that are deemed ``natural" because they preserve one property or another of modular forms. 

\sm

One choice of regularization is to adopt the \textit{Eisenstein summation}  which consists in cutting off the infinite sums in a symmetrical way, as was already used in (\ref{2c2}),
\bea
\label{G2Eis}
\HE_2 (\tau) = \lim_{N\to \infty} \, \lim _{M \to \infty} \, { 1 \over 2 \zeta(2)} \sum _{n = -N  } ^N \sum_{{m=-M \atop (m,n) \not= (0,0)}} ^M  { 1 \over (m + n \tau)^2} 
\eea
For finite $M$ and $N$ the sum is holomorphic in $\tau$ and the limit $M,N \to \infty$ produces a finite holomorphic  function $\HE_2(\tau)$. However, $\HE_2(\tau)$ fails to transform as a modular form under  $SL(2,\ZZ)$ transformations. This is not surprising as Theorem \ref{theorem2} tells us there are no modular forms of weight 2. Instead the transformation law of $\HE_2(\tau)$  is as follows,
\bea
\label{E2transfs}
\HE_2  ( \gamma \tau) = (c \tau +d)^2 \HE_2(\tau) + { 12 \over 2 \pi i} \, c (c \tau +d) 
\eea
To show this, we decompose the sum of (\ref{G2Eis}) into contributions from $n=0$ and $n \not=0$, 
\bea
\label{G2Eist}
\HE_2 (\tau) = 1 + \lim _{N \to \infty} \lim _{M \to \infty} { 1 \over  \zeta(2)} \sum _{n =1  } ^N \sum_{m=-M } ^M  { 1 \over (m + n \tau)^2} 
\eea
and perform the sum over $m$ using (\ref{lif}) in the convergent limit $M \to \infty$. The remaining sum over $n$ admits a finite limit as $N \to \infty$ and evaluates as follows, 
\bea
\label{3.E2sig}
\HE_2 (\tau) =  1 - 24 \sum _{n=1}^\infty \sum _{\ell=1}^ \infty \ell \, e^{2 \pi i \ell n \tau } 
=1 - 24 \sum _{\ell=1}^\infty \sigma _1(\ell) q^ \ell
\eea
Alternatively, the sum over $n$ in (\ref{3.E2sig}) may be performed first and the remaining sum over $\ell$ may be  rearranged in terms of the discriminant $\Delta (\tau)$, 
\bea
\label{3.E2Del}
\HE_2(\tau)  = { 1 \over 2 \pi i} \, \p_\tau \ln \Delta(\tau)
\eea
Using the modular transformation law of the weight 12 modular form $\Delta$ under $SL(2,\ZZ)$,
we deduce the transformation law of $\HE_2(\tau)$ by differentiating $\ln (\Delta)$ to obtain (\ref{E2transfs}).
Remarkable relations to the Weierstrass $\zeta$ and $\wp$ functions include the following,
\bea
\label{4.WPE}
\zeta(\thalf | \tau) = { \pi^2 \over 6} \, \HE_2(\tau)
\hskip 1in 
\wp(z|\tau) = - \p_z^2 \ln \tet _1 (z|\tau) -{\pi^2 \over 3 } \, \HE_2(\tau)
\eea
The latter demonstrates how $\HE_2(\tau)$ acts as a modular connection to guarantee the proper modular transformation of $\wp$ given that of the $\tet_1$-function.

\sm

Another choice of regularization was introduced by Siegel by defining the following sum,
\bea
\label{3.Estar}
\HE_2^* (\tau) = \lim _{\ep \to 0^+} { 1 \over 2 \zeta (2)} \sum _{\om \in \Lambda '} { 1 \over \om^2 \, |\om|^\ep} 
\eea
Finite $\ep>0$ renders the sum over $\Lambda'$ absolutely convergent and the limit $\ep \to 0$ turns out to be finite. 
The result is a weight 2 modular form under $SL(2,\ZZ)$, 
\bea
\HE_2^*(\gamma \tau)  = (c \tau +d)^2 \HE^*_2(\tau)
\eea
However, $\HE_2^*(\tau)$ fails to be holomorphic in $\tau$. The regularization factor $|\om|^\ep$ introduces non-holomorphic behavior for $\ep >0$ which leaves a non-holomorphic remnant as $\ep \to 0$,\footnote{Throughout, we shall use the notation $\tau = \tau_1 + i \tau_2$ with $\tau_1, \tau_2 \in \RR$.}
\bea
\label{eq:2E2rel}
\HE_2^*(\tau) = \HE_2 (\tau) - { 3 \over \pi \tau_2}
\eea
where the holomorphic $\HE_2(\tau)$ was given in (\ref{3.E2Del}). One verifies that the modular transformation laws of $\HE_2$ and $\HE_2^*$ map into one another under this relation.

\sm

In summary, while the Eisenstein series $\HE_k$ for $k \geq 4$ are holomorphic modular forms of weight $k$, the regularization of the infinite sum for $k=2$ allows for two ``natural" choices: one  is holomorphic but not modular, while the other  is modular but not holomorphic. One cannot enforce  both modularity and holomorphicity simultaneously.  The incompatibility between holomorphicity and modular invariance may be viewed as perhaps the first example  of what physicists refer to as an \textit{anomaly}. Eisenstein's result dates back to before  1850 !

\subsubsection{The rings of almost-holomorphic modular  and quasi-modular forms}
\label{sec:quasialmostmod}
A function $f: \cH \to \CC$ is an \textit{almost-holomorphic form of weight $k$} with respect to the full modular group $SL(2,\ZZ)$  if it transforms as follows under $\gamma \in SL(2,\ZZ)$, 
\bea
f(\gamma \tau) = (c \tau +d)^k f(\tau)
\eea
and is a polynomial in $\tau_2^{-1} $ whose coefficients $f_n(q)$ are holomorphic functions of $q=e^{ 2 \pi i \tau}$,
\bea
f(\tau) = \sum _{n=0}^N f_n(q) \tau_2^{-n}
\eea 
By this definition, every modular form of weight $k$ is an almost-holomorphic modular form of weight $k$, and $\HE_2^*(\tau)$ is an almost-holomorphic modular form of weight  2. Clearly, any linear combination of almost-holomorphic modular forms of weight $k$ is an almost-holomorphic modular form of weight $k$, and the product of two almost-holomorphic modular forms of weights $k$ and $\ell$ is an almost-holomorphic modular form of weight $k+\ell$, so that the space of almost-holomorphic forms a ring graded by the weight.

\sm

A function $g: \cH \to \CC$ is a \textit{quasi-modular form of weight $k$} if it is the holomorphic part $f_0(q)$ of some almost-holomorphic form $f$ of weight $k$, as defined above. By this definition, every modular form of weight $k$ is a quasi-modular form of weight $k$, and  $\HE_2(\tau)$ is a quasi-modular form of weight  2. The space of quasi-modular forms is a ring graded by the weight. More precisely, we have the following  theorem, presented here without proof.

{\thm 
\label{thm:almost}
The space of almost-holomorphic modular forms is a polynomial ring, graded by the weight $k$, and generated by $\HE_2^*, \HE_4$, and $\HE_6$.  The space of quasi-modular forms is a polynomial ring, graded by the weight $k$, and generated by $\HE_2, \HE_4$, and $\HE_6$. }

\subsubsection{Differential equations for quasi-modular forms}
\label{sec:introtodiffeq}

The inhomogeneous nature of the transformation law of $\HE_2(\tau)$ under $SL(2,\ZZ)$ suggests that $\HE_2(\tau)$  may be viewed geometrically as a connection in the space of modular forms $\cM$. The connection $E_2$ produces a covariant derivative $D_k$ acting on $\cM_k$, 
\bea
D_k : \cM_k \to \cM_{k+2} 
\hskip 1in
D_k = { 1 \over 2 \pi i} { d \over d\tau} - { k \over 12} \HE_2(\tau)
\eea
To show this, we use the fact that a modular form $f$ of weight $k$, namely $f \in \cM_k$, and its $\tau$-derivative $f'$ satisfy the following transformation rules,
\bea
f(\gamma \tau) & = & (c\tau+d)^k f(\tau) 
\no \\
f'(\gamma \tau) & = & (c \tau+d)^{k+2} f'(\tau) + k c (c\tau+d)^{k+1} f(\tau)
\eea
Combining (\ref{E2transfs}) with the above relation, and the definition of $D_k$, we obtain,
\bea
D_kf(\gamma \tau) = (c \tau +d)^{k+2} D_k f(\tau)
\eea
so that $D_k f$ is a modular form of weight $k+2$, as announced. The Leibnitz formula for the product of two modular forms $f_1$ and $f_2$  of respective weights $k_1$ and $k_2$ is given as follows,
\bea
D_{k_1+k_2} \left ( f_1 f_2 \right ) = \left ( D_{k_1} f_1 \right ) f_2 + f_1 \left ( D_{k_2} f_2 \right )
\eea
When the space $\cM_{k+2}$ is one-dimensional, one easily derives a differential equation for modular forms of weight $k$.  The precise normalizations may be deduced from the asymptotic behavior at the cusp. For example, we have, 
\begin{align}
D_4 \, \HE_4 & = - \tfrac{1}{3} \HE_6 & D_8 \, \HE_8 & = - \tfrac{2}{3} \HE_{10} & D_{12} \, \Delta & =0
\no \\
D_6 \, \HE_6 & = - \tfrac{1}{2} \HE_8 && & D_{12} \, \HE_{12}& = - \HE_{14}
\end{align}
Equivalent expressions may be obtained by using $\HE_8=\HE_4^2$, $\HE_{10}=\HE_4 \HE_6$, and $\HE_{14}=\HE_4^2 \HE_6$ together with Leibnitz's rule. The derivative of $\HE_{10}$ is, however, more involved since the space $\cM_{12}$ is two-dimensional.

\sm

Given that $\HE_2$ may naturally be interpreted as a \textit{connection} in $\cM_k$ there should be a natural notion of \textit{curvature}. The associated curvature should transform tensorially, namely it should be a modular form, and should be obtained from the derivative and the square of the connection. Given the transformation law (\ref{E2transfs}) for $\HE_2$, we obtain the transformations for its derivative with respect to $\tau$ and its square,
\bea
\HE_2' ( \tau'  ) & = & (c \tau +d)^4 \HE_2'(\tau) + 2 c (c \tau+d)^3 \HE_2 (\tau) + \tfrac{ 12 c^2}{2 \pi i}  (c\tau+d)^2 
\no \\
\HE_2  ( \tau' )^2 & = & (c \tau +d)^4 \HE_2(\tau)^2 + \tfrac{ 24 }{ 2 \pi i} \, c (c \tau +d) ^3 \HE_2 (\tau) 
+ \left ( \tfrac{ 12 c}{2 \pi i}\right )^2 (c\tau+d)^2
\eea
As a result, the following combination transforms as a modular form of weight 4,
\bea
{ 1 \over 2 \pi i} \HE_2' ( \tau'  ) -{1 \over 12} \HE_2(\tau')^2 = 
(c \tau +d)^4 \left (  { 1 \over 2 \pi i} \HE_2' ( \tau ) -{1 \over 12} \HE_2(\tau)^2\right ) 
\eea
Since this combination is both holomorphic in $\tau$ and a modular form of weight 4, it must belong to $\cM_4$. But by Theorem \ref{theorem2}, the space $\cM_4$ is dimension 1 and is generated by $\HE_4$. Matching the value at $q=0$, we obtain the relation, 
\bea
{ 1 \over 2 \pi i} { d \over d \tau} \HE_2 ( \tau ) -{1 \over 12} \HE_2(\tau)^2 = - { 1 \over 12} \HE_4(\tau)
\eea
It is natural to interpret $\HE_4$ as the curvature, given indeed by a linear combination of the derivative and the square of $\HE_2$.
We note that the coefficient multiplying the square is half of what one would have expected if the derivative had been applied to a modular form of weight 2. This is as usual when obtaining the curvature from the connection.

\subsection{Non-holomorphic Eisenstein series}
\label{sec:4.2}

We next introduce non-holomorphic Eisenstein series, which may be viewed as generalizing the Riemann $\zeta$-function, modular forms, and almost-holomorphic modular forms. Given a lattice $\Lambda = \ZZ + \tau \ZZ$, a non-holomorphic Eisenstein series may be defined by the following lattice sum,\footnote{The notation $E_s(\tau)$ used for the non-holomorphic Eisenstein series should not be confused with the notation $\HE_s(\tau)$ used for the holomorphic Eisenstein series.} or so-called \textit{Kronecker-Eisenstein sum},
\bea
\label{3.nhE}
E_s(\tau) = \sum _{\om \in \Lambda '} { \tau_2^s \over \pi^s |\om|^{2s}} = \sum _{m,n \in \ZZ}' { \tau_2^s \over \pi^s |m + n \tau|^{2s}}
\eea
where $\tau_2=\Im(\tau)$ and the prime on the sum indicates that the contribution $m=n=0$ is to be omitted from the sum. The sum is absolutely convergent for $\Re(s)>1$. The lattice sum definition of $E_s(\tau)$ may be used to show that $E_s$ is modular invariant,
\bea
E_s(\gamma \tau) = E_s(\tau) \hskip 1in \gamma = \left ( \bma a & b \cr c & d \ema \right ) \in SL(2,\ZZ)
\eea
It may also be used to factor the non-trivial common divisors of $m$ and $n$ in  the double sum,
\bea
E_s(\tau) = \zeta(2s) \sum_{{c,d \in \ZZ \atop \gcd(c,d)=1}}' { \tau_2^s \over \pi^s |c \tau +d|^{2s}}
\eea
where $\zeta(2s)$ is the Riemann $\zeta$-function and the remaining sum is over pairs $(c,d) \in \ZZ^2$ that have no common divisors strictly greater than 1. This expression in turn allows us to write $E_s(\tau)$ as a Poincar\'e series over the coset $\Gamma_\infty \backslash SL(2,\ZZ)$ as follows,
\bea
\label{eq:nonholoPoincare}
E_s(\tau) = { \zeta (2s) \over \pi^s} \sum _{\gamma \in \Gamma _\infty \backslash SL(2,\ZZ)} \left ( \Im \gamma \tau \right )^s
\eea
This relation implies that $E_s(\tau)$ vanishes for all $\tau$ whenever $2s$ is a zero of the Riemann $\zeta$-function, including at its zeros on the negative real axis, as well as at its non-trivial zeros when $\Re(4s)=1$. 
Finally, the non-holomorphic Eisenstein series $E_s(\tau)$ is an eigenfunction of  the Laplace equation, 
\bea
\label{eq:EisLapl}
\Delta E_s(\tau) = s(s-1) E_s(\tau) 
\hskip 1in
\Delta = 4  \tau_2^2 \p_\tau \p _{\bar \tau}=  \tau_2^2 ( \p_{\tau_1}^2 + \p_{\tau_2}^2)
\eea
as may be verified by applying the Laplace operator term-wise in the sum of (\ref{3.nhE}).

\subsubsection{Analytic continuation of $E_s(\tau)$}

While $E_s(\tau) $ is defined by the series (\ref{3.nhE}) for $\Re(s)>1$, it may be analytically continued in~$s$ throughout the complex $s$-plane to produce a meromorphic function, just as was possible for the Riemann $\zeta$-function. To exhibit the analytic structure of $E_s(\tau) $ in $s$, we use the following integral representation for the summands, 
\bea
\label{4.d6}
\Gamma (s) E_s(\tau)  = \int _0 ^\infty { dt \over t} \, t^s \left ( \sum _{m,n\in \ZZ} e^{ - \pi t |m +n\tau|^2 /\tau_2} -1 \right )
\eea
The subtraction  of $1$ eliminates the zero mode. The non-analyticity in $s$ arises from the small $t$ regime. As we did in the case of the Riemann $\zeta$-function, we split the integral at $t=1$ and Poisson re-sum the integrand for $0 < t < 1$ in both $m$ and $n$ to obtain, 
\bea
\label{4.d6a}
\Gamma (s) E_s(\tau)  & = & 
\int _1 ^\infty { dt \over t} \, t^s \left ( \sum _{m,n\in \ZZ} e^{ - \pi t |m +n\tau|^2 /\tau_2} -1 \right )
\no \\ &&
+ \int _0 ^1 { dt \over t} \, t^s \left ( {1 \over t} \sum _{m,n\in \ZZ} e^{ - \pi   |m +n\tau|^2/(t\tau_2)} -1 \right )
\eea 
Next, we change variable in the second integral $t \to 1/t$, collect the resulting integrals for $t \in [1,\infty [$, 
and analytically continue a trivial remaining integral. The result is as follows, 
\bea
\label{4.d7}
\Gamma (s)  E_s(\tau) =  - {1 \over s(1-s) }+ 
 \int _1 ^\infty { dt \over t} \, \Big ( t^s+ t^{1-s} \Big )  \left ( \sum _{m,n\in \ZZ} e^{ - \pi t |m +n\tau |^2 /\tau_2} -1 \right )
\eea
The first term on the right side is meromorphic in $s \in \CC$, while the integral is manifestly holomorphic throughout  $\CC$. As a result, $E_s(\tau) $ admits a meromorphic analytic continuation to $s \in \CC$ with a simple pole at $s=1$ with residue 1. Moreover, the expression is manifestly invariant under $s \to 1-s$ which implies  the functional equation, 
\bea
\label{4.d8}
\Gamma (s) E_s(\tau) = \Gamma (1-s) E_{1-s}(\tau) 
\eea
In particular, the functional relation requires $E_s(\tau)=0$ whenever $s=-1,-2, \cdots$ in view of the fact that the right side is finite for these values and that $\Gamma (s)$ on the left side  is infinite. This observation reproduces our earlier observation that $E_s(\tau)=0$ whenever $2s$ is a zero of the Riemann $\zeta$-function.

\subsubsection{Fourier series of $E_s(\tau)$}

The Fourier series of $E_s(\tau)$ for all $\Re(s)>1$ may be derived from the integral representation in (\ref{4.d6}), 
this time by Poisson re-summation  in the variable $m$ only, using the formula, 
\bea
\sum _{m,n \in \ZZ}  e^{- \pi t |m+n \tau|^2/\tau_2} 
= \sqrt{ { \tau _2 \over t}} \sum _{m,n \in \ZZ} e^{2 \pi i m n \tau_1} \, e^{ - \pi m^2\tau_2/t - \pi n^2 t \tau_2}
\eea
In the mathematics literature this procedure is referred to as the Chowla-Selberg method.
Substituting this result into (\ref{4.d6}) and  splitting the sum over $m,n$ into contributions for $m n \not=0 $ and $mn=0$, we have 
\bea
\Gamma (s) E_s(\tau)=  A_s(\tau)+B_s(\tau)
\eea
 where,
\bea
\label{4.d9}
A_s(\tau) & = & \sqrt{\tau_2} \sum _{m \not=0} \sum _{n \not=0}  e^{2 \pi i mn \tau_1} \int _0^\infty { dt \over t} \, 
t^{s-\half}  \, e^{- \pi \tau_2 (m^2/t + n^2 t)}
\no \\
B_s(\tau) & = &   \int _0^\infty { dt \over t} \, 
 t^s \left ( -1 + 2 \sum _{n =1}^\infty   \sqrt{{\tau_2 \over t}} \,  e^{ - \pi n^2 t \tau_2}
 +  \sum _{m \in \ZZ}  \sqrt{{\tau_2 \over t}} \,  e^{ - \pi m^2  \tau_2/ t } \right )
 \eea
 The integral $A_s(\tau)$ is manifestly holomorphic in $s$ throughout $\CC$, and may be computed by performing the $t$-integral in terms of a $K$-Bessel function, 
\bea
\label{4.d10}
A_s(\tau) = 2 \sqrt{\tau_2} \sum _{m,n \not=0} \left | { m \over n} \right |^{s-\half} e^{2 \pi i mn \tau_1} K_{s-\half} ( 2 \pi \tau_2  |mn| )
\eea
The integral $B_s(\tau)$ may be computed by  Poisson re-summing the $m$-sum in the integrand,
\bea
B_s(\tau) =  \sum _{n\not = 0} \int _0^\infty { dt \over t} \, 
 t^s \left (  \sqrt{{\tau_2 \over t}} \,  e^{ - \pi n^2 t \tau_2}
 +  e^{ - \pi n^2  t/ \tau_2 } \right )
 \eea
 which is readily evaluated in terms of the Riemann $\zeta$-function or associated $\xi$-function, 
 \bea
 \label{4.d11}
 B_s(\tau) & = & 2 \Gamma (s) \pi^{-s} \tau_2^s \zeta (2s) + 2 \Gamma (s-\thalf ) \pi^{\half -s} \tau_2^{1-s} \zeta (2s-1)
 \no \\
 & = & 
 \tau_2^s \xi (2s) + \tau_2^{1-s} \xi (2s-1)
 \eea
 where $\xi(s)$ is the function computed in (\ref{xi}). We may now verify the functional relation (\ref{4.d8}) directly in terms of the transformation properties of $A_s(\tau)$ and $B_s(\tau)$. For $A_s(\tau)$ this is manifest upon using the symmetry under interchange of $m$ and $n$, while for $B_s(\tau)$ it is a consequence of  the relation $\xi(1-s)=\xi(s)$.   Assembling both contributions, and recasting $A_s(\tau)$ in terms of a summation over a single variable $N=mn$, we obtain the celebrated Fourier series decomposition for  $ E_s(\tau)$, 
 \bea
 \label{4.d12}
\Gamma (s) E_s(\tau) & = & 
 2 \Gamma (s) \pi^{-s} \, \zeta (2s) \, \tau_2^s  + 2 \Gamma (s-\thalf ) \pi^{\half -s}  \, \zeta (2s-1) \, \tau_2^{1-s}
\no \\ &&
+ 4 \sqrt{\tau_2} \sum _{N \not=0} |N| ^{\half-s} \sigma _{2s-1}(|N|) \, e^{2 \pi i N \tau_1} K_{s-\half} ( 2 \pi \tau_2  |N| )
\eea
where $\sigma _\alpha (n)$ is defined to be the sum  $\sigma _\alpha (n) = \sum _{d |n} d^\alpha$ of positive divisors $d$ of $n$.  One verifies that each Fourier mode, including the power-behaved terms on the first line, individually satisfy the Laplace eigenvalue equation, as required by the fact that $\Delta$ commutes with the translation operator $\p_{\tau_1}$. 

\sm

For integer $s=k$ the Bessel function in the Fourier series becomes a spherical Bessel function $K_{k-\half}(x)$, which is a combination of an exponential in $x$ and a polynomial in $1/x$. As a result, the  Fourier series of $E_k(\tau)$ simplifies and may be recast as follows, 
\bea
\label{Eas}
E_k(\tau) &=& (- )^{k+1} { B_{2k} \over (2k)!} \, y^k 
+ { 4 \, (2k-3)! \, \zeta (2k-1) \over (k-2)! \, (k-1)! \, y^{k-1}}
\no \\ && \quad
+{2\over (k-1)! }\sum_{N=1}^\infty N^{k-1}\sigma_{1-2k}(N)P_k(Ny) \left(q^N + \bar q^N \right) 
\eea
where $q=e^{2 \pi i \tau}$ and $y = 4 \pi \tau_2$. Furthermore,  $B_{2k}$ are the Bernoulli numbers  and $P_k(x)$ is a polynomial in $1/x$ given by, 
\bea
P_k(x) = \sum_{m=0}^{k-1}{(k+m-1)! \over m! \, (k-m-1)! \, x^m}
\eea 
Non-holomorphic Eisenstein series with integer $s$ will play a crucial role in the theory of modular graph functions and forms, to be developed in section \ref{sec:MGF}.

\subsection{Maass forms}
\label{sec:4.3}

The non-holomorphic Eisenstein series discussed in the previous subsection are special examples of a more general class of non-holomorphic objects known as Maass forms. A  non-holomorphic function $f(\tau)$ is called a Maass form if it is $SL(2,\ZZ)$-invariant, satisfies a Laplace eigenvalue equation, 
\bea
\Delta f(\tau) = \lambda f(\tau) \hspace{0.8 in} \Delta =4 \tau_2^2 \p_\tau \p_{\bar \tau} = \tau_2^2 (\p_{\tau_1}^2 + \p_{\tau_2}^2)
\eea
for $\lambda \in \CC$, and has at most polynomial growth at the cusps. Without loss of generality, we will write $\lambda = s(s-1)$ and denote the space of such Maass forms by $\cN(s)$, where $\cN(s) = \cN(1-s)$. From (\ref{eq:EisLapl}) we see that the non-holomorphic Eisenstein series $E_s(\tau)$ are elements of $\cN(s)$. Indeed, below we will see that for generic $s\in \CC $ the non-holomorphic Eisenstein series are the \textit{only} elements of $\cN(s)$. 

\sm

The Fourier expansions of generic Maass forms admit surprisingly simple closed form expressions. In general, non-holomorphic modular forms admit Fourier expansions in separate powers of $q$ and $\bar q$, of the type,
\bea
f(\tau) = \sum_{M,N \in \ZZ} \widetilde{c}_{M,N}(\tau_2) q^M \bar q^N
\eea
with $\widetilde{c}_{M,N}(\tau_2)$ being functions of $\tau_2$ but not of $\tau_1$, since the latter would violate invariance under $T: \tau \rightarrow \tau + 1$. It is often useful to absorb all $\tau_2$ dependence, including that of $q$ and $\bar q$, into the Fourier coefficients, giving an alternative form for the Fourier expansion as,
\bea
 f(\tau)= \sum_{N\in \ZZ} c_N(\tau_2) \,e^{2 \pi i N \tau_1}
\eea
We will now see that, for Maass forms, the coefficients $c_N(\tau_2)$ can be fixed uniquely, up to overall constants. Indeed, inverting the Fourier expansion gives, 
\bea
c_N(\tau_2) = \int_0^1 d \tau_1 \,e^{-2 \pi i N \tau_1 } \,f(\tau)
\eea
and we may make use of the eigenvalue equation to obtain, 
\bea
c_N(\tau_2) 
& = & {1\over s(s-1) } \int_0^1 d\tau_1\, e^{-2 \pi i N \tau_1 } \,\Delta f(\tau)
\no \\
& = & {\tau_2^2 \over s(s-1)} \Big [ (2 \pi i N)^2 c_N(\tau_2) + \p_{\tau_2}^2 c_N(\tau_2) \Big ]
\eea
This gives a second order differential equation for the coefficients $c_N(\tau_2)$, 
\bea
\tau_2^2 \p_{\tau_2}^2 c_N(\tau_2) = \Big ( s(s-1) +(2 \pi N \tau_2)^2 \Big ) c_N(\tau_2)
\eea
For $N=0$ the general solution is of the form, 
\bea
c_0(\tau_2) = a \,\tau_2^s+ b \,\tau_2^{1-s}
\eea
For $N\neq 0$ we obtain a modified Bessel equation with general solutions $I_{s-1/2}$ and $K_{s-1/2}$. Discarding the solutions $I_{s-1/2}$ since they grow exponentially at the cusp, we obtain, 
\bea
c_N(\tau_2) = \alpha_N \,\sqrt{\tau_2} \,K_{s - 1/2}\left(2 \pi \tau_2|N|\right) \hspace{0.5in} N \neq 0
\eea
We conclude that a general element of $\cN(s)$ has a Fourier expansion of the form, 
\bea
\label{eq:Maassfourier}
f(\tau) = a \,\tau_2^{s}+ b\, \tau_2^{1-s} + \sum_{N \neq 0} \alpha_N \,\sqrt{\tau_2} \,e^{2 \pi i N \tau_1 }K_{s-1/2}\left(2 \pi \tau_2 |N|\right)
\eea
Note that the Fourier expansions of the non-holomorphic Eisenstein series were obtained already in (\ref{4.d12}), and match with the above result upon substituting, 
\bea
a = {2 \zeta(2s) \over \pi^{s}}
\hspace{0.4 in} 
b= {2 \Gamma(s-\half) \pi^{\half - s} { \zeta(2s-1)} \over \Gamma(s)}
\hspace{0.4 in} 
\alpha_N = {4 |N| ^{\half - s} \sigma_{2s-1}(|N|) \over \Gamma(s)}
\eea 

\sm

The constant part of the Fourier expansion, i.e. the term $a \,\tau_2^s+ b\, \tau_2^{1-s}$, is known as the \textit{Laurent polynomial} of the Maass form. Though it may at first sight seem that the space of Laurent polynomials is two-dimensional and spanned by $a$ and $b$, in fact this space is only one-dimensional. This may be shown by considering Maass forms on the truncated fundamental domain, 
\bea
F_L = \left\{\tau \in \cH, \,\,|\tau| \geq 1,\,\, |\mathrm{Re}(\tau)| \leq \half,\,\, \tau_2 \leq L\right\}
\eea
Indeed, consider two Maass forms $f,g \in \cN(s)$ with Fourier expansions 
\bea
f(\tau) = \sum_{N \in \ZZ} c_N(\tau_2) e^{2 \pi i N \tau_1}\hspace{0.6 in}g(\tau) = \sum_{N \in \ZZ} d_N(\tau_2) e^{2 \pi i N \tau_1}
\eea
Green's theorem on the region $F_L$ states that if ${\p\over \p n}$ is the normal derivative on $\p F_L$ and $ds$ is the differential arc length, then we have 
\bea
\int_{\p F_L} ds \left(f {\p g \over \p n} - d {\p f \over \p n} \right) =  \int_{F_L} {d \tau_1 d \tau_2 \over \tau_2^2} (f \Delta g - g \Delta f) =0
\eea
where the last equality follows from the eigenvalue equation. The left side may be evaluated in terms of the Fourier coefficients. Noticing that modular invariance reduces the boundary integral to an integral over $\tau_2 = L$ (since the remaining boundaries of the fundamental domain give equal and opposite contributions), we have 
\bea
\int_{\p F_L} ds \left(f {\p g \over \p n} - f {\p f \over \p n} \right) & =& \int_{-\half}^{\half} d \tau_1 \sum_{N,M\in \ZZ} \left(c_N(L) d'_M(L) - c'_N(L) d_M(L) \right)e^{2 \pi i (N+M) \tau_1} 
\no\\
&=& \sum_{N\in \ZZ} \left[c_N(L) d'_{-N}(L) - c'_N(L) d_{-N}(L) \right] 
\eea
where the derivative is defined by,
\bea
c'_N(L) = {\p c_N(\tau_2) \over \p \tau_2} \Big |_{\tau_2 = L}
\eea
and similarly for $d'_{-N}(L)$. In particular, this implies that $c_0 d'_0 - c'_0 d_0 = 0$, and if we write, 
\bea
c_0(\tau_2) = a \,\tau_2^s+ c\, \tau_2^{1-s} \hspace{0.7 in} d_0(\tau_2) = b \,\tau_2^s+ d\, \tau_2^{1-s}
\eea
then for $s \neq  {1\over 2}$ we obtain $ad-bc=0$, which indeed means that the space of Laurent polynomials has dimension one. In the case that $s = \half$ this space is also obviously of dimension one, since then the two factors of $\tau_2$ have the same power.

\subsubsection{The space of Maass forms}

As was mentioned above, at generic values of $s\in \CC$ the space of Maass forms $\cN(s)$ consists only of non-holomorphic Eisenstein series $E_s(\tau)$.  More generally, at non-generic values of~$s$, the space of Maass forms can include cusp forms. In the current context, cusp forms are defined to be the subset of elements of $\cN(s)$ such that the Laurent polynomial vanishes, that is to say $a=b=0$ in (\ref{eq:Maassfourier}). We will denote the space of cusp forms by $\cS(s)$. The non-holomorphic Eisenstein series are clearly not cusp forms for any $s$. For generic values of~$s$, the set $\cS(s)$ will be empty.

\sm

The structure of $\cN(s)$ can be summarized as follows:
{\thm 
\label{thm:Maassspace}
The space of Maass forms $\cN(s)$ is given by 
\bea
\cN(s) = \left\{ \begin{matrix} \CC E_s(\tau) & & \mathrm{Re} \,s > \half \,\,\mathrm{and}\,\, s \notin [0, 1]  \\ 
\CC  E_s(\tau)  \oplus \cS(s) &&  \mathrm{Re}\, s = \half\,\,\,\mathrm{or}\,\,\, s \in [0, 1 )  \\ \CC &&s = 1  \end{matrix} \right.
\eea}

\sm

The proof of this theorem follows almost immediately from the fact, proven above, that the space of Laurent polynomials has dimension 1. This means that for any $f(\tau) \in \cN(s)$, there exists a constant $c$ such that the combination $f(\tau) - c E(\tau,s)$ has trivial Laurent polynomial, i.e. is a cusp form. 

\sm

To complete the proof, we need only prove that the space of cusp forms is empty for all values of $s$ which  satisfy neither $\mathrm{Re}\, s = \half$ nor $s \in [0, 1 )$. The non-existence for $s=1$ is immediate: in this case the cusp form would be harmonic, i.e. satisfy a Laplace eigenvalue equation with zero eigenvalue, and hence would have to vanish by the maximum modulus principle. More generally we may argue as follows. Above we have seen that it is possible to find a constant $c$ such that for any $f(\tau) \in \cN(s)$, the combination  $h(\tau) := f(\tau) - c E_s(\tau)$ is a cusp form. Since cusp forms are square-integrable over the fundamental domain $F = SL(2, \ZZ)\backslash\cH$ equipped with the Poincar\'e measure $d^2\tau / \tau_2^2$,
\bea
 \int_{F}  {d^2 \tau \over \tau_2^2 }\,|h(\tau)|^2  < \infty
\eea
we may manipulate the following absolutely convergent integral by using the differential equation $\Delta h = s(s-1) h$  satisfied by $h$, and Green's theorem,
\bea
s(s-1) \int_{F}  {d^2 \tau \over \tau_2^2 }|h(\tau)|^2 = 
\int_F  {d^2 \tau \over \tau_2^2 }\Delta h(\tau)\, \overline{h(\tau)}
= -  4\int_{F}  {d^2 \tau}  \left | \p_\tau h \right |^2 
\eea
Since the right side is real and non-positive,  cusp forms must have real $s(s-1) \leq 0$, which restricts us to $s \in [0,1]$ or $\mathrm{Re}\, s = \half$. 

\sm 

In closing, let us note that the triviality of the space of cusp forms for $\mathrm{Re} \,s \neq \half$ and $s \not \in \RR$ implies that for these values the space $\cN(s)$ is one-dimensional and hence there must be a functional relation relating $E_s(\tau)$ and $E_{1-s}(\tau)$, since both have Laplace eigenvalue $s(s-1)$. Indeed, precisely such a relation was identified in (\ref{4.d8}).

\subsubsection{Cusp forms}

 We have just shown that the space $\cS(s)$ of cusp forms is empty unless $\mathrm{Re}\, s = \half\,\,\,\mathrm{or}\,\,\, s \in [0, 1 )$. Actually, one can further show that $\cS(s)$ vanishes for $s \in [0,1)$, and hence that all non-trivial cusp forms must exist at $\mathrm{Re}\, s = \half$ and away from the real axis: we refer to \cite{LangSL2R} for the proof. Denoting $s = \half + i t$ for $ t \in \RR$, the corresponding eigenvalue is $\lambda = s(s-1) = - \left({1\over 4} + t^2\right)$. 

\sm

For any $t \in \RR$, the space $\cS\left(\half + i t\right)$ is finite-dimensional. To show this, begin with $M$ elements $v_1(\tau), \dots, v_M(\tau) \in \cS\left(\half + i t\right)$. Given an additional element $\widetilde v(\tau)$, it is possible to choose coefficients $c_i \in \CC$ such that the linear combination, 
\bea
 v(\tau) :=\widetilde v(\tau) - \sum_{i=1}^M c_i v_i(\tau) 
\eea
has vanishing Fourier coefficients up to order $M$. We will now show that for $M$ sufficiently large but finite, the vanishing of Fourier coefficients up to order $M$ implies vanishing to all orders. In other words we will have $v(\tau) =0$ exactly, and hence all elements $\widetilde v(\tau)\in \cS\left(\half + i t\right)$ can be written as a linear combination of a finite number $M$ of other elements. 

\sm

To prove the desired result, begin by recalling that a cusp form $v(\tau)$ is bounded on the upper-half plane, and hence that $|v(\tau)|$ must have a maximum at some location $\tau^{(0)} = \tau_1^{(0)} + i \tau_2^{(0)}$. From the form of the Fourier expansion (\ref{eq:Maassfourier}) and orthogonality of exponentials, we have,
\bea
\a_N \,\sqrt{\tau_2^{(0)}}\, K_{it}(2 \pi \tau_2^{(0)} |N|) = \int_{-1/2}^{1/2} d \tau_1 \,v(\tau_1 + i {\tau_2^{(0)}/2})\, e^{- 2 \pi i N \tau_1}\left({\sqrt{2} K_{it} (2 \pi \tau_2^{(0)} |N|) \over K_{it} ( \pi \tau_2^{(0)} |N|) } \right) 
\eea 
It then follows that, 
\bea
\left|v(\tau^{(0)}) \right| < 2 \sqrt{2}\, \left|v(\tau^{(0)})\right|\, \sum_{|N| > M} {K_{it}(2 \pi \tau_2^{(0)} N) \over K_{it}( \pi \tau_2^{(0)} N) } 
\eea
If $M$ is sufficiently large (depending on $t$), we may estimate the Bessel functions by exponentials, giving the inequality, 
\bea
\left|v(\tau^{(0)})\right| \leq \left|v(\tau^{(0)})\right| \,C e^{- \pi M  \tau_2^{(0)}}
\eea
for some proportionality constant $C$. Hence if $M\geq (\ln C) / (\pi  \tau_2^{(0)}) $ then we conclude that $|v(\tau^{(0)})|=0$, and hence that $v(\tau)$ is identically zero.

\sm

We have just shown that $\cS\left(\half + i t\right)$ is finite-dimensional for any $t$. Furthermore, though we will not reproduce the proof here, it has been shown that there exist an infinite number of values of $t$ for which the space of cusp forms is non-empty.  However, there is no exact value of $t$ for which this is known to be the case. The smallest approximate values of $t$ for which cusp forms are believed to exist are, 
\bea
\label{eq:discreteseriesvalues}
t \approx\, 9.53369, \,\, 12.17300, \,\, 13.77975, \,\, 14.35850,\,\,16.13807,\,\,16.64425,\,\,17.73856, \, \dots
\eea
In particular, we see that there is only a \textit{discretum} of such values.

\subsubsection{Comments on the physical interpretation of the spectrum}

The fundamental domain for $SL(2,\ZZ)$ is non-compact as it has a puncture at the cusp. From the point of view of scattering theory, one may view the cusp as an asymptotic region from which \textit{a free wave} in the continuous spectrum  can be sent in. This wave is indeed oscillatory and of the form $\tau_2^{\half + it}$ for real $t$,  from the first constant term in the Fourier expansion. Since the fundamental domain is otherwise bounded, the wave hits a wall at $|\tau|=1$ and must be reflected out through the puncture and is asymptotically of the form $\tau_2^{\half -i t}$. This phenomenon is analogous to one-dimensional scattering on a semi-infinite line and explains in physical terms why the Eisenstein solutions have $E_{\half+it}(\tau) = E_{\half -it} (\tau)$. 

\sm

The cusp forms are wave functions of bound states, as is confirmed by the fact that they decay exponentially towards the cusp. In most quantum mechanics problems, the number of bound states in the presence of a wall would be finite. But there are exceptions, such as the Coulomb problem. Here, it is the special properties of the hyperbolic metric on the  fundamental domain of $SL(2,\ZZ)$ that produce an infinite number of cusp forms, or bound states, with a spectrum that rises all the way to infinity.

\subsection{Spectral decomposition}

We have now discussed various aspects of the spectrum of the Laplacian on the fundamental domain $SL(2,\ZZ) \backslash \cH$.
It is possible to extract from this data the following orthonormal basis of eigenfunctions of the Laplacian: 
\begin{itemize} 
\item \textbf{Discrete series:} The constant function $v_0 := \sqrt{\mathrm{Area}(F)} = \sqrt{3/ \pi}$ (which is trivially an eigenfunction of the Laplacian) together with an orthonormal basis of cusp forms $v_n(\tau)$ with corresponding eigenvalues $s_n(s_n-1) = - \left({1\over 4} + t_n^2 \right)$ for $t_n \in \RR$. The first few values of $t_n$ are given in (\ref{eq:discreteseriesvalues}).
\item \textbf{Continuous series:} The non-holomorphic Eisenstein series $E_s(\tau)$ with $s= \half + i t$ and $t\in \RR$ taking a continuum of values.
\end{itemize}

\sm

Orthogonality here is defined relative to the Petersson inner product with the Poincar\'e measure $d^2 \tau / \tau_2^2$ on the fundamental domain, i.e.
\bea
\langle f, g \rangle = \int_{F} {d^2 \tau \over \tau_2^2} f(\tau) \overline{g(\tau)}
\eea 
Let us use this inner product to verify the orthogonality claimed above. Consider the inner product between a non-holomorphic Eisenstein series $E_s(\tau) $  and a cusp form $v_n(\tau)$, 
\bea
\langle E_s(\tau) , \, v_n( \tau) \rangle = \int_{F} {d^2 \tau \over \tau_2^2} E_s(\tau) \overline{v_n( \tau)}
\eea
This inner product converges since cusp forms have exponential decay at the cusps, whereas non-holomoprhic Eisenstein series have at most polynomial growth. For $\mathrm{Re}\, s >1$, we may proceed by using the Poincar{\'e} series for $E_s(\tau) $ given in (\ref{eq:nonholoPoincare}), which gives
\bea
\langle E_s(\tau) , \, v_n( \tau) \rangle &=& {\zeta(2s) \over \pi^s}\int_{F} {d^2 \tau \over \tau_2^2} \,  \sum_{\gamma \in \Gamma_\infty \backslash SL(2, \ZZ)}(\mathrm{Im}\, \gamma \tau)^s \,\overline{v_n(\tau)}
\no\\
&=&  {\zeta(2s) \over \pi^s} \int_{-\half}^\half d\tau_1 \int_0^\infty  {d \tau_2 \over \tau_2^2}\,\tau_2^s\, \overline{v_n( \tau)}
\eea
where in the second equality we have used modular invariance to unfold the integral over the fundamental domain, i.e. we have replaced the integral over $SL(2,\ZZ)$ orbits in $F$ with an integral over the strip $\Gamma_\infty \backslash \cH$, 
\bea
\int_{SL(2,Z)\backslash \cH} ~~ \sum_{\Gamma_\infty \backslash SL(2,\ZZ)} \boldsymbol{\cdot} \,\,\,\,\,=\,\,\,\,\, \int_{\Gamma_\infty \backslash \cH} \boldsymbol{\cdot}
\eea
This is the analog of the unfolding trick introduced for periodic functions in section \ref{sec:unfolding}.
 The integral over $\tau_1$ may now be performed and isolates the Laurent polynomial of $v_n( \tau)$, which is vanishing by definition. This gives the desired orthogonality for $\mathrm{Re}\, s >1$. The analogous result for $\mathrm{Re}\, s <1$ may be obtained by analytic continuation. 
 
 \sm

An arbitrary square integrable modular function can now be decomposed into this orthogonal basis. We state without proof the following spectral decomposition theorem, 
{\thm [Roelcke-Selberg]
Let $v_0 = \sqrt{3 / \pi}$ be the normalized constant function  and $\{v_n(\tau)\}_{n \geq 1}$ be an orthonormal basis of cusps forms. Then any square integrable modular function $f(\tau)$ admits the following decomposition, 
\bea
f(\tau) = \sum_{n \geq 0} \langle f(\tau), v_n(\tau)\rangle\, v_n(\tau) + {1 \over 4 \pi} \int_{-\infty}^\infty dt \, \left\langle f(\tau),  \,E_{\thalf + i t} (\tau) \right\rangle E_{\thalf + i t} (\tau)\,\,
\eea 
}

The reader may be worried about the convergence properties of the continuous portion of this decomposition. Indeed, $E_s(\tau)$ is not square-integrable because of its Laurent polynomial. However, when $\mathrm{Re}\, s = \half$, one can show that $|E_s(\tau)| \ll \sqrt{\tau_2}$ when $\tau_2 \rightarrow \infty$, and the extra integration over $t$ in the decomposition formula saves a factor of $\log \tau_2$, which is sufficient to turn the right side into a square-integrable function.

\subsection{Maass forms of general weight}

Thus far we have focused on Maass forms of weight $(0,0)$. We now discuss Maass forms of arbitrary (integer) weight $k$, i.e. those satisfying 
\bea
f\left({a \tau + b \over c \tau + d}\right) = (c \tau + d)^k f(\tau)\hspace{0.5 in} \left(\begin{matrix} a & b \\ c & d\end{matrix} \right) \in SL(2, \ZZ)
\eea
The Maass property in this case must be recast in terms of a weight-$k$ Laplacian 
\bea
\Delta_k = 4 \tau_2^{2-k} \p_{\tau} \tau_2^k \p_{\bar{\tau}} = \Delta - 2 i k \tau_2 \p_{\bar \tau}
\eea
in order to get a modular covariant object. 

Of particular interest are the \textit{harmonic} Maass forms of weight $k$, i.e. those satisfying an eigenvalue equation $\Delta_k f = 0$. Unlike in the case of weight 0, for non-zero weight the space of harmonic Maass forms is non-trivial. 
\sm
A simple example of a weight-2 harmonic Maass form is the almost-holomorphic Eisenstein series $\HE_2^*(\tau)$ introduced in section \ref{E2section}. Indeed, from (\ref{eq:2E2rel}) and the fact that $\HE_2(\tau)$ is holomorphic we have 
\bea
\Delta_2 \HE^*_2(\tau)=  - {3 \over \pi} (\Delta - 4 i \tau_2 \p_{\bar \tau}) \tau_2^{-1} = - {3 \over \pi} ( 2 \tau_2^{-1} -2 \tau_2^{-1} )  = 0
\eea

\sm

The space of weight-$k$ harmonic Maass forms $\cI_k$ admits an interesting decomposition which will feature prominently in the subsequent section. This decomposition can be captured by the following exact sequence,
\bea
\label{eq:weightkMaassspace}
0 \rightarrow \cM^!_k \rightarrow \cI_k \rightarrow \cM^!_{2-k} \rightarrow 0 
\eea
with $\cM^!_k$ the space of weight-$k$ modular functions. The content of this exact sequence is two-fold. First, it says that the space $\cM^!_k$ of weight-$k$ modular functions is a subset of the space $\cI_k$ of weight-$k$ harmonic Maass forms. This follows trivially from the fact that modular functions are annihilated by $\Delta_k$.  Second, it says that the space $\cM^!_{2-k}$ of weight-$(2-k)$ modular functions is a quotient of $\cI_k$ by $\cM^!_k$. This follows from the fact that, given an element $f(\tau) \in \cI_k$, the function $g(\tau):=\tau_2^{k} \p_{ \tau} \overline{f(\tau)}$ is holomorphic and of weight $2-k$, and vanishes if and only if $f(\tau)$ is in $\cM^!_k$.

\subsection{Mock modular forms}

In section \ref{E2section} we saw that there exist two variants of the weight-two holomorphic Eisenstein series, namely the function $\HE_2(\tau)$ defined in (\ref{G2Eis}) which is holomorphic but not modular, and the function $\HE_2^*(\tau)$ defined in (\ref{3.Estar}) which is modular but not holomorphic. The former is an example of a quasi-modular form, whereas the latter is an example of an almost-holomorphic modular form. Section \ref{sec:quasialmostmod} described an isomorphism between the sets of quasi-modular forms and almost-holomorphic modular forms. 

\sm

Having introduced non-holomorphic modular forms in the previous subsections, it is natural to ask if their holomorphic parts lead to interesting analogs of quasi-modular forms. In particular, we will consider the holomorphic parts of harmonic  ``weak" Maass forms,\footnote{The adjective ``weak" indicates the potential for exponential growth at the cusps; in contrast, a ``Maass form" is required to have at most polynomial growth at the cusp.} 
which are known as \textit{mock modular forms}.\footnote{Mock modular forms made their first appearance in a letter of Ramanujan to Hardy, in which they were referred to as \textit{mock theta functions}. In the modern definition, a mock theta function is a $q$-series $H(q) = \sum_{n \geq 0} a_n q^n$ such that for some $\lambda \in \QQ$ the series $q^\lambda H(q)$ gives a mock modular form of weight $\half$, with the shadow (defined below) being a weight-$3\over 2$ function of the form $\sum_{n \in \ZZ} \eps(n) n q^{|\kappa| n^2}$ with $\kappa \in \QQ$ and $\eps$ an odd periodic function. 
} Given a harmonic Maass form $F^*(\tau)$, we may extract its holomorphic part as follows. We begin by defining, 
\bea
g(\tau) = \tau_2^k {\p \over \p  \tau} \overline{F^*(\tau)}
\eea
By the condition that $F^*(\tau)$ be harmonic, it follows that $g(\tau)$ is holomorphic, since $\Delta_k = \tau_2^{2-k} \p_\tau \tau_2^k \p_{\overline \tau}$.  
If one can identify a new function $g^*(\tau)$ such that its image under $\tau_2^k \p_{\overline \tau}$ is $\overline{g(\tau)}$, then it is clear that,
\bea
F(\tau) := F^*(\tau) - g^*(\tau)
\eea
will be holomorphic. Such a function $g^*(\tau)$ can be obtained by simply inverting the differential operator $\tau_2^k \p_{\overline \tau}$, giving the expression, 
\bea
g^*(\tau) = \left({i\over 2} \right)^{k-1} \int_{- \overline \tau}^{i\infty}d z\,\,  {\overline{g(-\overline{z})}\over (z+\tau)^k}
\eea
 known as a non-holomorphic Eichler integral. This integral is independent of the path since the integrand is holomorphic in $z$. 

\sm

The function $F(\tau)$ is holomorphic, but will in general fail to be modular, since $g^*(\tau)$ is not modular. The function $g(\tau)$ in terms of which $F(\tau)$ is defined is referred to as the \textit{shadow}, and is a weight-$(2-k)$ modular function. The pair of $F(\tau)$ and $g(\tau)$ are often together referred to as a mock modular form, though in some contexts the function $F(\tau)$ alone is given that name. 

\sm

Given a harmonic Maass form $F^*(\tau)$, it is always possible to extract from it a unique mock modular form by taking the holomorphic part in the way described above. Conversely, given a mock modular form, it is always possible to complete it to a harmonic Maass form by taking the sum of $F(\tau)$ with the function $g^*(\tau)$ obtained from the shadow. This establishes an isomorphism between the two spaces, exactly analogous to the isomorphism between the spaces of almost-holomorphic modular forms and quasi-modular forms given in section \ref{E2section}. The space of mock modular forms then fits into an exact sequence of the form in (\ref{eq:weightkMaassspace}). By nature of it being an exact sequence, the map from the space of mock modular forms to the space of shadows (known as the \textit{shadow map}) is surjective.

\subsubsection{Examples of mock modular forms}

We now give some concrete examples of mock modular forms. 

\sm

A first example is provided by $\HE_2(\tau)$. In the previous section we showed that $\HE_2^*(\tau)$ is a weight-2 harmonic Maass form, and hence by definition the holomorphic part of $\HE_2^*(\tau)$, namely $\HE_2(\tau)$, is a weight-2 mock modular form.  The shadow $g(\tau)$ in this case is simply a constant, namely $g(\tau) = - {12 \over \pi}$. It is easy to check that this constant gives $g^*(\tau) = - {3\over \pi \tau_2}$, reproducing the result in (\ref{eq:2E2rel}). Thus we see that $\HE_2(\tau)$ is both quasi-modular and mock modular. This is not in general the case:  quasi-modular forms are generically not mock modular. For example, $(\HE_2(\tau))^2$ is a quasi-modular form, but fails to be mock modular since its modular completion is not a harmonic Maass form. 

\sm

A second example, which appears rather unexpectedly in physics via the elliptic genus of strings propagating on a K3 manifold, is the following function, 
\bea
F^*(\tau) = F(\tau) + {6 \over \sqrt{ i}}\int_{- \overline{\tau}}^\infty dz \, { \overline{\eta(-\overline{z})^3}\over (z+\tau)^{1/2}}
\eea
with $F(\tau)$ defined as, 
\bea
\label{eq:mmf2}
F(\tau) = {1 \over \eta(\tau)^3}\bigg ( - \HE_2(\tau) 
+ 24 \sum_{\substack{r > s > 0 \\ r - s = 1 \,\, \mathrm{mod}\,\, 2}} (-1)^r \, s \, q^{rs/2} \bigg)
\eea
It is known that $F^*(\tau)$ is modular with weight $\half$. Hence its holomorphic part $F(\tau)$ must be a mock modular function. The shadow in this case is $g(\tau) = 3 \sqrt{2} \,\eta(\tau)^3$. 

\sm

A curious fact is that the first few terms in the Fourier expansion of $F(\tau)$, 
\bea
F(\tau) = q^{-{1\over 8}} \left( -1 + 45 q + 231 q^2 + 770 q^3 + 2277q^4 + 5796 q^5 + \dots\right) 
\eea
give dimensions of irreducible representations of the Mathieu group $M_{24}$, a sporadic simple group of order 244,823,040. This correspondence goes under the name of \textit{Mathieu moonshine}, and has been the subject of much research. 

\sm

As a third and final example, we mention with a classic result of Zagier, which states that the following two functions are harmonic with $k={3\over 2}$, 
\bea
F_0^*(\tau) &=& 3 \sum_{n \geq 0} H(4 n) \, q^n 
+ 6 \tau_2^{-1/2} \sum_{n \in \ZZ} \beta(4 \pi n^2 \tau_2) \, q^{-n^2}
\no\\
F_1^*(\tau) &=& 3 \sum_{n>0} H(4n-1) \, q^{n- {1\over 4}} 
+ 6 \tau_2^{-1/2} \sum_{n \in \ZZ} \beta\left(4 \pi (n+ 1/2)^2 \tau_2\right) q^{-\left(n+\half\right)^2}
\eea
where $\beta(t)$ is an incomplete gamma function 
\bea
\beta(t) = {1\over 16 \pi} \int_1^\infty du\, u^{-3/2} e^{- ut }
\eea
and $H(n)$ are the Hurwitz class number of imaginary quadratic fields. The latter are defined for $n>0$ to be the number of $PSL(2, \ZZ)$ equivalence classes of integral binary quadratic forms of discriminant $-n$, weighted by the reciprocal of the number of their automorphisms. The first few non-zero entries are given by
\bea
H(3) = {1\over 3} \hspace{0.5 in} H(4) = \half \hspace{0.5 in} H(7) = H(8) = H(11) = 1
\eea
 For $n=0$ we have $H(0) = -{1\over 12}$ and for $n<0$ we have $H(n) = 0$.
 
 \sm
 
 The holomorphic parts $F_{0,1}(\tau)$ of $F^*_{0,1}(\tau)$ are easy to read off, 
 \bea
 \label{eq:mmf3}
 F_0(\tau) &=& 3 \sum_{n \geq 0} H(4 n) q^n 
 \no\\
 F_1(\tau) &=&3 \sum_{n>0} H(4n-1) q^{n- {1\over 4}} 
 \eea
 and have corresponding shadows 
 \bea
 g_0(\tau) = - {3\over 4 \pi} \vartheta_3(0 | 2 \tau) \hspace{0.5 in} g_1(\tau) = - {3\over 4 \pi} \vartheta_2(0 | 2 \tau) 
 \eea
 It is important to note that these functions are not mock modular forms in the strict sense discussed above, since under $\tau \rightarrow - 1/\tau$ the original harmonic functions  $F^*_{0,1}(\tau)$  are not individually invariant, but rather transform into one another as 
 \bea
 \binom{F^*_0(-1/\tau)}{F^*_1(-1/\tau)} = - {(-i \tau)^{3/2}\over \sqrt{2}} \left(\begin{matrix} 1 & 1 \\ 1 & - 1 \end{matrix}  \right)  \binom{F^*_0(\tau)}{F^*_1(\tau)} 
 \eea
 There are two ways of interpreting this fact. The first is to think of $F^*_{0,1}(\tau)$  as forming a \textit{vector-valued} harmonic Maass form. In this interpretation the functions $F_{0,1}(\tau)$ likewise form a vector-valued mock modular form. Vector-valued modular forms will be discussed in more detail in section \ref{sec:MDEsvvmfs}. 
 
\sm

The second approach is to notice that $F^*_{0}(\tau)$ is invariant under $T$ and $F^*_{1}(\tau)$ is invariant under $T^4$. Hence we can say that $F^*_{0}(\tau)$ individually is modular invariant under the subgroup of $SL(2, \ZZ)$ generated by $T$ and $ST^4S$. This subgroup is denoted by $\Gamma_0(4)$, and is one of the so-called \textit{congruence subgroups} of $SL(2,\ZZ)$, discussed in detail in section \ref{sec:Cong}. In this interpretation one would say that $F^*_{0}(\tau)$ is a harmonic Maass form for the congruence subgroup $\Gamma_0(4)$, and that $F_{0}(\tau)$ is likewise a mock modular form for this subgroup. Modular forms for congruence subgroups will be discussed in detail in section \ref{sec:Forms}.

\subsection{Quantum modular forms}
 
Finally, let us briefly mention the notion of a \textit{quantum modular form}.\footnote{As far as we are aware, there is no direct connection between the term ``quantum" used here and the usual notion of ``quantum" in Physics. }  Unlike for standard modular forms for which the domain is the upper half-plane, quantum modular forms are defined on the space of cusps $\QQ \cup \{i \infty\}$ (or some subset thereof). Since this domain is a discrete space, quantum modular forms do not obey the usual analyticity properties of modular forms. Furthermore, they are not modular in the standard sense. Instead, a weight-~$k$ quantum modular form $f(\tau)$  is defined such that for all $\gamma \in SL(2, \ZZ)$, the difference 
 \bea
h_\gamma(\tau):=  f(\tau) - \eps_\gamma (c \tau + d)^k f\left({a\tau+b \over c \tau + d}\right)
 \eea
 is an analytic function on the upper half-plane. Here we have allowed for a multiplier phase $\eps_\gamma$, similar to that appearing in the transformation rules for the Dedekind $\eta$-function in section \ref{sec:Dedekindeta}.
 
\sm
 
 It is simplest to illustrate this definition by proceeding directly to an example. Consider the following $q$-expansion, 
 \bea
 F(q) = \sum_{n=0}^\infty \prod_{i=0}^{n-1}(1-q^{i+1})
 \eea
 known as Kontsevich's strange function. This function does not converge on any open subset of $\CC$, but is well-defined at roots of unity, i.e. when $\tau \in \QQ$. We now define the closely related object,
 \bea
 \phi(\tau) = q^{1\over 24 } F(q)
 \eea
It follows from this definition that $\phi(\tau+1) = e^{2\pi i /24} \phi(\tau)$. 

\sm

To understand the behavior under $S$, we first note that the $T$ transformation just given implies $\phi(n) = e^{2 \pi i n /24} \phi(0) =  e^{2 \pi i n /24}$ for $ n \in \ZZ$. On the other hand, Kontsevich and Zagier have shown that for $\tau = -{1\over n}$, one has,
\bea
\phi\left(-{1\over n} \right) \sim n^{3/2}e^{2 \pi i {(3+ n) \over 24}} + \sum_{j=0}^\infty c_j \left({2\pi i \over n} \right)^j
\eea
as $n \rightarrow \infty$, with the first few coefficients in the expansion being,
\bea
c_0 = 1 \hskip 0.6in c_1 = {23\over 24} \hskip 0.6in c_2 = {1681 \over 1152}
\eea 
Using the $T$ transformation, the first term above can be rewritten as,
\bea
n^{3/2}e^{2 \pi i \over 8} \phi(n) = (-in)^{3/2} \phi(n)
\eea 
and hence we conclude that the function
\bea
h_S(n) :=\phi(n) + (-i n)^{3/2} \phi(-1/n) 
\eea
has a well-defined Taylor expansion at $\tau = 0$, and so is smooth there. Indeed, the $n$-th derivative of $h_S(n)$ at $\tau = 0$ is given by $c_n \left({2\pi i / n} \right)^n$. Though we do not give details here, the function $h_S(n)$ can furthermore be extended to a real-analytic function on $\CC$ (except for at $\tau=0$), so $\phi(\tau)$ is an example of a quantum modular form.

\subsection*{$\bullet$ Bibliographical notes}
 
Classic introductions to spectral analysis on symmetric spaces with applications to Maass forms and non-holomorphic Eisenstein series  may be found in the books by Kubota~\cite{kubota}, Terras~\cite{terras1,terras2}, and Iwaniec~\cite{iwan2,iwan3}. {A discussion aimed at physicists, and focused on the definition over the adeles, can by found in the book by Fleig, Gustafsson, Kleinschmidt, and Persson \cite{Fleig:2015vky}.} Group-theoretic aspects of the spectral decomposition of automorphic forms are expounded on in the book by Gel'fand, Graev, and Pyatetskii-Shapiro~\cite{Gelfand}. Faddeev's approach, which is based on operator perturbation theory, and perhaps closer to physics, is reviewed in detail and commented upon in  Lang~\cite{LangSL2R}.  A construction of cusp forms which are not Maass forms was given in  \cite{BrownpartIII}.   The spectral decomposition theorem has recently been used in a physics context in \cite{Benjamin:2021ygh,Collier:2022emf,Paul:2022piq}. Generalizations to automorphic forms under the group $GL(n,\RR)$ are developed in the book by Goldfeld \cite{Goldfeld}. 
 
 \sm
 
A fundamental role was played by quasi-modular forms in Viazovska's proof of the sphere packing problem in eight dimensions, namely that no packing is more dense than the one corresponding to the root lattice of $E_8$. The original paper may be found in \cite{Via}, while the generalization to sphere packing in dimension 24 was settled in \cite{MV}.  
 
 \sm
 
Reviews on mock modular forms include the lecture notes by Zagier~\cite{Zagiernotes} in the mathematics literature and \cite{Dabholkar:2012nd,Anagiannis:2018jqf} in the physics literature. The second example of a mock modular form, given in (\ref{eq:mmf2}), appears in physics as the elliptic genus of a K3 manifold \cite{Eguchi:1988vra,Eguchi:2009cq}. The connection with dimensions of representations of the Monster group $M_{24}$ was observed in \cite{Eguchi:2010ej}, beginning the program of Mathieu moonshine. Reviews of moonshine aimed at physicists may be found in \cite{Anagiannis:2018jqf,GannonMoonshine1,GannonMoonshine2,Harrison:2022zee} and references therein. 
The third example of mock modular forms, given in (\ref{eq:mmf3}), also arises in at least one physics context, as the partition function of $\cN=4$ super Yang-Mills theory with gauge group $SO(3)$ on $\CC \mathbb{P}^2$ \cite{Vafa:1994tf}. 
 
 \sm 
 
Quantum modular forms were introduced in mathematics  in \cite{zagier2001vassiliev,zagier2010quantum,ono2013harmonic} but their role in physics remains unexplored to date, despite the word ``quantum" in their name.

 \newpage

\section{Quantum fields on a torus}
\setcounter{equation}{0}
\label{sec:TorusQFT}

In this section, we shall present some immediate applications of the theory of elliptic and modular functions to problems of physical interest, including the construction of the Green functions and the functional determinants for the $bc$ system, the scalar,  and the spinor fields on the torus. In particular, we shall show how the singular terms in the operator product expansion of holomorphic fields for the $bc$ system  essentially determine arbitrary correlation functions on the torus in terms of elliptic functions.

\subsection{Quantum fields}

A quantum field is a map from a manifold $\Sigma$ to a linear operator acting on a vector space~$\mF$, or Fock space,  which is equipped with a Hermitian inner product. If the inner product is positive definite then $\mF$ is a Hilbert space familar from standard quantum theory, but we shall consider quantum field theories where the inner product is not necessarily definite  as well. We shall be interested in the special case where the manifold $\Sigma$ is a (connected) Riemann surface which may be compact or non-compact, with or without a boundary. Quantum fields on the complex plane $\Sigma =\CC$ describe the continuum limit and critical phenomena  of two-dimensional statistical mechanics systems, such as the Ising, Potts, XY, and non-linear sigma models. Quantum fields on a general Riemann surface $\Sigma$ provide the proper setting for string perturbation theory, where probability amplitudes are obtained via an expansion in powers of the string coupling constant. The expansion is obtained by summing contributions from Riemann surfaces of all genera and, for each genus, by integrating  over the corresponding moduli space of Riemann surfaces, as will be explained in more detail in section \ref{sec:SA}.

\subsubsection{Conformal fields}
\label{sec:conffields}

A conformal field $\phi$ on $\Sigma$ may be defined as follows. We consider two arbitrary  coordinate charts $\cU, \cU' \subset \Sigma$ with local complex coordinate systems  $(z,\bar z)$ and $(z',\bar z')$. Assuming that these coordinate charts have a non-empty intersection $\cU \cap \, \cU' \not = \emptyset$, the coordinate systems in the overlap $\cU \cap \, \cU'$ are related to one another by a holomorphic transition function $z'=f(z)$ whose derivative $f'(z)$ is non-zero throughout $\cU \, \cap \, \cU'$. Any such transformation of the form $z \to z'=f(z)$ is a locally conformal transformation, namely it preserves all angles. For more details see appendix \ref{sec:RS}. In the coordinate systems $(z,\bar z)$ and $(z',\bar z')$ the field $\phi$ is given by $\phi(z,\bar z)$ and $\phi'(z', \bar z')$, respectively. The field $\phi$ is a conformal (primary) field of conformal weight $(h, \tilde h)$,   provided $\phi$ and $\phi'$ are related as follows,
\bea
\phi'(z', \bar z') (dz')^h (d\bar z')^{\tilde h} = \phi(z,\bar z) (dz)^h (d \bar z)^{\tilde h}
\eea
or  in view of the holomorphicity of $f(z)$,
\bea
\label{5.conf}
f'(z) ^{h} \,  \overline{f'(z)} ^{\tilde h} \, \phi'(z', \bar z') =    \phi(z,\bar z)
\eea
Equivalently, a conformal field may be viewed as a section of an operator-valued  holomorphic line bundle over $\Sigma$. Clearly, $h$ and $ \tilde h$  must be the same for all charts $\cU \subset \Sigma$.  

\sm

Local conformal transformations form an infinite-dimensional  Lie algebra, referred to as the \textit{Virasoro algebra}. Its generators $L_m$ for $m \in \ZZ$ obey the following structure relations, 
\bea
\label{eq:Virasoro}
{} [L_m, L_n] = (m-n) L_{m+n} + { \mc \over 12} m(m^2-1) \delta_{m+n,0}
\eea
Here $\mc$ is a real-valued constant referred to as the \textit{central charge} of the Virasoro algebra. The last term is proportional to the identity operator in the representation space of the Virasoro algebra, and therefore commutes with all $L_m$. When $\mc \not=0$, the Virasoro algebra is a central extension of the \textit{Witt algebra}, which has $\mc=0$. 

\sm

Actually, conformal transformations may be allowed to act independently on the holomorphic and anti-holomorphic coordinates, in which case we have two copies of the Virasoro algebra, whose generators are denoted $L_n$ acting on $z$ and $\tilde L_n$ acting on $\bar z$. These two Virasoro algebras commute so that $[L_m, \tilde L_n]=0$ for all $m,n \in \ZZ$. The conformal weights $h, \tilde h \in \RR$ characterize the representation of the Virasoro algebra under which the field $\phi$ transforms, 
\bea
\label{5.Lphi}
{} [L_m, \phi(z,\bar z) ] & = & z^{m+1} \p_z \phi (z, \bar z) + h(m+1) z^m \phi(z, \bar z)
\no \\
{} [\tilde L_m, \phi(z, \bar z ) ] & = & \bar z^{m+1} \pbz \phi (z,\bar z) + \tilde h(m+1) \bar z^m \phi(z, \bar z)
\eea
The {\sl minimal models} correspond to the irreducible representations of the Virasoro algebra for $0 < \mc <1$, for which the spectrum of allowed conformal weights $h$ (of primary conformal fields) is finite. Minimal models have rational conformal weights $h, \tilde h$ and a rational central charge $\mc$. They are special cases of the more general class of ``rational conformal field theories," discussed further in section \ref{eq:CMandNLSM}. For $0 < \mc < 1$, the unitary minimal models constitute a further subclass of conformal field theories that are so-called \textit{unitary}, and in particular are associated with $\mF$ being a genuine Hilbert space. Many conformal field theories may be realized in terms of free fields.

\subsubsection{Free conformal fields}

We shall limit our examples in the sequel  to free conformal field theories. A free quantum field satisfies a linear partial differential equation on $\Sigma$, given by a differential operator $\cD$. The order of the operator $\cD$ is usually of order no larger than 2. All correlation functions of a free quantum field may be evaluated in terms of the Green function $\cD^{-1} $ and the functional determinant of $\cD$.  For free conformal fields on an arbitrary genus-one Riemann surface~$\Sigma$,  the correlators and determinants may be expressed in terms of elliptic functions and modular forms. It turns out that the quantization of free conformal fields already provides a sufficiently rich structure for all of  string perturbation theory on an arbitrary flat space-time manifold, including Minkowski space-time with or without partial toroidal compactification, as will be further detailed in sections \ref{sec:SA} and \ref{sec:Toroidal}.

\subsection{The $bc$ system}
\label{sec:5.bc}

An important example of a free conformal field theory on a Riemann surface $\Sigma$, that has an immediate connection with elliptic functions and modular forms,  is provided by a system of two quantum fields  $b$ and $c$  governed by a first order action $S[b,c]$, 
\bea
S[b,c] = { 1 \over 2 \pi} \int _\Sigma d^2 z \, b(z,\bar z) \pbz c(z,\bar z) 
\hskip 1in 
d^2 z = { i \over 2} dz \wedge d\bar z
\eea
Invariance of $S[b,c]$ under local conformal transformations requires  both fields to have $\tilde h=0 $ and their remaining conformal weights to be conjugate: if $b$ has conformal weight $(h,0)$ then~$c$ must have conformal weight $(1-h,0)$. The field equations for $b$ and $c$ are obtained using the variational principle for $S[b,c]$, and are given by,
\bea
\pbz b = \pbz c=0
\eea
so that $b$ and $c$ are locally holomorphic functions of $z$ on $\Sigma$. The quantum field theory with these assignments is referred to as the $bc$ system of weight $h$. Below we shall consider the $bc$ system on the annulus and on the torus. The $bc$ system on a Riemann surface of arbitrary genus will be discussed in appendix \ref{app.bc}.

\subsubsection{The $bc$ system on an annulus}

An annulus is conformally isomorphic to a cylinder. Choosing the local complex coordinate $z'=x+iy$ on the cylinder with $x,y \in \RR$ and the periodic identification $x \approx x+1$, the coordinate $z$ on the annulus is given by the conformal map $z = e^{2 \pi i z'} = e^{2 \pi i (x+iy)}$.  Holomorphicity and periodicity of the fields $b$ and $c$ on the cylinder produce the following Fourier expansion in the coordinate $z'=x+iy$,\footnote{We use proportionality signs here in order to leave the precise normalization to be fixed on the annulus.}
\bea
b'(z') \sim \sum _{n \in \ZZ} b_n e^{- 2 \pi i n (x+i y)} 
\hskip 1in
c'(z') \sim \sum _{n \in \ZZ} c_n e^{- 2 \pi i n (x+i y)} 
\eea
If the fields $b$ and $c$ have conformal weights $(h,0)$ and $(1-h,0)$ respectively, then the fields on the annulus are obtained by the conformal transformation (\ref{5.conf}), and the Fourier expansions become  Laurent expansions in $z=e^{2 \pi i z'}$ given as follows,
\bea
b(z) = \sum_{n \in \ZZ} b_n \, z^{-n-h} 
\hskip 1in
c(z) = \sum_{n \in \ZZ} c_n z^{-n-1+h}
\eea
Considering the fields $b$ and $c$ as quantum operators, the coefficients $b_n$ and $c_n$ are operators. The closest connection with holomorphic and meromorphic functions will be obtained by requiring the operators to satisfy anti-commutation relations,
\bea
\label{5.ACom}
\{ b_m, b_n \} = \{ c_m, c_n \} =0
\hskip 1in
\{ b_m, c_n\} = \delta _{m+n,0}
\eea
where $\{A,B\} = AB+BA$ is the anti-commutator and $\delta_{m+n,0}$ is the Kronecker  symbol. We choose a polarization in which $b_n$ and $c_n$ are annihilation operators  for $n >0$ while $b_n^\dagger = b_{-n}$ and $c_n^\dagger = c_{-n}$  are creation operators. The ground state $|0\>$  is annihilated by all $b_n, c_n$ for $n >0$. The anti-commutation relations for $b_0, c_0$ define a Clifford-Dirac algebra whose unique irreducible representation has dimension 2. We shall define the ground state by, 
\begin{align}
b_0|0\> & = \< 0 | c_0 = 0    & b_n |0\>&=0 & c_n |0\> &=0 & n \geq 1
\end{align}
The Fock space $\mF$ is then generated by applying arbitrary polynomials in $b_n^\dagger$ and $c_n^\dagger$ for all possible values of $n >0$  to  the ground state $|0 \>$.

\subsubsection{The operator product expansion}

The fields $b(z)$ and $c(z)$ are holomorphic in $z$ on the annulus. The product $c(z) b(w)$ of operators, however, is singular as $z \to w$. This may be seen  by applying the operator to the ground state and using the fact that $|0\>$ is  annihilated by all $b_n$ with $n >0$, 
\bea
\label{5.OPE1}
c(z) b(w) |0\> = \sum_{m \in \ZZ} c_m z^{-m-1+h} \sum _{n \leq 0} b_n w^{-n-h} |0 \>
\eea
The contributions from $m \leq 0$ are non-singular  as $z \to w$ while for $m>0$ we may replace the product $c_m b_n$ by the anti-commutator $\{ c_m, b_n \} = \delta_{m+n,0}$  in view of $c_m |0 \>=0$. The resulting series is geometric in $w/z$, absolutely convergent for $|z/w|<1$, and may be analytically continued in $z$ and $w$ throughout the annulus, 
\bea
\label{5.OPE2}
c(z) b(w) |0\> = \sum_{m>0} z^{-m-1+h} w^{m-h} |0 \> + \hbox{ regular } 
= { 1 \over z-w} \, |0\> + \hbox{ regular } 
\eea
where ``regular" stands for all contributions that have a finite limit as $z \to w$.  The pole in $z-w$ arises due to the infinite nature of the series and is universal in the following sense.  If we apply the operator $c(z) b(w)$ to an arbitrary Fock space state, obtained by acting on $|0\>$ with an arbitrary polynomial in $b_n^\dagger $ and $c_n^\dagger $ with $n>0$, then only a finite number of terms in the infinite series (\ref{5.OPE2}) will be affected, which leaves the singularity unchanged. For this reason, one promotes the singularity to an operator statement,
\bea
\label{5.OPE3}
c(z) b(w) = { I_\mF \over z-w} + \hbox{ regular } 
\eea
where $I_\mF$ is the identity operator in the Fock space $\mF$. The Laurent series for the operator $b(z)c(w)$ is absolutely convergent for $|w/z|<1$, produces the pole $I_\mF/(z-w)$, and may be analytically continued to arbitrary $z,w $ in the annulus. Comparing with the pole in $c(z)b(w)$, we observe that the sign reversal is consistent with the fact that $b$ and $c$ satisfy anti-commutation relations. Actually, the pole term is only the first in an infinite Laurent series expansion of the operator $c(z)b(w)$ in powers of $z-w$. The general expansion is referred to as the \textit{operator product expansion} of the operators $c(z)$ and $b(w)$, 
\bea
c(z) b(w) = {I_\mF \over z-w} + \sum _{n=0}^\infty (z-w)^n j^{(n+1)}(w)
\eea
The coefficients $j^{(n+1)}(w)$ for $n >0$ are conformal primary fields of weight $(n+1,0)$. The transformation law for $j^{(1)}$ is more complicated and will be discussed later. The operators $b(z) b(w)$ and $c(z) c(w)$ similarly have operator product expansions. In view of the anti-commutation relations (\ref{5.ACom}) we have $b(z)^2=c(z)^2=0$ and,  
\bea
\label{5.OPE5}
b(z) b(w) &= & (z-w) (\p_w b(w)) \, b(w) + \cO\big ( (z-w)^2\big )
\no \\
c(z) c(w) & = & (z-w) (\p_w c(w)) \, c(w) + \cO\big ( (z-w)^2\big )
\eea
so that the operator product expansion is entirely regular.

\subsection{Correlators of the $bc$ system on the torus}

We now consider the $bc$ system on an arbitrary torus $\CC/\Lambda$ with $\Lambda = \ZZ \oplus \tau \ZZ$ and $\Im(\tau)>0$. The fields $b$ and $c$ are meromorphic and doubly periodic, and are thus elliptic functions. On the torus the $bc$ systems for different integer values of $h$ are all isomorphic to one another, in view of the fact that the torus has a holomorphic differential $dz$ which is nowhere vanishing. As a result, a field $\phi(z) (dz)^h$ of conformal weight $(h,0)$  is isomorphic to a scalar field $\phi(z)$ of weight $(0,0)$.  This isomorphism allows us for a  direct evaluation of all correlators of $b,c$ fields on the torus.  A correlator is the matrix element in the vacuum state $|0\>$ of a product of $b$ and $c$ fields  evaluated at distinct points on $\Sigma$, 
\bea
A= \< 0 | c(z_1) \cdots c(z_N) b(w_1) \cdots b(w_M) |0\>
\eea
The correlators are meromorphic in $z_n$ for all $n=1,\cdots, N$ and in $w_m$ for all $m=1,\cdots, M$. 
We shall now show that the positions of their zeros and poles in each variable are known from the operator product expansions discussed earlier. In view of (\ref{5.OPE5}), the correlator has a simple zero when two points $z_n$ and $z_{n'}$ coincide, as well as when two points $w_m$ and $w_{m'}$ coincide. In view of (\ref{5.OPE3}), the correlator has a simple pole whenever a point $z_n$ approaches a point $w_m$, whose residue is given by a correlator with one fewer $b$ and one fewer $c$,
\bea
A = { (-)^{N-1+m-n} \over z_n - w_m} 
\< 0 | c(z_1) \cdots \widehat{c(z_n)} \cdots c(z_N) b(w_1) \cdots \widehat{b(w_m)} \cdots  b(w_M) |0\>
+ \hbox{ regular }
\quad
\eea
The sign factor arises from permuting the positions of the fields, the hatted operators are to be omitted,    and ``regular" includes all terms that are non-singular as $z_n \to w_m$. 

\sm

Considered as a function of $z_n$,  the correlator  has precisely $M$ simple poles at $w_1, \cdots, w_M$. In view of Theorem \ref{thm:1} it must have $M$ zeros in $z_n$. These zeros must include the $N-1$ points $z_{n'}$ with $n'\not= n$, so we must have $M\geq N-1$. Similarly, as a function of $w_m$, the correlator has precisely $N$ simple poles at $z_1, \cdots, z_N$. In view of Theorem \ref{thm:1} it must have $N$ zeros in $w_m$  which must include the $M-1$ points $w_{m'}$ with $m' \not= m$, so we must also have $N \geq M-1$. Combing both inequalities allows for either $M=N$ or $M-N=\pm 1$. The latter is excluded by symmetry under reversing the sign of both $b$ and $c$. Thus only $M=N$ gives a non-vanishing correlator. We shall now  evaluate the correlators $A$ for $M=N$ in terms of $\tet$-functions using Theorems \ref{thm:1},  \ref{thm:2.5}, and \ref{thm:2.7}.

\subsubsection{The 4-point correlator}

We begin with the correlator $\<0| c(z_1) c(z_2) b(w_1) b(w_2) |0\>$. As a function of $z_1$, it has a simple zero at $z_2$, simple poles at $w_1$ and $w_2$, and no other poles. In view of the second relation in Theorem \ref{thm:1}, it must have a second zero, whose position is determined by the third relation in Theorem \ref{thm:1} to be given by $-z_2+w_1+w_2$ (mod $\Lambda$). Proceeding analogously for $z_2$ using Theorem \ref{thm:2.5} we obtain the following expression for the correlator, 
\bea
\<0| c(z_1) c(z_2) b(w_1) b(w_2) |0\> = A_0 \, { \tet_1(z_1+z_2-w_1-w_2) \tet_1(z_1-z_2) \tet_1(w_1-w_2) \over 
\tet_1(z_1-w_1) \tet_1(z_1-w_2) \tet_1(z_2-w_1) \tet_1(z_2-w_2) }
\eea
Here, $A_0$ is independent of $z_1,z_2,w_1,w_2$ and all dependence on $\tau$  has been suppressed. We may use this expression to obtain the correlator $\<0 |c(z_1) b(w_2)|0\>$ by letting $z_2\to w_1$, 
\bea
\label{5.res}
\<0| c(z_1) c(z_2) b(w_1) b(w_2) |0\> = { 1 \over z_2-w_1} \, { A_0 \over \tet_1'(0)} + \hbox{ regular}
\eea
The two-point correlator is constant and saturated by the zero modes of the $c$ and $b$ fields, 
\bea
\<0| c(z_1) b(w_2) |0\> ={ A_0 \over \tet_1'(0)} 
\eea
consistently with the fact that no function on a torus can have a single simple pole. Alternatively, we may use Theorem \ref{thm:2.7} to express the correlator in terms of the Weierstrass $\zeta$-function and its derivatives. Since the poles in $z_1, z_2$ of the correlator at $w_1, w_2$ are simple, and must have opposite residue, we have,
\bea
\<0| c(z_1) c(z_2) b(w_1) b(w_2) |0\> = A_3+ \sum _{i=1}^2 A_i \big \{ \zeta (z_i-w_1) -  \zeta(z_i-w_2) \big \}
\eea
where $A_1, A_2, A_3$ are independent of $z_1, z_2,w_1, w_2$. Matching the residues with (\ref{5.res}), we find $A_1=-A_2=-A_0/\tet_1'(0)$. This condition also shows that the correlator has zeros at $z_1=z_2$ and $w_1=w_2$ provided $A_3=0$. We leave it as an exercise to the reader to express the correlator in terms of the Weierstrass $\wp$ function and to show that these three expressions may be related to one another with explicit identities.

\subsubsection{Arbitrary correlators}

Correlators for  arbitrary $N$ may be evaluated iteratively.  As a function of $z_1$ the correlator has $N$ simple poles at $w_1, \cdots, w_N$, and $N-1$ known zeros at $z_2,\cdots, z_N$. The missing zero is given by Theorem \ref{thm:2} by the sum $-\sum_{n=2}^N z_n+ \sum_{n=1}^N w_n$. Putting all together, we have,
\bea
A= \hat A_1 \, { \tet_1 (D)  \, \prod _{n=2}^N \tet_1(z_1-z_n) 
\over \prod _{n=1}^N \tet_1(z_1-w_n)}
\hskip 1in 
D= \sum _{n=1} ^N (z_n-w_n) 
\eea
where $\hat A_1$ is independent of $z_1$, but depends on all other $z_n$ and $w_n$. Again, all dependence on $\tau$ has been suppressed. Proceeding iteratively on the points $z_n$ we find that the factor $\tet_1(D)$ always accounts for the missing zero. Putting all together, we obtain, 
\bea
A = A_0 \,  { \tet_1 (D)  \, \prod _{m<n} \tet_1(z_m-z_n) \tet_1 (w_m-w_n) 
\over \prod _{m,n} \tet_1(z_m-w_n)}
\eea
where $A_0$ depends only on $\tau$. We leave it as an exercise for the reader to express the correlator also in terms of Weierstrass $\wp$- and $\zeta$-functions using Theorems \ref{thm:2} and \ref{thm:2.7} respectively.

\subsection{The scalar field }

A real-valued scalar field $\f$  is governed  by a second order differential operator on the torus, namely the Laplace-Beltrami operator. In local complex coordinates $(z,\bar z)$ on the torus $\Sigma =\CC/\Lambda$ with $\Lambda = \ZZ \oplus \tau \ZZ$, its action may be expressed as follows,
\bea
S[\f] = { 1 \over 2 \pi} \int _\Sigma d^2 z \, \p_z \f \, \pbz \f
\eea
The corresponding field equation  $\p_z \pbz \f=0$ implies that $\p_z \f$ is a locally holomorphic field while $\pbz \f$ is a locally anti-holomorphic field. On an annulus, these quantum fields admit  Laurent expansions,
\bea
i \p_z \f (z, \bar z) = \sum_{m \in \ZZ} \f_m \, z^{-m-1} \hskip 1in
i \pbz \f (z, \bar z) = \sum_{m \in \ZZ} \tilde \f_m \, \bar z^{-m-1}
\eea
where the coefficient operators $\f_m$ and $\tilde \f_m$  satisfy commutation relations,
\bea
{} [ \f_m, \f_n] = m \delta _{m+n,0} \hskip 1in {} [ \tilde \f_m, \tilde \f_n] = m \delta _{m+n,0}
\eea
We choose a polarization such that $\f_{-n} = \f_n^\dagger $ and $\tilde \f_{-n} = \tilde \f_n^\dagger$, and define the ground state $|0\>$ to be  annihilated by $\f_n|0 \>= \tilde \f_n |0\>=0$ for $n \geq0$. The Fock space is constructed by applying arbitrary polynomials in $\f_n^\dagger$ for $n >0$ to $|0\>$. Each such state has positive norm so that the Fock space for a scalar field is actually a  Hilbert space, which we denote by $\mF$.  

\sm

The scalar field theory is conformal and has two commuting Virasoro algebras generated by $L_m$ and $\tilde L_m$ whose expressions in terms of $\f_m$ and $\tilde \f_m$ are given as follows,\footnote{The offset $-\tfrac{1}{24}$ in $L_0$ arises from the conformal transformation of the expression on the cylinder (which has no such offset) to the annulus. It may also be viewed as the sum of the zero-point energies of the harmonic oscillators, using the fact that the sum of all positive integers $m$ gives $\zeta(-1)=-\tfrac{1}{12}$.}
 \bea
 L_0 = - { 1 \over 24} + \half \f_0^2 + \sum _{n=1}^\infty \f_{-n} \f_n
 \hskip 1in 
 L_m = \half \sum_{n\in \ZZ} \f_{m-n} \f_n \quad \hbox{ for } m \not=0
 \eea 
 and similarly for $\tilde L_m$.  The operators $\p_z \f$ and $\pbz \f$ are conformal primary fields of weights $(1,0)$ and $(0,1)$ respectively, and satisfy the corresponding transformations given in (\ref{5.Lphi}).
 
 \sm
 
 The operator product expansion of two holomorphic fields is given by,
 \bea
 \label{5.OPEf}
 \p_z \f(z) \p_w \f(w) = { 1 \over (z-w)^2} - 2 \, T(w) + \cO(z-w)
 \eea
 where the Laurent expansion of the stress tensor $T(w)$ for the holomorphic component of the scalar field is given in terms of the Virasoro generators $L_m$,
 \bea
 T(w) = \sum _{n\in \ZZ} L_n \, w^{-n-2}
 \eea
 and similarly for the anti-holomorphic field $\pbz \f$ and the generators $\tilde L_m$. 
 
 \sm 
 
We may attempt to use the operator product expansion (\ref{5.OPEf}) to evaluate the correlator of holomorphic fields $\p_z\f(z)$, just as we had in the case of the $bc$ system. For simplicity, we choose matrix elements  between the ground state $|0\>=|0,0\>$ on the torus $\Sigma$, 
\bea
\< 0 | \p_z \f(z_1) \cdots \p_z \f (z_N) |0\>
\eea
The correlator vanishes for odd $N$ and is meromorphic in all its points $z_n$ for even $N$. As a function of $z_n$,  the correlator has a double pole at every point $z_m$ with $m \not=n$. For example, for the 2-point correlator, the double pole shows that the correlator must be of the form, 
\bea
\< 0 | \p_z \f(z_1)  \p_z \f (z_2) |0\> = \wp(z_1-z_2) + A_4
\eea
where $A_4$ is independent of $z_1, z_2$. In contrast with the $bc$ system, the zeros of the correlator are not manifest now, and the constant $A_4$ cannot be fixed by using known zeros. It may be determined, however, by using not just the pole term in the operator product expansion, but also the term proportional to the stress tensor $T$, and we find $A_4=0$. For higher point correlators, however, the corresponding constant terms will not be so easy to evaluate. Instead we shall turn to evaluating the full correlators of the field $\f$ itself.

\subsection{The scalar Green function on the torus}

The scalar Laplace-Beltrami operator on the torus has a zero mode and is not invertible on all scalar functions. 
However, an inverse may be defined on its range, namely on functions orthogonal to constant functions on $\Sigma$. To construct this inverse, we shall present two different approaches, the first based on Fourier series, and the second using $\tet$-functions.

\subsubsection{The Laplacian on the torus}
\label{5.LapTor}

An arbitrary Riemannian metric on the torus $\Sigma$ may be rescaled to a flat metric using a $\Sigma$-dependent Weyl transformation, as shown in more detail in appendix \ref{sec:RS}. A further rescaling by a constant Weyl transformation allows us to obtain a flat metric with unit area. This reduced metric depends on a single complex modulus parameter $\tau \in \cH$, and we represent the torus  by the quotient $\Sigma = \CC/\Lambda$ for the lattice $\Lambda = \ZZ + \ZZ \tau$. The points on $\Sigma$ may be parametrized  by a complex coordinate $z= x + y \tau$ for $x,y \in \RR/\ZZ$, for example in the range $0 \leq x,y \leq 1$ (see Figure \ref{2.fig:2}). In these coordinates,  the flat metric $\mg$  and associated volume form $d \mu_\mg$ of unit area are given by,
\bea
\mg = { |dz|^2 \over \tau _2} = { |dx+ \tau dy|^2 \over \tau_2} 
\hskip 1in 
d\mu_\mg = {i \over 2} { d z \wedge d \bar z \over \tau_2} =  dx \wedge dy 
\eea
We define an inner product on the space of complex-valued functions $f_1,f_2 : \Sigma \mapsto \CC$ by,
\bea
\label{5.f1f2}
\< f_1 | f_2 \>  = \int _\Sigma d\mu _\mg \, \bar f_1 \, f_2
\eea
making the space of  bounded functions on $\Sigma$ into an $L^2 (\Sigma)$ Hilbert space.  The Cauchy-Riemann operators in these coordinates take the form, 
\bea
\p_z = { 1 \over 2 i \tau_2} ( \p_y - \bar \tau \p_x) 
\hskip 1in
\pbz = - { 1 \over 2 i \tau_2} ( \p_y - \tau \p_x) 
\eea
while the Laplace-Beltrami operator, or simply \textit{Laplacian},  is given by,
\bea
\Delta_\mg = - 4 \tau _2 \pbz \p_z = - { 1 \over \tau_2} (\p_y - \tau \p_x ) (\p_y - \bar \tau \p_x )
\eea
The Laplacian is self-adjoint with respect to the inner product (\ref{5.f1f2}) on $L^2 (\Sigma)$ and positive. It is invariant under translations in $x,y$ and thus commutes with the operators $i \p_x$ and $i \p_y$ that generate infinitesimal translations in the $x$ and $y$ directions respectively. For fixed $\tau $, the Laplacian $\Delta_\mg$  has a discrete spectrum with a single zero mode, namely the constant function on $\Sigma$, so that ${\rm Ker \Delta_\mg} = \CC$. Its range is the set of  functions orthogonal to constant functions.

\subsubsection{The scalar Green function via Fourier series}

Since the self-adjoint operators $i \p_x$ and $i \p_y$ mutually commute and commute with $\Delta _\mg$, the Laplacian is diagonal in a basis where $i \p_x$ and $i \p_y$  are simultaneously diagonal. Such a basis is given by the following orthonormal basis of doubly periodic functions on $\Sigma$,
\bea
\label{10.orthon}
f_{m,n}(x,y) & = & e^{2 \pi i (nx-my)} 
\no \\
\< f_{m',n'} | f_{m,n} \> & = & \delta _{m',m} \, \delta _{n',n}
\hskip 1in m,m',n,n' \in \ZZ
\eea
They also satisfy the completeness relation, 
\bea
\label{10,comp}
\sum_{m,n \in \ZZ} \overline{ f_{m,n}}(u,v) f_{m,n}(x,y) = \delta (x-u) \delta (y-v)
\eea
where the Dirac $\delta$ functions on the right are periodic in their argument with period 1.
One verifies that the Laplace operator is indeed diagonal in this basis, and we find, 
\bea
\Delta_\mg \, f_{m,n}= \lambda _{m,n}(\tau)   \, f_{m,n}
\hskip 1in 
\lambda _{m,n} (\tau) = 4 \pi^2 \, { |m+  n\tau |^2 \over \tau_2}
\eea
The zero eigenvalue corresponds to the constant function with $m=n=0$, while all other eigenvalues are strictly positive, as expected. The Green function is customarily defined as the inverse of $\Delta_\mg /4\pi$ on the space of functions orthogonal to constants. In view of translation invariance, we have $G(z,w|\tau)=G(z-w|\tau)$ with $z= x + y \tau$,  and $G(z|\tau)$ is given by,
\bea
\label{10.green1}
G(z|\tau) = \sum _{(m,n) \not= (0,0)} { \tau_2  \over \pi |m+n \tau|^2} \, e^{ 2\pi i (nx - my)}
\eea
Applying the Laplacian, and using the completeness relation (\ref{10.orthon}), one verifies,
\bea
\label{10.green2}
\Delta _\mg G(z-w|\tau) = 4 \pi \tau_2 \, \delta^{(2)} (z-w) - 4 \pi 
\eea
where the 2-dimensional $\delta$-function, with $w=u+\tau v$, is defined by,
\bea
\tau_2 \, \delta^{(2)} (z-w) =\delta (x-u) \delta (y-v)
\eea 
By construction, the right side of (\ref{10.green2})  is orthogonal to constant functions. Actually, the Green function obtained in (\ref{10.green1})  itself is orthogonal to constant functions,
\bea
\label{10.norm}
\int _\Sigma d \mu _\mg(z) \, G(z-w|\tau) =0
\eea
since the zero mode $(m,n)=(0,0)$ has been removed from the sum in (\ref{10.green1}).

\subsubsection{The scalar Green function via $\tet$-functions}

We may take equations (\ref{10.green2}) and (\ref{10.norm}) as defining $G$ and express both in terms of the complex variables $z,w$,
\bea
\label{10.green3}
-  \p_z \pbz G(z-w|\tau) =  \pi \delta^{(2)} (z-w) - { \pi \over \tau_2}
\eea
 The solution for $G$ is easy to construct in terms of $\tet$-functions, as we shall now show. This method will generalize to constructing the Green function on Riemann surfaces of arbitrary genus, whereas the method of Fourier series of the preceding section does not generalize since higher genus Riemann surfaces have no continuous isometries.
 
 \sm
 
We proceed as follows for fixed $\tau \in \cH$. For $|z-w|\ll 1$ and  $|z-w|\ll\tau_2$,  the global structure of the torus is negligible and the Green function must approach its expression on $\CC$  which is given by $G(z-w|\tau) \approx - \ln |z-w|^2$. The function $\tet_1(z-w|\tau)/\tet_1'(0|\tau)$ behaves as $z-w$ for small $|z-w|$ and is therefore a reasonable candidate to play the role of $z-w$ on the torus. However, $\tet_1(z-w|\tau)$ cannot be a doubly periodic function as it is holomorphic and would then have to be constant by Liouville's theorem. Still, given the transformation laws of $\tet_1(z-w|\tau)$ under $z \to z+1$ and $z \to z+\tau$ given in (\ref{2.theta-shift}), one readily finds a simple additive term to restore double periodicity, and $G(z,w|\tau)=G(z-w|\tau)$ is given by,
\bea
G(z|\tau) = - \ln \left | { \tet _1 (z|\tau) \over \tet _1 '(0|\tau)} \right |^2 + 2 \pi {(\Im z)^2 \over \tau_2 } + G_0 (\tau)
\eea
Upon applying the Laplace operator, the first term produces the $\delta$-function, while the second term produces the constant on the right side of (\ref{10.green3}).  The $z$-independent term $G_0(\tau)$ may be determined  by imposing the normalization condition (\ref{10.norm}). 

\sm

To compute $G_0(\tau)$ it is convenient to choose the range of integration $-\half \leq x,y \leq \half$. The integrals of the last two terms in $G$ are readily evaluated. To compute the   integral of the first term in $G$ we use the product formula for $\tet_1$ given in (\ref{tetprod}),  expand the logarithm of the factors $(1-q^n e^{ \pm 2 \pi i z})$ in powers of $q$, observe that their integral vanishes term-by-term, and finally compute the integral of $\ln |\sin (\pi z)|^2$ by the same expansion method. Putting all together, we obtain the properly normalized scalar Green function,
 \bea
 \label{10.green-theta}
G(z|\tau) = - \ln \left | { \tet _1 (z|\tau) \over \eta (\tau)} \right |^2 + 2 \pi {(\Im z)^2 \over \tau_2 }
\eea
The expression (\ref{10.green1}) for $G$ may be related to (\ref{10.green-theta}) by carrying out the sum over $m$ first  in (\ref{10.green1}) using the following Lemma.
{\lem
\label{Lipschitz}
The Lipschitz formula for $\Re(z)>0$, $\Re(s)>0$ and $0< \alpha \leq 1$ is given by,
\bea
\sum_{n \in \ZZ} { e^{2 \pi i n \a} \over (z+in)^s} = { (2 \pi)^s \over \Gamma(s)} \sum_{m=0} ^\infty (m+\a)^{s-1} \, e^{-2 \pi z(m+\a)}
\eea}
The proof may be found in Rademacher \cite{Rad2}.

\subsubsection{Modular properties}

A modular transformation $ \gamma \in SL(2,\ZZ)$ leaves the metric $\mg$ and the Laplacian $\Delta _\mg$  invariant provided $\tau$ and the complex coordinate $z$ transform according to the following rules,
\bea
 \tau' = \gamma \tau = { a \tau + b \over c \tau + d} 
\hskip 0.7in
z' = \gamma z= { z \over c \tau +d}
\hskip 0.7in
\gamma = \left ( \bma a & b \cr c & d \ema \right )
\eea
The transformations on the real coordinates are as follows, 
\bea
\left ( \bma  x' \cr  y' \ema \right ) = \left ( \bma a & -b \cr -c & d \ema \right ) \left ( \bma x \cr y \ema \right )
\eea
The spectrum of the Laplacian is invariant under $SL(2,\ZZ)$, as the eigenvalues $\lambda_{m,n}(\tau)$ of $\Delta_\mg$ transform as follows, 
\bea
\lambda _{m,n} (x,y) =  \lambda _{m',n'}(x',y')
\hskip 1in 
\left ( \bma  m' \cr  n' \ema \right ) = \left ( \bma a & -b \cr -c & d \ema \right ) \left ( \bma m \cr n \ema \right )
\eea
in view of the relation $n'x'-m'y'=nx-my$. It readily follows that also the normalized Green function, given in (\ref{10.green1}), is modular invariant,
\bea
\label{10.G-mod}
G\left ( {z-w \over c \tau + d} \bigg | { a \tau + b \over c \tau +d} \right ) = G(z-w|\tau)
\eea 
Modular invariance may be verified directly on the expression for the Green function in terms of $\tet$-functions in (\ref{10.green-theta}) by using the transformation laws of $\tet_1$ given in (\ref{4.mod2}).

\subsection{Scalar determinant and the first Kronecker limit formula}

The functional determinant of $\Delta= \Delta_\mg/4 \pi$, considered as an operator on the subspace of functions orthogonal to constants,  is formally given by the product of the eigenvalues $\lambda_{m,n} (\tau)/4 \pi $. However, this product does not converge.  To define the determinant we introduce the following $\zeta$-function associated with $\Delta$, 
\bea
\label{4d5}
\zeta _{\Delta} (\tau, s) = \sum _{(m,n ) \not = (0,0)}   \left ( {4 \pi \over \lambda_{m,n}} \right )^s= E_s(\tau) 
\eea
where $E_s(\tau)$ is the non-holomorphic Eisenstein series introduced in section \ref{sec:4.2}. The series for $E_s(\tau)$ is absolutely convergent for $\Re(s) >1$. Its analytic continuation to $s \in \CC$ has a simple pole at $s=1$, is modular invariant, and satisfies the functional relation, 
\bea
\label{4d8}
\Gamma (s) E_s(\tau) = \Gamma (1-s) E_{1-s}(\tau)
\eea
Taking the limit $s \to 0$ of (\ref{4.d12}) gives  $ \zeta_\Delta(\tau,0)=-1$, while the determinant is given by,
\bea
\label{4d5a}
\ln \Det ' \Delta = - \zeta_{\Delta}  '(\tau, 0) 
=
- \p_s E_s(\tau) \Big |_{s=0} 
\eea
To compute the determinant, we use the expressions for $E_s(\tau)$ in terms of the functions $A_s(\tau)$ and $B_s(\tau)$ defined in (\ref{4.d9}) and (\ref{4.d10}), 
\bea
\zeta'_\Delta(\tau, 0) = A_0(\tau)  + {\pi \tau_2 \over 3} - \ln \tau_2 - \ln (4 \pi) 
\eea
The last three terms on the right side arise from differentiating $B_s(\tau)/\Gamma(s)$ at $s=0$. To compute $A_0(\tau)$, we simplify the summand in (\ref{4.d10}) by using the expression for the modified Bessel function in terms of elementary functions,  $K_{-\half}(x)= \sqrt{\pi/2x} \, e^{-x}$. We discover that $A_0(\tau)$ becomes a sum over holomorphic and anti-holomorphic dependence on $\tau$, 
\bea
A_0(\tau) = \sum _{m,n \not=0} { 1 \over |m| } \, \exp \Big \{ 2 \pi i mn  \tau _1 -2 \pi |mn| \tau_2  \Big \} 
\eea
which may be compactly expressed as follows, 
\bea
A_0(\tau) = -2 \ln \prod _{m=1}^\infty (1-q^m)(1-\bar q^m)
\eea
We may express the final combined result in terms of the Dedekind $\eta(\tau)$ function, 
\bea
\zeta'_\Delta(\tau, 0) = - \ln \left ( 4 \pi \tau _2  |\eta (\tau)|^4 \right ) 
\eea
Combining this with $\zeta _\Delta (\tau, 0)=-1$ and (\ref{4d5a}), we obtain, 
\bea
 \Det ' \Delta =   4 \pi \tau _2  |\eta (\tau)|^4 = 4 \pi  \Det ' \Delta _\mg
\eea
which is manifestly modular invariant.

\subsubsection{The first Kronecker limit formula}

The first Kronecker limit formula gives the residue and the finite part of $\zeta_\Delta(\tau, s)$ at its pole $s=1$ by the following expression,
\bea
\zeta_\Delta (\tau, s) = { a_{-1} \over s-1} + a_0 + \cO (s-1)
\eea
Using the functional relation (\ref{4d8}), it is immediate that $a_{-1}=-1$. From the same formula we get also $a_0$ by,
\bea
a_0 = { d \over ds}  \Big ( (s-1) E_s(\tau) \Big ) \Big |_{s=1} = - \ln \left ( 4 \pi \tau _2  |\eta (\tau)|^4 \right )  + 2 \gamma
\eea
 where $\gamma $ is Euler's constant as it arises in $\Gamma '(1)=-\gamma$.

 \subsubsection{The scalar partition function}
\label{sec:partitions}

The scalar determinant is intimately related to the partition function for the free scalar field in statistical mechanics. A partition function $Z(\beta)$  for a Hamiltonian $H$ acting in a Hilbert space $\mF$, evaluated at an inverse temperature $\beta$ is given by, 
\bea
Z(\beta)= \Tr _\mF \left ( e^{-\beta H} \right )
\eea
where $\Tr_\mF$ is the trace evaluated in the Hilbert space $\mF$. The scalar field has two independent sets of oscillators, namely $\f_m$ and $\tilde \f_m$, with independent Virasoro generators. In particular, the generators $L_0$ and $\tilde L_0$ are given by a sum of decoupled harmonic oscillator Hamiltonians plus the term due to the zero mode. 
Instead of the real-valued inverse temperature $\beta>0$, we shall complexify $\beta$ and  let $\beta \to -2 \pi i \tau$ whose real part is still positive in view of $\Im(\tau) >0$. The partition function for $L_0$ is then given as follows,
\bea
\Tr \left ( q^{ L_0} \right ) = q^{-\tfrac{1}{24} + \tfrac{1}{2} \f_0^2} \prod _{n=1}^ \infty { 1 \over (1-q^n)}
= q^{\tfrac{1}{2} \f_0^2} \, \eta (\tau)^{-1}
\eea 
where the trace is taken only over the oscillator part of the Hilbert space $\mF$. Proceeding similarly for $\tilde L_0$ and complexifying its inverse temperature $\beta \to 2 \pi i \bar \tau$, we obtain the combined partition function, 
\bea
\Tr  \left ( q^{L_0} \bar q ^{\tilde L_0} \right )
= q^{\tfrac{1}{2} \f_0^2} \bar q^{\tfrac{1}{2} \tilde \f_0^2} { 1 \over |\eta (\tau)|^2}
\eea
The zero mode operators $\f_0$ and $\tilde \f_0$ are self-adjoint and have a continuous spectrum of real eigenvalues. Physically, these eigenvalues are the momenta of left- and right-movers. Periodicity in $\tau_1 \to \tau_1 +1$ requires $\f_0^2 - \tilde \f_0^2 \in 2 \ZZ$. When the field $\f$ takes values in $\RR$, we require the eigenvalues of $\f_0$ and $\tilde \f_0$ to be equal to one another, and to range over $\RR$. The full trace is then given by,
\bea
\Tr _\mH \left ( q^{L_0} \bar q ^{\tilde L_0} \right ) = 
{ 1 \over |\eta (\tau)|^2} \int _\RR dp \, | q | ^{p^2} = { 1 \over \sqrt{ \tau_2} \, |\eta (\tau)|^2} = \left ( \Det ' \Delta _\mg \right )^{-{1 \over 2}}
\eea
which is precisely the modular invariant partition function evaluated using the functional integral.

\subsection{Spinor fields}

The orthogonal Lie algebra $SO(d)$ has spinor representations in addition to its tensorial representations. A spinor representation may be viewed either as  a double-valued representation of the group $SO(d)$ or as a single-valued representation of the double cover $Spin(d)$. A manifold supports spinor fields provided it is orientable and its second Stiefel-Whitney class vanishes. These conditions are satisfied in $d=2$ for Riemann surfaces.  In terms of bundles, one decomposes the cotangent bundle $T^*\Sigma$ into a direct sum of its holomorphic and anti-holomorphic part $T^*\Sigma = K \oplus K^*$ where $K$ is referred to as the canonical bundle. A spin bundle $S$ is a holomorphic line bundle whose square is isomorphic to the canonical bundle $S^{\otimes 2} \approx K$, and a spinor field is then an operator-valued section of $S$. For more details on Riemann surfaces see appendix \ref{sec:RS}, and for the general construction of line bundles on an arbitrary Riemann surface see appendix \ref{sec:LB}.

\sm

Different realizations of spinor fields, corresponding to different spin bundles,  on a given Riemann surface $\Sigma$,  are referred to as  \textit{spin structures}. They are labeled by  the first homology class of the surface with $\ZZ_2$ coefficients, namely $H_1(\Sigma, \ZZ_2)$. For example, the homology group of a compact genus $g$ Riemann surface is isomorphic to $\ZZ^{2g}$ and therefore the different spin structures are labeled by $\ZZ_2$ on each one of the $2g$ independent homology cycle. 

\sm

For the annulus, there is a single homology cycle, and thus two different spin structures, referred to as Neveu-Schwarz (NS) or Ramond (R). Denoting a holomorphic spinor field of conformal weight $(h,0)$ by $\psi$, its Laurent series on the cylinder and on the annulus, with coordinates $z'=x+iy$ and   $z=e^{2 \pi i z'}$ respectively,  are given as follows, 
\begin{align}
\hbox{NS} && \psi '(z')  & = \sum _{r \in \ZZ+\half} \psi_r \, e^{-2 \pi i r(x+iy)} & \psi (z) & = \sum_{r \in \ZZ+\half} 
\psi_r \, z^{-r -h}
\no \\
\hbox{R} \, && \psi '(z')  & = \sum _{n \in \ZZ} \psi_n \, e^{-2 \pi i n(x+iy)} & \psi (z) & = \sum_{n \in \ZZ} 
\psi_n \, z^{-r -h}
\end{align}
A free spinor field on a Riemann surface  is governed by a first order differential operator, namely by a generalized  $bc$ system in which the fields $b$ and $c$ are allowed either NS or R identifications. An important special case is that of a spinor field of conformal weight $(\half, 0)$, in which case the fields $b$ and $c$ have the same conformal weight $(\half, 0)$ and are quantized using anti-commutation relations as in the $bc$ system we discussed earlier.\footnote{Another important case is the superghost $\beta \gamma$ system where $b=\beta$ has weight $({3 \over 2}, 0)$ and $c=\gamma$ has weight $(-\half,0)$ and the oscillators obey commutation relations.}  In this case the NS spin structure makes the field $\psi(z)$ single-valued on the annulus, while the R spin structure makes the field $\psi(z)$ double-valued on the annulus. 

\sm

For the torus, a canonical choice of homology cycles $\mA$ and $\mB$ is given by the identifications $z\to z+1$ and $z \to z+\tau$ respectively. The four spin structures are distinguished according to whether the field is periodic or anti-periodic around each cycle. The signs are inherited from the cylinder, 
\bea
\psi (z+1) & = & - e^{2 \pi i \a} \psi(z) 
\no \\
\psi (z+\tau) & = & - e^{2 \pi i \beta} \psi(z) 
\eea
for $\a, \b \in \{ 0 , \thalf \}$ (mod 1) with the values 0 and $\thalf$ corresponding to NS and R respectively. Under the modular transformations $S$ and $T$ the four spin structures transform as follows, 
\begin{align} 
S: &  \left [ \bma \a \\ \b \ema \right ] \rightarrow \left [ \bma \b \\ \a \ema \right ]
& T: & \left [ \bma \a \\ \b \ema \right ] \rightarrow \left [ \bma \a \\ \a+\b +\thalf  \ema \right ]
\end{align}
where the addition $\a+\b+\thalf$ is evaluated mod 1. The spin structure $\a=\b=\thalf $ is referred to as the \textit{odd spin structure}, while the three other spin structures are referred to as \textit{even spin structures}. Under modular transformations, the odd spin structure maps to itself, while the three even spin structures are permuted amongst one another. This transformation rule precisely corresponds to the transformation rule for Jacobi $\tet$-functions with half-integer characteristics given in (\ref{4.mod2}). This is no accident, as will be further clarified below. Furthermore, in view of the existence of the holomorphic and nowhere vanishing differential $dz$, spinor fields on the torus for a given spin structure but different conformal weights $(h,0)$ are all isomorphic to one another. Therefore, we may focus attention on the case $h=\half$. For two functions $f_1$ and $f_2$ with the same spin structure (or more accurately for two sections of a given spin bundle $S$), we define a Hermitian inner product,
\bea
\<f_1 |f_2\> = \int _\Sigma d\mu_\mg \bar f_1 f_2
\eea
which has a positive definite norm and produces a Hilbert space $L^2(\Sigma, S)$.

\subsection{Spinor Green functions on the torus}

In this subsection, we shall derive the Green function for the spinor field on a torus $\Sigma = \CC/\Lambda$ with $\Lambda = \ZZ \oplus \tau \ZZ$ for  each one of the four spin structures. Just as we did for the scalar field, we offer two perspectives: a calculation via Fourier analysis and a calculation directly in terms of $\tet$-functions.

\subsubsection{Odd spin structure Green function}

For the odd spin structure $\a=\b=\half $, the Green function may be obtained from the results of the scalar case, because the spinor field  is doubly periodic and has the constant spinor as a zero mode. We define the Green function $S_1(z|\tau)$ that satisfies the following equation, 
\bea
\pbz S_1(z-w|\tau) = \pi \delta ^{(2)}(z-w) - {\pi \over \tau_2}
\eea
by analogy with the Laplace equation satisfied by the scalar Green function $G(z-w|\tau)$ in (\ref{10.green3}). 
The solution is given in terms of the scalar Green function by,
\bea
S_1(z-w|\tau) = - \p_z G(z-w|\tau) = \p_z \ln \tet_1(z-w|\tau) + { 2 \pi i \over \tau_2} \, \Im (z-w)
\eea
up to an additive constant which amounts to adding a term proportional to the Dirac zero mode.
Since $S_1(z-w|\tau)$ has a single pole in $z$ at $w$, and is doubly periodic, it cannot possibly be meromorphic
as no elliptic functions with a single simple pole exist. The second term on the right is present to compensate. 
Alternatively, a term-by-term differentiation of the Kronecker-Eisenstein series of $G(z-w|\tau)$ in (\ref{10.green1}) gives the following expression for $S_1$,
\bea
S_1(z-w|\tau) = \sum _{(m,n) \not= (0,0)} {  e^{ 2 \pi i (nx-my)} \over m + n \tau} 
\eea
The sum is only conditionally convergent and requires regularization. An alternative Green function $\tilde S_1(z;w_1,w_2|\tau) $ allows for two poles in $z$ at $w_1$ and $w_2$ with opposite residue. Requiring anti-symmetry in $w_1, w_2$ produces a unique Green function, 
\bea
\tilde S_1(z;w_1,w_2|\tau ) & = & S_1(z-w_1|\tau) + S_1(w_1-w_2|\tau) + S_1 (w_2-z|\tau)
\no \\
& = & \zeta (z-w_1|\tau) + \zeta (w_1-w_2|\tau) + \zeta(w_2-z|\tau)
\eea
where the expression is recast in terms of the Weierstrass $\zeta$-function on the second line above.

\subsubsection{Even spin structure Green functions via Fourier analysis}

The chiral Dirac operator $\pbz$ is not self-adjoint. However, it is a linear combination with constant coefficients of two self-adjoint operators $i \p_x$ and $i \p_y$,
\bea
\label{5.pbz}
\pbz = { 1 \over 2 \tau_2} ( i \p_y - \tau\,  i \p_x)
\eea
Since the operators $i\p_x$ and $i \p_y$ mutually commute, the operator $\pbz$ may be diagonalized in the basis in which $i\p_x$ and $i \p_y$ are diagonal. However, since the coefficients of the linear combination in (\ref{5.pbz}) are complex, the eigenvalues of $\pbz$ are complex. 

\sm

For a given even spin structure $[\kappa ]= \ss$, the operators $i \p_x$ and $i \p_y$ are diagonalized in the following basis, 
\bea
f [\kappa]  _{m,n} (x,y)  =  \exp \Big \{ 2 \pi i ( (n+\beta+\thalf )x - (m+\alpha +\thalf ) y) \Big \} 
\eea
The eigenvalues of the operator $\pbz$ are given by, 
\bea
\lambda [\kappa]  _{m,n} = { \pi \over \tau_2} \Big (m+\a + \thalf  + (n + \b + \thalf ) \tau \Big )
\eea
Since $\Im (\tau) >0$ and $\a, \b$ do not both equal $\thalf$ mod 1, none of the eigenvalues vanishes. The Green function for the operator $\tau_2 \pbz$ is then given by,
\bea
S[\kappa]  (z|\tau) = { 1 \over \tau_2} \sum_{m,n \in \ZZ} { f[\kappa]  _{m,n} (x,y) \over \lambda [\kappa] _{m,n}}
=\sum_{m,n \in \ZZ} { e^{2 \pi i \{ (n +\b +\half ) x - (m+\a+\half ) y \} } \over m+ \a+ \half  + (n +\b+\half ) \tau}
\eea
so that $S[\kappa]$ satisfies, 
\bea
\pbz S[\kappa] (z-w|\tau) = \pi \delta (z-w)
\eea
where the integral of $\delta(z-w)$ over $\Sigma$ is normalized to $\tau_2$.

\subsubsection{Even spin structure Green functions via $\tet$-functions}

We seek a meromorphic function $S[\kappa](z-w|\tau)$, which has a single simple pole  in a fundamental parallelogram for $\Lambda$, is doubly periodic with periods in $2 \Lambda$, and behaves as follows under translations in $\Lambda /(2 \Lambda)$, 
\bea
S[\kappa ] (z+1|\tau ) & = & - e^{2 \pi i \alpha} \, S [\kappa ] (z|\tau) \hskip 1in [\kappa] = \ss
\no \\ 
S[\kappa ] (z+\tau|\tau ) &  = & - e^{2 \pi i \beta} \,  S [\kappa ] (z|\tau)
\eea
For even spin structures $\kappa$, the simple pole in $\CC/\Lambda$ and the above translations are realized uniquely by the ratio $\tet [\kappa] (z|\tau)/\tet_1(z|\tau)$, and the normalization of the residue at $z=w$ gives, 
\bea
S [\kappa]  (z-w|\tau) = { \tet [\kappa]  (z-w|\tau) 
\, \tet _1 '(0|\tau) \over \tet [\kappa]  (0|\tau) \, \tet _1 (z-w|\tau) }
\eea
As an elliptic function in $\CC/(2 \Lambda)$, the \textit{Szeg\"o kernel} $S[\kappa](z-w|\tau)$ has four simple poles, namely at $z\equiv 0, \tau+1$ with residue 1, and at $z \equiv 1, \tau$ with residue $-1$.

\subsection{Spinor determinant and the second Kronecker limit formula}

The eigenvalues of the chiral Dirac operator $\pbz$ are complex, rendering the $\zeta$-function approach to evaluating its determinant inapplicable. Instead we shall make use of the fact that the dependence of the operator $\tau_2 \pbz$  on the modulus $\tau$ is holomorphic and we shall seek to construct an object $\Det (\pbz) $ that is holomorphic in $\tau$. 

\sm

What we do know how to evaluate using $\zeta$-function methods  is the determinant of the Laplace operator $\Delta _\kappa=  -4 \tau_2 \p_z \pbz$ on a spinor with even spin structure $\kappa$. The Laplace operator is self-adjoint, has positive spectrum, and no zero modes,
\bea
\lambda [\kappa] _{m,n} = { 4 \pi^2 \over \tau_2} \big | m+\a +(n+\beta) \tau \big |^2
\hskip 1in
\kappa = \left [ \bma \half +\a \\ \half +\b \ema \right ]
\eea
In fact we may generalize the problem by taking the characteristics $\a, \b$ to be arbitrary with the sole restriction that $(\a,\b) \not \equiv (0,0) $ mod 1.  The corresponding $\zeta$-function is defined by the Kronecker-Eisenstein sum,
\bea
\zeta _{\Delta_\kappa}  (\tau, s) = \sum _{m,n \in \ZZ} { \tau_2^s \over \pi^s |m+\alpha + (n + \beta )\tau|^{2s}}
\eea
which is absolutely convergent for $\Re(s)>1$. We shall show below that $\zeta_{\Delta_\kappa}(\tau, s)$ may be analytically continued in $s$ throughout the complex plane. The determinant is obtained from this analytic continuation as usual, 
\bea
\ln \Det \Delta _\kappa = - \zeta _{\Delta _\kappa} '(\tau, 0)
\eea
We recast the Kronecker-Eisenstein sum in terms of the following  heat-kernel representation, 
\bea
\label{4d6}
\Gamma (s) \zeta_{\Delta _\kappa}  (\tau, s) = \int _0 ^\infty { dt \over t} \, t^s  \sum _{m,n\in \ZZ} 
\exp \left \{  - { \pi t \over \tau_2}  \big |m+\alpha  +(n+\beta) \tau \big |^2  \right \}  
\eea
and Poisson re-sum either in the summation variable  $m$ if $\beta \not=0$ or in $n$  if $\a\not=0$ (but not in both). We shall work out the case $\beta \not=0$,  the other case being analogous, using the formula, 
\bea
\sum _{m,n \in \ZZ}  e^{- \pi t |m+\alpha +(n+\beta) \tau|^2/\tau_2} 
= \sqrt{ { \tau _2 \over t}} \sum _{m,n \in \ZZ} e^{2 \pi i m (\alpha + (n+\beta)  \tau_1} \, e^{ - \pi m^2\tau_2/t - \pi (n+\beta)^2 t \tau_2}
\eea
Substituting this result into (\ref{4d6}) and  splitting the sum over $m,n$ into contributions for $m  \not=0 $ and $m=0$, we have $\Gamma (s)  \zeta_{\Delta _\kappa} (\tau,s) =  A_s(\tau)+B_s(\tau)$ with,
\bea
\label{4d9}
A_s(\tau) & = & \sqrt{\tau_2} \sum _{m \not=0} \sum _n e^{2 \pi i m(\alpha + (n+\beta)  \tau_1)} 
\int _0^\infty { dt \over t} \, 
t^{s-\half}  \, e^{- \pi \tau_2 (m^2/t + (n+\beta) ^2 t)}
\no \\
B_s(\tau) & = &  \sqrt{\tau_2} \int _0^\infty { dt \over t} \, 
 t^{s-\half}  \sum _{n}   \,  e^{ - \pi (n+\beta) ^2 t \tau_2}
 \eea
 The integral $A_s(\tau)$ is manifestly holomorphic in $s$ throughout $\CC$, and may be evaluated  by performing the $t$-integral in terms of a Bessel function, 
\bea
\label{4d10a}
A_s(\tau) = 2 \sqrt{\tau_2} \sum _{m \not=0} \sum _n 
\left | { m \over n+\beta } \right |^{s-\half} e^{2 \pi i m (\alpha + (n+\beta)  \tau_1)} 
K_{s-\half} \big ( 2 \pi \tau_2  |m(n+\beta) | \big )
\eea
To evaluate the determinant, we only need $A_0(\tau)$, which may be read off from the above formula by setting $s=0$
and using the expression for the spherical Bessel $K_{-\half} (x) = \sqrt{\pi/2x} \, e^{-x}$, 
\bea
A_0(\tau) = \sum _{m=1}^ \infty \sum _n { 1 \over m} \, e^{-2 \pi \tau_2 m |n+\beta|} \left ( 
e^{ 2 \pi i m (\alpha + (n+\beta) \tau_1) } + e^{ - 2 \pi i m (\alpha + (n+\beta) \tau_1) } \right )
\eea
Without loss of generality we may take $0< \beta < 1$ to carry out the sum and obtain,
\bea
A_0(\tau) = - \sum _{n=0}^\infty  \ln \left | 1 - q^{n+\beta} \, e^{2 \pi i \alpha} \right |^2
- \sum _{n=1}^\infty \ln \left | 1 - q^{n-\beta} \, e^{- 2 \pi i \alpha} \right |^2
\eea
Finally, we compute $B_s(\tau)$ by performing the integrals under the sum,
\bea
B_s(\tau) =  \Gamma (s-\thalf) \pi ^{\half -s} \tau_2 ^{1-s} \sum _n { 1 \over |n+\beta|^{2s-1}}
 \eea
The sum is absolutely convergent for $\Re(s)>1$ and may be evaluated in terms of Hurwitz $\zeta$-functions. The analytic continuation of the Hurwitz $\zeta$-function is well-known, and the final result is given by,
\bea
\zeta _{\Delta_\kappa}  '(\tau, 0) = - \ln \left | { \tet [\alpha \, \beta ] (0|\tau) \over \eta (\tau) } \right |^2
\eea
This result is a version of the \textit{second Kronecker limit formula}.

\sm

The determinant obtained above is for the $bc$ system in which both $b$ and $c$ are conformal spinor fields of weight $(\thalf, 0)$, along with their complex conjugate fields. The natural candidate for the determinant of the chiral operator $\pbz$ is obtained by retaining only the part with holomorphic dependence on $\tau$, 
\bea
\Det \, \pbz \Big |_{bc} = { \tet [\alpha \, \beta ] (0|\tau) \over \eta (\tau) }  
\eea
The spinor fields that are required in string theory for the worldsheet matter fermions are further restricted to $\psi=b=c$, so that the $b$ and $c$ fields are actually identical fields. The functional chiral determinant of the operator $\pbz$  in this case is the square root of the determinant of the $bc$ system, 
\bea
\Det \, \pbz \Big |_\psi = { \tet [\alpha \, \beta ] (0|\tau)^\half  \over \eta (\tau)^\half }  
\eea
Neither  determinant is modular invariant or even covariant. Physicists say that there is a global anomaly, which manifests the incompatibility between maintaining holomorphicity and modular invariance. We may try a combination similar to the one we established in the non-chiral case, 
\bea 
\sum _{i=2}^4 { \tet _i (0|\tau)^{{N \over 2}}  \over \eta (\tau)^{{N \over 2}}  } =
{ \tet [00 ] (0|\tau)^{{N \over 2}}  \over \eta (\tau)^{{N \over 2}}  } + 
{ \tet [0\half  ] (0|\tau)^{{N \over 2}}  \over \eta (\tau)^{{N \over 2}}  } + 
{ \tet [\half 0 ] (0|\tau)^{{N \over 2}}  \over \eta (\tau)^{{N \over 2}}  } 
\eea
The smallest value of $N$ for which we have good modular transformations is $N=16$, and more generally any multiple of $16$. In those cases, the sum of the numerators transforms as a modular form of weight $N/4$.
The first modular invariant is encountered as $N=24$ for 48 chiral fermions.

\newpage

\subsection*{$\bullet$ Bibliographical notes}

A systematic introduction to quantum field theory and more specifically  conformal field theory in two dimensions is presented in the book by Di Francesco, Mathieu, and S\'en\'echal~\cite{DMS}, which also has an extensive bibliography. The lecture notes by Ginsparg~\cite{Ginsparg} provide an excellent introduction. 

\sm

The development of two-dimensional conformal field theory and the use of radial quantization date back to~\cite{Fubini:1972mf}. Modern conformal field theory,  based on  the representation theory of the Virasoro algebra, was developed in \cite{Belavin:1984vu}, where minimal models were also introduced. The $bc$ system was discussed in the context of string perturbation theory in \cite{Friedan:1985ge}.  Functional determinants of Laplace and Cauchy-Riemann operators on the torus  are evaluated there as well as in the classic paper by Polchinski~\cite{Polchinski:1985zf}. The Kronecker limit formulas are discussed in detail in the lecture notes by Siegel~\cite{Siegel5}, as well as in \cite{iwan2}. 

\sm

On higher genus Riemann surfaces, functional determinants of Laplace operators  were considered in \cite{RS} in connection with Ray-Singer torsion, in \cite{Hej}, and in \cite{DF} in connection with Reidemeister torsion, and were evaluated in terms of the Selberg zeta function in \cite{DHoker:1986eaw, Sarnak}. An overview is presented in the booklet by Jorgensen and Lang \cite{JL}.  Functional determinants of Laplace and Cauchy-Riemann operators were evaluated in terms of Riemann $\tet$-functions via chiral bosonization in \cite{Verlinde:1986kw} and \cite{Alvarez-Gaume:1987wwg} by building on the properties of the holomorphic anomaly established in  \cite{Belavin:1986cy} and \cite{Bost:1986uy}.

\newpage

\section{Congruence subgroups and modular curves}
\setcounter{equation}{0}
\label{sec:Cong}

The modular group $SL(2,\ZZ)$ possesses an infinite number of non-Abelian subgroups. Various objects encountered in the previous sections, such as the Jacobi $\tet$-functions,  $\tet$-constants, and the Dedekind $\eta$-function,  do not transform as modular forms under $SL(2,\ZZ)$, but they are modular forms under certain subgroups of $SL(2,\ZZ)$. These subgroups belong to the general class of \textit{congruence subgroups} of $SL(2,\ZZ)$, which we shall now define and study. 

\subsection{Definition of congruence subgroups of $SL(2,\ZZ)$}

The \textit{principal congruence subgroup} of level $N\geq 1$ is defined as follows, 
\bea
\label{6.Gam}
\Gamma (N) =  \left \{  \left ( \begin{matrix} a & b \cr c & d \cr \end{matrix} \right ),  ~ ~ a,d \equiv 1 ~ ({\rm mod} \, N)\,,  ~~b,c \equiv 0 ~ ({\rm mod} \, N) \right \}
\eea
Clearly, we have $\Gamma (1) = SL(2,\ZZ)$ and $\Gamma(N)$ is a subgroup of $SL(2,\ZZ)$ for any $N$. Actually, $\Gamma (N)$ is a normal subgroup of $SL(2,\ZZ)$ so that the quotient is itself a group. The quotient group is isomorphic to $SL(2,\ZZ_N)$ (to physicists, isomorphism is synonymous with equality),  
\bea
SL(2,\ZZ)/\Gamma(N) = SL(2,\ZZ_N)
\eea
namely the group of $2 \times 2$ matrices whose entries are integers mod $N$ and whose determinant is 1 mod $N$.  
Equivalently, the group $\Gamma(N)$ may be identified with the kernel of the homomorphism $\ZZ \to \ZZ_N$ of reduction mod $N$. The index $ [SL(2,\ZZ): \Gamma(N)]$ of $\Gamma(N)$ equals the order of $SL(2,\ZZ_N)$, which is finite. It will be determined below shortly. 

\sm   

By definition, a group $\Gamma$  is a \textit{congruence subgroup of $SL(2,\ZZ)$  of level $N$} provided there exists an integer $N \geq 1$ such that $\Gamma (N) \subset \Gamma \subset SL(2,\ZZ)$.  Since the index of $\Gamma(N)$ is finite, it follows that the index $[SL(2,\ZZ): \Gamma]$ is  finite for any congruence subgroup $\Gamma$. Note
that $\Gamma$ need not be a normal subgroup of $SL(2,\ZZ)$ so that $SL(2,\ZZ)/\Gamma$ need not be a group.

\subsection{The classic congruence subgroups}

Besides the principal congruence subgroups $\Gamma(N)$, two other fundamental congruence subgroups of level $N$ are defined as follows,\footnote{The groups of transposes are denoted by, 
$\Gamma^1(N) = \{ \gamma  | \gamma ^t \in \Gamma_1(N) \}$ and 
$\Gamma^0(N) = \{ \gamma | \gamma ^t \in \Gamma_0(N) \}$. }
\bea
\label{6.Gam01}
\Gamma_1(N) &=&  \left \{  \left ( \begin{matrix} a & b \cr c & d \cr \end{matrix} \right ),  ~ ~ a,d \equiv 1 ~ ({\rm mod} \, N)\,,  ~~c \equiv 0 ~ ({\rm mod} \, N) \right \}
\no\\
\Gamma_0 (N) &=& \left \{  \left ( \begin{matrix} a & b \cr c & d \cr \end{matrix} \right ),  ~ ~ c \equiv 0 ~ ({\rm mod} \, N) \right \}
\eea
with  the following manifest subgroup  inclusions,
\bea
\Gamma (N) \subset \Gamma _1(N) \subset \Gamma_0(N) \subset SL(2,\ZZ)
\eea
For $N \geq 2$, none of the groups $\Gamma(N), \Gamma _1(N), \Gamma _0(N)$ contains the generator $S$ of $SL(2,\ZZ)$ and, while $\Gamma_1(N)$ and $\Gamma_0(N)$ contain the translation generator~$T$,  the principal congruence subgroup $\Gamma(N)$ contains $T^N$ but not $T$.

\sm

We recall that $\Gamma(N)$ is a normal subgroup of $SL(2,\ZZ)$ induced by the homomorphism $\ZZ \to \ZZ_N$ on all its entries. Similarly, $\Gamma(N)$ is a normal subgroup of $\Gamma _1(N)$ induced by $\ZZ \to \ZZ_N$ on the entry~$b$ whose quotient is isomorphic to $\ZZ_N$. Finally, $\Gamma _1(N)$ is a normal subgroup of $\Gamma_0(N)$ whose quotient is isomorphic to the group $\ZZ_N^*$ of all multiplicatively invertible elements in $\ZZ_N$. In summary, we have the following quotients and  isomorphisms,
\bea
SL(2,\ZZ)/\Gamma(N) & = & SL(2,\ZZ_N)
\no \\
\Gamma_1(N) / \Gamma(N) & = & \ZZ_N
\no \\
\Gamma_0(N)/\Gamma _1(N) & = & \ZZ_N^*
\eea
The order of $SL(2,\ZZ_N)$ is denoted $2d_N$ and will be evaluated in section \ref{5.order}.  The order of $\ZZ_N$ is $N$. The order of $\ZZ_N^*$ equals the Euler function $\phi(N)$, which counts the number of integers in $\ZZ_N$ that are relatively prime to $N$, and is given by the following formula,
\bea
|\ZZ_N^*| = \phi(N) = N \prod _i \left ( 1 - { 1 \over p_i} \right )
\hskip 1in 
N = \prod _i p_i^{e_i} ~ \hbox{ for } ~ e_i \geq 1
\eea
where the products are over distinct primes $p_i$. For the special case where $N$ is prime, we have $\phi(N)=N-1$. The indices of these subgroups are related to the orders  of the quotient groups as follows, 
\bea
\label{6.index1}
{} [SL(2,\ZZ):\Gamma(N)] & = &  |SL(2,\ZZ_N)| = 2d_N
\no \\
{} [\Gamma _1(N): \Gamma (N)] & = & |\ZZ_N| = N
\no \\
{} [\Gamma_0(N) : \Gamma _1(N)] & = & |\ZZ_N^*| = \phi(N)
\eea
We obtain the  remaining indices,
\bea
\label{6.index2}
{} [SL(2,\ZZ):\Gamma_0(N)]  & = &  2 d_N / (N \phi(N))
\no \\
{} [SL(2,\ZZ):\Gamma_1(N)]  & = &  2 d_N / N 
\eea
using the multiplicative property $[G:H] \times [H:K]=[G:K]$ for groups $K \subset H \subset G$.

\subsection{Computing the order of $SL(2,\ZZ_N)$}
\label{5.order}

We shall now evaluate the order $2d_N$ of $SL(2,\ZZ_N)$, following \cite{DS}. The order will enter crucially in the valence formulas for congruence subgroups. To compute $d_N$, we shall make use of the Chinese remainder theorem (reviewed and proven in appendix \ref{sec:modN}), which provides a fundamental isomorphism for $\ZZ_N$ in terms of its prime decomposition factors. 

\sm

Let $N$ have the following prime number decomposition $N = \prod _i p_i ^{e_i}$ where the $p_i$ are distinct primes and $e_i$ are positive integer exponents. Then by the Chinese remainder theorem, we have the following isomorphism,
\bea
SL(2,\ZZ_N) = \prod _i SL(2,\ZZ_{p_i^{e_i}})
\eea
Thus, the order $|SL(2,\ZZ_N)|$ of the group $SL(2,\ZZ_N)$ is given by,
\bea
|SL(2,\ZZ_N)| = \prod _i |SL(2,\ZZ_{p_i^{e_i}})|
\eea
To compute the order of each component $SL(2,\ZZ_{p^{e}})$ we proceed by induction on $e$. We begin by computing the order of $GL(2,\ZZ_p)$. To do so, consider the set of all matrices $\gamma$, 
 \bea
\gamma = \left ( \begin{matrix} a & b \cr c & d \cr \end{matrix} \right )
\eea
whose entries $a,b,c,d \in \ZZ_p$ may be parametrized by $a,b,c,d  \in \{ 0,1, \cdots , p-1\}$.  To belong to $GL(2,\ZZ_p)$, the entries must satisfy $ad-bc \not=0 ~ (\mod p)$. To enumerate all its elements we proceed by listing the number of elements in the four subsets of possible values of the doublet $(a, b)$. No matrices with  $a=b=0$ belong to $GL(2,\ZZ_p)$; there are $p-1$ doublets with $a \not=0, b=0$ and for each non-zero value of $a$ the $p-1$ non-zero values of $d$ and the $p$ arbitrary values $c$ give  $(p-1)^2p$ elements 
$\gamma \in GL(2,\ZZ_p)$. For the case $a=0, b \not=0$ the roles of $c$ and $d$ are reversed and we obtain another $(p-1)^2p$ elements. Finally, for each one of the $ (p-1)^2$ doublets with $a,b \not=0$, the entry $c$ may take all $p$ values, while $d$ may take all $p$ values except the one that gives zero determinant, producing  $(p-1)^3 p$ contributions. Putting all together, we find, 
\bea
|GL(2,\ZZ_p)| = (p-1)^3 p + 2 (p-1)^2 p  = (p-1) p^3 \left ( 1 - { 1 \over p^2 } \right )
\eea
Any $\gamma \in GL(2,\ZZ_p)$ satisfies $\det \gamma \not=0$, and taking the  quotient by the determinant normal subgroup isomorphic to the multiplicative group $(\ZZ_p)^*$ with $p-1$ elements, we obtain, 
\bea
|SL(2,\ZZ_p)| = p^3 \left ( 1 - { 1 \over p^2 } \right )
\eea
To proceed by induction on the exponent $e$, we consider an arbitrary  matrix $\gamma '$ $(\mod p^{e+1})$  and parametrize it uniquely  $(\mod p^e)$,
\bea
\gamma' = \gamma + p^e \left ( \begin{matrix}  a'  &   b'  \cr  c'  &  d'  \cr \end{matrix} \right )
\hskip 1in 
\gamma = \left ( \begin{matrix} a   & b   \cr c   & d  \cr \end{matrix} \right )
\eea
where $a,b,c,d \in \ZZ_{p^e}$ and  $a',b',c',d' \in \ZZ_p$. Requiring $\det \gamma \equiv 1 ~ (\mod p^{e+1})$
requires $ad-bc \equiv 1 ~ (\mod p^e)$ so that  $\gamma \in SL(2,\ZZ_{p^e})$. Now for an arbitrary matrix $\gamma \in SL(2,\ZZ_{p^e})$, we obtain $\gamma ' \in SL(2,\ZZ_{p^{e+1}})$ if and only if,
\bea
\label{pe}
a d' + d a' - b c' - c b' \equiv 0 ~ (\mod p)
\eea
At least two entries in $\gamma$ are non-zero $(\mod p)$ since $ad-bc \equiv 1 ~ (\mod p)$. If $a \not=0$, we let $a',b',c'$  take independent arbitrary values in $\ZZ_p$ and determine the number $d'$ uniquely by solving equation (\ref{pe}). If $a=0$, then we must have $b\not \equiv 0 ~ (\mod p)$, take the values of $a',b',d'$ arbitrary and solve uniquely for $c'$ using (\ref{pe}). In each case, to every $\gamma \in SL(2,\ZZ_{p^e})$ there correspond $p^3$ elements $\gamma ' \in SL(2,\ZZ_{p^{e+1}})$. Hence we have,
\bea
|SL(2,\ZZ_{p^e})| = p^{3e} \left ( 1 - { 1 \over p^2 } \right )
\eea
Putting all together, for any $N\geq 2$ with prime number decomposition $N = \prod _i p_i^{e_i}$ we have, 
\bea
\label{6.dN}
2 d_N = |SL(2,\ZZ_N)| = N^3 \prod _i  \left ( 1 - { 1 \over p_i^2 } \right )
\eea
The order of the simplest non-trivial group  $SL(2, \ZZ_2)$ is 6 by the above formula. In fact, we can write its generators in terms of the generators $S$ and $T$ of $SL(2,\ZZ)$ and we find,
\bea
\label{5.sl2}
SL(2,\ZZ_2) = \big \{ I, \, S, \, T, \, ST, \, TS, \, STS \big \} \quad (\mod 2)
\eea
Opposite elements are not included in the set since  $I \equiv -I ~ (\mod 2)$.

\subsection{Modular curves}

Recall that the quotient $SL(2,\ZZ) \backslash \cH$ of the upper half-plane $\cH$ by $SL(2,\ZZ)$ is topologically a sphere with one puncture corresponding to the unique cusp. The quotient $SL(2,\ZZ) \backslash \cH$ may be represented  by a choice of fundamental domain in $\cH$ with boundary sides identified pairwise under generators of $SL(2,\ZZ)$. The standard choice $F$ of fundamental domain for $SL(2,\ZZ)$ was defined in (\ref{3.fund}) and depicted in Figure \ref{3.fig:2}. 

\sm

For an arbitrary congruence subgroup $\Gamma$ the quotient is the \textit{modular curve} $Y(\Gamma) = \Gamma \backslash \cH$ which is  again a Riemann surface. However, $Y(\Gamma)$ may be a sphere or a surface of genus  greater than zero and may have several punctures corresponding to the several cusps of $\Gamma$.

\subsubsection{Fundamental domain for a congruence subgroup $\Gamma$}

The modular curve $Y(\Gamma)$ may again be represented by a choice of fundamental domain $F_\Gamma$  in $\cH$ with opposite sides identified pairwise under generators of $\Gamma$. Given the fundamental domain $F$ for $SL(2,\ZZ)$, the fundamental domain $F_\Gamma$ may be obtained by applying the cosets of $SL(2,\ZZ)/\Gamma$ to $F$,
\bea
SL(2,\ZZ)/\Gamma :  F  \to F_\Gamma 
\eea
Therefore $F_\Gamma$ is the union of a number of copies of $F$ equal to the index $[SL(2,\ZZ): \Gamma]$. 
The elliptic points of $\Gamma$ are the images of the elliptic points $i$ and $\rho$ of $SL(2,\ZZ)$ under the cosets $SL(2,\ZZ)/\Gamma$ and, similarly,  the cusps of $\Gamma$ are the images of the cusp $i \infty$ of $F$ under the cosets $SL(2,\ZZ)/\Gamma$. Thus, the number of cusps in $Y(\Gamma)$ equals the index $[SL(2,\ZZ):\Gamma]$. For a non-trivial congruence subgroup $\Gamma \not= SL(2,\ZZ)$, the index is strictly greater than 1, as may be verified for the index  for $\Gamma = \Gamma(N), \Gamma_1(N),\Gamma_0(N)$ evaluated  in (\ref{6.index1}) and (\ref{6.index2}).
Since $Y(\Gamma)$ has at least one puncture, it is always non-compact. 

\sm

There is a useful alternative way of looking at the cusps of a congruence subgroup $\Gamma$ of $SL(2,\ZZ)$. Recall that the Borel subgroup $\Gamma _\infty$ of $SL(2,\ZZ)$  is defined by $\Gamma _\infty= \{ \pm T^n, n \in \ZZ\}$ and that  $SL(2,\ZZ)/\Gamma _\infty $ acts transitively on $\QQ \cup \{ \infty \}$. This property results from the fact that every rational number  $r \in \QQ$ is the image of $i \infty$ under a unique coset in $SL(2,\ZZ)/ \Gamma _\infty$.  To show this, we distinguish two cases: $r=0$ is the image of $i \infty$ under the transformation  $S \in SL(2,\ZZ)/\Gamma _\infty $, while for $r \not=0$ we write $r=a/c$ uniquely  with $a,c \in \ZZ^*$,  $\gcd (a,c)=1$, and $c >0$. By B\'ezout's theorem (reviewed in appendix \ref{sec:modN}) there exists a particular solution $(b_0,d_0)$ for $b$ and $d$ to the equation $ad-bc=1$.  The general solution is given by $b=b_0+ka$ and $d=d_0+kc$, which is equivalent to the particular solution by right multiplication by $T^k$, which proves the assertion. 
The cusps of $Y(\Gamma)$ may now be viewed as those points in $\QQ \cup \{ \infty\}$ which are inequivalent under the action of the congruence group $\Gamma$, or equivalently,\footnote{Throughout, it will be convenient to use the notation $\Gamma z = \{ \gamma z \hbox{ for all } \gamma \in \Gamma\}$ for the set of all images of the point $z$ under the action of the group $\Gamma$. }
\bea
\label{5.cusps}
\{ \hbox{cusps of } \Gamma \} = \big ( \Gamma \backslash SL(2,\ZZ)/\Gamma_\infty  \big ) \, \{ \infty \}
\eea
The number of cusps, counted with multiplicity, is again given by the index $[SL(2,\ZZ):\Gamma]$. 

\sm

\begin{figure}[htb]
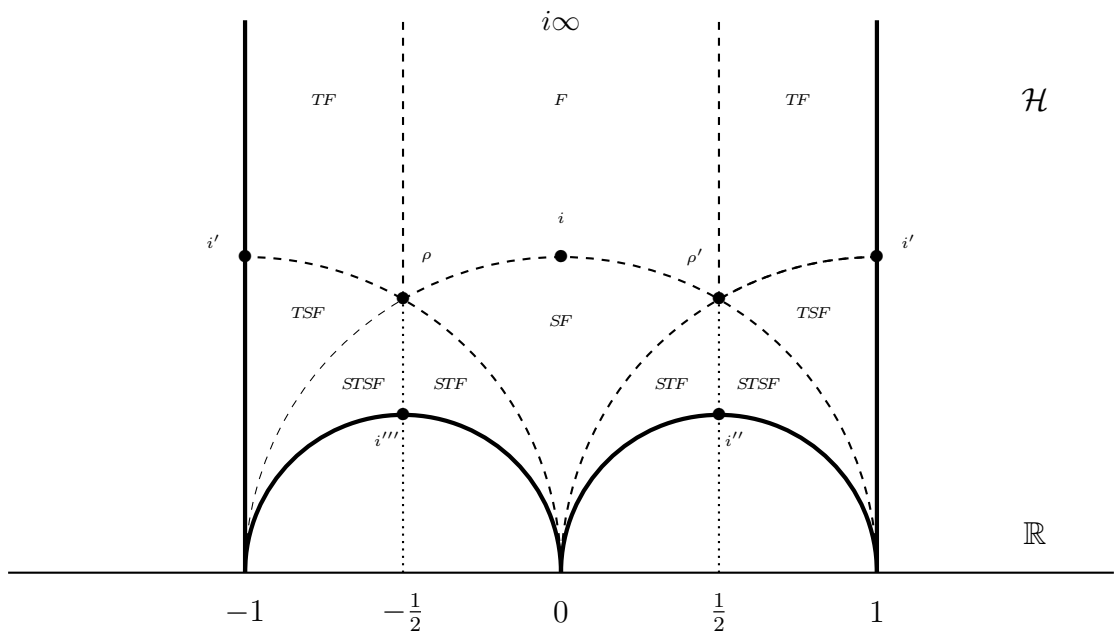

\begin{center}
\tikzpicture[scale=1.05]
\scope[xshift=-5cm,yshift=0cm]
\draw[thick] (-7,0) -- (7,0);
\draw[ultra thick] (-4, 0) -- (-4, 7);
\draw[ultra thick] (4, 0) -- (4, 7);
\draw [ultra thick] (0,0) arc (0:180:2 and 2);
\draw [ultra thick] (0,0) arc (180:0:2 and 2);

\draw[thick, dotted] (-2,0) -- (-2,3.464);
\draw[thick, dotted] (2,0) -- (2,3.464);

\draw[thick, dashed] (-2, 3.464) -- (-2, 7);
\draw[thick, dashed] (2 , 3.464) -- (2, 7);

\draw [thick, dashed] (-2,3.464) arc (120:60:4 and 4);
\draw [dashed] (-4,0) arc (180:120:4 and 4);
\draw [thick, dashed] (4,0) arc (0:60:4 and 4);
\draw [thick, dashed] (2,3.464) arc (120:90:4 and 4);
\draw [thick, dashed] (0,0) arc (0:90:4 and 4);
\draw [thick, dashed] (0,0) arc (180:90:4 and 4);
\draw (0,-0.5) node{$0$};
\draw (-2,-0.5) node{$-\half $};
\draw (2,-0.5) node{$\half $};
\draw (-4,-0.5) node{$-1$};
\draw (4,-0.5) node{$1$};
\draw (0,6) node{\tiny $F$};
\draw (3,6) node{\tiny $T\! F$};
\draw (-3,6) node{\tiny $T\! F$};
\draw (0,3.2) node{\tiny $S\! F$};
\draw (3.2,3.3) node{\tiny $T\! S\! F$};
\draw (-3.2,3.3) node{\tiny $T\! S\! F$};
\draw (1.4,2.4) node{\tiny $S\! T\! F$};
\draw (2.5,2.4) node{\tiny $S\! T\! S \! F$};
\draw (-1.4,2.4) node{\tiny $S\! T\! F$};
\draw (-2.5,2.4) node{\tiny $S\! T\! S \! F$};
\draw (6,6) node{$\cH$};
\draw (6,0.5) node{$\RR$};
\draw (0,4.5) node{\tiny $i$};
\draw (-4.4,4.2) node{\tiny $i'$};
\draw (4.4,4.2) node{\tiny $i'$};
\draw (2.2,1.7) node{\tiny $i''$};
\draw (-2.2,1.7) node{\tiny $i'''$};

\draw (0,4) node{$\bullet$};
\draw (4,4) node{$\bullet$};
\draw (-4,4) node{$\bullet$};
\draw (2,2) node{$\bullet$};
\draw (-2,2) node{$\bullet$};
\draw (2,3.464) node{$\bullet$};
\draw (-2,3.464) node{$\bullet$};
\draw (1.7,4) node{\tiny $\rho'$};
\draw (-1.7,4) node{\tiny $\rho$};
\draw (0,7) node{\small $i \infty$};
\endscope
\endtikzpicture
\end{center}
\caption{\textit{The fundamental domain $\Gamma _{\Gamma (2)}$ for the congruence subgroup $\Gamma(2)$, its elliptic points $i, i', i'', i''', \rho, \rho'$, and its cusps $0, \pm 1, i \infty$ are obtained  as the images of the fundamental domain $F$, the elliptic points $i, \rho$, and the cusp $i \infty$ of $SL(2,\ZZ)$, respectively,   under the action of the group of cosets $SL(2,\ZZ_2) = \{ I, S, T, ST, TS, STS \}$.} \label{5.fig:1}}
\end{figure}

For  the  example $\Gamma = \Gamma (2)$ with $SL(2,\ZZ)/\Gamma (2) = SL(2,\ZZ_2)$, the fundamental domain $F_{\Gamma (2)}$ is the union of the 6 images of $F$ under the elements of $SL(2,\ZZ_2)$ listed in (\ref{5.sl2}). The resulting fundamental domain $F_{\Gamma(2)}$  is depicted in Figure  \ref{5.fig:1}. Using the area formula (\ref{3.area}) we have ${\rm area}(F_{\Gamma (2)}) = 2 \pi$ or ${\rm area}(F_{\Gamma (2)}) = 6 \, {\rm area}(F)$, given by the order of $SL(2,\ZZ_2)$, as expected. The  $\Gamma(2)$-inequivalent elliptic points in $F_{\Gamma(2)}$ are $i, i', i'', i''', \rho, \rho'$ and its $\Gamma(2)$-inequivalent cusps are $0,1, i \infty$. The elliptic points $i$ and $i'$ have multiplicity 2;  $i''$ and $i'''$ have multiplicity 1; $ \rho$ and $\rho'$ have multiplicity 6, while the cusps $0,1,i \infty$ all have multiplicity 2.

\subsection{Compactification of the modular curves}

To compactify $Y(SL(2,\ZZ))$, one adjoins the point $\{i \infty \}$ to the fundamental domain $F$ of $SL(2,\ZZ)$. To compactify  $Y(\Gamma)$ for an arbitrary congruence subgroup $\Gamma$, one similarly adjoins $\{ \hbox{cusp of } \Gamma\} = \Gamma \backslash SL(2,\ZZ)/ \Gamma _\infty \, \{ i\infty \}$  to $Y(\Gamma)$. For an arbitrary congruence subgroup $\Gamma$, the resulting \textit{modular curve} $X(\Gamma)$ is a connected and compact Riemann surface.

\sm

\begin{figure}[tp]
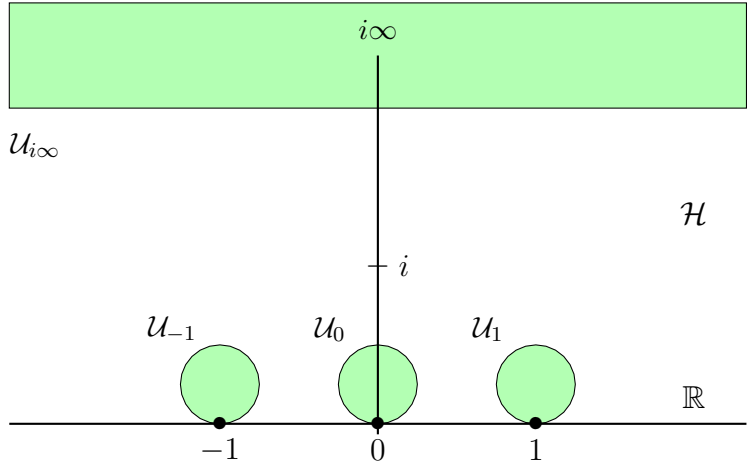

\begin{center}
\tikzpicture[scale=0.7]
\scope[xshift=0cm,yshift=0cm]
\draw[ fill=green!30!] (-7,6) rectangle (7,8);
\draw (0,7.5) node{\small $i \infty$};

\draw  [domain=0:360,fill=green!30!] plot ({3/4*cos(\x)},{3/4+3/4*sin(\x)});
\draw  [domain=0:360,fill=green!30!] plot ({3+3/4*cos(\x)},{3/4+3/4*sin(\x)});
\draw  [domain=0:360,fill=green!30!] plot ({-3+3/4*cos(\x)},{3/4+3/4*sin(\x)});
\draw [thick] (-7,0) -- (7,0);
\draw [thick] (0,-0.2) -- (0,7);
\draw (0,3) node{$-$};
\draw (0,0) node{$\bullet$};
\draw (-3,0) node{$\bullet$};
\draw (3,0) node{$\bullet$};
\draw (0,-0.5) node{$0$};
\draw (-3,-0.5) node{$-1$};
\draw (3,-0.5) node{$1$};
\draw (6,4) node{$\cH$};
\draw (6,0.5) node{$\RR$};
\draw (0.5,3) node{ $i$};
\draw (-6.5,5.3) node{\small  $\cU_{i \infty}$};
\draw (-3.9,1.8) node{\small $\cU_{-1}$};
\draw (-0.9,1.8) node{\small  $\cU_0$};
\draw (2.1,1.8) node{\small  $\cU_1$};
\endscope
\endtikzpicture
\end{center}
\caption{\textit{The green rectangle $\cU_{i \infty}$ represents a neighborhood of the cusp $i \infty$ of $SL(2,\ZZ)$ while the green discs $\cU_{-1}, \cU_0, \cU_1$ represent the neighborhoods of the additional cusps of $\Gamma (2)$. } \label{5.fig:2}}
\end{figure}

An alternative but equivalent way to obtain the compact modular curve $X(\Gamma)$ is to start from the union $\cH^* = \cH \cup \{i \infty\} \cup \QQ$ of the upper half-plane with the rational numbers and infinity. One chooses a topology for  $\cH^*$ whose open sets include all the standard open sets of $\cH$ (which are generated by the open coordinate discs whose closure lies entirely in $\cH$),  as well as the open sets generated by discs that are open in $\cH$ and intersect $\RR$ at a rational number. The latter provide an open neighborhood for every point in $\{i \infty\} \cup \QQ$ and are depicted in green in Figure \ref{5.fig:2} for the discs at infinity $\cU_{i \infty}$ and the discs $\cU_{\pm 1}$ and $\cU_0$ intersecting the real axis at the rational points $\pm 1$ and $0$. The topological space $\cH^*$ obtained this way is not Haussdorff. Recall that a topological space is Haussdorff if for any two distinct points of $\cH^*$ there exist neighborhoods of each point that are disjoint from each other.

\sm

The modular curve $X(\Gamma)$ may be obtained from $\cH^*$ by taking the \textit{extended quotient}, which includes the action of $\Gamma$ on $\QQ \cup \{i \infty \}$, 
\bea
X(\Gamma) = \Gamma \backslash \cH^*
\eea
The modular curve $X(\Gamma)$ is  Haussdorff even though $\cH^*$ is not. The compact modular curves for the classic congruence subgroups are denoted as follows,
\bea
X(N) & = & X \big (\Gamma(N) \big ) 
\no \\ 
X_1(N) & = & X \big ( \Gamma_1(N) \big ) 
\no \\
X_0(N) & = & X \big ( \Gamma_0(N) \big ) 
\eea
A modular form $f(\tau)$ of weight $k$ for a congruence subgroup $\Gamma \subset SL(2,\ZZ)$  corresponds to a $\Gamma$-invariant differential form $\mf= f(\tau) d\tau^{{k \over 2}}  $ on $\cH^*$, and thus pulls back to a well-defined differential on $X(\Gamma) = \Gamma \backslash \cH^*$. 
To obtain the dimensions of the spaces $\cM_k(\Gamma)$ and $\cS_k(\Gamma)$ of modular forms and cusp forms of weight $k$, we can now use the machinery of line bundles on the compact Riemann surfaces $X(\Gamma)$, reviewed in appendix \ref{sec:LB}.

\subsection{The genus of $X(\Gamma)$}

To study $X(\Gamma)$, we shall view it as a branched covering of the sphere $X(SL(2,\ZZ))$ by a non-constant holomorphic function $f : X(\Gamma) \to X(SL(2,\ZZ))$.  We provide a summary of various relevant topological aspects of Riemann surfaces in appendix \ref{sec:RS}.

\subsubsection{The Hurwitz formula for branched coverings}

Consider a non-constant holomorphic map  $f:X \to Y$ between two compact Riemann surfaces $X,Y$. The degree of $f$ is the cardinality of $f^{-1}$ denoted $d=|f^{-1}(y)|$ for all but finitely many points $y \in Y$. A formula valid for all points in $Y$ is obtained in terms of the \textit{degree of ramification} $e_x$ of $f$ at a point $x\in X$ which,  in local coordinates, is given by $f(z)-f(x) \sim (z-x)^{e_x}$.  The degree  of $f$ may then be defined for all $y \in Y$ by the  formula, 
\bea
\label{6.degreea}
d= \sum_{x \in f^{-1}(y)} e_x 
\eea
valid for an arbitrary point $y \in Y$. 
{\thm
\label{5.thm:1}
The Riemann-Hurwitz formula relates the genera $g_X$ and $g_Y$  of  $X$ and $Y$ to the degree of $f$ and the deviations from 1 of the degree of ramification, 
\bea
\label{6.RH}
2g_X -2 = d(2g_Y-2) + \sum _{x \in X} (e_x-1)
\eea}
A proof of the Riemann-Hurwitz formula is given in appendix \ref{sec:RS}.

\subsubsection{Calculating the genus of $X(\Gamma)$}

Returning now to the map $f : X(\Gamma) \to X(SL(2,\ZZ))$ we obtain an alternative formula for its degree in terms of the index of the congruence subgroup $\Gamma \subset SL(2,\ZZ)$,
\bea
\label{6.deg}
d=  \left \{ \bma [SL(2,\ZZ): \Gamma]  & \hbox{ if } & -I  \in \Gamma \cr
\thalf [SL(2,\ZZ): \Gamma]  & \hbox{ if } & -I  \not \in \Gamma \cr \ema \right .
\eea
We shall denote  the elliptic points of $SL(2,\ZZ)$ of order $2h$ by $y_h\in SL(2,\ZZ) \{e^{\pi i /h}\}$ for $h=2,3$ and the cusps by $y_\infty \in SL(2,\ZZ) \{i\infty\}$. The number of elliptic points of order $2h$  in $X(\Gamma)$ will be denoted by $\ep_h$, and the number of cusps in $X(\Gamma)$ by $\ep_\infty$. 

{\thm 
\label{5.thm:2}
The genus $g$ of $X(\Gamma)$ is given by,
\bea
\label{6.genus}
g = 1+{d \over 12} -{\ep_2 \over 4} -{ \ep_3 \over 3} -{\ep_\infty \over 2}
\eea
where $d$ is given in (\ref{6.deg}),  $\ep_h$ denotes the number of elliptic points of order $2h$, and $\ep_\infty$ denotes the number of cusps in $X(\Gamma)$.}

To prove this result using the Riemann-Hurwitz formula (\ref{6.RH}), we evaluate the contributions from the points $x\in X(\Gamma) $ for which $e_x >1$, namely from the points $f^{-1}(y_h)$ and the cusps of $X(\Gamma)$. If $f^{-1}(y_h)$ is a ramification point of $X(\Gamma)$ then  $e_{y_h}=1$, while if it is not a ramification point of $X(\Gamma)$ then $e_{y_h}=h$. Using the degree formula (\ref{6.degree}) at the points $y_h$, we have,
\bea
\sum_{x \in f^{-1}(y_h)} e_x = \ep_h + h (|f^{-1} (y_h)| -\ep_h)
\eea
where the first term arises from those points $y_h$ that are ramification points of $X(\Gamma)$ while the second term arises from those points in $f^{-1}(y_h)$ that are not ramification points of $X(\Gamma)$, in each case weighted by their respective ramification degrees 1 and $h$. Using the degree formula again to eliminate the number $|f^{-1}(y_h)|$, one obtains,
\bea
\label{6.yh}
\sum_{x \in f^{-1} (y_h) } (e_x-1) = { h-1 \over h} (d-\ep_h)
\eea
The formula holds for the cusps by setting $h = \infty$. Using the Riemann-Hurwitz formula (\ref{6.RH}), we set $g_Y=0$, $g=g_{X(\Gamma)}$, and substitute the contributions to the sum of $(e_x-1)$ obtained in (\ref{6.yh}) to obtain (\ref{6.genus}).

\subsection{Formulas for $d$, $\ep_2, \ep_3 $ and $ \ep_\infty$ for $\Gamma(N), \Gamma_1(N)$ and $\Gamma_0(N)$}

In order to obtain formulas for the genus of $X(\Gamma)$, and later on the dimensions of the spaces of modular and cusp forms on $X(\Gamma)$, we need to obtain the degree $d$, and formulas for $\ep_2, \ep_3, \ep_\infty$ for an arbitrary congruence subgroup $\Gamma$. But first, it will be useful to obtain these formulas  for the special cases of the congruence subgroups $\Gamma(N), \Gamma_1(N)$, and $\Gamma_0(N)$. To do so, we begin by assembling all the partial results that are required.

\subsubsection{The absence of elliptic points}

Recall  that the elliptic points of $SL(2,\ZZ)$ are the points of order four, namely $SL(2,\ZZ) \{i\}$, and the points of order six, namely $SL(2,\ZZ) \{e^{ 2 \pi i /6}\}$. In this subsection and the next, we shall count the number of elliptic points of orders four and six of the congruence subgroups $\Gamma(N), \Gamma_1(N)$, and $\Gamma _0(N)$.  We begin with the following theorem.

{\thm The following congruence groups have no elliptic points, 
\begin{enumerate}
\itemsep =-0.05in
\item $\Gamma(N)$ for $N \geq 2$;
\item $\Gamma_1(N)$ for $N \geq 4$;
\item $\Gamma _0(N)$ for any $N$ divisible by a prime $p$ which obeys $p \equiv -1 ~ (\mod 12)$. 
\end{enumerate}}
To prove 1, we parametrize $\gamma \in \Gamma (N) $ as follows,
\bea
\gamma 
= \left ( \begin{matrix} a &   b \cr c & d \cr \end{matrix} \right )
= \left ( \begin{matrix} 1+ N a' &  N b' \cr Nc' & 1+N d' \cr \end{matrix} \right )
\eea
For an elliptic $\gamma$ we must have $|a+d|<2$, following the analysis of subsection \ref{sec:3.3a}. The condition $|a+d| = |2+N(a'+d')|<2$ allows for three possibilities, namely $N(a'+d') =-1,-2,-3$, which are all ruled out for all $N \geq 4$. For both $N=2$ and $N=3$, the only solution is given by $a'+d'=-1$. Combining this condition with $\det \gamma =1$, we find that we must have $N(a'd'-b'c')=1$ which has no solutions for $N \geq 2$.  To prove 2, we parametrize $\gamma \in \Gamma _1(N) $ as follows,
\bea
\gamma = \left ( \begin{matrix} 1+ N a' &  b \cr Nc' & 1+N d' \cr \end{matrix} \right )
\eea
The condition $|a+d| = |2+N(a'+d')|<2$ for $\gamma$ to be elliptic again allows for three possibilities, namely $N(a'+d') =-1,-2,-3$, which are all ruled out for $N \geq 4$.  To prove~3, we parametrize $\gamma \in \Gamma _0(N)$ as follows,
\bea
\gamma = \left ( \begin{matrix} a &  b \cr Nc' & d \cr \end{matrix} \right )
\eea
The condition $|a+d| <2$ again leaves three possibilities, namely $a+d =0, \pm 1$. For $a+d=0$, we furthermore have
$-a^2-Nbc'=1$, so that $a^2 \equiv -1 ~ (\mod N)$. For $a+d=\pm 1$, we have $-a^2 \pm a -Nbc'=1$ so that $a^2 \pm a \equiv -1 ~ (\mod N)$. Now if $p \equiv -1 ~ (\mod 12)$ then $p \equiv -1 ~ (\mod 3)$  and $p \equiv -1 ~ (\mod 4)$. But $a^2 \equiv -1 ~ (\mod p)$ has no solutions if $p \equiv -1 ~ (\mod 4)$ in view of (\ref{quadcon.1}) while $a^2 + a \equiv -1 ~ (\mod p)$ has no solutions in view (\ref{quadcon.2}), which completes the proof of point 3 and of the theorem.

\subsubsection{Counting elliptic points for $\Gamma _0(N)$}

We now count elliptic points for $\Gamma _0(N)$, at the values of $N$ for which they exist. There are two cases, namely elliptic points of orders 4 and 6, for which the transformation in $SL(2,\ZZ)$ is conjugate to either,
\bea
S = \left ( \bma 0 & -1 \cr 1 & 0 \cr \ema \right )
\hskip 1in 
ST = \left ( \bma 0 & -1 \cr 1 & 1 \cr \ema \right )
\eea
with $S^4= (ST) ^6 = I$, $\tr (S)=0$, and $\tr (ST)=1$. Given the conditions on the traces, the representatives of $S$ and $ST$  in $\Gamma _0(N)$ are of the form,
\bea
\gamma_4 = \left ( \bma a & b \cr Nc' & -a \ema \right )
\hskip 1in
\gamma_6 = \left ( \bma a & b \cr Nc' & 1-a \ema \right )
\eea
with $-a^2-bc'N=1$ and $a-a^2-bc'N=1$ respectively. We shall now assume the prime decomposition $N= \prod _i p_i ^{\a_i}$ for distinct primes $p_i$ and $\a_i \geq 1$.
For the elliptic points of order~4  the congruence equation mod $N$, 
\bea
a^2 \equiv -1 ~ (\mod N)
\eea
is equivalent to the congruence equations $a^2 \equiv -1 ~ (\mod p_i ^{\a_i})$. Using the results of subsection \ref{sec:1.5.1} the number of solutions is given by (\ref{quadcon.1}),
\bea
\label{6.ep2}
\ep_2(\Gamma _0(N)) = \left \{ \bma 0 && 4 | N \cr 
\prod _i \Big ( 1+ (-1|p_i) \Big ) && 4 \nmid  N \cr \ema \right .
\eea
where $(-1|p) = \pm 1$ for $p \equiv \pm 1 ~ (\mod 4)$ and vanishes for $p=2$. 
For the elliptic points of order 6, the congruence equation, $a^2+a \equiv -1 ~ (\mod N)$ is equivalent to, 
\bea
(2a+1)^2+3 \equiv 0 ~ (\mod 4N)
\eea
The number of solutions was obtained in subsection \ref{sec:1.5.2}, and is given by,
\bea
\label{6.ep3}
\ep_3 (\Gamma _0(N)) = \left \{ \bma 0 && 9 | N \cr 
\prod _i \Big ( 1+ (-3|p_i) \Big ) && 9 \nmid  N \cr \ema \right .
\eea
where $(-3|p)=\pm 1$ for $p \equiv \pm 1 ~ (\mod 3)$ and vanishes for $p=3$.

\subsubsection{Counting the number of cusps}
\label{sec:numbercusps}

The cusps of a modular curve $X(\Gamma)$ for a congruence subgroup $\Gamma \subset SL(2,\ZZ)$ are given by the cusps $\{ \infty\} \cup \QQ = SL(2,\ZZ) \infty $ modulo equivalence under $\Gamma$. In this subsection, we shall count the number $\ep_\infty$ of cusps of $X(\Gamma)$ for $\Gamma = \Gamma (N), \Gamma _1(N)$, and $ \Gamma _0(N)$. 

\sm

We begin by considering the case $\Gamma = \Gamma(N)$. Two cusps $r=m/n$ and $r'=m'/n'$ with $\gcd(m,n)=\gcd(m',n')=1$ are equivalent under $\Gamma(N)$ provided there exists an $\gamma \in \Gamma (N)$ such that $r' = \gamma r$, or more explicitly, 
\bea
{m'\over n'} = { (1+Na') m + Nb'n \over Nc'm +(1+Nd') n}
\hskip 1in 
\gamma = \left ( \bma 1+Na' & Nb' \cr Nc' & 1+Nd' \ema \right )
\eea
The equation on the left side is equivalent to $mn' \equiv m'n ~ (\mod N)$, whose solution is given by $(m',n') = \pm (m,n) ~ (\mod N)$ in view of the fact that both pairs $(m,n)$ and $(m',n')$ are relatively prime. Therefore, the cusps for $\Gamma (N)$ are labelled by all pairs $(m,n) ~ (\mod N)$ with $\gcd(m,n)=1$. 

\sm 

An alternative perspective on this problem uses \textit{double cosets} and will lend itself more directly to the cases $\Gamma_1(N)$ and $\Gamma _0(N)$. Double cosets will be used in the construction of Hecke operators in section \ref{sec:Hecke}, where a more extensive discussion will be provided. Suffice it here to say that if $G$ is a group and $H_1, H_2$ are subgroups of $G$, then a double coset is an orbit under the left action of $H_1$ and the right action of $ H_2$ containing a given element $g \in G$, and is of the form $H_1 g H_2$. The set of all double cosets is denoted,
\bea
H_1 \backslash G / H_2 = \{ H_1 \, g \, H_2, ~ g \in G \}
\eea
Applying the construction to computing the cusps of $\Gamma (N)$, we set $G=SL(2,\ZZ)$, $H_1=\Gamma (N)$, $H_2= \Gamma _\infty = \{ \pm \left ( \begin{smallmatrix} 1 & b \cr 0 & 1 \end{smallmatrix} \right ), b \in \ZZ \}$. The simple coset $SL(2,\ZZ) /\Gamma _\infty$ provides a one-to-one map from $\infty$ to all the cusps of $SL(2,\ZZ)$ while the further left cosets identified cusps that are equivalent under $\Gamma(N)$. Hence we have, 
\bea
\ep_\infty (\Gamma (N)) = | \Gamma (N) \backslash SL(2,\ZZ) / \Gamma _\infty |
\eea
Since $\Gamma (N)$ is a normal subgroup of $SL(2,\ZZ)$ and $\Gamma (N) \backslash SL(2,\ZZ) = SL(2,\ZZ_N)$, we have,
\bea
\Gamma (N) \backslash SL(2,\ZZ) / \Gamma _\infty = SL(2,\ZZ_N) / \bar \Gamma _\infty
\eea
where $\bar \Gamma _\infty = \{ \pm \left ( \begin{smallmatrix} 1 & b \cr 0 & 1 \end{smallmatrix} \right ), b \in \ZZ_N \}$, and therefore the number of cusps is given by,
\bea
\label{6.cusp1}
\ep _\infty (\Gamma(N)) = |SL(2,\ZZ_N) |/ |\bar \Gamma_\infty| = { d_N \over N} 
\eea
in view of (\ref{6.dN}) and the fact that  $|\bar \Gamma_\infty| = 2N$.

\sm

The case of an arbitrary congruence subgroup $\Gamma$ may similarly be treated with the double coset construction, and we have,
\bea
\ep_\infty (\Gamma) = | \Gamma \backslash SL(2,\ZZ) / \Gamma _\infty |
\eea
The difference with the case of $\Gamma(N)$ is that an arbitrary congruence subgroup $\Gamma$ is not a normal subgroup of $SL(2,\ZZ)$, and this is indeed the case for $\Gamma_1(N)$ and $\Gamma_0(N)$. The calculation of the number of cusps in these cases is more involved and we refer to \cite{Shimura,DS} for a detailed exposition and here only quote the result,
\bea
\label{6.cusp2}
N \geq 5 & \hskip 0.3in & \ep_\infty (\Gamma _1(N)) = \half \sum _{ad=N} \phi(a) \phi(d) 
\no \\
N \geq 3 & \hskip 0.3in & \ep_\infty (\Gamma _0(N)) =  \sum _{ad=N} \phi \big ( \gcd(a,d) \big )  
\eea
where the sums are over $a,d >0$ and furthermore,
\bea
\label{6.cusp3}
\ep_\infty(\Gamma_1(2)) = \ep_\infty(\Gamma_1(3)) = \ep_\infty(\Gamma_0(2)) = 2 
\hskip 1in 
\ep_\infty(\Gamma_1(4)) = 3
\eea
This concludes our calculation of $d, \ep_2, \ep_3$ and $ \ep_\infty$, from which the genus of $X(\Gamma)$ may be computed using (\ref{6.genus}).

\begin{table}[htp]
\begin{center}
\begin{tabular}{|c||c|c|c|c|c|c|c|c|c|c|c|c|}
\hline
$N$ & 2 & 3 & 4 & 5 & 6 & 7 & 8 & 9 & 10 & 11 & 12 & 13  \\ \hline \hline
$d_N/N$ & ${3\over 2}$ & 4 & 6 & 12 & 12 & 24 & 24 & 36 & 36 & 60 & 48 & 84  \\ \hline
$g \big (X(N) \big )$ & 0  & 0 & 0 & 0 & 1 & 3 & 5 & 10 & 13 & 26 & 25 & 50  \\ \hline
$g \big (X_1(N) \big )$ &  0 & 0 & 0 & 0 & 0 & 0 & 0 & 0  & 0 & 1 & 0 & 2  \\ \hline
$g \big (X_0(N) \big )$ & 0 & 0 & 0 & 0 & 0 & 0 & 0 & 0 & 0 & 1 & 0 & 0  \\ 
$\ep_2,\ep_3$ & $1,0$ & $0,1$& $0,0$& $2,0$& $0,0$& $0,2$& $0,0$& $0,0$& $2,0$ & $0,0$& $0,0$& $2,2$ \\ \hline
\end{tabular}
\end{center}
\end{table}

\begin{table}[htp]
\begin{center}
\begin{tabular}{|c||c|c|c|c|c|c|c|c|c|c|c|c|}
\hline
$N$                               & 14 	& 15 		& 16 		& 17 		& 18 		& 19 		& 20 		& 21 		& 22 		& 23 		& 24		& 25   \\ \hline \hline
$d_N/N$                       & 72  	& 96 		&  96		&  144	&  108	&  180	&  144	&  192	&  180	&  264	&  192  	& 300 \\ \hline
$g \big (X(N) \big )$      & 49  	& 73		&  81		& 133  	&  109	&  196	&  169	&  241	&  241	&  375	&  289   	& 476 \\ \hline
$g \big (X_1(N) \big )$  & 1  	&  1		&  2		&  5		&  2		&  7		&  3		&   5		&  6		&  12		&  5  		& 12 \\ \hline
$g \big (X_0(N) \big )$  & 1 	&  1		&  0		&  1		&  0		&  1		&  1		&  1		&  2		&  2		&  1  		& 0 \\ 
$\ep_2,\ep_3$              & $0,0$ & $0,0$	& $0,0$	& $2,0$	& $0,0$	& $0,2$	& $0,0$	& $0,2$	& $0,0$ 	& $0,0$	& $0,0$ 	& 2,0 \\ \hline
\end{tabular}
\end{center}
\caption{\textit{Genera for $X(N)$, $X_1(N)$, $X_0(N)$ for $2\leq N \leq 25$ and the number of elliptic points $\ep_2$ and $\ep_3$ in $X_0(N)$.\label{tab:6.1}}}
\label{tab:generamodcurve}
\end{table}

\subsection{Explicit expressions for the genus}

The formulas for the genus of $X(\Gamma)$ are particularly simple for the cases when no elliptic points occur, as specified in Theorem \ref{5.thm:1}. In these cases the genus is obtained by combining (\ref{6.genus}) with the formula for the degree in (\ref{6.degreea}) with the expressions of the indices given in (\ref{6.index1}), (\ref{6.index2}), and the formulas for the number of cusps in (\ref{6.cusp1}), (\ref{6.cusp2}), (\ref{6.cusp3}),
\bea
N \geq 3: & \hskip 0.2in & g (X(N)) = 1+ { d_N (N-6) \over 12 N}
\\
N \geq 5: & & g(X_1(N)) = 1 + { d_N \over 12 N} - {1  \over 4} \sum_{ad=N} \phi(a) \phi(d)
\no \\
N\geq 2: & & g(X_0(N)) = 1 + { d_N \over 6 N \phi(N)} -{\ep_2(\Gamma_0(N)) \over 4} 
-{\ep_3(\Gamma_0(N)) \over 3} - {1  \over 2} \sum_{ad=N} \phi \big ( \gcd(a,d) \big ) 
\no
\eea
where $\ep_2(\Gamma_0(N))$ and $\ep_3(\Gamma_0(N))$ were given in (\ref{6.ep2}) and (\ref{6.ep3}) respectively. For low values of $N$, we obtain the expressions for $d_N$ and the genera given in Table \ref{tab:generamodcurve}.

\subsection*{$\bullet$ Bibliographical notes}

The books by Shimura~\cite{Shimura} and by Diamond and Shurman~\cite{DS} stand out for their clear and useful introductions to congruence subgroups and modular curves. Our summary here closely follows their presentation.

\newpage

\section{Modular forms for congruence subgroups}
\setcounter{equation}{0}
\label{sec:Forms}

In this section, we shall discuss the structure and the counting of modular forms  for an arbitrary congruence subgroup $\Gamma$, and present the corresponding Eisenstein series for $\Gamma$. These modular forms are maps from the modular curves $X(\Gamma)$ to spaces of meromorphic differentials. The modular curves are compact Riemann surfaces of arbitrary genus, and the spaces of meromorphic differentials may be viewed as spaces of meromorphic sections of holomorphic line bundles of the Riemann surface. We refer to appendix \ref{sec:LB} for a review of holomorphic line bundles on compact Riemann surfaces, the Riemann-Roch theorem,  various vanishing theorems, and dimension formulas for the spaces of meromorphic sections. We shall then give, without proof, the general formula for the dimensions of modular forms and cusp forms for arbitrary $\Gamma$ and provide concrete examples for the cases of the standard congruence subgroups.

\subsection{Modular forms and cusp forms with respect to $\Gamma$}

One defines modular functions, modular forms, and cusp forms  for an arbitrary  congruence subgroup $\Gamma$ by analogy with $SL(2,\ZZ)$.  Let $f(\tau)$  be a meromorphic function  on the upper half-plane $\cH$ which obeys the following transformation law, 
\bea
\label{6.gamma}
f(\gamma \tau) = (c \tau+d)^k f(\tau) 
\hskip 1in 
\gamma = \left ( \bma a & b \cr c & d \cr \ema \right )
\eea
for some integer $k \in \ZZ$, and all $\gamma \in \Gamma$.
Since $\Gamma$ is a congruence subgroup, there exists an integer $N$ such that $\Gamma (N) \subset \Gamma$, and hence a minimum integer $h$ which divides $N$ such that $T^h \in \Gamma$. As a result, $f$ is invariant under $\tau \to \tau + h$ and  admits a Fourier expansion,
\bea
f(\tau) = \sum _{n\in \ZZ} a_n q_h^n 
\hskip 1in 
q_h = e^{2 \pi i \tau/h}
\eea
\begin{enumerate}
\itemsep -0.02in
\item \textit{$f$ is a modular function of weight $k$} if $f$ is meromorphic at every cusp of $\Gamma \backslash \cH$, namely if the Fourier expansion of $f(\gamma \tau)$  has only a finite number of non-vanishing Fourier coefficients $a_n$ with negative $n$ for every $\gamma \in SL(2,\ZZ)$;
\item \textit{$f$ is a modular form of weight $k$} if it is holomorphic in $\cH$ and at every cusp of $\Gamma \backslash \cH$, namely the Fourier coefficients $a_n$ of $f(\gamma \tau)$  vanish for all negative $n$ for every $\gamma \in SL(2,\ZZ)$, and thus $f$ is finite at all cups;
\item \textit{$f$ is a cusp form of weight $k$} if it is a modular form  and vanishes at all cusps.
\end{enumerate}
The spaces of modular and cusp forms of weight $k$ are denoted $\cM_k(\Gamma)$ and $\cS_k(\Gamma)$ respectively. Clearly, the smaller the subgroup, the more modular forms it will have, 
\bea
\Gamma' \subset \Gamma \hskip 0.5in 
\left \{ \bma \cM_k(\Gamma) \subset \cM_k (\Gamma') \cr  \cS_k(\Gamma) \subset \cS_k (\Gamma') \ema \right .
\eea 
In particular, a congruence subgroup $\Gamma$ of level $N$ contains $\Gamma(N)$ so that $\cM_k(\Gamma) \subset \cM_k(\Gamma(N))$ and similarly for cusp forms.

\subsubsection{Modular forms of odd weight}

For the full modular group $SL(2,\ZZ)$ there are no non-zero modular forms of odd weight $k$ because the transformation law (\ref{6.gamma}) for the element $-I \in SL(2,\ZZ)$ shows that $f(\tau)=0$. The same conclusion holds for any congruence subgroup $\Gamma$ that contains $-I$. The congruence subgroups $\Gamma(2), \Gamma_1(2)=\Gamma_0(2)$ and $\Gamma_0(N)$ for every $N$  contain $-I$, and thus have no modular forms of odd weight $k$, while $\Gamma = \Gamma_1(N), \Gamma_0(N)$ for $N >2$ do not contain $-I$ so that the corresponding spaces $\cM_k(\Gamma)$ and $\cS_k(\Gamma)$ for $k$ odd are not necessarily trivial.

\subsubsection{Comments on terminology}
\label{sec:terminology}

The terminology for automorphic functions and automorphic forms and its relation with the terminology for modular functions and modular forms is not entirely standard in the Mathematics literature, and may lead to some confusion, which we shall eliminate here with the help of precise definitions.

\sm

Automorphic functions and forms may be defined  generally in terms of functions with specific transformation properties under a discrete group acting on a complex manifold. Here, we shall limit attention to the case where the complex manifold is the Poincar\'e upper half-plane $\cH$ and the discrete group $\Gamma $ is  a Fuchsian subgroup of $SL(2,\RR)$, acting on $\cH$ by M\"obius transformations (see appendix \ref{sec:RS} for the  definition of Fuchsian groups).

\sm

In the complex analytic category, following Shimura \cite{Shimura} and Diamond and Shurman \cite{DS}, an \textit{automorphic function} $f$ with respect to a Fuchsian group $\Gamma$ is a meromorphic  function on $\cH$ that is invariant under all $\gamma \in \Gamma$,
\bea
f(\gamma \tau) = f(\tau) 
\hskip 1in 
\hbox{ for all } \gamma = \left ( \bma a & b \cr c & d \cr \ema \right ) 
\eea
and is meromorphic at the cusps of $\Gamma$ in $\cH$. An \textit{automorphic form of weight $k$} with respect to $\Gamma$  is a meromorphic function on $\cH$ that instead transforms under all $\gamma \in \Gamma$ with a non-trivial automorphy factor $(c \tau+d)^k$, 
\bea
\label{6.Gamma}
f(\gamma \tau) = (c \tau+d)^k f(\tau) 
\eea
Equivalently, we may view an automorphic form $f$ of weight $k$ as a \textit{differential $k/2$-form} 
\bea
\mf = f(\tau) (d\tau)^{\tfrac{k}{2}} 
\eea
which is invariant under $\Gamma$, in analogy with the approach taken in subsection \ref{3.modif}. The nomenclature of a \textit{form} is particularly suited here, and familiar to a physics audience. 

\sm

In the category of non-holomorphic functions, following Terras \cite{terras1}, we may define an automorphic function as a complex-valued function of $\cH$ that is invariant under $\Gamma$, thereby generalizing the notion from the meromorphic case. From the point of view of a physicist, it is natural to generalize the notion of an automorphic form to the non-holomorphic category by defining an automorphic form of weight $(k,\ell)$ to transform under all $\gamma \in \Gamma $ by,
\bea
f(\gamma \tau) = (c \tau+d)^k (c\bar \tau + d)^\ell f(\tau) 
\eea
corresponding to a differential form $\mf$ of weight $(\tfrac{k}{2}, \tfrac{\ell}{2})$,
\bea
\mf = f(\tau) (d\tau)^{\tfrac{k}{2}} (d \bar \tau)^{\tfrac{\ell}{2}}
\eea
which is strictly invariant under $\Gamma$. In these notes, unless the meaning is clear from the context,  we shall add qualifiers, such as \textit{holomorphic automorphic forms}, \textit{meromorphic automorphic forms}, or more generally \textit{non-holomorphic automorphic forms}. 

\sm

In the special case where $\Gamma = SL(2,\ZZ)$ or a congruence subgroup thereof, one refers to modular functions and forms instead of automorphic functions or forms. Again, unless the meaning is clear from the context,  these terms will be accompanied by qualifiers, \textit{holomorphic modular forms} such as the holomorphic Eisenstein series $\HE_k(\tau)$, \textit{meromorphic modular functions} such as the $j(\tau)$-function, \textit{meromorphic modular forms} such as the inverse discriminant $\Delta (\tau)^{-1}$, and \textit{non-holomorphic modular functions} such as the \textit{non-holomorphic Eisenstein series} $E_s(\tau)$ and \textit{modular graph functions}.

\subsection{Holomorphic Eisenstein series for $\Gamma(N)$}

The holomorphic Eisenstein series of integer weight $k \geq 3$ was defined in (\ref{EisenG})  by,
\bea
G_k (\tau) = \sum  _{m,n \in \ZZ} ' { 1 \over ( m \tau + n)^k} 
\eea
where the prime on the sum instructs us to omit any contribution that makes the summand singular, in this case just $(m,n) = (0,0)$. Clearly $G_k(\tau)$ vanishes  for odd $k$, and is a modular form of weight $k$ for even $k$, transforming as follows, 
\bea
G_k (\gamma \tau)  = (c \tau+d)^k G_k(\tau) 
\hskip 0.7in 
\gamma = \left ( \bma a & b \cr c & d \cr \ema \right ) \in SL(2,\ZZ)
\eea
The ring of modular forms of $SL(2,\ZZ)$ is polynomial and generated by $G_4$ and $G_6$. 

\sm

To construct modular forms for congruence subgroup $\Gamma (N)$ with $N \geq 2$ we parametrize the elements of the group $\Gamma(N) \backslash SL(2,\ZZ) = (\ZZ_N)^2$ by a row vector  $v=(\mu ~  \nu) $ whose entries satisfy $\mu, \nu \in \{ 0,1, \cdots N-1\} = \ZZ_N$,  and introduce the following Kronecker-Eisenstein sum,\footnote{When no confusion is expected to arise we shall sometimes abbreviate $(\mod N)$ as simply $(N)$.} 
\bea
G _k (v | \tau) = \sum _{m,n\in \ZZ}' { 1 \over \big ( (mN+\mu)\tau + nN+\nu \big )^k}
= \sum_{(m,n) \equiv v \, (N)} ' { 1 \over (m\tau+n)^k}
\eea
For $v \not \equiv 0 ~ (\mod N)$ none of the denominators vanish and the prime on the sum may be omitted. 
The double sum is absolutely convergent for integer $k \geq 3$, while for $k=1,2$ it may be defined by the Eisenstein summation convention,
\bea
G _k (v | \tau) = \sum _{n \equiv \nu \, (N)} ' { 1 \over (\mu \tau + n)^k} + \sum _{{m \not =0  \atop m \equiv \mu \, (N)}}
\sum_{n \equiv \nu \, (N)} { 1 \over (m \tau + n)^k}
\eea
Under a transformation $ \gamma \in SL(2,\ZZ)$, we have,
\bea
G_k ( v \gamma | \g \tau) = (c \tau +d)^k G_k (v | \tau)
\eea
where the ingredients are defined as follows with $a,b,c,d \in \ZZ$ and $ad-bc=1$,
\bea
\gamma = \left ( \bma a & b \cr c & d \cr \ema \right )
\hskip 0.8in 
\g \tau = { a \tau + b \over c \tau +d} 
\hskip 0.8in 
v \gamma  = (a\mu+c \nu, b \mu + d \nu)
\eea
Under the action of $SL(2,\ZZ)$ a row vector $v$ may be mapped into itself or into different row vector. For example, $(0 ~ 0)$ maps to itself under any $\gamma \in SL(2,\ZZ)$, while the row $(0 ~ \nu_0)$ for $\nu_0 \not \equiv 0 ~ (\mod N)$  maps to $(\mu ~ \nu) = (c \nu_0 ~ d \nu_0) ~ (\mod N)$. Conversely,  given a row $v=(\mu ~ \nu)$ with $\gcd(\gcd(\mu,\nu), N) =1$, then there exists a $\g \in SL(2,\ZZ)$ which maps $v$ to $(0,\nu_0)$ with $\nu_0 = \gcd(\mu, \nu) ~ (\mod N)$.

\subsubsection{Invariance under $\Gamma (N)$}

A transformation $\gamma \in SL(2,\ZZ)$ maps  $G_k(v| \tau)$ to a multiple of itself provided $v$ satisfies $v \gamma - v \equiv 0 ~ (\mod N)$, or more explicitly, 
\bea
(a-1) \mu + c \nu & \equiv & 0 ~ (\mod N)
\no \\
b \mu + (d-1) \nu & \equiv & 0 ~ (\mod N)
\eea 
This condition is realized for an arbitrary $\gamma \in \Gamma(N)$ and an arbitrary pair $(\mu ~ \nu) \in (\ZZ_N)^2$.  A given pair  $v=(\mu ~ \nu)$ may be invariant under a subgroup of $SL(2,\ZZ)$ that is larger than $\Gamma(N)$. For example $(0 ~ 0)$ is invariant under all of $SL(2,\ZZ)$. But if we require that all $v \in (\ZZ_N)^2$ be invariant then the largest invariance group is $\Gamma (N)$. To see this, it suffices to choose $(\mu ~ \nu) = (1,0) ~ (\mod N)$ and $(\mu ~ \nu) = (0,1) ~ (\mod N)$ to obtain the conditions $a,d \equiv 1 ~ (\mod N)$ and $b,c \equiv 0 ~ (\mod N)$ so that $\gamma \in \Gamma (N)$.

\subsubsection{Asymptotics near the cusps}

The cusps of $\Gamma(N)$ are given by the double coset,
\bea
\Gamma(N) \backslash SL(2,\ZZ)/ \Gamma _\infty
\eea
acting on $i \infty$, namely  by all the images under $SL(2,\ZZ)$ of the cusp at infinity modulo equivalence under $\Gamma (N)$. The number of cusps is denoted $\ep_\infty (\Gamma (N))$ and was given in (\ref{6.cusp1}). Since every cusp  may be mapped to $\infty$ under  a suitable $\gamma \in SL(2,\ZZ)$, under which $G_k(v|\tau)$ maps to $G_k(v\gamma | \gamma \tau)$, it suffices to examine the asymptotics of $G_k(v|\tau)$ at the cusp $\infty$, the asymptotics at the other cusps being deduced by mapping back with $\gamma ^{-1}$.

\sm

The asymptotics of $G_k(v|\tau)$ is qualitatively different for the cases where $\mu \equiv 0 ~ (\mod N)$ or $\mu \not \equiv 0 ~ (\mod N)$.  For $\mu \not\equiv0 ~ (\mod N)$ the limit at the cusp $\tau \to i\infty$ vanishes, 
\bea
\lim _{\tau \to \infty} G_k(v| \tau) = 0 
\eea
Of course, $G_k(v|\tau)$ will in general be non-vanishing at cusps other than $i\infty$. 
If $\mu \equiv 0 ~ (\mod N)$ and $\nu \not \equiv 0 ~ (\mod N)$, then the limit at the cusp $\tau \to i\infty$ is given by, 
\bea
\lim _{\tau \to i\infty} G_k(v|\tau)  = \sum_{n \equiv \nu \, (\mod N)} { 1 \over n ^k}  = { 1 \over N^k} \Big ( \zeta (k, \tfrac{\nu}{N}) + (-)^k \zeta (k, 1-\tfrac{\nu}{N}) \Big )
\eea
It is shown in \cite{DS} that the Eisenstein series form a basis  for the Eisenstein space, which is defined to be the quotient, 
\bea
\cE_k \big ( \Gamma(N) \big ) = \cM_k \big ( \Gamma(N) \big )/\cS_k \big ( \Gamma(N) \big )
\eea 
where $ \cM_k(\Gamma(N))$ is the vector space of all holomorphic modular forms of weight $k$ and $\cS_k(\Gamma(N))$ is the sub-space of cusp forms. For even $k$, the dimension of the Eisenstein space is given by the number of cusps $\ep_\infty (\Gamma (N))$, 
\bea
\dim \cE_k \big ( \Gamma(N) \big )  = \ep _\infty \big ( \Gamma (N) \big )
\eea 
with $\ep _\infty (\Gamma (N))$ given in section \ref{sec:numbercusps}. One may consider  linear combinations of the forms $G_k(v|\tau)$ which produce  normalized Eisenstein series for $\Gamma(N)$, defined by,
\bea
\HE_k(v|\tau) = \sum _{(c,d) \equiv v \, (N) \atop \gcd(c,d)=1} { 1 \over (c \tau +d)^k}
\eea
For even $k$ the limit of $\HE_k(v|\tau)$ vanishes at all the cusps of $\Gamma(N)$ except for the cusp corresponding to $v$.

\subsection{Holomorphic Eisenstein series for $\Gamma_1(N)$ and $\Gamma _0(N)$}

The construction of holomorphic Eisenstein series for arbitrary congruence subgroups uses Dirichlet characters to symmetrize the Eisenstein series for $\Gamma(N)$. As discussed in more detail in appendix \ref{sec:Lseries}, a Dirichlet character $\chi_a$ for $a \in \NN$ is a map $\chi_a:\ZZ \rightarrow \CC$ which is periodic (i.e. $\chi_a(n) = \chi_a(n+a)$), completely multiplicative (i.e. $\chi_a(nm) = \chi_a (n) \chi_a(m)$), and which is non-vanishing only when $a$ and the argument $n$ are coprime. 

\sm

For any positive integer $N$ and integer $k \geq 3$, we define the set $\cC_{N,k}$ of triples $(\chi_{a_1}, \widetilde{\chi}_{a_2}, t)$ with $\chi_{a_1}$ and $\tilde{\chi}_{a_2}$ (primitive) Dirichlet characters and $t \in \NN$,
\bea
\cC_{N,k} := \left\{(\chi_{a_1}, \widetilde{\chi}_{a_2}, t) \, \big| \, \chi_{a_1}(-1)  \widetilde\chi_{a_2}(-1) = (-1)^k \,\,\,\,\mathrm{and}\,\,\,\, t a_1 a_2 \, | N \right\} 
\eea
In terms of Dirichlet characters, we also define the generalized divisor sums 
\bea
\sigma_m^{\chi_{a_1}, \widetilde \chi_{a_2}}(n) = \sum_{d|n} \chi_{a_1}(n/d) \widetilde\chi_{a_2}(d) d^{m}
\eea
as well as the delta function and Dirichlet $L$-series,
\bea
\delta(\chi_{a_1}) = \left\{ \begin{matrix} 1 && \chi_{a_1} = 1~, \\ 0 && \chi_{a_1} \neq 1~, \end{matrix} \right. \hspace{0.8 in}L(\widetilde \chi_{a_2}, s) = \sum_{n=1}^\infty {\widetilde \chi_{a_2}(n)\over n^s}~
\eea
In terms of these ingredients, we may then construct the following functions 
\bea
\HE_k^{\chi_{a_1}, \chi_{a_2},t}(q) = \delta(\chi_1) + {2 \over L(\widetilde{\chi}_{a_2}, 1-k)}\sum_{n=1}^\infty \sigma_{k-1}^{\chi_{a_1}, \widetilde\chi_{a_2}}(n) q^{tn}
\eea
When $\chi_{a_1}=\widetilde \chi_{a_2}= 1$ and $k \geq 4$, the generalized divisor sums reduce to the usual divisor sums, and the $L$-series reduces to the Riemann-zeta function. In this case the functions $\HE_k^{\chi_{a_1}, \chi_{a_2},t}(q)$ reduce to the usual Eisenstein series for $SL(2,\ZZ)$ at shifted values of $q$, i.e. $\HE_k^{1,1,t}(q) = \HE_k(q^t)$. It is clear that for any $t |N$ the functions $\HE_k(q^t)$ should be Eisenstein series for $\Gamma_1(N)$. What is less obvious is the following theorem, which we quote without proof,
{\thm
\label{thm:Gamma1basis}
The space of weight-$k$ holomorphic Eisenstein series for $\Gamma_1(N)$ is generated by $\HE_k^{\chi_{a_1}, \chi_{a_2},t}(q)$ for $({\chi_{a_1}, \widetilde\chi_{a_2},t}) \in \cC_{N,k}$, i.e.
\bea
\cE_k(\Gamma_1(N)) = \left\langle \HE_k^{\chi_{a_1}, \widetilde \chi_{a_2},t} \,\,\,\, \mathrm{for}\,\,\mathrm{all}\,\,\,\, ({\chi_{a_1}, \widetilde\chi_{a_2},t}) \in \cC_{N,k}  \right\rangle
\eea
for $k\geq 3$. For $k=2$, we have 
\bea
\label{eq:E2Gamma1Nres}
\cE_2(\Gamma_1(N)) = \left\{\begin{matrix} \left\langle \HE_2(q) - t \HE_2(q^t) \,\,\,\, \mathrm{for}\,\,\mathrm{all}\,\,\,\, t |N\right\rangle&& \chi_{a_1}= \widetilde \chi_{a_2}=1 \\  \left\langle \HE_2^{\chi_{a_1}, \widetilde \chi_{a_2},t} \,\,\,\, \mathrm{for}\,\,\mathrm{all}\,\,\,\, ({\chi_{a_1}, \widetilde\chi_{a_2},t}) \in \cC_{N,2}  \right\rangle && \mathrm{otherwise}  \end{matrix} \right. 
\eea
}
The proof can be found in \cite{DS,miyake}.
Note the qualitative distinction between the cases of $k=2$ and $k>2$, which results from the fact that there is no weight-2 modular function for $SL(2,\ZZ)$. The first line of (\ref{eq:E2Gamma1Nres}) is indeed consistent with the triviality of the space of weight-2 modular forms for $N=1$ (and hence $\chi_{a_1}= \widetilde \chi_{a_2}=t=1$). 

\sm

We now briefly discuss the case of $\Gamma_0(N)$. First note that if $\chi_{a_1}$ and $\widetilde \chi_{a_2}$ are two Dirichlet characters and $A$ is a common multiple of $a_1$ and $a_2$, then it is possible to define a modulo $A$ Dirichlet character $\widehat \chi_A = \chi_{a_1}^{(A)} \widetilde \chi_{a_2}^{(A)}$, where 
\bea
\chi_a^{(A)}(n) := \left\{ \begin{matrix} \chi_a(n) && (n,A) = 1 \\ 0 && (n,A)>1 \end{matrix} \right.
\eea
We may now further define the following subgroups of $\cE_k(\Gamma_1(N)) $,
\bea
\cE_k(N, \widehat \chi_N) = \left\langle \HE_k^{\chi_{a_1}, \chi_{a_2},t} \,\,\,\, \mathrm{for}\,\,\mathrm{all}\,\,\,\,  ({\chi_{a_1}, \widetilde \chi_{a_2},t}) \in \cC_{N,k}\,\,\,\,\mathrm{and}\,\,\,\, \chi_{a_1}^{(N)} \widetilde \chi_{a_2}^{(N)} = \widehat \chi_N  \right\rangle
\eea
such that $\cE_k(\Gamma_1(N)) = \bigoplus_{\widehat \chi_N}\cE_k(N, \widehat \chi_N) $. The quantity $\widehat \chi_N$ is referred to as the ``nebentypus."  The space of modular forms of $\Gamma_0(N)$ is then simply obtained by restricting to the sector with trivial nebentypus, i.e.
\bea
\label{eq:Gamma0Nbasis}
\cE_k(\Gamma_0(N))  = \cE_k(N, 1)
\eea

Let us close with a note for the practical researcher. Holomorphic Eisenstein series for various congruence subgroups, along with many of their properties, can be straightforwardly implemented in the computer program SageMath \cite{sagemath}. For example, the simple code 
\bea
&\vphantom{.}&> \mathtt{E=EisensteinForms(Gamma0(9),4,prec=10)}\no\\
&\vphantom{.}&>\mathtt{E.eisenstein\_series()}\no
\eea
can be used to output the Fourier expansions to order $O(q^{10})$ for all generators of weight-4 Eisenstein series of $\Gamma_0(9)$. The triplets $(\chi_{a_1}, \widetilde{\chi}_{a_2}, t)$ labelling each generator can be further obtained using 
\bea
&\vphantom{.}&> \mathtt{e=E.eisenstein\_series()}\no\\
&\vphantom{.}&>\mathtt{for\,\,\,e\,\,\,in\,\,\,E.eisenstein\_series():}\no\\
&\vphantom{.}&\qquad\mathtt{print\,\,\,e.parameters()}\no
\eea
In practice, using such computer algebra programs to generate the spaces of modular forms at a given level can be much simpler than working through the definitions above.

\subsection{Jacobi's theorem on sums of four squares}

We now provide a classic application of modular forms for congruence subgroups, here associated with $\Gamma _0(4)$. For any positive integer $n$, we denote by $r_k(n)$ the number of ways in which $n$ can be written as the sum of $k$ positive, zero, or negative integers squared,
\bea
r_k(n) = \# \left \{ m_1 , \cdots, m_k \in \ZZ ~ \hbox{such that} ~ n =  m_1^2 +\cdots + m_k^2  \right \}
\eea
The generating function for $r_k(n)$ in powers of $q= e^{2 \pi i \tau}$  is the function $\tet (0|2\tau)^k$, 
\bea
\tet (0 |2 \tau)^k = \sum _{m_1, \cdots, m_k \in \ZZ}  q^{ m_1^2 + \cdots + m_k^2}
= \sum _{n=0} ^\infty r_k(n) q^n
\eea
Note that the argument of $\tet$ has been taken to be $ 2 \tau$  for convenience, so that the expansion is in terms of $q$ raised to integer powers. The function $\tet (0 |2 \tau)^k$ is invariant under $\tau \to \tau+1$ but it is not invariant under $\tau \to -1/\tau$ for $k \not=0$ and, therefore,  is not a modular form under $SL(2,\ZZ)$. However, it is a modular form under a certain congruence subgroup of $SL(2,\ZZ)$. We specialize to the case $k=4$, and propose to prove Jacobi's theorem, 
\bea
\label{Jacobi}
r_4 (n) = 
8 \sum _{d \,  | \, n, ~ 4 \, \nmid \, d } d 
\eea
 with the help of modular forms.
The sum is over positive divisors $d$ of $n$ which are not divisible by 4. Jacobi's theorem  implies the famous theorem of Lagrange  that every positive integer can be written as the sum of four squares, but Lagrange did not evaluate the number of ways in which this can be done.

\sm

We begin by identifying the modular subgroup under which $\tet (0|2 \tau)^4$ is a modular form. One generator may be taken to be $T : \tau \to \tau+1$. Under $\tau \to -1/\tau$ we have,
\bea
\tet (0|-2/\tau)^4 = (- i \tau/2)^2 \tet (0|\tau/2)^4
\eea
so that $\tet (0|2 \tau)^4$ clearly is not a modular form under $SL(2,\ZZ)$. Instead, we look for more general modular transformations $\gamma = \left(\begin{smallmatrix}a & b \\ 0 & d \end{smallmatrix}\right) \in SL(2,\ZZ)$ under which we obtain good transformation laws. A general $\gamma \in SL(2, \ZZ)$ may be simplified by using the fact that we have invariance under $T$, and this allows us to set $b=0$. But then we have $ad=1$ which is solved by $a=d=1$ and $a=d=-1$. These cases are equivalent to one another upon reversal of the sign of $c$, so that we consider the modular transformation, 
\bea
\tau ' = \gamma \tau = { \tau \over c \tau +1}
\eea
under which we have,
\bea
\tet \left ( 0 |  2 \tau'  \right )^4  
=  \left ( i \, { c \tau +1 \over 2 \tau} \right )^2 \tet \left ( 0 \Big | - { c \over 2 } - {1 \over 2 \tau} \right )^4
\eea
The shift by $-c/2$ cancels in view of $T$-invariance if and only if $c \equiv 0 ~ ({\rm mod} \, 4)$, which may be established by recalling that $\tet(z|\tau+1) = \tet_2(z|\tau)$ from (\ref{tetplusone}). Assuming now that $c \equiv 0 ~ ({\rm mod} \, 4)$ the  remaining argument $-1/(2\tau)$ of $\tet$ is  then  independent of $c$. Using $\tet(0|-1/(2\tau))^4 = - 4 \tau^2 \tet (0|2 \tau)^4$, we obtain, 
\bea
\tet ( 0 | 2 \tau')^4  =  ( c \tau +1)^2 \tet (0  | 2 \tau)^4
\eea
We have now identified two transformations under which $\tet (0|2\tau)^4$  transforms as a modular form of weight 2, namely,
\bea
T: \tau \to \tau +1 
\hskip 1in 
\tilde S : \tau \to { \tau \over 4\tau+1}
\eea
These transformations may be used to generate the group of their compositions, which is $\Gamma _0 (4)$.
Therefore $\tet(0|2 \tau)^4$ is a weight-2 modular form under the congruence subgroup $\Gamma _0(4)$. Of course there are no modular forms of weight two under the full modular group $SL(2, \ZZ)$. So this is a simple example of the fact that, the ``smaller" the modular subgroup of $SL(2,\ZZ)$ is, the larger the number of modular forms it will admit for a given weight.

\sm

Now we would like to identify $\tet(0|2 \tau)^4$ with an Eisenstein series whose Fourier expansion we can compute directly in terms  of divisor functions $\sigma _\alpha (n)$. What are the modular forms of weight two? Let's start from $\HE_2(\tau)$ which is not a modular form under $SL(2,\ZZ)$, but rather a modular connection. Since the difference of two connections is a tensor, we investigate how $\HE_2 ( n \tau)$ transforms for various values of $n$. Under $T : \tau \to \tau +1$, the form  $\HE_2(n \tau)$ is invariant for any $n\in \NN$. Next, examine its transformation properties under the transformation 
\bea
\tau \to \tau' = {\tau \over N \tau+1}
\eea 
Since $\HE_2$ is a derivative of $\ln \Delta$ given by (\ref{3.E2Del}), we examine the transformation law of $\Delta$,
\bea
\Delta (n \tau') = \left ( - { N \tau +1 \over n \tau} \right )^{12} \Delta \left ( - {N \over n}  - {1 \over n \tau} \right )
\eea
When $n | N$, the combination $-N/n$ in the argument of $\Delta$ cancels by $T$-symmetry. Taking the transform of this simplified result, we obtain after some simplifications, 
\bea
\Delta (n \tau') = ( N \tau +1)^{12}  \Delta (n \tau)
\eea
Thus, for $n | N$, the function $\Delta (n\tau)$ is a weight 12 modular form under $\Gamma _0(N)$. Applying this general result to the sum of four squares problem we set $N=4$, so that $\Delta (\tau), \Delta (2\tau), \Delta (4\tau)$ are modular forms of $\Gamma _0(4)$. Taking the derivatives of the logarithms on both sides, and expressing the result in terms of $\HE_2$, we find, 
\bea
\HE_2 (n \tau') = ( 4 \tau +1)^2 \HE_2(n\tau) + { 4 \over n} \times { 12 \over 2 \pi i} (4 \tau +1)
\eea
It follows that the two linear combinations,
\bea
\HE_2 (\tau) - 2\,  \HE_2 (2 \tau)
\hskip 1in 
\HE_2 (2 \tau) - 2\,  \HE_2 (4 \tau)
\eea
are linearly independent modular forms under $\Gamma _0 (4)$ of weight 2. This matches with the results from (\ref{eq:Gamma0Nbasis}) and Theorem \ref{thm:Gamma1basis}. 

\sm

We can alternatively count the number of modular forms of weight two under $\Gamma _0(4)$ by the same arguments we used for $SL(2,\ZZ)$. The difference is that the fundamental domain for $\Gamma _0(4)$ is four times larger than the fundamental domain for $SL(2,\ZZ)$, so we would need to derive an extension of our previous result (\ref{thm1a}) to $\Gamma _0(4)$. We shall do so soon, but for the time being will simply invoke Theorem \ref{thm:Gamma1basis} to claim that the space has dimension two. The above combinations generate this space completely, and we must have,
\bea
\tet (0|2 \tau)^4 = \alpha \Big ( \HE_2 (\tau) - 2\, \HE_2 (2 \tau) \Big ) 
+ \beta \Big ( \HE_2 (2 \tau) - 2\, \HE_2 (4 \tau) \Big )
\eea
for some rational coefficients $\alpha, \beta$. Using the $q$-series representation for $\HE_2$,
\bea
\HE_2 (n \tau) = 1 - 24 \sum _{\ell=1}^\infty \sigma _1(\ell) q^{n \ell}
\eea
and matching the $q^0$ and $q^1$ terms requires $ 1 = -\alpha - \beta$ and $ 1 =  - 3 \alpha$ respectively so that, 
 \bea
 \tet (0|2 \tau)^4 = -{ 1 \over 3} \, \HE_2 (\tau)  + { 4 \over 3}  \, \HE_2 (4 \tau) 
 = 1 + 8 \sum _{\ell=1}^\infty \sigma _1(\ell) q^\ell - 32 \sum _{\ell=1}^\infty \sigma _1(\ell) q^{ 4 \ell}
\eea
Jacobi's theorem (\ref{Jacobi}) now readily follows.

\subsection*{$\bullet$ Bibliographical notes}

The books by Shimura~\cite{Shimura} and by Diamond and Shurman~\cite{DS} provide  useful introductions  for  modular forms for congruence subgroups. We also refer to the book by Miyake~\cite{miyake} and the software data base \cite{sagemath}. A clear account of Jacobi's sum of four squares theorem may be found in Apostol's book~\cite{Apostol}. Our summary  closely follows their presentation. Essays geared towards applications to the proof of Fermat's Last Theorem are collected in \cite{Fermat}.

\newpage

\section{Modular differential equations and vector-valued \\ modular forms}
\setcounter{equation}{0}
\label{sec:MDEsvvmfs}

In this section, we construct differential equations in the modular parameter $\tau$, and find solutions to these equations in simple cases. The solutions can generically be assembled into ``vector-valued modular forms," which have proven fruitful in recent works in Mathematics and Physics. In fact, we will see that in general the components of a vector-valued modular forms are modular forms for congruence subgroups, discussed in the previous section.

\subsection{Modular covariant derivatives}

In section \ref{E2section} we encountered the holomorphic function  $\HE_2(\tau)$, whose transformation under $SL(2,\ZZ)$ is that of a modular connection, as given in (\ref{E2transfs}). We may use this object to define a modular covariant differential operator, acting on modular forms of weight $k$,
\bea
D_k = {1 \over 2 \pi i}{d \over d \tau} - {k \over 12 } \HE_2(\tau)
\eea
The operator $D_k$ maps a modular form of weight $k$ to a modular form of weight $(k+2)$,
\bea
D_k : \cM_k \rightarrow \cM_{k+2}
\eea
as was already shown in section \ref{sec:introtodiffeq}. 

\sm

Note that the first order differential operators introduced above can be composed to define higher-order modular covariant derivatives,
\bea
\cD^{(d)}_k = \prod_{s=1}^d D_{k+2 s -2}~,\hspace{0.6 in}\cD^{(d)}_k : \cM_k \rightarrow \cM_{k+2 d}
\eea

For example, the first order differential operators on Eisenstein series give,
\bea
\label{eq:DonEk}
D_4 \HE_4(\tau) = - {1 \over 3} \HE_6 \hspace{0.8 in} D_6 \HE_6(\tau) = - {1 \over 2} \HE_4^2
\eea
while for the second order differential operators we have 
\bea
\cD^{(2)}_4 \HE_4(\tau) ={1 \over 6} \HE_4^2\hspace{0.8 in}\cD^{(2)}_6 \HE_6(\tau) ={1 \over 3} \HE_4 \HE_6
\eea
The second order derivatives follow directly from the first order ones. To prove the former, we note that the spaces of weight $6$ and $8$ modular forms are generated by $\HE_6$ and $\HE_4^2$ respectively, so the results are fixed up to normalization. To determine the normalization, one can compare the first terms in the $q$-expansion using (\ref{Ektau}).

\subsection{Modular differential equations}
\label{sec:MDEintro}

Consider now an arbitrary linear differential equation of order $d$ in $\tau$ on a modular form $f(\tau)$ of weight $k$, referred to as a modular invariant differential equation (MDE),
\bea
\label{basicMDE}
\left[ \cD^{(d)}_k + \sum_{r = 1}^d h_{r}(\tau) \cD^{(d-r)}_k\right] f(\tau) = 0
\eea
In the above equation, $h_{r}(\tau)$ is an arbitrary meromorphic modular form of weight $2 r$ and $f(\tau)$ has weight $k$. The simplest MDEs are obtained by taking the $h_{r}(\tau)$ to be holomorphic, in which case they are given by polynomials in $\HE_4(\tau)$ and $\HE_6(\tau)$. 

\subsubsection{First order MDE}

The generic first order MDE for modular weight $k$ takes the form, 
\bea
D_k f + h_1 f=0
\eea
for a modular function $h_1$ of weight 2 and a modular function $f$ of weight $k$. Using the relation,
\bea
\HE_2 (\tau) = { 1 \over 2 \pi i } \p_\tau \ln \Delta (\tau)
\eea
we obtain a simplified equation for the modular function $f_0$ defined by,  
\bea
f(\tau) = \Delta (\tau)^{\tfrac{k}{12}} f_0(\tau)
\hskip 1in
{ 1 \over 2 \pi i } { d f _0 \over d \tau} + h_1 (\tau) f_0(\tau)=0
\eea
Furthermore, since $h_1$ is a meromorphic function of weight 2 it may be expressed as $h_1(\tau) = {\HE_6(\tau)\over \HE_4(\tau)}  h_0(j)$, where $h_0$ is a modular function of $\tau$ which may be expressed as a function of $j(\tau)$. 

Using the definition of $j(\tau)$ in terms of $\HE_4$ and $\HE_6$ given by $j= 12^3 \HE_4^3/(\HE_4^3-\HE_6^2)$, together with (\ref{eq:DonEk}), we obtain the derivative of $j(\tau)$ respect to $\tau$,
\bea
\label{10.jder}
{1 \over 2 \pi i} {dj \over d \tau} = - {\HE_6 \over \HE_4} \, j
\eea
which by the chain rule also gives
\bea
{ 1 \over 2 \pi i } { d f _0 \over d \tau} = - {\HE_6 \over \HE_4} \, j \, { d f_0(j) \over dj}
\eea
 Cancelling a common factor of $\HE_6/\HE_4$, the equation for $f_0$ now becomes, 
 \bea
 j \, { d \ln f_0(j) \over d j} = h_0(j)
 \eea
 which may be solved by quadrature. It remains to ensure that the local solution $f(\tau)$ obtained via this procedure is a proper modular form of weight $k$.  This requirement may place conditions on the allowed values of $k$.

\subsubsection{Second order MDE}
\label{sec:secondorderMDE}

Consider now the example of a second order MDE with holomorphic coefficient functions, acting on a weight-zero modular function. The most general such MDE takes the form,
\bea
[D_2 D_0 + \g \HE_4(\tau)] f(\tau) = 0
\eea 
for some free constant parameter $\g$, or more explicitly, 
\bea
\label{eq:explicit2OMDE}
\left[\left({1 \over 2 \pi i}{d \over d \tau} \right)^2  - {1 \over 12 \pi i} \HE_2(\tau) {d \over d \tau} + \g \HE_4(\tau)\right] f(\tau) = 0
\eea
 The general solution to this equation may be obtained as follows. We begin by trading the derivatives with respect to $\tau$ for derivatives with respect to the $j$-function, as we had done for the first order MDE, using (\ref{10.jder}). It follows that 
\bea
{df \over dj} = - {\HE_4 \over j \, \HE_6} {1 \over 2 \pi i} {df \over d \tau}
\eea
and by similar means, 
\bea
{d^2f \over dj^2} = \left[\left({\HE_4 \over j \, \HE_6} \right)^2 \left(\left( {1 \over 2 \pi i} {d \over d \tau} \right)^2 - {\HE_2 \over 6} {1 \over 2 \pi i} {d \over d \tau} \right) - \left({7j - 6912  \over 6 j (j-1728)} \right) {d \over d j}\right] f
\eea
With these results the equation (\ref{eq:explicit2OMDE}) can be rewritten as, 
\bea
\left[j (j-1728){d^2 \over dj^2} +{1 \over 6} (7j - 6912) {d \over dj} + \g \right] f(\tau(j)) = 0
\eea 
In terms of the variable $j(\tau)/1728$, this equation is of standard hypergeometric type. For generic choice of $\g$, there exist two distinct solutions given in closed form by,\footnote{When $\mu \in 6 \, \NN$, the second solution is divergent, since the third argument of the hypergeometric function is a negative integer. In this case, the second solution should be replaced by the Meijer G-function, 
\bea
f_2(\tau) = G^{2,0}_{2,2}\left(\begin{matrix} {2 \over 3} & 1 \\{1- \mu \over 12} & {1+\mu \over 12}\end{matrix}\,\, \Big| 1728 \, j(\tau)^{-1}\right)
\eea
} 
\bea
\label{eq:2dexplicitsols}
f_1(\tau) &=& j(\tau)^{- \tfrac{(1+\mu)}{12}} {}_2 F_1\left(\tfrac{1 + \mu}{12}, \tfrac{5 + \mu}{12}\,;\, 1 + \tfrac{\mu}{6}\,; \, 1728 \, j(\tau)^{-1} \right)
\no\\
f_2(\tau) &=& j(\tau)^{-{(1-\mu) \over 12}} {}_2 F_1\left( \tfrac{1 - \mu}{12}, \tfrac{5 - \mu}{12}\,;\, 1 - \tfrac{\mu}{ 6}\,; \, 1728 \, j(\tau)^{-1} \right)
\eea
where we have defined $\mu = \sqrt{1 - 144 \g}$.

\subsubsection{Third order MDE}
\label{sec:thirdorderMDE}

We may similarly analyze the case of a third order modular differential equation with holomorphic coefficient functions and with $f$ of weight 0. In this case the most general modular differential equation depends on two free parameters $a$ and $b$, 
\bea
\label{3rdorderMDE}
\left[D_4 D_2 D_0 + a \HE_4(\tau) D_0 + b \HE_6(\tau) \right] f(\tau) = 0
\eea
By switching to the local coordinate $j(\tau)$ as before, we can again re-express this as a hypergeometric equation. For generic choices of parameters, there are three distinct solutions given by
 \bea
f_1(\tau) &=& j^{- \tfrac{\m_1 + 1}{6} } \, {}_3 F_2 \left( \tfrac{\m_1 + 1}{6}, \tfrac{\m_1 + 3}{6}, \tfrac{\m_1 + 5}{6}; \tfrac{\m_1 - \m_2 }{6}+1, \tfrac{\m_1 - \m_3}{6}+1 ; 1728 \, j^{-1} \right) 
 \no\\
  f_2(\tau) &=& j^{-{\m_2 + 1 \over 6} } \, {}_3 F_2 \left( \tfrac{\m_2 + 1}{6}, \tfrac{\m_2 + 3}{6}, \tfrac{\m_2 + 5}{6}; \tfrac{\m_2 - \m_1 }{6}+1, \tfrac{\m_2 - \m_3}{6}+1 ; 1728\, j^{-1} \right) 
   \no\\
f_3(\tau) &=& j^{-{\m_3 + 1 \over 6 }} \, {}_3 F_2 \left( \tfrac{\m_3 + 1}{6}, \tfrac{\m_3 + 3}{6}, \tfrac{\m_3 + 5}{6}; \tfrac{\m_3 - \m_2 }{6}+1, \tfrac{\m_3 - \m_1}{6}+1 ; 1728 \, j^{-1} \right) \,\,\,\,\,\,\,\,\,\,\,
 \eea
 Here we have defined $\m_i$ as 
 \bea
 \m_1 &=& 4 x_1 - 2 x_2 -2 x_3 
 \no\\
 \m_2 &=& 4 x_2 - 2 x_3 -2 x_1
  \no\\
 \m_3 &=& 4 x_3 - 2 x_1 -2 x_2
 \eea
and $x_i$, $i=1,2,3$ as the three roots of the cubic equation 
\bea
x^3 - \thalf  x^2 + \left(a + \tfrac{1}{18} \right) x + b =0
 \eea
 
 \subsubsection{Example}
 
One interesting example is when the parameters in (\ref{3rdorderMDE}) are taken to be
\bea
a = - {107 \over 2304}\hspace{0.8 in}b = {23 \over 55296}
\eea
In this case the exact solutions given above have the following $q$-expansions 
\bea
\label{Isingcharacterqexp}
f_1(\tau) &=& q^{23 \over 48}\left(1+q+q^2 + q^3 + 2 q^4 + 2 q^5+ \dots \right)
\no\\
f_2(\tau) &=& q^{-{1 \over 48}}\left(1+q^2 + q^3 + 2 q^4 +2 q^5 + \dots \right)
\no\\
f_3(\tau) &=& q^{1 \over 24}\left(1+q+q^2 + 2q^3 + 2 q^4 +3 q^5 + \dots \right)
\eea
To all order in $q$, these match with the following alternative closed-form expressions, 
\bea
\label{concretedim3sols}
f_1(\tau) & = & \half\left( \sqrt{\tet_3(0|\tau) \over \eta(\tau)} - \sqrt{\tet_4(0|\tau) \over \eta(\tau)}  \right)
\no\\
f_2(\tau) & = & \half\left( \sqrt{\tet_3(0|\tau) \over \eta(\tau)} + \sqrt{\tet_4(0|\tau) \over \eta(\tau)}  \right)
\no\\
f_3(\tau) & = &\sqrt{\tet_2(0|\tau) \over \eta(\tau)} 
\eea
Indeed, these can be checked to be the solutions to the MDE in question.

\subsubsection{Modular invariance of solution spaces}

Because MDEs are by definition modular invariant, so too are their solution spaces. This means that under $SL(2, \ZZ)$ the solutions must transform into linear combinations of themselves. It is useful to organize the solutions of an order-$d$ MDE into a $d$-vector, in which case the $S$ and $T$ generators of $SL(2,\ZZ)$ can be represented by $d \times d$ matrices. For example, in the concrete case of  (\ref{concretedim3sols}) we can organize the solutions into a three-dimensional vector ${\bf f}(\tau) = (f_1(\tau), \, f_2(\tau), \, f_3(\tau))^t$, and then from the known transformation properties of $\tet_i(0|\tau)$ and $\eta(\tau)$ we obtain $T$ and $S$ as 
\bea
T = \left(\begin{matrix} e^{23 \pi i \over 24}&0&0 \\ 0 & e^{- {\pi i \over 24}} & 0 \\ 0 & 0 & e^{\pi i \over 12} \end{matrix} \right)\hspace{0.5 in} S =\half \left(\begin{matrix}1 & \,1 &-1 \\1 &\, 1 & \,1 \\ -2&\, 2& \,0 \end{matrix} \right)
\eea
This leads us to a fruitful generalization of the notion of modular forms, known as \textit{vector-valued modular forms}.

\subsection{Vector-valued modular forms}

 A $d$-dimensional vector-valued modular form ${\bf f}(\tau)=(f_1(\tau), \dots, f_d(\tau))^t$ of weight $k$ is defined by its modular transformation properties, 
\bea
{\bf f}\left(\gamma \tau \right) =(c \tau + d)^k \rho (\gamma) \, {\bf f}(\tau) \hspace{0.5 in}  
\gamma = \left(\begin{matrix}a\,\, & b \\ c\,\, & d \end{matrix} \right) \in SL(2, \ZZ)
\eea
where $\rho: SL(2, \ZZ) \rightarrow GL(d,\CC)$ is a $d \times d$ matrix representation. ${\bf f}(\tau)$ is further required to be holomorphic in the upper-half plane and to have at most finite-order poles at the cusps. 

\subsubsection{Examples}

We begin with some simple examples of vector-valued modular functions. We have already seen that solutions to an order-$d$ MDE form a $d$-dimensional vector-valued modular form, and in the section \ref{sec:MDEvvmf} we will show that the converse is also true, i.e. the components of a $d$-dimensional vector-valued modular form are always solutions to an order-$d$ MDE. In this subsection however we will momentarily ignore the connection to MDEs. 

\sm

The Jacobi $\tet$-functions at $z=0$ form a 3-dimensional vector-valued modular form $(\tet_2^8(0|\tau), \tet_3^8(0|\tau), \tet_4^8(0|\tau))$. We have taken the eighth power to eliminate potential phases from the modular transformations given in (\ref{eq:Jactetmodtransf}). Under $S$ and $T$ transformations one has, 
\bea
\tet_2^8(0|\tau+1) &=& \tet_2^8(0|\tau)  \hspace{0.8 in}\tet_2^8(0|-1/\tau) \,\,=\,\, \tau^4\tet_4^8(0|\tau)
\no\\
\tet_3^8(0|\tau+1) &=& \tet_4^8(0|\tau)  \hspace{0.8 in}\tet_3^8(0|-1/\tau) \,\,= \,\,\tau^4\tet_3^8(0|\tau)
\no\\
\tet_4^8(0|\tau+1) &=& \tet_3^8(0|\tau)  \hspace{0.8 in}\tet_4^8(0|-1/\tau) \,\,=\,\,\tau^4 \tet_2^8(0|\tau)
\eea
Clearly these act as permutations of the three elements. In the usual permutation notation, the $T$ transformation corresponds to the transformation $(132)$, while the $S$ transformation corresponds to $(321)$. Together, these generate the group   $S_3$ of permutations of three elements.  One way of saying this is that there is a surjective homomorphism $\rho: SL(2, \ZZ)\rightarrow S_3$. It is interesting to ask what the kernel of the homomorphism $\rho$ is. That is to say, we are interested in the group defined by, 
\bea
{\rm Ker} (\rho) = \left\{ \gamma = \left(\begin{matrix} a & b \\ c & d \end{matrix} \right) \in SL(2, \ZZ) \,\,
\Big | \,\,\tet_i^8\left(0| \gamma \tau \right) = (c \tau + d)^4\tet_i^8(0|\tau), \quad i=2,3,4 \right\}
\eea
Recalling the notation in terms of half-integer characteristics $\tet_2 = \tet\left[\begin{smallmatrix}1/2 \\ 0 \end{smallmatrix} \right],\tet_3 = \tet\left[\begin{smallmatrix}0 \\ 0 \end{smallmatrix} \right],$ and $\tet_4 = \tet\left[\begin{smallmatrix}0 \\ 1/2 \end{smallmatrix} \right]$, from (\ref{eq:Jactetmodtransf}) we see that elements of  ${\rm Ker} (\rho)$ must have,
\bea
a \a + c \b \equiv \a \,\,\,({\rm mod}\,\,1) \hspace{0.8 in}b \a + d \b \equiv \b \,\,\,({\rm mod}\,\,1)
\eea
for any choice $(\a,\b)\in \left\{(0,0), \left(\half, 0\right), \left(0,\half\right) \right\}$. For $\a = \b = 0$ the equations are trivially satisfied. Taking $\a= \half$ and $\b= 0$, the two equations above require, 
\bea
a \equiv 1 \,\,\,({\rm mod}\,\,2)\hspace{0.8 in}b \equiv 0 \,\,\,({\rm mod}\,\,2)
\eea
while taking $\a= 0$ and $\b= \half$ require, 
\bea
c \equiv 0 \,\,\,({\rm mod}\,\,2)\hspace{0.8 in}d \equiv 1 \,\,\,({\rm mod}\,\,2)
\eea
We thus conclude that  ${\rm Ker}( \rho)$ is given by 
\bea
{\rm Ker} (\rho) = \left\{ \left(\begin{matrix} a & b \\ c & d \end{matrix} \right) \in SL(2, \ZZ) \,\,\Big |\left(\begin{matrix} a & b \\ c & d \end{matrix} \right) \equiv  \left(\begin{matrix} 1 & 0 \\ 0 & 1 \end{matrix} \right)\,\,\,({\rm mod}\,\,2) \right\}
\eea
In other words, we find that ${\rm Ker} (\rho)$ is equal to the principal congruence subgroup $\Gamma(2)$. 
Thus while the vector $(\tet_2^8(0|\tau), \tet_3^8(0|\tau), \tet_4^8(0|\tau))$ transforms as a vector-valued modular form in a non-trivial permutation representation of $SL(2,\ZZ)$, we see that each of the individual components of this vector is itself a  genuine modular form under $\Gamma(2)$. 

\sm

Proceeding to a four-dimensional example, we consider the vector ${\bf f}(\tau)$ with components, 
\bea
f_1(\tau) &=&3^{12}\, \eta(3 \tau)^{24}\hspace{0.98 in} f_2 (\tau)\,\, =\,\, \eta\left({\tau \over 3} \right)^{24}
\no\\
f_3(\tau) &=&\eta\left({\tau +1\over 3} \right)^{24}\hspace{0.8 in} f_4 (\tau) \,\,=\,\, \eta\left({\tau +2\over 3} \right)^{24}
\eea
Under $S$ and $T$ modular transformations, we have the following 
\bea
f_1(\tau+1 ) &=& f_1(\tau)  \hspace{0.8 in} f_1(-1/\tau)\,\,=\,\, \tau^{12} f_2(\tau)
\no\\
f_2(\tau+1)&=& f_3(\tau)\hspace{0.8 in} f_2(-1/\tau)\,\,=\,\, \tau^{12} f_1(\tau)
\no\\
f_3(\tau+1)&=& f_4(\tau) \hspace{0.8 in} f_3(-1/\tau)\,\,=\,\, \tau^{12} f_4(\tau)
\no\\
f_4(\tau+1) &=& f_2(\tau)\hspace{0.8 in} f_4(-1/\tau)\,\,=\,\, \tau^{12} f_3(\tau)
\eea
The generators $S$ and $T$ again permute elements of the vector; in permutation notation, they correspond to $(1342)$ and $(2143)$, respectively. However, now these transformations do not generate the full group $S_4$ of permutations of four elements. Instead, we only obtain the even permutations, i.e. we generate the alternating group $A_4$. So in this case we obtain a surjective homomorphism $\rho: SL(2,\ZZ) \rightarrow A_4$. We may again ask about the kernel of $\rho$. Demanding invariance of $f_1(\tau)$ under $\left(\begin{smallmatrix}a& b \\ c & d \end{smallmatrix} \right)\in SL(2, \ZZ)$ is equivalent to requiring, 
\bea
\left( \begin{matrix}3 & 0 \\ 0 & 1 \end{matrix}\right)\left( \begin{matrix}a & b \\ c & d \end{matrix}\right)\left( \begin{matrix}3 & 0 \\ 0 & 1 \end{matrix}\right)^{-1} = \left( \begin{matrix}a & 3b \\ c/3 & d \end{matrix}\right) \in SL(2, \ZZ)
\eea
which in particular requires that $c \in 3 \ZZ$. Similarly, demanding invariance of $f_2(\tau)$ requires, 
\bea
\left( \begin{matrix}1 & 0 \\ 0 & 3 \end{matrix}\right)\left( \begin{matrix}a & b \\ c & d \end{matrix}\right)\left( \begin{matrix}1 & 0 \\ 0 & 3 \end{matrix}\right)^{-1} = \left( \begin{matrix}a & b/3 \\ 3c & d \end{matrix}\right) \in SL(2, \ZZ)
\eea
which requires $b \in 3 \ZZ$. Any permutation in $A_4$ which fixes $f_1(\tau)$ and $f_2(\tau)$ also fixes $f_3(\tau)$ and $f_4(\tau)$. We conclude that the kernel consists of all matrices for which  $b \equiv c \equiv 0\,\,\,({\rm mod}\,\,3)$. This includes in particular the element $-I \in SL(2,\ZZ)$. Hence the kernel is not quite $\Gamma(3)$, but rather $\pm\Gamma(3)$, where we use the shorthand $\pm \Gamma(N) = \Gamma(N) \times \{\pm {I}\}$. So we have found that while the vector ${\bf f}(\tau)$ transforms as a vector-valued modular form in an alternating representation of $SL(2,\ZZ)$, each of the individual components of this vector is itself modular under the group $\pm\Gamma(3)$. 

\sm

For completeness, let us record here some of the isomorphisms which we have seen, as well as some we have not yet seen, 
\bea
SL(2, \ZZ) / \Gamma(2) &\cong&S_3 \hspace{0.8 in} SL(2, \ZZ) /\pm \Gamma(3) \,\,\cong\,\, A_4
\no\\
SL(2, \ZZ) /\pm \Gamma(4) &\cong&S_4 \hspace{0.8 in} SL(2, \ZZ) / \pm\Gamma(5) \,\,\cong \,\,A_5
\eea
These identities can be checked to be consistent with the index $[SL(2,\ZZ); \Gamma(N)]$, or equivalently with the order $|SL(2,\ZZ_N)|$, as given in (\ref{6.index1}) and (\ref{6.dN}). Indeed, from those formulas one finds
\bea 
[SL(2,\ZZ); \Gamma(2)]& = &6 \hspace{0.8 in}[SL(2,\ZZ); \Gamma(3)] \,\,= \,\,24 
\no\\
\vphantom{,}[SL(2,\ZZ); \Gamma(4)] &=& 48\hspace{0.8 in}[SL(2,\ZZ); \Gamma(5)]\,\,=\,\, 120
 \eea
which can be compared with the orders of the relevant permutation and alternating groups, 
\bea
|S_3| = 6 \hspace{0.5 in} |A_4| = 12 \hspace{0.5 in} |S_4 | =24 \hspace{0.5 in} |A_5 | = 60
\eea
For $N=2$ we see a match, whereas for $N>2$ there is a factor of two difference, stemming from the fact that we are quotienting by $\pm \Gamma(N)$ as opposed to $\Gamma(N)$. 

\subsubsection{Integrality and $\Gamma(N)$}
\label{sec:Intconj}

In both of the above examples, we found that the components of the vector-valued modular forms were themselves modular under $\Gamma(N)$ for some appropriate $N$. This is not true in general, and it is interesting to ask when it holds. The following  result has been proven for 2- and 3-dimensional vector-valued modular forms, and is conjectured to hold more generally:\footnote{In the context of vector-valued modular functions realized by the characters of RCFTs, which is the case of principal interest in Physics, this conjecture has been proven in \cite{Calegari:2021fwl}.}

\begin{conj}[Integrality Conjecture]
Consider a component $f_i(\tau)$ of a vector-valued modular form with Fourier expansion 
\bea
\label{vvmfFourier}
f_i(\tau) =q^{n^{(i)}_{0}} \sum_{n \geq0} a_n^{(i)} q^n 
\eea
with $n^{(i)}_{0} = p_i/N_i$ and $(p_i, N_i)=1$.  If all coefficients $a_n^{(i)}$ are algebraic integers, then $f_i(\tau)$ is a modular form  for $\Gamma(N_i) $.  
 \end{conj}
\noindent

The Integrality Conjecture states that a vector-valued modular form transforming in a representation $\rho$ is integral (i.e. each component of the vector has integer Fourier coefficients) only if $\mathrm{Ker}( \rho)$ contains a principal congruence subgroup $\Gamma(N)$. Indeed, the two examples given above had integer Fourier coefficients, and are directly in line with this conjecture.  As another example, the three-dimensional vector-valued modular function given in (\ref{concretedim3sols}) has all integer Fourier coefficients, and indeed the components can be seen to be modular functions for $\Gamma(48)$. 

\sm

One common place in Physics where vector-valued modular forms appear is as characters of CFTs. Since the Fourier coefficients in that case are interpretable as physical degeneracies, they must of course be algebraic integers (and in fact rational integers). The conjecture then tells us that CFT characters are always modular functions for some $\Gamma(N)$. Going back to the example of (\ref{concretedim3sols}), the functions in this case are actually characters for the Ising CFT, 
\bea
f_1(\tau) = \chi_\eps(\tau) \hspace{0.5 in} f_2(\tau) = \chi_1(\tau)\hspace{0.5 in} f_3(\tau) = \chi_\sigma(\tau)
\eea

\subsubsection{Relation to MDEs}
\label{sec:MDEvvmf}

We now show that the components of a $d$-dimensional vector-valued modular form are always solutions to an order-$d$ MDE. To see this, we begin by constructing the Wronskians $W_r$ associated to ${\bf f}(\tau)$,
\bea
\label{Wronskian1}
W_r =\det \left(\begin{matrix} 
f_1 & \dots & f_d 
\\ \cD^{(1)}_k f_1 & \dots & \cD^{(1)}_k f_d 
\\\vdots & & \vdots \\
\cD^{(r-1)}_k  f_1 & \dots & \cD^{(r-1)}_k f_d
\\\cD^{(r+1)}_k  f_1 & \dots & \cD^{(r+1)}_k f_d
\\\vdots & & \vdots \\
\cD^{(d-1)}_k  f_1 & \dots & \cD^{(d-1)}_k f_d
\\\cD^{(d)}_k  f_1 & \dots & \cD^{(d)}_k f_d 
\end{matrix}\right)
\eea
for $r=1, \dots, d$. From the trivial identity $W_{d-1} - \left({W_{d-1} \over W_d} \right) W_d = 0$, we obtain the following equation, 
\bea
\det  \left(\begin{matrix} 
f_1 & \dots & f_d 
\\ \cD^{(1)}_k f_1 & \dots & \cD^{(1)}_k f_d 
\\\vdots & & \vdots \\
\cD^{(d-2)}_k  f_1 & \dots & \cD^{(d-2)}_k f_d
\\\cD^{(d)}_k  f_1- {W_{d-1} \over W_d} \cD^{(d-1)}_k  f_1 & \dots & \cD^{(d)}_k  f_d- {W_{d-1} \over W_d} \cD^{(d-1)}_k  f_d
\end{matrix}\right) = 0
\eea
For this determinant to vanish, the bottom row must be a linear combination of the other rows, which tells us that 
\bea
\label{vvmfMDE1}
\cD^{(d)}_k  f_i- {W_{d-1} \over W_d} \cD^{(d-1)}_k  f_i + \sum_{r=2}^{d} h_r(\tau) \cD^{(d-r)}_k f_i = 0
\eea
This is of the usual MDE form (\ref{basicMDE}), where in particular we have the coefficient function
\bea
h_1(\tau) = - {W_{d-1} \over W_d}
\eea
In fact, by plugging the expression for $\cD^{(d)}_k  f_i$ obtained from (\ref{vvmfMDE1}) into (\ref{Wronskian1}), we may express all of the coefficient functions in terms of Wronskians, giving 
\bea
h_r(\tau) = (-1)^{r} {W_{d-r} \over W_d}
\eea

\sm

Say that the function $f_i(\tau)$ is bounded in the interior of the fundamental domain, and has zeroes at locations $p$ of order $\mathrm{ord}_{f_i}(p)$. We define the so-called \textit{Wronskian index} by
\bea
\label{Wronskianindexdef}
\ell := 6 \left(\half \mathrm{ord}_{W_d}(i) + {1 \over 3} \mathrm{ord}_{W_d}(\rho) + \sum_{p\in F} \mathrm{ord}_{W_d}(p) \right)
\eea
so that $\ell/6$ is the number of zeroes of $W_d$. The Wronskian index can take any non-negative integer value except $1$. We note for future use that the valence formula (\ref{thm1a}) implies 
\bea
\label{vvmfvalence}
\mathrm{ord}_{W_d}(i \infty) + {\ell \over 6} = {d(k+d-1) \over 12}
\eea
where we have used the fact that the Wronskian has modular weight $d(k+d-1)$. 

\sm

For $\ell=0$, the coefficient functions $h_r(\tau)$ are holomorphic, which is the case analyzed in detail for $d=2,3$ in the previous section. For $\ell >0$ on the other hand, the coefficient functions will generically be singular. For example, consider the case for which $d=\ell = 2$. By definition (\ref{Wronskianindexdef}) we conclude that $ \mathrm{ord}_{W_2}(\rho)  = 1$, and hence $W_2 \sim E_4(\tau)$. The coefficient functions then take the form $h_r(\tau) \sim {1 \over E_4(\tau)} W_{2-r}$. The remaining $\tau$ dependence can be fixed by simply requiring that one has the correct modular weight, giving $h_1(\tau) \sim E_4(\tau)$ and $h_2(\tau) \sim {E_6(\tau) \over E_4(\tau)}$. 

Finding a solution to an MDE with meromorphic coefficients is often significantly more difficult than for its holomorphic counterpart (though for $d=2$ and $\ell < 6$ it turns out that the solutions can again be written as hypergeometric functions). As we discuss in section \ref{sec:Heckevvmf}, Hecke operators can be used to generate such solutions.

\subsection*{$\bullet$ Bibliographical notes}

Solving modular differential equations by recasting them as hypergeometric equations has been discussed in \cite{KanekoKoike,BantayGannon2,FrancMasonHyper}. Vector-valued modular forms grew out of work of Gunning \cite{Gunning}, as well as Selberg \cite{Selbergvvmf}, who used them as a way of studying Fourier coefficients of modular forms for non-congruence subgroups. For a modern review of various properties of vector-valued modular forms see \cite{Gannon:2013jua}, and  for a discussion of modular forms and differential operators see \cite{Zagier7}. The integrality conjecture given in section \ref{sec:Intconj} was first put forward in \cite{AtkinSD}, and has since been proven for two-dimensional  and three-dimensional   vector-valued modular forms in \cite{Mason,FrancMason2} and \cite{Marks3,FrancMason3}, respectively. It has also been proven for vector-valued modular forms whose entries can be interpreted as the characters for RCFTs \cite{Calegari:2021fwl}. Its validity in a more general context has been discussed in e.g. \cite{BantayGannon2,FrancMason4}. 

\sm

Finally, we note that modular differential equations have been applied to the classification of rational conformal field theories, starting with the seminal work in \cite{Mathur:1988na,Mathur:1988gt}, together with follow-up papers \cite{Kiritsis:1988kq,Naculich:1988xv} shortly thereafter. In subsequent works modular differential techniques were applied to the classification of bosonic rational conformal fields theories with two characters at Wronskian index $\ell =2,4$ \cite{Hampapura:2015cea,Hampapura:2016mmz,Chandra:2018pjq}, and to the classification of theories with three characters at $\ell=0$ \cite{Mukhi:2020gnj}; for a recent review of some of this progress, see \cite{Mukhi:2019xjy}. 
As of the writing of these lecture notes, the state-of-the-art for the classification of bosonic rational conformal field theories is for theories with up to five characters (subject to restrictions on the Wronskian index), obtained in \cite{Kaidi:2021ent}. For fermionic theories on the other hand, there is a complicating factor that the relevant modular differential equations need not be invariant under $SL(2, \ZZ)$, but only an appropriate index 2 congruence subgroup. Works towards the classification of fermionic theories with small numbers of characters have appeared in \cite{Bae:2020xzl,Bae:2021mej}.

\newpage

\section{Modular graph functions and forms}
\setcounter{equation}{0}
\label{sec:MGF}

Modular graph functions and forms map  certain (decorated) graphs to complex-valued functions on the Poincar\'e upper half-plane $\cH$ with definite transformation properties under $SL(2,\ZZ)$. The terminology of modular graph \textit{functions} versus \textit{forms} follows the terminology adopted for automorphic functions and forms in the holomorphic and non-holomorphic categories discussed  in subsection \ref{sec:terminology}. Thus, modular graph functions may be viewed as $SL(2,\ZZ)$-invariant functions on $\cH$, while modular graph forms may be identified with $SL(2,\ZZ)$-invariant differential forms. 

\sm

To be more precise, we introduce the spaces $\cM_{k,\ell}$ of complex-valued smooth functions $f: \cH \to \CC$ with the following $SL(2,\ZZ)$ transformation properties,  
\bea
\label{6.sl2}
\cM_{k, \ell} = \left \{  f(\gamma \tau)  = (c \tau + d) ^k (c \bar \tau +d)^\ell f(\tau)
\hbox{ for all } 
\gamma = \left ( \bma a & b \cr c & d \ema \right ) \in SL(2,\ZZ) \right \}
\eea
for $k, \ell \in \CC$, subject to the condition $k-\ell \in \ZZ$. This condition is  imposed to ensure that the functions are single-valued. The functions of interest to us here will have at most polynomial growth at the cusp $\tau _2 \to \infty$, and we shall impose this property also on the functions in $\cM_{k,\ell}$. A modular graph form of modular weight $(k,\ell)$ is then a map from a certain (decorated) graph to $\cM_{k,\ell}$. For the special case $k=\ell$, one refers to such maps as modular graph functions. For $k, \ell \in \ZZ$, we may equivalently view the functions in $\cM_{k,\ell}$ as differential forms of weight $(\tfrac{k}{2},\tfrac{\ell}{2})$. 

\sm 

We shall establish that the non-holomorphic Eisenstein series introduced in section \ref{sec:4.2} may be associated with simple one-loop graphs and thus represent a special class of modular graph functions. Eisenstein series and modular graph functions and forms beyond Eisenstein series occur naturally and pervasively in the study of the low energy expansion of superstring amplitudes, as will be made clear in section \ref{sec:SA}. Here we shall present a purely mathematical approach with only minimal reference to Physics.

\subsection{One-loop modular graph functions are Eisenstein series}

We begin by considering the scalar Green function $G(z-w|\tau)$ on the torus $\Sigma=\CC/\Lambda$ for a lattice $\Lambda = \ZZ + \tau \ZZ$ with modulus $\tau \in \cH$ and $z, w \in \Sigma$. The scalar Green function  was defined  as the inverse of the scalar Laplacian on the torus on the space of functions orthogonal to constant functions in (\ref{10.green3}), solved in terms of Jacobi $\tet$-functions in (\ref{10.green-theta}), and shown to be invariant under $SL(2,\ZZ)$ in (\ref{10.G-mod}). Here we shall use its expression in terms of a double Fourier series or Kronecker-Eisenstein series, given in (\ref{10.green1}) and expressed here in terms of a sum over the lattice $\Lambda ' = \Lambda \setminus \{ 0 \}$,  
\bea
G(z|\tau) = \sum_{(m,n) \not = (0,0)} { \tau_2 \over \pi |m + n \tau|^2} \, e^{2 \pi i (nx-my)}
\eea
where $z=x+\tau y$ and $x, y$ are real ``co-moving coordinates" valued in $\RR/\ZZ$. By concatenating the Green function repeatedly using convolution,  we may construct an infinite family of modular invariant functions, which may be defined recursively as follows,
\bea
G_s (z|\tau) = \int _\Sigma {d^2 u \over \tau_2} \, G (z-u|\tau ) G_{s-1} (u|\tau)
\eea 
where we set $G_1=G$ and $s$ is an arbitrary positive integer.  To the Green function $G$ and its generalizations $G_s$ for positive integer $s$ it is natural to associate a graph, which physicists refer to as a Feynman graph,
\bea
\tikzpicture[scale=1.5]
\scope[xshift=0cm,yshift=0cm]
\draw [thick] (0,0) -- (1,0) ;
\draw (0,0) [fill=white] circle(0.05cm) ;
\draw (1,0) [fill=white] circle(0.05cm) ;
\draw(-1,0) node{$G(z-w|\tau) ~ = ~$};
\draw (0,-0.2) node{\small $z$};
\draw (1,-0.2) node{\small $w$};
\endscope
\scope[xshift=0cm,yshift=-0.8cm]
\draw [thick] (0,0) -- (3,0) ;
\draw [thick, dotted] (3,0) -- (4,0) ;
\draw [thick]  (4,0) -- (5,0) ;
\draw (0,0) [fill=white] circle(0.05cm) ;
\draw (1,0) [fill=black] circle(0.05cm) ;
\draw (2,0) [fill=black] circle(0.05cm) ;
\draw (3,0) [fill=black] circle(0.05cm) ;
\draw (4,0) [fill=black] circle(0.05cm) ;
\draw (5,0) [fill=white] circle(0.05cm) ;
\draw(-1,0) node{$G_s(z-w|\tau) ~ = ~$};
\draw (0,-0.2) node{\small $z$};
\draw (1,-0.2) node{\small $u_1$};
\draw (2,-0.2) node{\small $u_2$};
\draw (3,-0.2) node{\small $u_3$};
\draw (4,-0.2) node{\small $u_{s-1}$};
\draw (5,-0.2) node{\small $w$};
\endscope
\endtikzpicture
\label{6.fig:1}
\eea
The black dots represent points $u_1, \cdots u_{s-1}$  that are integrated over $\Sigma$, while the white dots represent given points in $\Sigma$, that are not integrated.  Upon Fourier transformation, convolution acts by multiplication, and we obtain the following representation for $G_s$, 
\bea
\label{6.Gs}
G_s(z|\tau) = \sum_{(m,n) \not= (0,0)} { \tau_2^s \over \pi^s |m + n \tau|^{2s}} \, e^{2 \pi i (nx-my)}
\eea
where $z=x+\tau y$. As long as $s \geq 2$, we may set the point $z=0$ which graphically is equivalent to closing the open chain in (\ref{6.fig:1}) into the closed loop of (\ref{6.fig:2}),
\bea
\tikzpicture[scale=1.5]
\scope[xshift=0cm,yshift=0cm]
\draw [thick]  (1,0) -- (1/2,0.866);
\draw [thick]  (1,0) -- (1/2,-0.866);
\draw [thick, dotted]  (-1,0) -- (-1/2,-0.866);
\draw [thick]  (-1,0) -- (-1/2,0.866);
\draw [thick]  (-1/2,-0.866) -- (1/2,-0.866);
\draw [thick]  (-1/2,0.866) -- (1/2,0.866);
\draw (1,0) [fill=black] circle(0.05cm) ;
\draw (-1,0) [fill=black] circle(0.05cm) ;
\draw (1/2,0.866) [fill=black] circle(0.05cm) ;
\draw (1/2,-0.866) [fill=black] circle(0.05cm) ;
\draw (-1/2,0.866) [fill=black] circle(0.05cm) ;
\draw (-1/2,-0.866)[fill=black] circle(0.05cm) ;
\draw(-2,0) node{$E_s(\tau) ~ = ~$};
\draw (1.2,0.15) node{\small $u_1$};
\draw (-1.2,0.15) node{\small $u_4$};
\draw (1/2,1.05) node{\small $u_2$};
\draw (1/2,-1.05) node{\small $u_s$};
\draw (-1/2,1.05) node{\small $u_3$};
\draw (-1/2,-1.05) node{\small $u_{s-1}$};
\endscope
\endtikzpicture
\label{6.fig:2}
\eea
Comparing to (\ref{3.nhE}), we see that $G_s(0|\tau) = E_s(\tau)$, the non-holomorphic Eisenstein series.  Actually, the double sum in (\ref{6.Gs}) converges absolutely for any $s \in \CC $ with $\Re(s) >1$, but the graphical interpretation no longer makes sense when $s$ is not a positive integer. 

\sm

When $s$ is integer and satisfies $s \geq 2$, the number $s$ counts the number of edges in the graph (\ref{6.fig:2}) and is referred to as the \textit{weight} of the graph (not to be confused with the modular weight of the corresponding modular graph function). In subsection \ref{sec:12.7} we shall identify this definition of the weight with that of \textit{transcendental weight}.

\subsection{Maass operators and Laplacians}
\label{sec:9.Maass}

Before studying higher-loop modular graph functions in detail, let us give a brief intermission on differential operators on the space $\cM_{k,\ell}$. There exist several natural maps between the spaces $\cM_{k,\ell}$ which we shall now exhibit.  

\sm

One noteworthy element of $\cM_{-1,-1}$ is the imaginary part $\tau_2$ of the modulus $\tau$, whose powers satisfy $\tau_2 ^s \in \cM_{-s,-s}$. Multiplication by a suitable power of $\tau_2$ therefore provides a map between the different spaces $\cM_{k,\ell}$. Since a function $f(\tau)$ and its image $\tau_2^s f(\tau)$ under this map are essentially equivalent to one another, we may  introduce the equivalence relations, 
\bea
\label{6.equiv}
\cM_{k,\ell} ~ \approx ~ \cM_{k-\ell, 0} ~ \approx ~ \cM_{0,\ell-k}
\eea
This equivalence is particularly convenient when it comes to the action of partial differential operators. The Cauchy-Riemann operator $\p_{\bar \tau}$ acts covariantly on the spaces $\cM_{k,0}$ without the need for a connection and maps to the space $\cM_{k,2}$. Similarly, the operator $\p_\tau$ maps $\cM_{0,\ell}$ covariantly to the space $\cM_{2,\ell}$ without the need for a connection. One may, of course, include a $k$- and $\ell$-dependent connection to obtain the Maass operators (see appendix \ref{sec:C4}). We shall circumvent doing so here by applying  $\p_{\bar \tau} $ to $\cM_{k,0}$ and $\p_\tau$ to $\cM_{0,\ell}$ only, using the equivalence of (\ref{6.equiv}). Actually, we can multiply the Cauchy-Riemann operators by suitable powers of $\tau_2$ so that the operators $\nabla = 2 i \tau_2^2 \p_\tau$ and $\bar \nabla = - 2 i \tau_2^2 \p_{\bar \tau}$ map the spaces as follows,
\bea
\bar \nabla : \cM_{k, 0} & \to & \cM_{k-2,0}
\no \\
\nabla : \cM_{0,\ell} & \to & \cM_{0,\ell-2}  
\eea
Laplace operators may be defined as follows,\footnote{Here we define the Laplacians as negative operators, while in appendix \ref{sec:C4} they are positive operators.}
\bea
\Delta _k ^+ : \cM_{k,0} \to \cM_{k,0} & \hskip 0.6in & 
\Delta _k^+ = \tau_2^{-k} \, \nabla \, \tau_2^{k-2} \, \bar \nabla 
\no \\
\Delta _\ell ^- : \cM_{0,\ell} \to \cM_{0,\ell} & \hskip 0.6in & 
\Delta _\ell^- = \tau_2^{-\ell} \, \bar \nabla \, \tau_2^{\ell-2} \, \nabla 
\eea
On modular forms of weight $(0,0)$ the Laplace operators $\Delta_0^+$ and $\Delta_0^-$ coincide and are denoted by $\Delta = 4 \tau_2^2 \p_\tau \p_{\bar \tau}$. Famously, the Eisenstein series satisfies the Laplace eigenvalue equation,
\bea
\Delta E_s  = s(s-1) E_s
\eea
Its expansion at the cusp consists of two power behaved terms, plus an infinite number of exponentially suppressed terms, given in (\ref{4.d12}), and which we recall here for convenience, 
 \bea
E_s(\tau)  =  
 2  \, { \zeta (2s) \over \pi^s} \, \tau_2^s  + 2 { \Gamma (s-\thalf ) \over \pi^{s-\half} \Gamma (s)}  \, \zeta (2s-1) \, \tau_2^{1-s} +\cO(e^{-2 \pi \tau_2}) 
\eea
In terms of the independent variables $\tau, \bar \tau$ and the real co-moving coordinates  $x,y$, the Laplace operator $\Delta$ also has $G_s(x+\tau y|\tau)$ as an eigenfunction with eigenvalue independent of $x,y$,
\bea
\Delta G_s(x+\tau y|\tau) = s(s-1) G_s(x+\tau y|\tau) 
\eea
Application of the operator $\nabla^k$ to $E_s$ produces modular graph forms of weight $(0,-2k)$, 
\bea
\nabla ^k E_s (\tau) = 
{ \Gamma (s+k) \over \Gamma (s)}  \sum_{(m,n) \not= (0,0)} { \tau_2^{s+k} \over \pi^s (m + n \tau)^{s+k} (m+n \bar \tau)^{s-k}} 
\eea
These functions coincide with Zagier's elliptic polylogarithm, up to an overall normalization.

\subsection{Two-loop modular graph functions}

We now investigate the special class of modular graph functions in $\cM_{0,0}$ associated with arbitrary (decorated) two-loop graphs. A connected two-loop graph has two trivalent vertices and an arbitrary number of bivalent vertices, distributed on the three concatenated edges connecting the trivalent vertices at $z$ and $w$ according to three positive integers $a_1, a_2, a_3$. The most general such graph is  depicted in (\ref{6.fig:3}), 
\bea
\tikzpicture[scale=3]
\scope[xshift=0cm,yshift=0cm]
\draw (1.9,0) node{$\bullet$} ;
\draw (2.4,0) node{$\bullet$} ;
\draw (2.9,0) node{$\bullet$} ;
\draw (3.4,0) node{$\bullet$} ;
\draw (3.9,0) node{$\bullet$} ;
\draw [thick] (1.9,0) -- (2.4,0) ;
\draw [thick]  (2.4,0) -- (2.9,0) ;
\draw[dotted, thick] (2.9,0) -- (3.4,0) ;
\draw [thick]  (3.4,0) -- (3.9,0) ;
\draw (2.4,0.25) node{$\bullet$} ;
\draw (2.9,0.30) node{$\bullet$} ;
\draw (3.4,0.25) node{$\bullet$} ;
\draw [thick]  (1.9,0) -- (2.4,0.25) ;
\draw [thick]  (2.4,0.25) -- (2.9,0.3) ;
\draw[dotted, thick] (2.9,0.3) -- (3.4,0.25) ;
\draw [thick]  (3.4,0.25) -- (3.9,0) ;
\draw (2.4,-0.25) node{$\bullet$} ;
\draw (2.9,-0.30) node{$\bullet$} ;
\draw (3.4,-0.25) node{$\bullet$} ;
\draw  [thick]  (1.9,0) -- (2.4,-0.25) ;
\draw [thick]  (2.4,-0.25) -- (2.9,-0.3) ;
\draw[dotted, thick] (2.9,-0.3) -- (3.4,-0.25) ;
\draw [thick]  (3.4,-0.25) -- (3.9,0) ;
\draw(1.3,0) node{$C_{a_1,a_2,a_3}(\tau) ~ = $};
\draw(2.9,0.39) node{$a_1$};
\draw(2.9,0.11) node{$a_2$};
\draw(2.9,-0.19) node{$a_3$};

\draw(1.9,0.1) node{$z$};
\draw(3.9,0.1) node{$w$};
\endscope
\endtikzpicture 
\label{6.fig:3}
\eea
The associated modular graph function may be defined in terms of a double integral over $\Sigma$ of a triple product of concatenated Green functions $G_s$, 
\bea
C_{a_1, a_2, a_3} (\tau) = \int _\Sigma { d^2 z \over \tau_2} \int _\Sigma { d^2 w \over \tau_2} G_{a_1}(z-w|\tau) 
 G_{a_2}(z-w|\tau)  G_{a_3}(z-w|\tau) 
 \eea
 Translation invariance on the torus renders one of the integrals redundant, and we have the following equivalent expression, 
 \bea
C_{a_1, a_2, a_3} (\tau) = \int _\Sigma { d^2 z \over \tau_2} G_{a_1}(z|\tau) 
 G_{a_2}(z|\tau)  G_{a_3}(z|\tau) 
 \eea
Upon Fourier transforming, the corresponding modular graph function is given by the following Kronecker-Eisenstein sums, 
\bea
C_{a_1, a_2, a_3} (\tau) = \sum _{{ m_r, n_r \in \ZZ \atop r=1,2,3}} ' 
\delta \Big ( \sum _{r=1}^3 m_r \Big ) 
\delta \Big ( \sum _{r=1}^3 n_r \Big ) 
\prod _{r=1}^3 \left (  {\tau _2 \over \pi |m _r + n_r \tau |^2} \right )^{a_r}
\eea
The sums are absolutely convergent provided $a_r + a_s > 1$ for all pairs $r \not= s$. The weight is given by $w=a_1+a_2+a_3$ and is required to satisfy  $w \geq 3$ for convergence. The Kronecker-Eisenstein sums are absolutely convergent for complex values of $a_1, a_2, a_3$ as long as $\Re(a_1), \Re(a_2), \Re(a_3) >1$, but contrarily to the case of the Eisenstein series,  the problem of the existence of analytic continuations in these variables remains unexplored. 

\sm

Two key properties of the non-holomorphic Eisenstein series, namely that they satisfy a differential equation, and have a simple expansion at the cusp, have more complicated counterparts in two-loop modular graph functions. 
{\thm 
\label{6.thm1}
Two-loop modular graph functions $C_{a_1,a_2,a_3}$  of weight $w=a_1+a_2+a_3$ obey a system of differential equations of uniform weight $w$ 
\bea
2 \Delta C_{a_1, a_2, a_3} & = & 
2 a_1 a_2 \, C_{a_1+1,a_2-1,a_3} + a_1a_2 \, C_{a_1+1,a_2+1,a_3-2} - 4 a_1a_2 \, C_{a_1+1,a_2, a_3-1} 
\no \\ &&
+  a_1 (a_1-1) \, C_{a_1,a_2,a_3}  +  \hbox{ 5 permutations of } (a_1,a_2,a_3)
\no 
\eea
where $\Delta = 4 \tau_2^2 \p_\tau \p_{\bar \tau}$. When $a_1, a_2, a_3 \in \NN$, the system of differential equations 
for a given weight truncates to a finite-dimensional linear system of inhomogeneous equations where the inhomogeneous part consists of a  linear combination of $E_w$ and $E_{w_1} E_{w_2} $ with $w=w_1+w_2$ and $w_1, w_2 \geq 2$, as a consequence of the following relations, 
\bea
C_{w_1,w_2,0} & = & E_{w_1} E_{w_2} - E_w
\no \\
 C_{w_1+1,w_2,-1} & = & E_{w_1} E_{w_2} + E_{w_1+1} E_{w_2-1}
\eea
The eigenvalues of the linear system are of the form $s(s-1)$ where $1 \leq s \leq w-2$ and $w-s \in 2\NN$ and have multiplicity $ [ (s+2)/3 ]$. } 
\sm

For example, up to weight 5, the relations of Theorem \ref{6.thm1} are given as follows,
\bea
\label{6.exC}
\Delta  C_{1,1,1} & = & 6 E_3
\no \\
(\Delta -2) C_{2,1,1} & = & 9 E_4 - E_2^2
\no \\
(\Delta -6) C_{3,1,1} - 3 C_{2,2,1} & = & 16 E_5 - 4 E_2 E_3
\no \\
\Delta C_{2,2,1} & = & 8 E_5
\eea
Note that the first and last identities imply $C_{1,1,1}-E_3$ is constant, as must be $5C_{2,2,1}-2E_5$. These constants may be determined from the asymptotic behavior of these functions at the cusp, a problem to which we now turn.

{\thm 
\label{6.thm2}
The expansion of $C_{a_1, a_2, a_3}(\tau)$ near the cusp $\tau \to i \infty$ is given by a Laurent polynomial in  $\tau_2$ of degree $(w,1-w)$
\bea
C_{a_1, a_2, a_3} (\tau) = c_w (-4 \pi \tau_2)^w + { c_{2-w} \over (4 \pi \tau_2)^{w-2}}
+ \sum _{k=1}^{w-1}  { c_{w-2k-1} \, \zeta (2k+1) \over (4 \pi \tau_2)^{2k+1-w}} + \cO(e^{ - 2 \pi \tau_2})
\eea
where  $c_w, c_{w-2k-1} \in \QQ$ and $c_{2-w}$ is a linear combination with integer coefficients of products of odd zeta-values,
\bea 
c_{2-w} = \sum_{k=1}^{w-2} \gamma _k \, \zeta (2k+1) \zeta (2w-2k-3) 
\hskip 1in \gamma _k \in \ZZ
\eea} A consequence of this theorem is that no multiple zeta-values occur in the Laurent polynomial of two-loop modular graph functions. It was shown by Zerbini that higher loop modular graph functions do involve irreducible multiple zeta values.  

\sm

At every odd weight, there is one linear combination of the above differential identities in the subspace of vanishing  eigenvalue whose inhomogeneous part does not involve bilinears in the Eisenstein series. This property may be verified in the identities for $C_{1,1,1} $ and $C_{2,2,1}$ in (\ref{6.exC}), and holds for all odd weights. Each one of these identities may  be integrated up to an additive constant which may be fixed by the asymptotics near the cusp, and we find,
\begin{align}
\label{eq:MGFidentities}
C_{1,1,1}  & =  E_3 + \zeta (3) & 
C_{3,3,1} + C_{3,2,2} & = \frac{3}{7} E_7 +\frac{\zeta (7)}{252} 
\no \\
C_{2,2,1} & =  {2 \over 5} E_5 +{\zeta (5) \over 30} &
9 C_{4,4,1} + 18 C_{4,3,2} + 4 C_{3,3,3}  &= 4 E_7 +\frac{ \zeta (9)}{240}
\end{align}
The equation for $C_{1,1,1}$ was derived by Zagier in an unpublished note  by direct summation of the Kronecker-Eisenstein series which defines $C_{1,1,1}$.

\subsection{General modular graph functions and forms}

At higher loops, it is no longer true that the Laplace operator in $\tau$, (acting on the immediate generalization of two-loop modular graph functions to higher loops) closes onto functions defined by Kronecker-Eisenstein series in which the exponents of holomorphic momenta $m_r + n _r \tau$ and their complex conjugate coincide. It is natural to also lift the restriction to modular functions, and consider more generally  \textit{modular graph forms} which have non-trivial modular weights. The required generalization may be defined as follows. 

\sm

The connectivity matrix $\Gamma$ of an arbitrary graph  has components $\Gamma _{v \, r}$ where the index $v=1,\cdots, V$ labels the vertices of the graph and the index $r=1,\cdots, R$ labels its edges. No edge is allowed to begin and end on the same vertex. When edge $r$ contains vertex $v$ we have $\Gamma _{v \, r}=\pm 1 $,  while otherwise we have $\Gamma _{v\, r} = 0$. We shall view the pair of exponents $(a_r, b_r)$ of the momenta $p_r=m_r+n_r \tau$ and its complex conjugate $ \bar p_r$ through edge $r$ as providing a \textit{decoration} of the graph. The decorated graph is then specified by the connectivity matrix $\Gamma$ and the pair of arrays  $(A,B)$ defined by,
\bea
\label{ab}
A = [a_1, \cdots, , a_R] & \hskip 1in & \ma=a_1+ \cdots + a_R
\no \\
B = [b_1, \cdots, , b_R] && \mb=b_1+ \cdots + b_R
\eea
where $a_r, b_r \in \CC$ with $a_r-b_r \in \ZZ$ for all $r = 1, \cdots , R$.  The pair of total exponents $(\ma,\mb)$ generalizes the weight to the full decorated graph.

\sm

To a decorated graph  $(\Gamma, A, B)$ we associate a complex-valued function  on  $\cH$, defined by the following Kronecker-Eisenstein sum, whenever this sum is absolutely convergent, 
\bea
\label{def:Cgamma}
\cC_\Gamma \left [ \begin{matrix} A \cr B \cr \end{matrix} \right ]  (\tau) 
=   \left ( { \tau_2 \over \pi} \right ) ^{ \half \ma+ \half \mb} \sum_{p_1,\dots,p_R \in \Lambda '}  ~ \prod_{r =1}^R
  { 1 \over  (p_r) ^{a_r} ~ (\bar p _r) ^{b_r} }\, \prod_{v =1}^V
  \delta \left ( \sum_{s =1}^R \Gamma _{v \, s} \, p_s \right )
\eea 
The Kronecker $\delta$-symbol equals 1 when its argument vanishes as an element of the lattice $\Lambda$, and equals 0  otherwise.  The number of loops $L$  is the number of independent momenta, given by  $L=R-V+1$.  For any given decorated graph $(\Gamma, A,B) $, the domain of absolute convergence of the sums in (\ref{def:Cgamma}) is given by a system of inequalities on the combinations $\Re (a_r+b_r)$. Outside of the domain of convergence, it is an open question as to whether and how the functions $\cC$ of (\ref{def:Cgamma}) may be defined by analytic continuation in the variables $a_r+ b_r$.
The function $\cC$ in (\ref{def:Cgamma}) vanishes whenever the integer $\ma-\mb$ is odd, or whenever it is  associated with a graph $\Gamma$ which becomes disconnected upon severing a single edge. A function $\cC$ associated with a graph $\Gamma$ which is the union of two graphs $\Gamma = \Gamma _1 \cup \Gamma_2$, such that the intersection of $\Gamma _ 1$ and $\Gamma_2$ consists of a single vertex, factorizes into the product of  functions $\cC$ for $\Gamma_1$ and $\Gamma_2$, with the corresponding partitions of the exponents.

\subsubsection{Modular properties}

Under $SL(2,\ZZ)$ the  functions defined in (\ref{def:Cgamma})  transform as follows, 
\bea
\label{mod}
\cC_\Gamma  \left [ \begin{matrix} A \cr B \cr \end{matrix}\right ] \left  ({ a \tau + b \over c \tau + d} \right ) 
= \left ( { c \tau+d \over c  \bar \tau +d } \right ) ^{\half \ma - \half \mb}
\cC_\Gamma  \left [ \begin{matrix} A \cr B \cr \end{matrix} \right ] (\tau) 
\eea
where $a,b,c,d \in \ZZ$ and $ad-bc =1$. One defines the  \textit{modular weight} of  $\cC$  by the pair $(\half (\ma-\mb), \half (\mb-\ma))$ which is integer-valued for $\ma-\mb$  even ($\cC$ vanishes for odd $\ma-\mb$).  As we have explained above, we refer to $\cC$ as a  \textit{modular graph form} when $\ma\not= \mb$ and a \textit{modular graph function}  when  $\ma=\mb$.  
For $\ma\not= \mb$ the normalization factor of $\tau_2$ is not canonical, and we  introduce the forms $\cC^+$ and $\cC^-$ of modular weight $(0,\mb-\ma)$ and $(\ma-\mb,0)$ respectively as follows,
\bea
\label{Cplus}
\cC_\Gamma^\pm  \left [ \begin{matrix} A \cr B \cr \end{matrix}\right ]
= (\tau_2)^{\pm {\ma-\mb \over 2}} \cC_\Gamma  \left [ \bma A \cr B \ema \right ] (\tau) 
\eea
$\cC^+_\Gamma $ transforms with a factor $(c \bar \tau +d)^{-\ma +\mb}$ while $\cC^-_\Gamma $ transforms with a factor $(c \tau +d)^{\ma-\mb}$.

\subsubsection{Examples: Eisenstein series and dihedral graphs}

In this newfound notation, we may represent and generalize some of the examples of modular graph functions and forms encountered earlier. The Eisenstein series and its successive derivatives take the form, 
\bea
\label{9.Eisen}
E_s = \cC_\Gamma \left [ \bma s & 0 \cr s & 0 \ema \right ]
\hskip 0.8in
\nabla ^k E_s = {\Gamma (s+k) \over \Gamma (s)} \, \cC^+_\Gamma  \left [ \bma s +k & 0 \cr s-k  & 0 \ema \right ]
\eea
where $\Gamma$ is a connected one-loop graph  with only bivalent vertices. The two-loop modular graph functions $C_{a_1, a_2, a_3}$ may be generalized to $(n-1)$-loop modular graph functions, 
\bea
C_{a_1, \cdots, a_n} = \cC_\Gamma \left [ \bma a_1 & \cdots & a_n  \cr a_1 & \cdots & a_n \ema \right ]
\eea
where the ``dihedral" graph $\Gamma$ contains two vertices of valence $n+1$ in addition to the remaining vertices which are all bivalent. For $n=3$ we recover the two-loop modular graph functions studied earlier.

\subsubsection{Examples: holomorphic modular graph forms}
\label{sec:9.4.7}

When all anti-holomorphic exponents $b_i$ vanish identically the corresponding modular graph form reduces to the following expression, 
\bea
\cC_\Gamma^-  \left [ \bma a_1 & \cdots & a_n  \cr 0 & \cdots & 0 \ema \right ](\tau) 
=   \sum_{p_1,\dots,p_R \in \Lambda '}  ~ \prod_{r =1}^R
  { 1 \over  (p_r) ^{a_r} }\, \prod_{v =1}^V
  \delta \left ( \sum_{s =1}^R \Gamma _{v \, s} \, p_s \right )
\eea 
The multiple sum is absolutely convergent provided $a_r \geq 2$ for all $r=1,\cdots,  R\geq 2$. In this case the result is 
holomorphic in $\tau$ and a modular form of weight $(\ma,0)$ with $\ma=a_1 + \cdots, + a_R$. When $\cC^-_\Gamma$ is a modular form of modular weight $(\ma,0)$, it may be expressed as a polynomial in the holomorphic modular forms $\HE_4$ and $\HE_6$, and therefore produces a map from graphs $\Gamma$ to holomorphic modular forms, or equivalently to polynomials in $\HE_4$ and $\HE_6$. The nature of this map, which is much simpler than the map to non-holomorphic modular forms, remains to be investigated. Under weaker conditions of conditional convergence  the sum must be suitably regularized. Maintaining holomorphicity leads to a violation of full modular invariance, such as is the case with $a_1=2, a_2=0$ for $R=2$ which yields the quasi-modular form $\HE_2$.

\subsubsection{Algebraic relations}

Under complex conjugation, any of these functions have their arrays of holomorphic and anti-holomorphic exponents swapped, 
\bea
\label{3d2}
\cC_\Gamma  \left [ \begin{matrix} A \cr B \cr \end{matrix}\right ] (\tau) ^* 
= \cC_\Gamma  \left [ \begin{matrix} B \cr A \cr \end{matrix}\right ] (\tau)
\eea
In addition, modular graph forms obey momentum conservation identities, 
\bea
\label{3d3}
\sum _{r=1}^R \Gamma _{k\, r} \, \cC_\Gamma  \left [ \begin{matrix} A - S_r \cr B \cr \end{matrix}\right ]=
\sum _{r=1}^R \Gamma _{k\, r} \, \cC_\Gamma  \left [ \begin{matrix} A \cr B - S_r \cr \end{matrix}\right ]=0
\eea
where the $R$-dimensional row-vector $S_r$ is defined to have zeroes in all slots except for the $r$-th, which instead has value 1, 
\bea
\label{Sr}
 S_r = [\underbrace{0,\ldots,0}_{r-1},1,\underbrace{0,\ldots,0}_{R-r}]
 \eea
These identities simply express the fact that the lattice momenta entering each vertex $v$ add up to zero or, as physicists would state, that momenta is conserved at each vertex as a consequence of translation invariance on the torus.

\subsubsection{Differential relations}

The action of the Cauchy-Riemann operator $\nabla = 2 i \tau_2^2 \p_{\tau}$ on modular graph forms $\cC^+$ defined and normalized  in (\ref{Cplus})  for arbitrary exponents $A,B$ is given by, 
\bea
\label{nab}
\nabla \cC_\Gamma^+  \left [ \begin{matrix} A \cr B \cr \end{matrix}\right ]
= \sum_{r=1}^R a_r \, \cC^+ \left [  \bma A + S_r \cr B - S_r \ema \right] 
\eea
where  $S_r$ was defined in (\ref{Sr}).  The action of the Laplace operator $\Delta = 4 \tau _2^2 \p_{\bar \tau} \p_\tau$ on modular graph functions (for which $\ma=\mb$) is given by,
\bea
\label{Lap1}
(\Delta + \ma ) \, \cC_\Gamma  \left [ \begin{matrix} A \cr B \cr \end{matrix}\right ]
= 
\sum _{r,s=1}^R a_r b_s \, \cC \left [ \bma A+S_r - S_s  \cr B -S_r + S_s  \ema \right ]
\eea
It is similarly possible to define the action of the Laplace operator on modular graph forms of arbitrary modular weight with $\ma \not= \mb$, but we shall not need them here.

\subsubsection{Higher loop algebraic identities}

Higher-loop modular graph functions and forms satisfy systems of algebraic and differential identities which generalize those found for two-loops. For example, one has the following algebraic identities involving modular graph functions of weight 4 and 5, 
\bea
\label{9.ids}
C_{1,1,1,1} & = & 24 C_{2,1,1} + 3 E_2^2 - 18 E_4
\no \\
C_{1,1,1,1,1} & = & 60 C_{3,1,1} + 10 E_2 \, C_{1,1,1} -48 E_5 + 16 \zeta (5)
\no \\
40 C_{2,1,1,1} & = & 300 C_{3,1,1} + 120 E_2 E_3 - 276 E_5 + 7 \zeta (5)
\eea
These identities were proven  using \textit{holomorphic subgraph reduction} and a \textit{sieve algorithm}. We shall discuss the techniques used in  the proof of these identities in the next subsection for the example of the first identity. We refer to the bibliographical notes for the papers where all algebraic identities between modular graphs of weight up to six, and some selected identities of weight 7 are proven, and where a Mathematica package is now available.  

\subsection{Relating modular graph functions to holomorphic forms}

The Maass operator $\nabla = 2 i \tau_2^2 \p_\tau$ maps a modular graph form of modular weight $(0,\ell)$ to a modular graph form of modular weight $(0,\ell-2)$, as was shown in subsection \ref{sec:9.Maass}. When the derivative is applied to a non-holomorphic Eisenstein series $E_n$, for example, and repeated $n$ times, one obtains the following form, using (\ref{9.Eisen}),
\bea
\label{9.Edif}
\nabla ^n E_n = {\Gamma (2n) \over \Gamma (n)} \, \cC^+_\Gamma  \left [ \bma 2n ~ 0 \cr 0  ~~\, 0 \ema \right ]
= {\Gamma (2n) \over \Gamma (n) \pi^n } \, \tau_2^{2n} \, G_{2n}(\tau)
\eea
where $G_{2n}(\tau)$ is the holomorphic Eisenstein series of modular weight $(2n,0)$. 

\sm

Applying the Maass operator to more complicated modular graph functions is not as simple as in the case of non-holomorphic Eisenstein series, but nonetheless a similar pattern emerges: $\nabla$ decreases the anti-holomorphic exponents and increases the holomorphic exponents. Let's see how this works in the case of the modular graph function $C_{1,1,1,1}$. Using the rules of (\ref{nab}) for $\nabla$, we obtain for the first derivative,
\bea
\nabla C_{1,1,1,1} = 
\nabla \cC  \left [ \bma 1 ~ 1 ~ 1 ~ 1 \cr 1  ~ 1 ~ 1 ~ 1  \ema \right ]
=
4 \, \cC^+  \left [ \bma 2 ~ 1 ~ 1 ~  1 \cr 0  ~ 1 ~ 1 ~ 1  \ema \right ]
\eea
The form on the right has modular weight $(0,-2)$. Applying $\nabla$ again gives, 
\bea
\label{9.del2}
\nabla^2 C_{1,1,1,1} = 
12\, \cC^+  \left [ \bma 2 ~ 2 ~ 1 ~ 1 \cr 0  ~ 0 ~ 1 ~ 1  \ema \right ]
-24 \, \cC^+  \left [ \bma 3 ~ 1 ~ 1 ~  1 \cr 0  ~ 0 ~ 1 ~ 1  \ema \right ]
\eea
where to obtain the last term we have a momentum conservation identity to convert a negative entry on the lower row, 
\bea
\cC^+  \left [ \bma 3 & 1 ~ 1 ~  1 \cr -1  & 1 ~ 1 ~ 1  \ema \right ]
= -3 \, \cC^+  \left [ \bma 3 ~ 1 ~ 1 ~  1 \cr 0  ~ 0 ~ 1 ~ 1  \ema \right ]
\eea
Further application of $\nabla$ to (\ref{9.del2}), in its present form,  would yield modular graph functions with some negative anti-holomorphic exponents that cannot be removed simply by using momentum conservation identities. However, the subgraph corresponding to the first two entries forms a closed loop with only holomorphic momenta. 
The momentum summations in such holomorphic subgraphs may be evaluated explicitly in terms of holomorphic modular forms. This procedure is dubbed \textit{holomorphic subgraph reduction} and, in the present case, leads to the following identity,
 \bea
 \label{9.holo}
 \cC^+  \left [ \bma 2 ~ 2 ~ 1 ~ 1 \cr 0  ~ 0 ~ 1 ~ 1  \ema \right ]
-2 \, \cC^+  \left [ \bma 3 ~ 1 ~ 1 ~  1 \cr 0  ~ 0 ~ 1 ~ 1  \ema \right ]
= 
2 \, \cC^+  \left [ \bma 4 ~ 1 ~ 1  \cr 0  ~ 1 ~ 1   \ema \right ] 
+ 3 \, \cC^+  \left [ \bma 4 ~ 0  \cr 0  ~ 0   \ema \right ] \cC^+  \left [ \bma 2 ~ 0  \cr 2  ~ 0   \ema \right ] 
\eea
Collecting all contributions to $\nabla ^2 C_{1,1,1,1}$ and proceeding analogously for the entries $C_{2,1,1}$ and $E_2^2$ in the first identity in (\ref{9.ids}), we obtain, 
\bea
\label{9.del3}
\nabla^2 C_{1,1,1,1} & = & 
24 \, \cC^+ \! \left [ \bma 4 ~ 1 ~ 1  \cr 0  ~ 1 ~ 1   \ema \right ] 
+ 36 \, \cC^+ \!  \left [ \bma 4 ~ 0  \cr 0  ~ 0   \ema \right ] \cC^+ \! \left [ \bma 2 ~ 0  \cr 2  ~ 0   \ema \right ] 
\no \\
\nabla^2 C_{2,1,1} & = & 
6 \, \cC^+ \! \left [ \bma 4 ~ 1 ~ 1  \cr 0  ~ 1 ~ 1   \ema \right ] 
+ 6 \, \cC^+ \! \left [ \bma 3 ~ 2 ~ 1  \cr 1  ~ 0 ~ 1   \ema \right ] 
+ 4 \, \cC^+ \!  \left [ \bma 6 ~ 0  \cr 2  ~ 0   \ema \right ] 
\no \\
\nabla ^2 E_2^2 & = & 
8 \, \cC^+ \!  \left [ \bma 3 ~ 0  \cr 1  ~ 0   \ema \right ]^2
+ 12 \, \cC^+ \! \left [ \bma 4 ~ 0  \cr 0  ~ 0   \ema \right ] \cC^+ \! \left [ \bma 2 ~ 0  \cr 2  ~ 0   \ema \right ] 
\eea
Applying $\nabla$ to either $\nabla^2 C_{1,1,1,1}$ or to $E_2^2$ produces modular graph functions with negative entries in the lower row that cannot be removed by using momentum conservation identities. This occurs when $\nabla$ is applied to one of the modular graph functions that is proportional to a holomorphic modular form, such as the first factors in the last terms on the right of $\nabla^2 C_{1,1,1,1}$ and $\nabla^2 E_2^2$. However, $C_{2,1,1}$ and the linear combination $C_{1,1,1,1}-3E_2^2$ (that cancels the last terms discussed above) produces the following $\nabla^3$ derivatives,
\bea
\label{9.nab3}
\nabla ^3 (C_{1,1,1,1}-3E_2^2) & = & 432 \, \cC^+ \!  \left [ \bma 7 ~ 0  \cr 1  ~ 0   \ema \right ] 
-288 \, \cC^+ \!  \left [ \bma 4 ~ 0  \cr 0  ~ 0   \ema \right ] \cC^+ \!  \left [ \bma 3 ~ 0  \cr 1  ~ 0   \ema \right ] 
\no \\
\nabla^3 C_{2,1,1} & = & 108 \, \cC^+ \!  \left [ \bma 7 ~ 0  \cr 1  ~ 0   \ema \right ] 
-12 \, \cC^+ \!  \left [ \bma 4 ~ 0  \cr 0  ~ 0   \ema \right ] \cC^+ \!  \left [ \bma 3 ~ 0  \cr 1  ~ 0   \ema \right ] 
\eea
Proceeding analogously for the fourth $\nabla$-derivative, we obtain, 
\bea
\nabla ^4 ( C_{1,1,1,1} - 24 C_{2,1,1} - 3 E_2^2) =- 15120 \, \cC^+ \!  \left [ \bma 8 ~ 0  \cr 0  ~ 0   \ema \right ]
\eea
The right side is proportional to $\tau_2^8 G_8$, where $G_8$ is the holomorphic Eisenstein series of weight $(8,0)$. 
But this form is proportional to $\nabla ^4 E_4$, so that we obtain,
\bea
\label{9.D4}
\nabla ^4\big ( C_{1,1,1,1} - 24 C_{2,1,1} - 3 E_2^2 +18 E_4 \big )=0
\eea
Next, we invoke the following lemma.

{\lem 
\label{9.lem1}
Let $F$ be a non-holomorphic modular function with polynomial growth near the cusp as $\tau_2 \to \infty$. If $F$ satisfies the differential equation,
\bea
\label{9.F}
\nabla ^n F=0
\eea
for an arbitrary integer $n \geq 1$, then $F$ is constant as a function of $\tau$.}

\sm

A simple proof of the lemma proceeds by making use of the following formula,
\bea
\bar \nabla ^n (\tau_2)^{-2n} \nabla ^n = \prod _{s=1}^n \Big ( \Delta -s(s-1) \Big ) 
\eea
This formula immediately implies that $F$ must satisfy the weaker condition,
\bea
\prod _{s=1}^n \Big ( \Delta -s(s-1) \Big ) F=0
\eea
Being a modular graph function,  $F$ has at most polynomial growth at the cusp. The solution to the equation $\Delta f_s=s(s-1)f_s$ for positive integer $s$ is either a constant for $s=1$ or an Eisenstein series $E_s$ for $s \geq 2$. This implies that $F$ must be a linear combination of a constant and the Eisenstein series $E_s$ for $s=2, \cdots, n$. Returning now to the original equation (\ref{9.F}) in the formulation of the lemma and using the asymptotics of the Eisenstein series at the cusp of (\ref{4.d12}), we conclude that the coefficients of the Eisenstein series in $F$ must all vanish so that $F$ is constant. This completes the proof of the lemma. 

\sm

Applying the lemma to  (\ref{9.D4}), it follows that,
\bea
F_4 = C_{1,1,1,1} - 24 C_{2,1,1} - 3 E_2^2 +18 E_4
\eea
must be constant. To calculate the constant, we investigate the asymptotics at the cusp. 
Clearly, all $\tau$-dependent terms must cancel on the right side as one may verify using the expansions of $C_{1,1,1,1}$ and $C_{2,1,1}$ in terms of the variable $y = \pi \tau_2$,
\bea
C_{1,1,1,1}(\tau) & = & { y^4 \over 945} + { 2 \zeta(3) y \over 3} + {10 \zeta(5) \over y}  -{ 3 \zeta(3)^2 \over y^2} +{9 \zeta(7) \over 4 y^3} + \cO(e^{-2y})
\no \\
C_{2,1,1} (\tau) & = & { 2y^4 \over 14175} +{\zeta(3) y \over 45} +{5 \zeta(5) \over 12 y} -{ \zeta (3)^2 \over 4 y^2} +{9 \zeta (7) \over 16 y^3} + \cO(e^{-2y})
\eea
and the expansion of the Eisenstein series in (\ref{4.d12}). The constant $y^0$ term has vanishing coefficient in each of the the modular graph functions in $F_4$, so that $F_4=0$. This completes the proof of the first identity in (\ref{9.ids}). The other identities in (\ref{9.ids}) may be established with the same methods, and holomorphic subgraph reduction formulas that generalize (\ref{9.holo}).

\subsection{Modular graph functions and iterated modular integrals}

The Fourier expansion of non-holomorphic Eisenstein series $E_k(\tau)$ for integer $k \geq 2$, presented in (\ref{Eas}), shows that the Fourier non-zero modes are of the form $(q^N+\bar q^N)$ times a polynomial  in $1/\tau_2$. 
Such a simple form is no longer shared by modular graph functions and forms at higher loops since in that case terms of the form $q^M \bar q^N$ also occur. Nonetheless, modular graph forms are related to holomorphic modular forms, even at higher loop, as demonstrated by the fact that certain repeated derivatives $\nabla$ map to holomorphic modular forms. Another way to exhibit this connection concretely is via iterated modular integrals. 

\sm

The basic building blocks of iterated modular integrals are the functions $\cE(k_1, \cdots, k_r |\tau)$ and $\cE^0 (k_1, \cdots, k_r|\tau)$ defined iteratively by the following relations, 
\bea
\cE(k_1, \cdots, k_r|\tau) & = & 2 \pi i \int ^{i \infty} _\tau { d \tau_r \over (2 \pi i)^{k_r}} \, G_{k_r}(\tau_r) \, \cE(k_1, \cdots, k_{r-1}|\tau)
\no \\
\cE^0(k_1, \cdots, k_r|\tau) & = & 2 \pi i \int ^{i \infty} _\tau { d \tau_r \over (2 \pi i)^{k_r}} \, G^0_{k_r}(\tau_r) \, \cE^0(k_1, \cdots, k_{r-1}|\tau)
\eea
The holomorphic modular form $G_k(\tau)$  of weight $k$ is defined for $k \geq 3$ by the by now familiar absolutely convergent Kronecker-Eisenstein series, 
\bea
G_k (\tau) = \sum _{{ m,n \in \ZZ \atop (m,n) \not = (0,0)}} { 1 \over (m + n \tau)^k} = 2 \zeta (k) + G^0 _k (\tau)
\eea
The function $G_k(\tau)$ vanishes for all odd $k \geq 3$ and evaluates to $2 \zeta(k)$ at the cusp, while $G^0_k(\tau)$ consists of the non-constant Fourier modes of $G_k(\tau)$, and decays exponentially at the cusp. We extend the definition to $k=0$ and supply initial conditions for the iterated integrals, 
\bea
G_0(\tau) = ~ G^0_0(\tau) & = & -1
\no \\
\cE(\cdot |\tau) = \cE^0(\cdot |\tau)  & = & 1
\eea
In view of the exponential decay of $G^0_k(\tau)$ at the cusp for $k \geq 4$, the corresponding integrals are convergent, while all other integrations are defined via the tangential base-point prescription at the cusp, which essentially amounts to the following prescription,
\bea
\int _\tau ^{i \infty} d \tau_r \to - \tau
\eea 
The iterated integrals satsify the following differential equations, 
\bea
\label{9.it1}
\nabla \cE(k_1, \cdots, k_r|\tau) & = & { 4 \pi \tau_2^2 \over (2 \pi i)^{k_r}} \, G_{k_r}(\tau) \, \cE(k_1, \cdots, k_{r-1} |\tau) 
\no \\
\nabla \cE^0(k_1, \cdots, k_r|\tau) & = & { 4 \pi \tau_2^2 \over (2 \pi i)^{k_r}} \, G^0_{k_r}(\tau) \, \cE^0(k_1, \cdots, k_{r-1} |\tau) 
\eea
For example, we have,
\bea
\label{9.itdif}
\nabla \cE^0(2k|\tau) & = & { 4 \pi \tau_2^2 \over (2 \pi i)^{2k}} \, G^0_{2k}(\tau) 
\no \\
\nabla  \cE^0(2k, 0_n |\tau) & = & - 4 \pi \tau_2^2 \, \cE^0(2k, 0_{n-1} |\tau) 
\eea
where the notation $0_n$ stands for an array of $n$ zeros.  

\sm

We begin by showing how the simplest Eisenstein series $E_2$ may be expressed in terms of these iterated integrals.  The starting point is the differential equation obtained in (\ref{9.Edif}),
\bea
\pi^2 \nabla ^2 E_2 = 6 \tau_2^4 \, G_4 = { 2 \pi^4 \tau_2^4 \over 15} + 6 \tau_2^4 \, G_4^0
\eea
Using the relation $\nabla \tau_2^n = n \tau_2^{n+1}$ and eliminating $G_4^0$ in favor of $\nabla \cE^0(4|\tau)$ using the first equation in (\ref{9.itdif}), we obtain, 
\bea
\pi^2 \nabla ^2 E_2 = { 2 \pi^4  \over 45} \nabla \tau_3^3  + 24 \pi^3  \tau_2^2 \, \nabla \cE^0(4|\tau)
\eea
Successive use of the second relation in (\ref{9.itdif}) for $n=1,2$ allows us to recast the right side in the form of a total 
derivative in terms of the iterated integrals $\cE^0(4|\tau), \cE^0(4,0|\tau), \cE^0(4,0_2|\tau)$,
\bea
\nabla (\pi \nabla E_2 ) = \nabla \Big ( { 2 \pi^3 \tau_2^3 \over 45} + 24 \pi^2 \tau_2^2 \cE^0(4|\tau) + 12 \pi \tau_2 \cE^0(4,0|\tau) + 3 \cE^0(4,0_2|\tau) \Big  )
\eea
The equation is readily solved, and the general solution is given as follows,
\bea
\label{9.nabE2}
\pi \nabla E_2  =  { 2 \pi^3 \tau_2^3 \over 45} + 24 \pi^2 \tau_2^2 \, \cE^0(4|\tau) 
+ 12 \pi \tau_2 \, \cE^0(4,0|\tau) + 3 \, \cE^0(4,0_2|\tau) + \overline{f_1(\tau)} 
\eea
where $f_1(\tau)$ is holomorphic. Making use anew of the equations (\ref{9.itdif}) we can integrate this equation once more, and find, 
\bea
E_2 (\tau) = { \pi^2 \tau_2^2 \over 45} - 6 \, \cE^0(4,0|\tau) - { 3 \over \pi \tau_2} \cE^0(4,0_2|\tau) - { \overline{f_1(\tau)} \over \pi \tau_2} + \overline{f_2 (\tau)}
\eea
where $f_2(\tau)$ is another holomorphic function. The reality of $E_2$ determines $f_1$ and $f_2$ uniquely  up to an additive constant in $f_1$, which may be fixed by the known asymptotics  at the cusp. Using again the shorthand $y=\pi \tau_2$, we have, 
\bea
E_2 (\tau) = { y^2 \over 45} + {\zeta (3) \over y} - 12 \, \Re \big [ \cE^0(4,0|\tau) \big ]  
- { 6 \over y} \Re \big [ \cE^0(4,0_2|\tau) \big ] 
\eea
The above expression reflects the structure of the Fourier decomposition of the Eisenstein series mentioned at the beginning of this section. 

\sm

Next, we provide an example of the decomposition into iterated integrals for the simplest  modular graph function that is not an Eisenstein series, namely $C_{2,1,1}$. The differential equations for $C_{2,1,1}$, given in the second line of (\ref{9.nab3}), may be recast in the following form,
\bea
\pi^2 \nabla ^3 \big (C_{2,1,1} - \tfrac{9}{10} E_4 \big )= - 6 \tau_2^2 G_4 (\tau) \nabla E_2
\eea
The subtraction of the term in $E_4$ is convenient as it readily cancels the first term on the right side of the second line in (\ref{9.nab3}). One may begin by using expression given in (\ref{9.nabE2}) to eliminate $\nabla E_2$ from the above equation in favor of iterated integrals. Successive use of  (\ref{9.it1}) again allows us to proceed by the same methods as used for $E_2$, and one obtains,
\bea
C_{2,1,1}& = & - \tfrac{y^4}{20250} + \tfrac{\zeta(3) y}{45} + \tfrac{5 \zeta(5)}{12 y} - \tfrac{\zeta(3)^2}{4 y^2} - \left ( \tfrac{ 2 y}{15} - \tfrac{ 3 \zeta(3)}{ y^2 } \right ) \Re \big [ \cE^0(4,0_2|\tau) \big ] 
\\
&&- \tfrac{9}{y^2} \, \Re\big[ \cE^0(4,0_2|\tau) \big ]^2 -72\, \Re \big [ \cE^0(4_2,0_2|\tau) \big ] -\tfrac{1}{5} \, \Re \big [ \cE^0(4,0_3|\tau) \big ]
\no\\
&&-\tfrac{36}{y} \, \Re \big [ \cE^0(4,0,4,0_2|\tau) \big ]-\tfrac{108}{y} \, \Re \big [ \cE^0(4_2,0_3|\tau) \big ]-\tfrac{1}{10 y} \, \Re \big [ \cE^0(4,0_4|\tau) \big ]\no
\eea
In this formulation, the first term on the second line reveals the presence of terms of the form $q^M \bar q^N$ in the Fourier series of $C_{2,1,1}$, as expected.

\subsection*{$\bullet$ Bibliographical notes}

The study of modular graph functions originated with explorations into the low energy effective interactions produced by one-loop superstring amplitudes \cite{Green:2008uj}, as will be described in sections \ref{sec:SA} and \ref{sec:IIB}.  Systematic investigations of modular graph functions were initiated in \cite{DHoker:2015gmr,DHoker:2015sve,Zerbini:2015rss}. The proof of  Theorem \ref{6.thm1} was given in \cite{DHoker:2015gmr}, while the proof of Theorem \ref{6.thm2} was given in \cite{d2018fourier}, where explicit expressions for all the coefficients in the expansion were also provided.  The Poincar\'e series for $\Gamma _\infty \backslash SL(2,\ZZ)$, and Fourier series were computed in  \cite{DHoker:2019txf}.  The relation between modular graph functions and single-valued elliptic polylogarithms was exhibited in \cite{DHoker:2015wxz}. Introductions to elliptic polylogarithms may be found in \cite{Zagier9, BL, Enriquez,Broedel:2017kkb}.  

\sm

The systematic construction of the algebraic and differential identities satisfied by modular graph functions and forms was carried out in  \cite{DHoker:2016mwo,DHoker:2016quv,Gerken:2018zcy}
using the method of holomorphic subgraph reduction. A variety of other methods were used in \cite{Basu:2015ayg,Basu:2016xrt,Basu:2016kli,Kleinschmidt:2017ege,Basu:2019idd}
to obtain identities within special classes of modular graph functions. An efficient Mathematica package for 
algebraic and differential relations was developed in \cite{Gerken:2020aju, Gerken:2020xte}. A class of related modular functions is presented in \cite{Brown:2017qwo,Brown:2020mhp}.

\sm

Fourier series expansions were obtained in \cite{d2018fourier,Ahlen:2018wng,DHoker:2019txf,Dorigoni:2019yoq}, while Poincar\'e series were derived in \cite{Ahlen:2018wng,DHoker:2019txf,Dorigoni:2019yoq}.  The role of modular graph functions in genus-one amplitudes for open and Heterotic  strings was elucidated in \cite{Broedel:2018izr,Gerken:2018jrq}.  The solutions to the differential identities obtained from holomorphic subgraph reduction may be obtained in terms  of iterated elliptic integrals and systematically represented in terms of  generating functions \cite{Broedel:2018izr,Gerken:2019cxz,Gerken:2020yii}. Additional discussions of the relation between modular graph functions and iterated Eisenstein integrals can be found in \cite{Dorigoni:2022npe,Hidding:2022vjf}. 
 
\sm

The generalization of modular graph functions to higher genera was initiated in \cite{DHoker:2017pvk} and is directly motivated by the study of the low energy expansion of superstring amplitudes, as will be discussed extensively in section \ref{sec:SA}. The natural generalization of modular graph forms to higher genus is provided by modular graph tensors, which were introduced in \cite{ DHoker:2020uid}. Various degenerations of the higher genus Riemann surface, studied in  \cite{DHoker:2018mys}, naturally lead one to introduce elliptic modular graph functions, whose study was begun in  \cite{DHoker:2020aex}.

\newpage

\part{Extensions and Applications}

In the second part of these lecture notes, we consider mathematical and physical extensions and applications of the material discussed in the first part. We shall also exhibit the connections with the second part of the title, namely various topics in string theory and related fields, wherever appropriate.

\sm

We start off in section \ref{sec:Hecke} with a review of Hecke operators, Hecke eigenforms, and the action of Hecke operators on Maass forms and vector-valued modular forms. We close with a physical application to the study of two-dimensional conformal field theory.   In section \ref{sec:CM}, we consider the topic of complex multiplication and singular moduli. We prove that the modular $j$-function is an algebraic integer at complex multiplication points $\tau$ and evaluate elliptic and $\tet$-functions at these points. 

\sm

In section \ref{sec:SA}, we present a quick introduction to string amplitudes. We review their perturbative expansion in the genus which corresponds to the loop expansion in quantum field theory, and the importance of conformal symmetry in decoupling of negative norm states. We summarize results for amplitudes to tree-level, one-loop, and two-loop orders without derivation.  We emphasize the crucial role played by modular invariance in the UV finiteness of string theory, and by modular graph functions in the low energy expansion in effective interactions. As an  application, we analyze toroidal compactifications in section \ref{sec:Toroidal}. 

\sm

In section \ref{sec:IIB} we review Type IIB supergravity and superstring theory, discuss its S-duality symmetry under $SL(2,\ZZ)$, spell out the modular properties of the low energy effective interactions, and make contact with the low energy effective interactions predicted earlier from superstring amplitudes in section \ref{sec:SA}. 

\sm

In section \ref{sec:SW} we present a brief introduction to Yang-Mills theories with extended supersymmetry in four space-time dimensions. We review  Goddard-Nuyts-Olive duality, its concrete realization as Montonen-Olive dualities in  $\cN=4$ theories, and then discuss the Seiberg-Witten solution with  emphasis on the role played by $SL(2,\ZZ)$. We close with a discussion of  dualities of $\cN=2$ superconformal Yang-Mills theory.

\sm

In the last section \ref{sec:Galois}, we address Galois theory, along with some applications to rational conformal field theory.

\newpage

\section{Hecke Theory}
\setcounter{equation}{0}
\label{sec:Hecke}

A natural set of operators acting on the space of modular forms are the Hecke operators $\Tt_n$, labeled by  positive integers~$n$. They map holomorphic functions to holomorphic functions, weight-$k$ modular forms to weight-$k$ modular forms, and weight-$k$ cusp forms to weight-$k$ cusp forms. For the group $SL(2,\ZZ)$ the Hecke operators are endomorphisms that map the space of holomorphic modular forms $\cM_k$ into itself, and map the subspace of cusp forms $\cS_k$ into itself. For congruence subgroups, the Hecke operators map weight-$k$ modular forms of one congruence subgroup into those of another congruence subgroup. Hecke operators commute with the Laplace-Beltrami operator on the upper half-plane so that non-holomorphic Eisenstein series and cusp forms are simultaneous eigenfunctions of all Hecke operators. Finally, given a modular form with positive integer Fourier coefficients, the Hecke transforms also have positive integer Fourier coefficients. For this reason, Hecke operators are relevant in a number of physical problems, as we shall see shortly.

\subsection{Definition of Hecke operators}
\label{sec:Heckeprops}

Given a lattice $\Lambda= \ZZ \omega_1 + \ZZ \omega_2$ with modulus $\tau=\om_2/\om_1 \in \cH$, a modular form $f(\tau)$ of weight $k$ under $SL(2,\ZZ)$ may be represented by a degree $k$ homogeneous function of the lattice $\Lambda$, denoted by the same letter, $f(\Lambda) = \omega_1^{-k} f(\omega_2/ \omega_1)$. Modularity of $f(\tau)$ is then equivalent to the fact that $f(\Lambda)$ is intrinsic and depends only on the lattice $\Lambda$, but not on the particular basis vectors  $\om_1, \om_2$ used to represent $\Lambda$. The Hecke operator $\Tt_n: \cM_k \rightarrow \cM_k$ may now be defined on $f(\Lambda)$ as a sum over all index $n$ sub-lattices $\Lambda_n \subset \Lambda$, 
\bea
\Tt_n f(\Lambda) = n^{k-1} \sum_{\substack{\Lambda_n \subset \Lambda \\ [\Lambda : \Lambda_n]=n}} f(\Lambda_n)
\eea
where the sub-lattice $\Lambda_n$ is understood to be a subgroup of $\Lambda$. The definition of $\Tt_n$ given above is intrinsic, in that it depends only on the lattices $\Lambda$ and $\Lambda_n$, and not on any particular choice of the basis vectors for each lattice. An arbitrary multiplicative normalization factor, which depends only on $k$ and $n$, is allowed in the definition of $\Tt_n$, and has been chosen here for later convenience. 

\sm

To make contact with modular forms expressed as functions of $\tau \in \cH$, we now choose specific basis vectors  for both $\Lambda$ and $\Lambda_n$. In particular, the set of index $n$ sub-lattices $\Lambda_n$ may be parametrized by choosing basis vectors $\omega_1^{(n)}$, $\omega_2^{(n)}$ for $\Lambda_n$  which are integer linear combinations of the basis vectors $\om_1$ and $\om_2$ of $\Lambda$, 
\bea
\omega_2^{(n)} = a\, \omega_2 + b\, \omega_1 \hspace{0.5 in} \omega_1^{(n)} = c\, \omega_2 + d \,\omega_1
\eea
where $a,b,c,d \in \ZZ$. Since $\Lambda$ and $\Lambda_n$ are Abelian groups, $\Lambda_n $ is a normal subgroup of $\Lambda$ and the quotient $\Lambda/\Lambda_n$ is a finite group. One may represent $\Lambda/\Lambda_n$ as the set of the points in $\Lambda$ modulo $\Lambda_n$, equipped with addition modulo $\Lambda_n$. The cardinality of $\Lambda / \Lambda_n$ is the index $[\Lambda: \Lambda _n]$ which is given by the number of points in $\Lambda$ that belong in the fundamental parallelogram of $\Lambda_n$, and this number is $n=ad-bc$, a number that is positive assuming that $\Lambda_n$ inherits its orientation from $\Lambda$.   The above parametrization of $\Lambda_n$  motivates us to introduce the set $M_n$ of all $2\times 2$ matrices with integer coefficients of determinant $n>0$, 
\bea
\label{eq:Mndef}
M_n = \left\{  \left( \begin{matrix} a & b \\ c & d\end{matrix} \right)  \Big | \,\,a,b,c,d \in \ZZ;\,\, ad - bc = n >0 \right\}
\eea
Given two matrices $A,B \in M_n$, one introduces an equivalence relation $A \sim B$ whenever $A= \mu B$ for some  element $\mu \in SL(2,\ZZ)$.  The quotient $SL(2, \ZZ)\backslash M_n$ is the set of equivalence classes of elements of $M_n$, subject to the equivalence relation. This relation is precisely the equivalence relation between lattices $\Lambda_n$ under $SL(2,\ZZ)$ transformations of the underlying lattice $\Lambda$. Therefore, the  cosets in $SL(2, \ZZ)\backslash M_n$ are in one-to-one correspondence with the lattices $\Lambda_n$ and the  Hecke operator $\Tt_n$ may equivalently be defined to act as follows,
\bea
\label{Heckedef}
\Tt_n f(\tau) = n^{k-1} \sum_{\g \in SL(2, \ZZ)\backslash M_n} (c\tau+d)^{-k} f \left(\g \tau \right)
\eea
The above definition makes it clear that $\Tt_n$ maps holomorphic functions to holomorphic functions. Below we shall show that it maps modular forms of $SL(2,\ZZ)$ to modular forms.

\subsection{Explicit parametrization of equivalence classes $SL(2, \ZZ)\backslash M_n$}

The action of the Hecke operators $\Tt_n$ can be made even more explicit by using the following  convenient explicit parametrization of the equivalence classes $SL(2, \ZZ)\backslash M_n$. 
{\lem 
\label{9.lem1a}
Every equivalence class in $SL(2, \ZZ)\backslash M_n$ contains an upper triangular representative of the form $\left(\begin{smallmatrix} a  & b \\ 0 & d \end{smallmatrix}\right) \in M_n$, with $d>0$. A complete set of inequivalent elements of $SL(2, \ZZ)\backslash M_n$ is given by $b \in \{ 0, 1, \cdots, d-1\}$. }

\sm

To prove the first part of the Lemma,  we show that for every matrix $A = \left(\begin{smallmatrix}a & b \\ c & d \end{smallmatrix}\right) \in M_n$ there exists a $\mu \in SL(2,\ZZ)$ such that $\mu A$ is of the  form given in the Lemma. To do so we parameterize $\mu$ as follows,
 \bea
\mu =  \left(\begin{matrix}p & q \\ r & s \end{matrix}\right)
\hskip 1in
\mu A = \left( \begin{matrix} p a + q c & p b + q d \\ r a + s c& r b + s d\end{matrix} \right) 
\eea
If $c=0$ then $d \not=0$ in view of $ad-bc=n>0$ by the definition of $M_n$ and we set $r=q=0$ and $p=s={\rm sign} (d)$ to obtain a representative of the desired form.  For the remaining cases $c\neq 0$  we choose  mutually prime $r$ and $s$ such that $s/r = - a/c$ to render the matrix $\mu A$ upper triangular. By B\'ezout's theorem (reviewed in appendix \ref{sec:modN}) there exist integers $p$ and $q$ such that $p s - q r = 1$ guarantees  $\mu \in SL(2,\ZZ)$, and  either $\mu$ or $-\mu$ gives the desired matrix.

\sm

To prove the second part of the Lemma, we show that the set given by $b \in \{ 0, 1, \cdots, d-1\}$ is complete and that all such elements are inequivalent under $SL(2, \ZZ)$. To see that the set is complete, we take two matrices $A_1 = \left(\begin{smallmatrix}a & b_1 \\ 0 & d \end{smallmatrix}\right)$ and $A_2= \left(\begin{smallmatrix}a & b_2 \\ 0 & d \end{smallmatrix}\right)$ with $b_1 \equiv b_2 \,\,\,({\rm mod}\,\,d)$. Then if $b_1 = b_2 + k d$ we define $\mu =  \left(\begin{smallmatrix}1 & k \\ 0 & 1 \end{smallmatrix}\right) \in SL(2, \ZZ)$ so that $A_1 = \mu A_2$, and hence $A_1 \sim A_2$. Conversely, to see that all elements are inequivalent we proceed as follows. If $A_1$ and $A_2$ are such that $A_1 \sim A_2$, there exists $\mu =  \left(\begin{smallmatrix}p & q \\ r & s \end{smallmatrix}\right) \in SL(2, \ZZ)$ such that $A_1 = \mu A_2$, namely,
\bea
\left(\begin{matrix}a_1 & b_1 \\ 0 & d_1 \end{matrix}\right) = \left(\begin{matrix}p & q \\ r & s \end{matrix}\right) \left(\begin{matrix}a_2 & b_2 \\ 0 & d_2 \end{matrix}\right) = \left(\begin{matrix}p a_2 & pb_2+q d_2 \\ r a_2 & r b_2 + s d_2 \end{matrix}\right) 
\eea
Since $a_2 d_2 = n$, clearly $a_2 \neq 0$, and hence by equating the bottom-left entries on both sides we conclude that $r=0$. Then since $p q - r s = 1$, we have $p q = 1$ and thus $p=q =\pm1$. We make the choice $p=q=+1$, which we can always do by replacing $\mu$ with $-\mu$ if needed. Then equating entries on both sides, we see that $a_1 = a_2$, $d_1 = d_2$, and  $b_1 \equiv b_2 \,\,\,({\rm mod}\,\,d)$. Hence if we restrict to $b \in \{0,\dots, d-1\}$, there are no equivalences.

\sm

We shall also make use of the following simple but important lemma.
{\lem 
\label{9.lem2}
For every pair $(A_1,\mu_1)$ with $A_1 \in M_n$ and $\mu_1 \in SL(2,\ZZ)$, there exists a pair $(A_2, \mu_2)$ with $A_2 \in M_n$ and $\mu_2 \in SL(2,\ZZ)$ such that $A_1 \mu_1= \mu_2 A_2$. As a result, the left and right cosets $SL(2,\ZZ)\backslash M_n$ and $M_n / SL(2,\ZZ)$ are isomorphic.}

\sm

To prove this lemma, we may choose $A_1$ and $A_2$ of the form provided by Lemma \ref{9.lem1a} without loss of generality, and parametrize the matrices $A_1,A_2,\mu_1, \mu_2$  involved as follows,
\bea
A _1= \left(\begin{matrix}a_1 & b_1 \\ 0 & d_1 \end{matrix}\right) 
\hspace{0.4 in}
A_2 = \left(\begin{matrix}a_2 & b_2 \\ 0 & d_2 \end{matrix}\right) 
\hspace{0.4 in}
\mu_1 = \left(\begin{matrix}p_1 & q_1 \\ r_1 & s_1 \end{matrix} \right)
\hspace{0.4 in}
\mu_2 = \left(\begin{matrix}p_2 & q_2 \\ r_2 & s_2 \end{matrix} \right)
\eea 
Given $\mu_1$ and $A_1$, one proceeds to solve for $A_2$ and $\mu_2$ using $\mu_2 A_2= A_1 \mu_1$ in components, 
\begin{align}
\label{9.Hecka}
p_2a_2 & =  a_1p_1+b_1r_1  & p_2 b_2 + q_2 d_2 & = a_1q_1 + b_1 s_1 
\no \\
r_2 a_2  & =  d_1r_1 & r_2 b_2 + s_2 d_2  & =  d_1 s_1
\end{align}
From the equations on the left one obtains $p_2/r_2 = (a_1p_1+b_1r_1)/(d_1 r_1)$, which determines the pair $(p_2, r_2)$ uniquely up to a common sign in view of the fact that $p_2, r_2$ must be relatively prime. The entry $a_2$ is then determined by $a_2=d_1 r_1/r_2$. Having determined the pair $p_2, r_2$,   B\'ezout's theorem guarantees the existence of a solution $q_2, s_2$ to $p_2s_2-q_2r_2=1$.  In terms of these solutions, we readily solve the equations on the right to obtain,
\bea
b_2 & = & s_2 a_1 q_1 +s_2 b_1 s_1 -q_2 d_1 s_1 
\no \\
d_2 & = & p_2d_1s_1- r_2 a_1 q_1 - r_2 b_1 s_1
\eea

\subsection{Hecke operators map $\cM_k$ to $\cM_k$}

Lemma \ref{9.lem1a} allows us to recast the action of the Hecke operators more explicitly as follows,
\bea
\label{Heckealtdef}
\Tt_n f(\tau) = n^{k-1}\sum_{\substack{a d = n\\ d >0}} {1 \over d^k}\sum_{b =0}^{d-1} f\left({a \tau + b \over d} \right)
\eea
or by solving for $a=n/d$, 
\bea
\Tt_n f(\tau) =  n^{k-1} \sum_{{d|n \atop d>0}}{1 \over d^k} \sum_{b=0}^{d-1}f\left({n \tau + b d \over d^2} \right)
\eea
In terms of these expressions for $\Tt_n$, it is now straightforward to prove the modularity of $\Tt_n f(\tau)$. We shall use the notations and parametrizations of Lemma \ref{9.lem2}.  Indeed, considering a transformation by $\mu_1 \in SL(2, \ZZ)$, we have, 
\bea
\Tt_n f(\mu_1 \tau) &=& n^{k-1}\sum_{A_1} {1 \over d_1^k}\, f\left(A_1 \mu_1 \tau \right)
\no\\
 &=& n^{k-1}\sum_{A_2} {1 \over d_1^k}\, f\left(\mu_2 A_2 \tau \right)
 \no\\
 &=& n^{k-1} \sum_{A_2} {1 \over d_1^k} \,(r_2 A_2 \tau + s_2)^kf\left( A_2 \tau \right)
\eea
where we have introduced a shorthand notation $\sum_{A}$ for the summations in (\ref{Heckealtdef}), and have used the fact that if $A_1$ runs through a complete set of inequivalent elements of $SL(2, \ZZ)\backslash M_n$, so too does $A_2$. Substituting the two equations  on the second line of (\ref{9.Hecka}), we then obtain \bea
\Tt_n f(\mu_1 \tau) =n^{k-1} \sum_{A_2} {1 \over d_2^k} (r_1 \tau + s_1)^k f\left( A_2 \tau \right) = (r_1 \tau + s_1)^k \,\Tt_n f( \tau)
\eea
which is the desired result.

\subsection{Fourier expansions}

For many applications in Physics and Mathematics, we will often be interested in the Fourier expansions of Hecke transforms. Given a weight-$k$ modular function $f(\tau)$ with Fourier expansion $f(\tau)= \sum_{m} a_m q^m$, the Fourier expansion of the Hecke transform can be computed explicitly as follows.
We begin by noting that,
\bea
\Tt_n f(\tau)  =  n^{k-1} \sum_{d|n}{1 \over d^k}\sum_{b=0}^{d-1} 
\sum_{m \in \ZZ} a_m \, q^{n m \over d^2}  e^{2 \pi i {b m \over d}} 
\, = \, \sum_{m \in \ZZ} \, \sum_{d | \gcd (n,m)} \left({n \over d} \right)^{k-1} a_m \, q^{n m \over d^2}
\eea
The second equality follows from the fact that $\sum_{b=0}^{d-1} e^{2 \pi i {b m \over d}}$ equals $d$ if ${m \over d}\in \ZZ$, and vanishes otherwise. Then we simply make repeated changes to the summation index to get, 
\bea
\Tt_n f(\tau)  &=& \sum_{\ell \in \ZZ} \sum_{d| n} \left({n \over d} \right)^{k-1} a_{\ell d} \, q^{n \ell \over d}
= \sum_{\ell \in \ZZ} \sum_{t | n} t^{k-1} a_{\ell n / t} \, q^{\ell t}
\no\\
&=& \sum_{m \in \ZZ} \, \sum_{t | {\gcd} (m,n)} t^{k-1}a_{mn / t^2} \, q^m
\eea
We thus conclude that the Fourier expansion of the Hecke transform is given by, 
\bea
\label{HeckeFourier}
\Tt_n f(\tau) = \sum_{m\in \ZZ} b^{(n)}_m q^m \hspace{1 in} b^{(n)}_m = \sum_{d | {\gcd}(m,n)} d^{k-1}a_{mn/d^2}
\eea
If the Fourier coefficients $a_m$ of $f(\tau)$ are integers, then by construction so are the Fourier coefficients of the Hecke transform $\Tt_n f$ for all $n >0$. Furthermore, we see from  (\ref{HeckeFourier}) that if $a_0 = 0$ then $b^{(n)}_0 = 0$ as well, and thus $\Tt_n$ acts as an endomorphism on the space $\cS_k$ of weight-$k$ cusp forms. 

\sm

In the particular case of $n=p$ being prime, the Fourier coefficients $b^{(n)}_m$ of the Hecke transform take a particularly simple form,
\bea
\label{eq:bnprime}
b^{(p)}_m= \left\{ \begin{matrix}\,\, p^k a_{pm}  & \hspace{0.3 in} p \nmid m \\ \,\,p^{k-1}(p \,a_{p m } + a_{m/p}) & \hspace{0.3 in} p\, |\, m \end{matrix}\right.
\eea
If $k \geq 1$, we see that if $a_m \in \ZZ$, then $b^{(p)}_m \in \ZZ$ as well. For $k = 0$ though this is no longer the case. For this reason, in the case of $k=0$ one usually drops the conventional factor of $n^{k-1}$ in (\ref{Heckedef}).

\subsection{Example: the Ramanujan tau function}
\label{sec:RamtauHecke}

We illustrate the utility of this machinery by means of a simple example. In particular, let us prove the product formulas (\ref{taumult}) and (\ref{taumult2}) for the Ramanujan tau function, defined in (\ref{3.tau}) of section \ref{sec:Discriminant}, and repeated here for convenience, 
\bea
\Delta (\tau) = (2 \pi)^{12} \sum _{n=1}^\infty \tau (n) q^n
\eea 
To do so, we apply the Hecke operators to the discriminant $\Delta(\tau)$ to obtain, 
\bea
\Tt_n \Delta(\tau) = \sum_m b_m^{(n)} q^m \hspace{0.8 in} b_m^{(n)} 
= (2\pi)^{12} \sum_{d | {\gcd} (m,n)}d^{11}\, \tau\left({n m \over d^2} \right)
\eea
Since $\Delta (\tau)$ and $\Tt_n \Delta (\tau)$ are both holomorphic cusp forms of weight 12, and  the space $\cS_{12}$ is one-dimensional, we conclude that we must have, 
\bea
\Tt_n \Delta(\tau) = \alpha_n \Delta(\tau)
\eea
for some sequence of constants $\alpha_n \in \CC$. Equating Fourier coefficients using (\ref{HeckeFourier}), we obtain the following relation, 
\bea
\label{9.mult}
\alpha_n  \tau(m) = \sum_{d | {\gcd} (m,n)}d^{11}\, \tau\left({n m \over d^2} \right)
\eea
For  the particular case of $m=1$ we have $\tau(1) = 1$ and only the term with $d=1$ contributes to the sum on the right side, so that $\a_n = \tau(n)$. Substituting this expression for $\a_n$ into (\ref{9.mult}) readily reproduces (\ref{taumult2}).

\subsection{Multiplicative properties of Hecke operators}

An important property of Hecke operators is their behavior under multiplication,
\bea
\label{Heckemult}
\Tt_{n}\Tt_m &=& \sum_{d| (n,m)} d^{k-1} \, \Tt_{m n / d^2} 
\eea
The full proof of this property is straightforward and standard (see e.g. \cite{Serre}), but somewhat lengthy. Here we settle for a proof of a special case of this relation,
\bea
\Tt_n \Tt_m = \Tt_{n m} \hspace{0.8 in}\gcd (m,n) = 1
\eea
To prove this, we begin by explicitly applying the Hecke operators on a weight-$k$ modular form $f(\tau)$ to obtain
\bea
\label{TnTmf}
\Tt_n \Tt_m f(\tau) = (nm)^{k-1}\sum_{\substack{\a \d = n \\ \d>0}}\,
\sum_{\substack{a d = m\\ d >0}} \,
\sum_{\b = 0}^{\d-1}\,
\sum_{b=0}^{d-1} {1 \over (d \d)^k} f\left({\a a \tau + (\a b + \b d) \over d \d} \right)
\eea
Now we make use of the fact that $(m,n)=1$, which tells us that since $d$ and $\delta$ run through positive divisors of $n$ and $m$ respectively, $d \delta$ should run through the positive divisors of the product $nm$. 
Furthermore, since $b$ $(\mod d)$ runs through a complete set of representatives for $SL(2, \ZZ)\backslash M_m$ and $\b$ $(\mod \delta)$ runs through a full set of representatives for $SL(2, \ZZ)\backslash M_n$, then $\a b + \b d$ $(\mod d \delta)$ runs through a complete set of inequivalent representatives for $SL(2, \ZZ)\backslash M_{mn}$. Hence by the definition given in (\ref{Heckedef}), the right side of (\ref{TnTmf}) is exactly the action of $\Tt_{m n}$ on $f(\tau)$.

\subsection{Hecke eigenforms}

The multiplication identity (\ref{Heckemult}) implies that the Hecke operators commute. As such, we may consider modular forms $f(\tau) \in \cM_k$ which are simultaneous eigenfunctions of all Hecke operators, 
\bea
\label{eigenformdef}
\Tt_n f(\tau) = \lambda_n f(\tau) \hspace{0.5 in}\forall \,n \in \NN
\eea
Such forms are called Hecke eigenforms, or just ``eigenforms" for short.\footnote{In some literature the term eigenform refers to eigenfunctions of only a single Hecke operator, whereas our definition of eigenforms would be referred to as ``simultaneous eigenforms."} Comparing the $\cO(q)$ terms in the Fourier expansions on both sides of (\ref{eigenformdef}), the Fourier coefficients of eigenforms are seen to satisfy, 
\bea
\label{Heckecoeffrel}
b_m^{(n)} = \lambda_n a_m
\eea
using (\ref{HeckeFourier}). Considering in particular $m=1$, we see that $a_n = \lambda_n a_1$, and thus that any non-trivial eigenform must have $a_1 \neq 0$. One often finds it useful to normalize $a_1=1$ so that the Hecke eigenvalues coincide with the Fourier coefficients of $f(\tau)$, 
\bea
\Tt_n f(\tau) = a_n f(\tau)
\eea
It is a general fact that normalized Hecke eigenforms have Fourier coefficients which are real (by Hermiticity of $\Tt_n$, as will be discussed momentarily) and multiplicative, 
\bea
\label{Fouriercoeffsmult}
a_m a_n = \sum_{ d|{\gcd} (m,n)} d^{k-1}a_{mn/d^2}
\eea

 The existence of eigenforms is obvious for certain weights. For example, for weights $k=4,6,8,10,$ and $14$, we saw in Theorem \ref{theorem2} that the space $\cM_k$ is one-dimensional, being generated by the relevant Eisenstein series $\HE_k(\tau)$. Hence for such values of $k$ the Hecke transform $\Tt_n \HE_k(\tau)$ must clearly be a multiple of $\HE_k(\tau)$ for any $n$, making $\HE_k(\tau)$ a Hecke eigenform. More surprisingly, the Eisenstein series are Hecke eigenforms even when ${\rm dim}\, \cM_k>1$, which is shown as follows. 
 Consider a generic non-cuspidal modular function satisfying 
\bea
\label{HeckeEisen}
\Tt_n f = \lambda_n f
\eea 
with $f$ having even weight $k$. For non-cuspidal $f$ the Fourier coefficient $a_0$ is non-zero, and by extension $b_0^{(n)}$ is non-zero as well. 
In particular, for $m=0,1$ we have from (\ref{HeckeFourier}) that
\bea
b_0^{(n)} &=& \sum_{d|n} d^{k-1}a_0 \,\,\,=\,\,\, a_0 \, \sigma_{k-1}(n)
\no\\
b_1^{(n)}&=& \sum_{d|{\gcd}(1,n)} d^{k-1}a_{n / d^2} \,\,\,=\,\,\, a_n
\eea
with $ \sigma_*(n)$ the sum of divisors function.  Comparing the first of these to (\ref{Heckecoeffrel}) allows us to conclude that $\lambda_n = \sigma_{k-1}(n)$, and then from the second of these we obtain $a_n = a_1 \sigma_{k-1}(n)$. Choosing normalization $a_0 = 1$, the Fourier expansion of the Hecke eigenform $f(\tau)$  is required to take the following form,
\bea
f(\tau) = 1+ a_1 \sum_{N=1}^\infty \sigma_{k-1}(N) \, q^N
\eea
The normalized Eisenstein series $\HE_k(\tau)$ take exactly this form, as in (\ref{Ektau}). Incidentally, note that the divisor function indeed satisfies the multiplicative property (\ref{Fouriercoeffsmult}). 

\sm

Besides Eisenstein series, all Hecke eigenforms are cuspidal. We have already encountered one example of a cuspidal eigenform in section \ref{sec:RamtauHecke}, where we saw that the modular discriminant $\Delta(\tau)$ is an eigenfunction of the Hecke operators, with the eigenvalues being precisely the values of the Ramanujan tau function.  We also saw that the Ramanujan tau function satisfied a multiplicativity property consistent with (\ref{Fouriercoeffsmult}). 

\sm

The fact that $\Delta(\tau)$ was an eigenform followed simply from the fact that ${\rm dim}\,\cS_{12} = 1$. The same argument holds for cusp forms of weight $k=16, 18, 20, 22,$ and $26$, where ${\rm dim}\,\cS_k = 1$, and is generated by $\Delta (\tau) \HE_k(\tau)$. However, it is a remarkable fact, due to Hecke, that there exist cuspidal eigenforms for other $k$ as well. In fact, Hecke showed that \textit{eigenforms provide a basis of $\cM_k$ for every $k$}. In order to prove this, he found it necessary to introduce an inner product on the space of cusp forms, as we now briefly review.

\subsubsection{Petersson inner product}

Weight-$k$ cusp forms admit an inner product known as the Petersson inner product,
\bea
\langle f, g \rangle = \int_{SL(2, \ZZ)\backslash \mathbb{H}} {d\tau_1 d\tau_2 \over \tau_2^2}\,(\tau_2)^k f(\tau) \overline{g(\tau)}
\eea
which is convergent due to the vanishing of $f$ or $g$ towards the cusp.   
A key fact is that Hecke operators are self-adjoint under this inner product,
\bea
\langle \Tt_n  f, g\rangle = \langle   f,  \Tt_ng\rangle
\eea
This together with commutativity of $\Tt_n$ implies that a generic $ f \in \cS_k$ can be decomposed into a basis of Hecke eigenforms, which is the result quoted before. This also tells us that the eigenvalues of $\Tt_n$ (and hence the Fourier coefficients of normalized eigenforms) are all real---another result quoted earlier.

\subsection{Hecke operators acting on Maass forms}

We now review the properties of Hecke operators acting on the space of Maass forms of weight 0, which was introduced in section \ref{sec:4.3}. Recall that a Maass form $f(\tau)$  is a complex-valued function of $\tau$ that satisfies a Laplace eigenvalue equation,
\bea
\Delta f(\tau) = s(s-1) f(\tau)
\eea
is invariant under $SL(2,\ZZ)$, and has at most polynomial growth at the cusp. The Fourier series for general Maass forms was obtained in (\ref{eq:Maassfourier}), and we repeat it here for convenience,
\bea
\label{9.Maass}
f(\tau) = a \, \tau_2 ^{s} + b \, \tau_2^{1-s} + \sum _{N \not= 0} \a_N \, \sqrt{\tau_2} \, K_{s-1/2} (2 \pi  |N| \tau_2) \, e^{2 \pi i N \tau_1} 
\eea
The space of all Maass forms for the full modular group $SL(2,\ZZ)$ and with eigenvalue $s(s-1)$ is denoted $\cN(s) = \cN(SL(2,\ZZ), s)$ with the understanding that it  satisfies $\cN(1-s)=\cN(s)$.  Non-holomorphic Eisenstein series $E_s(\tau)$, defined and normalized in (\ref{3.nhE}),  are Maass forms whose Fourier series were obtained in (\ref{4.d12}), 
\bea
a = { 2 \zeta (2s) \over \pi^s} 
\hskip 0.5in
b= { 2 \Gamma (s-\thalf) \zeta (2s-1) \over \Gamma (s) \, \pi^{s-\thalf}} 
\hskip 0.5in
\a_N = { 4 |N|^{\thalf -s} \sigma _{2s-1}(|N|) \over \Gamma (s) } 
\eea
A Maass form $f(\tau)$ is  a cusp form provided the constant Fourier mode in (\ref{9.Maass}) vanishes, $a=b=0$. Cusp forms decay exponentially at the cusp and have finite norm with respect to the Petersson inner product. The space of cusp forms is denoted $\cS(s)$. The space $\cS(s)$ is empty except for an infinite number of discrete eigenvalues of the form $s \in \thalf + i \RR$. Theorem~\ref{thm:Maassspace} provides a basis for $\cN(s)$ in terms of Eisenstein series and cusp forms. 

\sm

Hecke operators are linear operators that may be defined on Maass forms in $\cN(s)$ similarly to how  they were defined on holomorphic modular forms,
\bea
\Tt_n f(\tau) = {1 \over \sqrt{n}} \sum _{\gamma \in SL(2,\ZZ)\backslash M_n} f(\gamma \tau)
\eea
In contrast to (\ref{Heckedef}), there is no power of $(c \tau+d)$ multiplying the summand as the form $f(\tau)$ has weight 0. The definition allows for an arbitrary normalization prefactor that depends only on $n$ and is completely multiplicative and has been chosen to be $1/\sqrt{n}$ for later convenience. Using Lemma \ref{9.lem2}, and the fact that a Maass form is invariant under $SL(2,\ZZ)$, it follows immediately that $\Tt_n f(\tau)$ is also invariant under $SL(2,\ZZ)$. Furthermore, since $f(\tau)$ has at most polynomial growth at the cusp, so must $\Tt_n f(\tau)$. This is because it is defined to be  a finite sum of terms, each of which has at most polynomial growth. By manipulations analogous to those used for the holomorphic case, one may use an explicit parametrization of the coset $SL(2,\ZZ)\backslash M_n$ to recast the action of the Hecke operators in the following form,
\bea
\Tt_n f(\tau) = {1 \over \sqrt{n}} \sum _{{ad=n, d>0 \atop b=0,\cdots, d-1}} f \left ( { a \tau +b \over d} \right )
\eea
The Laplace-Beltrami operator $\Delta = 4 \tau_2^2 \p_\tau \p_{\bar \tau}$ is invariant under $\Tt_n$ because the transformation $\tau \to \tau'=(a \tau+d)/d$ implies $\tau_2 \to \tau_2'= a \tau_2/d$ and $\p_\tau \to \p_{\tau'} = d/a \,  \p_\tau$. These results may be summarized and further extended in the following theorem \cite{terras1}:

{\thm
\label{9.thm:Hecke}
The Hecke operators $\Tt_n$ satisfy the following properties. 
\begin{enumerate}
\itemsep=-0.05in
\item $\Tt_n$ map $\cN(s)$ to $\cN(s)$ and $\cS(s)$ to $\cS(s)$. 
\item The Fourier series of a Maass form $f(\tau)$ in $\cN(s)$, given in (\ref{9.Maass}),  and its Hecke image $\Tt_n f(\tau)$ expressed in the form given in (\ref{9.Maass}) with $a\to a', b \to b', \a_N \to \a_N'$ are related as follows,
\bea
\left \{ \bma a' = n^{1/2-s} \sigma_{2s-1}(n) \, a \cr b' = n^{s-1/2} \sigma_{1-2s}(n) \, b \ema \right . 
\hskip 1in
\a_N' = \sum_{d | {\gcd} (N,n)} \a_{Nn/d^2}
\eea
\item The product of operators $\Tt_m$ and $\Tt_n$ on $\cN(s)$  is commutative  and  given as follows,
\bea
\Tt_m \Tt_n = \sum_{d |{\gcd} (m,n)} \Tt_{mn/d^2}
\eea
\item The operators $\Tt_n$ are self-adjoint linear operators with respect to the Petersson inner product on the space of cusp forms, and may be simultaneously diagonalized on $\cS(s)$.
\item If $f \in \cN(s)$ is a simultaneous eigenfunction of all Hecke operators with eigenvalues $\Tt_n f = \lambda _n f$ then the coefficients $\a_N$  in its Fourier expansion  (\ref{9.Maass}) satisfy, 
\bea
\a_{\pm N} = \lambda _N \, \a_{\pm 1}  \hskip 1in N >0
\eea
Therefore, up to the normalizations $\a_{\pm 1}$, the coefficients of the non-constant Fourier modes are the eigenvalues of the Hecke operators. 
\end{enumerate}
}

\sm

Item 1 was already established in the preamble to the theorem, while Item 2 readily follows by identifying terms in the Fourier series for $f$ and $\Tt_nf$. Item 3 may be established along the same lines as the product formula (\ref{Heckemult}) was for holomorphic modular forms. The proof of Item 4 is given in \cite{terras1}. Finally, Item 5 is established using the second relation in Item 2 for $\a_N ' = \lambda _n \a_N$, which implies,
\bea
\lambda _n \, \a_N = \sum_{d | {\gcd} (N,n)} \a_{Nn/d^2}
\eea
for all $n >0$ and all $N \not=0$. Setting $N=\pm 1$ collapses the sum on the right to just the term $d=1$, which produces the result announced.

\subsection{Hecke operators on vvmfs}
\label{sec:Heckevvmf}

In section \ref{sec:MDEsvvmfs} we introduced modular differential equations (MDEs), and discussed how they were usefully organized according to the so-called Wronskian index $\ell$, defined in (\ref{Wronskianindexdef}), which gives a rough measure of the number of poles allowed in the coefficient functions. For the case of $\ell=0$ the coefficient functions must be holomorphic, and for low orders the MDEs can be recast as hypergeometric equations and solved exactly, as was done in sections \ref{sec:secondorderMDE} and \ref{sec:thirdorderMDE} for the cases of second- and third order, respectively.

Obtaining exact solutions to MDEs for general $\ell$ is more difficult. One way to obtain (a subset of) solutions to MDEs with non-zero $\ell$ is to make use of Hecke operators. Here we will give only a flavor of how this is done.  Above we have defined the action of Hecke operators on modular forms for $SL(2,\ZZ)$. In fact, there exists a straightforward generalization to modular forms for congruence subgroups. Assuming the Integrality Conjecture of section \ref{sec:Intconj}, this then leads to a component-wise action of the Hecke operators on \textit{integral} vector-valued modular forms. For a vector-valued modular form ${\bf f}(\tau)$ with components $f_i(\tau)$ transforming covariantly under $\Gamma(N)$, the action of $\Tt_p$ for $p$ prime such that $(p,N)=1$  is given by
\bea
(\Tt_p f)_i(\tau) = \sum_j \rho_{ij}(\sigma_p) f_j(p \tau) + \sum_{b=0}^{p-1} f_i\left(  {\tau + b N \over p}\right)
\eea
where $\rho(\sigma_p)$ is a matrix representation of
\bea
\sigma_p  = T^{\overline p} S^{-1} T^p S T^{\overline p} S
\eea
and $\overline p$ is the modular inverse of $p$, i.e. $p \overline p \equiv 1$ (mod $N$). This definition of the Hecke operators is chosen to lead to a simple action on the Fourier coefficients. Indeed, if we write the Fourier expansion of the components of a vector-valued modular function as, 
\bea
f_i(\tau) =\sum_{n} a_n^{(i)} q^{n/N}
\eea
then one has that 
\bea
(\Tt_pf)_i(\tau) =\sum_{n } b_{n}^{(i)}(p) q^{n/N} 
\eea
with 
\bea
\label{eq:vvmfHeckecoeffs}
b_{n}^{(i)}(p)= \left\{\begin{matrix} p a_{p n}^{(i)} & \,\,\,\,p \nmid n \\ p a_{p n}^{(i)} + \sum_j \rho_{ij}(\sigma_p) a_{n/p}^{(j)} &\,\,\,\, p \,|\, \,n \end{matrix} \right.
\eea
 This action of the Hecke operator gives a new vector-valued modular form with components which have the same weight $k$ under $\Gamma(N)$ as before, but now with different leading exponents $n_0^{(i)}$ in the notation of (\ref{vvmfFourier}). Because $\mathrm{ord}_{W_d}(i \infty)  = \sum_i n_0^{(i)}$, we expect from (\ref{vvmfvalence}) that the Hecke image $\Tt_p {\bf f}(\tau)$ generically satisfies an MDE with different Wronskian index $\ell$. The Hecke operators can thus be used to transform solutions to MDEs with $\ell=0$ to solutions to those with $\ell >0$. 

\sm

We content ourselves with a single example here. Consider the three-dimensional vector-valued modular function made up of characters of the Ising CFT, given in (\ref{concretedim3sols}). From (\ref{Isingcharacterqexp}), the leading exponents are seen to be
\bea
n_0^{(1)} = {23 \over 48}\hspace{0.5 in} n_0^{(2)} = -{1 \over 48}\hspace{0.5 in} n_0^{(3)} = {1 \over 24}
\eea
in the notation of (\ref{vvmfFourier}). Upon application of the Hecke operator $\Tt_p$, these change to 
\bea
\tilde{n}_0^{(1)} = {23p\,\,{\rm mod}\,48 \over 48}\hspace{0.5 in} \tilde{n}_0^{(2)} = -{p \over 48}\hspace{0.5 in} \tilde{n}_0^{(3)} = {2p \,\,{\rm mod}\,48 \over 48}
\eea
from (\ref{eq:vvmfHeckecoeffs}). Then using the formula (\ref{vvmfvalence}), we conclude that for e.g. $\Tt_{49}, \Tt_{53}$, and $\Tt_{55}$ we have $\ell = 6$, and hence application of these Hecke operators to the Ising characters gives solutions to an MDE of order $3$ with meromorphic coefficient functions. These solutions are exact in the sense that their full Fourier expansions can be obtained, though they are not known to be expressible in terms of any standard functions.

\sm

Another interesting case is that of $\Tt_{47}$ acting on the Ising characters. In that case we obtain 
\bea
(\Tt_{47} \, \chi_1)(\tau) &=& q^{-{47 \over 48}}\left( 1 + 96256 \, q^2 + 9646891 \, q^3 + 366845011 \, q^4 + \dots\right) 
\\
(\Tt_{47} \, \chi_\eps)(\tau) &=& q^{25 \over 48} \left( 4371 + 1143745 \, q + 64680601 \, q^2 + 1829005611 \, q^3 + \dots\right)
\no\\
(\Tt_{47} \, \chi_\sigma)(\tau) &=& q^{23 \over 24 } \left( 96256 + 10602496 \, q + 420831232 \, q^2 + 9685952512 \, q^3 + \dots\right)
\no
\eea
These turn out to be solutions to an order $3$ MDE with $\ell=0$, and can be shown to be precisely the characters of the so-called Baby Monster CFT, which has as its global symmetry group the Baby Monster $\mathbb{B}$, the second-largest finite sporadic group. Hence the Hecke operator $\Tt_{47}$ maps from the Ising CFT to the Baby Monster CFT, in some appropriate sense.

\subsection{Physics applications}
\label{sec:3dgrav}

We now give a physical application of Hecke theory. Before doing so, we will need to recall the basic notion of Virasoro characters.  In section \ref{sec:charLfunc}, we introduced characters for finite groups. Characters can be introduced for Lie groups as well, in which case, given a  representation $\cR$ of the group, we have, 
\bea
\chi_{\cR}(g) = {\rm Tr}_{\cR}\, (g )
\eea
with the trace $\Tr_\cR$ taken in  representation space of $\cR$. The general theory of characters is rich, but largely unnecessary for our purposes. The particular case of importance to us here is that of the characters of the Virasoro group. As we have discussed in section \ref{sec:conffields}, the Virasoro group is an infinite-dimensional Lie group which is the universal central extension of the group of diffeomorphisms of the circle. The corresponding Lie algebra is the Virasoro algebra, whose generators $L_n$ satisfy the structure relations given in (\ref{eq:Virasoro}), which we repeat here for convenience, 
\bea
[L_n, L_m] = (n-m)L_{n+m} + {\mc \over 12} (n^3 - n) \delta_{n,-m}
\eea
The parameter $\mc$ here is known as the central charge. 

\sm

A quantum field theory with Virasoro symmetry is known as a conformal field theory (CFT). By ``symmetry," one means that the state space of the theory is organized into Virasoro representations. In a unitary CFT, these representations can be constructed by beginning with a highest weight state $|h\rangle$ satisfying 
\bea
L_0 |h\rangle = h |h \rangle \hspace{0.5 in} L_n|h\rangle = 0 \hspace{0.3 in} n >0
\eea
and then generating other states in the representation by taking linear combinations of states of the form, 
\bea
L_{-n_1}L_{-n_2} \dots L_{-n_r}|h\rangle  \hspace{0.5in} n_i > 0
\eea
Such states are known as descendants, and will be collectively denoted by $|h, \{n\}\rangle$. The character associated to a representation labelled by $h$ is then defined to be, 
\bea
\chi_h(\tau) = q^{h - {\mc \over 24}} \sum_{\{n\}} q^{N}
\eea
with $q = e^{2 \pi i \tau}$ and $N = \sum_i n_i$. This is sometimes also written as, 
\bea
\chi_h(\tau) = {\rm Tr}_{\cV_h} \left ( q^{L_0 - {\mc \over 24}} \right )
\eea
with $\cV_h$ known as the \textit{weight-h Verma module}. 

\sm

Characters play an important role in the study of 2d CFTs. In particular, the torus partition function of many CFTs can be decomposed into a sum over conformal families and a sum within each family, giving 
\bea
\label{eq:CFTtoruspartfunct}
Z(\tau, \bar \tau) = \sum_{h, \tilde{h}} N_{h, {\tilde{h}}}\,\chi_h(\tau) \overline{\chi_{\tilde{h}}(\tau)}
\eea
for some gluing matrix $N_{h, {\tilde{h}}}$. Here we are assuming that the indices $h, \tilde{h}$ are discrete, but not necessarily that the sums over conformal families are finite.

\sm

The partition function is by definition a weight-0 modular function. But conversely, not every modular function corresponds to a valid CFT partition function. For example, since partition functions count degeneracies of states, any physically sensible partition function must have positive integer Fourier coefficients. The same is true for the constituent characters. In this context, it is useful to introduce a variant of the Hecke operators with alternative normalization, 
\bea
\cT_n := n \,\Tt_n
\eea
which has the virtue that if our original function $\chi(\tau)$ has positive integer Fourier coefficients $a_m \in \ZZ_{\geq 0}$, the Hecke transform $\cT_n \chi(\tau)$ has positive integer Fourier coefficients $n \,b_m^{(n)} \in \ZZ$ as well; c.f. the discussion after (\ref{eq:bnprime}). The application of Hecke operators $\cT_n$ to a CFT character thus produces another function, which itself has some  properties expected of a CFT character. In this sense, we might hope that the Hecke operators provide a ``map between CFTs," as in the example of the Ising and Baby Monster CFTs mentioned above.
We now study this phenomenon in the simplest examples of holomorphically factorizable CFTs, i.e. those for which the partition function is given simply as a product $Z(\tau, \bar \tau) = Z(\tau) \overline{Z(\tau)}$. 

\sm

Let us consider a class of two-dimensional CFTs arising as the holographic duals to three-dimensional gravity in anti-de Sitter space AdS$_3$. We will consider theories of pure gravity, which in three dimensions are known to not have any propagating gravitational waves. Indeed, three-dimensional gravity can be rewritten as a topological theory, namely an $SL(2, \RR) \times SL(2, \RR)$ Chern-Simons theory specified by levels $k_L, k_R$. Here we work with the case of $k_L = k_R = \kappa$. By comparing the Chern-Simons formulation of the theory with the Einstein-Hilbert formulation, one can show that the curvature radius of AdS$_3$ is related to the level $\kappa$ by $\ell = 16 \k G_N$, where $G_N$ is Newton's constant. In particular, we see that $\ell$ is quantized. 

\sm

What do the tentative dual CFTs look like? By the Brown-Henneaux formula \cite{Brown:1986nw}, the CFTs should have central charges,
\bea
\mc = {3 \ell \over 2 G_N} = 24 \kappa
\eea
The vacuum energy is given by $- {\mc \over 24} = - \k$. Additional states can be obtained from the vacuum state by application of the generators of the Virasoro algebra. In particular,   the generators $L_n$ for $n\geq -1$ annihilate the vacuum, while the generators $L_{-n}$ for $n>1$ act on the vacuum to give new states with energy $-\k + n$. For given energy $-\k + n$, the naive degeneracy would be given by $p(n)$, the number of integer partitions of $n$. For example, for $n=3$ we have $p(3)=3$, corresponding to the three states, 
\bea
L_{-1}^3 | 0 \rangle \hspace{0.6 in} L_{-1} L_{-2}|0\rangle \hspace{0.6 in} L_{-3} |0\rangle
\eea
This would give total vacuum contribution to the chiral partition function,
\bea
Z_\k^{\rm vac}(\tau) \stackrel{?}{=} q^{- \k} \sum_{n=0}^\infty p(n) q^n = q^{-\k} \prod_{n=1}^\infty{1 \over 1-q^n} 
\eea
as per the discussion in section \ref{sec:partitions}. However, in the current case $L_{-1}$ annihilates the vacuum, and the corresponding degeneracies must be removed. This is accomplished by including a factor of $(1-q)$, giving the actual vacuum contribution 
\bea
\label{eq:vaccontrib}
Z_\k^{\rm vac}(\tau) = q^{-\k} \prod_{n=2}^\infty{1 \over 1-q^n} 
\eea
The vacuum contribution to the partition function is not itself modular invariant, which means that there must be additional states beyond the vacuum and its descendants. Indeed, there must be some states in the CFT which capture the known black holes states of the holographic dual. These black holes are expected on general grounds to appear at order $O(q^1)$, giving a full partition function of the form
\bea
Z_\k(\tau) = q^{-\k} \prod_{n=2}^\infty{1 \over 1-q^n} + \cO(q)
\eea
Remarkably, there is a unique such partition function for every $\k$. 

\sm

For $\k=1$, i.e. pure AdS$_3$ gravity with the smallest value of the cosmological constant, the only possible candidate is
\bea
Z_1(\tau) &=& j(\tau) - 744
\no\\
&=& q^{-1}+ 196884 \, q + \dots
\eea
with $j(\tau)$ the $j$-function introduced in section \ref{sec:jfunction}. Physically, the 196884 states at level 1 can be interpreted as 1 descendant of the vacuum, and an additional 196883 black hole states.  The 2d CFT with this partition function was first constructed by Frenkel, Lepowsky, and Merman and is known as the ``Monster CFT," due to it having global symmetry group given by the Fischer-Greiss monster $\mathbb{M}$, the largest finite sporadic group. The smallest non-trivial representation of $\mathbb{M}$ has dimension 196883, and the black hole states indeed lie in this representation.

\sm

To obtain partition functions for higher $\k>1$, we now make use of Hecke operators. Thus far in this section we have focused on Hecke operators acting on holomorphic modular forms, but the extension to the meromorphic case is straightforward. The $q$-expansions of Hecke transforms can be obtained using (\ref{HeckeFourier}), giving e.g.
\bea
\cT_2 J(\tau) &=& q^{-2} + 42987520 \, q + \dots 
\no\\
\cT_3 J(\tau) &=& q^{-3} + 2592899910 \, q + \dots
\eea
where we have defined $J(\tau) = j(\tau) - 744$. For any $\k$, it is easy to show that one has 
\bea
\cT_\k J(\tau) = q^{-\k} + \cO(q)
\eea
Using this, we can now write down closed form expressions for $Z_\k(\tau)$ as follows. Say that 
\bea
Z_\k^{\rm vac}(\tau) = \sum_{r=-\k}^\infty c_r q^r
\eea
Of course, using (\ref{eq:vaccontrib}) all $c_r$ are known. Then, the full chiral partition function is given simply by 
\bea
\label{eq:fullHeckepartfunc}
Z_\k(\tau) = \sum_{r = 0}^\k c_{-r} \cT_r J
\eea
where we take $\cT_0 J(\tau) = 1$. Using the formulas above, we get explicitly
\bea
Z_2(\tau) &=& q^{-2}+ 1 + 42987520 \, q+\dots 
\no\\
Z_3(\tau) &=& q^{-3} + q^{-1} + 1 + 2593096794 \, q + \dots
\eea
and so on. Though no concrete construction of CFTs with these partition functions are known (assuming they actually exist), there exists much literature on the subject. Such tentative theories are referred to as ``extremal CFTs." We have thus seen that the Hecke operators act as maps between extremal CFTs. 

\sm

Furthermore, by using Hecke operators we have obtained the exact number of quantum black hole microstates in theories of pure gravity. In particular, this number is obtained by subtracting the number of vacuum descendants from the coefficient of the $\cO(q^1)$ term. It is interesting to compare these results to those predicted by the Bekenstein-Hawking entropy,
\bea
S_{BH} = 4\pi \sqrt{\k L_0}
\eea
For example, for $\k=L_0=1$ one finds $S_{BH} =4 \pi \approx  12.57$, to be compared to the exact result $S_{\rm exact} = \log 196883 \approx 12.19$. The agreement between the Bekenstein-Hawking entropy and the exact microstate count improves in the semiclassical limit $\k, L_0 \rightarrow \infty$ with $L_0/k$ fixed. The origin of this agreement is in fact a classic result of Petersson and Rademacher. Writing the Fourier expansion of $j(\tau)$ as
\bea
j(\tau) -744= \sum_{n=-1}^\infty c_n q^n
\eea 
with $c_0=0$, Petersson and Rademacher proved that as $n \rightarrow \infty$, one has 
\bea
\label{eq:asympjcn}
\log c_n \approx 4 \pi \sqrt{n} - {3 \over 4} \ln m - \half \ln 2 + \dots
\eea
We now combine this with the results for the Fourier coefficients of Hecke transforms. Since we are working in the limit $\k, L_0 \rightarrow \infty$, it suffices to consider only the $\cT_\k J$ contribution to (\ref{eq:fullHeckepartfunc}). The dominant contribution to the corresponding Fourier coefficients $b^{(\k)}_n$ are $b^{(\k)}_n \approx \k c_{\k n}$, which gives 
\bea
\log b^{(\k)}_n \approx 4 \pi \sqrt{\k n} + {1 \over 4} \ln \k - {3 \over 4} \ln n - \half \ln 2 + \dots
\eea
The first term reproduces the Bekenstein-Hawking result, while the others give the leading-order quantum corrections.

\subsection*{$\bullet$ Bibliographical notes}

The formal theory of Hecke operators was developed by Hecke in \cite{Hecke1,Hecke2}, but the operators had been introduced two decades earlier when Mordell proved the multiplication law of the Ramanujan $\tau$-function \cite{Mordelltau}, reproduced in section \ref{sec:RamtauHecke}. Most of the material discussed above is by now standard, and may be found in the books by Apostol \cite{Apostol}, Shimura \cite{Shimura}, Diamond and Shurman \cite{DS}, and Serre \cite{Serre}.   The theory of Hecke operators acting on Maass forms is described in the book by Terras \cite{terras1}. The double coset construction is described in detail in the book by Diamond and Shurman \cite{DS} and the book by Iwaniec \cite{Iwan1}. The action of Hecke operators on vector-valued modular forms was developed recently in both physics \cite{Harvey:2018rdc,Harvey:2019qzs} and mathematics \cite{Bouchard:2018pem} literature. 

\sm

The discussion of three-dimensional gravity given in section \ref{sec:3dgrav} follows closely the work of \cite{Witten:2007kt}. For standard introductions to two-dimensional quantum field theory and conformal field theory we refer again to  \cite{Ginsparg,DMS}, already mentioned in section \ref{sec:TorusQFT}. The basic tools of AdS/CFT are reviewed in \cite{Aharony:1999ti,DHoker:2002nbb} as well as in the book by Ammon and Erdmenger \cite{Ammon}. The non-existence of several extremal CFTs was recently proven in \cite{Lin:2021bcp}, using the theory of topological modular forms. The asymptotic form of the coefficients of the $j$-function quoted in (\ref{eq:asympjcn}) was obtained by Petersson in \cite{Peterssonj}, and independently by Rademacher \cite{Rademacherj}. 

\newpage

\section{Singular moduli and complex multiplication}
\setcounter{equation}{0}
\label{sec:CM}

In previous sections we introduced $SL(2,\ZZ)$ as the automorphism group for any lattice $\Lambda = \ZZ \om _1 + \ZZ \om_2$ with $\Im (\om_2/\om_1) >0$. Singular moduli correspond to special values of $\tau=\om_2/\om_1$ for which the lattice has  an automorphism group that is larger than $SL(2,\ZZ)$.  The mechanism by which this happens is referred to as \textit{complex multiplication}. 

\sm

A first rather obvious example is when $\tau=i$, in which case the lattice has an extra symmetry group $\{ \pm I, \pm S\}$ of order four.  A second obvious example is when $\tau=\rho=e^{ 2 \pi i /3}$, in which case the extra symmetry group $\{ \pm I, \pm (ST), \pm (ST)^2 \}$ is of order six. One way of looking at these automorphisms is by complex multiplication, namely in the first case we may take the periods to be $\om_1=1$ and $\om_2=i$, and the extra non-trivial symmetries correspond mapping $\om_1 \to i \om _1= \om_2$ and $\om_2  \to i \om_2=-\om_1$, leaving the lattice invariant.

\subsection{Conditions for complex multiplication}

We seek the lattices for which there exists a complex number $\alpha \in \CC \setminus \ZZ$ such that $\alpha \Lambda \subset \Lambda$. The condition may be written out explicitly on the periods as follows,
\bea
\alpha \om_2 & = & a \om_2 + b \om_1
\no \\
\alpha \om_1 & = & c \om_2 + d \om_1
\eea
with $a,b,c,d \in \ZZ$ and $ad-bc \not=0$. Note that we are not requiring $ad-bc=1$ since we only demand that $\alpha \Lambda$ be contained in $\Lambda$ without  necessarily being equal to $\Lambda$. Taking the ratio gives a condition on $\tau$, while given such a solution $\tau$, the value of $\alpha$ is then determined, 
\bea
\tau = { a \tau + b \over c \tau + d}
\hskip 1in
\alpha = c\tau  + d
\eea
The condition on $\tau$ is that it satisfies a quadratic polynomial equation,
\bea
\label{eq:tauquadCM}
c \tau^2 +(d-a)\tau-b=0
\eea 
with integer coefficients. The discriminant of this equation is denoted by $D$,
\bea
D=(a-d)^2+4bc=(a+d)^2 -4(ad-bc)
\eea 
The restriction $\Im (\tau) \not= 0$ imposes the condition $D<0$, which in turn requires $bc<0$ and $ad-bc \geq1$. Without loss of generality we may assume that $c>0$ and define $\sqrt{D}$ to be the square root of $D$ satisfying $\Im (\sqrt{D})>0$,  so that we have,
\bea
\tau = { a-d + \sqrt{D} \over 2c}
\hskip 1in
\alpha = { a+d + \sqrt{D} \over 2}
\hskip 0.8in 
|\a|^2= ad-bc
\eea
Extending the field of the rationals $\QQ$ by $\tau$ gives the  quadratic extension field $\QQ(\tau)$.

\subsection{Elliptic functions at complex multiplication points}

When $\alpha \Lambda \subset \Lambda$ and $\alpha \in \CC \setminus \ZZ$, elliptic functions and modular forms at $\tau$ satisfy remarkable properties. A first key result is for elliptic functions. It suffices to investigate the Weierstrass function $\wp (z)= \wp (z|\Lambda)$ for the lattice $\Lambda$. Since $\alpha \Lambda \subset \Lambda$, the function $\wp(\alpha z)$ is an elliptic function for the original lattice $\Lambda$ and may therefore be expressed as a rational function of $\wp(z)$. More specifically, we have,
\bea
\label{eq:compmultprel}
\wp (\alpha z) = { A(\wp(z)) \over B(\wp(z))}
\eea
with $A$ and $B$ relatively prime polynomials of degrees related by,
\bea
\deg A = 1+ \deg B = |\alpha|^2= ad-bc
\eea
The only assertion above that needs proof is the relation between the degrees. First, since $\wp(\alpha z)$ has a double pole at $z=0$ and since $A$ and $B$ will have poles at $z=0$ of respective degrees $2 \deg A$ and $2 \deg B$, we must have the first equality above between the degrees. Second, the lattice $\alpha \Lambda$ has a fundamental parallelogram which is scaled up from $\Lambda$  by a factor of $|\alpha|$, so that the area of the fundamental parallelogram is $|\alpha |^2$ times that of the fundamental parallelogram of $\Lambda$. Thus, the index is given by $[\Lambda : \alpha \Lambda] = |\alpha |^2$. While $\wp(z)$ has one double pole in the fundamental parallelogram of $\Lambda$, the total number of poles of $\wp (\alpha z)$ in the fundamental parallelogram of $\Lambda$ is  $2 |\alpha|^2$, which is also its number of zeros. It follows that $\deg A = |\alpha|^2$ giving the second equality.

\sm

Constructing the polynomials $A,B$ is achieved by matching the Laurent expansions of both sides at $z=0$ using the Laurent series of $\wp(z)$, 
\bea
\wp (z) = { 1 \over z^2} + \sum _{m=1}^\infty (2m+1) G_{2m+2} z^{2m}
\eea
where we have taken into account the fact that $G_m=0$ for $m$ odd. We shall also use the fact that $G_{2m}$ for $m \geq 4$ may be expressed as a polynomial in the generators $G_4,G_6$ of the ring of modular forms of $SL(2,\ZZ)$ of (\ref{moditer}), 
\bea
G_{2m+4} = \sum_{k=1}^{m-1} { 3 (2k+1)(2m-2k+1) \over (m-1)(2m+3)(2m+5)} \, G_{2k+2}G_{2m-2k+2}
\eea
Of course, the relation between $\wp(\alpha z)$ and $\wp(z)$, and therefore the matching of the Laurent series, will only be possible at the singular modulus $\tau$ for which we have the corresponding complex conjugation factor $\alpha$. Therefore, the matching will impose a relation between the Eisenstein series $G_4$ and $G_6$, and this will give a formula for the value of $j(\tau)$.

\subsection{Examples}

Consider the simple example $\alpha = i \sqrt{2}$ for which  $a+d=0$, the discriminant is given by $D=-8$, and $|\alpha |^2 =2$. We may take the representative $\tau = i \sqrt{2}$. In this case $A$ has degree two and $B$ degree one, so that we may express the relation (\ref{eq:compmultprel}) as,
\bea
\wp (i \sqrt{2} \, z) = c_1 \wp (z) + c_2 +{ c_3 \over \wp(z) + c_4}
\eea
Since $\wp(i \sqrt{2} \, z) = -1 /(2 z^2) + \cO(z^2)$ we must have $c_1=-\half$ and $c_2=0$.
Substituting the Laurent expansion we obtain the following expression, 
\bea
\left ( \wp (i \sqrt{2} \, z) + \half \wp (z) \right )^{-1}
& = &
- { 2 \over 9 G_4} \left ( { 1 \over z^2} + 5 {G_6 \over G_4} + 25 {G_6^2 \over G_4^2} z^2 - 5 G_4 z^2 + \cO(z^4) \right )
\no \\
& = &
{1 \over c_3}  \left ( \wp(z)  + c_4 + 25 {G_6^2 \over G_4^2} z^2 - 8 G_4 z^2 + \cO(z^4) \right )
\eea
To match the Laurent expansions we have set,
\bea
c_3 = -{ 9 \over 2} G_4
\hskip 1in 
c_4 = 5 {G_6 \over G_4}
\eea
But the relation must be exact to all orders in $z$ by our global analysis, so the terms of order $z^2$ and higher in the above parentheses must vanish. This can be achieved only at the singular modulus associated with the complex multiplication factor $\alpha$, and this gives us a relation between $G_4$ and $G_6$,
\bea
25\, G_6^2 = 8 G_4^3
\eea
It may be checked that all higher order terms  in the parentheses on the second line also vanish. We thus find that the $j$-invariant at the complex multiplication point is given by,
\bea
j (i \sqrt{2}) = { (12)^3 \, g_2^3 \over g_2^3 - 27 g_3^2} =  (12)^3 \left ( 1 - { 49 \, G_6^2 \over 20 G_4^3} \right )^{-1}
= (20)^3= 8000
\eea
Remarkably, $j(i \sqrt{2})$ is an integer, and in fact the cube of an integer !

\subsubsection{Heegner points}

A related set of points are those labelled by the so-called Heegner numbers. The Heegner numbers can be defined as numbers $d$ such that $j\left(\half (1 + i \sqrt{d}) \right) \in \ZZ$. There exist precisely eight such numbers, 
\bea
1,  \, 3, \, 7, \, 11, \, 19, \, 43, \, 67, \, 163
\eea
a fact first conjectured by Gauss and proven by Heegner, Baker, and Stark over a century later. It is a remarkable fact that the $j$-invariant evaluates to a cube on each of them, 
\bea
\label{8.tablej}
j\left(\thalf(1+ i) \right) &=& 12^3 \hspace{0.8 in}j\left(\thalf(1+ i\sqrt{3}) \right) = 0
\no\\
j\left(\thalf(1+ i\sqrt{7}) \right) &=& -15^3\hspace{0.65 in}j\left(\thalf(1+ i\sqrt{11}) \right) = -32^3
\no\\
j\left(\thalf(1+ i\sqrt{19}) \right) &=& -96^3 \hspace{0.65 in}j\left(\thalf(1+ i\sqrt{43}) \right) = -960^3
\no\\
j\left(\thalf(1+ i\sqrt{67}) \right) &=& -5280^3 \hspace{0.5 in}j\left(\thalf(1+ i\sqrt{163}) \right) = -640320^3
\eea
These are precisely the discriminants $D=-d$ for which the class number is one and the corresponding ring of integers has unique factorization.

\sm

Incidentally, the Heegner numbers are related to some well-known examples of ``almost integers." Perhaps the most famous almost integer is Ramanujan's constant, 
\bea
e^{\pi \sqrt{163}} = 262537412640768743.99999999999925\dots 
\eea
whose near-integrality can be explained by using the value of $j\left(\half(1+ i\sqrt{163}) \right) $ given above together with the $q$-expansion of $j(\tau)$ to write 
\bea
e^{\pi \sqrt{163}} = 640320^3 + 744 + O(e^{-\pi \sqrt{163}})
\eea

\subsection{$j(\tau)$ as an algebraic integer}
\label{sec:jalgebraicCM}
An \textit{algebraic integer} is a complex number that is a root of a monic polynomial with integer coefficients, namely of a monic polynomial in $\ZZ[x]$. Recall that a \textit{monic} polynomial is one whose leading monomial has coefficient 1.  A general result regarding the arithmetic properties of $j(\tau)$ at points $\tau$ of complex multiplication is the following:

{\thm{
\label{thm:jalgint} For a lattice $\ZZ + \tau \ZZ$ with complex multiplication,  $j(\tau)$ is an \textit{algebraic integer}.}}
\newline

To prove this result,  we define the following monic polynomial in $x$ for given $\tau$, 
\bea
\label{8.Pn}
P_n(x|\tau) = \prod_{\g \in SL(2, \ZZ) \backslash M_n} \left(x - j(\g \tau)\right)
\eea
where $M_n$ is the set of $2 \times 2$ matrices with determinant $n$, introduced in (\ref{eq:Mndef}). Since $j(\tau)$ is invariant under $SL(2,\ZZ)$ the product in $\gamma$ is taken over the left cosets $SL(2, \ZZ) \backslash M_n$ discussed above (\ref{Heckedef}). The function $P_n(x|\tau)$ is invariant under $\tau \to \gamma \tau$ for all $\gamma \in SL(2,\ZZ)$ because the right action of $SL(2,\ZZ)$ on the left cosets  $SL(2, \ZZ) \backslash M_n$ simply permutes the cosets. 

\sm

Because the set of matrices,
\bea
\left\{\left(\begin{smallmatrix}a & b \\ 0 & d \end{smallmatrix}\right)\, \big| \, ad = n, \, 0 \leq b \leq d-1\right\}
\eea 
provides a complete set of representatives of the quotient $SL(2, \ZZ) \backslash M_n$, as was explained in section \ref{sec:Heckeprops},  we obtain the following equivalent  product formula,
\bea
\label{8.Pn1}
P_n(x|\tau)  = \prod_{ad= n} ~ \prod_{b=0}^{ d-1} \left(x - j \left({a \tau + b \over d} \right) \right)
\eea
This expression confirms  that $P_n(x|\tau+1) = P_n(x|\tau)$ since the shift in $\tau$ amounts to a shift $b \to b+a ~ (\mod d)$ under which the product over $b$ is invariant.  The degree of $P_n$ is given by,
\bea
\deg(P_n) = \# (SL(2, \ZZ) \backslash M_n) = \sigma _1(n)
\eea
where $\sigma_1(n)=\sum_{d|n} d$ is the divisor sum.
The full structure of the $\tau$-dependence of $P_n(x|\tau)$ is given by the following Lemma.

{\lem{$P_n(x|\tau)$ is a polynomial in $x$ and $j(\tau)$ with integer coefficients, so that there exists a polynomial $Q_n(x,y) \in \ZZ[x,y]$ which satisfies $P_n(x|\tau) = Q_n(x,j(\tau))$.}}
\newline

To prove the Lemma, we begin by focussing on the product over $b$ for given $d$. Writing the $q$-expansion of $j(\tau)$ as  $j(\tau) = q^{-1} + 744 + \sum_{k \geq1} c_k q^k$ with $c_k \in \NN$, we have, 
\bea
\label{eq:secondprod}
\prod_{0 \leq b \leq d-1}\left(x - j \left( \tfrac{a \tau + b } {d} \right) \right) &=& (-1)^d \prod_{0 \leq b \leq d-1} \left(e^{-{2 \pi i b\over d}} q^{-{a \over d}} + 744 +\sum_{k \geq1} c_k e^{ {2 \pi i k b\over d}}q^{ka \over d}-x\right) 
\no\\
&=& (-1)^d q^{-a} + \sum_{\ell \geq 1 -d} a_\ell (x) q^{\ell a \over d}
\eea
By construction, the coefficients $a_\ell(x)$ are polynomials in $x$ and the root of unity $e^{2 \pi i /d}$, namely $a_\ell(x) \in \ZZ(x, e^{2 \pi i \over d})$. Actually, (\ref{eq:secondprod}) implies that the $a_\ell(x)$ are invariant under all transformations $\sigma_r: e^{2 \pi i \over d} \rightarrow e^{2 \pi i r \over d}$ for $0 \leq r \leq d-1$; that is, the so-called ``Galois automorphisms" of $\QQ(e^{2 \pi i \over d})/\QQ$, which will be further discussed in section \ref{sec:Galois}.  As a result, $a_\ell(x)$ must be independent of $e^{2 \pi i \over d}$ and therefore belongs to $\ZZ[x]$. Furthermore, the left side of (\ref{eq:secondprod}) is invariant under $\tau \mapsto \tau+1$, so that we must have $a_\ell(x) =0$ for all $d \nmid \ell$.  The function $P_n(x|\tau)$ is obtained by taking the product of  (\ref{eq:secondprod}) over $ad=n$, which we
reorganize as follows,
\bea
P_n(x|\tau) = \prod_{ad = n} \left[(-1)^d q^{-a} + \sum_{m \geq 0} a_{md}(x) q^{ma} \right]
= \sum_{r=0}^{\deg(P_n)} h_r(\tau) x^r
\eea
where the coefficient functions $h_r(\tau)$ admit a Fourier expansion $h_r(\tau) = \sum_\ell a_\ell^{(r)} q^\ell$ with $a_\ell^{(r)} \in \ZZ$. Recall that any meromorphic modular function with integer Fourier coefficients can be written as a polynomial in $j(\tau)$ with integer coefficients; that is, we can write $h_r(\tau) = \sum_\ell \tilde{a}_\ell^{(r)} j(\tau)^\ell$ with $\tilde{a}_\ell^{(r)} \in \ZZ$. This allows us to conclude that the same is true of $P_n(x|\tau)$, thereby proving that $P_n(x|\tau)$ is a polynomial in $x$ and $j(\tau)$ with integer coefficient.  This result may be summarized by the statement that there exists a polynomial $Q_n(x,y)$ with integer coefficients such that,
\bea
P_n(x|\tau) = Q_n(x, j(\tau))
\eea
Actually $Q_n$ satisfies $Q_n(y,x) = \pm Q_n(x,y)$ and is of degree $\sigma _1(n)$ in $x$ and thus also in $y$.

\sm

To complete the proof of Theorem \ref{thm:jalgint}, we now specialize to a point $\tau=\tau_{\rm CM} \in \cH$ of complex multiplication of order $n$, namely satisfying $\g_0 \tau_{\rm CM} = \tau_{\rm CM}$ for a matrix $\g_0 =\left(\begin{smallmatrix}a & b \\ c & d \end{smallmatrix}\right) \in M_n$ with determinant $n=ad-bc$. The polynomial $P_n(x|\tau_{{\rm CM}})= Q_n(x,j(\tau_{{\rm CM}}) $   must vanish at the point $x=j(\tau_{{\rm CM}})$  since, up to an $SL(2,\ZZ)$ transformation, $\gamma_0$ will coincide with one of the $\gamma$-matrices in the product in (\ref{8.Pn}). Thus, the polynomial $Q_n(x,x) \in \ZZ[x]$  with integer coefficients vanishes at $x=j(\tau_{{\rm CM}})$. 

\sm

To complete the proof that $j(\tau_{{\rm CM}})$ is an algebraic integer, it remains to show that it is a root of a non-vanishing polynomial with integer coefficients. To this end, we evaluate the leading behavior at the cusp of the polynomial $Q_n(x,x) \in \ZZ[x]$ from (\ref{8.Pn1}) with the help of  the factorization formula,
\bea
\prod_{b=0}^{d-1}  (x- e^{2 \pi i b/d}  y) = x^d-y^d
\eea
and we obtain, 
\bea
Q_n(j(\tau), j(\tau)) = \prod _{ad=n}  \Big ( q^{-d} - q^{-a} + \hbox{ lower order terms } \Big )
\eea
There are now two cases to be considered, depending on whether $n$ is a perfect square or not. We first consider the case when $n$ is not a perfect square. Every factor in the product has $a \not = d$ since otherwise $n=ad=a^2$ is a perfect square and the leading behavior at the cusp is given by the power of $q$ obtained by summing the maxima of $(d,a)$,
\bea
Q_n(j(\tau), j(\tau)) = \pm q^{- \sigma _1^+(n)} + \hbox{ lower orders }
\hskip 0.6in 
\sigma _1^+ (n) = \sum_{ad=n} \max(a,d)
\eea
Therefore, the polynomial $Q_n(x,x)$ is non-vanishing and monic (up to a sign) when $n$ is not a perfect square. 
When $n=\nu^2$ is a perfect square, then one element in $SL(2, \ZZ) \backslash M_n$ is of the form $\nu I$ so that $a=d=\nu$ and $Q_n(j(\tau), j(\tau)) =0$. The vanishing is caused by the fact that $Q_n(x,y)$ has a factor $(x-y)$. Adapting the above arguments to this case, one shows that the polynomial,
\bea
\tilde Q_n (x,y) = { Q_n(x,y) \over (x-y)}
\eea 
evaluated at $x=y=j(\tau)$ is an integer multiple of a monic polynomial of degree $\sigma_1^+(n)-n$. This completes the proof of the Theorem.

\subsubsection{The examples of $n=2,3,4,5$}

Zagier worked out the example of $n=2$ in detail. The cases $n=3,4,5$ may be obtained analogously, and we find the following factorizations of $Q_n(x,x)$,
\bea
Q_2(x,x) & = & - (x-12^3)(x-20^3)(x+15^3)^2
 \\
Q_3(x,x) & = & - x(x-20^3)(x+32^3)(x-54000)
\no \\
\tilde Q_4(x,x) & = & 2 (x+15^3)^2(x-54000)^2(x-66^3)(x^2+191025 x-121287375)^2
\no \\
Q_5(x,x) & = & (x-12^3)^2(x+32^3)^2(x-66^3)^2(x+96^3)^2(x^2-1264000x-681472000)
\no
\eea
The second order polynomials factor as follows,
{\small \bea
x^2+191025 x-121287375 & = & (x-\thalf( -191025 + 85995 \sqrt{5} ) )(x-\thalf( -191025 - 85995 \sqrt{5} ) )
\no \\
x^2-1264000x-681472000 & = & (x- 632000 - 282889 \sqrt{5}) (x- 632000+ 282889 \sqrt{5}) 
\eea}
In addition to the entries listed in (\ref{8.tablej})  we also recognize,
\begin{align}
j(i) & = 12^3 &  j(i \sqrt{3}) &= 54000 
\no \\
j(2i) & = 66^3 &  j(i\sqrt{5}) & = 632000 + 282880 \sqrt{5}
\no \\
j(i \sqrt{2}) &= 20^3  &  j \left ( \thalf (1+i \sqrt{15}) \right )  & = -\thalf \big ( 191025 + 85995 \sqrt{5} \big ) 
\end{align}

\subsection{$\tet$-functions at complex multiplication points}
\label{sec:tetCM}

We now examine Jacobi theta functions at some arguments $\tau =i \sqrt{n}$ with $n \in \NN$. 
To do so, we begin by defining the function, 
\bea
\label{eq:Kkdef}
K(k) \,\,=\,\, \int_0^{\pi \over 2} {dx \over \sqrt{1-k^2 \sin^2 x}}\,\,=\,\,{\pi \over 2} \, {}_2F_1\left(\half, \half; 1; k^2 \right)
\eea
known as a complete elliptic integral of the first kind. A valuable result involving elliptic integrals and Jacobi theta-functions is that
\bea
\label{eq:maintetCM}
\tet_3(0|\tau)= \sqrt{{2 \over \pi} K(k)}
\eea
where the parameter $\tau$ is defined in terms of the elliptic integrals as, 
\bea
\label{eq:qFdict}
\tau = i F(k)~, \hspace{0.5 in}  F(k) = {K(\sqrt{1-k^2}) \over K(k)} = \, { {}_2 F_1(\thalf, \thalf; 1; 1-k^2) \over {}_2 F_1(\thalf, \thalf; 1; k^2)}
\eea
In this presentation, the action of the generators $T$ and $S$ of $SL(2,\ZZ)$ corresponds to,
\begin{align} 
T: \hskip 0.5in \tau & \to \tau+1 & k^2 & \to -{k^2 \over 1-k^2}
\no \\
S: \hskip 0.5in \tau & \to - {1 \over \tau} & k^2 & \to 1-k^2
\end{align}
which may be readily verified using the transformation formulas for the hypergeometric function
(see Bateman \cite{Bateman2}, Vol II page 318).

\sm

As we will now describe, the above relations between $\tet$-functions and hypergeometric functions can be used to evaluate $\tet_3(0|\tau) $ at special values of $\tau$. Having obtained $\tet_3(0|\tau) $, one can then obtain the values of other theta functions by means of, 
\bea
\tet_2(0|\tau) = \sqrt{k}\, \tet_3(0|\tau) \hspace{0.5 in}\tet_4(0|\tau) = (1-k^2)^{1/4} \,\tet_3(0|\tau)
\eea

\subsubsection{The point $\tau = i$}

Let us begin by evaluating $\tet_3(0|i)$.  In this case, (\ref{eq:qFdict}) tells us to consider $F(k)=1$, i.e.
\bea
K(\sqrt{1-k^2}) = K(k)
\eea
This is solved by $k = {1 \over \sqrt{2}}$. Using this in (\ref{eq:maintetCM}) then gives
\bea
\tet_3(0|i )  = \sqrt{{2 \over \pi} K\left({1\over \sqrt{2}}\right)} 
\eea
In order to evaluate the remaining elliptic integral, we convert it to a hypergeometric function as per the second equality of (\ref{eq:Kkdef}) and make use of the following standard result, 
\bea
\label{eq:hypergeomGammas}
 {}_2F_1\left(a,b; \half(a+b+1); \half \right) = { \sqrt{\pi} \,  \Gamma(\half(a+b+1)) \over \Gamma(\half(1+a)) \Gamma(\half(1+b))}
\eea
valid for arbitrary $a,b$. Using this, we conclude that,
\bea
\label{eq:tet3pi}
\tet_3(0|i)  = \frac{\pi^{\frac{1}{4}} }{ \Gamma(\tfrac{3}{4})}
\eea

\subsubsection{The point $\tau = 2 i$}

Next consider $\tet_3(0|2i)$. We begin by searching for $k$ such that $F(k) = 2$. To do so, we make use of the elliptic integral identity,
\bea
F(k) = 2F\left( {2 \sqrt{k} \over 1+k}\right)
\eea
which follows from a similar functional identity involving hypergeometric functions, 
(see for example Bateman \cite{Bateman1}, Vol I, page 110)
\bea
\label{eq:useful1}
  {}_2F_1\left( \thalf, \thalf; 1; \frac{4 x}{ (1+x)^2}\right) =(1+x)\, {}_2F_1\left( \thalf, \thalf; 1; x^2\right) 
\eea
Using this identity, we may recast our equation for $k$ as follows, 
\bea
F\left( {2 \sqrt{k} \over 1+k} \right) =1 
\eea
The solution to this equation was identified in the previous subsection---we require, 
\bea
{2 \sqrt{k} \over 1+k} = {1 \over \sqrt{2}}\hspace{0.5 in} \Rightarrow \hspace{0.5in} k = (\sqrt{2}-1)^2
\eea
and thus we have 
\bea
\tet_3(0|2i) =  \sqrt{{}_2F_1\left( \half, \half; 1;(\sqrt{2}-1)^4\right)} 
\eea
Making use of duplication and reflection identities of hypergeometric functions, one shows, 
\bea
{}_2F_1\left( \thalf, \thalf; 1;(\sqrt{2}-1)^4\right) = \frac{1}{4 - 2 \sqrt{2}}\,\,{}_2F_1\left( \thalf, \thalf; 1;\thalf \right)
\eea
and then using (\ref{eq:hypergeomGammas}) we obtain, 
\bea
\tet_3(0|2 i) =  \frac{\pi^{\frac{1}{4}} }{ 2\Gamma(\tfrac{3}{4})} \sqrt{2 + \sqrt{2}}
\eea

\subsubsection{The point $\tau = \sqrt{2}\, i$}

Finally we consider $\tet_3(0 | \sqrt{2} i)$. We would like to solve $F(k) =\sqrt{2}$, or equivalently,
\bea
\sqrt{2}\, {}_2F_1\left( \thalf, \thalf; 1; k^2\right)  ={}_2F_1\left( \thalf, \thalf; 1; 1-k^2 \right) 
\eea
Comparing this to (\ref{eq:useful1}), we see that $k= \sqrt{2}-1$ is a solution. Thus we have, 
\bea
\theta_3(0| \sqrt{2} i) =  \sqrt{{}_2F_1\left( \thalf, \thalf; 1;(\sqrt{2}-1)^2\right)} 
\eea
To proceed, we again use duplication and reflection identities of hypergeometric functions to show that,
\bea
 {{}_2F_1\left( \thalf, \thalf; 1;(\sqrt{2}-1)^2\right)} = 
 \frac{1}{\sqrt{4 - 2 \sqrt{2}}} \,\,{}_2F_1\left( \tfrac{1}{4}, \tfrac{3}{4}; 1;\thalf\right)
\eea
from which we can evaluate, 
\bea
\tet_3(0 | \sqrt{2} i) = \frac{\Gamma(\tfrac{9}{8})}{\Gamma(\tfrac{5}{4})}\sqrt{\frac{\Gamma(\tfrac{1}{4})}{2^{1/4} \pi}} 
\eea

\subsubsection{The points $\tau=i n$ and $\tau = i/n$ with $n \in \NN$}

To close, we quote without proof the results for some other $\tet$-values at imaginary complex multiplication points $\tet_3(0 | i n)$. It is a general fact that one can write, 
\bea
\tet_3(0 | i n) =  \frac{\tet_3(0 | i)}{n^{1/4} h_n} 
\eea
with $\tet_3(0 | i)$ given in (\ref{eq:tet3pi}). The first few values of $h_n$ are found to be,
\bea
\label{eq:tethnlist}
h_1 &=&1
\no\\
h_2 &=&\sqrt{2\sqrt{2}-2}
\no\\
h_3 &=& (2 \sqrt{3}-3)^{1/4}
\no\\
h_4&=& = \frac{2^{3/4}}{2^{1/4} + 1}
\no\\
h_5&=& \sqrt{5 - 2 \sqrt{5}}
\no\\
h_6 &=& \frac{2^{3/4} 3^{1/8} ((\sqrt{2}-1)(\sqrt{3}-1))^{1/6}}{(-4+3 \sqrt{2} + 3^{5/4} + 2 \sqrt{3} - 3^{3/4} + 2^{3/2} 3^{3/4})^{1/3}}
\eea
From the above results, one can also obtain expressions for $\tet_3(0 | {i / n})$ by means of the following useful relation, 
\bea
\tet_3\left(0  | {i / \sqrt{n}} \right)= n^{1/4}\, \tet_3(0 | i \sqrt{n})
\eea

\subsection{The values of $\HE_2, \, \HE_4, \, \HE_6$ at the points $\tau = i, \rho$}

The quasi-modular transformation property of $\HE_2(\tau)$, 
\bea
\label{9.E2transfs}
\HE_2  ( \gamma \tau) = (c \tau +d)^2 \HE_2(\tau) + { 12 \over 2 \pi i} \, c (c \tau +d) 
\eea
allows one to evaluate $\HE_2(\tau)$ at the points $\tau=i$ and $\tau=\rho=e^{2 \pi i /3}$. These points are fixed points of the modular transformations $\gamma = S$, for which $(c,d)=(1,0)$,  and $\gamma = ST$, for which $(c,d)=(1,1)$,  respectively.  Substituting these expressions into the transformation law (\ref{9.E2transfs}), we readily obtain,
\bea
\HE_2 (i) = { 3 \over \pi}
\hskip 1in
\HE_2(\rho) = { 2 \sqrt{3} \over \pi}
\eea
The modular forms $\HE_4$ and $\HE_6$ vanish at the points $\rho$ and $i$ respectively, as was already shown in  (\ref{3.G4G6}) in terms of the closely related modular forms $G_4$ and $G_6$. Evaluating $\HE_4$ and $\HE_6$ at the points $i$ and $\rho$ respectively requires expressing elliptic integrals in terms of hypergeometric functions, and we shall quote here the result without proof,
\begin{align} 
\HE_4(i) & = { 48 \, \Gamma( \tfrac{5}{4})^4 \over \pi^2 \, \Gamma (\tfrac{3}{4})^4} &
\HE_4(\rho) & =0  &
\HE_6 (i) & = 0 &
 \HE_6(\rho) & = { 729 \, \Gamma( \tfrac{4}{3})^6 \over 2 \pi^3 \, \Gamma (\tfrac{5}{6})^6}
\end{align}
For a derivation of the first and fourth relations, see Diamond and Shurman \cite{DS}.

\newpage

\subsection*{$\bullet$ Bibliographical notes}
 
The algebraic nature of $j(\tau)$ at complex multiplication points is a classic result; the discussion given in section \ref{sec:jalgebraicCM} follows the one given by Zagier in section 6.1 of \cite{Zagier123}. Another useful account may be found in the beautiful book by Cox~\cite{Cox}.

\sm

The method used to evaluate $\tet$-functions at special complex multiplication points in section \ref{sec:tetCM} was pioneered by Ramanujan, and can be found in his third notebook \cite{Ramanujan3}; see Chapter 17, Entry 6. It was systematized by Chowla and Selberg in \cite{ChowlaSelberg1,ChowlaSelberg2}. The explicit values of $h_n$ listed in (\ref{eq:tethnlist}) were  obtained in \cite{yi2004theta}. 

\sm

Generally speaking, points of enhanced symmetry tend to play a special role in physics, and singular moduli are precisely the values of the moduli where such enhancements occur. As such, singular moduli and complex multiplication have appeared in various places in physics. Concrete examples can be found in Seiberg-Witten theory \cite{Seiberg:1994rs,Seiberg:1994aj,Argyres:1995jj,Minahan:1996fg,Minahan:1996cj}, where points on the Coulomb branch corresponding to singular moduli generically have extra massless degrees of freedom, as well as in string theory where the additional discrete symmetries at singular values of the axion-dilaton can be gauged to give so-called ``S-folds" \cite{Garcia-Etxebarria:2015wns,Aharony:2016kai,Garcia-Etxebarria:2016erx,Evtikhiev:2020yix,Kaidi:2022lyo}. In section \ref{eq:CMandNLSM} we will introduce another physical application: namely we will show that a non-linear sigma model CFT with $T^2$ target space is rational if and only if the torus admits complex multiplication \cite{Gukov:2002nw}. 

\sm

We close by noting that another context in which singular moduli appear is in the study of orbifolds, in which one quotients by the emergent symmetry at such points. In the context of perturbative string theory on a target space torus, orbifolds were first introduced in \cite{Dixon:1985jw,Dixon:1986jc}, and were further developed  in \cite{Dijkgraaf:1989hb} among many others publications. A large number of references (with a particular emphasis on string phenomenology) can be found in the bibliography of \cite{Cvetic:2022fnv}. The study of singular moduli at genus-2 was carried out in the 1960s by Gottschling in \cite{gottschling1961fixpunkte,gottschling1961fixpunktuntergruppen,gottschling1967uniformisierbarkeit}, though the application to genus-2 orbifolds is fairly recent \cite{Nilles:2021glx}.
Some more general comments at arbitrary genus, from a two-dimensional conformal field theory perspective, can be found in \cite{Robbins:2019ayj}.

\newpage

\section{String amplitudes}
\setcounter{equation}{0}
\label{sec:SA}

We now finally turn to the topic of string theory. The starting point for string theory is the assumption that elementary  particles  are one-dimensional objects, namely strings, in contrast to quantum field theory where elementary particles are assumed to be point-like.  The size of strings is set by the Planck length $\ell_P$, 
\bea
\ell_P = \sqrt{G_N \hbar / c^3} \approx 10^{-33} \, {\rm cm}
\eea
where $G_N$ is Newton's gravitational constant, $\hbar$ is Planck's constant, and $c$ is the speed of light. 
The Planck length is approximately  $10^{-19}$ times the size of a proton. All distance scales accessible to us to date  are enormously larger than $\ell_P$. At long distance scales or, equivalently, at low energy scales strings effectively behave as point-like particles. 

\subsection{Overview}

\begin{figure}[h]
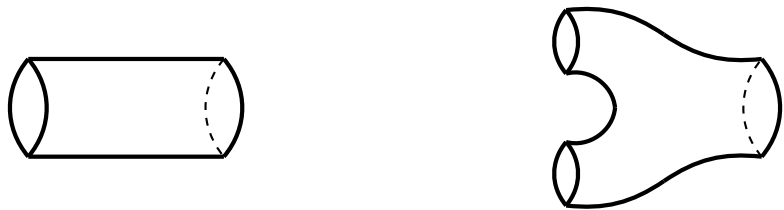

\begin{center}
\tikzpicture[scale=0.65]
\begin{scope}[xshift=10cm]
\draw [black, ultra thick] (-1,1) to (3,1) ;
\draw [black, ultra thick] (-1,-1) to (3,-1);
\draw [black, ultra thick] (-1,1) to [bend left=40] (-1,-1);
\draw [black, ultra thick] (-1,1) to [bend left=-40] (-1,-1);
\draw [black, ultra thick] (3,1) to [bend left=40] (3,-1);
\draw [black, thick, dashed] (3,1) to [bend left=-40] (3,-1);
\draw [black, ultra thick] (10,2) to [bend left=40] (10,0.7);
\draw [black, ultra thick] (10,2) to [bend left=-40] (10,0.7);
\draw [black, ultra thick] (10,-2) to [bend left=40] (10,-0.7);
\draw [black, ultra thick] (10,-2) to [bend left=-40] (10,-0.7);
\draw [black, ultra thick] (10,-0.7) to [bend left=-50] (11,0);
\draw [black, ultra thick] (10,0.7) to [bend left=50] (11,0);
\draw [black, thick, dashed] (14,-1) to [bend left=40] (14,1);
\draw [black, ultra thick] (14,-1) to [bend left=-40] (14,1);
\draw [black, ultra thick] (10,2) to [bend left=20] (12,1.5);
\draw [black, ultra thick] (12,1.5) to [bend left=-20] (14,1);
\draw [black, ultra thick] (10,-2) to [bend left=-20] (12,-1.5);
\draw [black, ultra thick] (12,-1.5) to [bend left=20] (14,-1);
\end{scope}
\endtikzpicture
\caption{\textit{Time evolution of a free closed string is represented in the left panel.  Interaction by joining and splitting of closed strings is represented in the right panel. }\label{13.fig:1}} 
\end{center}
\end{figure}

Being a one-dimensional object, a connected string may have two different topologies: that of a circle or \textit{closed string}, or that of an interval or \textit{open string}. The time evolution of a closed string sweeps out a two-dimensional surface with the topology of a cylinder, represented in the left panel of  Figure \ref{13.fig:1}, while an open string  sweeps out a  rectangle. The surface swept out by a string is referred to as the \textit{worldsheet}. One of the most striking features of string theory is that all interactions are governed by the joining and splitting of strings, as represented in the right panel of Figure \ref{13.fig:1} for closed strings, without the need for the point-like interactions required in quantum field theory. The interaction by joining and splitting is unique once the free propagation of the string is known, since locally on the worldsheet there is no way to distinguish free propagation from interaction.

String theory may be promoted into a consistent quantum theory provided certain conditions are met on the space-time in which strings propagate and interact. Superstrings can propagate in flat space-time provided the dimension of the space-time is ten.  The spectrum of superstring theory then automatically contains a graviton, and  string theory is automatically a theory of quantum gravity, albeit in space-time dimension greater than four. Supersymmetry, which swaps boson and fermion states,  is a key ingredient in making superstring theory mathematically and physically consistent. The center of attention in these lectures  is modular invariance which is responsible for making superstring theory UV finite. 

\sm

 The ultimate goal of string theory is the unification of gravity with the Standard Model of Particle Physics into a single theory that is consistent with quantum mechanics and general relativity. Successful unification will require principles and mechanisms for selecting our four-dimensional universe from all the solutions of string theory,  for breaking space-time supersymmetry, and for maintaining a small cosmological constant in the process. Satisfactory answers to these questions remain largely outstanding today.

\sm

In this section we shall review  some of the string amplitudes derived from \textit{superstring perturbation theory} and use those to obtain corrections to supergravity and super-Yang-Mills theory in the form of local \textit{effective interactions}. Superstring perturbation theory uses a series expansion in powers of the string coupling $g_s$ to evaluate \textit{quantum mechanical probability amplitudes}, or simply \textit{amplitudes}. The absolute value squared of an amplitude gives the quantum mechanical probability for a given  process to occur. A process is specified by the data of the incoming and the outgoing string states, and the amplitude for the process is schematically given by the Feynman functional integral prescription of summing over all possible configurations of the string worldsheet given the initial and final state data. 

\sm

\begin{figure}[hbt]
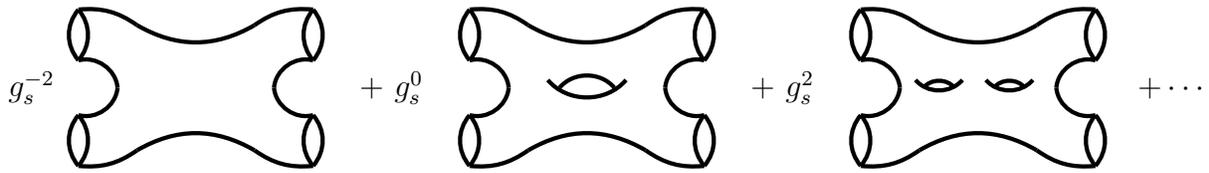

\begin{center}
\tikzpicture[scale=0.52]
\begin{scope}[xshift=10cm]
\draw [black, ultra thick] (0,2) to [bend left=40] (0,0.7);
\draw [black, ultra thick] (0,2) to [bend left=-40] (0,0.7);
\draw [black, ultra thick] (0,-2) to [bend left=40] (0,-0.7);
\draw [black, ultra thick] (0,-2) to [bend left=-40] (0,-0.7);
\draw [black, ultra thick] (0,-0.7) to [bend left=-50] (1,0);
\draw [black, ultra thick] (0,0.7) to [bend left=50] (1,0);
\draw [black, ultra thick] (0,2) to [bend left=20] (1.5,1.6);
\draw [black, ultra thick] (1.5,1.6) to [bend left=-30] (4.5,1.6);
\draw [black, ultra thick] (0,-2) to [bend left=-20] (1.5,-1.6);
\draw [black, ultra thick] (1.5,-1.6) to [bend left=30] (4.5,-1.6);
\draw [black, ultra thick] (4.5,1.6) to [bend left=20] (6,2);
\draw [black, ultra thick] (4.5,-1.6) to [bend left=-20] (6,-2);
\draw [black, ultra thick] (6,2) to [bend left=40] (6,0.7);
\draw [black, ultra thick] (6,2) to [bend left=-40] (6,0.7);
\draw [black, ultra thick] (6,-2) to [bend left=40] (6,-0.7);
\draw [black, ultra thick] (6,-2) to [bend left=-40] (6,-0.7);
\draw [black, ultra thick] (6,-0.7) to [bend left=50] (5,0);
\draw [black, ultra thick] (6,0.7) to [bend left=-50] (5,0);
\draw [black, ultra thick] (10,2) to [bend left=40] (10,0.7);
\draw [black, ultra thick] (10,2) to [bend left=-40] (10,0.7);
\draw [black, ultra thick] (10,-2) to [bend left=40] (10,-0.7);
\draw [black, ultra thick] (10,-2) to [bend left=-40] (10,-0.7);
\draw [black, ultra thick] (10,-0.7) to [bend left=-50] (11,0);
\draw [black, ultra thick] (10,0.7) to [bend left=50] (11,0);
\draw [black, ultra thick] (10,2) to [bend left=20] (11.5,1.6);
\draw [black, ultra thick] (11.5,1.6) to [bend left=-30] (14.5,1.6);
\draw [black, ultra thick] (10,-2) to [bend left=-20] (11.5,-1.6);
\draw [black, ultra thick] (11.5,-1.6) to [bend left=30] (14.5,-1.6);
\draw [black, ultra thick] (14.5,1.6) to [bend left=20] (16,2);
\draw [black, ultra thick] (14.5,-1.6) to [bend left=-20] (16,-2);
\draw [black, ultra thick] (16,2) to [bend left=40] (16,0.7);
\draw [black, ultra thick] (16,2) to [bend left=-40] (16,0.7);
\draw [black, ultra thick] (16,-2) to [bend left=40] (16,-0.7);
\draw [black, ultra thick] (16,-2) to [bend left=-40] (16,-0.7);
\draw [black, ultra thick] (16,-0.7) to [bend left=50] (15,0);
\draw [black, ultra thick] (16,0.7) to [bend left=-50] (15,0);
\draw [black, ultra thick] (12, 0.2) to [bend left=-50] (14, 0.2);
\draw [black, ultra thick] (12.3,0) to [bend left=50] (13.7,0);
\draw [black, ultra thick] (20,2) to [bend left=40] (20,0.7);
\draw [black, ultra thick] (20,2) to [bend left=-40] (20,0.7);
\draw [black, ultra thick] (20,-2) to [bend left=40] (20,-0.7);
\draw [black, ultra thick] (20,-2) to [bend left=-40] (20,-0.7);
\draw [black, ultra thick] (20,-0.7) to [bend left=-50] (21,0);
\draw [black, ultra thick] (20,0.7) to [bend left=50] (21,0);
\draw [black, ultra thick] (20,2) to [bend left=20] (21.5,1.6);
\draw [black, ultra thick] (21.5,1.6) to [bend left=-30] (24.5,1.6);
\draw [black, ultra thick] (20,-2) to [bend left=-20] (21.5,-1.6);
\draw [black, ultra thick] (21.5,-1.6) to [bend left=30] (24.5,-1.6);
\draw [black, ultra thick] (24.5,1.6) to [bend left=20] (26,2);
\draw [black, ultra thick] (24.5,-1.6) to [bend left=-20] (26,-2);
\draw [black, ultra thick] (26,2) to [bend left=40] (26,0.7);
\draw [black, ultra thick] (26,2) to [bend left=-40] (26,0.7);
\draw [black, ultra thick] (26,-2) to [bend left=40] (26,-0.7);
\draw [black, ultra thick] (26,-2) to [bend left=-40] (26,-0.7);
\draw [black, ultra thick] (26,-0.7) to [bend left=50] (25,0);
\draw [black, ultra thick] (26,0.7) to [bend left=-50] (25,0);
\draw [black, ultra thick] (21.4, 0.2) to [bend left=-50] (22.6, 0.2);
\draw [black, ultra thick] (21.6,0) to [bend left=50] (22.4,0);
\draw [black, ultra thick] (23.2, 0.2) to [bend left=-50] (24.4, 0.2);
\draw [black, ultra thick] (23.4,0) to [bend left=50] (24.2,0);
\draw (-1.2, 0) node{$ g_s^{-2}$};
\draw (8, 0) node{$+ \, \, g_s^0$};
\draw (18, 0) node{$+ \, \, g_s^2$};
\draw (28, 0) node{$+ \cdots$};
\end{scope}
\endtikzpicture
\end{center}
\caption{\textit{Schematic representation of the perturbative expansion in the string coupling $g_s$ of the string amplitude $\cA$ for a process of four incoming and/or outgoing string states.} \label{13.fig:2}}
\end{figure}

The only topological information that remains, once the initial and final string data have been fixed by specifying the process, is the genus $h\geq 0$ of the surface, namely the number of handles on the worldsheet. The string coupling $g_s$ provides the weight given to each topology, so that the perturbative string amplitude $\cA$ for a given process is provided by the following topological expansion in powers of $g_s$,
\bea
\cA = \sum _{h=0}^\infty g_s ^{2h-2} \cA^{(h)} + \hbox{ non-perturbative effects}
\eea
where $\cA^{(h)}$ is the  \textit{genus $h$} (or $h$-loop) contribution to the string amplitude $\cA$. The expansion  is schematically represented in Figure \ref{13.fig:2}. The prescription of the initial and final data for the incoming and/or outgoing string states will be discussed in subsection \ref{sec:25}.

The construction of string amplitudes that we shall sketch below contains only bosonic states and the corresponding string theory  is referred to as the \textit{closed bosonic string}. To include fermions introduces many technical complications and has led to the elaboration of several possible different formulations, which are mostly beyond the scope of these lectures. Instead we shall focus on the bosonic string, and some mild extensions thereof, to illustrate the questions and results that relate to modular invariance of string perturbation theory in this section, to T-duality for toroidal compactifications in the next section \ref{sec:Toroidal}, and to  S-duality in Type IIB superstring theory in section \ref{sec:IIB}.

\subsection{String amplitudes as integrals over moduli space}

Space-time is assumed to be a manifold $M$  (or an orbifold) with a space-time metric~$G$. Physically, the signature of $G$ is Minkowskian. However,  as is familiar from quantum field theory, it will often be convenient to construct string amplitudes in terms of a metric of Euclidean signature, so that $M$ is a Riemannian manifold,  and then analytically continue to Minkowski signature. String amplitudes are formulated in terms of two-dimensional orientable surfaces~$\Sigma$ and  continuous maps $X : \Sigma \to M$. The Riemannian metric $G$ on $M$  induces a Riemannian metric $X^* (G)$ on $\Sigma$ under the pull-back $X^*$ of the map $X$. Since~$\Sigma$ is orientable and carries a Riemannian metric it is automatically a Riemann surface. 

\sm

The Polyakov formulation of string theory also invokes an intrinsic Riemannian metric~$g$ on $\Sigma$, which is independent  of  the induced metric $X^*(G)$. The map $X$ and the intrinsic metric $g$ on $\Sigma$ are governed by a worldsheet action. String amplitudes are required to be  independent of the coordinates used to parametrize $\Sigma$ and $M$, which leads us to require invariance under the diffeomorphism groups ${\rm Diff}(\Sigma)$ and ${\rm Diff}(M)$. A natural candidate satisfying these conditions  is given by  the following action,\footnote{Throughout,  $\xi^m$ with $m=1,2$ are real  local coordinates on $\Sigma$ and $\p _m = \p / \p \xi^m$ are the derivatives with respect to $\xi^m$. The Riemannian metric on $\Sigma$ takes the form $g = g_{mn} d\xi^m \otimes d\xi^n$ while the volume form   is given by $d \mu _g =  \half  \sqrt{\det g} \, \eps_{mn} d \xi ^m \wedge d \xi^n$ where $\eps_{mn}=-\eps_{nm}$ specifies the orientation of $\Sigma$  with $\eps_{12}=1$. Throughout, the Einstein summation convention on a pair of matching upper and lower indices will be implied, both for the worldsheet indices $m,n=1,2$ as well as for the space-time indices $\mu, \nu = 1, \cdots, \dim(M)$.}
\bea
\label{2a2}
I_G [X,g] = { 1 \over 4 \pi \alpha'} \int _\Sigma d\mu _g \, g^{mn} \p_m X^\mu \p_n X^\nu G_{\mu \nu}(X)
\eea
Feynman's functional integral prescription for probability amplitudes  gives the amplitudes as integrals over the spaces  ${\rm Map} (\Sigma)$ of maps $X$ and ${\rm Met} (\Sigma)$ of metrics $g$. The corresponding functional measures of integration $Dg$ and $DX$ may be constructed from suitable $L^2$ norms on the tangent spaces to ${\rm Map} (\Sigma)$ and ${\rm Met} (\Sigma)$. Denoting their tangent vectors   by $\delta X$ and $\delta g$ respectively, the norms are dictated by ${\rm Diff}(\Sigma)$ and ${\rm Diff}(M)$ invariance, 
\bea
\label{12.measure}
\| \delta X \|^2 = \int _\Sigma d \mu_g \, G_{\mu \nu}(x) \, \delta X^\mu \delta X^\nu
\hskip 0.8in
\| \delta g \|^2 = \int _\Sigma d \mu_g \, g^{mn} \, g^{pq} \, \delta g_{mp} \, \delta g_{nq}
\eea
The partial integral over ${\rm Map} (\Sigma)$ is given as follows,
\bea
e^{-W_G[g]} = \int _{{\rm Map}(\Sigma)} DX \, e^{- I_G [X,g]}
\eea
The two-dimensional quantum field theory defined by $I_G[X,g]$,  for fixed metrics $g$ and $G$, is referred to as a  \textit{non-linear sigma model}.   The quantum field theory on $\Sigma$ defined by the functional integral of $DX$ depends on the space-time metric $G$, which is a function of $X$ and therefore may depend on an infinite number of couplings.  This may be seen explicitly  by Taylor expanding $G$ in powers of the field $X$ in the neighborhood of a flat metric $G^{(0)} _{\mu \nu}$,
\bea
G_{\mu \nu}(X) = G^{(0)} _{\mu \nu} + G^{(1)} _{\mu \nu; \rho} X^\rho + \half G^{(2)} _{\mu \nu; \rho \sigma} X^\rho X^\sigma + \cO(X^3)
\eea
The sets of coefficients $G^{(0)} _{\mu \nu}$, $G^{(1)} _{\mu \nu; \rho}$, $G^{(2)} _{\mu \nu; \rho \sigma}, \cdots$  may be viewed as independent coupling parameters of the quantum field theory corresponding to the non-linear sigma model.  This theory is renormalizable in the sense of Friedan namely that, upon insisting on invariance under ${\rm Diff}(M)$, the number of counter-terms is finite at each order in $\alpha '$. 

\sm

The genus $h$ contribution $\cA^{(h)}$ to the string amplitude for metric $G$ at genus $h$ is \textit{formally} given by the functional  integral over ${\rm Met} (\Sigma)$,  
\bea
\label{2a1}
\cA^{(h)} \sim \int _{{\rm Met}(\Sigma)} Dg \, e^{-W_G[g]}
\eea
where the symbol $\sim$  indicates that this relation is formal. Taken literally, the integral over~$g$ would be divergent as the measure and the integrand  are both invariant under ${\rm Diff}(\Sigma)$ so that ${\rm Met} (\Sigma)$ contains an infinite number of images of each metric $g$ under ${\rm Diff}(\Sigma)$. Factoring out ${\rm Diff}(\Sigma)$ amounts to a reduction of the integration to the quotient  ${\rm Met} (\Sigma)/{\rm Diff}(\Sigma)$ of orbits, and may be carried out using the Faddeev-Popov gauge fixing procedure familiar from Yang-Mills gauge theory. The Faddeev-Popov  determinant produced in the process may be represented by a functional integral over ghost fields $b$ and $c$ with action $I_{{\rm gh}}[b,c,g]$,
\bea
e^{-W_{{\rm gh}} [g]} = \int D(bc) e^{- I_{{\rm gh}} [b,c,g]}
\hskip 0.8in
I_{{\rm gh}}[b,c,g] = { 1 \over 2 \pi } \int _\Sigma d\mu_g \left ( b \pbz c + {\rm c.c.  } \right )
\eea 
The resulting amplitude is then given by,
\bea
\cA^{(h)} = \int _{{\rm Met}(\Sigma)/{\rm Diff}(\Sigma)} Dg \, e^{-W_G[g]- W_{{\rm gh}}[g]}
\eea 
The reduction removes two of the three functional degrees of freedom of a two-dimensional metric, leaving a single real field. Such theories are referred to as \textit{non-critical strings}. 

\sm

\textit{Critical string theories} are obtained by requiring that the combination of the measure $Dg$ and the integrand are invariant also under Weyl transformations.  The group of Weyl transformations ${\rm Weyl}(\Sigma)$ leaves the metric $G$ and the map $X$ invariant, and multiplies the metric  $g$ by a positive scalar function,   $g_{mn} (\xi) \to e^{2 \sigma (\xi)} g_{mn} (\xi)$, where $\sigma : \Sigma \to \RR$. The quotient of the space ${\rm Met} (\Sigma)$ of metrics by the semi-direct product of  ${\rm Diff}(\Sigma) $ and ${\rm Weyl}(\Sigma)$ may be identified with the moduli space $\cM_h$  of conformal structures on $\Sigma$, and is isomorphic to the space of complex structures on a Riemann surface $\Sigma$ of genus $h$,
\bea
\cM_h = {\rm Met} (\Sigma) / \big [ {\rm Diff}(\Sigma) \ltimes {\rm Weyl}(\Sigma) \big ]
\eea
Thus, our final expression for the genus $h$ contribution $\cA^{(h)}$  to the amplitude for the critical string is given by a finite-dimensional integral over moduli space, 
\bea
\label{12.intA}
\cA^{(h)} = \int _{\cM_h} D\hat g \, e^{-W_G[g]- W_{{\rm gh}}[g]}
\eea 
where $D\hat g$ is the integration measure on moduli induced on the quotient. 
The resulting amplitude $\cA^{(h)}$ is a function of the metric $G$ on the space-time $M$.

\subsection{Conformal invariance and decoupling negative norm states}
\label{sec:22}

 The classical actions  $I_G[X,g]$ and $I_{{\rm gh}}[b,c,g]$  are Weyl-invariant, but the measures $DX$,  $Dg$, and $D(bc)$ are not Weyl invariant, thereby producing  Weyl anomalies. In particular, the functional $W_G[g]$, defined in (\ref{2a2}), is not invariant under ${\rm Weyl}(\Sigma)$. For general metric $G$, the Weyl transformation law of $W_G$ is complicated and not known explicitly. However, when the transformation law of $W_G$ takes the following special form,
\bea
\label{weyl}
\delta W_G[g]= { \mc \over 24 \pi } \int _\Sigma d\mu_g \, R_g \, \delta \sigma
\eea
where $R_g$ is the scalar curvature of the metric $g$ and $\mc$ is a constant, then $I_G$ produces a \textit{conformal field theory with central charge $\mc$}. For example, when $G$ is the flat metric on $\RR^D$, then the action $I_G$ is  quadratic  in $X$, and defines a conformal field theory with central charge $\mc=D$, and $I_{{\rm gh}}$ is a conformal field theory with central charge $\mc=-26$. In the critical dimension $D=26$ the combined field theory of $X,b,c$ is therefore Weyl-invariant.   

\sm

When the metric $G$ is not flat, the Weyl transformation of $W_G$ may be evaluated in an expansion in powers of $\alpha'$, which is equivalent to an expansion in powers of the Riemann tensor of the metric $G$ and its derivatives.  To leading order in $\alpha'$, one finds,
\bea 
\delta W_G [g]=  { 1 \over 4 \pi} \int _\Sigma d\mu_g \, \left ( { D \over 6} R_g + 
\half g^{mn} \p_m \bar X^\mu \p_n \bar X^\nu R_{\mu \nu}(\bar X) \right ) \delta \sigma + \cO(\alpha ')
\eea
The first term is familiar from (\ref{weyl}) for $\mc=D$. The second term involves the Ricci tensor $R$ of the metric $G$ and an arbitrary background field $\bar X$ used to evaluate the quantum corrections in the background field method.  The term in $R_{\mu \nu}$  cancels when the metric $G$ is Ricci flat. Thus we find the remarkable fact that the requirement of Weyl symmetry on the worldsheet metric $g$ imposes Einstein's equations on the space-time metric $G$. 

\sm

Conformal symmetry plays a fundamental role  in decoupling the negative norm and null  states which arise in the Lorentz-covariant formulation of string theory in flat Minkowski space-time $M=\RR^{26}$. The maps $X : \Sigma \to M$ satisfy the Laplace equation $\p_z \pbz X^\mu=0$, so that the field $\p_z X^\mu$ is holomorphic, while the field $\pbz X^\mu$ is anti-holomorphic. Both fields may be expanded in a Laurent series, 
\bea
i \p_z X^\mu =  \sum _{m \in \ZZ} z^{-m-1} X^\mu _m  
\hskip 1in
-i \pbz X^\mu =  \sum _{m \in \ZZ} \bar z^{-m-1} \tilde X^\mu _m  
\eea
 which upon canonical quantization satisfy, 
\bea
(X^\mu _m)^\dagger = X_{-m} ^\mu 
& \hskip 0.7in &
{}[X^\mu _m, X^\nu _n] = m \delta _{m+n,0} 
\no \\
(\tilde X^\mu _m)^\dagger = \tilde X_{-m} ^\mu 
& \hskip 0.7in &
{}[\tilde X^\mu _m, \tilde X^\nu _n] = m \delta _{m+n,0} 
\eea
while $X^\mu_m$ and $\tilde X^\nu _n$ commute with one another.\footnote{We note that the fields $\p_z X^\mu$ and $\pbz X^\mu$ are treated as independent of one another, and not as complex conjugates of one another, a property that is inherited from the worldsheet with Minkowskian signature where the fields correspond to the data on the two light-cone directions.}
The operators $X_0^\mu$ and $\tilde X_0^\mu$ are self-adjoint. They commute with one another and with all other modes and correspond to left and right moving momentum operators. (They will play a key role in toroidal compactification and will be discussed in greater detail in section \ref{sec:Toroidal}.) Thus the free fields $\p_z X^\mu$ and $\pbz X^\mu$ decompose into $X_0^\mu, \tilde X_0^\mu$ in addition to an infinite number of decoupled harmonic oscillators $X_m^\mu, \tilde X_m^\mu$ for $m \not=0$, each of which satisfies a Heisenberg algebra. 

\sm

The Fock space of the closed bosonic string in flat Minkowski space-time $\RR^{26}$  is constructed as follows. The operators $X_0^\mu$ and $\tilde X_0^\mu$ may be diagonalized simultaneously; their eigenvalues are real and are denoted $k^\mu$ and $\tilde k^\mu$; their eigenvectors $|0;k\>$ and $|0;\tilde k\>$ are ground states of the string excitation spectrum of momenta $k^\mu$ and $\tilde k^\mu$ respectively. In flat Minkowski space-time, the left and right momenta must be equal $\tilde k^\mu = k^\mu$. (This condition will be relaxed for toroidal compactifications in order to accommodate winding modes, as will be discussed in section \ref{sec:Toroidal}.)  The Fock space is the sum over string momenta $k \in \RR^{26}$ of the tensor product of  left and right chiral Fock spaces, 
\bea
\mF_{{\rm closed}} = \bigoplus _{k \in \RR^{26}} \mF_k \otimes \tilde \mF_k
\eea
The \textit{chiral Fock space} $\mF_k$ is the infinite tensor product of the Fock spaces of the harmonic oscillators $X_n^\mu$ for $n \not=0$.  Concretely, we define the ground state $|0; k\>$ to satisfy,
\bea
X^\mu_0 |0;k\>= k^\mu |0;k\> 
\hskip 1in
X^\mu _n |0;k\>=0 \hskip 0.3in n >0
\eea 
The ground state is then a scalar under the Lorentz group and $\mF_k$ is obtained as follows, 
\bea
\mF_k ~ = \bigoplus _{{\ell \geq 0 \atop  n_1, n_2, \cdots, n_\ell >0}} 
\Big \{ x_{-n_1}^{\mu_1} \cdots x_{-n_\ell} ^{\mu_\ell} |0;k\> \Big \} 
\eea
with the analogous construction for $\ti \mF_k$. We define a norm on these spaces by normalizing the string ground state for given momentum by $ \| |0;k \> \|^2 =1$. The lowest excited state in $\mF_k$ with polarization vector $\ep^\mu(k)$ is given by $\ep _\mu (k) X^\mu _{-1} |0;k\>$ and has the following norm,
\bea
\| \ep _\mu (k) X^\mu _{-1} |0;k\> \|^2 = \ep_\mu (k) \ep ^\mu(k) 
\eea
The norm is respectively positive, null, or negative when  $\ep^\mu(k)$ is space-like,  light-like, or time-like.  \textit{Negative norm states}  are inconsistent with the principles of quantum mechanics, but they are unavoidable byproducts of the Lorentz-covariant quantization of the string, just as they are unavoidable in the covariant quantization of gauge theories.

\sm

In flat 26-dimensional space-time negative norm states and null states  may be decoupled with the help of conformal symmetry, whose Virasoro generators take the form, 
\bea
L_m = \sum _{n \in \ZZ} \half X_{m-n} \cdot X_n 
\hskip 1in 
L_0 = \half X_0^2 + \sum _{n \in \NN} X_{-n} \cdot X_n
\eea
The conformal transformations  by the Virasoro algebras act as follows on the oscillators,
\bea
{} [L_m, X_n^\mu]  =  - n X^\mu _{m+n}
\eea
and similarly for the tilde operators. One defines the subspace of \textit{physical states} $ \mF^{{\rm phys}} _k\subset \mF_k$ by imposing  the following conditions on the states $| \psi\> \in \mF_k$ ,
\bea
(L_0 -1) |\psi\> = L_m |\psi \> =0 \hskip 1inm >0
\eea 
All negative norm states are eliminated by these physical state conditions, and all null states decouple, just as is the case when imposing Gauss's law in gauge theory. The ground state is physical provided $k^2= 2$ which makes it a tachyon. The tachyon renders the bosonic string vacuum unstable but it is eliminated in the superstring, and we shall omit its discussion in the sequel. Returning to the example of the first excited state, $|\psi \> = \ep _\mu (k) X^\mu _{-1} |0;k\>$, we see that the physical state conditions $L_m |\psi \> =0$ for $m \geq 2$ are automatically satisfied, while those for $m=1, 0$ impose the constraints $k \cdot \ep(k)=0$ and the massless condition $k^2=0$ on the momenta. The resulting physical states constitute the spectrum of the open bosonic string and include a massless vector particle, such as the photon.

\sm

The above construction of the Fock space also shows that the closed bosonic string spectrum contains the massless states $\ep_\mu (k) \tilde \ep _\nu (k) X_{-1}^\mu \tilde X_{-1} ^\nu |0; k\>$ where $k^2=0 $ and $k \cdot \ep = k \cdot \tilde \ep=0$. The traceless symmetric part of $\ep_\mu (k) \tilde \ep _\nu (k)$ corresponds to the graviton state as expected from the presence of the metric $G$ in the worldsheet action. However, there is also a trace-part which corresponds to the scalar dilaton state, and an anti-symmetric part which corresponds to a rank-two anti-symmetric state. These additional states and particles are not required by general relativity, but their presence in the spectrum will, in fact, be mandated by supergravity, as we shall see in section \ref{sec:IIB}. 

\sm

In view of the presence of the extra states of the dilaton and rank two anti-symmetric tensor field, we may generalize the non-linear sigma model action (\ref{2a2}) to include a field $\Phi$ for the dilaton as well as a rank two anti-symmetric field $B_{\mu \nu}$,\footnote{The factor of $i$ multiplying the $B_{\mu \nu}$ term is present for worldsheet metric $\mg$ of Euclidean signature, but is absent when Minkowski signature is used.}
\bea
\label{2k2}
I_{G,\Phi, B}  [X,g]  & = & { 1 \over 4 \pi \alpha'} \int _\Sigma d\mu _g  \bigg  \{ \Big ( g^{mn}   G_{\mu \nu}(X)
\! - i \ep ^{mn}  B _{\mu \nu}(X) \Big ) \p_m X^\mu \p_n X^\nu 
\no \\ && \hskip 1in
 +  2 \alpha ' \, R_g \, \Phi (X) \bigg \}
\qquad
\eea
where $R_g$ is the scalar curvature of the metric $g$, already encountered in (\ref{weyl}). In the presence of the fields $G, \Phi, B$ conformal invariance requires, up to lowest non-trivial order in the $\alpha'$-expansion, a generalization of Einstein's equations that includes the dynamics of the  fields $\Phi$ and $B_{\mu \nu}$. Those field equations may be derived from the following action,
\bea
S[G,\Phi, B]= { 1 \over 2 \kappa^2} \int d \mu_G \, e^{-2\Phi} \left ( R_G + 4 D_\mu \Phi D^\mu \Phi -{ 1 \over 12} H_{\mu \nu \rho} H^{\mu \nu \rho} \right )
\eea
where $\kappa$ is a constant, $D_\mu$ is the covariant derivative, and $H_{\mu \nu \rho} = D_\mu B_{\nu \rho} + D_\nu B_{\rho \mu}  + D_\rho B_{\mu \nu}$.

\sm

In summary, conformal symmetry is the local gauge symmetry that eliminates negative norm states and decouples null states. In this section we have illustrated this mechanism when space-time is flat, but the result is expected to hold in  space-times which are asymptotically flat or anti-de Sitter.

\subsection{String amplitudes in terms of vertex operators}
\label{sec:25}

\begin{figure}[htb]
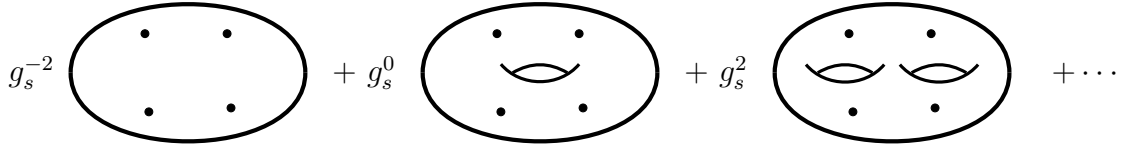

\begin{center}
\tikzpicture[scale=0.52]
\begin{scope}[xshift=0cm]
\draw [black, ultra thick] (-3,0) to [bend left=90] (3,0);
\draw [black, ultra thick] (-3,0) to [bend right=90] (3,0);
\draw (-1.1,1) [fill=black] circle(0.1cm) ;
\draw (-1,-1) [fill=black] circle(0.1cm) ;
\draw (1,1) [fill=black] circle(0.1cm) ;
\draw (1.1,-0.9) [fill=black] circle(0.1cm) ;
\draw (-4, 0) node{$ g_s^{-2}$};
\end{scope}
\begin{scope}[xshift=9cm]
\draw [black, ultra thick] (-3,0) to [bend left=90] (3,0);
\draw [black, ultra thick] (-3,0) to [bend right=90] (3,0);
\draw [black, very thick] (-1,0.22) to [bend right=50] (1,0.22);
\draw [black, very thick] (-0.7,0) to [bend left=30] (0.7,0);
\draw (-1.1,1) [fill=black] circle(0.1cm) ;
\draw (-1,-1) [fill=black] circle(0.1cm) ;
\draw (1,1) [fill=black] circle(0.1cm) ;
\draw (1.1,-0.9) [fill=black] circle(0.1cm) ;
\draw (-4.5, 0) node{$+ \, \, g_s^0$};
\end{scope}
\begin{scope}[xshift=18cm]
\draw [black, ultra thick] (-3,0) to [bend left=90] (3,0);
\draw [black, ultra thick] (-3,0) to [bend right=90] (3,0);
\draw [black, very thick] (-2.2,0.21) to [bend right=50] (-0.2,0.21);
\draw [black, very thick] (-1.9,0) to [bend left=30] (-0.5,0);
\draw [black, very thick] (0.2,0.21) to [bend right=50] (2.2,0.21);
\draw [black, very thick] (0.5,0) to [bend left=30] (1.9,0);
\draw (-1.1,1) [fill=black] circle(0.1cm) ;
\draw (-1,-1) [fill=black] circle(0.1cm) ;
\draw (1,1) [fill=black] circle(0.1cm) ;
\draw (1.1,-0.9) [fill=black] circle(0.1cm) ;
\draw (-4.5, 0) node{$+ \, \, g_s^2$};
\draw (5, 0) node{$+ \cdots$};
\end{scope}
\endtikzpicture
\end{center}
\caption{\textit{Schematic representation of the perturbative expansion in the string coupling $g_s$ formulated in terms of vertex operator insertions on the worldsheet.} \label{13.fig:3}}
\end{figure}

Conformal symmetry also allows us to simplify the formulation of string scattering processes. Instead of representing each incoming and outgoing string in a scattering process by a boundary of the string worldsheet, as was presented in Figure \ref{13.fig:2}, we can encode all the physical information of the incoming or outgoing strings in terms of vertex operators. The data for these vertex operators is specified at vertex insertion points on the worldsheet, as illustrated in Figure \ref{13.fig:3}. One may view these insertion points as the limit in which the radii of the boundary discs of Figure \ref{13.fig:2} are shrunk to zero by a conformal transformation. 

Consider the scattering of $N$ on-shell gravitons in flat space-time $\RR^{26}$, with momenta $k_i$ and polarization tensors $\ep_{i; \mu \nu}(k)$ for $i=1,\cdots, N$. A graviton of momentum $k_i$ introduce a small ripple on the metric of space-time, given in terms of a plane monochromatic wave. Adding up the contributions of the $N$ gravitons the space-time metric is perturbed as follows to lowest order in the amplitude of the wave given by the polarizations tensors $\ep_i$, 
\bea
\label{2e1}
G_{\mu \nu} (X) = \eta _{\mu \nu} + \sum _{i=1}^N \ep_{i; \mu \nu} (k) \, e^{i k_{i} \cdot X} + \cO(\ep_i^2)
\eea 
Conformal invariance requires $G$ to satisfy Einstein's equations, as was discussed in the second paragraph of subsection \ref{sec:22}. Einstein's equations apply in linearized form, because we consider the metric only to first order in each $\ep_i$,  and impose the conditions $k_i^2=0$ and $k_i^\mu \ep _{i; \mu \nu}(k)=0$. Substituting the expression (\ref{2e1}) for the metric fluctuations into the action $I_G[X,g]$ of (\ref{2a2}) and the functional integral of (\ref{12.intA}), and retaining the contribution which is linear in each $\ep_i$ gives the genus $h$ contribution to the amplitude for  the scattering of $N$ gravitons in terms of vertex operators, 
\bea
\label{2e2}
\cA^{(h)} = \int _{\cM_h } D\hat g \int _{{\rm Map}(\Sigma)} DX \, 
\cV_1[X,g] \cdots \cV_N [X,g] \, e^{- I_G [X,g] - W_{{\rm gh}}[g] }
\eea
The vertex operators are read off the expansion in $\ep_i$ to linear order and are given by,
\bea
\cV_i [X,g] = \ep_{i ; \mu \nu} (k) \int _\Sigma d \mu_g \, g^{mn} \p_m X^\mu \p_n X^\nu \, e^{i k_i \cdot X}
\eea
The physical state conditions for the graviton, namely its massless condition  $k_i^2=0$, and the transversality of its polarization tensors $k^\mu _i \ep_{i; \mu \nu}(k)=0$, and $\ep_{i;\mu \nu} \ep_i^{\mu \nu}=0$,  guarantee that the unintegrated vertex operator $g^{mn} \p_m X^\mu \p_n X^\nu \, e^{i k_i \cdot X}$ has conformal dimension $(1,1)$ and may be integrated over $\Sigma$ consistently with conformal symmetry. One may proceed analogously for the fields $\Phi$ and $B_{\mu \nu}$.

\subsection{Superstring amplitudes}

Scattering amplitudes for the superstrings may similarly be formulated in terms of functional integrals   over the string degrees of freedom, such as $X^\mu, b, c$,  involving vertex operators. Unlike for the case of the bosonic string, several different formulations have been developed over time for the case of the superstring, each with certain advantages and disadvantages. Of central concern to us here is not so much how the amplitudes have been obtained but rather what their structure is. Thus, we shall proceed below by simply quoting the final expressions for the amplitudes we need, and refer to the literature for their detailed derivations, which are often quite involved. We shall concentrate on Type II theories here, but a similar analysis may be carried out for Type I and Heterotic superstrings. 

\sm

For definiteness, we shall consider the scattering process that involves four gravitons, whose momenta we denote by $k_i$ and whose polarization tensors are chosen in factorized form $\ep_i^\mu (k) \tilde \ep _i^\nu (k)$ with $k_i^2= k_i \cdot \ep_i = k_i \cdot \tilde \ep_i=0$ and $i=1,2,3,4$. It will be useful to introduce the dimensionless kinematic Lorentz-invariants defined by,
\bea
s_{ij} = - { \alpha ' \over 4} (k_i+k_j)^2
\eea
These variables satisfy a number of kinematic relations as a result of momentum conservation and the massless conditions $k_i^2=0$, which may be solved as follows,
\begin{align}
s&=s_{12}=s_{34}
\no \\
t&=s_{14}=s_{23}
\no \\
u&=s_{13}=s_{24} & s+t+u & =0
\end{align}
Closed superstring perturbation theory produces the following on-shell four-graviton amplitude in Type II superstring theory,
\bea
\label{8a1}
\cA (\ep_i; k_i) = \kappa _{10}^2 \, \cR^4 \sum _{h=0}^\infty g_s^{2h-2} \cA^{(h)}  (s_{ij})
\eea
where $\cR$ stands for the on-shell linearized Riemann tensor, $\cR^4$ for a particular scalar invariant constructed out of $\cR$, and $\cA^{(h)}$ is a Lorentz scalar function that depends only on $s,t,u$. The fact that a single kinematic combination $\cR^4$ is shared by the amplitudes at all loop order is a consequence of the space-time supersymmetry of the Type~II superstring, and does not hold for other superstring theories such as Type~I or Heterotic strings. The particular structure of the Lorentz contractions needed to form the scalar $\cR^4$ is also a direct consequence of supersymmetry. In the factorized basis for the polarization tensors for the gravitons, the combination $\cR^4$ is itself factorized,
\bea
\label{8a2}
2^6 \, \cR^4 = \cK \, \tilde \cK
\eea
where $\cK$ depends only on the polarization vectors $\ep_i^\mu$ and $\tilde \cK$ depends only on the polarization vectors $\tilde \ep_i^\mu$.  In terms of the linearized field strengths,
\bea
f_i^{\mu \nu} & = & k_i^\mu \ep _i ^\nu - k_i ^\nu \ep _i ^\mu
\no \\
\tilde f_i^{\mu \nu} & = & k_i^\mu \tilde \ep _i ^\nu - k_i ^\nu \tilde \ep _i ^\mu
\eea 
the factor $\cK$ is given  as follows,
\bea
\label{8a3}
\cK & = & (f_1 f_2) (f_3 f_4) + (f_1 f_4) (f_2 f_3) + (f_1 f_3) (f_2 f_4)
\no \\ &&
- 4 (f_1 f_2 f_3 f_4) - 4 (f_1 f_2 f_4 f_3) - 4 (f_1 f_3 f_2 f_4)
\eea
where the parentheses are defined by $(f_i f_j) = f_i ^{\mu \nu} f_j ^{\nu \mu}$ and $(f_i f_j f_k f_\ell) = f_i ^{\mu \nu} f_j ^{\nu \rho} f_k ^{\rho \sigma} f_\ell ^{\sigma \mu}$. The expression for  $\tilde \cK$ is given by the above expression with $f \to \tilde f$.

\sm

Explicit formulas for $\cA^{(h)}$ have been established from first principles for tree-level $h=0$, one-loop $h=1$, and two loops $h=2$, and are given as follows,
\bea
\label{12.amps}
\cA^{(0)} (s_{ij}) & = &  { 1 \over stu} { \Gamma (1-s) \Gamma (1-t) \Gamma (1-u) \over 
\Gamma (1+s) \Gamma (1+t) \Gamma (1+u)}
\no \\
\cA^{(1)}  (s_{ij}) & = & { \pi \over 16} \int _{\cM_1} { |d \tau|^2 \over (\Im \tau)^2} \, \cB^{(1)}  (s,t,u |\tau) 
\no \\
\cA^{(2)}  (s_{ij}) & = & {\pi \over 64} \int _{\cM_2} { |d \Omega|^2 \over (\det \Im \Omega)^3} \, \cB^{(2)}  (s,t,u |\Omega) 
\eea
The integrands $\cB^{(1)} $ and $\cB^{(2)} $ are dimensionless Lorentz scalar functions of $s,t,u$ given by,
\bea
\label{12.bamps}
\cB^{(1)}  (s,t,u |\tau)  & = & 
\int _{\Sigma ^4} \prod _{i=1}^4 { d^2 z_i \over \Im \tau} \, \exp \left \{ \sum_{i<j} s_{ij} G(z_i, z_j|\tau) \right \}
\no \\
\cB^{(2)}  (s,t,u |\Omega)  & = & 
\int _{\Sigma ^4} { |\cY|^2 \over (\det \Im \Omega)^2}  \, \exp \left \{ \sum_{i<j} s_{ij} \cG(z_i, z_j|\Omega) \right \}
\eea
Here, $z_i$ are the vertex insertion points on the surface $\Sigma$, $\tau$ is the modulus used here as a local complex coordinate for the genus-one moduli space $\cM_1$, and $\Omega$ is the period matrix used here as a set of local coordinates for genus-two moduli space $\cM_2$, both of which are defined in appendix \ref{sec:RS}. Furthermore, in the genus-two amplitude, $\cY$ is a holomorphic $(1,0)$ form in each one of the vertex points $z_i$ given by,
\bea
\cY  =  (t-u) \Delta(1,2) \Delta (3,4) + (s-t) \Delta(1,3) \Delta(4,2) + (u-s) \Delta(1,4)\Delta(2,3)
\eea
where $\Delta$ is a holomorphic $(1,0)$ form in its entries defined by, 
\bea
\Delta (i,j)  =  \om_1(z_i) \om_2(z_j) - \om_2(z_i) \om_1(z_j)
\eea
and $\om_I(z)$ are the canonically normalized holomorphic Abelian differentials on a Riemann surface of genus 2 (see appendix \ref{sec:RS}).  In $\cB^{(1)}(s,t,u|\tau)$, the Green function $G(x_i,z_j|\tau)$ is the scalar Green function on the torus of modulus $\tau$  defined in (\ref{10.green2}) and given explicitly either in terms of  Kronecker-Eisenstein series in (\ref{10.green1}) or Jacobi $\tet$-functions in (\ref{10.green-theta}).  In $\cB^{(2)}  (s,t,u |\Omega) $, the Green function $\cG(z_i,z_j|\Omega)$ is the Arakelov Green function on the genus-two Riemann surface $\Sigma$ with period matrix $\Omega$, which we shall now construct.  

\sm

Analogous formulas hold for amplitudes with more than four external states, as well as for Type I and Heterotic strings.

\subsubsection{The Arakelov Green function for genus 2 Riemann surfaces}

To construct the Arakelov Green function on  a genus-two Riemann surface $\Sigma$, we begin by defining the canonical K\"ahler form $\kappa$ on $\Sigma$ which is given by the pull-back under the Abel map of the canonical K\"ahler form on the Jacobian $J(\Sigma)$ (see appendix \ref{sec:Theta}),
\bea
\kappa = { i \over 4} \sum_{I,J=1}^2 \big (Y^{-1}) ^{IJ} \om_I \wedge \bar \om_J \hskip 1in \int _\Sigma \kappa =1
\eea
where $Y=\Im \Omega$.  In terms of local complex coordinates $z, \bar z$ on $\Sigma$, we have $\kappa = {i \over 2} \kappa_{z \bar z} dz \wedge d \bar z$, and the Arakelov Green function is defined by, 
\bea
\p_{\bar z} \p_z \cG(z,w|\Omega) = - \pi \delta(z,w) + \pi \kappa_{z \bar z} (z) 
\hskip 0.5in 
\int _\Sigma \kappa(z) \cG(z,w|\Omega) =0
\eea
It may be constructed explicitly in terms of the prime form and Abelian integrals, 
\bea
\cG(z,w|\Omega) & = & G(z,w|\Omega) - \gamma (z|\Omega) - \gamma (w|\Omega) + \gamma ^1 (\Omega)
\no \\
G(z,w|\Omega) & = & - \ln |E(z,w|\Omega)|^2 + 2 \pi \Im \left ( \int _w^z \om_I \right ) (Y^{-1} )^{IJ} \Im \left ( \int _w ^z \om_J \right )
\eea
where, 
\bea
\gamma (z|\Omega) = \int _\Sigma \kappa(w) G(z,w|\Omega) 
\hskip 1in 
\gamma ^1(\Omega) = \int _\Sigma \kappa (z) \gamma (z|\Omega)
\eea

\subsubsection{Physical singularity structure of the four-graviton amplitudes}

The singularity structure of the genus-one and genus-two amplitudes is as follows.
For fixed moduli, the singularities of $\cB^{(h)} $ as a function of $s,t,u$ are governed by the operator product expansion of the graviton vertex operators, namely the conformal field theory of free scalar fields on $\Sigma$. The Green functions $G(z,w|\tau)$ and $\cG(z,w|\Omega)$ are smooth throughout $\Sigma$, except for a logarithmic singularity at $z=w$ where they both behave as $ - \ln |z-w|^2$ plus regular terms. As a result, the integrations which define $\cB^{(h)} $ are absolutely convergent for $\Re(s_{ij}) <1$.  The analytic continuation of $\cB^{(h)} $ may be carried out to the complex plane in each variable and produces simple poles at positive integers $s, t, u \in \NN$. 

\sm

The further integration over moduli, required to obtain the physical amplitude $\cA^{(h)} $ for $h=1,2$,  is absolutely convergent only when $\Re (s_{ij}) =0$. Analytic continuation in $s_{ij}$ to the complex plane may be carried out using a decomposition of moduli space into subregions in each of which the domain of convergence is enlarged. This procedure has been carried out explicitly only for genus one. The resulting analytic continuation  has branch cuts in $s_{ij}$ starting at any non-negative integer, which signals two-particle intermediate states. The analytic continuation process will be illustrated in the low energy expansion, near the massless branch cut starting at $s_{ij}=0$ below.

\subsection{Ultraviolet finiteness from modular invariance}

An alternative presentation of the genus-one and genus-two amplitudes  may be given by introducing internal loop momenta. We shall illustrate this here for the case of genus one. One begins by choosing a canonical homology basis of cycles $\mA, \mB$ on the torus $\Sigma$, and chooses the loop momentum $p$ to flow through the cycle $\mA$. The amplitude is then given in terms of a Hermitian pairing between left and right chiral amplitudes,
\bea
\cA^{(1)}  (s_{ij})= \int _{\RR^{10}} d p \int _{\cM_1} \int _{\Sigma ^4} 
\cF(z_i, k_i, p|\tau) \, \overline{\cF(z_i, - k_i, p|\tau)}
\eea
where we have suppressed the dependence on the polarization tensors.
The chiral amplitude $\cF(z_i, k_i, p |\tau)$ is locally holomorphic in $\tau$ and in $z_i$ for $i=1,\cdots, 4$. Its explicit form is given by the following top holomorphic differential form on $\cM_1 \times \Sigma ^4$, 
\bea
\cF(z_i, k_i, p |\tau) 
=
e^{ i \pi \tau p^2 + 2 \pi i \sum _i k_i z_i } \prod _{i<j} \tet _1 (z_i-z_j|\tau)^{-s_{ij}} d \tau \prod _{i=1}^4 dz_i
\eea
The price to pay for the local holomorphicity is that $\cF$ has non-trivial monodromy when a point $z_\ell$ is taken around one of the homology cycles of the surface, and we have,
\bea
\cF(z_i + \delta _{i, \ell} \, \mA, k_i, p |\tau) & = & e^{2 \pi i k_\ell \cdot p} \, \cF(z_i, k_i, p |\tau) 
\no \\
\cF(z_i + \delta _{i, \ell} \, \mB, k_i, p |\tau) & = &  \cF(z_i, k_i, p+k_\ell |\tau) 
\eea
When the point $z_\ell$ is taken around the point $z_i$ with $\ell \not= i$, the chiral amplitude $\cF$ is multiplied by a phase factor $e^{- 2\pi i s_{ij}}$. Modular transformations of $\cF$ involve a change of homology basis $\mA, \mB$ and thus a change of momentum routing through the torus. The Hermitian pairing of $\cF$ and $\bar \cF$ is familiar from two-dimensional conformal field theory where the  loop momentum $p$ labels the conformal blocks of ten copies of a $c=1$ theory.

\sm

Thanks to modular invariance, all string amplitudes are ultraviolet (UV) finite. The result was obtained by Shapiro for the bosonic string at genus one, but holds for all modular invariant superstrings to all genera. 

\sm

We can illustrate the mechanism by which the genus-one string amplitude is UV finite by  inspection of the chiral amplitude derived in the previous paragraph. For any string theory, the chiral amplitude, with loop momentum $p$ chosen to flow through the cycle $\mA$,  has the following universal pre-factor which is Gaussian in the loop momentum,
\bea
\cF(z_i, \ep_i, k_i, p|\tau) = e^{ i \pi \tau p^2} \times \hbox{ exponentials and powers in } p
\eea
The Gaussian factor is universal in the sense that it is independent of the momenta $k_i$ and polarization tensors $\ep_i$ of the external states and actually independent of the specific string theory under consideration. Modular invariance allows one to choose the fundamental domain $\cM_1$ for $\cH / SL(2,\ZZ)$ that is adapted to the particular choice for the flow of loop momentum. For our choice of loop momentum traversing the $\mA$-cycle, the choice of fundamental domain is the standard one, $\cM_1 = \{ \tau \in \cH, ~ |\tau|\geq 1, ~ |\Re(\tau)| \leq \half \}$. The high energy behavior of the loop momentum is governed by the magnitude of $\Im (\tau)$, which is uniformly bounded away from zero by our choice of fundamental domain. The uniform Gaussian suppression at large loop momenta implies UV finiteness as all other factors are either only exponential or polynomial in $p$.
Higher genus amplitudes have a generalization of the above factor which involves the period matrix of the corresponding higher genus surface, but the structure is otherwise analogous to the one-loop case.

\subsection{Effective interactions from the four-graviton amplitude}
\label{sec:12.7}

The tree-level, one-loop, and two-loop string amplitudes listed in (\ref{12.amps})  for four gravitons have important applications to the evaluation of string theory corrections to supergravity in the low energy limit. Type~IIB supergravity and its relation to Type~IIB superstring theory will be reviewed in section \ref{sec:IIB}. In preparation for those discussions and the comparison of perturbative results with the implications from supersymmetry and $SL(2,\ZZ)$-duality of Type~IIB string theory, we shall here evaluate the behavior of the partial amplitudes in (\ref{12.bamps}) and the full amplitudes in (\ref{12.amps}) in the low energy limit.

\sm

To illustrate the emergence of low energy effective interactions from superstring amplitudes, we recall the expansion in powers of the string coupling $g_s$ of the four-graviton amplitude, given in (\ref{8a1}) and repeated here for convenience, 
\bea
\label{8a1a}
\cA (\ep_i; k_i) = \kappa_{10}^2 \cR^4 \sum _{h=0}^\infty g_s^{2h-2} \cA^{(h)}   (s,t,u)
\eea
where $\cR$ stands for the on-shell linearized Weyl tensor, whose expression was given in (\ref{8a2}) and (\ref{8a3}), while the expressions for $\cA^{(0)} , \cA^{(1)} , \cA^{(2)} $ were given in (\ref{12.amps}).

\subsubsection{Low energy expansion at tree-level}

We concentrate first on the tree-level contribution $\cA^{(0)} $ and use the following series expansion for the ratio of $\Gamma$-functions, 
\bea
\label{12.exp}
{ \Gamma (1-s) \Gamma (1-t) \Gamma (1-u)  \over \Gamma (1+s) \Gamma (1+t) \Gamma (1+u)} 
= \exp \left \{ \sum _{n=1} ^\infty { 2 \zeta (2n+1) \over 2n+1} (s^{2n+1} + t^{2n+1} + u^{2n+1}) \right \} 
\eea
The argument of the exponential must clearly be odd in $s,t,u$, since the left side is mapped to its inverse under simultaneous sign reversal of $s,t,u$. The linear term cancels by $s+t+u=0$. The amplitude $\cA^{(0)} $ is a symmetric function of $s,t,u$ and may therefore be expanded in powers of the two remaining symmetric polynomials in $s,t,u$,
\bea
\sigma _2 = s^2 + t^2 + u^2
\hskip 1in 
\sigma _3 = s^3 + t^3 + u^3 = 3 stu 
\eea
The first few terms in the expansion in powers of $\sigma _2$ and $\sigma _3$ are given by, 
\bea
\label{8a5}
\cA^{(0)}  = { 1 \over stu} + 2 \zeta (3) + \sigma _2 \, \zeta (5) 
+ {2 \over 3}  \sigma _3 \, \zeta (3)^2 +\half \sigma _2^2 \, \zeta (7) + \cO(s_{ij})^5
\eea
The first term on the right side  in the expansion is non-analytic at low energy and corresponds to the exchange of massless gravitons and other massless bosons in Type IIB supergravity.  The remaining terms on the right side  are all local and successively correspond to effective interactions which are schematically of the form $D^{2k} \cR^4$ for $k=0, 2, 3$ and $4$. The effective interaction $D^2 \cR^4$ is missing in this list because of the relation $s+t+u=0$.

\subsubsection{Transcendental weight}
\label{sec:12.7.2}

An important notion that has arisen over the past few years, starting in the study of quantum field theory amplitudes, is that of \textit{transcendental weight}. A natural starting point for its definition is with zeta-values and multiple zeta-values, 
\bea
\label{12.MZV}
\zeta(k) = \sum _{n=1}^\infty { 1 \over n^k} 
\hskip 1in 
\zeta(k_1, \cdots, k_p) = \sum _{n_1\geq  n_2 \geq \cdots \geq n_p \geq 1} { 1 \over n_1^{k_1} \cdots n_p^{k_p}}
\eea
for integer $k, k_1 \geq 2$ and $k_2, \cdots, k_p \geq 1$. When $k$ is even, $\zeta(k) \in \pi^k \QQ$. Transcendental weight provides a grading upon multiplication and one assigns weight 0 to any algebraic number, weight 1 to $\pi$, and thus weight $k$ to $\zeta (k)$ for $k$ even. By extension, one assigns weight $k$ to $\zeta (k)$ for all $k$ and weight $k=k_1+\cdots +k_p$ to the multiple zeta value $\zeta(k_1, \cdots, k_p)$. 

\sm

With these assignments of transcendental weight, we observe that the argument of the exponential on the right of (\ref{12.exp}) has transcendental weight 0 provided we assign weight $-1$ to $s,t,$ and $u$. Each term in the expansion of the tree-level amplitude in (\ref{8a5}) thus has weight 3. The amplitude is said to exhibit \textit{uniform transcendentality}. 

\sm

Remarkably, string theory has non-trivial transcendentality properties already at tree-level. This is in contrast with the case of quantum field theory where tree-level amplitudes have trivial transcendentality and it is only through the expansion of the $\Gamma$-functions that arise in dimensional regularization at loop level that non-trivial transcendentality arises. Uniform transcendentality, observed here for tree-level Type~II string amplitudes, is shared in quantum field theory by the maximally supersymmetric $\cN=4$ Yang-Mills theory in four dimensions. A thorough understanding of how this property arises, and for precisely which correlators it holds, remains an open problem.

\subsection{Genus one in terms of modular graph functions}

The genus-one amplitude $\cA^{(1)}$ is obtained in (\ref{12.amps}) by integrating the partial amplitude $\cB^{(1)}(s,t,u|\tau)$, which  is a modular function in $\tau$, over the genus-one moduli space $\cM_1$.  The partial amplitude in turn is given by the  integration over four copies of the torus $\Sigma$ with modulus $\tau$ in (\ref{12.bamps}).
For fixed $\tau \in \cH$, the scalar Green function $G(z|\tau)$ on the torus  is smooth throughout the torus except for the logarithmic singularity at $z=0$ given by $G(z|\tau) \approx - \ln |z|^2$. As a result, for fixed $\tau$, the integrals which define $\cB^{(1)}(s,t,u|\tau)$ as a function of $s, t, u$  are absolutely convergent for $\Re(s), \Re (t), \Re (u) <1$, and may be expanded in a Taylor series in powers of $s,t,u$ with radius of convergence $|s|, |t|, |u| <1$. Since the function $\cB^{(1)}(s,t,u|\tau)$ is symmetric in $s,t,u$, we may organize this expansion in powers of the symmetric polynomials $\sigma _2 $ and $\sigma _3$ in $s,t,u$ as we did for the tree-level amplitude, 
\bea
\label{8c4}
\cB^{(1)} (s,t,u|\tau) = \sum _{p, q =0} ^\infty \cB _{(p,q)}^{(1)} (\tau) \, {\sigma _2 ^p \, \sigma _3^q \over p ! \, q !}
\eea
Since $\cB^{(1)} (s,t,u|\tau)$ is a modular function in $\tau$ each coefficient $\cB_{(p,q)}^{(1)}(\tau)$ is a modular function of $\tau$. The coefficients may be computed by expanding the exponential in the integrand of (\ref{12.bamps}) in powers of its argument to a given order $w$, which is the overall degree in the variables $s,t,u$, given by $w=2p+3q$, so that we have, 
\bea
\label{8c5}
\sum _{{p, q \geq 0 \atop 2p+3q=w}}  \cB _{(p,q)}^{(1)} (\tau) \, {\sigma _2 ^p \, \sigma _3^q \over p ! \, q !} = { 1 \over w!} \prod _{i=1}^ 4 \int _\Sigma { d^2 z_i \over \Im \tau} \,  \left (
\sum _{i < j} s_{ij} G(z_i-z_j |\tau) \right )^w
\eea
The coefficients $\cB_{(p,q)}^{(1)}(\tau)$ for different values of $p,q$ satisfying $w=2p+3q$ may then be sorted out by further expanding the $w$-th power on  the right side. The expansion may be presented  graphically, and the modular function corresponding to a given graph is precisely one of the  \textit{modular graph functions} discussed in section \ref{sec:MGF}.  Each graph has four vertices, namely the points $z_i$ for $i=1,2,3,4$, and $w$ edges. An  edge may connect any pair of distinct vertices, but is not allowed to begin and end on the same vertex. 

\sm

\begin{figure}[htb]
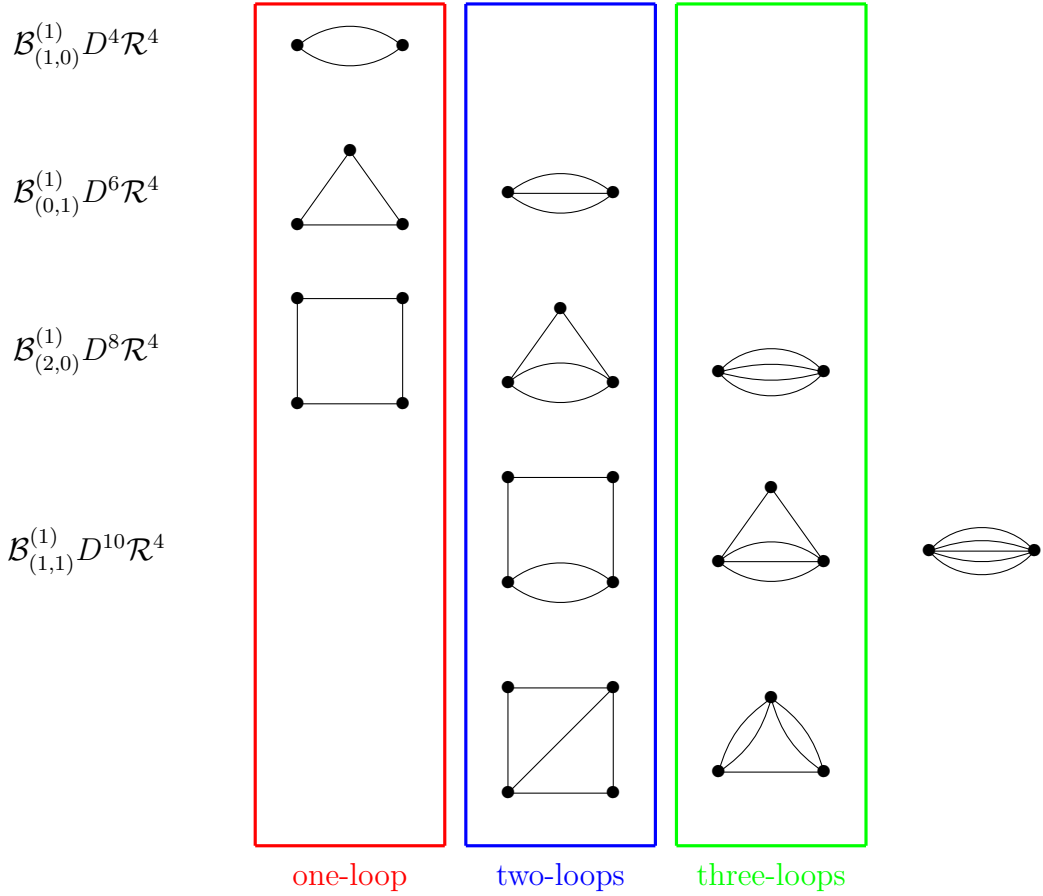

\begin{center}
\tikzpicture[scale=1.4]
\begin{scope}[xshift=10cm]
\draw (0,5.1) node{$\cB_{(1,0)}^{(1)} D^4\cR^4$} ;
\draw (2,5.1) node{$\bullet$} ;
\draw (3,5.1) node{$\bullet$} ;
\draw (2,5.1) to [bend left=40] (3,5.1) ;
\draw (2,5.1) to [bend right=40] (3,5.1) ;
\draw (0,3.7) node{$\cB_{(0,1)}^{(1)} D^6 \cR^4 $} ;
\draw (2,3.4) node{$\bullet$} ;
\draw (2.5,4.1) node{$\bullet$} ;
\draw (3,3.4) node{$\bullet$} ;
\draw (2,3.4) to (2.5, 4.1) ;
\draw (3,3.4) to  (2.5, 4.1) ;
\draw (2,3.4) to  (3,3.4) ;
\draw (4,3.7) node{$\bullet$} ;
\draw (5,3.7) node{$\bullet$} ;
\draw (4,3.7) to [bend left=40] (5,3.7) ;
\draw (4,3.7) to  (5,3.7) ;
\draw (4,3.7) to [bend right=40] (5,3.7) ;
\draw (0,2.2) node{$\cB_{(2,0)}^{(1)} D^8\cR^4$} ;
\draw (2,1.7) node{$\bullet$} ;
\draw (2,2.7) node{$\bullet$} ;
\draw (3,1.7) node{$\bullet$} ;
\draw (3,2.7) node{$\bullet$} ;
\draw (2,1.7) to (3,1.7) ;
\draw (2,1.7) to (2,2.7) ;
\draw (3,1.7) to (3,2.7) ;
\draw (2,2.7) to (3,2.7) ;
\draw (4,1.9) node{$\bullet$} ;
\draw (4.5,2.6) node{$\bullet$} ;
\draw (5,1.9) node{$\bullet$} ;
\draw (4,1.9) to (4.5,2.6) ;
\draw (5,1.9) to  (4.5,2.6) ;
\draw (4,1.9) to [bend left=40] (5,1.9) ;
\draw (4,1.9) to [bend left=-40] (5,1.9) ;
\draw (6,2) node{$\bullet$} ;
\draw (7,2) node{$\bullet$} ;
\draw (6,2) to [bend left=50] (7,2) ;
\draw (6,2) to [bend left=15] (7,2) ;
\draw (6,2) to [bend right=15] (7,2) ;
\draw (6,2) to [bend right=50] (7,2) ;
\draw (0,0.3) node{$\cB_{(1,1)}^{(1)} D^{10}\cR^4$} ;
\draw (4,0) node{$\bullet$} ;
\draw (4,1) node{$\bullet$} ;
\draw (5,0) node{$\bullet$} ;
\draw (5,1) node{$\bullet$} ;
\draw (4,0) to [bend left=40]  (5,0) ;
\draw (4,0) to [bend left=-40]  (5,0) ;
\draw (4,0) to (4,1) ;
\draw (5,0) to (5,1) ;
\draw (4,1) to (5,1) ;
\draw (6,0.2) node{$\bullet$} ;
\draw (6.5,0.9) node{$\bullet$} ;
\draw (7,0.2) node{$\bullet$} ;
\draw (6,0.2) to (6.5,0.9) ;
\draw (7,0.2) to  (6.5,0.9) ;
\draw (6,0.2) to [bend left=40] (7,0.2) ;
\draw (6,0.2) to  (7,0.2) ;
\draw (6,0.2) to [bend left=-40] (7,0.2) ;
\draw (8,0.3) node{$\bullet$} ;
\draw (9,0.3) node{$\bullet$} ;
\draw (8,0.3) to [bend left=50] (9,0.3) ;
\draw (8,0.3) to [bend left=20] (9,0.3) ;
\draw (8,0.3) to (9,0.3) ;
\draw (8,0.3) to [bend right=20] (9,0.3) ;
\draw (8,0.3) to [bend right=50] (9,0.3) ;
\draw (4,-2) node{$\bullet$} ;
\draw (4,-1) node{$\bullet$} ;
\draw (5,-2) node{$\bullet$} ;
\draw (5,-1) node{$\bullet$} ;
\draw (4,-2) to (5,-2) ;
\draw (4,-2) to (5,-1) ;
\draw (4,-2) to (4,-1) ;
\draw (5,-2) to (5,-1) ;
\draw (4,-1) to (5,-1) ;
\draw (6,-1.8) node{$\bullet$} ;
\draw (6.5,-1.1) node{$\bullet$} ;
\draw (7,-1.8) node{$\bullet$} ;
\draw (6,-1.8) to [bend left=20] (6.5,-1.1) ;
\draw (6,-1.8) to [bend left=-20] (6.5,-1.1) ;
\draw (7,-1.8) to [bend left=20] (6.5,-1.1) ;
\draw (7,-1.8) to [bend left=-20] (6.5,-1.1) ;
\draw (6,-1.8) to  (7,-1.8) ;
\draw [red, very thick] (1.6,-2.5) to (3.4,-2.5);
\draw [red, very thick] (1.6, -2.5) to (1.6, 5.5);
\draw [red, very thick] (3.4,-2.5) to (3.4,5.5);
\draw [red, very thick] (1.6, 5.5) to (3.4, 5.5);
\draw [blue, very thick] (3.6,-2.5) to (5.4,-2.5);
\draw [blue, very thick] (3.6, -2.5) to (3.6, 5.5);
\draw [blue, very thick] (5.4,-2.5) to (5.4,5.5);
\draw [blue, very thick] (3.6, 5.5) to (5.4, 5.5);
\draw [green, very thick] (5.6,-2.5) to (7.4,-2.5);
\draw [green, very thick] (5.6, -2.5) to (5.6, 5.5);
\draw [green, very thick] (7.4,-2.5) to (7.4,5.5);
\draw [green, very thick] (5.6, 5.5) to (7.4, 5.5);
\draw (2.5, -2.8) [red] node{one-loop};
\draw (4.5, -2.8) [blue] node{two-loops};
\draw (6.5, -2.8) [green] node{three-loops};
\end{scope}
\endtikzpicture
\end{center}
\caption{\textit{Modular graph functions contributing up to the genus-one four-graviton amplitude, up to  weight $w=5$ included. }\label{fig:8}}
\end{figure}

The expansion is simplified by the following observations. Any modular graph function corresponding to 
\begin{itemize}
\itemsep=0in
\item a graph containing a vertex on which only a single edge begins or ends vanishes;
\item a graph which becomes disconnected upon removing one edge vanishes;
\item a graph which becomes disconnected upon removing a single vertex factorizes into the modular graph functions of the resulting connected components.
\end{itemize}
The graphs corresponding to the modular graph functions which contribute to the modular functions $\cB_{(p,q)}(\tau)$ up to weight $w=5$ are presented in Figure \ref{fig:8}. Instead of organizing the graphs by weight $w$, they may alternatively be organized by loop order, which is indicated in Figure \ref{fig:8} by the colored rectangular boxes.

\sm

The weight $w$ of the graphs contributing to $\cB_{(p,q)} (\tau)$ for $w=2p+3q$ may now be seen to coincide with the  definition of \textit{transcendental weight} given in subsection \ref{sec:12.7.2}, provided we continue to assign weight $-1$ to $s,t,u$ and weight 0 to $\tau$.

\sm

By expanding (\ref{8c5}) and using identities such as the one in (\ref{eq:MGFidentities}), one finds, 
\bea
\label{12.calB}
\cB_{(0,0)}^{(1)} &=& 1 
\no \\
\cB_{(1,0)}^{(1)} &=& E_2 
\no\\
\cB_{(0,1)}^{(1)} &=& \tfrac{5}{3} E_3 + \tfrac{1}{3} \zeta(3) 
 \no\\
\cB_{(2,0)}^{(1)} &= &  2 C_{2,1,1} + E_2^2 - E_4 
\no \\
\cB_{(1,1)} ^{(1)} & = &  \tfrac{7}{3} C_{3,1,1} +\tfrac{5}{3} E_2 E_3 +\tfrac{1}{3} \zeta(3)E_3 -\tfrac{34}{15}E_5 +\tfrac{1}{5} \zeta(5)
\eea
up to weight $w=5$ included. The graphs that contribute are shown in Figure \ref{fig:8}.

\subsection{Genus two in terms of modular graph functions}

The genus-two amplitude  $\cA^{(2)}$ is obtained in (\ref{12.amps}) by integrating $\cB^{(2)}(s,t,u|\Omega)$ over the genus-two moduli space $\cM_2$.  The partial amplitude $\cB^{(2)}(s,t,u|\Omega)$ in turn is given by the  integration over four copies on the genus-two surface $\Sigma$ with moduli $\Omega$ in (\ref{12.bamps}). For fixed $\Omega \in \cH_2$, the scalar Arakelov Green function $\cG(z,w|\Omega)$ is smooth throughout $\Sigma$ except for the logarithmic singularity at $z=w$ given by $\cG(z,w|\Omega) \approx - \ln |z-w|^2$. As a result, for fixed~$\Omega$, the integrals that define $\cB^{(2)}(s,t,u|\Omega)$ as a function of $s, t, u$  are absolutely convergent for $\Re(s), \Re (t), \Re (u) <1$ and may be expanded in a Taylor series in powers of $s,t,u$ with radius of convergence $|s|, |t|, |u| <1$. This situation is completely analogous to the genus-one case. The presence of the additional  measure factor $|\cY|^2$ is inconsequential for these results to hold as $\cY$ is holomorphic on $\Sigma$. 

\sm

Since the function $\cB^{(2)}(s,t,u|\Omega)$ is symmetric in $s,t,u$, we may organize its Taylor expansion in powers of $s,t,u$ via the symmetric polynomials   $\sigma _2 $ and $\sigma _3$   as we did for the tree-level amplitude and the genus-one partial  amplitude, 
\bea
\label{8c4a}
\cB^{(2)} (s,t,u|\Omega) = \sum _{p, q =0} ^\infty \cB _{(p,q)}^{(2)} (\Omega) \, {\sigma _2 ^p \, \sigma _3^q \over p ! \, q !}
\eea
Since $\cB^{(2)} (s,t,u|\Omega)$ is a modular function of $\Omega$, namely it is invariant under the genus-two modular group $Sp(4,\ZZ)$, each coefficient $\cB_{(p,q)}^{(2)}(\Omega)$ is itself a modular function of $\Omega$. The coefficients may be computed by expanding the exponential in the integrand of (\ref{12.bamps}) in powers of its argument to a given order $w$, which is related to the overall degree in the variables $s,t,u$, given by $w+2=2p+3q$, so that we have, 
\bea
\label{8c5a}
\sum _{{p, q \geq 0 \atop 2p+3q=w+2}}  \cB _{(p,q)}^{(2)} (\Omega) \, {\sigma _2 ^p \, \sigma _3^q \over p ! \, q !} = { 1 \over w!}  \int _{\Sigma^4}  { |\cY|^2 \over (\det \Im \Omega)^2}   \left (
\sum _{i < j} s_{ij} \cG(z_i, z_j |\Omega) \right )^w
\eea
Note that the shift to $w+2$ included  in the range of summation on the left side is due to the fact that $\cY$ is linear in the variables $s,t,u$.

\subsubsection{Contributions to low weight}
\label{sec:12.9.1}

Linearity of $\cY$ in $s,t,u$ immediately implies that the genus-two contribution to the $\cR^4$ interaction without derivatives must vanish $\cB_{(0,0)}^{(2)}=0$. As was already mentioned earlier, the $D^2 \cR^4$ contribution vanishes in view of the momentum conservation identity $s+t+u=0$. 

\sm

The leading low energy contribution is the $D^4 \cR^4$ interaction obtained from the zero-th order expansion term of the exponential in (\ref{8c5}). The remaining integrations required to evaluate $\cB^{(2)} _{(1,0)}$ may be carried out using the Riemann bilinear relations, reviewed in appendix~\ref{sec:RS}, and yield an $\Omega$-independent contribution. 

\sm

The coefficient $\cB^{(2)}_{(0,1)}$ has non-trivial dependence on $\Omega$, denoted by $64 \, \f(\Omega)$. Assembling these results to order $D^6\cR^4$ we have, 
\bea
\cB^{(2)}(s,t,u|\Omega) = 32 (s^2+t^2+u^2) + 192 \, stu \, \f (\Omega)+ \cO(s_{ij}^4 )
\eea
The genus-two modular function $\f(\Omega)$  is given in terms of the Arakelov Green function by,
\bea
\f(\Omega) = { 1 \over 8} 
\int _{\Sigma ^2} \om_I(x) \om_J(y)  \Big ( \bar \omega^I(x) \bar \omega^J(y) - 2 \bar \om ^J(x) \bar \om ^I(y) \Big )
\cG(x,y|\Omega) 
\eea
where $\bar \om^I= (Y^{-1})^{IJ} \, \overline{\om_J}$. This object was identified with the spectral invariant introduced by Kawazumi and Zhang. Higher order contributions to $\cB^{(2)}(s,t,u|\Omega) $ may similarly be expressed in terms of integrals of products of Arakelov Green functions, and may be viewed as higher genus generalizations of the genus-one modular graph functions investigated in section \ref{sec:MGF}. Their general structure has, however, been less well-studied and is much less well-understood than that of their genus-one counterparts.

\subsection{Integration over the genus-one moduli space}

When considering one-loop or two-loop amplitudes we have, so far,  focused attention on the structure of the partial 
amplitudes $\cB^{(1)}(s,t,u|\tau)$ and $\cB^{(2)}(s,t,u|\Omega)$, for which we have obtained a systematic low energy expansion in terms of modular graph functions, both at genus one and genus two. For fixed $\tau$ and $\Omega$, this expansion is absolutely convergent for $\Re(s), \Re(t), \Re(u)  <1$, and admits an analytic continuation in $s,t,u$ whose sole singularities are simple poles at positive integers in $s,t,u$, corresponding to the exchanges of massive string states, as may be seen from the operator product expansion.   

\sm

Obtaining the corresponding physical string  amplitudes $\cA^{(1)}(s_{ij})$ and $\cA^{(2)}(s_{ij})$ further requires the integrations over the moduli spaces of genus-one and genus-two Riemann surfaces shown in (\ref{12.amps}).  These integrations are absolutely convergent only when $s,t,u$ are purely imaginary, in which case $\cB^{(1)}(s,t,u|\tau)$ and $\cB^{(2)}(s,t,u|\Omega)$ may be uniformly bounded on their respective moduli spaces. For the case of genus one, the lack of convergence originates from the integration in the region near the cusp. This may be seen by expressing the genus-one Green function $G(z|\tau)$ in the co-moving coordinates $z=x+y \tau$ for $x,y \in \RR/\ZZ$ first introduced in section \ref{5.LapTor} and given by (\ref{10.green-theta}),
\bea
G(x+y \tau | \tau) = - \ln \left | { \tet_1(x+y \tau |\tau) \over \eta(\tau) } \right |^2 + 2 \pi \tau_2 y^2
\eea
For $\tau_2 \to \infty$ and fixed $x$ and $|y| < \thalf$, the Green function behaves as follows,
\bea
G(x+y \tau | \tau) \approx 2 \pi \tau_2 (y^2 -|y|) 
\eea 
which diverges as $\tau_2 \to \infty$. 

\sm

The analytic continuation in $s,t,u$ of the integrals over moduli now produces double and additional simple poles at all positive integers, as well as branch cuts in $s,t,u$ that originate at every non-negative  integer, including zero. 
These singularities are fully expected on physical grounds. The poles produce  mass renormalization and non-zero decay widths of massive string states, while the branch cuts correspond to the decay of a single string into a two-string state. The presence of the branch cuts implies that the string amplitudes $\cA$ do not admit a convergent low energy Taylor expansion. We now review the derivation of the branch cuts that arise in the low energy expansion for the case of one-loop amplitudes.

\subsubsection{Partitioning the genus-one moduli space}

\begin{figure}[htb]
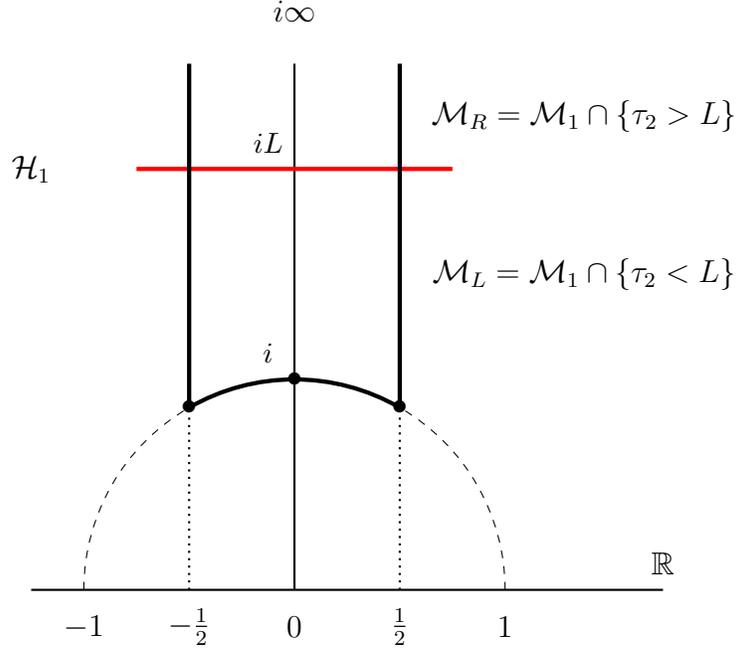

\begin{center}
\tikzpicture[scale=0.7]
\scope[xshift=-5cm,yshift=0cm]
\draw[thick] (-5,0) -- (7,0);
\draw[thick] (0,0) -- (0,10);
\draw[ultra thick, color=red] (-3,8) -- (3,8);
\draw[thick, dotted] (-2,0) -- (-2,3.464);
\draw[thick, dotted] (2,0) -- (2,3.464);
\draw[ultra thick] (-2, 3.464) -- (-2, 10);
\draw[ultra thick] (2 , 3.464) -- (2, 10);
\draw [ultra thick] (-2,3.464) arc (120:60:4 and 4);
\draw [dashed] (-4,0) arc (180:120:4 and 4);
\draw [dashed] (4,0) arc (0:60:4 and 4);
\draw (0,-0.7) node{$0$};
\draw (-2,-0.7) node{$-\half $};
\draw (2,-0.7) node{$\half $};
\draw (-4,-0.7) node{$-1$};
\draw (4,-0.7) node{$1$};
\draw (5.5,6) node{$\cM_L=\cM_1 \cap \{ \tau_2<L\} $};
\draw (5.5,9) node{$\cM_R=\cM_1 \cap \{ \tau_2>L\} $};
\draw (7,0.5) node{$\RR$};
\draw (-5,8) node{$\cH_1$};
\draw (-0.5,4.5) node{\small $i$};
\draw (-0.5,8.5) node{\small $iL$};
\draw (0,4) node{$\bullet$};
\draw (2,3.464) node{$\bullet$};
\draw (-2,3.464) node{$\bullet$};
\draw (0,11) node{$i \infty$};
\endscope
\endtikzpicture
\end{center}
\caption{\textit{Partition of the genus-one moduli space  $\cM_1$ into $\cM_R$ and $\cM_L$. }\label{fig:12.4}}
\end{figure}

To proceed with the analysis of the branch cuts produced in the low energy limit, we partition the moduli space $\cM_1$ of genus-one Riemann surfaces into a region near the cusp $\cM_R$ and its complement $\cM_L$, as illustrated in the Figure \ref{fig:12.4}, and given as follows for $L >1$,
\bea
\cM_R & = & \cM_1 \cap \{ \tau_2 >L \} 
\hskip 1in \cM_L \cap \cM_R = \emptyset
\no \\
\cM_L & = & \cM_1 \cap \{ \tau_2 \leq L \}
\hskip 1in \cM_L \cup \cM_R = \cM_1
\eea

We shall think of $L$ as a free parameter, but in practice take $L \gg 1$ so that $\cM_R$ is a small neighborhood of the cusp. The amplitude $\cA^{(1)}$ is  a sum of the corresponding integrals, 
\bea
\cA^{(1)} (s_{ij} ) = \cA^{(1)}_L (s_{ij} ) + \cA^{(1)}_R (s_{ij} )
\eea
where
\bea
\cA^{(1)}_{L,R} (s_{ij} ) = { \pi \over 16} \int _{\cM_{L,R}}   { |d \tau|^2 \over (\Im \tau)^2} \, \cB^{(1)}  (s,t,u |\tau) 
\eea
The region $\cM_L$ is compact and the integrand $\cB^{(1)}  (s,t,u |\tau) $ admits a uniform low energy expansion in the region of convergence $|s|, |t|, |u|<1$. Therefore, the resulting contribution $\cA^{(1)}_L (s_{ij} )$ will be analytic in $s,t,u$ but dependent on $L$. 

\sm

The region $\cM_R$ is not compact and $\cB^{(1)}  (s,t,u |\tau) $ cannot be handled  with a uniformly convergent low energy expansion. Instead we need a treatment of this contribution that is exact up to contributions exponentially suppressed in $L$. The resulting $\cA^{(1)}_R (s_{ij} )$ will exhibit branch cuts in $s,t,u$ and its $L$-dependence will cancel the $L$-dependence of $\cA^{(1)}_L (s_{ij})$.

\subsubsection{Integrals involving Eisenstein series and modular graph functions}

To carry out the integrations over the fundamental domain of the partial amplitudes $\cB^{(1)}_{(p,q)} (\tau)$, one invariably encounters integrals of Eisenstein series, products of Eisenstein series, and other modular graph functions,  over the truncated fundamental domain $\cM_L$. These integrals are of interest in their own right and we dedicate this subsection to deriving some general results and applying these results to the simplest cases, and then close by quoting the integral of the product of three Eisenstein series from Zagier's work. We begin with the following straightforward  application of Stokes's theorem to an arbitrary modular graph function $\cC$ of weight $(0,0)$, or to an arbitrary modular graph form $\cC^+$  of weight $(0,2)$,
\bea
\label{12.ints}
\int _{\cM_L} { d^2 \tau \over \tau_2^2} \, \Delta \cC & = & \int _0^1 d\tau _1 \, \p_{\tau_2} \cC(\tau_1, \tau_2) \Big | _{\tau_2 =L}
\no \\
\int _{\cM_L} { d^2 \tau \over \tau_2^2} \, \nabla \cC^+ & = & \int _0^1 d\tau _1 \, \cC^+(\tau_1, \tau_2) \Big | _{\tau_2 =L}
\eea
where the result on the right side may be read off from the constant Fourier mode part of $\cC$ and $\cC^+$. The result may be readily used to evaluate the integral of a non-holomorphic Eisenstein series or products thereof. It turns out that the simplest expressions are obtained in terms of the normalized Riemann zeta function $\zeta^*(s)$ and the normalized non-holomorphic Eisenstein series $E^*_s(\tau)$, defined by,
\bea
\zeta ^*(s) =  { \Gamma (s/2) \zeta (s) \over \pi^{{s \over 2}}} 
\hskip 1in 
E_s^*(\tau) = \thalf \Gamma (s) E_s(\tau)
\eea
The normalization guarantees the functional equations $\zeta^*(1-s) = \zeta^*(s)$ and $E^*_{1-s} (\tau) = E^*_s(\tau)$, in addition to the following simple Fourier expansion, 
\bea
\label{12.Eisas}
E^*_s(\tau) = \zeta ^*(2s) \, \tau_2^s + \zeta ^*(2s-1) \, \tau_2^{1-s} + \cO(e^{-2 \pi \tau_2})
\eea
In terms of these functions, we have the following result.

{\lem
\label{12.Eisint}
The following integrals involving Eisenstein series are given, up to exponentially decaying terms $\cO(e^{-2 \pi \tau_2})$ at the cusp, by,
\bea
\int _{\cM_L} { d^2 \tau \over \tau_2^2} \, E_s^* & = & \sum _{x=s,1-s} \zeta ^*(2x) { L^{x-1} \over x-1} 
 \\
\int _{\cM_L} { d^2 \tau \over \tau_2^2} \, E_s^*E_t^* & = & \sum _{x=s,1-s} \, \sum_{y=t,1-t} \zeta ^*(2x) \zeta ^*(2y) { L^{x+y-1} \over x+y-1} 
\no \\
\int _{\cM_L} { d^2 \tau \over \tau_2^2} \, E_s^*E_t^* E^*_u & = & 
\zeta^*(w-1) \zeta^*(w-2s) \zeta^*(w-2t) \zeta ^*(w-2u)
\no \\ &&
+ \sum _{x=s,1-s} \, \sum_{y=t,1-t} \sum _{z=u,1-u} \zeta ^*(2x) \zeta ^*(2y) \zeta^*(2z) { L^{x+y+z-1} \over x+y+z-1} 
\no
\eea
where $w=s+t+u$ in the last line. }

\sm

The first line in Lemma \ref{12.Eisint} may be proven by substituting $E_s^*= \Delta E_s^*/s(s-1)$ in the integrand  and then  using the first line on (\ref{12.ints}). The second line in Lemma \ref{12.Eisint} may be proven by expressing the combination $\Delta E_s^* \, E^*_t - E_s^* \Delta E^*_t $ in two different ways, 
\bea
\Delta E_s^* \, E^*_t - E_s^* \Delta E^*_t 
& = & 
\Big  ( s(s-1) - t(t-1) \Big ) E^*_s E^*_t
 \\ & = &
\nabla \left ( { E^*_s \bar \nabla E_t^* - E^*_t \bar \nabla E_s^* \over 2 \tau_2^2}     \right ) 
+ \bar \nabla \left ( { E^*_s \nabla E_t^* - E^*_t  \nabla E_s^* \over 2 \tau_2^2}     \right )
\no
\eea
and then using the expression on the right side of the first line to eliminate $E_s^* E_t^*$ from the integrand of \ref{12.Eisint} in favor of the total derivative terms in the second line. Finally, to integrate the total derivative terms, we use the second line in (\ref{12.ints}) and its complex conjugate relation along with the asymptotics (\ref{12.Eisas}) of the Eisenstein series.  The third line in Lemma \ref{12.Eisint} was proven by Zagier using different methods, as will be indicated in the bibliographical notes.

\subsubsection{The non-analytic contribution $\cA_R^{(1)}$}

To evaluate $\cB^{(1)}(s,t,u|\tau)$ for $\tau \in \cM_R$, we cannot use  (\ref{8c4}) since this expansion is not uniformly convergent throughout $\cM_R$. Instead, our starting point is the original expression in (\ref{12.amps}), which we shall evaluate exactly up to terms that are exponentially suppressed in~$L$. To do so, we use co-moving coordinates, $z_i = x_i + y_i \tau$ with $i=1,2,3,4$ and partition the integration region $\Sigma ^4$ into the six possible orderings of $y_1,y_2, y_3, y_4$ up to cyclic permutations.  Doing so allows us to decompose $\cA^{(1)}_R(s_{ij})$ into a sum of six more elementary contributions, which are pairwise equal to one another, 
\bea
\cA^{(1)}_R(s_{ij}) = 2\cA_*(L; s,t) + 2\cA_*(L; t,u) + 2\cA_*(L; u,s) \hskip 0.5in s+t+u=0
\eea
The function $\cA_*(L; s,t) $ may be computed exactly, up to exponentially suppressed terms, 
\bea
\cA_*(L; s,t)  =  \mC (s,t;0) \, \ln(-4 \pi Ls) + { \p \over \p \ep} \mC (s,t;\ep) \Big |_{\ep=0} + \cO(e^{ - 2 \pi L}) 
\eea
where $\mC$ is conveniently expressed as the sum of two integral representations, 
\bea
\mC(s,t;\ep) & = & - { 8 \pi s^4 \over \Gamma (3+\ep)} 
\int _0^1 dx \int _0^1 dy \, x^{5+2\ep} y^{2+\ep} (1-y)^{2+\ep} W(s, -sxy+t(1-x)) 
\no \\ &&
- 4 \pi s^7 \sum_{k=0}^\infty s^k t^k \int _0^1 \! dx \, { x^{k+3+\ep} (1-x)^{k+3+\ep} \over 2 \Gamma (k+4+\ep) \, k!} 
\left [ { \p^k \over \p \mu^k} W(s,\mu-sx) \Big |_{\mu=0} \right ]^2
\qquad
\eea
The function $W$ is closely related to the Virasoro-Shapiro amplitude, 
\bea
W(s,t) = { 1 \over stu} \left ( { \Gamma (1-s) \Gamma (1-t) \Gamma (1-u) \over \Gamma (1+s) \Gamma (1+t) \Gamma (1+u)} -1 \right ) 
\hskip 1in u=-s-t
\eea
Although the above expression may look daunting at first, it analytical structure is manifest. The function $W(s,t)$ is meromorphic throughout $s,t \in \CC$, holomorphic in discs of unit radius near the origin, and has only simple poles at positive integer values of $s,t,u$, and thus admits a Taylor expansion in $s,t$. Substituting this Taylor expansion into the integrals defining $\mC$, we see that the integrals preserve the holomorphicity in $s,t,u$ near the origin, so that $\mC$ itself has a convergent Taylor expansion  near the origin, which starts at order $s^4$. To this lowest order we have,
\bea
\label{ARtot}
\mC(s,t;\ep) = -  16 \pi \zeta(3) s^4 { \Gamma (3 + \ep) \over \Gamma (7+2 \ep)} + \cO(s^5, s^4t)
\eea
The coefficients of the expansion to all orders  inherit contributions linear and bilinear in odd zeta-values from the expansion of the Virasoro-Shapiro amplitude,  with rational coefficients in the coefficient $\mC(s,t;0)$  of the $\ln(-4 \pi Ls)$ term, and additional harmonic sums in the $\p_\ep \mC(s,t;\ep) \big |_{\ep=0}$ analytic contribution to $\cA_*(L;s,t)$. The discontinuity of $\ln(-4 \pi Ls)$  near the origin reproduces the square of the tree-level amplitude by unitarity.

\subsubsection{The analytic contribution $\cA_L^{(1)}$}

The contribution $\cA^{(1)}_L(s_{ij})$ is analytic in $s,t,u$ near the origin and we may use the expansion of $\cB^{(1)}  (s,t,u |\tau)$ in terms of modular graph functions of (\ref{8c4}) to evaluate $\cA_L^{(1)}(s_{ij})$ order by order in powers of $s,t,u$,
\bea
\cA^{(1)} _L(s_{ij}) = \sum_{p,q=0}^\infty \cA_{(p,q)}^{(1)} \, { \sigma _2^p \sigma _3^q \over p! \, q!}
\eea
  Since the non-analytic part $\cA_R^{(1)} (s_{ij})$ starts contributing to $\cA^{(1)}(s_{ij}) $ only at order $s^4$, all its lower orders are given entirely by $\cA^{(1)}_L(s_{ij})$.  The corresponding amplitudes are readily evaluated, and we find,
\begin{align}
\label{12.loen}
\cA_{(0,0)}^{(1)} & = { \pi \over 3} & \cA_{(0,1)}^{(1)} & = {\pi  \over 9} \zeta (3)
&
\cA_{(1,0)}^{(1)} & = 0 & \cA_{(1,1) }^{(1)} & = {29 \pi  \over 540}\zeta(5)
\end{align}
The remaining coefficients at weight four are given as follows,
\bea
\label{ALtot}
\cA_{(2,0)} ^{(1)}=  
 { 4 \pi \zeta(3)  \over 45} \left [ 
\ln (2 L) + {\zeta '(4) \over \zeta (4) } - {\zeta '(3) \over \zeta (3) }  -{ 1 \over 4} \right ]
 \eea
 
 \subsubsection{Assembling analytic and non-analytic contributions}
 
Assembling the contributions to $\cA_L^{(1)}$ in (\ref{ALtot}) and to $\cA_R^{(1)}$ in (\ref{ARtot}) gives $\cA^{(1)}(s_{ij})$, the total amplitude,  to order $\cO(s_{ij}^4)$ included. We see explicitly that all $L$-dependence cancels as required for the consistency of the calculation. It is instructive to rearrange  the total amplitude as the sum 
of ``analytic" and ``non-analytic" pieces,  
\bea
\label{suman}
\cA^{(1)}(s_{ij}) = 2\pi \Big ( \cA_L ^{(1)} (L;s_{ij}) + \cA_R ^{(1)} (L;s_{ij}) \Big ) = 
\cA_{{\rm an}}(s_{ij}) +\cA_{{\rm non-an}}(s_{ij}) 
\eea
The reason for the quotation marks on  analytic and non-analytic is that the non-analytic piece actually contains also analytic contributions, so that the nomenclature is natural and suggestive but not entirely precise.
 The ``analytic'' piece $ \cA_{{\rm an}} (s_{ij})$   is given by, 
\bea
\label{Aan}
 \cA_{{\rm an}} (s_{ij}) = { 2 \pi^2 \over 3} \left ( 1 + {\zeta (3) \sigma _3 \over 3}  + \cO(s_{ij}^5)\right ) 
 \eea
 and the  ``non-analytic'' piece $ \cA_{{\rm non-an}} (s_{ij})$ has the form, 
 \bea
\label{Anonan}
 \cA_{{\rm non-an}} (s_{ij}) = {2\pi^2 \over 3}  \left ( \cA_{{\rm sugra}} + \cA_4   +  \cO(s_{ij}^5) \right )
 \eea
 The lowest-order term $ \cA_{{\rm sugra}}$ is a regularized version of the ten-dimensional one-loop supergravity amplitude.  The contribution $\cA_4$ is given by, 
  \bea
\label{Ahatfour}
 \cA_4  =  { 4 s^4 \over 15} \, \zeta (3)  \left ( - \ln (- 2 \pi s) +  {\zeta '(4) \over \zeta (4) } - {\zeta '(3) \over \zeta (3) } - \gamma  +{63 \over 20} \right ) +
\hbox{ cycl } (s,t,u)
\eea
where $\gamma$ is the Euler constant. The discontinuity of $ \cA_{{\rm non-an}} (s_{ij}) $, namely the  coefficient of the $\ln(-2\pi s)$ terms in (\ref{Ahatfour}), reproduce those obtained in \cite{Green:2008uj}.

\subsection{Integration over the genus-two moduli space}
\label{12.11}

The function $\cB^{(2)} _{(1,0)} (\Omega)$ needed to obtain the coefficient of the $D^4\cR^4$ interaction was found to be independent of moduli $\Omega$ in subsection \ref{sec:12.9.1}. Thus, its integral over moduli space $\cM_2$ is given by the volume of $\cM_2$ with the Siegel metric on the Siegel upper half space,
\bea
ds^2 = \sum_{I,J,K,L} Y^{-1}_{IJ} \, Y^{-1}_{KL} \, d\bar \Omega_{JK} \, d\Omega_{IL}
\eea
 of constant negative curvature.  The volume form $d \mu_2$ and the value of the volume, computed by Siegel, are given by,
\bea
\int _{\cM_2} d \mu_2  = { 8 \pi^3 \over 270}
\hskip 1in 
d \mu_2 = { |d \Omega_{11} \wedge d \Omega_{12} \wedge d \Omega _{22}  |^2 \over (\det \Im \Omega)^3}
\eea
As a result, the low energy effective interaction to this order is given by,
\bea
\label{12.gen2a}
\cA^{(2)}_{(1,0)} = { 4 \pi^4 \over 270} (s^2+t^2+u^2) \cR^4 
\eea

The function $\cB^{(2)} _{(0,1)} (\Omega)$ is proportional to the Kawazumi-Zhang invariant $\f$. Its integral over $\cM_2$ is convergent and yields the contribution $\cA^{(2)}_{(0,1)} $ to the low energy effective action. Evaluation of the integral is made possible by the fact that $\f$ satisfies an inhomogeneous Laplace eigenvalue equation, 
\bea
(\Delta - 5) \f = - 2 \pi \delta_{SN}
\eea
where $\Delta$ is the Laplace-Beltrami operator on the Siegel upper half space for the Siegel metric, 
\bea
\Delta = \sum_{I,J,K,L} 4 \, Y_{IK} \, Y_{JL} \bar \p_{IJ} \p_{KL} 
\hskip 1in 
\p_{IJ} = \half (1+\delta_{IJ}) { \p \over \p \Omega_{IJ}}
\eea
and $\delta_{SN}$ is the Dirac $\delta$-function supported on the separating degeneration node. 
The coefficient $\cB_{(0,1)}^{(2)}$ is then given by the integral of the Kawazumi-Zhang invariant over $\cM_2$, 
\bea
\cB_{(0,1)} ^{(2)} = \pi \int _{\cM_2} d\mu_2 \,  \f = {\pi \over 5} \int _{\cM_2} d \mu_2 \Big ( \Delta \f + 2 \pi \delta_{SN} \Big ) = {2 \pi^3 \over 45} 
\eea
The integral may be evaluated using Stokes's theorem and the asymptotic behavior for the Kawazumi-Zhang invariant, and we find, 
\bea
\label{12.gen2b}
\cA_{(0,1)} ^{(2)} =  {2 \pi^3 \over 45} (s^3+t^3+u^3) \cR^4
\eea
For possible verifications to genus-three order, we refer to the references of this section.

\subsection*{$\bullet$ Bibliographical notes}

Classic books  by Green, Schwarz, and Witten \cite{GSW1,GSW2} and by Polchinski \cite{Polchinski:1998rq,Polchinski:1998rr} present  comprehensive perspectives on string theory. A pedagogical introduction, aimed at undergraduate and graduate students, is given in the book by Zwiebach \cite{Zwiebach}. Other accounts may be found in the books by Johnson \cite{Johnson}, by Becker, Becker, and Schwarz \cite{Becker}, by Blumenhagen, L\"ust, and Theissen  \cite{Blumenhagen}, and by Kiritsis  \cite{Kiritsis}. {A beautiful discussion of the connection between modular/automorphic forms and string amplitudes can be found in the book by Fleig, Gustafsson, Kleinschmidt, and Persson \cite{Fleig:2015vky}.}

\sm

The early development of string theory was dominated by the study of string amplitudes. The Ramond and Neveu-Schwarz sectors of the RNS formulation were introduced in \cite{R} and \cite{NS}, respectively. The different spin structure sectors and the Gliozzi-Scherk-Olive (GSO) projection onto supersymmetric string spectra were introduced in \cite{GSO}. 
The Polyakov formulation of string theory, along with many other key problems in modern theoretical physics, is presented in the book by Polyakov \cite{Polyakov}. Friedan's renormalization of the non-linear $\sigma$-model and the relation between Einstein's equations and conformal invariance may be found in \cite{Friedan:1980jm}. Generalizations with worldsheet fermions are given in \cite{Alvarez-Gaume:1981exa}, and including torsion in \cite{Braaten:1985is} and the dilaton in \cite{Osborn:1991gm}. The $\alpha'$ expansion of the effective action including the metric, anti-symmetric tensor field, and dilaton is presented in detail in the book of Princeton Institute for Advanced Study lecture notes \cite{EDias}.

\sm

The standard treatment of the decoupling of negative norm states for both the bosonic and superstrings  is clearly reviewed in  \cite{GSW1}. The modern BRST-based approach to the covariant quantization of superstrings in the RNS formulation was developed in \cite{Friedan:1985ge}, and reviewed in \cite{GSW1,GSW2} and \cite{Polchinski:1998rq,Polchinski:1998rr}. The structure of gauge and gravitational anomalies, relevant to supergravity and string theories in various dimensions, was obtained in \cite{Alvarez-Gaume:1983ihn}. The cancellation of anomalies in the Type I theory with gauge group $SO(32)$ was famously discovered in \cite{Green:1984sg}. Useful lecture notes on anomalies may be found in \cite{Harvey:2005it}.

\sm

Reviews of superstring perturbation theory may be found in \cite{RMP,Lectures} for the RNS formulation and in  \cite{Berkovits:2004px} for the pure spinor formulation.  Lectures accessible to a more mathematically oriented readership are in \cite{EDias}, while an elaboration of  the mathematical underpinning of the RNS formulation is provided in \cite{Witten:2012ga,Witten:2012bh}. A broad overview of the current status of superstring perturbation theory, and open problems,  is given in \cite{Berkovits:2022ivl}, which contains an extensive bibliography. 

\sm

The tree-level and one-loop four graviton amplitudes were derived in the original papers  in which the Type II \cite{Green:1981yb}  and Heterotic strings \cite{Gross:1985fr,Gross:1985rr}  were discovered. The existence of the analytic continuation of the one-loop four graviton amplitude was shown in  \cite{DHoker:1993hvl,DHoker:1994gnm}. 

\sm

 The fact that the low energy expansion of string theory leads to $\cN=4$ super-Yang-Mills theory and $\cN=8$ supergravity in four space-time dimensions was discovered  in \cite{Green:1982sw}.  The systematic analysis of the low energy effective interactions induced by the tree-level four-graviton amplitude was initiated in \cite{Gross:1986iv,Grisaru:1986px}, 
while the effects at genus-one were obtained in  \cite{Green:1997as,Green:1999pv,Green:2008uj}. The low energy expansion, up to order $D^8\cR^4$, including the logarithmic branch cuts and the transcendentality properties of the amplitude  discussed in section \ref{sec:12.7.2} were obtained in \cite{DHoker:2019blr}. 

\sm

The original Rankin-Selberg method for integrating cusp forms over the fundamental domain for $SL(2,\ZZ)$ 
was developed in \cite{Rankin, Selberg}, and was extended to automorphic functions of non-rapid decay in \cite{Zagier8}. { It was applied to string amplitudes in \cite{Florakis:2016boz} and subsequent works \cite{Angelantonj:2011br,Angelantonj:2012gw}. } The proof of the last integral identity in Lemma \ref{12.Eisint} is in \cite{Zagier8}, while methods to carry out integrals of modular graph functions may be found  in \cite{DHoker:2021ous} and references therein.

\sm

Considerable progress has been made recently towards evaluating tree-level and genus-one string amplitudes with arbitrary numbers of external states in \cite{Mafra:2011nv,Azevedo:2018dgo,Mafra:2018nla,Mafra:2018pll,Mafra:2018qqe,Mafra:2022wml} and references therein. The role of motivic and multiple zeta-values played in the form of the amplitudes was examined in \cite{Broedel:2013tta,Stieberger:2013wea,SCHLotterer:2012ny}.
Genus-one amplitudes have also recently been reproduced from ${\cal N}=4$ supersymmetric Yang--Mills theory by considering a flat-space limit of $AdS_5\times S^5$  \cite{Alday:2018pdi}.  

\sm

For genus-two amplitudes, the measure on supermoduli space was derived in the RNS formulation in \cite{DHoker:2001kkt, DHoker:2001qqx,  DP4} and reproduced using algebraic-geometric methods in \cite{Witten:2013tpa}. The full amplitude for four gravitons was obtained in \cite{DP6} in the RNS formulation, and extended to include external fermion states in   \cite{Berkovits:2005df,Berkovits:2005ng} using the pure spinor formulation. The overall normalization of the amplitude was obtained in  \cite{DGP,Gomez:2010ad}. 

\sm

Recently, the genus-two amplitudes with five massless states were constructed in \cite{DHoker:2020prr,DHoker:2020tcq} using an amalgam of methods from the RNS and pure spinor formulations and. For external NS boson states and even spin structure, these results were confirmed from first principles in the RNS formulation in \cite{DHoker:2021kks}. 

\sm

The generalization of  modular graph functions to genus two and higher genus was developed in \cite{DHoker:2017pvk} while the relation between the coefficient of the $D^6\cR^4$ term and the invariant of Kawazumi  \cite{kawazumi2008johnson} and Zhang \cite{zhang2008gross} was identified in \cite{DHoker:2013fcx} and used to derive the corresponding differential equation in  \cite{DHoker:2014oxd}. Degenerations of genus-two modular forms naturally produce non-holomorphic versions of Jacobi forms, systematic treatment of which may be found in the book by Eichler and Zagier \cite{Eichler}.  A $\tet$-lift representation of the genus-two Kawazumi-Zhang invariant was obtained in \cite{Pioline:2015qha}. For the  derivation of the results in section \ref{12.11} we refer to  \cite{DHoker:2014oxd}. 
 
 \sm
 
For amplitudes at genus three and beyond, no first principle derivations are available at the time of this writing. 
 A proposal for the measure on supermoduli space in the RNS formulation was advanced in \cite{Cacciatori:2008ay},
building on  some earlier unsuccessful attempts in \cite{DHoker:2004fcs}. The proposal of \cite{Cacciatori:2008ay} was critiqued, however, in the appendix of \cite{Witten:2015hwa}.  The leading low energy effective interaction in Type II string theory was obtained using the pure spinor formulation in \cite{Gomez:2013sla}. The corresponding expression for the amplitude, however, diverges beyond the leading low energy limit so that no independent checks on the unitarity of the amplitude is available. A proposal for the full genus-three four graviton amplitude has been advanced in \cite{Geyer:2021oox}, but details of the derivation are not yet available. Proposals for the measure on supermoduli space in the RNS formulation at genus four may be found in \cite{Cacciatori:2008pj,Grushevsky:2008zm}.

\newpage

\section{Toroidal compactification}
\setcounter{equation}{0}
\label{sec:Toroidal}

In this section, we discuss the modular properties of quantum field theory of scalar fields that take values in a $d$-dimensional torus with a flat metric and constant anti-symmetric tensor. The problem is of great interest in quantum field theory and string theory in view of the fact that such \textit{toroidal compactifications} admit solutions using free field theory methods on the worldsheet, preserve Poincar\'e supersymmetries, and may be used to relate different perturbative string theories via \textit{T-duality}. Toroidal compactifications produce large duality groups that generalize the modular group $SL(2,\ZZ)$ considered thus far.

\subsection{Conformal field theory on flat manifolds}

In section \ref{sec:TorusQFT} we considered the functional integral for a single free scalar field $\varphi$ that  takes values in $\RR$. The situation is readily generalized to the case of  $d$ free scalar fields taking values in a flat Euclidean $\RR^d$. As discussed in section \ref{sec:SA}, the generalization to $d$ scalar fields taking values in (the local coordinate systems of) an arbitrary space-time manifold in general leads to fully interacting quantum field theories on the worldsheet that cannot be solved exactly. 

\sm

There is one class of theories, however, where an exact quantization may be obtained, namely when all the fields $G,B,\Phi$ are constant, and the manifold $M$ is flat. Although flat, $M$ may  have non-trivial topology as in the case of a $d$-dimensional torus $M=T_d$, which is the case we shall consider here.  Constant $\Phi$ gives rise to a term proportional to the Euler number of $\Sigma$ and weighs the functional integral by a power of the string coupling which is independent on the quantum field $X$. Finally, the term involving $B_{\mu \nu}$ is a topological term when $B$ is constant. It does not contribute when the topology of $M$ is trivial, but when $M$ is a torus, it does have very interesting physical effects which we shall now examine. For constant $G,B,\Phi$, the $X$-dependent part of the action thus reduces to,
\bea
I[X, g]  =  
{ 1 \over 4 \pi \alpha ' } \int _\Sigma d \mu _g \, \Big  (  g^{mn} G_{\mu \nu} 
- i \, \ep^{mn}  B_{\mu \nu}   \Big ) \p_m X ^\mu  \p_n X ^\nu 
\eea
Henceforth, we shall restrict to the case where the worldsheet  surface is a torus of modulus~$\tau$, so that $\Sigma = \CC /\Lambda$ with $\Lambda = \ZZ + \tau \ZZ$. We shall choose a system of complex coordinates $z,\bar z$ on $\Sigma$ in which the metric $g$ takes the form,
\bea
g_{mn} \, d\xi ^m d \xi^n = { |dz|^2 \over \tau_2}
\eea
or in components we have $g_{zz}=g_{\bar z \bar z}=0$ and $g_{z \bar z} = 1/(2\tau_2)$. Furthermore,  the space-time torus $T_d$ is given by $\RR^d / L$ for some $d$-dimensional lattice $L$.

\subsection{Lattices and tori of dimension $d$}

A lattice $L$ of rank $d$, also referred to as a lattice of dimension $d$,  may be represented in terms of a basis of $d$ vectors $v_1 , \cdots, v_d $  in $\RR^d$,
\bea
L = \Big \{ \ell = \sum_{i=1}^d n_i v_i  ~~ n_i \in \ZZ \Big \}
\eea
It will be useful to  view the assignment of basis vectors as obtained by a $GL(d,\RR)$ transformation applied to a standard basis of $d$ vectors, such as for example $v_1^0=(1,0,\cdots, 0)$, $v_2^0 = (0,1,0,\cdots, 0), \cdots, v_n^0 =(0,\cdots, 0, 1)$. Decomposing $GL(d,\RR) = SL(d,\RR) \times \RR^+$, the factor $\RR^+$ corresponds to overall rescaling of the lattice.  An arbitrary  set of $d$ vectors $v_i' \in L$ for $i=1,\cdots, d$ may be decomposed in the basis $v_i$,
\bea
v'_i = \sum_{j=1}^d M_{ij} v_j \hskip 1in M_{ij} \in \ZZ
\eea 
and generates a sub-lattice $L' \subset L$. Lattices may be viewed as Abelian groups so that $L'$ is a subgroup of $L$.  The index of $L'$ as a subgroup of $L$ is given as follows, 
\bea
[L:L'] = |\det M|
\eea
Equivalently, the unit lattice cell of $L'$ contains $|\det M|$ copies of the unit lattice cell of $L$. In the special case where $\det M=1$, we have $L'=L$ and therefore $M$ is an automorphism of~$L$. The group of all automorphisms of $L$ is $SL(d,\ZZ)$, thereby generalizing to $d$ dimensions the result familiar from the two-dimensional torus. When $\det M=-1$ the lattice $L'$ is the mirror image of $L$, reversing the orientation of the lattice. 

\sm

Introducing the equivalence relation between two lattices that are related to one another by an overall rescaling in $\RR^+$, an overall rotation in $SO(d)$, and a lattice automorphism in $SL(d,\ZZ)$, we may identify the space of all inequivalent lattices as follows,
\bea
\hbox{space of inequivalent lattices } = SL(d,\ZZ) \backslash SL(d,\RR)/SO(d,\RR)
\eea
In the case of a two-dimensional lattice $L$ with $d=2$, we may identify $SL(2,\RR)/SO(2)$ with the Poincar\'e upper half-plane $\cH$ and the entire coset $SL(2,\ZZ) \backslash SL(2,\RR)/SO(2,\RR)$ with the moduli space of  toroidal Riemann surfaces.  The dynamics of point particles propagating on the torus $T_d=\RR^d/L$ is sensitive to the lattice geometry of this quotient. We shall see below that the dynamics of strings propagating on the torus $\RR^d/L$ is sensitive to a different equivalence class of lattices, which is ``smaller" than  $SL(d,\ZZ) \backslash SL(d,\RR)$ thanks to the quintessentially string-theoretic phenomenon of $T$-duality.

\subsection{Fields taking values in a torus}

We begin by reviewing the construction of the space of $d$ scalar fields $X(z)$ that take values in the torus $\RR^d / L$ and are functions on a worldsheet $\Sigma= \CC/ \Lambda$ with the topology of a two-dimensional torus, whose complex structure is given by its modulus $\tau \in \cH$ so that $\Lambda = \ZZ \oplus \tau \ZZ$.  The space of maps,
\bea
X : \, \Sigma = \CC/\Lambda \, \to \, T_d= \RR^d /L
\eea
is disconnected and may be decomposed into the union of an infinite number of connected components which are labelled by two copies of the lattice $L$. In a sector labelled by a pair $(\ell_A, \ell_B) \in L\times L$, the field $X^\mu$ satisfies the following monodromy conditions, 
\bea
X (z+1) & = & X (z) + \ell _A
\no \\
X (z+\tau) & = & X (z) + \ell  _B
\hskip 1in \ell _A, \ell_B \in L
\eea
The field $X$ may be decomposed into a purely periodic part $Y$, which satisfies,
\bea
Y(z+1)=Y(z+\tau)=Y(z)
\eea 
and a special solution to the monodromy conditions. The latter  is arbitrary but it will be convenient to choose it  linear in $z$ and $\bar z$. Since the field $X$ and the vectors $\ell_A, \ell_B$  are real-valued, we have the following decomposition, 
\bea
X _{\ell_A, \ell_B} (z) = Y  (z) + z \, { \ell _B - \bar \tau \ell  _A \over \tau - \bar \tau}
- \bar z \, { \ell  _B - \tau \ell  _A \over \tau - \bar \tau}
\eea
Such a configuration is sometimes referred to as a \textit{worldsheet instanton}. The lattice elements $\ell_A, \ell_B$ are related to the momentum and the winding modes of strings on $T_d$, as we shall clarify below. The integration over all $X : \Sigma \to T_d$ decomposes into a sum over sectors labelled by $L \times L$. The  functional integral takes the following form, 
\bea
\int _{{\rm maps}(\Sigma \to T_d)} D X e^{-I[X,g]}
= {\rm Vol} (T_d) \sum _{\ell _A, \ell_B \in L} \int _{{\rm maps}(\Sigma \to \RR^d)} D' Y \, e^{-I[X_{\ell_A, \ell_B}, g]}
\eea
For given worldsheet metric $g$, the measures $DX$ and $D 'Y$ are evaluated using the norm of (\ref{12.measure}). 
The factor ${\rm Vol } (T_d)$ arises from the normalizable zero mode of the field $Y$, and the integration $D' Y$ is over all  non-zero modes, i.e. all functions $Y$ orthogonal to constants. The contribution to the action of $Y$ decouples from the contribution of the momentum and winding modes, which may be evaluated explicitly, 
\bea
I[X_{\ell_A, \ell_B}, g] = I[Y, g] + { 1 \over 4 \pi \alpha ' \tau_2} (\ell ^\mu _B - \bar \tau \ell ^\mu _A) 
(\ell ^\nu _B -  \tau \ell ^\nu _A) (G_{\mu \nu} +  B_{\mu \nu} )
\eea
The integral over $Y$ factorizes, the dependence on $B_{\mu \nu}$ cancels out in this integral and reduces to the functional integrals computed earlier in section \ref{sec:TorusQFT}. The remaining factor is given by the following lattice sum,
\bea
\label{4h9}
Z (L, \tau) = {\rm Vol} (T_d)  \sum _{\ell _A, \ell_B \in L} \exp \left \{ - 
{ 1 \over 4 \pi \alpha ' \tau_2} (\ell ^\mu _B - \bar \tau \ell ^\mu _A) 
(\ell ^\nu _B -  \tau \ell ^\nu _A) (G_{\mu \nu} +  B_{\mu \nu} ) \right \}
\eea
The dependence on $B_{\mu \nu}$ reduces to $-i B_{\mu \nu} \, \ell ^\mu _A \, \ell ^\nu _B / (2 \pi \alpha ')$ and is independent of $\tau$, as is expected for a topological term. Before investigating the properties for a general torus $T_d$, we first look at the much simpler case of $d=1$.

\subsection{T-duality on a circle}

In this subsection, we study the simplest case of toroidal compactification when $d=1$ and exhibit the phenomenon of $T$-duality for a circle. We take $G=1$, and introduce the lattice $L$ associated with a circle of radius $R$, namely $L = 2 \pi R \, \ZZ$. We label the monodromy vectors $\ell_A$ and $\ell_B$, which are one-dimensional in this case,  as follows,
\bea
\label{4j0}
\ell _A & = & 2 \pi R \, n
\no \\
\ell_B & = & 2 \pi R \, m \hskip 1in m,n \in \ZZ
\eea
The sum over momentum and winding modes then simplifies to,
\bea
\label{4j1}
Z(R, \tau) = \sqrt{R^2 \over \alpha'} \sum _{m,n \in \ZZ}  \exp \left \{ - 
{ \pi R^2 \over \alpha ' \tau_2} | m - \tau n|^2  \right \}
\eea
The prefactor in $\sqrt{R^2/\alpha'}$ arises from the zero mode of the field $Y$ which, for a circle of radius $R$, gives a finite contribution. The above sum $Z(R,\tau)$  is a rather familiar looking object. To exhibit $T$-duality, we  perform a Poisson resummation on both $m,n$. To do so we compute the Fourier transform in both variables, 
\bea
\int _{\RR^2} dm \, dn  \, e^{ - 2 \pi i (m x + ny)} \, \exp \left \{ - { \pi R^2 \over \alpha ' \tau_2} | m - \tau n|^2  
\right \} = { \alpha' \over R^2} \, \exp \left \{ - { \pi \alpha ' \over R^2 \tau_2} | y+\tau x|^2 \right \}
\eea
so that Poisson resummation gives us,
\bea
Z(R, \tau) = \sqrt{\alpha '  \over R^2} \sum _{m,n \in \ZZ}  \exp \left \{ - 
{ \pi \alpha '  \over R^2 \tau_2} | m + \tau n|^2  \right \}
\eea
Noticing that Poisson resummation has inverted the dependence on $R^2/\alpha'$, we have,
\bea
Z(R,\tau) = Z\left ( { \alpha ' \over R}, \tau \right )
\eea
Thus, a free bosonic string on a circle of radius $R$ has exactly the same partition function as a free bosonic string on a circle of radius $\alpha ' / R$. While we have proven this above for the free string on a genus-one worldsheet only, the effect actually is valid on surfaces of arbitrary genus, and thus holds for fully interacting string theory. The effect is referred to as \textit{$T$-duality}.

\sm

We may organize the double sum in a slightly different way which makes the decomposition into momentum and winding modes transparent. Performing a Poisson resummation on (\ref{4j1}) in $m$ only, we obtain the representation, 
\bea
\label{eq:compbosZ}
Z(R,\tau) = (\tau_2)^\half \sum _{m,n \in \ZZ} e^{2 \pi i \tau p_L^2} \, e^{-2 \pi i \bar \tau p_R^2}
\eea
where the left and right momenta are given in terms of the integers $m,n$ by,
\bea
p_L = \half \left ( { \sqrt{\alpha'} \over  R}  m + { R \over \sqrt{\alpha'}} n \right )
& \hskip 1in & p_L+p_R = { \sqrt{\alpha'} \over  R}  m
\no \\
p_R = \half \left ( { \sqrt{\alpha'} \over  R}  m - { R \over \sqrt{\alpha'}} n \right )
& \hskip 1in & p_L-p_R = { R \over \sqrt{\alpha'}} n 
\eea
The lattice points $p_L+p_R \in \sqrt{\alpha'}/R \, \ZZ$ are clearly associated with the total momentum of the string, while the lattice points $p_L-p_R \in R/\sqrt{\alpha'} \, \ZZ$ are associated with the winding modes.  

\sm

The space of all possible compactifications on the circle is labeled by the radius $R$ modulo the identification $R \equiv \alpha '/R$, a space we may denote by $\RR^+/\ZZ_2$. Clearly, the lattices $L=\sqrt{\alpha'}/R \, \ZZ$ and $L^+=R/\sqrt{\alpha'} \, \ZZ$ are dual to one another, a fact that is reflected in the relation $(p_L+p_R)(p_L-p_R) = mn \in \ZZ$. Thus, the lattice sum may be written alternatively as,
\bea
Z(R,\tau) = (\tau_2)^\half \sum _{p \in L} \sum _{ w\in L^+} e^{ \pi i \tau  (p+w)^2/2} \, e^{- \pi i \bar \tau  (p-w)^2/2}
\eea
At the radius $R = \sqrt{\alpha'}$ the conformal field theory is self-dual and enjoys an enhanced symmetry to the $SU(2)$ Kac-Moody algebra. The radius $R = \sqrt{2 \alpha'}$ is also special in a different sense: it is at this point that the $\mc=1$ Virasoro algebra representation of the conformal field theory becomes reducible to two $\mc=\half$ fermions. 

\sm

$T$-duality has several important consequences for string theory, which include,
\begin{enumerate}
\item The fact that (bosonic) string theory on a circle of radius $R$ is physically equivalent to a string theory on a circle of radius $\alpha '/R$ means that we can never physically probe distances smaller than $\sqrt{\alpha '}$. Analogous arguments exist for the energy of scattering processes, namely probing a string with energy $E$ and with energy $1/(\alpha ' E)$ are physically equivalent. It means that string theory has built in its dynamics a {\sm smallest string length scale} $\sqrt{\alpha'}$.
\item Actually, the above picture is modified in the case of superstring theories. In Type II string theories, compactification of Type IIA on a circle of radius $R$ is physically equivalent to Type IIB theory on a circle of radius $\alpha '/R$. In Heterotic string theories, the $E_8 \times E_8$ theory is invariant under $R \leftrightarrow \alpha ' /R$ but the $Spin (32)/\ZZ_2$ theory is exchanged with the theory of open and closed strings Type I. 
In both cases, $T$-duality provides a mechanism for a partial unification of the five 10-dimensional superstring theories.
\item $T$-duality extends to string theories on curved manifolds under certain conditions. For example, on Calabi-Yau spaces, it is promoted to \textit{mirror symmetry}.
\end{enumerate}

\subsection{T-duality on a torus $T_d$}

For a general torus $T_d= \RR^d /L$ with $d \geq 2$ new effects arise. We consider the lattice sum of (\ref{4h9}) and introduce a basis $\{ v_1 ^\mu , \cdots, v_d ^\mu \}$ with $\mu =1, \cdots,d$ for the lattice $L$. Generalizing the conventions of (\ref{4j0}) used for the circle, we have,
\bea
\ell _A ^\mu & = & 2 \pi \sum _{i=1}^d n^i  \, v_i ^\mu \hskip 1in n^i \in \ZZ
\no \\
\ell _B ^\mu & = & 2 \pi \sum _{i=1}^d m^i \, v_i ^\mu  \hskip 1in m^i \in \ZZ
\eea
What enters $Z(L,\tau)$ is the combination, 
\bea
{1 \over \alpha '} (G_{\mu \nu} + B_{\mu \nu} ) v^\mu _i v^\nu _j = \cG _{ij}+ \cB_{ij}
\eea
where $\cG_{ij}$ and $\cB_{ij}$ are respectively symmetric and anti-symmetric under the interchange of the indices $i,j$.
Using matrix notation for $m,n,\cG, \cB$, we have,
\bea
\label{4k2}
Z (L, \tau) = {\rm Vol} (T^d)  \sum _{m, n \in \ZZ^d} \exp \left \{ - 
{ \pi \over \tau_2} (m + \tau n)^\dagger  (\cG +\cB) (m +  \tau n) \right \}
\eea
Since $v_i$ forms a basis of $L$, the volume of the torus is given by,
\bea
{\rm Vol}(T_d) = (\det \cG)^\half
\eea
up to some factor of $2 \pi$ which will be of no interest here.

\subsubsection{A first look at $T$-duality}

The argument of the exponential may be decomposed as follows, 
\bea
-  { \pi \over \tau_2} (m + \tau n)^\dagger  (\cG +\cB) (m +  \tau n)
& = & 
- { \pi \over \tau_2} (m + \tau_1 n)^t \cG (m + \tau_1 n) 
\no \\ &&
- \pi \tau_2 n^t \cG n - 2 \pi i m^t \cB n
\eea
As a result, each term in the exponential is invariant under a shift in $\cB$ by an arbitrary anti-symmetric matrix with integer entries, or in components, 
\bea
\cB_{ij}  \to \cB_{ij}  + b_{ij} \hskip 1in b_{ij}=-b_{ji}\in \ZZ
\eea
This invariance is an immediate consequence of the fact that, for constant $B$,  the $B$-term  is topological and its values on the field configurations for the torus $T_d$ are quantized. The corresponding symmetry is the analogue of the $T$-transformation $\tau_1 \to \tau_1 + 1$  for $SL(2,\ZZ)$, and will provide some of the generators of the full $T$-duality group for $T_d$.

\sm

Next, we wish to obtain the analogue of the $S$-transformation. To do so, we Poisson resum in both $m,n$.
It is helpful to express the quadratic form combining $m,n$ into a single column matrix, so that, 
\bea
- { \pi \over \tau_2} (m + \tau n)^\dagger  (\cG +\cB) (m +  \tau n)
= 
- \pi \left ( \begin{matrix}m \cr n \cr \end{matrix} \right ) ^t \cM \left ( \begin{matrix} m \cr n \cr \end{matrix} \right ) 
\eea
where $\cM$ is given by,
\bea
\cM = {1 \over \tau_2}  \left (  \begin{matrix}  \cG &  & \tau_1  \cG + i \tau_2 \cB \cr && \cr
 \tau_1 \cG - i \tau_2 \cB & & (\tau_1^2 + \tau_2^2) \cG \cr \end{matrix} \right ) 
\eea
The inverse matrix may be expressed as follows,
\bea
\cM^{-1} = {1 \over \tau_2} \left ( \begin{matrix} 0 && I \cr && \cr -I && 0 \cr \end{matrix} \right )^t 
 \left (  \begin{matrix}  \tilde \cG &  & \tau_1  \tilde \cG + i \tau_2 \tilde \cB \cr && \cr
 \tau_1 \tilde \cG - i \tau_2 \tilde \cB & & (\tau_1^2 + \tau_2^2) \tilde \cG \cr \end{matrix} \right ) 
  \left ( \begin{matrix} 0 && I \cr && \cr -I && 0 \cr \end{matrix} \right )
 \eea
 where $\tilde \cG$ and $\tilde \cB$ are given by the $S$-transformation, 
 \bea
 \tilde \cG = \Big ( \cG + \cB^t \, \cG^{-1} \, \cB \Big )^{-1} 
 \hskip 1in 
 \tilde \cB = \cG^{-1} \cB \, \tilde \cG
 \eea
The action of $S^2$ is obtained by repeating the transformation, 
\bea
\hat \cG = \Big ( \tilde \cG + \tilde \cB^t \, \tilde \cG^{-1} \, \tilde \cB \Big )^{-1} 
\hskip 1in
\hat \cB = \tilde \cG^{-1} \tilde \cB \, \hat \cG
 \eea
and eliminating $\tilde \cB$ and $\tilde \cG$. The result is $\hat \cB=\cB$ and $\hat \cG = \cG$, so that the action of $S^2$ on $\cG, \cB$ is the identity.\footnote{Although in $SL(2,\ZZ)$ the matrix $S$ squares to $S^2=-I$, its action on $\tau$ reduces to the identity; the above statement is the direct analogue thereof.} We shall show shortly that  $S$ and the integer translations of $\cB$ generate the duality group $SO(d,d, \ZZ)$.

\subsubsection{Holomorphic block decomposition}

As in the case of the scalar field theory valued in the circle, it is instructive to perform a Poisson resummation only on the integers $m$, and expose the holomorphic structure in $\tau$. To do so, we start from the decomposition of the quadratic form given earlier, and we need the following Fourier transform,
\bea
&&
\int _{\RR^d} d^d x \, e^{-2 \pi i x^t m} \, \exp \left \{ 
- { \pi \over \tau_2} (x + \tau_1 n)^t \cG (x + \tau_1 n)  - 2 \pi i x^t \cB n \right \}
\no \\ && \qquad =
{ (\tau_2)^{{ d \over 2}} \over (\det \cG)^\half } \exp \Big \{ 
2 \pi i \tau   p_L^2 - 2 \pi i \bar \tau  p_R^2 \Big \}
\eea
where we obtain after minimal simplifications,
\bea
4 p_L^2 & = & (m+\cB n)^t \cG^{-1} (m+\cB n) + n^t \cG n + 2 m^t n
\no \\
4 p_R^2 & = & (m+\cB n)^t \cG^{-1} (m+\cB n) + n^t \cG n - 2 m^t n
\eea
Note that while the integers $n$ were  defined with an upper index, as were the old $m$, by contrast the Poisson resummation variable $m$ have lower indices. So, it is instructive to write out the above relations in terms of indices,
\bea
4 p_L^2 & = &  \Big (m_i+ \cB_{ii'}  n^{i'} \Big  ) (\cG^{-1})^{ij} \Big (m_j+ \cB_{jj'} n^{j'} \Big ) + n^i \cG_{ij}  n^j  + 2  m_i  n^i
\no \\
4 p_R^2 & = &  \Big (m_i+ \cB_{ii'}  n^{i'} \Big  ) (\cG^{-1})^{ij} \Big (m_j+ \cB_{jj'} n^{j'} \Big ) + n^i \cG_{ij}  n^j  - 2  m_i  n^i
\no
\eea
In evaluating the partition function, the factor ${\rm Vol} (T^d)$ cancels out, and we find, 
\bea
Z(L,\tau) = (\tau_2)^{{d \over 2}} \sum _{m,n \in \ZZ^d} \exp \Big \{ 2 \pi i \tau \, p_L^2 - 2 \pi i \bar \tau \, p_R^2 \Big \}
\eea
with $p_L^2, p_R^2$ functions of $m,n$ as defined above. The invariance under integer shifts of the matrix $\cB$ is now manifested by the fact that such a shift may be compensated in the sum by a shift in the integers $m$.

\subsubsection{$T$-duality in terms of the lattice}

The description of the torus with both an arbitrary lattice $L$ and an arbitrary metric $G$ is in fact redundant. By general linear transformation, we may always make the metric equal to the identity. Geometrically, this is equivalent to expressing the metric $\cG_{ij}$  in terms of an orthonormal frame $e_\mu {}^a$ and its inverse $e_a {}^\mu$ by 
\bea
\cG_{ij} = e_i {}^a e_j{}^b \delta _{ab}
\eea 
where $i$ are ``Einstein indices" and $a$ are \textit{frame indices}. The lattice $L$ with identity metric is then generated by the vector $ e_a {}^i$, while the dual lattice $L^+$ is generated by the dual vector $e_i {}^a$. Defining also the $B$-field in frame indices,
\bea
\mB _{ab} = \cB_{ij} \, e_a{}^i e_b{}^j
\eea
and defining the lattice and dual lattice vectors, 
\bea
\mu _a & = & m_i \, e_a{}^i
\no \\
\nu ^a & = & n^i \, e_i {}^a
\eea
we find that $p_L^2$ and $p_R^2$ may be interpreted as square of vectors, respectively given by,
\bea
p_L & = & \half \Big (   \mu + \mB \, \nu + \nu   \Big )
\hskip 1in \mu \in L^+
\no \\
p_R & = & \half \Big ( \mu + \mB \, \nu -  \nu    \Big )
\hskip 1in \nu \in L
\eea
with,
\bea
p_L\cdot p_L - p_R \cdot p_R = \mu \cdot \nu = m^t n \in \ZZ
\eea
To parametrize the space of all lattices, we proceed as follows. Given a lattice, it may be deformed continuously provided we maintain the condition $p_L\cdot p_L - p_R \cdot p_R  \in \ZZ$. This condition is invariant under the group $SO(d,d;\RR)$. But rotations under $SO(d)$ on both $p_L$ and $p_R$ are physically indistinguishable from the original lattice. Therefore, any lattice and its dual may be parametrized by a point in the coset space,
\bea
SO(d,d;\RR) / (SO(d;\RR) \times SO(d;\RR))
\eea
The dimension of this space is,
\bea
\half (2d)(2d-1) - 2 \times \half d(d-1) = d^2
\eea
This dimension is easily understood in the original formulation with arbitrary metric $G$ and $B$-field in a square torus: together they have $d^2$ components. 

\sm

Two different points in $SO(d,d;\RR) / (SO(d;\RR) \times SO(d;\RR))$ may correspond to the same lattice. 
We already know that this is the case when $\cB_{ij}$ are shifted by integers, and when the $S$ transformation is applied to both $\cG$ and $\cB$. The shift is realized in terms of an anti-symmetric matrix $K$ with integer entries, 
\bea
\gamma = \left ( \begin{matrix} I + K & - K \cr K & I - K \cr \end{matrix} \right )
\hskip 1in 
\left ( \begin{matrix} \tilde p_L \cr \tilde p_R \cr \end{matrix} \right ) =  \gamma
\left ( \begin{matrix} p_L \cr  p_R \cr \end{matrix} \right )
\eea
whose effect is to shift $\cB \to \cB + 2 K$. The matrix $\gamma$ is easily seen to belong to $SO(d,d;\RR)$, and to be  a translation or \textit{parabolic} element, such that,
\bea
\gamma ^n = \left ( \begin{matrix} I + nK & - nK \cr nK & I - nK \cr \end{matrix} \right )
\eea
The $S$-transformation is given by,
\bea
S = \left ( \begin{matrix} 0 & I \cr -I & 0 \cr \end{matrix} \right )
\eea
The space of inequivalent tori $T_d$ from the string point of view are thus parametrized by the double coset space,
\bea
SO(d,d;\ZZ) \backslash SO(d,d; \RR) / SO(d;\RR) \times SO(d;\RR)
\eea
In the case of heterotic string, the result extends to include the left-moving 16 dimensional torus, and the duality group is $SO(d, d+16;\ZZ)$.

\subsection{Rationality and complex multiplication}
\label{eq:CMandNLSM} 

In section \ref{sec:3dgrav} we introduced the torus partition function of a two-dimensional conformal field theory, e.g.  in (\ref{eq:CFTtoruspartfunct}).  A CFT is said to be \textit{rational} if the sum over characters is finite. In a sense, a rational CFT is one which is exactly solvable, and great effort has been expended towards obtaining a classification of such theories.
In this section, we ask when a $T_2$ non-linear sigma model is rational.

\subsubsection{The case $T_1$}

 To begin, let us consider the simpler case of an $S^1$ non-linear sigma model, with the circle having radius $R$. In this case, the partition function may be computed to be
\bea
Z(R, \tau) = {1 \over |\eta(\tau)|^2} \sum_{m,n \in \ZZ} q^{\half \left({m \over R} + {n R \over 2}\right)^2}\bar q^{\half \left({m \over R} -{n R \over 2}\right)^2}
\eea
This comes from (\ref{eq:compbosZ}) by choosing $\alpha' = 2$ and dressing with a factor of $|\eta(\tau)|^2$ to account for Virasoro descendants. To address the question of rationality, we must ask when this can be written as a sum over a finite number of characters. It turns out that this is possible if $R^2 \in \QQ$. Indeed, say that $R^2 = {2k \over \ell}$ for $k, \ell \in \ZZ$. In this case we can write the partition function as,
\bea
Z\left(\sqrt{{2k \over \ell}},\, \tau\right) ={1 \over |\eta(\tau)|^2}\sum_{m,n\in \ZZ} q^{{1 \over 4 k \ell} (m \ell + n k)^2} \overline q^{{1 \over 4 k \ell} (m \ell - n k)^2}
\eea
It will be useful to write 
\bea
m = \m + 2 k r \hspace{0.8 in} n = \n + 2 \ell s 
\eea
where $\m \in \{0,\dots, 2k -1\}$, $\n \in \{0,\dots, 2\ell-1\}$, and $r, s \in \ZZ$. 
We may then split the sums as
\bea
Z\left(\sqrt{{2k \over \ell}},\, \tau\right) ={1 \over |\eta(\tau)|^2}\sum_{\m =0}^{2k-1} \sum_{\n =0}^{2 \ell-1} \sum_{r, s \in \ZZ}q^{{1 \over 4 k \ell} (2 k \ell (r+s) + \m \ell + \n k)^2} \overline q^{{1 \over 4 k \ell} (2 k \ell (r-s) +\m \ell - \n k)^2}
\eea
We now again change summation variables to,  
\bea
\tilde{\m} = \m \ell + \n k \hspace{0.5 in} \tilde{\n} = \m \ell - \n k \hspace{0.5in} \rho = r+s \hspace{0.5 in} \sigma = r - s
\eea
By definition we have,
\bea
\label{eq:rattorus}
\tilde{\m} + \tilde{\n} \equiv 0 \,\,\,({\rm mod}\,\,2 \ell) \hspace{0.8 in}\tilde{\m} - \tilde{\n} \equiv 0 \,\,\,({\rm mod}\,\,2k)
\eea
Hence we can rewrite the partition function as a finite sum,
\bea
Z\left(\sqrt{{2k \over \ell}},\, \tau\right)  = \sum'_{\substack{\tilde{\m}, \tilde{\n} = 0,\dots,2k \ell - 1}} \chi_{\tilde \m}(\tau) \overline \chi_{\tilde \n}(\bar \tau)
\eea
where the primed sum denotes a sum subject to restrictions (\ref{eq:rattorus}). The result has been written in terms of the characters, 
\bea
\chi_{\tilde \m}(\tau) = {1 \over \eta(\tau)}\sum_{\rho \in \ZZ}q^{k \ell \left(\rho + {\tilde \m \over 2 k\ell } \right)^2}
\eea
 That these are the correct characters for $T_1$ is confirmed by noting that the compact boson at radius $R^2 = {2k \over \ell}$ is equivalent to a $\ZZ_\ell$ orbifold of the $U(1)_k$ current algebra theory.

\subsubsection{The case $T_2$}

We now proceed to the case of $M=T_2$. For the moment we will consider only tori that are trivial products of circles, with radii $R_1$ and $R_2$.  Then the CFT factorizes into two $S^1$ CFTs studied above, and the total CFT is rational if and only if the two components CFTs are. In other words, we must take $R_{1,2}^2 \in \QQ$. Say concretely that, 
\bea
R_1^2 = {2k_1 \over \ell_1} \hspace{0.8 in}R_2^2 = {2k_2 \over \ell_2} \hspace{0.5 in} k_i, \ell_i \in \ZZ
\eea
The complex structure of the spacetime torus, which we will denote as $\rho$ to avoid confusion with the complex structure $\tau$ of the worldsheet torus, is given by $\rho = i R_1 / R_2$ up to $SL(2, \ZZ)$ transformations. Hence  $\rho$ satisfies a quadratic equation with integer coefficients,
\bea
\ell_1 k_2 \rho^2 + \ell_2 k_1 = 0
\eea
This is precisely the condition that the spacetime $T_2$ admit complex multiplication. Though we have restricted ourselves to tori which are trivial products of circles here, one can argue by similar methods that for generic elliptic curves $\CC/L$, rationality of the CFT is tantamount to the presence of complex multiplication on $\CC/L$.

\subsection*{$\bullet$ Bibliographical notes}

Toroidal compactification of higher space-time dimensions  goes back to \cite{Scherk:1979zr} and was generalized to the Heterotic string in \cite{Narain:1985jj,Narain:1986am}. It  was used to map out the moduli space of $\mc=1$ unitary conformal field theories in \cite{Dijkgraaf:1987vp}.  Comprehensive overviews of toroidal compactification of the bosonic string may be found in the first volume of  Polchinski's book \cite{Polchinski:1998rq}, while toroidal compactification of the superstring, T-duality, and related topics are discussed in detail in the second volume of Polchinski's book  \cite{Polchinski:1998rr} as well as in the book by Kiritsis \cite{Kiritsis}. Further useful references on correlation functions for compact scalars, representations of current and Kac-Moody algebras and related topics may be found in \cite{DiFrancesco:1987gwq} and \cite{diFrancesco:1987qf}. For the contents of section \ref{eq:CMandNLSM}, we refer again to  \cite{Gukov:2002nw}, and to section \ref{sec:CM}  on complex multiplication of these lecture notes.

\newpage

\section{S-duality of Type IIB superstrings}
\setcounter{equation}{0}
\label{sec:IIB}

In this section, we shall draw together a number of different strands of inquiry addressed earlier  in these lecture notes.  We shall study the interplay between superstring amplitudes, their low energy effective interactions, Type~IIB supergravity, and the S-duality symmetry of Type~IIB superstring theory. We begin with a brief review of Type~IIB supergravity which, in particular, provides the massless sector of Type IIB superstring theory. We then discuss how the $SL(2,\RR)$ symmetry of Type IIB supergravity is reduced to the $SL(2,\ZZ)$ symmetry of Type IIB superstring theory via an anomaly. We conclude with a discussion of how the low energy effective interactions induced by string theory on supergravity may be organized in terms of modular functions and forms under this $SL(2,\ZZ)$ symmetry, and match the predictions provided by perturbative calculations of section \ref{sec:SA}.

\subsection{Type IIB supergravity}
\label{sec:IIBSUGRA}
In section \ref{sec:SA} we discussed the expansion of bosonic and superstring amplitudes in powers of the string coupling constant~$g_s$. While this expansion produces only an asymptotic series in~$g_s$, it has the advantage of being valid for all energy ranges. A different, and complementary, expansion of string theory is in powers of $\alpha'$, also referred to as the \textit{low energy expansion}. This is the expansion that was used for the string amplitudes in subsection~\ref{sec:12.7}. To leading order in this approximation, one omits the effects of all the massive string states, whose mass-squares are set  by the string scale $1/\alpha'$. The remaining massless states precisely make up the contents of the corresponding supergravity. 

\sm

To each 10-dimensional perturbative string theory, namely Type~I, Type~IIA, Type~IIB, and Heterotic $E_8 \times E_8$ and ${\rm Spin}(32)/\ZZ_2$ there is a corresponding supergravity. The Type~II theories have the maximal number of 32 supersymmetries, while the other three have 16 supersymmetries. We shall focus here on Type~IIB as its corresponding Type~IIB superstring theory has $SL(2,\ZZ)$ duality symmetry, which is the focus of these lecture notes. 

\sm

The  contents of Type IIB supergravity consist of the following fields,
\bea
G \, & \hskip 1in & \hbox{ the metric, whose excitations are the graviton}
\no \\
C_2 & \hskip 1in & \hbox{ the complex 2-form potential}
\no \\
\tau \, & & \hbox{ the complex axion/dilaton field}
\no \\
C_4 && \hbox{ the real 4-form potential, whose field strength is self-dual}
\no \\
\psi \, && \hbox{ chirality $+$ Weyl spinor gravitino}
\no \\
\lambda \, && \hbox{ chirality $-$ Weyl spinor dilatino}
\eea
The fields $G, C_2, \tau$, and $C_4$ are bosonic fields, while $\psi, \lambda$ are fermions. Here we shall concentrate on the properties and field equations of the bosonic fields. The field strength of $C_2$ is $F_3=dC_2$, while the field strengths of $\tau$ and $C_4$ are given as follows,
\begin{align}
P & =  f^2 \, dB  & B & = (1+i\tau)/(1-i\tau)
\no \\
Q & =  f^2 \, \Im (B d \bar B) & f^2 & = (1-|B|^2)^{-1}
\no \\
F_5 & =  d C_4 + \tfrac{i}{16} \big ( C_2  \wedge \bar F_3 - \bar C_2 \wedge F_3 \big ) & \star F_5 & = F_5
\end{align}
The axion-dilaton field $\tau$ is valued in the hyperbolic upper half-plane $\tau_2=\Im \tau >0$.  The  field equation for $F_5$ coincides with its Bianchi identity $dF_5 =\tfrac{i}{8} F_3  \wedge \bar F_3$ in view of its self-duality relation. In terms of the components of the fields $P,Q,F_5$, and $K  = f(F_3 - B \bar F_3) $, 
\begin{align}
P & =P_\mu \, dx^\mu & K &  = \tfrac{1}{3!} K_{\mu \nu \rho} \, dx^\mu \wedge dx^\nu \wedge dx^\rho
\no \\
Q& =Q_\mu \, dx^\mu & F_5 & = \tfrac{1}{5!}F_{5\mu \nu \rho \sigma \tau} \, dx^\mu \wedge dx^\nu \wedge dx^\rho \wedge dx^\sigma \wedge dx ^\tau
\end{align}
the remaining field equations are conveniently expressed as follows,
\bea
0 & = & \nabla ^\mu P_\mu - 2 i Q^\mu P_\mu + \tfrac{1}{24} K_{\mu \nu \rho} K^{\mu \nu \rho}
\no \\
0 & = & \nabla ^\rho K_{\mu \nu \rho} - i Q^\rho K_{\mu \nu \rho} - P^\rho \bar K_{\mu \nu \rho} +\tfrac{2i}{3} F_{5 \mu \nu \rho \sigma \tau} K^{\rho \sigma \tau}
\no \\
0 & = & R_{\mu \nu} -P_\mu \bar P_\nu - \bar P_\mu P_\nu - \tfrac{1}{6} (F_5^2)_{\mu \nu} 
- \tfrac{1}{4} \Re( K_\mu {}^{\rho \sigma} \bar K_{\nu \rho \sigma} ) + \tfrac{1}{48} G_{\mu \nu} K^{\rho \tau \sigma} \bar K_{\rho \tau \sigma}
\eea
where $R_{\mu \nu}$ is the Ricci tensor for the metric $G_{\mu \nu}$.   The axion-dilaton field $\TAU$ may be related to the axion field $\chi$ and the dilaton field  $\Phi$,
\bea
 \TAU = \chi + i e^{- \Phi}
\eea
In view of the self-duality condition on $F_5$, there does not exist a Lorentz-covariant action from which the field equations of Type IIB supergravity may be deduced using the standard variational principle. Instead one may define the  following action,\footnote{Our notation for the contraction of two rank $n$ anti-symmetric tensors is $F_n \cdot F'_n = \tfrac{1}{n!} F_{n \mu_1 \cdots \mu_n} F_n ^{\prime \mu_1 \cdots \mu_n}$.} 
\bea
S_{{\rm IIB}} & = & { 1 \over 2 \kappa_{10}^2} \int d\mu_G \left ( 
R - { \p_\mu \TAU \p^\mu \bar \TAU \over 2 \TAU_2^2} 
-   { (F_3^1 - \TAU F_3^2 )\cdot (F_3^1 - \bar \TAU F_3^2 ) \over 2 \, \TAU_2} 
- {1 \over 4} F_5 \cdot F_5 \right )
\no \\ &&
- {2 \over \kappa_{10}^2} \int C_4 \wedge F_3^1 \wedge F_3^2
\eea
Here $d\mu_G$ is the invariant volume form for the metric $G$; $\kappa_{10}^2 $ is related to the 10-dimensional Newton constant; and  we have $F_3 = F_3^1 + i F_3 ^2$ and $\TAU=\TAU_1 + i \TAU_2$ for real field components $F_3^1, F_3^2, \TAU_1, \TAU_2$. Importantly, the action $S_{{\rm IIB}}$ should be considered for a field $F_5$ that  is \textit{unconstrained by the self-duality condition}. The field equations for Type IIB supergravity may be obtained from this action by varying the \textit{unconstrained field $F_5$}, along with all the other fields,  and only subsequently imposing the self-duality condition $\star F_5 = F_5$.

\subsection{From $SL(2,\RR)$ to $SL(2,\ZZ)$}

The Type IIB supergravity field equations are invariant under the group $SL(2,\RR)$,  under which the metric $G$ and the anti-symmetric fields $C_4$ and $F_5$ are invariant, and the  fields $P, Q, K, \tau$ and $F_3=F_3^1+iF_3^2$  transform  as follows,
\begin{align}
P & \to  P'= e^{2 i \theta} P  & \tau & \to   \tau' = (a \tau +b)/(c \tau+d)  
\no \\
Q & \to  Q' = Q + d \theta & F_3^1 & \to (F_3^1)'= a F_3^1 + b F_3^2
\no \\
K & \to  K' = e^{i \theta } K & F_3^2 & \to (F_3^2)'= c F_3^1 + d F_3^2
\end{align}
The variable $\theta$ is given by, 
\bea
e^{i \theta} = \left ( { (\tau+i)  ( a \bar \tau -i c \bar \tau +b-id  )  \over (\bar \tau-i) \big ( a \tau + ic \tau +b+id \big )}  \right )^\half
\eea
The axion-dilaton takes values in the coset $SL(2,\RR)/U(1)$ and the phase $e^{i \theta}$ performs a field-dependent $U(1)$ rotation. The gravitino field $\psi$ and the dilatino field $\lambda$ transform with $U(1)$ phase factors under $SL(2,\RR)$, given by $\psi \to \psi' = e^{ i \theta /2} \psi $ and $\lambda \to \lambda ' = e^{3i \theta /2} \lambda$. 

\sm

The $SL(2,\RR) $ symmetry discussed above is a symmetry of the classical supergravity field equations, as well as of the action $S_{{\rm IIB}}$ and the self-duality condition $\star F_5 = F_5$. A classical symmetry cannot always be promoted to a symmetry of the corresponding quantized field theory, in which case the symmetry is said to suffer an \textit{anomaly}. Phase rotations on chiral fermions often suffer anomalies, and this is also the case here with the field-dependent $U(1)$ transformations acting on the gravitino and dilatino chiral fermions. A treatment of anomalies is beyond the scope of this paper, so we shall just state the result. 

\sm

It turns out that the continuous $SL(2,\RR)$ symmetry is broken by the $U(1)$-anomaly, but the quantum theory retains a discrete symmetry, under which the axion field is shifted by an integer $ \chi \to \chi'= \chi + b$ with $b \in \ZZ$. The discrete nature of the remaining symmetry is due to the fact that the axion field couples to the topological charge density of the $D_{-1}$ instantons in Type IIB superstring theory, whose total charge is quantized. The shift in the axion field $ \chi \to \chi' = \chi + b$ with $b \in \ZZ$ that survives quantization also shifts the complex axion-dilaton field $\TAU \to \TAU' = \TAU+b$, and is supplemented by the transformation $\TAU \to -1/\TAU$. Together these transformations generate the discrete group $SL(2,\ZZ)$. This symmetry extends to a symmetry of the entire Type IIB superstring theory.

\sm

This phenomenon is actually familiar from an extension of the Standard Model in which an axion field is included. Continuous shifts in the axion field again suffer an anomaly but discrete shifts survive. In this case the discrete nature of the shift is due to the coupling of the axion field to the topological charge density of Yang-Mills instantons whose topological charge is quantized in terms of the second Chern class of the gauge group.

\subsection{Low energy effective interactions}

Supergravity, as a classical theory, is valid for all values of the string coupling. Thus, it permits  investigations into certain strong coupling phenomena such as solitonic states, NS branes and D-branes. Supergravity has been an invaluable tool in the search for semi-realistic compactifications of superstring theory, such as on tori, oribifolds, and Calabi-Yau spaces. 

\sm

However, supergravity alone is  not a consistent quantum theory. In 10 space-time dimensions, it exhibits UV divergences starting at one loop. The UV convergence situation is somewhat improved by lowering the dimension of space-time. In four space-time dimensions the four-graviton amplitude is UV convergent up to five loops, but it is likely that non-renormalizable UV divergences start occurring at seven loops, rendering the quantum theory inconsistent, or more accurately, incomplete. Its UV completion is precisely superstring theory. From the vantage point of superstring theory, supergravity should be thought of as an \textit{effective low energy  field theory} that captures the dynamics of the massless states of Type IIB string theory, to leading order  in the $\alpha'$ expansion. 

\sm

\begin{figure}[htb]
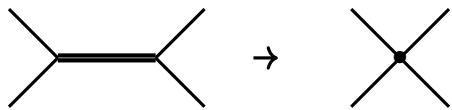

\begin{center}
\tikzpicture[scale=0.65]
\begin{scope}[xshift=10cm]
\draw [very thick] (-1,1) to (0,0) ;
\draw [very thick] (-1,-1) to (0,0) ;
\draw [ultra thick] (0,0.05) to (2,0.05) ;
\draw [ultra thick] (0,-0.05) to (2,-0.05) ;
\draw [very thick] (2,0) to (3,1) ;
\draw [very thick] (2,0) to (3,-1) ;
\draw [very thick,->] (4,0) to (4.5,0) ;
\draw [very thick] (6,1) to (7,0) ;
\draw [very thick] (6,-1) to (7,0) ;
\draw [very thick] (7,0) to (8,1) ;
\draw [very thick] (7,0) to (8,-1) ;
\draw (7,0) node{$\bullet$};
\end{scope}
\endtikzpicture
\caption{\textit{The exchange of a massive string state, indicated by the thick line in the left panel, induces a local effective interaction indicated by a thick dot in the right panel. }\label{13.fig:4}} 
\end{center}
\end{figure}

The validity of an effective field theory may be extended beyond its leading order contributions. Viewed as an effective field theory of Type IIB superstring theory, the corresponding supergravity may be supplemented with contributions of higher order in $\alpha'$. Such contributions are due to the effects of massive string states whose mass, we recall, is of order $1/\sqrt{\alpha'}$ and are referred to as \textit{effective interactions} In the approximation where the momenta used to probe string amplitudes are small compared to $1/\sqrt{\alpha'}$, the effective interactions are local, as is illustrated  in Figure \ref{13.fig:4} for the exchange of a massive string state. Although these string induced effective interactions are highly suppressed they provide systematic corrections to supergravity in an expansion in powers of $\alpha'$.

The space-time we observe has dimension four and is approximately flat when its curvature is measured in units of the Planck scale. Therefore, if the space-time of superstring theory is to be 10 then the six extra dimensions must have  radii that are smaller than the smallest length scales accessible to experiment today. Physically realistic superstring theories are often based on space-times of the form $\RR^{10-d} \times M_d$ where $M_d$ is a compact manifold or orbifold and the physically observed dimension of space time corresponds to $d=6$. For the case where $M_d$ is a flat torus, superstring perturbation theory continues to be well-understood.  When $M_d$ is an orbifold of a torus,  the space $M_d$ is flat away from isolated point-like singularities, and string theory on toroidal orbifold spaces  still lends itself to reasonably  explicit solutions. However, when $M_d$ is an arbitrary curved manifold, the predictions of superstring theory are much more difficult to obtain. This is the case even when space-time is a curved maximally symmetric space such as $AdS_5 \times S^5$, for which the quantization of the superstring is still largely an unresolved problem. An important exception is when the curvature of $M_d$ is uniformly small compared to the Planck scale, a case that is referred to as the large radius expansion, and that we shall discuss next.

\sm

Restricting, for example, to purely gravitational effects, the supergravity Lagrangian reduces to the Einstein-Hilbert term given by the Ricci scalar for the space-time metric. In Type II superstring theory, for example, the lowest order $\alpha'$ correction to the  Einstein-Hilbert action is given by a term which we symbolically represent by $\cR^4$ where $\cR$ stands for the Riemann tensor, and the contribution to the action is obtained via a special contraction of the four factors, consistent with space-time supersymmetry, as will be explained below. This effective interaction arises at order $(\alpha')^3$. Higher order effective interactions involve more derivatives and higher powers of $\alpha'$ and are symbolically represented by $D^{2k} \cR^4$ where $D$ represents a covariant derivative, again suitably contracted.

\subsection{$SL(2, \ZZ)$ duality in Type IIB superstring theory}

Type IIB superstring theory exhibits $SL(2, \ZZ)$ duality symmetry which leaves the space-time metric $G_E$  in Einstein frame and the self-dual five form invariant, and transforms the 3-form flux field strengths linearly into one another, 
\bea
\left ( \bma F_3^1 \cr F_3^2 \ema \right ) \to 
 \left( \begin{matrix} a & b \cr c & d \cr  \end{matrix} \right) \left ( \bma F_3^1 \cr F_3^2 \ema \right )
  \hskip 1in 
\left ( \begin{matrix} a & b \cr c & d \cr  \end{matrix} \right ) \in SL(2,\ZZ)
\eea
As for the axion-dilaton field obtained by combining the dilaton field $\Phi$ and the axion field $\chi$ into a complex scalar field $\TAU = \chi + i e^{-\Phi}$, transformations under $SL(2,\ZZ)$ occur as follows,
\bea
\label{8b1}
\TAU \to { a \TAU + b \over c \TAU + d} 
\eea
The transformation properties of the fermionic fields are more complicated, and will not be reproduced here. In fact, when one accounts for the fermionic fields the true duality symmetry of Type IIB is \textit{not} $SL(2, \ZZ)$, but rather an extension of it known as the meta-linear group $ML^+(2, \ZZ)$; see \cite{Tachikawa:2018njr} for details. This subtlety will not affect the purely bosonic analysis below.

Being a symmetry of Type IIB string theory, $SL(2,\ZZ)$ must be a symmetry of the low energy effective action of the theory. The effective interactions accessible from the four-graviton amplitude are of the form $D^{2k} \cR^4$, as explained above, but they have coefficients that depend upon the vacuum expectation value of the dilaton $g_s = e^\Phi$ and the axion $\chi$. Expressing the effective action in terms of the Einstein frame metric $G_E$, with associated Riemann tensor $\cR_E$, and the vacuum expectation value $\tau$, we find to  order $(\alpha ')^7$,
\bea
\label{4b2}
\int d\mu_G  \Big ( \cE_0 (\TAU) \cR_E^4 + \cE_4 (\TAU) D^4 \cR_E^4 +\cE_6 (\TAU) D^6 \cR_E^4 + \cE_8 (\TAU) D^8 \cR_E^4 + \cO((\alpha ')^8) \Big )
\eea
where the coefficients $\cE_{2k}(\TAU)$ are real-valued scalar functions of $\TAU$. Derivatives of $\tau$ may occur as well but will be systematically omitted here for simplicity.  Since the action must be $SL(2,\ZZ)$-invariant and, in Einstein frame, $G_E$ and $\cR_E$ are invariant as well, we see that each coefficient $\cE_{2k}(\TAU)$ must be invariant, 
\bea
\cE_{2k} \left (  { a \TAU + b \over c \TAU + d}  \right ) = \cE_{2k} (\TAU)
\hskip 1in 
\left ( \begin{matrix} a & b \cr c & d \cr  \end{matrix} \right ) \in SL(2,\ZZ)
\eea
a property which makes them into \textit{non-holomorphic modular functions}.

\sm

String perturbation theory calculations are carried out in the string frame metric which we shall denote by $G_{\mu \nu}$ and which is related to the Einstein frame metric by $G_{\mu \nu} = e^{\Phi /2} G_{E \mu \nu}$. Converting the expression for the effective action of (\ref{4b2}) to string frame may be done by using the following relations (for constant values of $\tau$), 
\bea
\sqrt{G_E} \, \cE_{2k} (\TAU) \, D_E ^{2k} \cR_E ^4 = e^{ (k-1)\Phi/2} \, \sqrt{G} \, \cE_{2k} (\TAU) \, D ^{2k} \cR ^4
\eea
where $D$ and $\cR$ now stand respectively for the covariant derivative and Riemann tensor in the string frame. Now let us carry out the following exercise: express combinations of the non-holomorphic Eisenstein series with effective interactions $D_E^{2k} \cR_E ^4$ in the Einstein frame such that the leading perturbative contribution in string frame is tree-level,
\bea
\half \pi^{3/2} \sqrt{G_E} \, E_{{ 3 \over 2}} (\TAU)  \cR_E ^4 & = &
e^{-2 \Phi} \zeta (3) \cR^4 +{ \pi^2 \over 3} \cR^4 + \cdots
\no \\
\half \pi^{5/2} \sqrt{G_E} \, E_{{ 5 \over 2}} (\TAU) D_E^4 \cR_E ^4 & = &
e^{-2 \Phi} \zeta (5) D^4 \cR^4 +{ 2 \pi^4 \over 135} e^{2 \Phi} D^4 \cR^4 + \cdots
\eea
Comparing the terms in the above table which have the $\Phi$-dependence of tree-level perturbative contributions with the known result of (\ref{12.loen}) at genus one and (\ref{12.gen2a}) and (\ref{12.gen2b})  at genus two, we see that their dependence on odd zeta-values is precisely reproduced. This observation would suggest that the coefficients $\cE_{2k}(\TAU)$ in (\ref{4b2}) may be given by the  non-holomorphic Eisenstein series.

\sm

\subsection{Eisenstein series from supersymmetry and S-duality}

We will now show how supersymmetry and S-duality can be used to prove the identifications between $\cE_{2k}(\tau)$ and the appropriate Eisenstein series. We will begin with the case of $\cE_{0}(\tau)$, namely the coefficient of $\cR^4$.  Because the metric $G$ is in the same supermultiplet as all of the other supergravity fields reviewed in section \ref{sec:IIBSUGRA}, supersymmetry requires that  the $\cR^4$ term in the low energy effective action be accompanied by a number of analogous terms related by supersymmetry. To make this more concrete, we first introduce some notation. As we have discussed above, the complex 2-form potential $C_2$, and consequently the corresponding field strength $F_3$,  transforms as a doublet under $SL(2, \ZZ)$. It is useful to package these field strengths into an $SL(2,\ZZ)$ singlet via, 
\bea
M = {e^{i \phi}\over \sqrt{- 2 i \tau_2}}(F_3^{1} - \tau F_3^{2}) 
\eea
 where the superscript on $F_3$ is a doublet index. We further define the ``super-covariantized" combination,
 \bea
 \widehat M_{\mu \nu \rho} = M_{\mu \nu \rho} - 3 \overline{\psi}_{[\mu} \gamma_{\nu \rho]}\lambda - 6 i \overline{\psi}^*_{[\mu}\gamma_\nu \psi_{\rho]}
 \eea
obtained by dressing $M$ with appropriate combinations of the gravitino and dilatino fields $\psi$ and $\lambda$. The combination $\widehat M$ has the property that under supersymmetry transformations it does not contain derivatives of the transformation parameter. 

\sm

In terms of this supercovariant field strength, the supersymmetric completion to the $\cR^4$ term can be written as \cite{Green:1998by}, 
\bea
\label{eq:S3def}
&\vphantom{.}&S^{(3)} = \int d \mu_G \left(\cE_{0}^{(12,-12)} \lambda^{16} + \cE_{0}^{(11,-11)} \widehat M \lambda^{14} + \dots + \cE_{0}^{(8,-8)} \widehat M^8 \right.
\no\\
&\vphantom{.}&\hspace{1.5in}\left. + \dots +   \cE_{0}^{(0,0)}\cR^4 + \dots + \cE_{0}^{(-12,12)} (\lambda^*)^{16}\right)
\eea
where $\lambda^{16}$ and $\widehat M \lambda^{14}$ (which will be the main terms of interest to us below) are short for the following contractions,
\bea
\lambda^{16} &=& {1 \over 16!} \eps_{a_1\dots a_{16}} \lambda^{a_1} \dots \lambda^{a_{16}}
\no\\
\widehat M \lambda^{14} &=& {1 \over 14!} \widehat M_{\mu \nu \rho} (\gamma^{\mu \nu \rho} \gamma^0)_{a_{15} a_{16}} \eps_{a_1\dots a_{16}} \lambda^{a_1} \dots \lambda^{a_{14}}
\eea
The $SL(2,\ZZ)$ symmetry of Type IIB demands that the coefficient functions $\cE_{0}^{(w,-w)}(\tau)$ be modular forms of holomorphic weight $w$ and anti-holomorphic weight $-w$. The coefficient function $\cE_{0}^{(0,0)}(\tau) = \cE_{0}(\tau)$ is the original modular invariant coefficient function of $\cR^4$. The coefficient functions are furthermore required by supersymmetry to satisfy
\bea
\label{eq:relbetweenEww}
\cE_{0}^{(w,-w)}(\tau) = D_{w-1}\dots D_0 \cE_{0}(\tau)
\eea
with $D_w = i \tau_2 \left( {\p \over \p \tau} - i {w \over 2 \tau_2} \right)$ the covariant derivative mapping weight $(w,w')$ modular forms to weight $(w+1, w'-1)$ modular forms. Because of these relations,  constraints on $\cE_{0}^{(w,-w)}$ for any $w$ can imply non-trivial constraints on $\cE_{0}$. We will now see how to use supersymmetry to obtain differential constraints on $\cE_{0}^{(12,-12)}$ and $\cE_{0}^{(11,-11)}$ in particular, which will suffice to prove that $\cE_{0}$ is, up to normalization, equivalent to $E_{3\over 2}(\tau)$. 

\sm

Let us begin by schematically writing the low-energy effective action as 
\bea
S = {1 \over (\alpha')^4} \left[ S^{(0)} + (\alpha')^3 S^{(3)} + (\alpha')^4 S^{(4)}+ (\alpha')^5 S^{(5)} + \dots  \right]
\eea
with $S^{(3)}$ in particular given in (\ref{eq:S3def}) above. Likewise, a generic SUSY transformation $\delta$ can be expanded in powers of $\alpha'$ as 
\bea
\label{eq:SUSYvarexp}
\delta = \delta^{(0)} +  (\alpha')^3 \delta^{(3)} + (\alpha')^4 \delta^{(4)} + (\alpha')^5 \delta^{(5)} + \dots
\eea
Requiring that the action has supersymmetry implies that we must have 
\bea
\delta^{(0)} S^{(0)} = \delta^{(0)} S^{(3)} + \delta^{(3)} S^{(0)}  = \delta^{(0)} S^{(5)} + \delta^{(5)} S^{(0)} = \dots = 0
\eea
on the equations of motion. 

\sm

We now consider the first two terms in the action, 
\bea
\cL^{(3)} = \cE_{0}^{(12,-12)} \lambda^{16} + \cE_{0}^{(11,-11)} \widehat M \lambda^{14} 
\eea
A crucial feature of these two terms is that they are related by a subset of the SUSY transformations that do not mix with any of the other terms at this order, though we will have to keep track of variations of the lowest order action $S^{(0)}$ by $\delta^{(3)}$. It is a straightforward exercise to find the lowest-order $\delta^{(0)}$ supersymmetry variation of $\cL^{(3)}$. Denoting the supersymmetry parameter by $\eps$ and keeping only terms proportional to $\lambda^{16} \psi_\mu^*\eps$ gives 
\bea
\delta^{(0)} \cL^{(3)} |_{\lambda^{16} \psi_\mu^* \eps} = - 8i \left( \cE_{0}^{(12,-12)}(\tau) + 108 D_{11} \cE_{0}^{(11,-11)}(\tau)\right)
\eea
On the other hand, the $\delta^{(3)}$ variation of $S^{(0)}$ cannot produce any such term. Hence we require $\delta^{(0)} \cL^{(3)}|_{\lambda^{16} \psi_\mu^* \eps} = 0 $ alone, which demands
\bea
\label{eq:SUSYconst1}
D_{11} \cE_{0}^{(11,-11)}(\tau) = - {1\over 108} \cE_{0}^{(12,-12)} (\tau)
\eea
Further constraints can be obtained by considering the term in the supersymmetry variation proportional to $\lambda^{16} \lambda^* \eps^*$. In this case the computation is slightly more involved since now there \textit{is} a contribution from the $\delta^{(3)}$ variation of $S^{(0)}$. The analysis will not be reproduced here, but the result is the following constraint
\bea
\label{eq:SUSYconst2}
\overline{D}_{-12}\, \cE_{0}^{(12,-12)}(\tau) + 3240\, \cE_{0}^{(11,-11)}(\tau) - 90 g = 0
\eea
where $g(\tau, \overline \tau)$ is an unknown function. A final constraint may be obtained by demanding closure of the SUSY algebra on $\lambda^*$, giving 
\bea
\label{eq:SUSYconst3}
32 D_{11} g = \cE_{0}^{(12,-12)}(\tau)
\eea
We have now obtained three constraints (\ref{eq:SUSYconst1}), (\ref{eq:SUSYconst2}), and (\ref{eq:SUSYconst3}). Combining them, we may derive a Laplace eigenvalue equation for $\cE_{0}^{(12,-12)}$, 
\bea
\Delta \,\cE_{0}^{(12,-12)}(\tau)= \left(-132 + {3\over 4}\right) \cE_{0}^{(12,-12)}(\tau)
\eea
where we have noted that, when acting on weight-$(12,-12)$ modular functions, the Laplacian is given by $\Delta = 4 D_{11} \overline{D}_{-12}$.

\sm

It remains only to translate this differential constraint on $\cE_{0}^{(12,-12)}(\tau)$ to a differential constraint on $\cE_{0}(\tau)$, which can be done using the relation (\ref{eq:relbetweenEww}) between the two. This  gives the following Laplace eigenvalue equation for $\cE_{0}(\tau)$,
\bea
\Delta \cE_{0}(\tau) = {3\over 4}\cE_{0}(\tau)
\eea
Together with the asymptotic expansion obtained from string perturbation theory, this equation uniquely fixes $\cE_{0}(\tau)$ to be the non-holomorphic Eisenstein series $E_{3\over 2}(\tau)$. Indeed, we saw already in (\ref{eq:EisLapl}) that the Eisenstein series satisfy precisely such eigenvalue equations. 

\sm

We next briefly outline the analogous calculation for the $D^4 \cR^4$ term. As before, the first step is to identify a subset of the terms appearing in $S^{(5)}$ which mix with each other, but not with other terms, under some subset of the supersymmetry transformations. In the current case, a judicious choice is the following, 
\bea
\cL^{(5)} = \lambda^{16}\widehat M^4 \cE_4^{(14,-14)} + \lambda^{15} \gamma^\mu \psi_\mu^* \widehat M^4 \cE_4^{(13,-13)} + \lambda^{16} \widehat M^2 \widehat M_{\rho_1 \rho_2 \rho_3}\widehat M^{\rho_1 \rho_2 \rho_3} \widetilde{\cE}_4^{(13,-13)}
\eea
As before, modular invariance demands that the coefficient functions $\cE_4^{(w,-w)}$ be modular forms of weight-$(w,-w)$, with $\cE_4^{(0,0)} = \cE_4$ the coefficient of $D^4 \cR^4$ itself. Note that there are two coefficient functions of weight-$(13,-13)$ above, which we have denoted by $\cE_4^{(13,-13)}$ and $\widetilde{\cE}_4^{(13,-13)}$. We now proceed by again checking invariance under supersymmetry. Considering the term proportional to $\lambda^{16} \psi_\mu^* \eps$, we find only the contribution from $\delta^{(0)} \cL^{(5)}$, and demanding that this vanishes gives the constraint
\bea
2 D_{13} \, \cE_4^{(13,-13)} - 11 \, \cE_4^{(14,-14)} = 0
\eea
Considering instead the term proportional to $\lambda^{16} \lambda^* \eps^*$, we obtain both a contribution from the $\delta^{(0)}$ variation of $\cL^{(5)}$ as well as the $\delta^{(5)}$ variation of $S^{(0)}$. Demanding the vanishing of the combination gives the constraint
\bea
2 \overline{D}_{-14} \, \cE_4^{(14,-14)}+ 15 \, \cE_4^{(13,-13)} 
- {9 i \over 16} \, \widetilde{\cE}_4^{(13,-13)} - 1080 i \, g_1 - {3\over 4} i \, g_2 = 0 
\eea
with $g_1(\tau, \overline{\tau})$ and $g_2(\tau, \overline{\tau})$ undetermined functions. Closure of the supersymmetry algebra on $\lambda^*$ further gives rise to the following three equations, 
\bea
-192 i \, D_{13} g_1 = \cE_4^{(14,-14)} \hspace{0.3 in} -108 \, g_1 
= \widetilde \cE_4^{(13,-13)} \hspace{0.3 in} i \, g_2 + 191 i g_1 = \half  \cE_4^{(13,-13)}
\eea
These constraints can then be combined to obtain a Laplace eigenvalue equation for $\cE_4^{(14,-14)}$, namely 
\bea
\Delta \cE_4^{(14,-14)} = \left(-182 + {15 \over 4} \right) \, \cE_4^{(14,-14)}
\eea
Recalling that $\cE_4^{(14,-14)} = D_{13} \dots D_0 \cE_4$, we conclude that $\cE_4$ satisfies the equation 
\bea
\Delta \cE_4(\tau) = {15 \over 4}\, \cE_4(\tau)
\eea
This eigenvalue equation, together with the asymptotics at the cusp, fix $\cE_4$ to be equal to $E_{5/2}(\tau)$ up to an overall constant factor. 

\sm

Finally, for the coefficient of the $D^6 \cR^4$ term, analogous steps can again be carried out, though one now obtains an \textit{inhomogeneous} Laplace eigenvalue equation, 
\bea
\label{eq:D6R4diffeq}
(\Delta - 12) \cE_{6}(\tau) = - 6 \pi^3 E_{3/2}(\tau)^2
\eea
the solution to which is no longer an Eisenstein series. No closed form solution is known to this differential equation, but it is possible to get a closed form expression for the Laurent polynomial $\cE_6^{(0)}$ of $\cE_6$, namely the portions of $\cE_6$ which are power-law in $\tau_2$ and independent of $\tau_1$. Fortunately, the Laurent polynomial pieces are the most interesting for the present purposes, since they are the ones which can be compared to calculations in string perturbation theory. 

\sm

To obtain the Laurent polynomial $\cE_6^{(0)}(\tau)$ we begin by inserting the Fourier series of $E_{3/2}(\tau)$, given in (\ref{4.d12}), into (\ref{eq:D6R4diffeq}) and keeping only the Laurent polynomial pieces,
\bea
(\Delta - 12) \cE_{6}^{(0)}(\tau) &=& - 6\left(\left(2 \zeta(3) \tau_2^{3/2} + {2 \over 3}\pi^2 \tau_2^{-1/2}\right)^2 \right.
\no\\
&\vphantom{.}& \hspace{0.5 in} \left.+ 64 \pi^2 \tau_2 \sum_{N \neq 0} |N|^{-2} \sigma_2(|N|)^2 K_1(2 \pi \tau_2 |N|)^2 \right) 
\eea
Because the Laplacian is given by $\Delta = \tau_2^2 (\p_{\tau_1}^2 + \p_{\tau_2}^2)$, and hence the derivatives $\p_{\tau_2}^2$ are dressed with factors of $\tau_2^2$, the function $\cE_6^{(0)}$ must be a polynomial in $\tau_2$ with powers being those appearing on the right side, namely $\tau_2^3$, $\tau_2$, and $\tau_2^{-1}$, together with the powers solving the homogeneous equation $(\Delta - 12) f = 0$, namely $\tau_2^{-3}$ and $\tau_2^4$. The term $\tau_2^4$ can be ruled out on physical grounds, since there should be no term more singular than the tree-level $\tau_2^3$ term. On the other hand, the $\tau_2^{-3}$ contribution corresponds to a three-loop contribution which can be (and indeed is) non-zero. 

\sm

The general Ansatz for $\cE_{6}^{(0)}(\tau)$ is then, 
\bea
\cE_{6}^{(0)}(\tau) = a_1 \tau_2^3 + a_2 \tau_2 + a_3 \tau_2^{-1} + a_4 \tau_2^{-3}
\eea 
By plugging into (\ref{eq:D6R4diffeq}) we may uniquely fix $a_1, a_2,a_3$, giving,
\bea
\cE_{6}^{(0)}(\tau) = 4 \zeta(3)^2 \tau_2^3 + {4\over 3}\pi^2 \zeta(3) \tau_2 + {4 \pi^4 \over 15} \tau_2^{-1} + a_4 \tau_2^{-3}
\eea
These match with the expectations from string perturbation given in (\ref{8a5}), (\ref{12.loen}), and (\ref{12.gen2b}) at tree-level, one-loop, and two-loops, respectively. As for the three-loop coefficient $a_4$, this cannot be fixed by simply inserting the Ansatz into (\ref{eq:D6R4diffeq}), since this term already solves the homogenous equation. Instead, this coefficient was fixed in \cite{Green:2005ba} by multiplying both sides of the Laplace equation by $E_4(\tau)$ and integrating over the fundamental domain, using the results presented in Lemma \ref{12.Eisint}. The final result is 
\bea
a_4 = {8 \pi^6 \over 8505}
\eea 
We note in closing that beyond $D^6 \cR_4$, the terms in the low energy effective action are no longer protected by supersymmetry, and thus we do not expect to be able to extend the above analysis to study them.

\sm

\subsubsection{Non-renormalization Theorems}
\label{sec:nonrenorm}

Perhaps the most striking implications of the modular structure of the $\cR^4$, $D^4 \cR^4$, and $D^6 \cR^4$ effective interactions is that they make predictions for the amplitudes at all orders of perturbation theory, i.e. all genera. From the Fourier series expansion of the functions $E_{3 \over 2}(\TAU) $ we see that, aside from the leading perturbative contribution, there is only one other perturbative contribution, given by the second term, and behaving as $\TAU_2 ^{-\half}$. In the string frame, this gives a contribution of order $\TAU_2^0$ which corresponds to genus one. This means that from superstring perturbation theory, the $\cR^4$ effective interaction receives contributions from genus zero and genus one only. There are no contributions to genus 2 and higher. Similarly, for the $D^4 \cR^4$ effective interaction, there is a genus zero contribution and the only other contribution behaves as $\TAU_2^{- {3 \over 2}} $ in the Einstein frame, or $\TAU_2^{-2}$ in the string frame. Hence there is no genus one correction to the $D^4 \cR^4$ effective interaction but there is a non-zero genus two contribution. There are no higher contributions.

\subsection*{$\bullet$ Bibliographical notes}

The existence of a web of dualities between different supergravities and string theories  was proposed in \cite{Hull:1994ys}, and led to the discovery of M-theory in \cite{Witten:1995ex}. The action of $SL(2,\ZZ)$ on branes in Type IIB was developed in \cite{Schwarz:1995dk}, while evidence of the duality between Type I and Heterotic string theory was given in \cite{Polchinski:1995df}. Useful lecture notes may be found in \cite{Schwarz:1996bh}, and an early account of the role played by automorphic functions in dualities is in \cite{Obers:1998fb,Obers:1999um}.
The textbooks by Polchinski \cite{Polchinski:1998rq,Polchinski:1998rr}, Johnson \cite{Johnson}, and by Becker, Becker and Schwarz \cite{Becker} provide excellent and comprehensive accounts of string dualities.

\sm

The use of S-duality to constrain the low-energy effective action goes back to \cite{Green:1997tv}, where the coefficient of the $\cR^4$ term was related to the non-holomorphic Eisenstein series $E_{3/2}(\tau)$ using results from tree-level string theory and D-instantons. A derivation  using supersymmetry, in the way reviewed above, was provided in \cite{Green:1998by} and extended  in \cite{Pioline:1998mn}. That the same techniques could also be applied to the $D^4 \cR^4$ term was proposed in \cite{Green:1998by}, and used to provide a full derivation in \cite{Sinha:2002zr}. {As we have mentioned in the text, there is no closed form expression for the $D^6 \cR^4$ term, though it is known to satisfy a Laplace eigenvalue equation; this property and others were discussed in \cite{Green:2014yxa,Green:2010wi,Green:2005ba,Pioline:2015yea,Green:2019rhz}. An implicit expression for the  $D^6 \cR^4$ term as a two-loop Schwinger integral can be found in \cite{Bossard:2020xod}. }
\sm

As mentioned in section \ref{sec:nonrenorm}, the results quoted in the previous paragraph imply a number of non-renormalization theorems, some of which were anticipated from superstring perturbation theory in \cite{Martinec:1986wa,Berkovits:2004px}.  For the $\cR^4$ term, the S-duality and supersymmetry analysis predict the absence of perturbative corrections beyond one-loop, and indeed the vanishing at genus two was proven in \cite{DP6}. Conversely, for the $D^4 \cR^4$ term there is expected to be no genus one correction, as was verified in \cite{Green:1999pv}, but there is a non-zero genus-two contribution which was successfully computed and matched in \cite{DGP}. 
 
 \sm
 
The systematic study of genus-one contributions to effective interactions of the form $D^{2 \ell} \cR^4$ with $\ell \geq 3$ was initiated in \cite{Green:2008uj,DHoker:2015gmr}. From this grew the notion of modular graph functions \cite{DHoker:2015wxz,DHoker:2016mwo}, introduced in section \ref{sec:MGF}.

\sm 

One of the hallmarks of superstring theory is the infinite tower of BPS states, which includes black hole microstates. If the string background under consideration contains a torus, then the functions which count these microstates often enjoy nice modular or automorphic properties. A particularly well-studied setup is string theory compactified to 4d with $\cN=4$ supersymmetry. This can be engineered by considering Type II superstrings on $K3 \times T^2$ or Heterotic strings on $T^6$. Either way, the theory is expected to have an $SL(2, \ZZ) \times SO(22,6,\ZZ)$ duality symmetry, with the corresponding counting functions expressible as Siegel modular forms \cite{Dijkgraaf:1996it}. The contribution of multi-centered black holes was studied in \cite{Denef:2017zxm}, while the single-centered black holes were shown to have interesting connections with mock modular and Jacobi forms in \cite{Dabholkar:2012nd}. Recent connections have been made with Hurwitz class numbers \cite{Kachru:2017eju} and class groups \cite{Benjamin:2018mdo} as well. More generally, the tools of Rademacher sums and Farey tail expansions have been applied to the study of microstate counting in \cite{Dijkgraaf:2000fq,deBoer:2006vg,Manschot:2007ha}. A pedagogical introduction to the relation between blackhole microstate counting and automorphic forms can be found in \cite{Pioline:2006ni}.

\newpage

\section{Dualities in $\cN=2$ super Yang-Mills theories}
\setcounter{equation}{0}
\label{sec:SW}

In this penultimate section, we shall  discuss dualities in Yang-Mills theories with extended supersymmetry in four-dimensional Minkowski space-time. We briefly review supersymmetry multiplets of states and fields and the construction of supersymmetric Lagrangian theories with  $\cN=1,2,4$ Poincar\'e supersymmetries. We then discuss the $SL(2,\ZZ)$ Montonen-Olive duality properties of the maximally supersymmetric $\cN=4$ theory and  the low energy effective Lagrangians for $\cN=2$ theories via the Seiberg-Witten solution. We shall close this section with a discussion of dualities of $\cN=2$ superconformal gauge theories, which possess interesting spaces of marginal gauge couplings. In some cases these spaces of couplings can be identified with the moduli spaces for Riemann surfaces of various genera.   

\subsection{Super Yang-Mills: states and fields}

The quantum field theories we shall consider here are invariant under the Poincar\'e algebra, whose generators of translations $P_\mu$ and Lorentz transformations $L_{\mu \nu}$ satisfy the following structure relations,\footnote{Our conventions are as follows. The non-zero components of the Minkowski metric $\eta_{\mu \nu}$ are  $-\eta_{00} =\eta_{11} =\eta_{22}=\eta_{33}=1$. The Clifford-Dirac algebra is $\{ \gamma ^\mu, \gamma ^\nu \} = -2 \eta^{\mu \nu} I_4$, where $I_d$ is the $d \times d$ identity matrix. In a basis where the chirality matrix $\gamma _5= \left ( \begin{smallmatrix} I_2 & 0 \cr 0 & -I_2 \end{smallmatrix} \right )$ is diagonal, we have  $\gamma ^\mu = \left ( \begin{smallmatrix} 0 & \sigma ^\mu \cr \bar \sigma ^\mu & 0 \end{smallmatrix} \right )$ where $\sigma ^0 = \bar \sigma ^0 =- I_2$ and $\sigma ^i= - \bar \sigma ^i$ are the standard Pauli matrices for $i=1,2,3$. Two-component spinor indices $\a, \dot \a$ are raised and lowered with the help of the anti-symmetric symbols $\epsilon^{\a \b},\epsilon ^{\dot \a \dot \b}$ and $\epsilon_{\a \b},\epsilon _{\dot \a \dot \b}$ normalized to  $\epsilon^{12}=\epsilon ^{\dot 1 \dot 2}=-\epsilon_{12}=-\epsilon _{\dot 1 \dot 2}=1$.  We use the notations $\psi \chi= \psi^\a \chi_\a$, $\bar \psi \bar \chi = \bar \psi _{\dot \a} \bar \chi ^{\dot \a}$, and $\chi \sigma ^\mu \bar \psi = \chi^\a \sigma ^\mu _{\a \dot \a} \bar \psi ^{\dot \a}$.  For more details on spinor index notation, see for example \cite{Wess:1992cp}.}
\begin{align}
[L_{\m \n}, L_{\rho \sigma}] &= \eta_{\n \rho} L_{\mu \sigma} + \eta_{\m \sigma} L_{\n \rho} - \eta_{\m \rho} L_{\n \sigma} - \eta_{\n \sigma} L_{\m \rho} 
\no\\
{[L_{\m \n}, P_\rho]} &= \eta_{\n \rho} P_\m - \eta_{\m \rho} P_\n 
\no \\
{} [P_\mu, P_\nu] & = 0
\end{align}
 In supersymmetric theories the Poincar\'e  algebra is extended to a super Poincar\'e algebra by supplementing $P_\mu$ and $L_{\mu \nu}$ with $\cN$ supercharges $Q_\a^I$ and their adjoints $\bar Q_{\dot \a I} = (Q^I_\a)^\dagger$ for spinor index $\a =1,2$ and $I=1,\cdots, \cN$.  The supercharges commute with $P_\mu$, are Weyl spinors under the Lorentz generators $L_{\mu \nu}$, and obey the following anti-commutation relations,
\bea
\label{15.QQ}
\{ Q_\a ^I, \bar Q_{\dot \b J} \} & = & 2 \, \sigma ^\mu _{\a \dot \b} \, P_\mu \, \delta ^I _J \hskip 0.6in I,J =1 , \cdots, \cN
\no \\
 \{ Q_\a ^I, Q_\b ^J \} & = & 2\,  \epsilon _{\a \b} \, Z^{IJ}
 \eea
The generators $Z^{IJ} = - Z^{JI}$ can arise only when $\cN\geq 2$. They are central generators in the Poincar\'e superalgebra since they commute with one another and with all other generators. When $Z^{IJ}=0$, the supersymmetry algebra is invariant under a $U(\cN)_R$ automorphism group which, in Physics, is referred to as the R-symmetry group.

\sm

Applying a supercharge to a state changes the spin  by $\pm \thalf$ so that the values of the spin in the multiplets of the Poincar\'e superalgebra must increase with $\cN$. When $\cN>4$, the multiplets necessarily involve states of spin greater than 1, which cannot be accommodated in Yang-Mills theory. For $\cN\leq 8$ they can, however, be accommodated in supergravity, but this requires states of spin 2. Thus $\cN=4$ is the maximal number of Poincar\'e supersymmetries allowed for supersymmetric Yang-Mills theories in four dimensions.

\subsubsection{States}

The spectrum of states in a supersymmetric theory exhibits a characteristic pattern. All states have positive or zero energy.   If the state of lowest energy in the spectrum---namely the ground state---has exactly zero energy then  supersymmetry is said to be \textit{unbroken} or \textit{manifest}. In this case  all positive energy states occur in boson-fermion  pairs of equal mass, momentum, and internal quantum numbers. If the ground state has strictly positive energy, however, the ground state is not supersymmetric and supersymmetry is said to be \textit{spontaneously broken}. Bosons and fermions of identical internal quantum numbers do not generally have the same mass.

\sm

In this section we shall consider only the case where supersymmetry is manifest.  For the case of $\cN=1$, this is the complete story. For $\cN\geq 2$, there is an additional interplay between  the mass of the state and its central charge. 

\sm

For $\cN=2$ there is a single complex-valued central charge $Z=Z^{12}$, which we decompose into its absolute value and its phase, $Z=e^{i \f} |Z|$ for $\f \in \RR$. We consider a state with mass $M \not= 0$ and use a Poincar\'e transformation to the rest frame of this state with momentum  $P^\mu = (M,0,0,0)$. In terms of the combinations $ \cQ^\pm _\a= \thalf (Q^1 _\a \mp e^{- i \f} \sigma ^0_{\a \dot \b} Q^{2 \dot \b})$  the anti-commutation relations of (\ref{15.QQ}) are equivalent to the relations,   
\bea
\big \{ \cQ^\pm _\a, (\cQ^\pm _\b)^\dagger \big \} =  \delta _{\a}^{\b} (M \pm |Z|)
\eea
while all other anti-commutators vanish identically. Since the operators $\big \{ \cQ^\pm _\a, (\cQ^\pm _\a)^\dagger \big \}$  are self-adjoint and positive in a unitary theory, we obtain the following so-called \textit{BPS bound} between the mass $M$ and the central charge $Z$ of any state,
\bea
M \geq |Z|
\eea
In the limit where the state becomes massless, its central charge must vanish. Of particular interest are the massive  states for which the BPS bound is saturated, say by $M=|Z|$. In these states, which are referred to as BPS states,   the supersymmetry algebra reduces to,
\bea
\big \{ \cQ^+ _\a, (\cQ^+ _\b)^\dagger \big \} = 2M \delta _\a ^\b
\hskip 1in 
\big \{ \cQ^- _\a, (\cQ^- _\b)^\dagger \big \} = 0
\eea
while all other anti-commutators vanish. The  operators $\cQ^-_\a$ and $(\cQ^-_\a)^\dagger$ both annihilate the BPS states which means that BPS states are invariant under half of the total number of Poincar\'e supersymmetries. More specifically they are referred to as $\thalf$-BPS states, and they make up a supermultiplet that is half as long as the non-BPS  supermultiplets.  

\sm

For $\cN=4$, the matrix of central charges $Z^{IJ}=-Z^{JI}$ has two complex-valued eigenvalues $Z_1$ and $Z_2$. The BPS bound is now $M \geq |Z_1|, |Z_2|$. When $|Z_1| \not= |Z_2|$ one obtains $\thalf$-BPS states by setting $M$ equal to the maximum of $|Z_1|$ and $|Z_2|$. But when $|Z_1|=|Z_2|$, a further shortening of the supersymmetry multiplet takes place  to $\tfrac{1}{4}$-BPS states. It is a general result of CPT invariance in four space-time dimensions that supersymmetric Yang-Mills theories  with $\cN=3$ automatically enjoy the full $\cN=4$ supersymmetry, so that the case $\cN=3$ need not be considered separately. 

\subsubsection{Fields}
\label{sec:supermults}

A super Yang-Mills theory with gauge group~$G$ and associated Lie algebra~$\mathfrak{g}$ is constructed in terms of the customary fields of four-dimensional quantum field theory absent gravity, namely spin-1 gauge fields $A_\mu$  in the  adjoint representation of~$G$,  together with spin-$\thalf $ left Weyl fermion fields $\psi_\a, \lambda_\a$ and spin-0 scalar fields $\phi,h$  transforming under various representations of $G$. We recall that, in even dimensions,  a Dirac spinor is the direct sum of two Weyl spinors of opposite chirality and that, in four dimensions, a Majorana spinor is equivalent to a Weyl spinor.  The field multiplets for $\cN=1,2,4$ are as follows: 

\begin{itemize}
\itemsep=0in
\item $\cN=1$ \textit{Gauge multiplet} $(A_\mu, \lambda_\a)$ in the adjoint representation of $G$,  where $A_\mu$ is the gauge field and $\lambda_\a$ is a left Weyl fermion referred to as the gaugino;
\item $\cN=1$ \textit{Chiral multiplet} $(\psi_\a, \phi)$ in a representation $\cR$ of $G$, where $\psi_\a$ is a left Weyl fermion and $\phi$ is a complex scalar field;\\
\item $\cN=2$ \textit{Gauge multiplet} $(A_\mu, \lambda_\a, \tilde \lambda_\a , \phi )$ in the adjoint representation of $G$, where $A_\mu$ and $\phi$ are singlets under $SU(2)_R$  while $(\lambda_\a , \tilde \lambda_\a)$ is a doublet;
\item $\cN=2$ \textit{Hypermultiplet} $(h, \tilde h, \psi_\a, \tilde \psi_\a)$ in a representation $\cR$ of $G$,
where $(h, \tilde h)$ is a doublet under $SU(2)_R$ while $\psi_\a, \tilde \psi _\a$ are singlets;\\
\item $\cN=4$ \textit{Gauge multiplet} $(A_\mu, \lambda _\a^A, \phi^I)$ in the adjoint representation of $G$ where $\lambda^A _\a$ with $A=1,\cdots,4$ are left Weyl fermions in the ${\bf 4}$ of $SU(4)_R$  and $\phi^I$ with $I=1,\cdots, 6$ are complex scalars in the ${\bf 6}$ of $SU(4)_R$ while $A_\mu$ is a singlet. 
\end{itemize}

The fields of an $\cN=1$ theory may be collected in fully off-shell superfields with linear transformations under the action of the super Poincar\'e algebra, as will be made explicit in the next subsection. The construction of superfields for higher numbers of supersymmetries is, however, much more complicated and will not be addressed here. For this reason it will be useful to decompose the field content of the $\cN=2$ and $\cN=4$ theories in terms of $\cN=1$ superfields. In this spirit, the $\cN=2$ gauge multiplet is the direct sum of an $\cN=1$ gauge multiplet and an $\cN=1$ chiral multiplet, both in the adjoint representation of $G$. The $\cN=2$ hypermultiplet is a sum of two $\cN=1$ chiral multiplets in complex conjugate representations of one another $\cR \oplus \bar \cR$. Similarly, the $\cN=4$ gauge multiplet is the direct sum of an $\cN=2$ gauge multiplet and an $\cN=2$ hypermultiplet, both in the adjoint representation of $G$. Note that for $\cN=4$ theories the  $U(1)_R$ factor of the automorphism group $U(\cN)_R$ of the supersymmetry algebra   is anomalous and fails to be  a quantum symmetry. The charge assignments of the $U(1)_R$ factor in the case of $\cN=1$ and $\cN=2$ will not be important in the sequel, and we shall not spell them out here.

\sm

Before moving on, we remark that when $\cR$ is a pseudoreal representation, and hence admits an antisymmetric invariant tensor $\eps_{IJ}$, the fields $h_I$ and $\tilde h^J$ in the $\cN=2$ hypermultiplet (with indices $I,J$ in the representation $\cR$ shown explicitly) can be constrained to satisfy, 
\bea
h_I = \eps_{IJ} \tilde h^J
\eea
This constraint is compatible with $\cN=2$ supersymmetry, and halves the number of degrees of freedom in the multiplet. The resulting multiplet is referred to as a \textit{half-hypermultiplet}. More generally, given a complex representation $\cR$ there is a standard pseudo-real structure on $\cR \oplus \overline \cR$, and hence we may define half-hypermultiplets in the representation $\cR \oplus \overline \cR$. These are precisely the same as full hypermultiplets in the complex representation $\cR$.

\subsection{Super Yang-Mills: Lagrangians}

Two types of supersymmetric Lagrangians will be of interest to us: those that correspond to renormalizable quantum field theories, and those that collect the low energy effective interactions of some arbitrary Lagrangian or non-Lagrangian quantum field theory. In either case, we shall restrict attention to Lagrangians in which bosonic fields have a total of at most two derivatives while fermions have a total of at most one derivative acting on any field. Furthermore, the Lagrangians for theories with $\cN=2$ and  $\cN=4$ supersymmetries will be expressed in terms of $\cN=1$ chiral and vector superfields, using the decomposition of the corresponding multiplets given in the penultimate paragraph of the preceding subsection. Throughout we shall assume that the gauge group $G$ is compact with Lie algebra $\mathfrak{g}$, and of the form of a semi-simple group times a certain number of $U(1)$ factors. 

\sm

To introduce $\cN=1$ superfields, we supplement the coordinates $x^\mu$ of Minkowski space-time with Grassmann 
spinor variables $\theta ^\a$ and $\bar \theta _{\dot \a}$ of odd grading.  The superfield $\Phi$ for the $\cN=1$ chiral multiplet containing the fields $(\phi, \psi _\a)$  is given by,
\bea
\Phi = \phi + 2 \psi_\a \theta^\a + F \theta_\a \theta^\a
\eea  
where $F$ is a non-dynamical complex-valued scalar \textit{auxiliary field}. The chiral superfield $\Phi$ is generally matrix-valued with components $\Phi^i$ and transforms under an arbitrary representation $\cR$ of the gauge algebra $\mathfrak{g}$ with $i = 1 ,\cdots, \dim \cR$. The superfield $V$ for the $\cN=1$ gauge multiplet satisfies a reality condition $V^\dagger = V$ and transforms in the adjoint representation of the gauge algebra~$\mathfrak{g}$. The general decomposition of the superfield $V$  is more complicated than that of the chiral multiplet and we present it here in \textit{Wess-Zumino gauge} only,
\bea
V = - \theta \sigma^\m \overline\theta _\m + i \theta \theta \bar \theta \bar \lambda -  i \bar \theta \bar \theta  \theta  \lambda + \half \theta\theta \bar\theta \bar \theta D
\eea
where $D$ is a real-valued scalar auxiliary field. The vector superfield $V$ is generally matrix-valued and may be  decomposed $V= \sum_a T^a V^a$ onto the generators  $T^a$ of $\mathfrak{g}$, with $a=1, \cdots, \dim \mathfrak{g}$, in the adjoint representation of the gauge algebra $\mathfrak{g}$.

\subsubsection{$\cN=1$ Lagrangians}

In terms of the $\cN=1$ gauge superfield $V$  and chiral superfield $\Phi$, the most general Lagrangian density, under the assumptions spelled out earlier,   is of the following form, 
\bea
\label{eq:4dN2Lag}
\cL = \int d^4 \theta \, K(e^V \Phi, \Phi^\dagger) + \mathrm{Re}\int d^2 \theta \Big (  U(\Phi) + \tau_{ab}(\Phi)  W^a W^b \Big )
\eea
in superspace notation. Here $K$ is the real-valued \textit{K{\"a}hler potential},  $U$ is the \textit{superpotential},
and $\tau_{ab}$ is the matrix of couplings and mixings of the field strengths $W_\a^a$ of the gauge fields $V^a$. The functions $U(\Phi)$ and $\tau_{ab}(\Phi)$ are locally holomorphic in $\Phi$.  The superfield $W_\a= \sum _a W_\a ^a T^a$ is obtained from the vector superfield $V$ by taking, 
\bea
W_\alpha  = -{1\over 4} \bar D \bar D \left ( e^{-V} D_\alpha e^V \right )
\hspace{0.7 in} 
D_\a = {\p \over \p \theta^\alpha} + i \sigma^\m_{\a \dot \a} \bar \theta^{\dot \a} \p_\m
\eea
The field $W_\a$ is a chiral superfield since $\bar D ^{\dot \a} W_\b=0$ by construction.

\subsubsection{Renormalizable $\cN=2$ Lagrangians}

A general four-dimensional  $\cN=2$ Lagrangian theory is specified by a gauge algebra $\mathfrak{g}$ which determines the vector multiplets, as well as a representation $\cR$ of $\mathfrak{g}$ for the hypermultiplets, or half-hypermultiplets if $\cR$ is pseudo-real.\footnote{This data is not quite complete: in general, additional global data such as the spectrum of line operators is necessary to fully specify the theory. We will discuss such additional data in the context of $\cN=4$ Yang-Mills in section \ref{sec:SL2ZN4YM}, but will otherwise largely ignore this subtlety. As long as one is concerned only with the local operator spectrum of the theory, no harm arises from this omission.}
The $\cN=2$ gauge multiplet consists of an $\cN=1$ gauge multiplet $V$ and an $\cN=1$ chiral multiplet $\Phi$ in the adjoint representation of $\mathfrak{g}$, while the $\cN=2$ hypermultiplet consists of two $\cN=1$ chiral multiplets $H, \tilde H$ in the representations $\cR$ and $\bar \cR$ of~$\mathfrak{g}$, respectively. The $\cN=2$ half-hypermultiplet consists of a single $\cN=1$ chiral multiplet $H$ in a pseudo-real representation $\cR$. 

\sm

Let us consider first a renormalizable  $\cN=2$  theory without hypermultiplets. We begin by arguing that the superpotential $U(\Phi)$  must vanish. Indeed, for the theory to be renormalizable, one requires that $\tau_{ab}$ be independent of $\Phi$ and that the superpotential $U(\Phi)$ be at most cubic in the chiral superfield $\Phi$. 
Upon integrating over superspace, the final two terms of the Lagrangian (\ref{eq:4dN2Lag}) give,
\bea
\mathrm{Re}\int d^2 \theta \,U(\Phi)&=& \mathrm{Re} \Big \{
F^a \p_a U(\phi) + \psi^a \psi^b \p_a \p_b U(\phi)  \Big \}
\\
\mathrm{Re} \int d^2 \theta \,\tau_{ab}\, W^a W^b  
&=& 
- \mathrm{Re}\left\{\tau_{ab} \left({i\over 2} F_{\m \n}^a F^{b \m \n} - \half F_{\m \n}^a \tilde F^{b \m \n} + \overline \lambda^a \overline \sigma^\m D_\m \lambda^b \right)  \right\}
\no
\eea
where $\p_a = \p / \p \phi^a$.  To obtain a Lagrangian with  $\cN=2$ supersymmetry, the combined contributions of the $\cN=1$ gauge and chiral multiplets must be invariant under the $SU(2)_R$ symmetry rotating $\psi_\a$ into $\lambda_\a$, which is possible only if $U(\Phi)$ is linear in $\Phi$. Upon integrating out the auxiliary field $F$, a linear $U(\Phi)$ would add a constant $|\p U|^2$ to the energy density and spontaneously break supersymmetry. Therefore, in a theory with supersymmetric vacua, the superpotential  $U$ is a constant that may set to zero without loss of generality. 

\sm

What this means is that the only potential for the scalar fields $\phi^a$ comes from the K{\"a}hler potential. To proceed, we shall assume for simplicity that the gauge group $G$ is a simple Lie group, so that there is a single $\cN=1$ vector superfield $V$ transforming under the adjoint representation of $\mathfrak{g}$. For a renormalizable theory, the K{\"a}hler potential must be \textit{canonical} i.e. $K(e^V \Phi, \Phi^\dagger) = \mathrm{Tr} \left(\Phi^\dagger e^{V} \Phi\right)$. In that case the auxiliary field $D$ may be integrated out using its field equations and the resulting scalar potential takes the form, 
\bea
\label{eq:vectmultscalarpot}
V_{{\rm gauge}}(\phi, \bar \phi) = \mathrm{Tr}\left( i [\phi, \bar \phi] \right)^2
\eea
The structure of this potential will be of central importance in disentangling the vacuum structure of the $\cN=2$ theories.

\sm

Next, we include an $\cN=2$ hypermultiplet, which  consists of $\cN=1$ chiral superfields $H=(h,\psi_\a)$ and $\tilde H = (\tilde h, \tilde \psi_\a)$ in representations $\cR$ and $\bar \cR$ of $\mathfrak{g}$, respectively.   The representation $\cR \oplus \bar \cR$ may be reducible into a direct sum of irreducible representations $\cR_I$ and $ \bar \cR_I$ of $\mathfrak{g}$, 
\bea
\cR = \bigoplus _I n_I \,  \cR_I \oplus \bar \cR_I \hskip 1in 
\sum_I n_I  =N_f
\eea
where $n_I$ is the number of hypermultiplets in representation $\cR_I$, and $N_f$ is the total number of hypermultiplets in all representations $\cR_I$. We decompose the hypermultiplet accordingly into hypermultiplets $H_I$ and $\tilde H_I$ in the representations $\cR_I$ and $\bar \cR_I$ respectively.  The scalar fields $h$ and $\tilde h$ in the hypermultiplets produce the following addition to the scalar potential, 
\bea
V_{{\rm hyper}} (h, \tilde h, \phi) = \sum_{i,j=1}^{\dim \cR} h_i  (\phi_a T^a )_{ij}  \tilde h_j + \sum _I \sum _{s,t =1}^{n_I} M_{I,s,t} h_{I,s}  \tilde h_{I,t}
\eea
Here $s,t$ are flavor indices running from $1, \dots, n_I$ in each irreducible representation $\cR_I$, while $i,j$ are the indices in the representation $\cR$, and $\phi$ is the scalar component of the superfield~$\Phi$. The matrix $M_{I,s,t}$ contains parameters that can be interpreted as masses and mixings of the hypermultiplets in a given representation $\cR_I$ of $\mathfrak{g}$. Gauge invariance precludes mixings between hypermultiplets $\cR_I$ and $\cR_J$ with $I \not= J$. Combining this scalar potential with that in (\ref{eq:vectmultscalarpot}) allows for interesting structure for the moduli space of vacua.

\subsection{Low-energy Lagrangian on the Coulomb branch}

In a supersymmetric Yang-Mills theory, the energy is bounded from below by zero. In the sequel we shall assume that the gauge group of the UV theory is a compact semi-simple Lie group $G$ with Lie algebra $\mathfrak{g}$. In particular, the potential energy $V_{{\rm gauge}}$ of (\ref{eq:vectmultscalarpot})  is non-negative since the Cartan-Killing form of $\mathfrak{g}$ is negative definite. Any supersymmetric vacuum state has exactly zero energy, which requires vanishing fermion fields, $D_\mu \phi=F_{\mu \nu}=0$ and, for the solutions to these equations in the gauge $A_\mu=0$, a constant field $\phi$ satisfying, 
\bea
\label{eq:Dtermconst}
[\phi, \bar \phi] = 0
\eea
Decomposing $\phi=\phi_1 + i \phi_2$ into Hermitian matrices $\phi_1,\phi_2$ gives the equivalent condition $[\phi_1, \phi_2]=0$ which guarantees that $\phi_1$ and $\phi_2$  may be simultaneously diagonalized, and gives  the following general solution to (\ref{eq:Dtermconst}) for the vacuum expectation value of the field $\phi$,
\bea
\label{15.vac}
 \phi  = \sum _{I=1}^r a_I \mh_I \hskip 1in r = {\rm rank} (\mathfrak{g})
\eea
where $\mh_I$ for $I=1,\cdots, r={\rm rank} (\mathfrak{g})$ are the mutually commuting generators of the Cartan subalgebra of $\mathfrak{g}$. For generic values of  $a_I\not= 0$ the vacuum is referred to as the \textit{Coulomb branch}. The standard Higgs mechanism breaks the gauge symmetry $G$ to the Cartan subgroup $U(1)^r$. Actually, the Cartan subgroup is invariant under the Weyl group $\cW(\mathfrak{g})$ group which is the residual symmetry remaining from the non-Abelian part of $G$. Thus, the precise symmetry breaking pattern is as follows,
\bea
G \longrightarrow U(1)^r/ \cW(\mathfrak{g})
\eea 
Therefore, the low energy effective theory consists of $r$ copies of  $\cN=2$ electro-magnetism with compact $U(1)$ gauge groups, up to  identification under the Weyl group $\cW(\mathfrak{g})$. 

\sm

The field contents of the low energy theory consists of $\cN=2$  gauge multiplets with component fields $(A_\mu ^I, \lambda _\a ^I , \tilde \lambda _\a^I, \phi^I)$ for gauge group $U(1)^r$ and $I=1,\cdots, r$ which may be decomposed into $\cN=1$ gauge superfields $V^I$ and chiral superfields $\Phi^I$. The theory is  governed by the Lagrangian constructed in (\ref{eq:4dN2Lag})  for gauge group $U(1)^r$,
\bea    
\cL = \int d^4 \theta \, K \big ( \Phi, \Phi^\dagger \big ) + \Re \int d^2 \theta \, \tau_{IJ} \, W^I W^J
\eea
The field strength superfields $W^I$ of the Abelian gauge multiplets simplifies,
\bea
W^I_\a = - {1 \over 4} \bar D \bar D D_\a V^I
\eea
and $\tau_{IJ} (\Phi)$ is holomorphic in the fields $\Phi^I$.  Since the chiral fields $\Phi^I$ are neutral under the gauge group $U(1)^r$ the factor $e^V$ multiplying $\Phi$ in  (\ref{eq:4dN2Lag}) is absent here. No superpotential $U(\Phi^I)$ is allowed in view of  $\cN=2$ supersymmetry, as may be shown by the same argument used earlier in the case of renormalizable Lagrangians. Expressed in terms of component fields the Lagrangian takes the form,\footnote{We use the standard notation for coordinates on  K\"ahler  manifolds where $\phi^{\bar I} = \bar \phi^I$ so that the metric is $g_{I \bar J} \, d\phi^I \otimes d\phi^{\bar J}$. The Levi-Civita connection $\Gamma$ for the metric $g_{I \bar J}$ is compatible with the complex structure so that its mixed components $\Gamma ^I_{\bar J \bar K} , \, \Gamma ^I_{J \bar K}$ and their complex conjugates vanish. The corresponding covariant derivatives  are given by $D_\mu f ^I = \p_\mu f^I + \Gamma ^I _{JK}  \p_\mu \phi^J\, f^K$ for the fields $f^I=\phi^I, \lambda ^I, \tilde \lambda ^I$.}
\bea
\cL & = & - g_{I \bar J}(\phi,\bar \phi)  \Big ( D_\mu \phi^I D^\mu \phi^{\bar J} + i \overline{ \tilde \lambda} ^{\bar J} \bar \sigma ^\mu D_\mu \tilde \lambda ^I \Big )
\no \\ &&
- \Re \left \{ \tau_{IJ}(\phi)  \left ( {i \over 2} F_{\mu \nu} ^I F^{J \mu \nu } - \half F^I _{\mu \nu} \tilde F^{J\mu \nu} + \bar \lambda ^I \bar \sigma^\mu  D_\mu \lambda ^J \right ) \right \} 
\eea
The $SU(2)_R$ symmetry of $\cN=2$ rotates $\lambda ^I$ into $\tilde \lambda^I$ so that invariance of $\cL$ requires the following relation between their coefficients $\tau_{IJ}$ and the K\"ahler metric $g_{I\bar J}$. 
\bea
\Im \tau_{IJ}(\phi) = g_{I \bar J} (\phi, \bar \phi) = { \p^2 K (\phi, \bar \phi) \over \p \phi^I \p \phi ^{\bar J}}
\eea
Since $\tau_{IJ}(\phi)$ is holomorphic, this relation implies that the K\"ahler potential $K$ must be of the \textit{special K\"ahler} type, so that $\tau_{IJ}(\phi)$ and $K$ are given by,
\bea
\tau_{IJ} (\phi) = { \p^2 \cF (\phi) \over \p \phi^I \p \phi^J} 
\hskip 0.6in
K(\phi, \bar \phi) = {1 \over 2i}  \left ( \bar \phi^I \,\phi_{DI} 
- \phi^I \,  \bar \phi_{DI}  \right )
\hskip 0.5in
\phi_{DI}  = {\p \cF(\phi) \over \p \phi ^I}
\eea
where $\cF(\phi)$ is a locally holomorphic function of $\phi^I$ referred to as the \textit{pre-potential}. Since the pre-potential $\cF$ depends only on the field $\phi$ and not on any derivatives of $\phi$, its expression is entirely determined by evaluating $\cF$ on the vacuum expectation values of $\phi$, given in (\ref{15.vac}).

\subsection{BPS states, monopoles, and dyons}

In addition to the massless states described by the canonical fields of the $\cN=2$ gauge multiplet, the theory also contains the massive gauge boson states as well as the `t Hooft-Polyakov magnetic monopoles produced by the Higgs mechanism $\mathfrak{g} \to \muu(1)^r$. These states are BPS states and carry non-zero electric and/or magnetic charges under the various $\muu(1)$ fields. In this subsection we shall review the properties of these states.

\sm

The massive vector bosons, and their superpartners,  that arise via the Higgs mechanism from the spontaneous symmetry breaking $\mathfrak{g} \to \muu(1)^r$ correspond to the root generators $\me_{\ab}$ in the decomposition\footnote{Our notation for Cartan generators and roots for a rank $r$ Lie algebra $\mg$ in the Cartan basis is as follows. The Cartan  generators will be denoted $\mh_I$ with $I=1,\cdots, r$, as we have already used in (\ref{15.vac}) when parametrizing the vacua of the Coulomb branch. The root generators are denoted $\me_\ab$, where the corresponding root vector is an $r$-dimensional vector with components $\a_I$. Their commutators may now be expressed as follows, $[\mh_I, \mh_J]=0$, $[\mh_I, \me_\ab] = \a_I \me_\ab$, and $[\me_\ab, \me_{-\ab}] = \sum _I \a_I \mh_I$.} of the generators of $\mathfrak{g}$ while the massless gauge bosons of $\muu(1)^r$ correspond to the Cartan generators $\mh_I$ identified in (\ref{15.vac}).  For each root $\ab$ there is a unique massive vector boson multiplet in the spectrum. Counting the number of states for each spin,  we see that these massive vector bosons must belong to  BPS multiplets.  As a result, the BPS formula gives their masses in terms of their central charges, 
\be
M^W _{\ab} = |\ab \cdot \ba |
\hskip 1in
\ba = (a_1 , \cdots , a_r)
\ee
a formula that is a direct consequence of the Higgs mechanism by  a scalar $\phi$ in the adjoint representation of the gauge algebra $\mathfrak{g}$.

\sm

The massive magnetic monopole states, and their superpartners, also arise via the Higgs mechanism $\mathfrak{g} \to \muu(1)^r$ as `t Hooft-Polyakov magnetic monopoles. In the classical limit, they arise as static solutions of finite mass that saturate the Bogomolnyi  bound,
\bea
M =  { 2 \over g^2} \left | \int d^3 x \, \boldsymbol{\nabla} \cdot \tr \big ( \phi \, \bB^I) \right |
\eea
where $\bB^I$ is the magnetic field component of the field strength $F_{\mu \nu}^I$ and the integral gives the magnetic charge of the field $F^I_{\mu \nu}$. Precisely one magnetic monopole arises for each possible embedding of $\mathfrak{su}(2)$ into $\mathfrak{g}$, i.e. for each root $\boldsymbol{\beta}$ of $\mathfrak{g}$. This construction will give us a BPS formula (at the semi-classical level) for
the mass of the `t Hooft-Polyakov magnetic monopole in terms of the vacuum expectation value of the dual gauge scalar $\phi_{DI}$. Applying this construction for each $SU(2)$ subgroup of $G$, namely for each root of $\mathfrak{g}$, we find the semi-classical magnetic monopole BPS mass formula, 
\be
M^M _{\betb} =  |\boldsymbol{\beta}  \cdot \ba_D  |
\hskip 1in
\ba_D = (a_{D1} , \cdots , a_{Dr})\, .
\ee   
Here, we have also included the effect of the $\theta$ angle, which is to produce an electric charge on the magnetic monopole and make it into a \textit{dyon}. 

\sm

More generally,  a dyon may be specified by two roots: one root $\ab$ for its electric charge as in the construction of vector boson masses, and one (co)root $\betb$ for its magnetic charge as in the construction of the magnetic monopole masses. The full quantum BPS mass formula for such a dyon is,
\be
M^D _{\ab, \betb} = \big |\ab  \cdot \ba + \betb \cdot \ba_D \big |
\ee  
The theory further contains an infinite number of states that may be neutral or carry arbitrary electric and magnetic charges, but that do not saturate the BPS bound and are thus non-BPS states. No simple formula for their mass in terms of the other quantum numbers is known to exist. Henceforth, we restrict attention to the BPS spectrum only.

\subsection{$SL(2,\ZZ)$-duality of the $\cN=4$ theory}
\label{sec:SL2ZN4YM}
For the $\cN=4$ theory, the pre-potential is given by,
\bea
\cF(\phi) = \half \tau \sum_I \phi^I \phi^I
\eea
so that $a_{DI} = \tau a_I$ and $\tau_{IJ} = \tau \delta_{IJ}$ with $\tau$ independent of $\phi$. The spectrum of  vector boson, magnetic monopole, and dyon BPS states is then given by,
\be
M^D _{\ab, \betb} (\tau)= \big |\ab  \cdot \ba + \tau \betb \cdot \ba \big |
\ee  
The roots $\ab$ and $\betb$ may be decomposed onto a basis of simple roots $\ab_I$, 
\bea
\ab = \sum _I n_I \ab_I 
\hskip 1in 
\betb = \sum_I m_I \ab_I
\eea
where the coefficients $m_I,n_I$ are integers. The masses of the BPS states are then given by,
\bea
M_{m,n} (\tau) = \big | Z \big | 
\hskip 1in Z =  \sum _I (n_I + m_I \tau) \, \ab_I \cdot \ba
\eea
One of the original indications of the $SL(2,\ZZ)$ Montonen-Olive duality symmetry of the maximally supersymmetric $\cN=4$ Yang-Mills theory was that the spectrum of BPS states is invariant under $SL(2,\ZZ)$ transformations of the complexified coupling $\tau$,
\bea
M_{m',n'}(\tau') = M_{m,n}(\tau) 
\hskip 0.6in
\tau \to \tau'= { a \tau + b \over c\tau +d} 
\hskip 0.6in 
\ba \to \ba' = { \ba \over c \tau +d} 
\eea
provided the states are mapped as follows, 
\bea
m_I & = & am_I' + cn_I'
\no \\ 
n_I & = & bm_I' + dn _I'
\eea 
We note, however, that $SL(2,\ZZ)$ does not in general map $\cN=4$ Yang-Mills with gauge group $G$ to itself. Instead, as we shall now explain, it transforms between theories with the same Lie algebra $\mathfrak{g}$ but different global structures.

\subsubsection{The global structure of $\cN=4$ theories}

The Lie algebra $\mathfrak{g}$ of an $\cN=4$ Yang-Mills theory does not uniquely specify the theory. Rather there remain different choices for the \textit{global structure} of the theory, including (but not limited to) the particular choice of gauge group $G$ corresponding to the algebra $\mathfrak{g}$.\footnote{Throughout this section, we shall restrict attention  to gauge groups that are connected and compact. An example in which the global structure of the gauge group alone  is not enough to fully specify the global structure is provided by the case of $SO(3)$ super Yang-Mills, which will be discussed below.} The various Lie groups associated with a  Lie algebra $\mg$ may be obtained by first identifying the unique simply-connected group $\cG$ whose Lie algebra is $\mg$, and then identifying the center $Z(\cG)$ of $\cG$. All of the other gauge groups corresponding to the algebra $\mathfrak{g}$ are then of the form $G_H=\cG/H$ for $H$ a subgroup of $Z(\cG)$. 
Whenever $H$ is non-trivial, i.e. contains at least one element other than the identity, the group $G_H$ is non-simply connected and a standard argument shows that its first homotopy group $\pi_1(G_H)$ is isomorphic to $H$. 

\sm

The weight lattice $\Lambda _w$ of a Lie algebra $\mg$ is the lattice spanned by the weights of all possible finite-dimensional representations of $\mg$, and coincides with the weight lattice of the simply connected Lie group $\cG$ so that $\Lambda_w = \Lambda _w ^\cG$. The weight lattice of a non-simply connected subgroup $G_H \subset \cG$ will be denoted $\Lambda _w ^{G_H}$.  Every representation of $G_H$ is a representation of $\cG$ so that $\Lambda _w ^{G/H} \subset \Lambda _w ^\cG$ but the converse is not true, so that the inclusion is a strict one whenever $H$ is non-trivial. The root lattice of $\mg$ is a sub-lattice of $\Lambda_w^{G/H}$ for every $H \subset Z(\cG)$. 

\sm 

We shall now consider the $\cN=4$ Yang-Mills theories with gauge algebra $\mg$ and gauge group $G=G_H$ for the different choices of $H$.  An $SL(2,\ZZ)$ transformation maps the gauge algebra $\mg$ to itself, and maps the gauge group $G_H$ to a gauge group $G_{H'}$ where $H'$ is not necessarily equal to $H$. This means that the allowed representations of $G_H$ and $G_{H'}$ will differ from one another when $H'\not = H$ and is most easily seen by analyzing the spectrum of line operators in each theory. These line operators come in the following types: 

\begin{itemize}
\item \textbf{Wilson lines:} To each center subgroup $H$ and corresponding gauge group $G_H$ is associated a set of \textit{Wilson lines}, which correspond physically to the worldlines of heavy probe particles that  transform under a representation $\cR$ of $G_H$. The Wilson line may be expressed as the holonomy of the gauge field $A$ in the representation $\cR$ of $G_H$,
\bea
W_\cR(\gamma) = \mathrm{Tr}_\cR \left (  P \exp \Big \{ i \oint_\gamma A \Big \} \right )
\eea
with $\gamma$ a 1-cycle in the four-dimensional  spacetime. Defined in this way, each Wilson line is manifestly invariant under $H$ since $A$ is invariant, but the selection of the representation $\cR$ of $G_H$ depends on $H$, as explained earlier.  Note that since Wilson lines are worldlines of \textit{probe} particles, their existence does not depend on the dynamical matter content of the theory: once the gauge group $G_H$ is fixed, Wilson lines will exist for every representation $\cR$ of $G_H$. In terms of the weight lattice $\Lambda_w$ of $\mathfrak{g}$, the Wilson lines correspond to points in $\Lambda^{G_H}_w / \cW(\mg)$, with $\Lambda_w^{G_H} \subset \Lambda_w$ the sub-lattice of weights of $G_H$ and $\cW(\mg)$ the Weyl group of $\mg$. 

\sm

\item \textbf{'t Hooft lines:} Gauge theories also come equipped with \textit{'t Hooft lines}, which are the \textit{magnetic} counterparts of the Wilson lines. Unlike Wilson lines, they cannot be written as local functionals of the fields of the gauge theory, but they are labelled in an analogous way. In particular, let us denote the magnetic dual (also referred to as the Langlands or Goddard-Nuyts-Olive (GNO) dual) algebra of $\mathfrak{g}$ by $\mathfrak{g}^\vee$. The algebra $\mathfrak{g}^\vee$ is dual to $\mathfrak{g}$ in the sense that the weight lattice of $\mathfrak{g}^\vee$, which we denote by $\Lambda_w^\vee$, is the dual of the root lattice of $\mathfrak{g}$. For simply-laced $\mathfrak{g}$ we have $\mathfrak{g} = \mathfrak{g}^\vee$, whereas for non-simply-laced algebras one has
\bea
 \mathfrak{b}_r^\vee = \mathfrak{c}_r\hspace{0.5 in} \mathfrak{c}_r^\vee = \mathfrak{b}_r\hspace{0.5 in} \mathfrak{g}_2^\vee = \mathfrak{g}_2\hspace{0.5 in} \mathfrak{f}_4^\vee = \mathfrak{f}_4
 \eea
 in the standard Cartan notation in which $\mb_r, \mc_r, \mg_2, \mf_4$ are the Lie algebras of the Lie groups  $SO(2r+1)$, $Sp(2r)$, $G_2$, $F_4$, respectively.  At the level of groups the correspondence is more subtle, with the results collected in Table \ref{tab:Langlandsduals}. Note that,
\bea
\cW(\mg^\vee) =\cW(\mathfrak{g}) \hskip 1in Z(\cG^\vee) = Z(\cG)
\eea  
The 't Hooft lines are then labelled by a particular subset of the points in $\Lambda_w^\vee/ \cW(\mathfrak{g})$, specified by the global structure of the magnetic gauge group $G^\vee$. The way to determine this subset will be explained below.
\sm
\item \textbf{Dyonic lines:} Finally, we can consider mixed Wilson and 't Hooft lines, labelled by a subset of the electric and magnetic charges, 
\bea
(q_e, q_m) \in \Lambda_w \times \Lambda_w^\vee
\eea
subject to the identification,
\bea
(q_e,\, q_m) \sim (w q_e,\, w q_m) \hspace{0.5 in} w \in \cW(\mathfrak{g})
\eea
Note that this set can be larger than the set of pairs of representations of $\mathfrak{g}$ and $\mathfrak{g}^\vee$, which is  labelled instead by $(\Lambda_w / \cW(\mathfrak{g})) \times (\Lambda_w^\vee / \cW(\mathfrak{g}))$. 
\end{itemize}

\begin{table}[tp]
\begin{center}
\begin{tabular}{|c|c|}
\hline
$G$ & $G^\vee$
\\\hline\hline
$SU(NM)/\ZZ_N$ & $SU(NM)/\ZZ_M$
\\\hline
$SO(2N)$ & $SO(2N)$
\\\hline
$Spin(2N)$ $N$ odd & $Spin(2N)/\ZZ_4$
\\\hline
$Spin(2N)$ $N$ even & $SO(2N)/\ZZ_2$
\\\hline
$SO(2N+1)$ & $USp(2N)$
\\\hline
$Spin(2N+1)$ & $USp(2N)/\ZZ_2$
\\\hline
$E_6$ & $E_6/\ZZ_3$
\\\hline
$E_7$ & $E_7/\ZZ_2$
\\\hline
\end{tabular}
\end{center}
\caption{\textit{Langlands or GNO dual groups $G^\vee$ for various gauge groups $G$, with $(G^\vee)^\vee = G$. The gauge groups which do not appear in this table, namely $G_2, F_4,$ and $E_8$, have trivial center and hence only a single global form, meaning that $G^\vee =G$.}}
\label{tab:Langlandsduals}
\end{table}%

\sm

We now explain how to determine the subset of allowed 't Hooft and dyonic lines. Having fixed a global form $G_H$ for the gauge group, we know that the spectrum of Wilson lines is the set of all lines $(q_e,0)$ with $q_e \in \Lambda_w^{G_H}/\cW (\mathfrak{g})$. The spectrum of 't Hooft and dyonic lines is then given by the full spectrum of lines which are \textit{mutually local} to the Wilson lines. By mutual locality, we mean that if one computes the correlation function of two lines supported on $\gamma$ and $\gamma'$, and then moves $\gamma'$ in a loop around $\gamma$, the correlation function is unchanged. 

\sm 

To make this more concrete, note that independent of the particular global group $G_H$, Wilson lines labelled by \textit{roots} of $\mathfrak{g}$ must always be present, since roots are elements of $\Lambda^{G_H}_w$ for any choice of subgroup $H$. Likewise, even without solving the mutual locality constraint, it is clear that we must have at least the 't Hooft lines labelled by roots of $\mathfrak{g}^\vee$. Because the charge lattice is closed under addition and inversion---i.e. if $(q_e, q_m)$ is present then so is $(-q_e, -q_m)$, and if furthermore $(q_e', q_m')$ is present then so is $(q_e + q_e', q_m+q_m')$---it is useful to organize lines into families labelled by the weight lattice modulo the root lattice. The weight lattice modulo the root lattice is nothing but the center, and recalling that $Z(\cG) = Z(\cG^\vee)$ we conclude that we may label families of lines by pairs
\bea
(z_e, z_m) \in Z(\cG) \times Z(\cG)
\eea
If one element in such a family exists, so do all of the other elements, obtained by adding arbitrary root vectors.

\sm

For charges valued in the center, the mutual locality condition mentioned above takes a simple form. For $Z(\cG) = \ZZ_N$, which is the case for all simple Lie groups except for $SO(4N+2)$,\footnote{In the case of $SO(4N+2)$ the center is $\ZZ_2 \times \ZZ_2$. This case is more complicated and will not be discussed here; see \cite{Aharony:2013hda} for details.} then for lines $(z_e, z_m)$ and $(z_e', z_m')$ the mutual locality constraint reads \cite{Aharony:2013hda},
\bea
z_e z_m' - z_m z_e' = 0 \,\,\, \mathrm{mod}\,\, N
\eea
We note that even upon fixing the global form of the gauge group $G_H$, and hence the lattice of allowed electric charges $z_e$, there can be multiple solutions to the mutual locality constraint, giving different lattices of dyonic lines. To fully specify the $\cN=4$ theory, one must specify all of this additional global data.

\subsubsection{Example: $\mathfrak{su}(2)$ SYM}

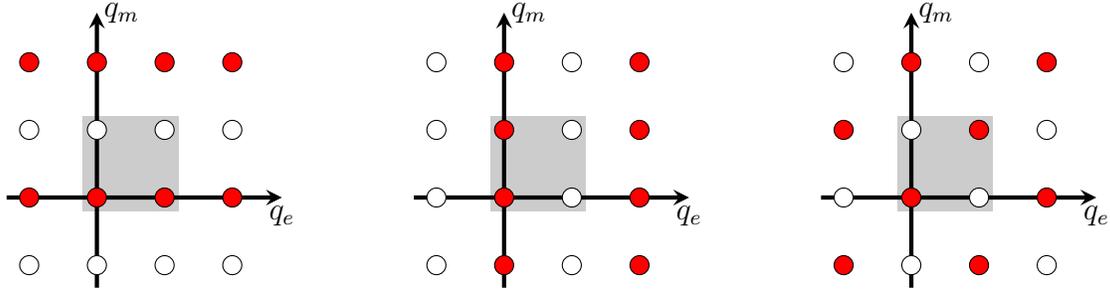
\begin{figure}[tp]
\begin{center}
\begin{tikzpicture}[baseline=0,scale = 0.6, baseline=0]

\shade[top color=black!20, bottom color=black!20]  (-0.3,1.8) -- (1.8,1.8) -- (1.8,-0.3) -- (-0.3,-0.3)--  (-0.3,1.8);
\draw [ultra thick,-stealth] (-2,0) to (4.1,0);
\draw [ultra thick,-stealth] (0,-2) to (0,4.1);
\node at (4.1,-0.4) {$q_e$};
\node at (0.55,4.1) {$q_m$};

\draw[fill = white] (-1.5,-1.5) circle (6pt);
\draw[fill = white] (0,-1.5) circle (6pt);
\draw[fill = white] (1.5,-1.5) circle (6pt);
\draw[fill = white] (3,-1.5) circle (6pt);

\draw[fill = red] (-1.5,0) circle (6pt);
\draw[fill = red] (0,0) circle (6pt);
\draw[fill = red] (1.5,0) circle (6pt);
\draw[fill = red] (3,0) circle (6pt);

\draw[fill = white] (-1.5,1.5) circle (6pt);
\draw[fill = white] (0,1.5) circle (6pt);
\draw[fill = white] (1.5,1.5) circle (6pt);
\draw[fill = white] (3,1.5) circle (6pt);

\draw[fill = red] (-1.5,3) circle (6pt);
\draw[fill = red] (0,3) circle (6pt);
\draw[fill = red] (1.5,3) circle (6pt);
\draw[fill = red] (3,3) circle (6pt);
\end{tikzpicture}
\hspace{0.45 in}
\begin{tikzpicture}[baseline=0,scale = 0.6, baseline=0]

\shade[top color=black!20, bottom color=black!20]  (-0.3,1.8) -- (1.8,1.8) -- (1.8,-0.3) -- (-0.3,-0.3)--  (-0.3,1.8);
\draw [ultra thick,-stealth] (-2,0) to (4.1,0);
\draw [ultra thick,-stealth] (0,-2) to (0,4.1);
\node at (4.1,-0.4) {$q_e$};
\node at (0.55,4.1) {$q_m$};

\draw[fill = white] (-1.5,-1.5) circle (6pt);
\draw[fill = red] (0,-1.5) circle (6pt);
\draw[fill = white] (1.5,-1.5) circle (6pt);
\draw[fill = red] (3,-1.5) circle (6pt);

\draw[fill = white] (-1.5,0) circle (6pt);
\draw[fill = red] (0,0) circle (6pt);
\draw[fill = white] (1.5,0) circle (6pt);
\draw[fill = red] (3,0) circle (6pt);

\draw[fill = white] (-1.5,1.5) circle (6pt);
\draw[fill = red] (0,1.5) circle (6pt);
\draw[fill = white] (1.5,1.5) circle (6pt);
\draw[fill = red] (3,1.5) circle (6pt);

\draw[fill = white] (-1.5,3) circle (6pt);
\draw[fill = red] (0,3) circle (6pt);
\draw[fill = white] (1.5,3) circle (6pt);
\draw[fill = red] (3,3) circle (6pt);
\end{tikzpicture}
\hspace{0.45 in}
\begin{tikzpicture}[baseline=0,scale = 0.6, baseline=0]

\shade[top color=black!20, bottom color=black!20]  (-0.3,1.8) -- (1.8,1.8) -- (1.8,-0.3) -- (-0.3,-0.3)--  (-0.3,1.8);
\draw [ultra thick,-stealth] (-2,0) to (4.1,0);
\draw [ultra thick,-stealth] (0,-2) to (0,4.1);
\node at (4.1,-0.4) {$q_e$};
\node at (0.55,4.1) {$q_m$};

\draw[fill = red] (-1.5,-1.5) circle (6pt);
\draw[fill = white] (0,-1.5) circle (6pt);
\draw[fill = red] (1.5,-1.5) circle (6pt);
\draw[fill = white] (3,-1.5) circle (6pt);

\draw[fill = white] (-1.5,0) circle (6pt);
\draw[fill = red] (0,0) circle (6pt);
\draw[fill = white] (1.5,0) circle (6pt);
\draw[fill = red] (3,0) circle (6pt);

\draw[fill = red] (-1.5,1.5) circle (6pt);
\draw[fill = white] (0,1.5) circle (6pt);
\draw[fill = red] (1.5,1.5) circle (6pt);
\draw[fill = white] (3,1.5) circle (6pt);

\draw[fill = white] (-1.5,3) circle (6pt);
\draw[fill = red] (0,3) circle (6pt);
\draw[fill = white] (1.5,3) circle (6pt);
\draw[fill = red] (3,3) circle (6pt);
\end{tikzpicture}

\caption{\textit{Lattice of line operators for the $SU(2)$, $SO(3)_+$, and $SO(3)_-$ theories. The red dots represent occupied lattice points. The grey square represents the sublattice of elements $(z_e, z_m) \in Z(SU(2)) \times Z(SU(2))$ valued in the center, which can be used to obtain the entire lattice by adding arbitrary root vectors.}}
\label{fig:chargelattices}
\end{center}
\end{figure}

Let us illustrate the above through the example of $\mathfrak{g} = \mathfrak{su}(2)$. In this case $\cG = SU(2)$ and $Z(\cG) = \ZZ_2$. If we take $H \subset Z(\cG) $ to be trivial, then $G_H = SU(2)$ itself. Aside from the identity line $(z_e, z_m) = (0,0)$, the spectrum of Wilson lines is generated by elements of the non-trivial center, which is just $(z_e, z_m) = (1,0)$. The mutual locality condition requires that any other family of lines labelled by $(z_e', z_m')$ satisfies $z_m' = 0\,\,\,\mathrm{mod}\,\,2$, and hence that no other non-trivial representation of the center is allowed. The lattice of charges for  $G_H = SU(2)$ is  shown in the gray square in the leftmost part of Figure \ref{fig:chargelattices}---the full lattice is obtained from this by adding elements of the root lattice. 

\sm

We can instead choose $H = \ZZ_2$, in which case $G_H = SU(2)/\ZZ_2 = SO(3)$. The Wilson lines are now labelled by representations of $SO(3)$, which have even electric charges, and hence which, when reduced to the center, are all trivial $(z_e, z_m)=(0,0)$. Because of this, demanding mutual locality with Wilson lines does not impose any constraints on the spectrum of 't Hooft and dyonic lines operators, and we may choose to have either $(z_e, z_m) = (0,1)$ or $(1,1)$ occupied (but not both, since the two are not mutually local). This gives the middle and right lattices in Figure \ref{fig:chargelattices}. We note here that even upon fixing the gauge group $G_H = SO(3)$, there are still two distinct choices for the spectrum of line operators, and one must specify the full lattice of lines before one completely specifies the theory. The two theories in this case are typically denoted by $SO(3)_+$ and $SO(3)_-$.

\sm 

Let us point out here that the only ingredient entering in the above analysis was the \textit{center} of $\cG$. We may thus conclude that the case of $\mathfrak{g} = \mathfrak{e}_7$, for which the center is again $\ZZ_2$, is exactly identical, again with three distinct global variants. For groups with larger centers the analysis is more involved, but conceptually straightforward. On the other hand, when the center is trivial as for $\mathfrak{g} = \mathfrak{e}_8$, there is only a single variant of the theory.

\subsubsection{$SL(2,\ZZ)$ on line operators}

Having understood the spectrum of line operators, we may now return to the question of the action of $SL(2,\ZZ)$. The $S$ and $T$ transformations act in a straightforward way on the spectrum of line operators, 
\bea
S: (q_e, q_m) \rightarrow (q_m, q_e) \hspace{0.5 in} T: (q_e, q_m) \rightarrow (q_e + q_m, q_m)
\eea
The first of these is the famous result that the $S$ transformation implements electro-magnetic duality or so-called \textit{Montonen-Olive duality} on the theory, while the second is the effect by which magnetic monopoles become dyonic upon shift of the theta angle. 

\sm

In the context of the example of $\mathfrak{g} = \mathfrak{su}(2)$, examining the action of $S$ and $T$ on the charge lattices in Figure \ref{fig:chargelattices} reveals that 
\bea
S: \hspace{0.2 in} SU(2) ~ &\rightarrow& SO(3)_+ \hspace{0.6 in} T:  \hspace{0.23 in}SU(2) \,\,\, \rightarrow\,\,\, SU(2)
\no\\
SO(3)_+ &\rightarrow& SU(2) \hspace{1.1 in}  SO(3)_+ \,\,\,\rightarrow\,\,\, SO(3)_-
\no\\
SO(3)_- &\rightarrow& SO(3)_- \hspace{1 in}  SO(3)_- \,\,\,\rightarrow\,\,\, SO(3)_+
\eea
Thus we see that no individual theory is actually self-dual under $SL(2,\ZZ)$. Instead, one can think of the set of three theories as transforming as a three-dimensional vector-valued modular function of $SL(2,\ZZ)$. Alternatively, each individual theory can be said to be invariant under an appropriate congruence subgroup. For example, for $G_H = SU(2)$ the theory is invariant under $T$ and $ST^2S$, which together generate the subgroup $\Gamma_0(2) \subset SL(2,\ZZ)$. Thus the $SU(2)$ theory is self-dual under $\Gamma_0(2)$. 

\sm

Let us close by mentioning that, in the cases in which there is only a single global variant of the theory, e.g. for $\mathfrak{g} = \mathfrak{e}_8$ mentioned above, then the full $SL(2,\ZZ)$ really does act as a duality group for the theory.

\subsection{The Seiberg-Witten Solution}

The Seiberg-Witten solution for an $\cN=2$ supersymmetric Yang-Mills theory in four space-time dimensions  with gauge group $G$ of rank $r$,  is given in terms of the vacuum expectation values   of the complex scalars $a_I$ of the gauge multiplet and their magnetic duals $a_{DI}$  in the Coulomb phase. When hypermultiplets are present in a representation $\cR \oplus \bar \cR$ of the gauge group $G$, we shall denote their $G$-invariant masses by $m_f$ where the index $f$ runs over the various irreducible components $\cR_f$ of $\cR$. The prepotential $\cF$ and the matrix $\tau_{IJ}$ for $I,J=1,\cdots, r$ of effective couplings and mixings of the unbroken $U(1)^r$ gauge multiplets are related to these expectation values by,
\bea
a_{DI} = { \p \cF \over \p a_I} 
\hskip 1in
\tau_{IJ} = { \p a_{DI} \over \p a_J} = { \p^2 \cF \over \p a_I \, \p a_J}
\eea
The original Seiberg-Witten solution was obtained for gauge group $SU(2)$, but the results were quickly generalized to other semi-simple compact gauge groups $G$. 

\sm

The building blocks of the Seiberg-Witten solution are a curve $\cC$ and a differential $\lambda$ on~$\cC$. More precisely, $\cC=\cC(u,m)$ is a family of genus $r$ Riemann surfaces parametrized by $G$-invariant  local complex coordinates $u_1, \cdots u_r$ and  hypermultiplet masses $m_f$. A convenient definition of the $u_I$ may be given in terms of the gauge-invariant traces $\tr (\phi^n)$ of the gauge scalar $\phi$.  The differential $\lambda$ is  meromorphic on $\cC$ and its  partial derivatives with respect to the parameters $u_I$ are holomorphic Abelian differentials, possibly up to the addition of exact differentials of meromorphic functions on $\cC$. In the presence of  hypermultiplets, $\lambda$ has poles whose residues are given by the masses $m_f$.

\sm

The Seiberg-Witten solution gives the vacuum expectation values of the gauge  scalars $a_I$ and their magnetic duals $a_{DI}$ in terms of  the periods of the differential $\lambda$ on a canonical basis of homology cycles $\mA_I$ and $\mB_I$ of the curve $\cC$,  
\bea
2 \pi i a_I = \oint _{\mA_I} \lambda 
\hskip 1in
2 \pi i a_{DI} = \oint _{\mB_I} \lambda 
\eea
By the very construction of this set-up, the matrix $\tau_{IJ}$ has positive imaginary part as required on physical grounds by the fact that it provides the couplings for the $U(1)^r$ gauge field strengths $F_{\mu \nu}^I$, which must be positive. To see this, we combine the following formulas, 
\bea
\tau_{IJ} = \sum _K { \p a_{DI} \over \p u_K} { \p u_K \over \p a_J}
\hskip 0.6in 
2 \pi i { \p a_I \over \p u_K} = \oint _{\mA_I} { \p \lambda  \over \p u_K} 
\hskip 0.6in
2 \pi i { \p a_{DI} \over \p u_K} = \oint _{\mB_I} { \p \lambda  \over \p u_K} 
\eea
to observe that $\tau_{IJ}$ is the period matrix of the curve $\cC(u)$, whose imaginary part is positive by the Riemann bilinear relations; see appendix \ref{sec:Riemannrelations}. 

\sm

To illustrate the general construction outlined above, we present the Seiberg-Witten curves and differentials for the gauge group $G=SU(N)$ in two examples.  The first has $N_f< 2 N$ hypermultiplets with masses $m_f$  in the defining  representation of $G$ and is asymptotically free. The second has one hypermultiplet with mass $m$ in the adjoint representation of $G$ and is referred to as the $\cN=2^*$ theory. The latter is UV finite and reduces to the $\cN=4$ super Yang-Mills theory upon setting the mass $m$ of the adjoint hypermultiplet to zero, and to the pure $\cN=2$ theory without hypermultiplets by sending the mass $m$ to infinity while suitably scaling the gauge-coupling to zero. 

\subsubsection{The $N_f < 2 N$ theory for gauge group $SU(N)$}

For the theory with $N_f<2N$ hypermultiplets in the fundamental representation of $SU(N)$ with masses $m_f$, the differential $\lambda$ and the curve $\cC$ are given by,
\bea
 \lambda  = \left (  A' - \half (A-y) { B' \over B} \right ) { x dx \over y}
\hskip 1in
y^2  = A(x)^2 - B(x) 
\eea
where the polynomials $A$ and $B$ are given as follows,
\bea
A(x)  = \prod _{i=1}^N (x-\bar a_i)  \hskip 1in B(x)  = \Lambda ^{2N-N_f} \prod_{f=1}^{N_f} (x-m_f)
\eea
and $\Lambda$ is the renormalization scale of the asymptotically free theory.  The gauge-invariant coordinates $u_I$ are given in terms of symmetric polynomials in the variables $\bar a_i$ which, since the gauge group is $SU(N)$, must satisfy $\sum_i \bar a_i=0$. The curve $\cC$ is hyperelliptic of genus $N-1$. The differential $\lambda$ has simple poles with residue $m_f$ at the points $(x,y)=(m_f, -A(m_f))$ on the second sheet of the hyper-elliptic curve, but is regular at the points $(x,y)=(m_f, A(m_f))$ on the first sheet of $\cC$.

\subsection{The $\cN=2^*$ theory for gauge group $SU(N)$}

The $\cN=2^*$ theory possesses a rich modular structure, in part because it may be obtained as a mass deformation from the $\cN=4$ theory, shares the same ultraviolet finiteness properties, and  has a marginal deformation controlled by the complex gauge coupling $\tau$.  Its Seiberg-Witten theory may be developed in terms of integrable mechanical systems, of the Hitchin or elliptic Calogero-Moser type. It is the latter approach that we shall follow here.

\subsubsection{The Seiberg-Witten solution}

To formulate the Seiberg-Witten solution for the $\cN=2^*$ theory for gauge group $SU(N)$, we introduce a torus $\CC/\Lambda$ for a lattice  $\Lambda = \om_1 \ZZ \oplus \om_2 \ZZ$ where the modulus  $\tau = \om _2/\om_1$ is related to the gauge coupling $g$ and the $\theta$ angle of the non-Abelian gauge theory  by,
\bea
\tau = { \theta \over 2 \pi} + { 4 \pi i \over g^2}
\eea
Denoting the mass of the adjoint hypermultiplet by $m$, the differential $\lambda$ and the curve $\cC$ are constructed in terms of a local complex coordinate $z$ on $\CC/\Lambda$,
\bea
\lambda = k \, dz  \hskip 1in  R_N(k,z |\Lambda) =0
\eea
The function $R_N(k,z|\Lambda)$ is given by the characteristic polynomial for the Lax operator of the elliptic Calogero-Moser system  for the root system $\mathfrak{a}_{N-1}$ of the Lie algebra $\mathfrak{su}(N)$,
\bea
R_N(k,z|\Lambda)= \det \big ( kI - L(z|\Lambda) \big )
\eea
The components of the matrix $L(z|\Lambda)$ and its conjugate Lax matrix $M(z|\Lambda)$  are given by,
\bea
L_{ij} (z|\Lambda) & = & p_i \delta_{ij} - m(1-\delta_{ij}) \Phi(x_i-x_j,z|\Lambda) 
\no \\
M_{ij} (z |\Lambda) & = & m (1-\delta_{ij} ) \Phi'(x_i-x_j,z|\Lambda) + m \delta_{ij} \sum_{k \not= i} \wp(x_i-x_k|\Lambda) 
\eea
where $\wp$ is the Weierstrass function and $\Phi$ is the Lam\'e function,
\bea
\Phi(x,z|\Lambda) = { \sigma (z-x |\Lambda) \over \sigma (z|\Lambda) \sigma (x|\Lambda )} \, e^{x\zeta(z|\Lambda ) }
\eea
given in terms of the Weierstrass $\zeta$- and $\sigma$-functions for the lattice $\Lambda$; see subsection \ref{wpZ} for definitions of $\zeta$ and $\sigma$. The variables $x_i$ and $p_i$ are the positions and momenta for the Calogero-Moser system. The integrability condition $\dot L = [M,L]$ guarantees that  $R_N(k,z)$ is a conserved quantity for all values of $z,k$, and $\Lambda$.

\subsubsection{The Seiberg-Witten curve}

\begin{figure}[tp]
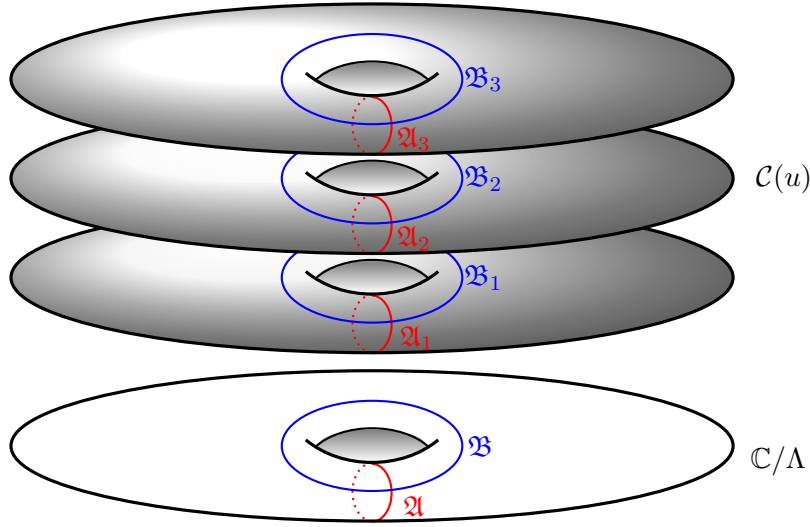

\begin{center}
\tikzpicture[scale=0.4, line width=0.35mm]
\begin{scope}[xshift=0cm, yshift=0cm]
\shade[ball color = gray!10, opacity = 1] (0,0) ellipse  (12cm and 2.5cm);
\shade[color=gray!0] (0,0) ellipse  (1.8cm and 0.545cm);
\draw[thick, red] (0,-2.52) arc (-90:90:0.65cm and 0.96cm);
\draw[thick, red,dotted] (0,-0.575) arc (90:270:0.65cm and 0.96cm);
\draw[very thick] (0,0) ellipse  (12cm and 2.5cm);
\draw[very thick] (-2.2,0.2) .. controls (-1,-0.8) and (1,-0.8) .. (2.2,0.2);
\draw[thick] (-1.9,-0.05) .. controls (-1,0.8) and (1,0.8) .. (1.9,-0.05);
\draw[thick, blue](0,0) ellipse  (3cm and 1.5cm);
\draw[red] (1.5,-2) node{\small $\mA_1$};
\draw[blue] (3.7,0) node{\small $\mB_1$};

\end{scope}
\begin{scope}[xshift=0cm, yshift=3.3cm]
\shade[ball color = gray!5, opacity = 1] (0,0) ellipse  (12cm and 2.5cm);
\shade[color=gray!0] (0,0) ellipse  (1.8cm and 0.545cm);
\draw[thick, red] (0,-2.52) arc (-90:90:0.65cm and 0.96cm);
\draw[thick, red,dotted] (0,-0.575) arc (90:270:0.65cm and 0.96cm);
\draw[very thick] (0,0) ellipse  (12cm and 2.5cm);
\draw[very thick] (-2.2,0.2) .. controls (-1,-0.8) and (1,-0.8) .. (2.2,0.2);
\draw[thick] (-1.9,-0.05) .. controls (-1,0.8) and (1,0.8) .. (1.9,-0.05);
\draw[thick, blue](0,0) ellipse  (3cm and 1.5cm);
\draw[red] (1.4,-2) node{\small $\mA_2$};
\draw[blue] (3.7,0) node{\small $\mB_2$};
\draw  (13.7,0) node{\small $\cC(u)$};
\end{scope}
\begin{scope}[xshift=0cm, yshift=6.6cm]
\shade[ball color = gray!0, opacity = 1] (0,0) ellipse  (12cm and 2.5cm);
\shade[color=gray!0] (0,0) ellipse  (1.8cm and 0.545cm);
\draw[thick, red] (0,-2.52) arc (-90:90:0.65cm and 0.96cm);
\draw[thick, red,dotted] (0,-0.575) arc (90:270:0.65cm and 0.96cm);
\draw[very thick] (0,0) ellipse  (12cm and 2.5cm);
\draw[very thick] (-2.2,0.2) .. controls (-1,-0.8) and (1,-0.8) .. (2.2,0.2);
\draw[thick] (-1.9,-0.05) .. controls (-1,0.8) and (1,0.8) .. (1.9,-0.05);
\draw[thick, blue](0,0) ellipse  (3cm and 1.5cm);
\draw[red] (1.4,-2) node{\small $\mA_3$};
\draw[blue] (3.7,0) node{\small $\mB_3$};
\end{scope}
\begin{scope}[xshift=0cm, yshift=-5.6cm]
\shade[color=gray!0] (0,0) ellipse  (1.8cm and 0.545cm);
\draw[thick, red] (0,-2.52) arc (-90:90:0.65cm and 0.96cm);
\draw[thick, red,dotted] (0,-0.575) arc (90:270:0.65cm and 0.96cm);
\draw[very thick] (0,0) ellipse  (12cm and 2.5cm);
\draw[very thick] (-2.2,0.2) .. controls (-1,-0.8) and (1,-0.8) .. (2.2,0.2);
\draw[thick] (-1.9,-0.05) .. controls (-1,0.8) and (1,0.8) .. (1.9,-0.05);
\draw[thick, blue](0,0) ellipse  (3cm and 1.5cm);
\draw[red] (1.4,-2) node{\small $\mA$};
\draw[blue] (3.6,0) node{\small $\mB$};
\draw (13.5,-0.5) node{\small $\CC/\Lambda$};

\end{scope}
\endtikzpicture
\caption{\textit{Schematic representation of the Seiberg-Witten curve $\cC(u)$ and the underlying torus $\CC/\Lambda$  for  gauge group $SU(3)$. Canonical homology cycles $\mA_1, \mA_2, \mB_1, \mB_2$ on $\cC(u)$ may be provided by uplifting the cycles $\mA$ and $\mB$ of a canonical homology basis on $\CC/\Lambda$. The identifications between the different sheets are not shown here. }
\label{15.fig:1}}
\end{center}
\end{figure}

The curve $R(k,z|\Lambda)$ admits an explicit presentation directly in terms of $\tet$-functions, 
\bea
\label{15.SUN}
R_N(k,z|\Lambda) = \tet_1 \left ({ z \over  \om_1} \bigg |\tau \right )^{-1}  \tet_1 \left ({ z \over  \om_1} - m { \p \over \p \tilde k}  \bigg |\tau \right ) P(\tilde k) \bigg |_{\tilde k = k + m h_1(z|\Lambda)}
\eea
Here $P(k)$ is a monic polynomial in $k$ of degree $N$  with vanishing $k^{N-1}$ term, reflecting the fact that the gauge group is $SU(N)$ as opposed to $U(N)$, 
\bea
P(k) = k^N + \sum_{n=0}^{N-2} u_n k^n
\eea
The $N-1$ remaining  independent coefficients $u_n$ are the moduli of the Coulomb branch. The functions $h_n$ are defined by, 
\bea
h_n(z|\Lambda) = \tet_1 \left ({ z \over  \om_1} \bigg |\tau \right )^{-1}  {\p ^n \over \p z^n}  \tet_1 \left ({ z \over  \om_1}  \bigg |\tau \right )
\eea
One may think of the roots of the polynomial as the vacuum expectation values of the gauge scalars $a_I$ when $m=0$.   The corresponding curves for $N=2,3,4$ are given by,\footnote{The Weierstrass function $\wp(z|\tau)$ differs from $h_2(z|\Lambda) - h_1(z|\Lambda)^2$  by a $z$-independent term given by 
$ \wp(z|\Lambda) = - \p_z^2 \ln \tet_1(z/\om_1|\tau) -\HE_2(\tau)/(3 \om_1^2)$, as may be established using (\ref{4.WPE}).  In the final expressions given in (\ref{15.curves})  we have used the freedom to redefine the moduli parameters $u_n$ by $\Lambda$-dependent but $z$-independent shifts, such as for example $u_0 \to u_0 - m^2 \, \HE_2/(3\om_1^2 )$ for $N=2$. These shifts are immaterial as the $u_n$ provide an arbitrary coordinate system for the moduli of the Coulomb branch. The resulting shifted versions of (\ref{15.curves})  have the advantage of enjoying transparent  modular transformation properties.}
\bea
\label{15.curves}
R_2(k,z|\Lambda) & = & k^2 + u_0 - m^2 \wp 
 \\
R_3(k,z|\Lambda) & = & k^3 + (u_1- 3 m^2 \wp)k + u_0 + m^3 \wp'
\no \\
R_4(k,z|\Lambda) & = & k^4 + (u_2 -6 m^2 \wp) k^2 +(u_1+4 m^3 \wp') k + u_0 -m^4 \wp'' + 3 m^4 \wp^2
\no
\eea
The SW curve may thus be constructed from $N$ copies, or sheets, of the underlying torus $\CC/\Lambda$, namely one for each solution $k_I$, glued together wherever two sheets $I \not= J$ intersect one another. A schematic representation of the underlying torus $\CC/ \Lambda$ and the curve $\cC(u)$ for gauge group $SU(3)$ is presented in Figure \ref{15.fig:1}.

\sm

A canonical choice for $\mA, \mB$ cycles on the torus $\CC/\Lambda$ may be uplifted to a canonical homology basis of cycles $\mA_I, \mB_I$ for $I=1, \cdots, N-1$ on the sheet  corresponding to the solution $k_I$, as illustrated for $N=3$  in Figure \ref{15.fig:1}. The modular group $SL(2,\ZZ)$ acts on the cycles $\mA,\mB$ by,
\bea
\left ( \bma \mB \cr \mA \ema \right ) \to \left ( \bma \mB' \cr \mA' \ema \right )
= \left ( \bma a & b \cr c & d \ema \right ) \left ( \bma \mB \cr \mA \ema \right )
\hskip 1in 
\left ( \bma a & b \cr c & d \ema \right ) \in SL(2,\ZZ)
\eea
and this transformation may be lifted to a transformation on the homology cycles $\mA_I$ and $\mB_I$ on the curves $\cC(u)$ by a subgroup $SL(2,\ZZ)$ of the full modular group of the curve $Sp(2N-2,\ZZ)$. However, independent of this induced action of $SL(2,\ZZ)$, we may act on the cycles $\mA_I, \mB_I$ of the curve $\cC(u)$ by the full modular group $Sp(2N-2,\ZZ)$.   Thus, the action of $SL(2,\ZZ)$ on $\mA,\mB$ and the action of $Sp(2N-2,\ZZ)$ on $\mA_I, \mB_I$ may be viewed as independent of one another, producing a larger modular structure governed by the group, 
\bea
SL(2,\ZZ) \ltimes Sp(2N-2,\ZZ)
\eea
The corresponding construction gives rise to the notion of \textit{bi-modular forms}, and may be extended to the superconformal $\cN=2$ theory with gauge group $SU(N)$ and $2N$ hypermultiplets in the fundamental representation of $SU(N)$.

\sm

\subsubsection{Gauge scalar vacuum expectation values and pre-potential}

In terms of these data, the expectation values $a_I$ and their magnetic duals $a_{DI}$ are given by,
\bea
2 \pi i a_I = \oint _{\mA_I} k \, dz = \oint _\mA k_I dz
\hskip 1in 
2 \pi i a_{DI} = \oint _{\mB_I} k \, dz = \oint _\mB k_I dz
\eea
The pre-potential $\cF$ satisfies the following renormalization-type differential equation in $\tau$,
\bea
\label{eq:RGtype}
{ \p \cF \over \p \tau} \bigg |_{m,a_I} = { 1 \over 4 \pi i} \sum_{I=1}^{N-1} \oint _{\mA_I} dz \, k^2  
= { 1 \over 4 \pi i} \oint _\mA dz \sum_{I=1}^{N-1}  k_I^2 
\eea
The partial derivative is considered at fixed $m$ and $a_I$.  Note that $\p \cF / \p \tau$ receives no perturbative contributions, as those are independent of $\tau$. The advantage of this formula is that we can now evaluate $a_I$ and $\p \cF / \p \tau$ entirely in terms of integrals over $\mA$-periods without having to carry out the rather complicated analysis of turning points and regularization required when evaluating the $\mB$-periods. 

\sm

In the limit $m \to 0$, we recover the $\cN=4$ theory and we have $R(k,z|\tau) = P(k)$, so that the roots of $P(k)$ correspond to the vacuum expectation values of the gauge scalar of the $\cN=2$ gauge multiplet. Since we have $\oint _\mA dz= \om_1$ by definition of these periods, we also have $2 \pi i a_I =  \om_1 k_I$ in the $m\to 0 $ limits, and thus we  set $\om_1 = 2\pi i$.

\subsection{The $\cN=2^*$ theory for gauge group  $SU(2)$ }
\label{sec:15.7}

It is instructive to analyze the action of the duality group $SL(2,\ZZ)$ for the case of gauge group $SU(2)$ where many formulas and transformation laws can be made completely explicit.  The curve $\cC$ of (\ref{15.curves})  and differential $\lambda$ may be parametrized as follows, 
\bea
k^2 = \kappa^2 + m^2 \wp (z|\Lambda)   \hskip 1in \lambda =  \sqrt{\kappa^2 + m^2 \wp(z|\Lambda )} \, dz
\eea
where we have set $u_0=- \kappa^2$ for later convenience. As $z \to 0$, we have $k \sim \pm m /z$ which produces the pole in the differential $\lambda$ whose residue is the hypermultiplet mass, as expected on the basis of the general construction outlined above. To evaluate the periods $a$ and $a_D$, we begin by evaluating the following elliptic integrals,  using the relation $\wp(z|\Lambda)=-\zeta'(z|\Lambda )$ and the monodromy relations of $\zeta(z|\Lambda )$ given in (\ref{zmon}), 
\bea
\label{15.AB1}
A_1 & = & { 1 \over 2 \pi i} \oint _\mA dz \, \wp(z|\Lambda) 
= -{ 2 \, \zeta (\thalf |\tau)  \over (2 \pi i)^2} = { 1 \over 12} \, \HE_2(\tau)
\no \\
B_1 & = & { 1 \over 2 \pi i} \oint _\mB dz \, \wp(z|\Lambda) 
= -{ 2 \, \zeta (\tfrac{\tau}{2}  |\tau) \over (2 \pi i)^2}  = { \tau \over 12} \, \HE_2(\tau) +{ 1 \over 2 \pi i } 
\eea
where the last equality was obtained by combining the formulas (\ref{2.eta1}), (\ref{eq:deletarel}),  (\ref{eq:tetpetarel}), and (\ref{3.E2Del}). This immediately allows us to compute the right side of the RG equation (\ref{eq:RGtype}), 
\bea
\label{15.RG2}
{ \p \cF \over \p \tau} \bigg |_{m,a} = { 1 \over 4 \pi i} \oint _\mA dz k_+^2
=  { 1 \over 4 \pi i} \oint _\mA dz \Big ( \kappa^2 + m^2 \wp(z|\Lambda) \Big )= \half \kappa^2  +  {m^2  \over 24} \HE_2
\eea

\subsubsection{Expansion in powers of the mass $m$}

To obtain the periods $a$ and $a_D$, we expand the SW differential $\lambda$ for $SU(2)$ in powers of $m$,  
\bea 
\lambda = \kappa \sum _{n=0}^\infty {\Gamma (n-\thalf) \over \Gamma (- \thalf) n!} \left ( - { m^2 \over \kappa^2} \right )^n   \wp(z|\Lambda)^n \, dz
\eea
This expansion will be uniformly convergent for small enough $m$ provided $z$ is kept away from $0$, which can always be achieved when we integrate over suitably chosen $\mA$ and $\mB$ cycles. To evaluate the periods $a$ and $a_D$, we compute the integrals, 
\bea
A_n = { 1 \over 2 \pi i} \oint _\mA dz \, \wp(z|\Lambda) ^n
\hskip 1in 
B_n = { 1 \over 2 \pi i} \oint _\mB dz \, \wp(z|\Lambda) ^n
\eea
Clearly, we have $A_0=1$ and $B_0=\tau$ while $A_1$ and $B_1$ were already evaluated in (\ref{15.AB1}). For $n \geq 2$, we make use of the following recursion relation,
\bea
\label{15.recur1}
(8n-4) \wp^n = (2n-3) g_2 \wp^{n-2} +(2n-4) g_3 \wp^{n-3} + 2 \big ( \wp ' \wp^{n-2} \big )'
\eea
which is readily established using the defining equation $(\wp')^2 = 4 \wp^3 - g_2 \wp -g_3$. Eliminating  $g_2, g_3$ in favor of $\HE_4, \HE_6$ for the period normalization $\om_1=2\pi i$, 
\begin{align}
g_2 & = { 4 \pi^4 \, \HE_4 \over 3  \om_1^4} = {\HE_4 \over 12}
&
 g_3 & = { 8 \pi^6 \, \HE_6 \over 27  \om_1^6} = - { \HE_6 \over 216}
\end{align}
and integrating the relation (\ref{15.recur1}), we obtain the following recursion relation, satisfied by both $A_n$ and $B_n$ with $n \geq 2$,  
\bea
\label{15.recur2}
(8n-4) A_n = { 2n-3 \over 12} \, \HE_4 A_{n-2} - { n-2 \over 108} \, \HE_6 A_{n-3}
\eea
The coefficients of this recursion relation are modular forms, but the initial conditions for $A_1$ and $B_1$ involve the quasi-modular form $\HE_2$. Linearity of (\ref{15.recur2}) guarantees that the solutions $A_n, B_n$ are linear in $\HE_2$ and may be cast in the following form, 
\bea
A_n = K_n + { \HE_2 \over 12} \, L_{n-1}
\hskip 1in 
B_n = \tau A_n + { L_{n-1} \over 2 \pi i}
\eea
where $K_n$ and $L_n$ are modular forms of weight $2n$. The function $A_n$  satisfies (\ref{15.recur2}) given for $A_n$ with initial conditions $A_0$ and $A_1$, while $L_n$ satisfies (\ref{15.recur2})  for $B_n-\tau A_n$ with initial conditions $B_0-\tau A_0=0$ and $B_1-\tau A_1 = (2 \pi i)^{-1}$.  Substituting these expressions into the expansion in powers of $m$ of the period integrals, we obtain  the $a$-period, 
\bea
\label{15.a1}
a = \kappa \sum _{n=0}^\infty {\Gamma (n-\thalf) \over \Gamma (- \thalf) n!} \left ( - { m^2 \over \kappa^2} \right )^n
\left ( K_n + { \bE_2 \over 12} \, L_{n-1}  \right )
\eea
and the following result for the combination $a_D - \tau a$ of the $a_D$ and $a$ periods, 
\bea
\label{15.a2}
a_D-\tau a = { \kappa \over 2 \pi i} \sum _{n=0}^\infty {\Gamma (n-\thalf) \over \Gamma (- \thalf) n!} 
\left ( - { m^2 \over \kappa^2} \right )^n  L_{n-1} 
\eea
We shall establish below that the combination $a_D-\tau a$ of the periods transforms as a modular form in the modulus $\tau$.

\subsubsection{Low orders and perturbative contribution}

The Eisenstein series $\HE_{2n}(\tau)$ for $n \geq 1$ are normalized to 1 at the cusp $\tau \to i \infty$ and admit a series expansion  at the cusp in powers of $q=e^{2 \pi i \tau}$. The following coefficients vanish, 
\bea
K_1=L_{-1}=L_1=0
\eea
 while the non-vanishing low order coefficients are given by,
\begin{align}
K_0& = 1 & K_2 & = { \HE_4 \over 144} & K_3 & = - {\HE_6 \over 2160} & K_4 &= { 5 \, \HE_4^2 \over 48384} & 
K_5 & = - {\HE_4 \HE_6 \over 77760} 
\no \\
L_0 & = 1 & L_2 & = {\HE_4 \over 80} & L_3 & = - {\HE_6 \over 1512} & L_4 & = { 7 \, \HE_4^2 \over 34560} 
& L_5 &= - { 29 \, \HE_4 \HE_6 \over 1330560}
\end{align}
To solve for the prepotential, we may either integrate the RG equation, or integrate $a_D = \p \cF / \p a$. In either case, we need to express $\kappa$ as a function of $\tau, m, a$, which can be done order by order in $m$ by inverting the series given in (\ref{15.a1}) for $a$ as a function of $\tau, m, \kappa$, 
\bea
\kappa  =  a \Bigg [ 1 - {\HE_2 \over 6} { m^2 \over (2a)^2} 
+ { \HE_4 - 2 \HE_2^2 \over 72} {m^4 \over (2a)^4}
- { 20 \HE_2^3 -11 \HE_2 \HE_4-4 \HE_6 \over 216} {m^6 \over (2a)^6}
+ \cO(m^{8}) \bigg ]
\quad
\eea
The expression for $\cF$ derived from the RG equation is given by,
\bea
\cF(\tau, m, a) = \half \tau a^2 + \cF_0(m,a) - { 1 \over 2 \pi i} \bigg  [ {\HE_2 \over 24} {m^4 \over (2a)^2}
+\left ( {\HE_4 \over 720} +{\HE_2^2 \over 144} \right ) {m^6 \over (2a)^4} \bigg ]
+\cO(m^{8})
\eea
where $\cF_0$ is independent of $\tau$ and undetermined by the RG equation. However, it may be determined from the expressions for $a_D$ and $a$ that we have already obtained, and we have,
\bea
\cF_0(m,a)  = c m^2 + { m^2 \over 4 \pi i}  \ln { a \over m}
\eea 
In the weak coupling limit, where $q \to 0$ and $E_2, E_4, E_6 \to 1$ we evaluate the effective gauge coupling $\tau_{{\rm IR}}$ as follows, 
\bea
\tau_{{\rm IR}} =  \tau - { 1 \over \pi i} \sum _{n=1}^\infty { 1 \over n} \left ( {m^2 \over  4a^2} \right )^n + \cO(q)
= \tau +{ 1 \over \pi i} \ln \left ( 1-{m^2 \over  4a^2} \right ) + \cO(q)
\eea
Thus, the perturbative logarithm at the mass of the massive vector boson $m^2 = 4 a^2$ is the result of re-summing an infinite series of genuine modular form contributions.  Note that the logarithmic term produced by $\cF_0$ is required to obtain the $n=1$ term needed to complete the series for the logarithm.

\subsubsection{Modular properties}

Exceptionally, in this subsection only,  we shall parametrize $SL(2,\ZZ)$ transformations with Greek letters in order to avoid a clash with the notation for the periods $a$ and $a_D$. Under modular modular transformations, 
\bea
\tilde \tau = { \a \tau + \b \over \g \tau + \delta} 
\hskip 0.6in 
\left ( \bma \tilde \om_2 \\ \tilde \om_1 \ema \right ) 
=  \left ( \bma \a & \b  \\ \g & \delta  \ema \right )  \left ( \bma \om_2 \\ \om_1 \ema \right ) 
\hskip 0.6in 
\left ( \bma \a & \b  \\ \g & \delta  \ema \right ) \in SL(2,\ZZ)
\eea
we have the following transformations of Eisenstein series for $n \geq 1$, 
\bea
\HE_{2n}(\tilde \tau) = (\g \tau + \delta)^{2n} \HE_{2n}(\tau) +   { 12 \over 2 \pi i} \, \g (\g \tau + \delta) \,  \delta_{n,1}
\eea
and the Weierstrass function, in which we keep the period $  \om _1 = 2 \pi i$ fixed, 
\bea
\wp( \tilde z |\tilde \tau) = (\g \tau + \delta)^2 \wp(z|\tau) 
\hskip 1in \tilde z = { z \over \g \tau + \delta}
\eea
The transformation laws of the SW data may be deduced as follows. First of all, modular invariance of the SW differential implies the transformation law for $k^2$, 
\bea
\lambda = k \, dz = \tilde k \, d \tilde z \hskip 1in \tilde k^2 = (\g \tau + \delta)^2k^2
\eea
Using the relation between $z, k,$ and $\kappa$, we readily obtain,
\bea
\tilde  \kappa = (\g \tau + \delta) \kappa
\eea
Here, the sign ambiguity introduced by taking the square root of $\kappa^2$ amounts to implementing the transformation $-I \in SL(2,\ZZ)$ on $\kappa$.  The transformation laws  on $a_D, a$ may be deduced from their explicit expressions in terms of $\kappa$ and $\tau$. We begin by observing in (\ref{15.a2}) that the combination $a_D - \tau a$ transforms as a modular form of weight $(-1,0)$, 
\bea
\tilde a_D - \tilde \tau \tilde a = (\g \tau +\delta)^{-1} ( a_D - \tau a)
\eea
Next, we use the explicit expression for $a$ from (\ref{15.a1}) along with the transformation laws of $\kappa$, $K_{2n},$ and $L_{2n-2}$ as modular forms of weight $(1,0), (2n,0)$, and $(2n-2,0)$ respectively, as well as the transformation law of $\HE_2$, and we obtain, 
\bea
\tilde a = (\g \tau + \delta) a + \gamma (a_D-\tau a) = \gamma \, a_D + \delta \, a
\eea
where the second term in the middle equality arises from the inhomogeneous transformation term of $\HE_2$. Assembling both transformation laws we obtain, 
\bea
\left ( \bma \tilde a_D \\ \tilde a \ema \right ) =  \left ( \bma \a & \b  \\ \g & \delta  \ema \right )  \left ( \bma a_D \\ a \ema \right ) 
\eea
which is precisely the transformation law induced by the action of the modular group on the cycles $\mA,\mB$ for a modular invariant SW differential.

\subsection{Linear quiver chains from a limit of $\cN=2^*$}

We now return to the $\cN=2^*$ theory for the gauge group $SU(N)$, whose independent free parameters are the UV gauge coupling $\tau$, the hypermultiplet mass $m$, and the vacuum expectation values $a_I$  of the gauge scalars, or equivalently  the roots of the polynomial $P(k)$.   The masses of the gauge boson BPS states are $|a_I - a_J|$ while the masses of the BPS hypermultiplet states are $|a_I-a_J-m|$. Several interesting limits may be taken in this set-up. 

\sm

A first interesting limit is obtained by keeping the vacuum expectation values $a_I$ fixed while sending the  hypermultiplet mass $m \to \infty$ and letting $\tau \to i \infty$ at the rate dictated by the one-loop renormalization group $\beta$-function with renormalization scale $\Lambda$ (not to be confused with the lattice $\Lambda$), $\Lambda^{2N} = (-)^N m^{2N} q$ where $q=e^{2 \pi i \tau}$. This gives the pure $\cN=2$ theory with gauge group $SU(N)$. 

\begin{figure}[tp]
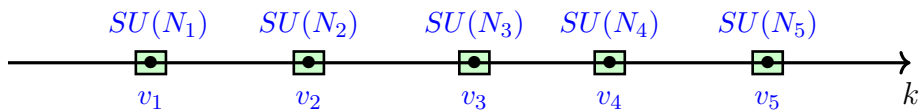

\begin{center}
\tikzpicture[scale=1, line width=0.35mm]
\begin{scope}[xshift=0cm, yshift=0cm]
\draw[ fill=green!20!] (-4.3,0.15) rectangle (-3.9,-0.15);
\draw[ fill=green!20!] (-2.2,0.15) rectangle (-1.8,-0.15);
\draw[ fill=green!20!] (0,0.15) rectangle (0.4,-0.15);
\draw[ fill=green!20!] (1.8,0.15) rectangle (2.2,-0.15);
\draw[ fill=green!20!] (4.3,0.15) rectangle (3.9,-0.15);

\draw[very thick,->] (-6,0) -- (6,0);
\draw (0.2,0) node{$\bullet$};
\draw (2,0) node{$\bullet$};
\draw (4.1,0) node{$\bullet$};
\draw (-2,0) node{$\bullet$};
\draw (-4.1,0) node{$\bullet$};
\draw (6,-0.4) node{$k$};
\draw[blue] (-4.1,-0.5) node{\small $v_1$};
\draw[blue] (-2,-0.5) node{\small $v_2$};
\draw[blue] (0.2,-0.5) node{\small $v_3$};
\draw[blue] (2,-0.5) node{\small $v_4$};
\draw[blue] (4.1,-0.5) node{\small $v_5$};
\draw[blue] (-4,0.5) node{\small $SU(N_1) $};
\draw[blue] (-2,0.5) node{\small $SU(N_2) $};
\draw[blue] (0.2,0.5) node{\small $SU(N_3) $};
\draw[blue] (2,0.5) node{\small $SU(N_4) $};
\draw[blue] (4.1,0.5) node{\small $SU(N_5) $};
\end{scope}

\endtikzpicture
\caption{\textit{The distribution of vacuum expectation values for a breaking from $SU(N)$ to $SU(N_1) \times \cdots \times SU(N_5)$ with $N=N_1+\cdots+N_5$ where the $v_p$ are separated from their nearest neighbors by $\pm m$ plus a term $\Lambda_p$ that is held fixed as $m, v_p \to \infty$. The green rectangles schematically represent the ranges in which the $x^p_{i_p}$ can vary.}
\label{15.fig:2}}
\end{center}
\end{figure}

A second interesting limit is obtained by letting $m \to \infty$ and $\tau \to \infty$, but now also sending some of the vacuum expectation values $a_I$ to infinity. Upon carefully tuning the correlations between these limits, one can send some of the hypermultiplet masses and some of the gauge boson masses to $\infty$. This limit will break the gauge symmetry to $
SU(N_1) \times \cdots \times SU(N_P) $
 with $N=N_1+\cdots + N_P$ and $N_p \geq 2$ for all $p=1,\cdots, P$, while any $U(1)$ factors in the gauge group that arise may be shown to decouple. To obtain the above breaking pattern, we arrange the classical values for the expectation values in a linear chain. We partition all the order parameters $k_I$ for $I=1, \cdots, N$ into $P$ groups with $I=i_1=1, \cdots, N_1$ labeling the indices in the first group, $I=N_1+i_2$ with $i_2=1,\cdots, N_2$ those in the second group; $I=N_1+\cdots +N_{p-1} +i_p $ with $i_p=1,\cdots, N_p$ those in group $p$, and so on until $I = N_1 + \cdots N_{P-1} +i_P$ with $i_P =1,\cdots, N_P$. We now set,
\bea
k_{N_1 + \cdots + N_p + i_p} = v_p + x^p _{i_p} \hskip 1in
p=1,\cdots, P
\eea
The decomposition is not unique as it stands since $v_p$ and all $x^p_{i_p}$ may be shifted by an $i_p$-independent amount such that the corresponding $k_I$ is unchanged. To make the decomposition unique, we impose the following further conditions, 
\bea
\sum_{p=1}^P N_p v_p =0 
\hskip 1in
\sum _{i_p=1}^{N_p} x^p _{i_p} =0
\eea
We then order the values of $v_p$ such that,
\bea
v_{p+1} - v_p = m + \Lambda _p
\eea 
and then send $v_p, m \to \infty$ along with $\tau \to i \infty$ while keeping all $\Lambda_p$ and $x^p_{i_p}$ fixed. The result is a linear quiver chain, depicted in Figure \ref{15.fig:2}. The gauge group is $SU(N_1) \times \cdots \times SU(N_P)$, and the theory contains  hypermultiplets in bi-fundamental representations,
\bea
\bigoplus _{p=1}^{P-1} \Big \{ (\bN_p, \overline{\bN}_{p+1}) \oplus ( \overline{ \bN}_p , \bN_{p+1}) \Big \}
\eea
Depending on the values of $N_p$, hypermultiplets in the fundamental of $SU(N)$ may also arise.

\subsection{$\cN=2$ dualities}

In the remainder of this section we discuss a set of dualities between $\cN=2$ theories which makes interesting contact with the moduli spaces of Riemann surfaces. The theories in question will be \textit{superconformal}, i.e. theories with both supersymmetry and conformal symmetry. We begin by briefly reviewing these notions.

\subsubsection{Conformal symmetry}

The conformal algebra in four dimensions is given by $\mathfrak{so}(4,2)$, generated by Lorentz transformations $L_{\m \n}$, translations $P_\m$, dilations $D$, and special conformal transformations $K_\m$. These generators satisfy the following commutation relations, 
\begin{align}
[L_{\m \n}, L_{\rho \sigma}] &= \eta_{\n \rho} L_{\mu \sigma} + \eta_{\m \sigma} L_{\n \rho} - \eta_{\m \rho} L_{\n \sigma} - \eta_{\n \sigma} L_{\m \rho} & {[K_\m, P_\n]} &= 2(\eta_{\m \n} D - L_{\m \n} )
\no\\
{[L_{\m \n}, P_\rho]} &= \eta_{\n \rho} P_\m - \eta_{\m \rho} P_\n & {[D, P_\m]}&= P_\m
\no\\
{[L_{\m \n}, K_\rho]} &= \eta_{\n \rho} K_\m - \eta_{\m \rho} K_\n
& {[D,K_\m]} &= - K_\m
\end{align}
while $[P_\mu, P_\nu]=[K_\mu, K_\nu]=[D, L_{\mu \nu}]=0$. Dilations $D$ generate an Abelian subalgebra $\mathfrak{so}(1,1) \subset \mathfrak{so}(4,2)$ that commutes with the Lorentz subalgebra $\mathfrak{so}(1,3)$ generated by $L_{\mu \nu}$. Therefore, we may choose the Cartan subalgebra of $ \mathfrak{so}(4,2)$ to be generated by $D$ and by the Cartan subalgebra of $\mathfrak{so}(1,3)$. The dilation generator can be used to assign weights to the other generators. The eigenvalue of $D$ is customarily  denoted by $\Delta$ and referred to as the \textit{conformal weight}. The final two commutation relations are the statement that the momenta $P_\m$ and special conformal generators $K_\m$ have conformal weights $1$ and $-1$, respectively. Furthermore, they imply that $P_\m$ is \textit{not} in the Cartan subalgebra, meaning that operators or states in a conformal field theory cannot be labelled by an energy/mass---indeed, such quantities would introduce a dimensionful scale to the theory and thereby violate conformal symmetry. 

\sm

Conformal fields have already been introduced in section \ref{sec:conffields}, where they were labelled by weights $(h, \overline h)$, related to the conformal weight $\Delta$ via $\Delta = h + \overline h$. An important difference between the discussion here and in section \ref{sec:conffields} is that in the latter case the discussion was specific to two spacetime dimensions, for which the ``global" conformal group discussed in this section is enhanced to (two copies of) an infinite-dimensional Virasoro symmetry, given in (\ref{eq:Virasoro}). In higher dimensions such an enhancement does not occur, though the notion of conformal primary introduced in section \ref{sec:conffields} persists. In the current language, a state $|[j_1, j_2], \Delta \rangle$ labelled by its conformal weight $\Delta$ and representation of the Lorentz group $\mathfrak{so}(3,1) \cong \mathfrak{su}(2) \oplus \mathfrak{su}(2)$ is a conformal primary if it satisfies,
\bea
\label{eq:confprimdef}
K_\m |[j_1, j_2], \Delta \rangle = 0
\eea
Starting from this highest weight state, a series of \textit{conformal descendants} can be obtained via repeated application of $P_\mu$.

\subsubsection{Superconformal symmetry}

When $\cN$ supercharges transforming in the Weyl spinor representation of $\mathfrak{so}(3,1)$ are added to the conformal algebra $\mathfrak{so}(4,2)$, one obtains the superalgebra $\mathfrak{su}(2,2| \cN)$. In the current section we will be mainly interested in the case of $\cN=2$, for which the maximal bosonic subalgebra of the superconformal algebra is, 
\bea
\mathfrak{so}(4,2) \oplus \mathfrak{u}(2)_R \subset \mathfrak{su}(2,2|2)
\eea 
The $\mathfrak{u}(2)_R$ factor is referred to as the \textit{R-symmetry}, and its generators will be denoted by $R^I_J$. They satisfy the standard commutation relations, 
\bea
[R^I_J, R^K_L] = \delta_J^K R^I_L - \delta^I_L R^K_J
\eea
As for the fermionic generators, in addition to the Poincar{\'e} supercharges $Q_\alpha^I$, $\overline Q_{{\dot \alpha} I}$, we now have the superconformal partners $S_{\alpha I}$, $\overline S_{{\dot \alpha}}^I$. The anti-commutation relations involving these generators take the form, 
\bea
\label{eq:SCAferms}
\{ Q_\alpha^I, Q_\beta^J\} &=& \{S_{\alpha I}, S_{\beta J}\} = \{Q_\alpha^I, \overline S_{\dot \beta}^I\} = 0
\no\\
\{Q_\alpha^I, \overline Q_{\dot \beta J}\} &=& 2 \sigma^\mu_{\alpha \dot \beta} P_\mu \delta^I_J
\no\\
\{S_{\alpha I}, \overline S_{\dot \beta}^J\} &=& 2 \sigma^\mu_{\alpha \dot \beta} K_\mu \delta^J_I
\no\\
\{Q_\alpha^I, S_{\beta J}\} &=& \eps_{\alpha \beta} (\delta_J^I \,D + R^I_J) + \half \delta^I_J \,L_{\m \n} \,\sigma^{\m \n}_{\alpha \beta}
\eea
Note that unlike in the non-conformal case, the superconformal algebra does not admit a deformation by central charges $Z$, since these would  introduce a scale and thus would break conformal symmetry.

\sm

The states of an $\cN=2$ superconformal field theory can be labelled by their conformal weight $\Delta$ and Lorentz quantum numbers $[j_1,j_2]$, as well as additional labels $(R,r)$ for the $\mathfrak{u}(2)_R \cong \mathfrak{su}(2)_R \times \mathfrak{u}(1)_r$ R-symmetry. We will denote such states by $|[j_1,j_2], \Delta\rangle^{(R,r)}$. In analogy to a conformal primary, one defines a \textit{superconformal primary} via 
\bea
S_{\alpha I} |[j_1,j_2], \Delta\rangle^{(R,r)} \,\,=\,\, \overline S_{\dot \alpha}^I |[j_1,j_2], \Delta\rangle^{(R,r)} \,\,=\,\, 0 
\eea
Note that these two conditions, together with the anti-commutation relation in (\ref{eq:SCAferms}), imply that all superconformal primaries satisfy $K_\m |[j_1,j_2], \Delta\rangle^{(R,r)}=0$, and hence are particular cases of conformal primaries in the definition of (\ref{eq:confprimdef}). A superconformal multiplet is obtained by beginning with a superconformal primary and applying arbitrary combinations of $Q_\alpha^I, \overline Q_{\dot \alpha I}$. By the anti-commutation relations in  (\ref{eq:SCAferms}), this also includes all applications of the momentum operator $P_\m$, and hence includes all conformal descendants.

\subsubsection{Superconformal field theories}

We may construct $\cN=2$ superconformal Lagrangians by combining the $\cN=2$ supermultiplets discussed in subsection \ref{sec:supermults} in such a way as to cancel any potential non-conformality. Setting all masses for the hypermultiplets to zero, the only parameter left in an $\cN=2$ Lagrangian is the coupling $\tau$, which is classically dimensionless. Thus any such Lagrangian would naively seem to have conformal invariance. While this is true at the classical level, quantum mechanically the coupling $\tau$---or rather only $g$, since $\theta$ is topologically protected---can receive radiative corrections  making it dimensionful. The dependence of the quantum-corrected coupling on the energy scale $\mu$ is captured by the $\beta$-function,
\bea
{\beta(g)} = \mu {d \over d \mu}\left({8 \pi^2 \over g^2} \right)
\eea
which, for a generic four-dimensional  Yang-Mills theory with $N_f$ Weyl fermions in a representation $\cR_f$ and $N_s$ scalars in a representation $\cR_s$, is given by,
\bea
{\beta(g)} = {11 \over 3} C(\mathfrak{g}) - {2 \over 3} N_f C(\cR_f) - {1 \over 3} N_s C(\cR_s) + \cO(g)
\eea
Here $C(\cR)$ is the Dynkin index of the representation $\cR$, defined via  $\mathrm{Tr}_{\cR}(T^a T^b) = C(\cR) \delta^{ab}$ with $T^a$ the generators of the gauge algebra. By a slight abuse of notation, we shall denote the adjoint representation of the algebra $\mathfrak{g}$ by the same symbol $\mg$. In particular, $C(\mathfrak{g})  = h^\vee$ is the dual Coxeter number. 

\sm

For an $\cN=1$ theory the vector multiplet, which contains one vector and one Weyl fermion in the adjoint representation, contributes a factor of, 
\bea
{11 \over 3 } C(\mathfrak{g}) - {2 \over 3} C(\mathfrak{g}) = 3\, h^\vee
\eea
to the $\beta$-function. The contribution for a chiral multiplet, containing a Weyl fermion and a scalar both in the representation $\cR$, is given by, 
\bea
-{2 \over 3}C(\cR) - {1\over 3} C(\cR) = - C(\cR)
\eea
For example,  an  $\cN=1$ super-Yang-Mills theory with gauge group $SU(N_c)$  (for which $h^\vee = N_c$) with $N_f$ chiral multiplets in the representation $\cR$ has the following $\beta$-function, 
\bea
{\beta(g)} = 3 N_c - N_f C(\cR) + \cO(g)
\eea
For conformality, we see that we must require $N_f C(\cR) = 3 N_c$. 

\sm

We now move on to the main case of interest to us, namely $\cN=2$ theories. The contribution to the $\beta$-function of an $\cN=2$ vector multiplet is, 
\bea
{11 \over 3} C(\mathfrak{g}) -2 \times {2 \over 3} C(\mathfrak{g}) - {1\over 3} C(\mathfrak{g}) = 2 \,h^\vee
\eea
while the contribution from a hypermultiplet in the representation $\cR$ is, 
\bea
- 2 \times {2\over 3} C(\cR)- 2 \times {1\over 3}C(\cR) = - 2\, C(\cR)
\eea
We thus see that for an $\cN=2$ $SU(N_c)$ gauge theory with $N_f$ hypermultiplets in the representation $\cR$, 
\bea
{\beta(g)} = 2 N_c - 2 N_f C(\cR) + \cO(g)
\eea 
Taking in particular the case in which all hypermultiplets are in the fundamental representation for which $C(\square) = \half$, we see that conformality can be achieved only if $N_f = 2 N_c$. 

\sm

The theories that we will study in the rest of this section will be exactly of this type, namely $\cN=2$ theories with $SU(N_c)$ gauge group and $N_f$ fundamental hypermultiplets. The data of such a theory can be captured in a so-called ``quiver diagram"---a diagram whose internal nodes are gauge groups, whose external nodes are fundamental hypermultiplets, and whose internal edges are bifundamental hypermultiplets, with the structure of the quiver chosen to be compatible with the vanishing of all beta functions.\footnote{One could ask why we allow for only bifundamental edges in the graph. It turns out that for more complicated edges, it is impossible to satisfy both anomaly cancellation and vanishing of the beta functions. The only exception is the case of trifundamentals in theories with $SU(2)$ gauge nodes, which we will return to below. } For example, $(N)-[2N]$ represents $SU(N)$ gauge theory with $2N$ fundamentals, and has a manifestly vanishing beta function. Likewise the quiver 
\bea
[N]-(N)-(N)-[N]
\eea
represents $SU(N)\times SU(N)$ gauge theory with one bifundamental hypermultiplet corresponding to the internal edge, and with $N$ additional fundamentals for each copy of $SU(N)$. Since the internal edge can be interpreted as giving $N$ fundamental hypers to the left gauge node and $N$ anti-fundamental hypers to the right gauge node, the beta functions for both gauge couplings indeed vanish.

We will focus on the case of $N=2$, for which something rather special happens. In particular, note that in this case each bifundamental has two $SU(2)$ gauges indices and one $SU(2)$ flavor index, i.e. we can write them as $\hat Q_{ij s}$ with indices as defined above. We can imagine weakly gauging the flavor symmetry, or conversely ungauging one of the gauge symmetries, which makes it clear that all three indices should be treated democratically. In other words, it is useful to think of the elementary building blocks as \textit{trifundamental} matter, which we will represent pictorially by 
\[\hat Q_{ijs}\hspace{0.35 in}\Rightarrow\hspace{0.35 in}
\begin{tikzpicture}[baseline=19,scale=0.6]
\draw[  thick] (-1,0)--(0,1);
\draw[  thick] (1,0)--(0,1);
\draw[  thick] (0,1)--(0,2);
  \node[left] at (-1,0) {$i$};
    \node[right] at (1,0) {$j$};
        \node[right] at (0,2) {$s$};
\end{tikzpicture}
\]
With this in mind, it is possible to replace any $SU(2)$ quiver diagram with a trivalent network, for example 
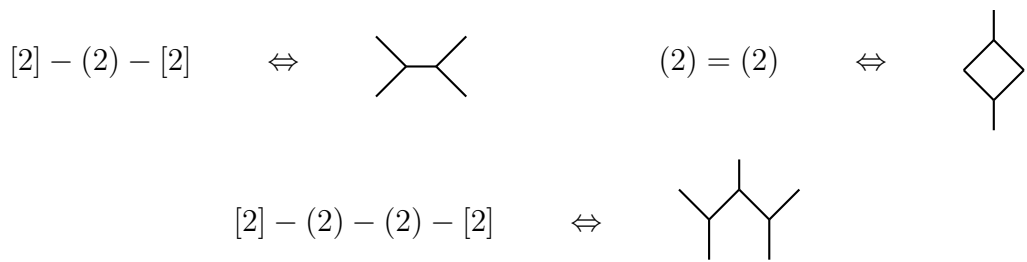
\[[2]-(2)-[2] \hspace{0.35 in}\Leftrightarrow\hspace{0.35 in}
\begin{tikzpicture}[baseline=10,scale=0.4]
\draw[  thick] (-1,0)--(0,1);
\draw[  thick] (-1,2)--(0,1);
\draw[  thick] (2,0)--(1,1);
\draw[  thick] (2,2)--(1,1);
\draw[  thick] (0,1)--(1,1);
\end{tikzpicture}
\hspace{1 in}
(2)=(2) \hspace{0.35 in}\Leftrightarrow\hspace{0.35 in}
\begin{tikzpicture}[baseline=0,scale=0.4]
\draw[  thick] (-1,0)--(0,1);
\draw[  thick] (1,0)--(0,1);
\draw[  thick] (0,1)--(0,2);
\draw[  thick] (-1,0)--(0,-1);
\draw[  thick] (1,0)--(0,-1);
\draw[  thick] (0,-1)--(0,-2);
\end{tikzpicture}
\]
\[[2]-(2)-(2)-[2] \hspace{0.35 in}\Leftrightarrow\hspace{0.35 in}
\begin{tikzpicture}[baseline=7,scale=0.4]
\draw[  thick] (-1,-1/3)--(-1,1);
\draw[  thick] (-1,1)--(-2,2);
\draw[  thick] (-1,1)--(0,2);
\draw[  thick] (0,2)--(0,3);
\draw[  thick] (1,-1/3)--(1,1);
\draw[  thick] (1,1)--(2,2);
\draw[  thick] (1,1)--(0,2);
\end{tikzpicture}
\]
Conversely, any trivalent graphs can be given an interpretation as a SCFT with $SU(2)$ nodes. 

\subsubsection{$SU(2)$ quivers and Riemann surfaces}
Consider a trivalent graph $\Gamma$ with $g$ faces and $n$ external legs. We will denote the theories obtained from such graphs by $\cT_{g,n}$. There is one gauge $SU(2)$ per internal edge, and it can be shown that the number of such internal edges is $3g-3+n$. To each of these internal nodes is associated a gauge coupling $\tau$. The quantity $3g-3+n$ appearing here may seem familiar---indeed, if $n=0$ then this precisely the dimension of the moduli space of genus $g$ Riemann surfaces, as given in (\ref{eq:modspacedim}). For non-zero $n$, this gives the dimension of the moduli space of genus $g$ Riemann surfaces \textit{with $n$ marked points}.

The mapping between the moduli space of 4d $\cN=2$ SU(2) quiver theories and that of punctured Riemann surfaces can be made more precise as follows. We identify each trivalent vertex with a three-punctured sphere, 
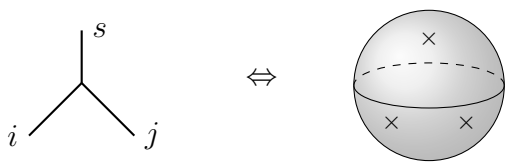
\[
\begin{tikzpicture}[baseline=19,scale=0.7]
\draw[  thick] (-1,0)--(0,1);
\draw[  thick] (1,0)--(0,1);
\draw[  thick] (0,1)--(0,2);
  \node[left] at (-1,0) {$i$};
    \node[right] at (1,0) {$j$};
        \node[right] at (0,2) {$s$};
\end{tikzpicture}
 \hspace{0.35 in}\Leftrightarrow\hspace{0.35 in}
\begin{tikzpicture}[baseline=0,scale=0.5]
  \shade[ball color = gray!40, opacity = 0.4] (0,0) circle (2cm);
  \draw (0,0) circle (2cm);
  \draw (-2,0) arc (180:360:2 and 0.6);
  \draw[dashed] (2,0) arc (0:180:2 and 0.6);
  \node[above,scale=0.8] at (0,0.8)  {$\times$};
   \node[left,scale=0.8] at (-0.5,-1)  {$\tiny\times$};
   \node[right,scale=0.8] at (0.5,-1)  {$\tiny\times$};
\end{tikzpicture}
\]
Gauging involves gluing two such trivalent vertices together, which in other words corresponds to plumbing together two three-punctured spheres. 

Indeed, let us consider gluing two trivalent vertices together to obtain the $SU(2)$ $N_f = 4$ theory.  In the corresponding Riemann surface picture, we would like to plumb together two spheres along their punctures. Say that the two spheres have complex coordinates $w$ and $z$, with the three punctures being located at $w,z=0,1,\infty$. We will glue the point at $w= \infty$ to the point at $z = 0$, 
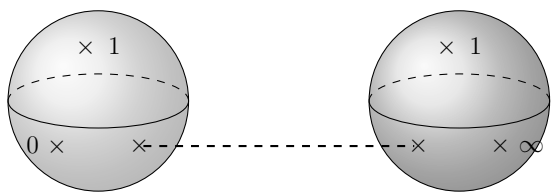
\[
\begin{tikzpicture}[baseline=0,scale=0.6]
  \shade[ball color = gray!40, opacity = 0.4] (0,0) circle (2cm);
  \draw (0,0) circle (2cm);
  \draw (-2,0) arc (180:360:2 and 0.6);
  \draw[dashed] (2,0) arc (0:180:2 and 0.6);
  \node[above,scale=0.8] at (0,0.8)  {$\times\,\,\,1$};
   \node[left,scale=0.8] at (-0.5,-1)  {$0\,\,\tiny\times$};
   \node[right,scale=0.8] at (0.5,-1)  {$\tiny\times$};
    \shade[ball color = gray!40, opacity = 0.6] (8,0) circle (2cm);
  \draw (8,0) circle (2cm);
  \draw (6,0) arc (180:360:2 and 0.6);
  \draw[dashed] (10,0) arc (0:180:2 and 0.6);
  \node[above,scale=0.8] at (8,0.8)  {$\times\,\,\,1$};
   \node[left,scale=0.8] at (7.5,-1)  {$\tiny\times$};
   \node[right,scale=0.8] at (8.5,-1)  {$\tiny\times\,\,\infty$};

\draw[thick,dashed](1,-1) --(7,-1)  ;
\end{tikzpicture}
\]
To glue them, we begin by changing to the local coordinate $\tilde w = {1\over w}$ which describes the region around the $w=\infty$ puncture, and then we identify 
\bea
z \tilde w = q 
\eea
with $q$ some complex parameter. This parameter is identified with the coupling in the field theory via $ q = e^{i \pi \tau}$. 

From this picture, it is clear that the conformal manifold of the $SU(2)$ $N_f = 4$ theory---that is to say the space in which $\tau$ can take values---can be identified with the moduli space of a four-punctured sphere. The latter is none other than the usual fundamental domain, shown in Figure \ref{3.fig:2}. The cusp at $\tau = i \infty$, i.e. $q \rightarrow 0$, corresponds to the separating limit in which the four-punctured sphere pinches off into two three-punctured spheres. 

\begin{figure}
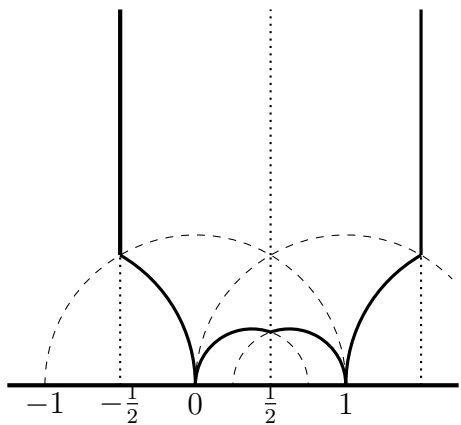

\begin{center}
\tikzpicture[scale=0.5]
\scope[xshift=-5cm,yshift=0cm]
\draw[ultra thick] (-5,0) -- (7,0);
\draw[thick, dotted] (-2,0) -- (-2,3.464);
\draw[thick, dotted] (2,0) -- (2,3.464);
\draw[thick, dotted] (6,0) -- (6,3.464);
\draw[ultra thick] (-2, 3.464) -- (-2, 10);
\draw[thick,dotted] (2 , 3.464) -- (2, 10);
\draw [dashed] (-4,0) arc (180:0:4 and 4);
\draw [dashed] (0,0) arc (180:45:4 and 4);
\draw [dashed] (0,0) arc (180:0:1.5 and 1.5);
\draw [dashed] (4,0) arc (0:180:1.5 and 1.5);

\draw[very thick] (6 , 3.464) -- (6, 10);
\draw [very thick] (0,0) arc (180:70:1.5 and 1.5);
\draw [very thick] (4,0) arc (0:110:1.5 and 1.5);
\draw [very thick] (0,0) arc (0:60:4 and 4);
\draw [very thick] (4,0) arc (180:120:4 and 4);
\draw (0,-0.5) node{$0$};
\draw (-2,-0.5) node{$-\half $};
\draw (2,-0.5) node{$\half $};
\draw (-4,-0.5) node{$-1$};
\draw (4,-0.5) node{$1$};
\endscope
\endtikzpicture
\end{center}
\caption{\textit{The fundamental domain for a sphere with four distinguishable marked points.} \label{fig:markedsphere}}
\end{figure}

It is also useful to consider the moduli space of a four-punctured sphere when the four punctures are taken to be \textit{distinguishable}. In that case, the moduli space takes the form shown in Figure \ref{fig:markedsphere}. There are now three cusps, at $\tau = 0,1,i \infty$, which correspond to configurations in which different pairs of marked points collide. It is interesting to consider the behavior of the field theory at each of these three cusps.

Let us begin with the weakly-coupled cusp at $\tau = i \infty$. At this point the theory is well-described by the usual $SU(2)$ $N_f = 4$ theory discussed before. This theory has an $SO(8)$ flavor symmetry, under which the hypermultiplets transform in the $\mathbf{8}_v$ representation. It will be useful to consider the following subgroup of $SO(8)$, 
\bea
SO(8) \supset SO(4) \times SO(4) \supset SU(2)_a \times SU(2)_b \times SU(2)_c \times SU(2)_d
\eea
This is the subgroup which is manifest in the trivalent diagram. In particular, each of the four external legs carries one copy of $SU(2)$. Let us assign them as follows,
\[\tau \rightarrow i \infty : \qquad\begin{tikzpicture}[baseline=10,scale=0.4]
\draw[  thick] (-1,0)--(0,1);
\draw[  thick] (-1,2)--(0,1);
\draw[  thick] (2,0)--(1,1);
\draw[  thick] (2,2)--(1,1);
\draw[  thick] (0,1)--(1,1);
\node[left] at (-1,0) {$a$};
\node[left] at (-1,2) {$b$};
\node[right] at (2,0) {$d$};
\node[right] at (2,2) {$c$};
\end{tikzpicture}
\]
In the corresponding four-punctured sphere, the points labelled by $a$ and $b$ are close to one another, and separated by a long thin tube from the points $c$ and $d$. We now move towards the point in moduli space with $\tau = 0$. In this case the four-punctured sphere changes to a configuration in which the points $a$ and $c$ approach one another, and are separated by a long tube from $b$ and $d$. Likewise as we approach the cusp at $\tau = 1$, the points $a$ and $d$ approach one another, and are separated by a long tube from $c$ and $b$. In terms of trivalent diagrams, this is 
\[ \tau \rightarrow 0 : \qquad\begin{tikzpicture}[baseline=10,scale=0.4]
\draw[  thick] (-1,0)--(0,1);
\draw[  thick] (-1,2)--(0,1);
\draw[  thick] (2,0)--(1,1);
\draw[  thick] (2,2)--(1,1);
\draw[  thick] (0,1)--(1,1);
\node[left] at (-1,0) {$a$};
\node[left] at (-1,2) {$c$};
\node[right] at (2,0) {$d$};
\node[right] at (2,2) {$b$};
\end{tikzpicture}\hspace{1 in }
\tau \rightarrow 1 : \qquad\begin{tikzpicture}[baseline=10,scale=0.4]
\draw[  thick] (-1,0)--(0,1);
\draw[  thick] (-1,2)--(0,1);
\draw[  thick] (2,0)--(1,1);
\draw[  thick] (2,2)--(1,1);
\draw[  thick] (0,1)--(1,1);
\node[left] at (-1,0) {$a$};
\node[left] at (-1,2) {$d$};
\node[right] at (2,0) {$b$};
\node[right] at (2,2) {$c$};
\end{tikzpicture}
\]

We see that in all cases, the quiver is of the same general form as before, suggesting that the strong-coupling regime of $SU(2)$ $N_f = 4$ is again described by $SU(2)$ $N_f = 4$. However, distinguishing between the marked points allows us to make an important observation about these strong-coupling dualities. In particular, let us recall that in the original weakly-coupled theory, the hypermultiplets transformed in the $\mathbf{8}_v$ of $SO(8)$. Decomposing this representation into those of the $SU(2)$ subgroups, we have 
\bea
\mathbf{8}_v = \mathbf{4} \oplus \mathbf{4} = (\mathbf{2}_a \otimes \mathbf{2}_b)\oplus (\mathbf{2}_c \otimes \mathbf{2}_d)
\eea
On the other hand, one has the following decompositions, 
\bea
\mathbf{8}_s &=&  (\mathbf{2}_a \otimes \mathbf{2}_c)\oplus (\mathbf{2}_b \otimes \mathbf{2}_d)\no\\
\mathbf{8}_c &=& (\mathbf{2}_a \otimes \mathbf{2}_d)\oplus (\mathbf{2}_b \otimes \mathbf{2}_c)
\eea
with $\mathbf{8}_s$ and $\mathbf{8}_c$ the spinor and conjugate spinor representations of $SO(8)$. Thus swapping $SU(2)_b$ with $SU(2)_c$ has the effect of switching from the vector to the spinor representation of $SO(8)$, and likewise for swapping $SU(2)_b$ with $SU(2)_d$. We thus see that upon approaching the strongly-coupled cusps in moduli space, we reobtain $SU(2)$ $N_f = 4$, but with the hypermultiplets transforming in spinor representations of the flavor symmetry group.

Similar statements hold for the $[2]-(2)-(2)-[2]$ quiver whose trivalent diagram was given above. Again, approaching any cusp gives rise to a quiver in the same form, but with a rearrangement of the $SU(2)$ subgroups of the total flavor symmetry. In general, though, the story is far richer. Indeed, it is often possible that the theory emerging at a cusp has a completely distinct quiver than that of the original theory. This for example happens in the case of the three node quiver $[2]-(2)-(2)-(2)-[2]$. In particular, if we consider the strong-coupling limit of the middle node, while keeping the outer nodes weakly coupled, the quiver transforms as
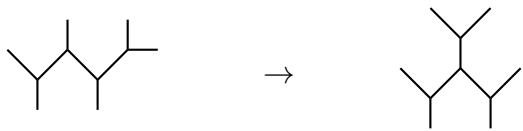
\[\begin{tikzpicture}[baseline=10,scale=0.4]
\draw[  thick] (1,0)--(1,1);
\draw[  thick] (1,1)--(0,2);
\draw[  thick] (1,1)--(2,2);
\draw[  thick] (2,2)--(2,3);
\draw[  thick] (2,2)--(3,1);
\draw[  thick] (3,1)--(3,0);
\draw[  thick] (3,1)--(4,2);
\draw[  thick] (4,2)--(4,3);
\draw[  thick] (4,2)--(5,2);
\end{tikzpicture}
\hspace{0.5 in}\rightarrow\hspace{0.5 in}
\begin{tikzpicture}[baseline=17,scale=0.4]
\draw[  thick] (1,0)--(1,1);
\draw[  thick] (1,1)--(0,2);
\draw[  thick] (1,1)--(2,2);

\draw[  thick] (2,2)--(2,3);
\draw[  thick] (2,3)--(1,4);
\draw[  thick] (2,3)--(3,4);

\draw[  thick] (2,2)--(3,1);
\draw[  thick] (3,1)--(3,0);
\draw[  thick] (3,1)--(4,2);
\end{tikzpicture}
\]
which gives a quiver of an entirely new type.

We close this subsection by noting that a physical origin of the correspondence between 4d $\cN=2$ theories and Riemann surfaces can be obtained via M-theory. In particular, if we wrap an M5-brane on the relevant Riemann surfaces, with the insertion of appropriate defect operators at the marked points, then reducing the worldvolume theory from six- to four-dimension gives precisely the 4d $\cN=2$ theories in question. Due to their six-dimensional origins, the 4d $\cN=2$ theories constructed in this way are referred to as ``Class $\cS$" theories.

\subsection*{$\bullet$ Bibliographical notes}

Comprehensive expositions of supersymmetric quantum field theories and supersymmetric Yang Mills theories may be found in the books by Wess and Bagger \cite{Wess:1992cp},  Weinberg \cite{Weinberg}, Freedman and Van Proeyen \cite{FVP}, and Shifman \cite{Shifman:2022shi}. 

\sm

Goddard-Nuyts-Olive (GNO) duality was introduced in  \cite{Goddard:1976qe} and realized in $\cN=4$ Yang-Mills theory as Montonen-Olive duality in \cite{Montonen:1977sn,Osborn:1979tq}. The effect of topological charges on the central charge of the supersymmetry algebra was obtained in \cite{Witten:1978mh}, while the effect by which magnetic monopoles in the presence of a $\theta$-angle become dyons was derived in \cite{Witten:1979ey}. Subsequent tests of S-duality in $\cN=4$ were carried out in \cite{Vafa:1994tf}. The relation between electric-magnetic duality and the geometric Langlands program was established in \cite{Kapustin:2006pk}. Integrated correlators in the superconformal phase of $\cN=4$ at fixed gauge coupling provided explicit realizations of S-duality \cite{Chester:2020vyz,Dorigoni:2021rdo,Dorigoni:2021guq,Dorigoni:2022cua}. For a SAGEX overview of modular invariance in $\cN=4$ Yang-Mills theory and Type IIB superstring theory, see \cite{Dorigoni:2022iem}.

\sm

Seiberg-Witten theory was originally formulated for gauge group $SU(2)$ in \cite{Seiberg:1994rs,Seiberg:1994aj}. Generalizations to other gauge groups with various hypermultiplet contents were soon thereafter produced, and are reviewed    in the lecture notes \cite{Intriligator:1995au,Alvarez-Gaume:1996ohl,DHoker:1999yni,Martone:2020hvy} and in the book by Tachikawa \cite{Tachikawa:2013kta}, where  extensive bibliographical references may also be found. 

\sm

Construction of the prepotential using localization techniques was initiated for the instanton part in \cite{Nekrasov:2002qd}, and generalized to include also the perturbative contributions in \cite{Pestun:2007rz}. An overview of localization techniques is provided by the collection of essays in \cite{Pestun:2016zxk}.

\sm

The relation between integrable systems and the Seiberg-Witten solution for the $\cN=2^*$ theory was put forward in \cite{Donagi:1995cf,Martinec:1995qn}. The implementation using the Calogero-Moser system for $SU(N)$ gauge group was carried out  in  \cite{DHoker:1997hut}, and for other gauge groups in \cite{DHoker:1998zuv,DHoker:1998xad} where the behavior under $SL(2,\ZZ)$ transformations was also analyzed. The derivations in section \ref{sec:15.7} closely follow  \cite{DHoker:1997hut,Manschot:2019pog}, and reproduce the hypermultiplet mass expansion of \cite{Minahan:1997if}. The modular properties of the Seiberg-Witten solution for gauge group $SU(2)$ and $N_f\leq 4$ fundamentals were analyzed in detail in  \cite{Aspman:2021vhs,Aspman:2021evt} and for $SU(3)$ in \cite{Aspman:2020lmf}. The role of bi-modular forms was discussed in those references,  in  \cite{Manschot:2021qqe}, and in the mathematical literature in \cite{SienZag}. The Seiberg-Witten curve for linear quivers and gauge group $SU(N)$ was constructed from M-theory in \cite{Witten:1997sc}.

\sm

The dualities of $\cN=2$ superconformal Yang-Mills theories were developed in \cite{Gaiotto:2009we} and are  reviewed in \cite{Akhond:2021xio}. The  global structure and spectrum of line operators in $\cN=4$ Yang-Mills is clearly presented in \cite{Aharony:2013hda}. Useful lecture notes on the dynamics of four-dimensional supersymmetric gauge theories may be found in \cite{Razamat:2022gpm}. 

\sm

Besides the beautiful Mathematics which arose from Seiberg-Witten theory, various exciting physical results also emerged, including the  discovery of a number of strongly-coupled, non-Lagrangian theories known as \textit{Argyres-Douglas theories} \cite{Argyres:1995jj}, as well as the later-discovered \textit{Minahan-Nemeschansky theories} \cite{Minahan:1996fg,Minahan:1996cj}. These bear close relation to complex-multiplication, as well as to Kodaira's theory of the degeneration of elliptic fibers.

\sm 

Finally, let us mention that in the last decade an interesting connection between four-dimensional  $\cN=2$ theories and 2d vertex operator algebras (VOAs) has been uncovered \cite{Beem:2013sza}, which has given even more direct connections between 4d $\cN=2$ theories and modularity. Indeed, by studying the modular differential equations satisfied by the corresponding VOAs \cite{Beem:2017ooy,Kaidi:2022sng}, it has been possible to provide a partial classification for low-rank 4d $\cN=2$ theories. The full power of these techniques remains to be explored.

\newpage

\section{Basic Galois Theory}
\setcounter{equation}{0}
\label{sec:Galois}

We close these notes with a brief introduction to Galois theory, largely following the discussions of \cite{escofier2012galois,hungerford2012abstract}. After having introduced the basic Mathematical ideas, we will illustrate their use in Physics through examples from conformal field theory.

\subsection{Fields}

In Mathematics, a field\footnote{Not to be confused with classical or quantum fields in Physics, such as the electro-magnetic field.} is a set $F$ equipped with two binary operations $F \times F \to F$ on elements $a,b \in F$, referred to as addition $a+b$ and multiplication $ab=a \cdot b$. Under addition $F$ forms a group with unit element 0, while under multiplication $F \setminus \{ 0 \}$  forms  a group with unit~1. The two operations are commutative and are  intertwined by the property of distributivity which requires $a \cdot (b+c) = a \cdot b + a \cdot c$ for all $a,b,c \in F$. 

\sm

A field $F$ is said to have characteristic 0 if there exists no integer $n$ such that the $n$-fold sum of the unit 1 vanishes. If there does exist a positive integer $n$ such that the $n$-fold sum of the unit 1 vanishes, then the smallest such integer may be shown to be a prime number $p$ and the characteristic of $F$ is defined to be $p$. If $F$ has characteristic $p$ then we have $p \cdot a=0$ for all $a \in F$. The simplest finite fields of characteristic $p$ prime are $\FF_p = \ZZ/(p \ZZ)$.

\sm

Well-known examples of fields of characteristic 0 include the rational numbers $\QQ$, the real numbers $\RR$, the complex numbers $\CC$, and the algebraic numbers (namely complex numbers that satisfy a polynomial equation with rational coefficients).  Rational functions of a single variable $x$ form fields of characteristic 0, namely $\QQ(x)$, $\RR(x)$, and $\CC(x)$ depending on whether the coefficients in the rational function are in $\QQ$, $\RR$, or $\CC$ respectively.  

\sm

A subset $F \subset K$ of a field $K$ is referred to as a subfield if $F$ is a field with respect to the binary operations of the field $K$. In the above examples, we have the subfield inclusions $\QQ \subset \RR \subset \CC$, and $\QQ(x) \subset \RR(x) \subset \CC(x)$, as well as $\QQ \subset \QQ(x)$, $\RR \subset \RR(x)$, and $\CC \subset \CC(x)$.

\subsection{Field Extensions}

A field $K$ is said to be an \textit{extension} of a field $F$ if  $F$ is a subfield of $K$, namely $F \subset K$. The \textit{degree} of the extension, denoted  $[K:F]$, is the dimension of $K$ viewed as a vector space over the field $F$. If $F$ is a subfield of a field $K$ which itself is a subfield of a field $L$, then the following relation holds between the degrees of extension, 
\bea
 F \subset K \subset L 
 \hskip 1in 
 [L:F] = [L:K][K:F]
\eea
Equivalently, these relations hold when $L $ is an extension of $K$ which in turn is an extension of $F$.  For example, $\CC$ is a field extension of $\RR$ by the element $i$, and the degree of the extension is $[\CC:\RR]=2$ since $\{1,i\}$ is a basis of $\CC$ over $\RR$. On the other hand, $\RR$ is a field  extension of $\QQ$ of infinite order. 

\sm

An element $y \in K$ is said to be \textit{algebraic} over $F$ if there exists a polynomial relation, 
\bea
a_n y^n + a_{n-1}y^{n-1} + \dots + a_0 = 0 \hspace{0.7in}a_n , \dots, a_0\in F
\eea
with $a_n \not=0$ for some $n\geq 1$. Given an extension $K$ containing an element $y$ that is algebraic over $F$, there exists a unique monic irreducible polynomial $f(x) \in F[x]$ that has $y$ as a root. This polynomial is known as the \textit{minimal polynomial of $y$ over $F$}. The uniqueness is a non-trivial result---it implies that once we have found any monic irreducible polynomial in $F[x]$ with $y$ as a root, it must be the minimal polynomial. 

\sm

As an example, consider $K = \QQ(\sqrt{3})$ and $F= \QQ$. The polynomial $f(x) = x^2 - 3$ is monic and irreducible in $\QQ[x]$, and has $\sqrt{3}$ as a root. It is thus the minimal polynomial of $\sqrt{3}$ over $\QQ$.  This simple example illustrates an important fact: the minimal polynomial is dependent on the choice of base field $F$. Indeed, if we  had chosen $F=\RR$, then $f(x)$ would have been reducible, splitting to $f(x) = (x- \sqrt{3})(x+ \sqrt{3})$. Thus the minimal polynomial of $\sqrt{3}$ over $\RR$ is instead $\tilde{f}(x) = x - \sqrt{3}$.

\subsubsection{Simple extensions}

Let $K$ be an extension of $F$ and $y$ be an element in $K$. We denote by $F(y)$ the intersection of all subfields of $K$ which contain both $F$ and $y$. This intersection is a field, since the intersection of subfields of $K$ is itself a field. Since $F(y)$ is contained in all subfields containing $F$ and $y$, it is by definition the smallest subfield containing $F$ and $y$. We refer to such a minimal extension $F(y)$ as a \textit{simple extension} of $F$ by $y$. In fact, \textit{any} finite extension of a field $F$ of characteristic 0 is a simple extension.  For algebraic $y$, namely $y$ satisfying $f(y)=0$ for the minimal polynomial $f$ of degree $n$  with coefficients in $F$, one has the following result. 

{\thm If $y$ is algebraic over $F$, and $K$ is a field extension of $F$ by the element $y$,  then the simple extension $F(y)$ does not depend on $K$, and satisfies 
\bea
F(y) \cong F[x]/f(x) \hspace{0.8 in }[F(y): F] = n
\eea
\label{thm:minextthm}
Thus, $F(y)$ is completely determined by $F[x]$ and the minimal polynomial $f$ of $y$.}
 
 \sm
 
  An analogous statement holds for finitely generated extensions $F(y_1, y_2, \dots, y_N)$, as long as $K$ is an \textit{algebraic extension} of $F$, i.e.  if every element in $K$ is algebraic over $F$. 

\sm

An immediate consequence of Theorem \ref{thm:minextthm} is that if $y$ and $w$ have the same minimal polynomial $f(x)$ in $F[x]$, then $F(y)$ is isomorphic to $F(w)$.

\subsubsection{Splitting fields}

So far we have considered extension fields of $F$ containing a root of a polynomial $f(x) \in F[x]$. We now consider extension fields that contain \textit{all} the roots of $f(x)$. In particular, let $K$ be an extension field of $F$ and $f(x)$ be a non-constant polynomial of degree $n$ in $F[x]$. If $f(x)$ factors in $K[x]$ as $f(x) = \a (x-y_1) (x-y_2) \dots (x-y_n)$, then we say that $f(x)$ \textit{splits} in $K$. If $K$ is the smallest extension containing all the roots of $f(x)$, then we refer to $K$ as the \textit{splitting field} for $f(x)$. 

\sm

By Theorem \ref{thm:minextthm}, any two splitting fields of a polynomial in $F[x]$ are isomorphic. Furthermore, every polynomial $f(x) \in F[x]$ admits a splitting field over $F$.

{\thm Let $F$ be a field and $f(x)$ be a non-constant polynomial of degree $n$ in $F(x)$. Then there exists a splitting field $K$ of $f(x)$ over $F$ such that $[K:F]\leq n!$.}
\newline\newline
\noindent Of course, we already know that,  if $K$ is a simple extension and the polynomial $f(x)$ has $n$ distinct roots, then $[K:F] = n$ by Theorem \ref{thm:minextthm}.

\sm

We have just noted that every polynomial admits a splitting field; it is then natural to ask if there exists a field extension of $F$ over which \textit{every} polynomial splits. If such a universal splitting field exists, the field $F$ is said to be \textit{algebraically closed}. A well-known example is $F=\CC$. Another is the field $\AA$ of all algebraic numbers over $\QQ$.

\subsubsection{Normal extensions}

We close this subsection by introducing the concept of a \textit{normal extension}. An algebraic extension $K$ of $F$ is said to be normal provided that whenever an irreducible polynomial has one root in $K$, it has all roots in $K$ (i.e. it splits over $K$). For example, $\QQ(\sqrt{3})$ is a normal extension of $\QQ$ since both of the roots $\pm \sqrt{3}$ of the minimal polynomial $x^2 - 3$ are in $\QQ(\sqrt{3})$. On the other hand, the extension $\QQ(\sqrt[3]{2})$ is not a normal extension of $\QQ$ since two roots of the minimal polynomial $x^3 - 2$ are complex and hence not contained in $\QQ(\sqrt[3]{2})$.

\subsection{Field automorphisms and the Galois Group}

An automorphism of a field $K$ is a bijection $\sigma : K \rightarrow K$ such that for all $a,b \in K$ we have, 
\bea
\sigma(a+b) & = & \sigma(a) + \sigma(b) 
\no \\
 \sigma(ab) & = & \sigma(a) \, \sigma(b)
\eea
Two automorphisms $\sigma, \tilde{\sigma}$ are distinct if $\sigma(a) \neq \tilde{\sigma}(a)$ for at least one element $a \in K$. Distinct automorphisms are linearly independent---that is, if $\sigma_1, \dots, \sigma_n$ are distinct automorphisms of $K$, then 
a linear relation valid for all $y \in K$,
\bea
a_1 \sigma_1 (y) + \dots + a_n \sigma_n(y)=0 
\eea
 implies that $a_1 = \dots = a_n = 0$. 

\sm

Let $K$ be an extension field of $F$. An \textit{$F$-automorphism of $K$} is an automorphism of the field $K$, namely $\sigma:K \rightarrow K$, that fixes each element of $F$. The set of all $F$-automorphisms will be denoted ${\rm Gal}(K,F)$ and is called the \textit{Galois group of $K$ over $F$}. The use of the word ``group" is justified since ${\rm Gal}(K,F)$ can be endowed with group structure via function composition. Indeed, if $\sigma_1, \sigma_2 \in {\rm Gal}(K,F)$ then clearly $\sigma_1 \circ \sigma_2$ is also an automorphism of $K$ leavings $F$ invariant.  We begin with the following simple result,

{\thm Consider a polynomial $f(x) \in F[x]$. If $y \in K$ is a root of $f(x)$ and $\sigma \in {\rm Gal}(K,F)$, then $\sigma(y) $ is also a root of $f(x)$. \label{eq:basicGalres}}
\newline\newline
To prove this theorem, let $f(x)$ be an arbitrary polynomial of arbitrary degree $n$, 
\bea
f(x) = c_n x^n + \dots + c_1 x + c_0
\eea
Since $f(y)=0$, and $\sigma$ is a homomorphism with $\sigma(c_i) = c_i$ for each $c_i \in F$, we have, 
\bea
0 = \sigma(0) &=& \sigma\left(c_n y^n + \dots c_1 y + c_0 \right) 
\no\\
&=&\sigma(c_n) \sigma(y^n) +\dots+ \sigma(c_1) \sigma(y) + \sigma(c_0)
\no\\
&=& c_n \sigma(y)^n +\dots+ c_1 \sigma(y) +c_0
\eea
and hence $\sigma(y)$ is a root of $f(x)$. 

\sm

Theorem \ref{eq:basicGalres} shows that every image of a root $y$ of $f(x)$ under an automorphism of the Galois group is also a root of $f(x)$. Conversely, we could ask if every root of $f(x)$ can be obtained as the image of some root $y$ under an automorphism $\sigma \in {\rm Gal}(K,F)$. In some instances, this is indeed the case:

\sm

{\thm If $K$ is a splitting field for a polynomial in $F[x]$ and $y_1,y_2 \in K$ are two roots, then there exists an element $\sigma \in {\rm Gal}(K,F)$ such that $\sigma(y_1) = y_2$  if and only if $y_1$ and $y_2$ have the same minimal polynomial over $F$. \label{thm:Galoisiff}}
\newline\newline
The forward direction follows immediately from Theorem \ref{eq:basicGalres}. For the converse, note from Theorem \ref{thm:minextthm} that if $y_1$ and $y_2$ have the same minimal polynomial, then there is an isomorphism $\sigma : F(y_1) \cong F(y_2)$. This isomorphism can be chosen such that $\sigma(y_1) = y_2$, while fixing $F$ elementwise. Since $K$ is a splitting field of some polynomial in $F[x]$, it is also a splitting field for the same polynomial over $F(y_1)$ and $F(y_2)$, and thus $\sigma$ extends to an $F$-automorphism of $K$, i.e. $\sigma\in {\rm Gal}(K,F)$.

\sm

A important fact of which we make repeated use  is the following, 

{\thm If $K = F(y_1, \dots, y_n)$ is an algebraic extension of $F$, then automorphisms in $ {\rm Gal}(K,F)$ are determined completely by their action on $y_1, \dots, y_n$. That is, if $\sigma_1, \sigma_2 \in {\rm Gal}(K,F)$ satisfy $\sigma_1(y_i) = \sigma_2(y_i)$ for all $i=1, \dots ,n$, then $\sigma_1 = \sigma_2$. \label{eq:Galonlyext}}
\newline\newline
Together with Theorem \ref{eq:basicGalres}, this result can be used to at least partially construct concrete Galois groups. For example, consider the Galois group of $\QQ(\sqrt{3}, \sqrt{5})$ over $\QQ$. 
 By Theorem \ref{eq:Galonlyext} the Galois automorphisms are determined completely by their action on $\sqrt{3}$ and $\sqrt{5}$. The minimal polynomial of $\sqrt{3}$ is $x^2 - 3$, and hence by Theorem \ref{thm:Galoisiff} we must have $\sqrt{3} \rightarrow \pm \sqrt{3}$. Similar statements hold for $\sqrt{5}$. Hence the Galois group $ {\rm Gal}(\QQ(\sqrt{3}, \sqrt{5}),\QQ)$ has at most four elements, 
\bea
\label{eq:Galsqrt35}
\sqrt{3} \xrightarrow[]{1} \sqrt{3} \hspace{0.5in}\sqrt{3} \xrightarrow[]{\a} -\sqrt{3} \hspace{0.5in}\sqrt{3} \xrightarrow[]{\b} \sqrt{3} \hspace{0.5in}\sqrt{3} \xrightarrow[]{\g} -\sqrt{3}
\no\\
\sqrt{5} \xrightarrow[]{} \sqrt{5} \hspace{0.5in}\sqrt{5} \xrightarrow[]{} \sqrt{5} \hspace{0.5in}\sqrt{5} \xrightarrow[]{} -\sqrt{5} \hspace{0.5in}\sqrt{5} \xrightarrow[]{} -\sqrt{5}
\eea
In fact, it turns out that in this case there are exactly four elements, and $ {\rm Gal}(\QQ(\sqrt{3}, \sqrt{5}),\QQ) \cong \ZZ_2 \times \ZZ_2$, though we will not prove this here. 

\sm

In the above example, every automorphism of the Galois group permuted the four roots $\pm \sqrt{3}, \pm \sqrt{5}$. This is a particular case of a more general result,

{\thm If $K$ is the splitting field of a separable polynomial $f(x)$ of degree $n$ in $F[x]$, then ${\rm Gal}(K,F)$ is isomorphic to a subgroup of $S_n$. \label{thm:Galperm}}
\newline\newline
That is, every element of ${\rm Gal}(K,F)$ produces a permutation of the roots of $f(x)$. Of course, the converse is not necessarily true. For example, in the case of $\QQ(\sqrt{3}, \sqrt{5})$, the permutation mapping $\pm \sqrt{3} \rightarrow \pm\sqrt{5}, \pm \sqrt{5} \rightarrow \pm \sqrt{3}$ does not correspond to an $F$-automorphism of $K$.

\subsubsection{Intermediate Fields}
\label{sec:IntermediateFields}

Let $K$ be an extension field of $F$. A field $E$ such that $F \subseteq E \subseteq K$ is called an \textit{intermediate field} of the extension. It can be shown that the Galois group ${\rm Gal}(E,F)$ associated to the intermediate field is a subgroup of ${\rm Gal}(K,F)$.  Hence to every  intermediate field we can associate a subgroup of the Galois group. Conversely, if $H$ is a subgroup of the Galois group, one can associate to it an intermediate field $E_H$ such that, 
\bea
E_H = \left\{ k \in K\, | \,\sigma(k) = k\quad \forall \,\sigma \in H\right\}
\eea
known as the \textit{fixed field of $H$}. That $E_H$ is indeed a field is easy to check, since for example the equalities $\sigma(0)= 0$ and $\sigma(1)=1$ imply that $0$ and $1$ are contained in $E_H$, and since, 
\bea
\sigma(c+d) = \sigma(c) + \sigma(d) = c+d \hspace{0.5 in} \sigma(cd) = \sigma(c) \sigma(d) = c d
\eea
imply that $E_H$ is closed under addition and multiplication. The presence of inverses follow simply as well.
A somewhat stronger result is the following, 

{\thm Let $K$ be a finite-dimensional extension field of $F$. If $H$ is a subgroup of ${\rm Gal}(K,F)$ and $E_H$ is the fixed field of $H$, then $H={\rm Gal}(K,E_H)$ and $|H| = [K:E]$. \label{thm:extGal}}
\newline\newline
We may now return to the example of $ {\rm Gal}(\QQ(\sqrt{3}, \sqrt{5}),\QQ)$. In this case $\QQ(\sqrt{3})$ is an intermediate field of the extension $\QQ(\sqrt{3}, \sqrt{5})$ of $\QQ$. Recall that $ {\rm Gal}(\QQ(\sqrt{3}, \sqrt{5}),\QQ) = \{ 1, \a, \b , \g \}$ as defined in (\ref{eq:Galsqrt35}). On the other hand, $ {\rm Gal}(\QQ(\sqrt{3}, \sqrt{5}),\QQ(\sqrt{3}))$ should consist of only the transformations leaving the base field  $\QQ(\sqrt{3})$ invariant. Hence we have that 
\bea {\rm Gal}(\QQ(\sqrt{3}, \sqrt{5}),\QQ(\sqrt{3})) = \{ 1, \b \}
\eea which is clearly a subgroup of $ {\rm Gal}(\QQ(\sqrt{3}, \sqrt{5}),\QQ)$. Conversely, the fixed field of $H=\{1,\b\}$ is precisely $E_H = \QQ(\sqrt{3})$.

\subsection{The Fundamental Theorem of Galois Theory}

Consider a finite-dimensional extension field $K$ of $F$. Let $S$ be the set of all intermediate fields, and $T$ be the set of all subgroups of the Galois group  $ {\rm Gal}(K,F)$. We introduce a function $\f:S\rightarrow T$ such that for each intermediate field $E$, 
\bea
\f(E) = {\rm Gal}(K,E)
\eea
The function $\f$ is referred to as a \textit{Galois correspondence}. In this correspondence $K$ is mapped to the identity subgroup of $ {\rm Gal}(K,F)$, while $F$ is mapped to the entire group  ${\rm Gal}(K,F)$.  By Theorem \ref{thm:extGal}, the Galois correspondence is surjective.

\sm

While surjectivity of the Galois correspondence is guaranteed, injectivity is not. For example, consider the extension field $\QQ(\sqrt[3]{2})$ over $\QQ$. Every automorphism in the Galois group maps $\sqrt[3]{2}$ to a root of the minimal polynomial $x^3 -2$ by Theorem \ref{eq:basicGalres}. Since $\sqrt[3]{2}$ is the only real root, and since all elements of $\QQ(\sqrt[3]{2})$ are real, every automorphism must map $\sqrt[3]{2}$ to itself. Thus by Theorem \ref{eq:Galonlyext} we conclude that ${\rm Gal}(\QQ(\sqrt[3]{2}), \QQ)$ is trivial. This means that under the Galois correspondence, both $\QQ(\sqrt[3]{2})$ and $\QQ$ must be associated to the same group, i.e. the trivial group $\{1\}$. Therefore the Galois correspondence is not injective.

\sm

In the cases in which the Galois correspondence \textit{is} injective, the extension $K$ of $F$ is called a \textit{Galois extension}. Given a Galois extension $K$, if $E_1$ and $E_2$ are intermediate fields such that ${\rm Gal}(K,E_1) = {\rm Gal}(K,E_2)$, then we can conclude that $E_1= E_2$. Note that in characteristic 0, a Galois extension is simply a splitting field. 

\sm

As an example, consider the extension $\QQ(\sqrt{3}, \sqrt{5})$ over $\QQ$. This extension is Galois because it is the splitting field of $f(x) = (x^2 - 3)(x^2 - 5)$. In this case the Galois correspondence $\f$ must be bijective, and indeed we have
\bea
\label{eq:Q35surj}
\f: \QQ(\sqrt{3}, \sqrt{5}) &\mapsto & {\rm Gal}(\QQ(\sqrt{3}, \sqrt{5}),\QQ(\sqrt{3}, \sqrt{5}))=1
\no\\
\QQ(\sqrt{3}) &\mapsto & {\rm Gal}(\QQ(\sqrt{3}, \sqrt{5}),\QQ(\sqrt{3}))=\{1,\b\}
\no\\
\QQ(\sqrt{5}) &\mapsto & {\rm Gal}(\QQ(\sqrt{3}, \sqrt{5}),\QQ(\sqrt{5}))=\{1,\a\}
\no\\
\QQ(\sqrt{15}) &\mapsto & {\rm Gal}(\QQ(\sqrt{3}, \sqrt{5}),\QQ(\sqrt{5}))=\{1,\g\}
\no\\
\QQ &\mapsto & {\rm Gal}(\QQ(\sqrt{3}, \sqrt{5}),\QQ)=\{1,\a,\b,\g\}
\eea
In this example we observe that all intermediate fields appearing are themselves Galois extensions of $\QQ$. For example, $\QQ(\sqrt{3})$ is a Galois extension since it is the splitting field of $x^2 - 3$. Furthermore, the corresponding subgroups of the Galois group are all normal. The general version of this statement is given by the Fundamental Theorem of Galois Theory, 

{\thm
Let $K$ be a Galois extension of $F$. Then
\begin{enumerate}
\item The map $\f:S\rightarrow T$ between the set $S$ of intermediate fields and the set $T$ of subgroups of the Galois group ${\rm Gal}(K,F)$ defined by $\f(E) = {\rm Gal}(K,E)$ is a bijection. One has 
\bea
[K:E] = |{\rm Gal}(K,E)|\hspace{0.8 in} [E:F] = [{\rm Gal}(K,F):{\rm Gal}(K,E)]
\eea
\item An intermediate field $E$ is a normal extension of $F$ if and only if the corresponding group ${\rm Gal}(K,E)$ is a normal subgroup of ${\rm Gal}(K,F)$, in which case ${\rm Gal}(E,F) \cong {\rm Gal}(K,F)/{\rm Gal}(K,E)$.
\end{enumerate}
\label{thm:fundthmGal}
}
 
 \subsection{Solvability by Radicals}
 
The most famous application of Galois theory, and the reason for its initial development, is to the study of roots of polynomial equations. In particular, Galois theory provides a criterion for a polynomial to be solvable by radicals, as we now discuss. Let us first define the notion of solvability by radicals. A field $K$ is said to be a \textit{radical extension} of a field $F$ if there exists a chain of fields, 
 \bea
 F = F_0 \subseteq F_1 \subseteq F_1 \subseteq F_2 \subseteq\dots \subseteq F_r = K
 \eea
 such that for each $i=1,2,\dots, r$ we have $F_i = F_{i-1}(y_i)$ and some power of $y_i$ is in $F_{i-1}$. 
 Now let $f(x) \in F[x]$. The equation $f(x) = 0$ is said to be \textit{solvable by radicals} if there exists a radical extension of $F$ that contains a splitting field of $f(x)$. 
 
 \sm
 
 On the other hand, a group $G$ is said to be solvable if there exists a chain of subgroups, 
 \bea
 G= G_0 \supseteq G_1 \supseteq G_2\supseteq\dots \supseteq G_{n-1}\supseteq G_n = \{1\}
 \eea
 such that $G_i$ is a normal subgroup of the preceding group $G_{i-1}$ and the quotient group $G_{i-1}/G_i$ is abelian. For example, every Abelian group $G$ is solvable since $G \supseteq \{1\}$ is a chain of the required form.
 As a more non-trivial example, consider the cyclic subgroup $\langle (123) \rangle$ of order $3$ in $S_3$. The chain, 
 \bea 
 S_3 \supseteq \langle (123) \rangle\supseteq \langle (1) \rangle
 \eea
 shows that $S_3$ is solvable. On the other hand, it can be shown that for $n \geq 5$ the group $S_n$ is \textit{not} solvable.
 
 \sm
 
We now state the famous Galois Criterion for solvability by radicals.
 {\thm Let $F$ be a field of characteristic 0 and $f(x) \in F[x]$. Then $f(x) = 0$ is solvable by radicals if and only if the Galois group of $f(x)$ is solvable. \label{thm:Galcriterion}}
 \newline\newline
As a simple example, consider the polynomial $f(x) = 2x^5 - 10x+5 \in \QQ[x]$. By Theorem \ref{thm:Galperm}, the Galois group must be a subgroup of $S_5$. We will now show that it is in fact \textit{equal} to $S_5$. To do so, note that the function $f(x)$ has a maximum at $x=-1$, a minimum at $x=1$, and an inflection point at $x=0$. This means that $f(x)$ crosses the $x$-axis exactly three times, and hence that $f(x)$ has exactly three real roots. If $K$ is a splitting field of $f(x)$ over $\QQ$, then $|{\rm Gal}(K,\QQ)| = [K:\QQ]$ by Theorem \ref{thm:fundthmGal}. Taking any root $r$ of $f(x)$, we have 
\bea
[K:\QQ] = [K:\QQ(r)][\QQ(r):\QQ] = 5 [K:\QQ(r)]
\eea by Theorem \ref{thm:minextthm}. From this it follows that $5$ divides $|{\rm Gal}(K,\QQ)|$, and thus that ${\rm Gal}(K,\QQ)$ contains an element of order 5. In other words,  ${\rm Gal}(K,\QQ)$ contains a 5-cycle of $S_5$. In addition, ${\rm Gal}(K,\QQ)$ also contains complex conjugation, which fixes the three real roots of $f(x)$ but interchanges the two non-real roots. Thus ${\rm Gal}(K,\QQ)$ contains a transposition. But in fact, the only subgroup of $S_5$ containing both a 5-cycle and a transposition is $S_5$ itself. We thus conclude that ${\rm Gal}(K,\QQ)=S_5$. Then using the fact that $S_n$ is not solvable for $n \geq 5$, we conclude from Theorem \ref{thm:Galcriterion} that $f(x)$ does not admit a solution in radicals. This implies the famous result that there is no universal formula (involving only field operations and roots) for the solutions of fifth-degree polynomial equations.

 \subsection{Galois Theory in rational conformal field theory}
 
We close with an application of Galois theory to Physics. As discussed in sections \ref{sec:Intconj} and \ref{sec:3dgrav}, the torus partition of a rational conformal field theory (RCFT) can be expressed in terms of a finite number of characters $\boldsymbol{\chi}(\tau) = (\chi_1(\tau), \dots, \chi_d(\tau))$, transforming in a vector representation of the modular group. The action of $S: \tau \rightarrow - 1/\tau$ in particular is implemented by a matrix known as the modular $S$-matrix, 
\bea
\boldsymbol{\chi}(-1/\tau) = S\, \boldsymbol{\chi}(\tau)
\eea
which, by slight abuse of notation, will also be denoted by $S$. A crucial constraint on $S$ is that it satisfy the Verlinde formula,
\bea
N_{pq}^r = \sum_{s \in \cI }{S_{p,s}S_{q,s}S^\dagger_{r,s}\over S_{0,s}}
\eea
where $s \in \cI$ labels the finite set of primary fields $\phi_s$ of the theory and $N_{pq}^r$ are fusion coefficients of primaries, namely they satisfy,
\bea
\phi_p \times \phi_q = \sum_r N_{pq}^r \phi_r
\eea
In particular, the operator product coefficients $N_{pq}^r$ are integers. Treating $N_p$ with entries $(N_p)_{q}^r=N^r_{pq}$ as matrices, we may write their characteristic polynomials as follows,
\bea
{\rm det}\left(\lambda_p \mathds{1} - N_p \right)=0
\eea
The polynomial in each $\lambda_p$ is a monic with integer coefficients. From the Verlinde formula, it follows that the roots are given by, 
\bea
\lambda_p^{(q)} = {S_{p,q} \over S_{0,q}}
\eea

Whereas the matrix entries of $N_p$ are integers,  the ratios of S-matrix elements $\lambda_p^{(q)}$ are generically not even rational. Consider the extension field $\QQ(S_{p,q})$ over $\QQ$, where $\QQ(S_{p,q})$ denotes extension by the matrix elements of the modular S-matrix. By Theorem \ref{eq:basicGalres}, the elements  $\sigma \in {\rm Gal}(\QQ(S_{p,q}), \QQ)$ act on the $\lambda_p^{(q)}$ by permuting them amongst themselves. In particular, we have, 
\bea
\sigma\left({S_{p,q} \over S_{0,q}} \right) = {\sigma(S_{p,q}) \over \sigma(S_{0,q})} = {S_{p, \sigma(q)} \over S_{0,\sigma(q)}}
\eea
Note that here we have associated the permutation of the roots $\lambda_p^{(q)}$ to a permutation of the underlying primary fields. Indeed, it is possible to prove that there exist a set of signs $ \eps_{\sigma}(q) \in \{\pm1\}$ such that
\bea
\sigma(S_{p,q}) = \eps_{\sigma}(q) S_{p,\sigma(q)}
\eea
i.e. Galois conjugation acts on the S-matrix by signed permutation. 

\sm

This so-called \textit{Galois symmetry} of the RCFT leads to the following physical constraints.  Recall that the partition function of a RCFT is obtained by gluing together the chiral and anti-chiral characters via,
\bea
Z(\tau,\bar\tau) = \sum_{(p, \bar p) \, \in \, \cI \, \times \, \overline{\cI}}\cN_{p,\bar p}\, \chi_p(\tau) \, \overline{\chi}_{\bar p}(\bar \tau)
\eea
By modular invariance the gluing matrix $\cN_{p,\bar p}$ is required to satisfy $ (\cN \overline{S})_{p, \bar q} =(S\cN)_{p, \bar q} $. We now apply a Galois automorphism $\sigma$ to this equation, giving, 
\bea
\sum_{\bar m} \cN_{p, \bar m} \bar\eps_{\sigma}(\bar m) \overline{S}_{\sigma(\bar m),\bar q} &=& \eps_{\sigma}(p) \sum_m S_{\sigma(p),m}\cN_{m, \bar q}
\no\\
&=& \eps_{\sigma}(p) \sum_{\bar m} \cN_{\sigma(p),\bar m} \overline{S}_{\bar m, \bar q}
\no\\
&=& \eps_{\sigma}(p) \sum_{\bar m} \cN_{\sigma(p), \sigma(\bar m)} \overline{S}_{\sigma(\bar m), \bar q}
\eea
Using invertibility of $S$, we then obtain,
\bea
\cN_{\sigma(p), \sigma(q)} = \eps_\sigma(p) \bar \eps_{\sigma}(\bar q) \cN_{p,\bar q}
\eea
Because $\cN\geq 0$, we conclude that  $\cN_{p,\bar q}$ can only be non-zero when $\eps_\sigma(p) =\bar \eps_{\sigma}(\bar q)$ for all $\sigma$. This turns out to be a very useful result, leading to powerful selection rules for possible modular invariant combinations of characters. In the literature on rational conformal field theory, it is often referred to as the ``parity rule".

\subsection*{$\bullet$ Bibliographical notes}

An all-time classic introduction to Galois theory may be found in the book by Artin~\cite{Artin}, while a general treatment of modern algebra pertinent to Galois theory is given in the book by van der Waerden~\cite{VDW}. The books by Escofier~\cite{escofier2012galois} and Cox~\cite{Cox2} provide clear and systematic introductions with many examples, problems sets, and historical notes. We also refer to the recent book by Hungerford \cite{hungerford2012abstract}. 

\sm

In the context of rational conformal field theory, Galois symmetry first appeared in \cite{de1991markov}, with numerous follow-up works including \cite{Coste:1993af,Coste:1999yc,Bantay:2001ni}. The connection to Hecke operators was discussed in \cite{Harvey:2018rdc,Harvey:2019qzs}.  Galois symmetry also plays a role in the study of three-dimensional  topological quantum field theories, as discussed in e.g. \cite{Buican:2019evc,Buican:2021axn,Kaidi:2021gbs}.

\newpage

\appendix

\part{Appendix}

Four relatively short appendices collect basic mathematical material that, while central to various parts of the core material in these lecture notes, is not concerned directly with either modular forms or string theory. 

\sm

Appendix \ref{sec:modN} is primarily included for the benefit of the reader with a physics training, and is presumably familiar to every reader with a mathematics training. It includes a basic review of arithmetic modulo~$N$, the Chinese remainder theorem, solving polynomial equations in modular arithmetic, quadratic residue calculations, and the quadratic reciprocity theorem. We conclude with a brief introduction to Dirichlet characters and $L$-functions.

\sm

Appendix \ref{sec:RS} provides a review of some of the basic definitions,   theorems, and general results on compact Riemann surfaces needed in the main text. This material is directly relevant to  the physical applications to conformal field theory discussed in section~\ref{sec:TorusQFT}; the structure of the modular curves in section~\ref{sec:Cong};  the perturbative string amplitudes in section~\ref{sec:SA}; the topic of Seiberg-Witten theory in section \ref{sec:SW}; and to the subsequent appendices \ref{sec:LB} and \ref{sec:Theta}. The topological, metrical, and complex structural aspects of Riemann surfaces are reviewed as are their uniformization and construction in terms of Fuchsian groups.   

\sm

Appendix \ref{sec:LB} provides a review of holomorphic line bundles on compact Riemann surfaces, the Riemann-Roch theorem, the vanishing theorems, and their relation to the dimension of moduli space and to the construction of the spaces of tensors and spinor fields on Riemann surfaces. This material is directly relevant to the physical applications discussed in section~\ref{sec:TorusQFT}, to perturbative string amplitudes in section \ref{sec:SA}, and to the material in appendix \ref{sec:Theta}.

\sm

Appendix \ref{sec:Theta} provides an overview of higher rank Riemann $\tet$-functions, the Riemann relations on $\tet$-functions, modular geometry of the Siegel upper half space, the Abel map,  Jacobian varieties, and their relation to higher genus Riemann surfaces via the Riemann vanishing theorem. Using these ingredients, the construction of holomorphic and meromorphic differentials of arbitrary weight on Riemann surfaces of arbitrary genus {is} carried out, using the prime form as an essential building block. As a Physics application, the correlation functions for the $bc$ system, which were already discussed in the special case of genus one Riemann surfaces in section \ref{sec:TorusQFT}, are evaluated explicitly using the properties of meromorphic differentials and the operator product expansion.

\newpage

\section{Some arithmetic}
\setcounter{equation}{0}
\label{sec:modN}

In this appendix we collect some basic results in number theory, including the Chinese remainder theorem, its application to solving polynomial equations, the Legendre and Jacobi symbols, quadratic reciprocity, its application to solving quadratic equations mod $N$, and a brief introduction to Dirichlet characters and $L$-functions. Throughout the set of positive integers is denoted $\NN = \{ 1, 2,3, \cdots \}$ and does not include 0.

\subsection{Arithmetic mod N}

We denote by $\ZZ_N= \ZZ/N \ZZ = \{ 0,1,2,\cdots, N-1\}$ the additive group of integers modulo $N$. Addition of two elements $a,b \in \ZZ_N$ is defined by $a+b ~ (\mod N) \in \ZZ_N$, the identity is $0$, and the inverse of $a \in \ZZ_N$  is $-a ~ (\mod N)$. The group $\ZZ_N$ is isomorphic to the multiplicative group of $N$-th roots of unity $\{ 1, \ep, \ep ^2, \cdots , \ep ^{N-1} \}$ where $\ep = e^{2 \pi i /N}$ and multiplication is in $\CC$.

\sm

For $p \in \NN$ a prime number, the set $\ZZ_p^*= \{ 1,2,\cdots, p-1\}$ forms a group under multiplication of integers $(\mod p)$, and $\FF_p= \{0,1,2,\cdots, p-1\}$ forms a \textit{finite field}.

\sm

For arbitrary $N \in \NN$, we denote by $\ZZ^*_N$ the maximal multiplicative subgroup of $\ZZ_N$. When $N$ is prime, the definition coincides with the one provided above. When $N$ is not prime, however, there are elements in $\ZZ_N$ other than 0 which are not invertible, and have to be removed to produce a multiplicative group. Consider an element $a \in \ZZ_N$ which is not relatively prime with $N$, namely such that $\gcd(a,N) >1$, then it follows that the element $b=N/\gcd(a,N) \in \ZZ_N$ satisfies $ab =N$ so that $ab \equiv 0 ~ (\mod N)$ and therefore the element $a$ is not invertible in $\ZZ_N$. Every invertible element $a \in \ZZ_N$ must be relatively prime to $N$, so that the multiplicative subgroup is given by,
\bea
\ZZ_N^* = \{ a \in \ZZ_N , ~ \gcd(a,N)=1 \} 
\hskip 1in 
| \ZZ_N^* | = \phi(N)
\eea
where $\phi(N)$ is Euler's totient function counting precisely the number of integers less than $N$ that are relatively prime to $N$, and $|S|$ denotes the cardinality of the set $S$. For example, we may represent the group $\ZZ_{15}^*$ by listing its elements in pairs of opposites modulo 15 as follows $\ZZ_{15}^* = \{ \pm 1, \pm 2, \pm 4, \pm 7\}$. The elements $\pm 1$ and $\pm 4$ square to 1 $(\mod 15)$, while   $7\times (-2) \equiv (-7) \times 2 \equiv 1 ~ (\mod 15)$.

\subsection{Chinese remainder theorem}

{\thm
\label{A.thm:1}  
Let $n_1, \cdots , n_r$ be positive integers which are relatively prime in pairs so that they satisfy $\gcd(n_i,n_j)=1$ for all $1\leq i \not = j \leq r$ and let $b_1, \cdots b_r$ be arbitrary integers. Then the system of congruences,
\bea
x \equiv b_i  ~ (\mod n_i) \qquad \hbox{ for all } ~ i=1,\cdots , r
\eea
has a unique solution modulo $N= n_1 \cdots n_r$. An equivalent statement is that the map,
\bea
x ~(\mod N) \, \mapsto \, \big ( x ~ (\mod n_1), \cdots , x ~ (\mod n_r) \big )
\eea
is a  ring isomorphism,
\bea
\ZZ_N \, \cong \, \ZZ_{n_1} \times \cdots \times \ZZ_{n_r}
\eea}
To prove the theorem, we begin by proving that the solution is unique, so that the map is injective. If we had  two solutions $x, y$ satisfying the same congruence conditions then $x-y \equiv 0 ~ (\mod n_i)$ for all $i$. Since the $n_i$ are pairwise co-prime their product $N$ also divides $x-y$ so that $x-y \equiv 0 ~ (\mod N)$. Since the cardinality of the left and right sides are equal to one another, the map must be surjective as well, and is thereby an isomorphism.

\subsection{Solving polynomial equations}

{\thm 
\label{A.thm:2}
Let $f(x)$ be a polynomial in $x$ with integer coefficients $f \in \ZZ[x]$, let $n_1, \cdots n_r$ be positive integers which are relatively prime in pairs $\gcd(n_i,n_j)=1$ for all $1\leq i \not= j \leq r$, and let $N=n_1 \cdots n_r$. Then the congruence,
\bea
\label{Ch.1}
f(x) \equiv 0 ~ (\mod N)
\eea
has a solution if and only if each one of the congruences,
\bea
\label{Ch.2}
f(x) \equiv 0 ~ (\mod n_i)  \qquad i=1,\cdots, r
\eea
has a solution. The number of solutions $\nu(N) $  to (\ref{Ch.1}) is given in terms of the number of solutions $\nu(n_i)$ for each of the congruences by,
\bea
\nu(N) = \nu(n_1) \cdots \nu(n_r)
\eea}
To prove the theorem we prove both implications. Clearly if $x$ satisfies (\ref{Ch.1}) then it satisfies (\ref{Ch.2}). Conversely, let $x_i ~ (\mod n_i)$ be a solution to each congruence (\ref{Ch.2}), which exists by the assumption of the theorem, then there exists $x ~ (\mod N)$ which solves,
\bea
x \equiv x_i ~ (\mod n_i)
\eea
by the Chinese remainder theorem.

\sm

Applying Theorem \ref{A.thm:2} to  polynomial equations modulo an integer $N$ whose prime factorization is given by $N = \prod _i p_i ^{\a_i}$ for $p_i$ all distinct primes and $\a_i\geq 1$, we have the equivalence,
\bea
f(x) \equiv 0 ~ (\mod N) \qquad \Leftrightarrow \qquad \big \{  f(x) \equiv 0 ~ (\mod p_i^{\a_i}) \hbox{ for all } i \big \}
\eea
The individual congruences for each prime power are given as follows.

\sm

{\thm 
\label{A.thm:3}
Assume $p$ prime, $\a\geq 2$, and let $r$ be a solution of the congruence,
\bea
f(x) \equiv 0 ~ (\mod p^{\a-1}) 
\eea
with $0 \leq r < p^{\a-1}$.  The existence of solutions proceeds as follows.
\begin{description}
\itemsep=0in
\item (a) If $f'(r) \not \equiv 0 ~ (\mod p)$ then $r$ lifts in a unique way  from $(\mod p^{\a-1})$ to $(\mod p^\a)$. 
\item (b) If $f'(r)\equiv 0 ~ (\mod p)$
\begin{itemize}
\item $f(r) \equiv 0 ~ (\mod p^\a) $ then $r$ lifts from $(\mod p^{\a-1})$ to $(\mod p^\a)$ in $p$ different ways.
\item $f(r) \not \equiv 0 ~ (\mod p^\a) $, then $r$ cannot be lifted from $(\mod p^{\a-1})$ to $(\mod p^\a)$.
\end{itemize}
\end{description}}
To prove the theorem, we consider all possible lifts of $r$ from $(\mod p^{\a-1})$ to $(\mod p^\a)$, which  are given by $r + q p^{\a-1} ~ (\mod p^\a)$. Taylor expanding the polynomial $f$ of degree $N$, 
\bea
f(r+qp^{\a-1}) = f(r) + f'(r) q p^{\a-1} + \half f''(r) q^2 p^{2\a-2} + \cdots + { 1 \over N!} f^{(N)}(r) q^N p^{N \a-N}
\eea
and reducing this equation $(\mod p^\a)$ we obtain, 
\bea
f(r+ qp^{\a-1}) \equiv f(r) + f'(r) q p^{\a-1} ~ (\mod p^\a)
\eea
Since $f(r) \equiv 0 ~ (\mod p^{\a-1})$, we must have $f(r) = k p^{\a-1}$ for some integer $k$. Therefore the number of solutions to $f(x) \equiv 0 ~ (\mod p^\a)$ equals the number of solutions to the  congruence
$k + q f'(r) \equiv 0 ~ (\mod p)$.  The statements of the Theorem now readily follows.

\subsection{Quadratic residues and quadratic residue symbols}
\label{sec:5.4}

If $m,N \in \NN$ and $n \in \ZZ$ with $\gcd(n,N)=1$, then $n$ is an $m$-th power residue mod $N$ if $x^m \equiv n ~ (\mod N)$ admits a solution, and a non-residue if it admits no solutions. In the special case of quadratic residues for $m=2$ with prime  $N=p$ an odd prime,  an integer $n \not \equiv 0 ~ (\mod p)$ is a quadratic residue if the equation,
\bea
\label{quad}
x^2 \equiv n ~ (\mod p)
\eea 
has a solution, and a quadratic non-residue if the equation has no solutions.  If $n \not \equiv 0 ~ (\mod p)$ is a quadratic residue and $x$ is a solution to (\ref{quad}), then  $-x$ is also a solution to the same equation.
When $n \equiv 0 ~ (\mod p)$ then equation (\ref{quad}) has one solution, namely $x \equiv 0 ~ (\mod p)$.  

\sm

The \textit{Legendre symbol} $(n|p)$ for $p$ prime combines all these cases and  is defined as follows,\footnote{Frequently used alternative notations for the Legendre symbol are $({ n \over p})$ and $(n/p)$.}
\bea
(n|p) = \left \{ \bma +1 & \hbox{ if }  n  \hbox{ is  a quadratic residue } \mod p \cr 
-1 & \hbox{ if }  n  \hbox{ is a quadratic non-residue } \mod p \cr
0 & \hbox{ if }  n \equiv 0 ~ (\mod p) \cr \ema \right .
\eea
The number of solutions $x$ to equation (\ref{quad}) is $1+(n|p)$ in all cases. 

\sm

{\prop 
\label{5.prop1}
Further properties of quadratic residues and the Legendre symbol for an arbitrary odd prime $p$ are as follows.
\begin{enumerate}
\itemsep=0in
\item There are $(p-1)/2$ residues and $(p-1)/2$ non-residues in $\ZZ_p^*$;
\item For any $n \in \ZZ$ we have $(n|p) = n^{(p-1)/2} ~ (\mod p)$;
\item The first supplement is  $(-1|p)=(-)^{(p-1)/2}$;
\item The second supplement is $(2|p)=(-)^{(p^2-1)/8}$;
\item For arbitrary $m,n \in \ZZ$ we have  $(mn|p)=(m|p)(n|p)$;
\end{enumerate}}
The proof proceeds as follows. Item 1 follows from the fact that the map $x \mapsto x^2$ is two-to-one in $\ZZ_p^*$. 
For Item 2, we first note that Fermat's little theorem, namely  that $n^{p-1} \equiv 1 ~ (\mod p) $ for any $n$ not divisible by $p$, implies $n^{(p-1)/2} \equiv \pm 1 ~ (\mod p)$. For $(n|p)=1$  there exists an $x$ so that $x^2 \equiv n ~ (\mod p)$ and hence $n^{(p-1)/2}= x^{p-1} \equiv 1 =(n|p) ~ (\mod p)$. For $(n|p)=-1$  the degree $(p-1)/2$ polynomial congruence equation $x^{(p-1)/2} -1 \equiv 0 ~ (\mod p)$ has $(p-1)/2$ solutions. The quadratic residues already provide $(p-1)/2$ solutions, so the non-residues cannot be solutions. As  a result,  we have $x^{(p-1)/2} \not \equiv 1 ~ (\mod p)$ and hence $x^{(p-1)/2} \equiv -1 = (n|p) ~ (\mod p)$. Finally, Items 3,  4, and 5 follow from Item 2. Item 3 may also be derived from Fermat's theorem that a prime $p$ is the sum of the squares of two integers if and only if $p\equiv 1 ~ (\mod 4)$.

\sm

The \textit{Jacobi symbol} $(n|N) $ generalizes the Legendre symbol to an arbitrary positive odd integer $N=\prod _i p_i ^{\a_i}$ where $p_i$ are distinct primes and $\a_i \geq 1$ and any $n \in \ZZ$ by,
\bea
\label{A.jacobi}
(n|N) = \prod _i (n|p_i)^{\a_i}
\eea
where in each factor $(n|p_i)$ is the Legendre symbol.  The \textit{Kronecker symbol} further generalizes the Jacobi symbol by lifting the restrictions that $N$ be odd and positive, and is defined as follows. If $N = u \prod _i p_i ^{\a_i}$ where $u = \mp 1$ and the product runs over all distinct primes including 2, then the Kronecker symbol is defined by,
\bea
(n|N) = (n|u) \prod _i (n|p_i)^{\a_i}
\eea
where $(n|p_i)$ stands for the Legendre symbol when $p_i$ is an odd prime, while the remaining data are supplied by,
\bea
(n|-1) = \left \{ \bma 1 & \hbox{ if } n \geq 0 \cr -1 & \hbox{ if } n <0 \cr \ema \right .
\hskip 0.9in 
(n|2) = \left \{ \bma 
0 & \hbox{ if } n \equiv 0  ~ (\mod 2) ~~ \cr 
+1 & \hbox{ if } n \equiv \pm 1 ~ (\mod 8) \cr 
-1 & \hbox{ if } n  \equiv \pm 3 ~ (\mod 8) \cr 
\ema \right .
\eea
The Legendre, Jacobi, and Kronecker symbols are referred to as \textit{quadratic residue symbols}.

\subsubsection{Quadratic residues $x^2 \equiv -1 ~ (\mod N)$}
\label{sec:1.5.1}
We now analyze the particular example of  quadratic residues of $-1$ modulo $N$. This example, as well as that in the following subsection, will be used extensively when we discuss the dimensions and genera of congruence subgroups. 

\sm

To obtain the number of solutions of the congruence $x^2 \equiv -1 ~ (\mod N)$ for an arbitrary integer $N$ with prime factorization $N = \prod _i p_i ^{\a_i}$ we consider the polynomial $f(x) = x^2 +1$ with $f'(x)=2x$. If $p$ is an odd prime, and $0\leq r < p$ solves $r^2+1 \equiv 0 ~ (\mod p)$ then $r \not \equiv 0 ~(\mod p)$. As a result, $f'(r) = 2 r \not \equiv 0 ~ (\mod p)$ so that there is a unique lift of $r$ to $(\mod p^2)$, and to any $(\mod p^\a)$. If $p=2$ then $r \equiv 1 ~ (\mod 2)$ so there is one solution $(\mod 2)$. Now using Theorem \ref{A.thm:3}, $f'(r) \equiv 0 ~ (\mod 2)$,  and the fact that every lift  $f(1+2q) \equiv 1 ~ (\mod 4)$ implies that there are no solutions $(\mod 4)$, and hence no solutions $(\mod 2^\a)$ for all $\a \geq 2$. Hence the number of solutions to $x^2 \equiv -1 ~ (\mod N)$ is given by,
\bea
\label{quadcon.1}
\# \Big \{ x \in \ZZ_N \,  \big | \, x^2 \equiv -1 ~ (\mod N) \Big \} = \left \{ \bma 0 && 4 | N \cr 
\prod _i \Big ( 1+ (-1|p_i) \Big ) && 4 \nmid  N \cr \ema \right .
\eea
where  the Legendre symbol $(-1|p)$ evaluates to $(-1)^{{ p-1 \over 2}}$ for an odd prime $p$. For $p=2$, the number of solutions is 1, so we set $(-1|2)=0$.

\subsubsection{Quadratic residues  $x^2+x \equiv -1 ~ (\mod N)$}
\label{sec:1.5.2}

To obtain the number of solutions of the congruence $x^2+x \equiv -1 ~ (\mod N)$ for an arbitrary integer $N$ with prime factorization $N = \prod _i p_i ^{\a_i}$ we consider the polynomial $f(x) = x^2+x +1$ with $f'(x)=2x+1$. 
Solving $f(x) \equiv 0 ~ (\mod N)$ is equivalent to solving,
\bea
(2x+1)^2 + 3 \equiv 0 ~ (\mod 4N)
\eea
If $9 |N$, then the congruence $y^2+3 \equiv 0 ~ (\mod 9)$ has no solutions as the quadratic residues $\mod 9$ are $0,1,4,7$, giving the residues of $y^2+3$ to be $3,4,7,1$ but not 0.  For an odd prime $p$, the congruence  $y^2 + 3 \equiv 0 ~ (\mod p)$ has $1+(-3|p)$ solutions. By quadratic reciprocity, if $p \not= 3$ then $(-3|p) = (-1|p)(3|p) =(p|3)$ which evaluates to $\pm 1$ when $p \equiv \pm 1 ~ (\mod 3)$. For $p \equiv -1 ~ (\mod 3)$ there are no solutions $y \in \ZZ$ and hence no solution of the form $y \equiv 2x+1~ (\mod p)$. For $p \equiv 1 ~ (\mod 3)$ there are two solutions $y ~ (\mod p)$. For any given $y$, there is then a unique solution $x$ to $y \equiv 2x+1~ (\mod p)$. 
 The number of solutions is thus given by,
\bea
\label{quadcon.2}
\# \Big \{ x \in \ZZ_N \, \big | \, x^2+x \equiv -1 ~ (\mod N) \Big \} = \left \{ \bma 0 && 9 | N \cr 
\prod _i \Big ( 1+ (-3|p_i) \Big ) && 9 \nmid  N \cr \ema \right .
\eea
For $p=3$ the number of solutions is 1, so we set $(-3|3)=0$.

\sm

\subsection{Gauss sums}

For an odd prime $p$, and an arbitrary $a \in \ZZ$, the Gauss sum $g_a$ is defined by,
\bea
\label{Gauss1}
g_a = \sum _{k=0}^{p-1} (k|p) \, \ep ^{ka}
\hskip 1in 
\ep = e^{2\pi i/p}
\eea
Gauss sums are directly related to the Legendre symbol, 
\bea
\label{Gauss2}
g_a = (a|p) g \hskip 1in g= g_1
\eea
Indeed, for $a$ divisible by $p$ we have $g_a=0$ in view of the fact that the number of residues in $\ZZ_p^*$ equals the number of non-residues. Hence the relation holds trivially when $p | a$ as both sides vanish. For $p \nmid a$, 
we use the fact that $(a|p)^2=1$, the definition of $g_a$, and the complete multiplication property of the Legendre symbol,
\bea
g_a =  (a|p) \sum _{k=0}^{p-1} (a|p) (k|p) \, \ep ^{ka} = (a|p) \sum _{k=0}^{p-1} (ka|p)  \, \ep ^{ka} = (a|p) g
\eea
The last equality follows,  for $p \nmid  a$, from the fact that the sum over $k$ is identical to the sum over $ka$. Furthermore, we have the following relation, 
\bea
\label{g2}
g^2 = ( -)^{(p-1)/2} p
\eea
which follows from $g_a g_{-a} = (-1|p) g^2$ and expressing its sum over $a$ in two different ways, 
\bea
\sum _{a=0}^{p-1} g_a g_{-a} = (p-1) (-1|p) g^2  =  \sum_{a,k,\ell=0}^{p-1} (k\ell |p) \ep^{(k-\ell)a}= p \sum_{k,\ell=0}^{p-1} (k\ell | p) \delta_{k,\ell} = p(p-1)
\eea
where $\delta_{k,\ell}=1$ when $k=\ell$ and zero otherwise. The result (\ref{g2}) follows upon using the first supplement $(-1|p)=(-)^{(p-1)/2}$.

\subsection{Quadratic reciprocity}

Quadratic reciprocity states that for $p,q$ distinct odd primes, we have, 
\bea
(p|q)(q|p) = (-)^{(p-1)(q-1)/4}
\eea
Many proofs exist of the law of quadratic reciprocity, of which Gauss produced at least 8. Here we shall use Gauss's sums as defined in the preceding subsection. For any odd prime $q$ not equal to $p$, we begin by using (\ref{g2}) to write,
\bea
\label{pq}
g^{q-1} = (g^2)^{(q-1)/2} = (g^2| q) = (-1|q)^{(p-1)/2} (p|q)= (-)^{(p-1)(q-1)/2} (p|q)
\eea
where we used the relation $(n|p) \equiv n^{(p-1)/2} ~ (\mod p)$ of property 2. of \ref{5.prop1}  in the second equality.  Next, we use the fact that the prime power of a multiple sum evaluates as follows, 
\bea
(a_1 + \cdots + a_n)^q = a_1^q + \cdots a_n^q ~ (\mod q)
\eea
This relation is valid for $a_1, \cdots, a_n \in \ZZ$, but we generalize it here to the ring of algebraic integers $\ZZ_p(\ep)$. As a result we have,
\bea
g^q = \left ( \sum_{k=0}^{p-1} (k|p) \ep^k \right )^q = \sum_{k=0}^{p-1} (k|p)^q \ep^{kq} 
= \sum ^{p-1}_{k=0} (k|p) \ep^{kq} = g_q ~ (\mod q)
\eea
By (\ref{Gauss2}), we have $g_q = (q|p) g$, and upon combining with (\ref{pq}), we get, 
\bea
g^{q-1} =  (-)^{(p-1)(q-1)|2} (p|q) = (q|p)
\eea
The last equality  implies the law of quadratic reciprocity upon using $(p|q)^2=1$.

\subsection{Characters}
\label{sec:charLfunc}

Consider a finite Abelian group $G$. A function $\chi: G \rightarrow \CC^*$ is said to be a \textit{character} if it is a group homomorphism, i.e. $\chi (g h) = \chi(g) \chi(h)$ for all $g,h \in G$. Characters form a group $\hat G = {\rm Hom}(G,\CC^*)$, known as the Poincar{\'e} dual of $G$.  We will denote the identity element of this group as $\chi_0$, such that $\chi_0(g) = 1$ for all $g$. The inverse elements are obtained by complex conjugation $\chi^{-1}(g) = \overline{\chi(g)}$. 

\sm

As an example, consider the case of a cyclic group $G = \ZZ_n$. If we denote the order-$n$ generator by $v$, then it is clear that $\chi(v)^n = \chi(v^n) = 1$ and hence we can take $\chi(v)$ to be an $n$-th root of unity. Conversely, every $n$-th root of unity defines a character of $\ZZ_n$. Thus we conclude that $\hat G$ is in fact isomorphic to the original group, $\hat G \cong \ZZ_n$. 

\sm

Characters satisfy the following relations 
\bea
\label{eq:preSchurorth}
\sum_{g \in G} \chi(g) & = & \left\{ \begin{matrix} |G| & \chi = \chi_0 \\ 0 & {\rm otherwise}\end{matrix}\right. 
\no \\
\sum_{\chi \in \hat G} \chi(g) & = & \left\{ \begin{matrix} |\hat G| & g = \mathds{1} \\ 0 & {\rm otherwise}\end{matrix}\right. 
\eea
The first line of each of these is obvious. To prove the second line of the first relation, begin by choosing $h \in G$ such that $\chi(h) \neq 1$. Then we have 
\bea
\chi(h) \sum_{g \in G} \chi(g)  = \sum_{g \in G} \chi(hg) = \sum_{g \in G} \chi(g)
\eea
where we have used the fact that $\chi$ is a homomorphism, as well as closure of the group $G$. Since $\chi(h) \neq 1$, we conclude $\sum_{g \in G} \chi(g)=0$ as per the claim. The second relation is proven exactly analogously---namely we multiply by $\chi'(g) \neq 1$ and used the fact that $\hat G$ is a group. 

\sm

From these results we may obtain the following Schur orthogonality relations 
\bea
\sum_{g \in G} \chi(g) \overline{\chi'(g)} & = & \left\{ \begin{matrix} |G| & \chi = \chi' \\ 0 & {\rm otherwise}\end{matrix}\right. 
\no \\
\sum_{\chi \in \hat G} \chi(g) \overline{\chi(h)} & = & \left\{ \begin{matrix} |\hat G| & g = h \\ 0 & {\rm otherwise}\end{matrix}\right. 
\eea
Indeed, for the former note that if $\chi, \chi' \in \hat G$ then $\chi (\chi')^{-1} \in \hat G$ as well. We may thus use the relation (\ref{eq:preSchurorth}) with $\chi$ replaced by $\chi (\chi')^{-1}$. Then noting that,
\bea
\sum_{g \in G}(\chi (\chi')^{-1})(g) = \sum_{g \in G}\chi(g) \overline{\chi'(g)}
\eea 
we obtain the desired result. Likewise for the latter relation, we use the fact that,
\bea
\sum_{\chi \in \hat G} \chi (g h^{-1}) = \sum_{\chi \in \hat G} \chi(g) \overline{\chi(h)}
\eea 
This completes the proof of both relations in (\ref{eq:preSchurorth}).

\newpage

\subsection{Dirichlet characters and L-functions}
\label{sec:Lseries}

A \textit{Dirichlet character} for $a \in \NN$ is a function $\chi_a: \ZZ \rightarrow \CC$ satisfying the following properties: 
\begin{enumerate}
\itemsep=0in
\item Periodicity:  $\chi_a(n) = \chi_a(n+a)$ for all $ n \in \ZZ$. 
\item $\chi_a(n) \neq 0$ if and only if $\gcd(n,a) = 1$.
\item Multiplicativity: $\chi_a(nm) = \chi_a(n) \chi_a(m)$ for all $n,m \in \ZZ$. 
\end{enumerate}
Dirichlet characters can be thought of as characters for $ G = \ZZ_a^*$, the multiplicative group of invertible elements of $\ZZ_a$. This is an Abelian group with $|G| = \phi(a)$, where $\phi$ is Euler's totient function. In this case our orthogonality relations read as follows, 
\bea
\label{eq:Dircharsortho}
\sum_{g =0}^{a-1} \chi_a(g) \overline{\chi'_a(g)} & = & \left\{ \begin{matrix} \phi(a) & \chi_a = \chi'_a \\ 0 & {\rm otherwise}\end{matrix}\right. 
\no\\
\sum_{\chi_a \in \hat G} \chi_a(g) \overline{\chi_a(h)} & = & \left\{ \begin{matrix}\phi(a) & g \equiv h ~(\mod a)\\ 0 & {\rm otherwise}\end{matrix}\right. 
\eea
We now introduce the Dirichlet L-function, 
\bea
L_a(s, \chi) = \sum_{n= 1 }^\infty{\chi_a(n) \over n^s}
\eea
For $\chi_a$ not equal to the identity character, $L_a(s,\chi)$ converges and is analytic for ${\rm Re}\,s >1$ \cite{iwan3,Serre}. In this region, we have the following Euler product formula, 
\bea
\label{eq:DirLprod}
L_a(s, \chi) = \prod_p {1\over 1 - {\chi_a(p) \over p^s}}
\eea
the product being over prime numbers. To prove this, note that for each prime $p$, we have, 
\bea
(1- \chi_a(p) p^{-s})^{-1} = \sum_{n=0}^\infty\chi_a(p)^n p^{-ns} = \sum_{n=0}^\infty \chi_a(p^n) p^{-ns}
\eea
Hence for fixed prime $q$, we have, 
\bea
\prod_{p \leq q}{1\over 1 - {\chi_a(p) \over p^s}} = \sum_{n \in S_q} {\chi_a(n) \over n^s}
\eea
with $S_q$ the set of all natural numbers whose prime factors are less than or equal to $q$. Then for any $N \in \NN$, we conclude that, 
\bea
\sum_{n=1}^N{\chi_a(n) \over n^s} = \prod_{p \leq q}{1 \over 1- {\chi_a(p) \over p^s}} - \sum_{\substack{n \in S_q\\ n \geq N}} {\chi_a(n) \over n^s}
\eea
with $q$ the largest prime less than or equal to $N$. Taking $N\rightarrow \infty$ gives (\ref{eq:DirLprod}). 

Using this product formula, it is possible to show that $L_a(s, \chi_0)$ extends to a meromorphic function on ${\rm Res} \,s>0$, with the only pole at $s=1$. We do not reproduce the proof here; it can be found in e.g. \cite{iwan3,Serre}.

\subsubsection{Dirichlet's Theorem on Arithmetic Progressions}
We now use Dirichlet L-functions to prove Dirichlet's Theorem on arithmetic progressions, 

{\theorem Let $m$ and $a$ be relatively prime natural numbers. Then there are infinitely many prime numbers $p$ such that $p \equiv m ~(\mod a)$. }
\newline

In order to prove this, we aim to prove the following stronger result, 
\bea
\sum_{p \equiv m ~(\mod a)} {1\over p}  \hspace{0.2 in} \rm{is \,\,\, divergent}
\eea
which clearly implies the theorem. To do so, begin by considering the logarithm of the Dirichlet L-function, which by the product formula takes the form
\bea
\log L_a(s, \chi) = \sum_p \left(- \log\left(1- {\chi_a(p) \over p^s}\right)\right)
\eea
valid for ${\rm Re}\, s >1$. Taylor expanding the logarithm, this gives
\bea
\label{eq:LTaylorexp}
\log L_a(s, \chi) = \sum_p {\chi_a(p) \over p^s} + \sum_p \sum_{n=2}^\infty {\chi_a(p)^n \over n p^{ns}}
\eea
where, for reasons to be clear in a moment, we have split the $n=1$ term from the rest. Using the orthogonality of Dirichlet characters in (\ref{eq:Dircharsortho}), we find that 
\bea
\label{eq:Dirmain}
\sum_{p \equiv m ~(\mod a)} {1\over p^s} &=& {1 \over \phi(a)} \sum_{\chi_a \in \hat G}\overline{\chi_a(m)} \sum_p {\chi_a(p) \over p^s} 
\\
&=& {1 \over \phi(a)} \left[\sum_{\chi_a \in \hat G} \overline{\chi_a(m)} \log L_a(s, \chi) - \sum_{\chi_a \in \hat G} \overline{\chi_a(m)} \sum_p \sum_{n=2}^\infty {\chi_a(p)^n \over n p^{ns}} \right] \no
\eea
where $(a,m)=1$ lest $\chi_a(m) = 0$. To prove the theorem, we must now show that the right-hand side is divergent as $ s \rightarrow 1$. To do so, we begin by showing that $ \sum_p \sum_{n=2}^\infty {\chi_a(p)^n \over n p^{ns}}$ is bounded as $ s\rightarrow 1$, and hence that the second term in the difference does not diverge. To see this, note that since $|\chi_a(p)| = 0$ or 1, we have $|\chi_a(p) p^{-s}| \leq |p^{-s}| \leq \half$. Then we have 
\bea
\left|\sum_{n=2}^\infty {\chi_a(p)^n \over n p^{ns}}\right| \leq \left|{\chi_a(p) \over p^s} \right|^2 \sum_{n=2}^\infty{1 \over n}\left| {\chi_a(p) \over p^s}\right|^{n-2} \leq \left|{\chi_a(p) \over p^s} \right|^2 \sum_{n=2 }^\infty \half {1 \over 2^{n-2}} = \left|{\chi_a(p) \over p^s} \right|^2 \leq {1 \over p^2}\,\,\,\,\,\,\,
\eea
Hence we conclude that 
\bea
\left|  \sum_p \sum_{n=2}^\infty {\chi_a(p)^n \over n p^{ns}}\right| \leq \sum_p{1 \over p^2}< \sum_{n=1}^\infty{1 \over n^2}
\eea
the latter of which is convergent. 

\sm

Having shown that the second term on the right-hand side of (\ref{eq:Dirmain}) is not divergent as $s \rightarrow 1$, we are now tasked with showing that the first term \textit{is} divergent. At the end of appendix \ref{sec:Lseries} we quoted the result that $L_a(s, \chi_0)\rightarrow \infty$ as $s \rightarrow 1$, and hence $\log L_a(s, \chi_0)$ diverges in this limit as well. To complete the proof it thus suffices to show that $L_a(1, \chi) \neq 0$ for all $\chi \neq \chi_0$. 

To show this, first consider a character $\chi_a$ such that $\chi_a^2$ is not the identity character. Then we can consider the function 
\bea
\lambda(s) = L_a(s, \chi_0)^3 L_a(s, \chi)^4 L_a(s, \chi^2) 
\eea
By (\ref{eq:LTaylorexp}) we have
\bea
|\lambda(s)| = \left|{\rm exp}\left[\sum_{p}\sum_{n=1}^\infty{3 + 4 \chi_a(p^n)+\chi_a^2(p^n) \over n p^{n s}} \right] \right|
\eea
Letting $\theta_{n,p} = {\rm Arg}\, \chi_a(p^n)$, we then obtain 
\bea
|\lambda(s)| = {\rm exp}\left| \sum_{p}\sum_{n=1}^\infty{3 + 4 \cos\theta_{n,p}+\cos 2\theta_{n,p} \over n p^{n\, {\rm Re}\,s}} \right|
\eea
But for any real $\theta$, we have 
\bea
3 + 4 \cos\theta+2 \cos\theta = 2(1+\cos\theta)^2 \geq 0
\eea
Thus we conclude that 
\bea
|\lambda(s) | \geq e^0 = 1
\eea
This gives the desired result. Indeed, if we had $L_a(1, \chi) = 0$, then since $L_a(s, \chi_0)$ has a simple pole at $s=1$ but $L_a(s, \chi^2)$ does \textit{not} (because $\chi_a^2 \neq \chi_0$ by assumption) the degree $\geq 4$ zero wins against the degree 3 pole, and we would have $\lambda(1) = 0$, in contradiction to the inequality just obtained. Thus we must have $L_a(1, \chi) \neq 0$.

\sm

For the case of $\chi_a^2 = \chi_0$, a similar argument can be run using $\lambda(s) = L_a(s, \chi)L_a(s, \chi_0)$. In this case one aims to show that $\log \lambda(s)  \geq \log \zeta(2 s)$. In total, one concludes that $L_a(1, \chi) \neq 0$ for all $\chi \neq \chi_0$, completing the proof of the theorem.

\subsubsection{Perfect squares and quadratic residues}
We close  by using the various tools of this section, including Dirichlet's theorem of arithmetic progressions, to prove the following statement:

{\theorem A number $N \in \NN$ is a perfect square if and only if it is a quadratic residue for all primes $p$ coprime to $N$.}
\newline\newline
One direction of this statement is obvious: if $N$ can be written as a perfect square, it is clear that it is also a perfect square modulo any prime. We now focus on the other direction. Assume that $N$ is a quadratic residue modulo all coprime $p$. Say that $N$ is \textit{not} a perfect square. We may then write $N$ in the form $N= j^2 p_1 \dots p_r$ for distinct primes $p_1, \dots p_r$. Since any odd power of 2 is a quadratic non-residue mod 3, we may assume that there is at least one odd factor. 
Take $p_1$ to be odd and consider non-zero integers $m$ and $n_i$ such that $(m|p_1) = -1$ and $(n_i|p_i) = 1$ for $i = 2, \dots, r$. By the Chinese remainder theorem, there exists a solution to the following set of congruences 
\bea
\label{eq:perfectsqcong}
x \equiv 1 ~(\mod 4)\hspace{0.5 in}x \equiv m ~(\mod p_1)\hspace{0.5 in}x \equiv n_i ~(\mod p_i) \hspace{0.5 in} p_i \neq 2
\eea
If one of the prime factors $p_i = 2$, then we replace the final congruence with $x \equiv 1~(\mod 8)$. Given a particular solution $x_0$ to these congruences, the general solution takes the form $x = x_0 + 4 k p_1 \dots p_r$ for integer $k$. By Dirichlet's theorem of arithmetic progressions this takes an infinite number of prime values, and we choose one, say $x = p$, such that $p \nmid j$. By definition, we have 
\bea
(p| p_1) = (m| p_1) = -1 \hspace{0.8 in} (p| p_i) = (n_i | p_i) = 1 \hspace{0.5 in} i = 2, \dots, r
\eea
Then using quadratic reciprocity, together with the first congruence of (\ref{eq:perfectsqcong}), we conclude that 
\bea
(p_1| p) = -1\hspace{0.8 in} {\rm and}\hspace{0.8 in}(p_i | p) = 1 \hspace{0.3 in} i = 2, \dots, r
\eea
Thus $(N|p) = (p_1|p)\dots(p_r| p) =  -1$, and we have identified a prime $p$ coprime to $N$ for which $N$ is not a quadratic residue, contrary to hypothesis. We conclude that $N$ must be a perfect square.

\subsection*{$\bullet$ Bibliographical notes}

An excellent overview of some classic problems in arithmetic is given in the book by Cox~\cite{Cox}, where a historical perspective is also provided. Two other classics are the books by Serre~\cite{Serre} and Hecke~\cite{Hecke}. Useful references for results in analytic number theory, including $L$-functions, may be found in the book by Iwaniec \cite{iwan3}, Rademacher \cite{Rad2}, and Siegel \cite{Siegel5}.

 \newpage
 
\section{Riemann surfaces}
\setcounter{equation}{0}
\label{sec:RS}

A Riemann surface is a connected complex manifold of two real dimensions or equivalently a connected complex manifold of one complex dimension, also referred to as a complex curve.  In this appendix, we shall review the topology of Riemann surfaces, their homology groups,  homotopy groups, uniformization, construction in terms of Fuchsian groups, as well as their emergence from two-dimensional orientable Riemannian manifolds. 

\subsection{Topology}

The topology of a Riemann surface $\Sigma$ is the topology of the underlying orientable two-dimensional manifold $\Sigma$. Throughout, we shall restrict to Riemann surfaces whose boundary $\p \Sigma$ is the union of a finite number $b$ of one-dimensional components, and a finite number $n$ of points referred to as punctures of $\Sigma$.   Given a triangulation of $\Sigma$ with $F_\Sigma$ faces, $E_\Sigma$ edges, and $V_\Sigma$ vertices, the Euler characteristic $\chi(\Sigma)$ is defined by,
\bea
\label{B.Euler}
\chi (\Sigma) = F_\Sigma - E_\Sigma + V_\Sigma
\eea
The Euler characteristic is a topological invariant in the sense that its value is independent of the triangulation chosen for a given $\Sigma$.  In suitably chosen triangulations, the operation of adding a puncture to $\Sigma$ (namely removing a point from $\Sigma$) leaves $F_\Sigma$ and $ E_\Sigma$ unchanged but diminishes $V_\Sigma$ by one while adding a one-dimensional boundary component leaves $E_\Sigma $ and $ V_\Sigma$ unchanged but diminishes $F_\Sigma$ by one. The remaining contribution to $\chi(\Sigma)$ derived from the genus $g$,  which is the number of handles on $\Sigma$, is,
\bea
\chi(\Sigma) = 2-2g -n-b
\eea
A Riemann surface with $b \not=0$ may be studied by considering its double, which is obtained by gluing $\Sigma$ and its mirror image $\Sigma '$ into a Riemann surface $\hat \Sigma$ without one-dimensional boundary components. The surface $\hat \Sigma$ is endowed with an involution $\iota$ which swaps $\Sigma$ and $\Sigma '$. Henceforth, we shall restrict attention to surfaces with $b=0$. 

\sm

Compact Riemann surfaces play a central role in both Mathematics and Physics. It will often be useful to consider the  generalization to a \textit{compact Riemann surface with $n$ punctures} defined to be a compact Riemann surface with a finite number $n$ of points removed, namely with $n$ punctures. Their study may be initiated by first studying the underlying compact surface and then removing $n$ points.

\subsubsection{The Riemann-Hurwitz formula}

Holomorphic maps between Riemann surfaces allow us to relate different Riemann surfaces. 
We begin by considering a non-constant holomorphic map  $f:X \to Y$ between two compact Riemann surfaces $X$ and $Y$. The degree of $f$ is the multiplicity of the function $f^{-1}$ given by the cardinality $d=|f^{-1}(y)|$ for all but finitely many points $y \in Y$. A formula valid for all points in $Y$ is obtained in terms of the \textit{degree of ramification} $e_x$ of $f$ at a point $x\in X$ which,  in local coordinates, is given by $f(z)-f(x) \sim (z-x)^{e_x}$.  The degree  of $f$ may then be defined for all $y \in Y$ by the following formula, 
\bea
\label{6.degree}
d= \sum_{x \in f^{-1}(y)} e_x 
\eea
and $d$ is independent of $y$. The Riemann-Hurwitz formula relates the genera $g_X$ and $g_Y$  of the surfaces $X$ and $Y$ to the degree of $f$ and the deviations from 1 of the degree of ramification, 
\bea
\label{6.RHa}
2g_X -2 = d(2g_Y-2) + \sum _{x \in X} (e_x-1)
\eea
A hyper-elliptic Riemann surface $X$, viewed as a double cover of  the Riemann sphere $Y$, provides an illuminating example of the Riemann-Hurwitz formula for $d=2$ since the presence of $2n$ branch points at which $e_x=2$ readily gives $g_X= n-1$. The formula (\ref{6.RHa}) may be proven by using the expression for the Euler characteristic $\chi(X)$  in (\ref{B.Euler}) of $X$ in terms of a triangulation by $F_X$ faces, $E_X$ edges, and $V_X$ vertices (including all  the ramification points),  and similarly for $\chi(Y)$. Given that the degree of the map is $d$, we have $F_X = d F_Y$ and $E_X=d E_Y$ but $V_X= d V_Y - \sum _{x \in X} (e_x-1)$ since a vertex with ramification degree $e_x$ counts for only one vertex of $X$. Expressing the result in terms of the Euler numbers we have $\chi(X) = d \chi(Y) - \sum _{x \in X} (e_x-1)$. For a compact Riemann surface without boundary or punctures, the Euler numbers are related to the genus by $\chi(X) = 2 - 2 g_X$ and $\chi(Y) = 2 - 2g_Y$ from which (\ref{6.RHa}) follows.\footnote{For the special case of an un-ramified covering the formula $\chi(X) = d \chi(Y)$  for $\chi<0$ the formula readily follows from the fact that the Euler number is proportional to the area for unit negative curvature metrics.}

\subsubsection{Homotopy}

The homotopy groups $\pi_0(\Sigma)$ and $\pi_2(\Sigma)$ for a compact Riemann surface are trivial, and the interesting homotopy resides entirely in the first homotopy group $\pi_1(\Sigma)$, which we now define.

\sm

A closed curve $\mC: [0,1] \to \Sigma$ is a continuous map such that $\mC(1)=\mC(0)$.  Two closed curves $\mC_0,\mC_1 $ are said to be homotopic (to one another) iff there exists a family of curves $\mF(s,t)$ parametrized by $t \in [0,1]$, such that $\mF(s,t) \subset \Sigma$ and is continuous in $s,t$  for all $s,t \in [0,1]$ and  interpolates as follows: $\mF(s,0)=\cC_0(s)$ and $\mF(s,1) = \mC_1(s)$. The homotopy relation amongst closed curves is clearly an equivalence relation, whose classes are the first homotopy classes.  To make the set of classes into a group under composition of maps, we choose a base-point $P \in \Sigma$ and restrict to curves in each class that pass through $P$. Thus, we may choose the parametrization of each closed curve $\mC$ that passes through $P$  so that $\mC(0)=\mC(1)=P$. The composition $\mC_1 \circ \mC_0$ of two curves $\mC_0,\mC_1$ is given by, 
\bea
\mC_1 \circ \mC_0 (s) = \left \{ \bma \mC_0(2s)  & \hbox{ for } & s \in [0,\half] \cr 
\mC_1(2s-1) & \hbox{ for } & s \in [\half, 1] \ema \right .
\eea 
The composition of closed curves induces a composition of first homotopy classes of curves, where 
the identity element $e$ is the class of contractile curves, and the inverse is the class of curves with opposite orientation. The resulting group of classes is independent of the base-point $P$ \footnote{Independence of $P$ requires $\Sigma$ to be path-connected as we shall assume throughout.}  and is referred to as the first homotopy group, or fundamental group $\pi_1(\Sigma)$.

\sm

{\prop The Riemann sphere $\hat \CC$, the complex plane $\CC$, the Poincar\'e upper half-plane  $\cH$, and the unit disc $\cD$ are simply connected Riemann surfaces, as is any compact Riemann surface $\Sigma _0$ of genus 0,  so that their first homotopy groups are given by,
\bea
\pi_1 (\hat \CC) = \pi_1(\CC)= \pi_1(\cH) = \pi_1(\cD) = \pi_1(\Sigma _0)= \{ e \}
\eea 
For a compact surface  $\Sigma$ of genus 1, namely a torus,  we have $\pi_1(\Sigma) = \ZZ^2$.} 

\sm

For surfaces of genus $g \geq 2$, the fundamental group is non-Abelian, and will be discussed concretely in appendix \ref{sec:B4} in terms of representations of $SL(2,\RR)$.

\begin{figure}[htb]
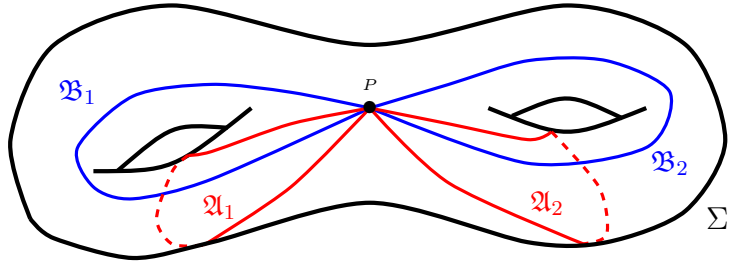

\begin{center}
\tikzpicture[scale=1.05]
\scope[xshift=0cm,yshift=0cm]

\draw[ultra thick] (1.5,1.2) .. controls (2.5, 1.2) .. (3.5,2);
\draw[ultra thick] (1.8,1.2) .. controls (2.5, 1.8) .. (3.2,1.75);
\draw[ultra thick] (6.5,2) .. controls (7.5, 1.6) .. (8.5,2);
\draw[ultra thick] (6.8,1.9) .. controls (7.5, 2.2) .. (8.2,1.87);
\draw[very thick, color=blue] plot [smooth] coordinates {(5,2) (4,2.2) (3, 2.3) (2, 2.15) (1.3,1.5) (1.5,1) ( 2, 0.85) (3,1.1) (5,2) };
\draw[very thick, color=blue] plot [smooth] coordinates {(5,2) (6,2.3) (7, 2.6) (8, 2.6) (8.8,2.2) (8.5,1.5)  (7,1.3) (5,2) };
\draw[very thick, color=red] plot [smooth] coordinates { (5,2) (4,1.8) (3,1.45) (2.7,1.4)};
\draw[very thick, color=red, dashed] plot [smooth] coordinates { (2.7,1.4) (2.5,1.2) (2.3,0.6) (2.5, 0.275) (2.9, 0.265)};
\draw[very thick, color=red] plot [smooth] coordinates { (2.9,0.265) (4,1) (5,2)};
\draw[very thick, color=red] plot [smooth] coordinates { (5,2) (6,1.8) (7,1.6) (7.3,1.7)};
\draw[very thick, color=red, dashed] plot [smooth] coordinates { (7.3,1.7) (7.9,1) (8,0.4) (7.7, 0.28)};
\draw[very thick, color=red] plot [smooth] coordinates { (7.7, 0.28) (6,1.05) (5,2)};
\draw[ultra thick] plot [smooth] coordinates {(0.7,0.6) (0.45,1.6) (1,2.8) (2.5, 3.3)  (5,2.8) (7.5, 3.3) (9,2.8) (9.5,1.6) (9,0.7) (8,0.3) (7, 0.3) (5,0.8) (3, 0.3) (2,0.1) (1,0.35) (0.7,0.6) };
\draw [color=red] (3.1,0.75) node{\small $\mA_1$};
\draw [color=red] (7.25,0.8) node{\small $\mA_2$};
\draw [color=blue] (1.3,2.2) node{\small $\mB_1$};
\draw [color=blue]  (8.8,1.3) node{\small $\mB_2$};
\draw (9.4,0.6) node{{$\Sigma$}};
\draw (5,2) node{$\bullet$};
\draw (5,2.3) node{\tiny $P$};
\endscope
\endtikzpicture
\caption{ \textit{A choice of homotopy generators $\mA_I$ and $\mB_I$ with common base point $P$ is illustrated for a compact genus-two Riemann surface $\Sigma$. }
\label{fig:B.1}}
\end{center}
\end{figure}

\subsubsection{Homology}

The top and bottom homology groups of a Riemann surface are trivial.  All the interesting homology resides in the first homology group which, for a compact Riemann surface of genus $g$ with $n$ punctures, is given by $H_1(\Sigma, \ZZ) = \ZZ^{2g+n}$. The primitive homology cycle associated with each puncture is homologous to a small circle centered at the puncture. Having chosen an orientation for the Riemann surface $\Sigma$, we introduce a binary intersection pairing $\mJ(\mC_1, \mC_2)$ on oriented 1-cycles $\mC_1$ and $\mC_2$,  which counts the number of intersections of the 1-cycles $\mC_1$ with $\mC_2$  weighed by $\pm 1$ factors for the relative orientation of the cycles at their intersection points. The pairing  $\mJ(\mC_1, \mC_2)$ is odd under swapping $\mC_1, \mC_2$. 

\sm

Specializing to the case of a compact Riemann surface $\Sigma$ without punctures, the intersection form $\mJ$ is non-degenerate, and we may choose a basis for $H_1(\Sigma, \ZZ) = \ZZ^{2g}$ consisting of homology cycles $\mA_I, \mB_J$ for $I,J=1,\cdots, g$ such that,
\begin{align}
\label{B.2.1}
\mJ (\mA_I, \mA_J) & = \, \mJ (\mB_I, \mB_J)  \, = 0 
\no \\
 \mJ (\mA_I, \mB_J) & =  - \mJ (\mB_I, \mA_J)  =  \delta _{IJ}
\end{align}
Two different canonical bases $(\mA, \mB)$ and $(\tilde \mA, \tilde \mB)$  are related by a linear transformation that may be represented by a matrix $M$  with integer entries, 
\bea
\label{B.2.2}
 \left ( \begin{matrix} \tilde \mB \cr \tilde \mA \cr \end{matrix} \right ) 
 = M \left ( \begin{matrix} \mB \cr \mA \cr \end{matrix} \right )
\eea
Here, $\mA$ and $\mB$ are column matrices with entries $\mA_I$ and $ \mB_I$, respectively.  The matrix $M$ is an element of  the rank $g$ modular group $Sp(2g,\ZZ)$ which preserves the intersection matrix~$\mJ$, 
\bea
\label{B.2.3}
 M^t \mJ M = \mJ 
 \hskip 0.7in 
\mJ = \left ( \begin{matrix}0 & -I_g \cr I_g & 0 \cr \end{matrix} \right )
 \hskip 0.7in
M = \left ( \begin{matrix} A & B \cr C & D \cr \end{matrix} \right )
\eea
 where $A,B,C,D$ are $g \times g$ matrices with integer entries. An important subgroup of $Sp(2g,\ZZ)$ is the group $GL(g,\ZZ)$ which  consists of those $M$ that transform $\mA$-cycles into linear combinations of $\mA$-cycles and $\mB$-cycles into linear combinations of $\mB$-cycles. It is obtained by setting $B=C=0$ and $D^{-1}= A^t$.

 \subsubsection{Cohomology}
 \label{sec:B.1}

By the de Rham theorem, the first de Rham cohomology group is given by $H^1(\Sigma, \RR) = \RR^{2g}$ and is generated by $2g$ real-valued harmonic 1-forms on $\Sigma$. Since $\Sigma$ is a complex manifold,  harmonic one-forms may be decomposed into holomorphic and anti-holomorphic one-forms, and the cohomology may be decomposed into the Dolbeault cohomology groups, 
\bea
H^1(\Sigma, \RR) =  H^{(1,0)} (\Sigma ) \oplus H^{(0,1)} (\Sigma)
\eea
generated by holomorphic $(1,0)$ forms and anti-holomorphic $(0,1)$ forms respectively. A canonical basis of the Dolbeault cohomology group  $H^{(1,0)} (\Sigma, \ZZ)= \ZZ^g$ consists of holomorphic $(1,0)$-forms  $\om_I$ with $I=1,\cdots, g$ whose periods are normalized on the cycles $\mA_I$ of the canonical homology basis $(\mA, \mB)$, as follows,
\bea
\label{B.3.1}
\oint _{\mA_I} \om_J = \delta _{IJ} 
\eea
The periods on the remaining 1-cycles are then completely determined and give the matrix elements of the period matrix $\Omega$ of $\Sigma$, 
\bea
\label{B.3.2}
\oint _{\mB_I} \om_J = \Omega _{IJ}
\eea
\sm
Under modular transformations $M \in Sp(2g,\ZZ)$, whose  parametrization in terms of $g \times g$ matrices $A,B,C,D$ is given in (\ref{B.2.3}), the matrix of holomorphic Abelian differentials $\om$, the period matrix $\Omega$, and its imaginary part $Y$,  transform as follows,
\bea
\label{B.3.4}
\tilde \om & = & \om (C \Omega+D)^{-1}
\no \\
\tilde \Omega & = & (A \Omega +B) (C \Omega +D)^{-1} 
\no \\
\tilde Y & = & (\Omega C^t+D^t)^{-1} Y (C\Omega^* +D)
\eea
Clearly, the transformation formula for $\Omega$ reduces to the standard M\"obius transformation formula  for the special case of $g=1$.

\subsubsection{Riemann bilinear relations}
\label{sec:Riemannrelations}

\begin{figure}[htb]
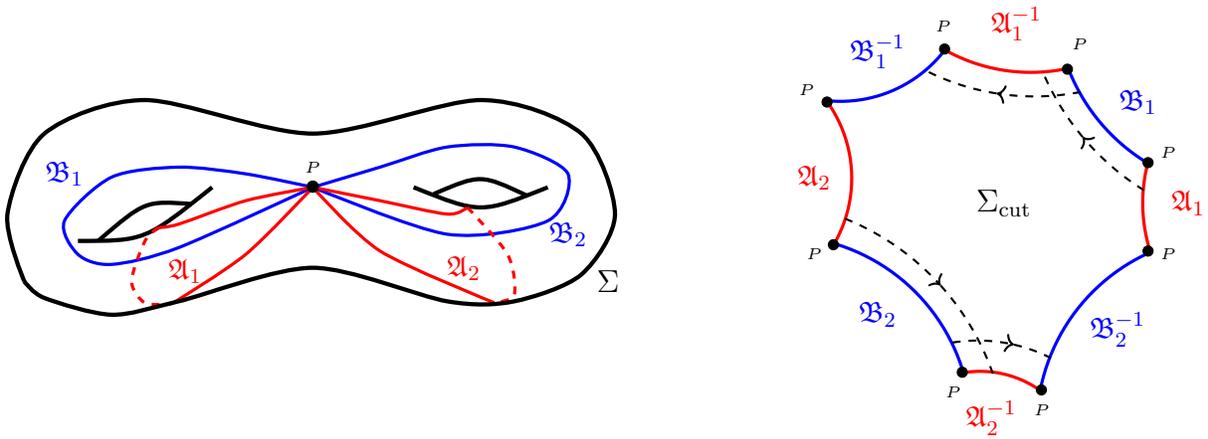

\begin{center}
\tikzpicture[scale=1.05]
\scope[xshift=0cm,yshift=0cm, scale=0.85]

\draw[ultra thick] (1.5,1.2) .. controls (2.5, 1.2) .. (3.5,2);
\draw[ultra thick] (1.8,1.2) .. controls (2.5, 1.8) .. (3.2,1.75);
\draw[ultra thick] (6.5,2) .. controls (7.5, 1.6) .. (8.5,2);
\draw[ultra thick] (6.8,1.9) .. controls (7.5, 2.2) .. (8.2,1.87);
\draw[very thick, color=blue] plot [smooth] coordinates {(5,2) (4,2.2) (3, 2.3) (2, 2.15) (1.3,1.5) (1.5,1) ( 2, 0.85) (3,1.1) (5,2) };
\draw[very thick, color=blue] plot [smooth] coordinates {(5,2) (6,2.3) (7, 2.6) (8, 2.6) (8.8,2.2) (8.5,1.5)  (7,1.3) (5,2) };
\draw[very thick, color=red] plot [smooth] coordinates { (5,2) (4,1.8) (3,1.45) (2.7,1.4)};
\draw[very thick, color=red, dashed] plot [smooth] coordinates { (2.7,1.4) (2.5,1.2) (2.3,0.6) (2.5, 0.275) (2.9, 0.265)};
\draw[very thick, color=red] plot [smooth] coordinates { (2.9,0.265) (4,1) (5,2)};
\draw[very thick, color=red] plot [smooth] coordinates { (5,2) (6,1.8) (7,1.6) (7.3,1.7)};
\draw[very thick, color=red, dashed] plot [smooth] coordinates { (7.3,1.7) (7.9,1) (8,0.4) (7.7, 0.28)};
\draw[very thick, color=red] plot [smooth] coordinates { (7.7, 0.28) (6,1.05) (5,2)};
\draw[ultra thick] plot [smooth] coordinates {(0.7,0.6) (0.45,1.6) (1,2.8) (2.5, 3.3)  (5,2.8) (7.5, 3.3) (9,2.8) (9.5,1.6) (9,0.7) (8,0.3) (7, 0.3) (5,0.8) (3, 0.3) (2,0.1) (1,0.35) (0.7,0.6) };
\draw [color=red] (3.1,0.75) node{\small $\mA_1$};
\draw [color=red] (7.25,0.8) node{\small $\mA_2$};
\draw [color=blue] (1.3,2.2) node{\small $\mB_1$};
\draw [color=blue]  (8.8,1.3) node{\small $\mB_2$};
\draw (9.4,0.6) node{{$\Sigma$}};
\draw (5,2) node{$\bullet$};
\draw (5,2.3) node{\tiny $P$};
\endscope
\scope[xshift=13cm,yshift=1.5cm, scale=3.2]
\draw  [very  thick, color=red, domain=166:198] plot ({cosh(0.6)*cos(0)+sinh(0.6)*cos(\x)}, {cosh(0.6)*sin(0)+sinh(0.6)*sin(\x)});
\draw  [very  thick, color=red, domain=241:282] plot ({cosh(0.65)*cos(85)+sinh(0.65)*cos(\x)}, {cosh(0.65)*sin(85)+sinh(0.65)*sin(\x)});
\draw  [very  thick, color=red, domain=330:396] plot ({cosh(0.5)*cos(175)+sinh(0.5)*cos(\x)}, {cosh(0.5)*sin(175)+sinh(0.5)*sin(\x)});
\draw  [very  thick, color=red, domain=54:100] plot ({cosh(0.4)*cos(265)+sinh(0.4)*cos(\x)}, {cosh(0.4)*sin(265)+sinh(0.4)*sin(\x)});
\draw  [very  thick, color=blue, domain=201:239] plot ({cosh(0.7)*cos(40)+sinh(0.7)*cos(\x)}, {cosh(0.7)*sin(40)+sinh(0.7)*sin(\x)});
\draw  [very  thick, color=blue, domain=264:323] plot ({cosh(0.5)*cos(125)+sinh(0.5)*cos(\x)}, {cosh(0.5)*sin(125)+sinh(0.5)*sin(\x)});
\draw  [very  thick, color=blue, domain=17:73] plot ({cosh(0.7)*cos(225)+sinh(0.7)*cos(\x)}, {cosh(0.7)*sin(225)+sinh(0.7)*sin(\x)});
\draw  [very  thick, color=blue, domain=114:169] plot ({cosh(0.7)*cos(315)+sinh(0.7)*cos(\x)}, {cosh(0.7)*sin(315)+sinh(0.7)*sin(\x)});
\draw  [thick, color=black, dashed, ->, domain=238:220] plot ({cosh(0.9)*cos(40)+sinh(0.9)*cos(\x)}, {cosh(0.9)*sin(40)+sinh(0.9)*sin(\x)});
\draw  [thick, color=black, dashed,  domain=220:203] plot ({cosh(0.9)*cos(40)+sinh(0.9)*cos(\x)}, {cosh(0.9)*sin(40)+sinh(0.9)*sin(\x)});
\draw (0.148,-0.74) [fill=black] circle(0.02cm) ;
\draw  [thick, color=black, dashed, ->, domain=68:42] plot ({cosh(0.9)*cos(225)+sinh(0.9)*cos(\x)}, {cosh(0.9)*sin(225)+sinh(0.9)*sin(\x)});
\draw  [thick, color=black, dashed,  domain=42:19] plot ({cosh(0.9)*cos(225)+sinh(0.9)*cos(\x)}, {cosh(0.9)*sin(225)+sinh(0.9)*sin(\x)});
\draw  [thick, color=black, dashed, ->, domain=281:261] plot ({cosh(0.85)*cos(85)+sinh(0.85)*cos(\x)}, {cosh(0.85)*sin(85)+sinh(0.85)*sin(\x)});
\draw  [thick, color=black, dashed, domain=261:243] plot ({cosh(0.85)*cos(85)+sinh(0.85)*cos(\x)}, {cosh(0.85)*sin(85)+sinh(0.85)*sin(\x)});
\draw  [thick, color=black, dashed, ->, domain=99:77] plot ({cosh(0.6)*cos(265)+sinh(0.6)*cos(\x)}, {cosh(0.6)*sin(265)+sinh(0.6)*sin(\x)});
\draw  [thick, color=black, dashed, domain=77:63] plot ({cosh(0.6)*cos(265)+sinh(0.6)*cos(\x)}, {cosh(0.6)*sin(265)+sinh(0.6)*sin(\x)});
\draw (0.57,0.16) [fill=black] circle(0.02cm) ;
\draw (0.251,0.53) [fill=black] circle(0.02cm) ;
\draw (-0.7,0.4) [fill=black] circle(0.02cm) ;
\draw (-0.235,0.61) [fill=black] circle(0.02cm) ;
\draw (-0.675,-0.165) [fill=black] circle(0.02cm) ;
\draw (-0.165,-0.67) [fill=black] circle(0.02cm) ;
\draw (0.57,-0.19) [fill=black] circle(0.02cm) ;

\draw (0.15, -0.82) node{\tiny $P$};
\draw (0.65, 0.2) node{\tiny $P$};
\draw (0.3, 0.63) node{\tiny $P$};
\draw (-0.78, 0.45) node{\tiny $P$};
\draw (-0.24, 0.7) node{\tiny $P$};
\draw (-0.75, -0.2) node{\tiny $P$};
\draw (-0.2, -0.75) node{\tiny $P$};
\draw (0.65, -0.2) node{\tiny $P$};
\draw [color=red] (0.73,0) node{\small $\mA_1$};
\draw [color=blue] (0.53,0.4) node{\small $\mB_1$};
\draw [color=red] (0.05,0.71) node{\small $\mA_1^{-1}$};
\draw [color=blue] (-0.5,0.6) node{\small $\mB_1^{-1}$};
\draw [color=red] (-0.75,0.1) node{\small $\mA_2$};
\draw [color=blue]  (-0.5,-0.44) node{\small $\mB_2$};
\draw [color=red] (-0.05,-0.85) node{\small $\mA_2^{-1}$};
\draw [color=blue]  (0.45,-0.5) node{\small $\mB_2^{-1}$};
\draw (0,0) node{{$\Sigma_{{\rm cut}}$}};
\endscope

\endtikzpicture
\caption{ \textit{Representation of the Riemann surface $\Sigma$ in terms of a simply connected domain $\Sigma_{{\rm cut}}$ in $\CC$ obtained by cutting $\Sigma$ along the cycles $\mA_I$ and $\mB_I$ with common base point $P$ in common, with inverse cycles pairwise identified under the dashed arrows,  illustrated here for a compact genus-two Riemann surface $\Sigma$. }
\label{fig:B.5}}
\end{center}
\end{figure}

A useful formula relates the integral over $\Sigma$ of a wedge product of closed forms $\varpi_1$ and $\varpi_2$ to their line integrals over canonical homology cycles,
\bea
\label{B.Riem}
\int _\Sigma \varpi_1 \wedge \varpi _2 = \sum _{K=1}^g \left ( \oint _{\mA_K} \varpi_1  \oint _{\mB_K} \varpi_2
- \oint _{\mB_K} \varpi_1  \oint _{\mA_K} \varpi_2 \right )
\eea
The formula may be established by cutting the surface $\Sigma$ along homology cycles $\mA_K$ and $\mB_K$ for $K=1,\cdots, g$, chosen so as to have a point in common, into a simply-connected domain $\Sigma_{{\rm cut}}$  in the plane, as illustrated in Figure \ref{fig:B.5}. Expressing one of the closed differentials as an exact differential of a local function  in $\Sigma_{{\rm cut}}$, for example, $\varpi _2 = d f_2$, the formula may be derived using Green's theorem for an arbitrary 1-form $\varpi$,
\bea
\int _{\Sigma_{{\rm cut}}} d \varpi = \oint _{\p \Sigma _{{\rm cut}}} \varpi
\eea
where the boundary consists of the union of cycles, 
\bea
\p \Sigma _{{\rm cut}}  = \bigcup _I \Big ( \mA_I \cup \mB_I \cup \mA_I^{-1} \cup \mB_I^{-1} \Big )
\eea
as illustrated in Figure \ref{fig:B.5} for genus $g=2$.  

\sm

The period matrix $\Omega $ is symmetric by the Riemann bilinear relations obtained by setting $\varpi_1 = \om_I$ and $\varpi _2=\om_J$ in (\ref{B.Riem}). Furthermore, the imaginary part of the period matrix,
\bea
\label{B.3.3}
Y= \Im \, \Omega
\eea 
is positive definite by the Riemann relations obtained by setting $\varpi_1 = \om_I$ and $\varpi _2= \bar \om_J$ in (\ref{B.Riem}), and using the fact that $\tfrac{i}{2} \int _\Sigma \varpi \wedge  \varpi^*$ is positive definite for holomorphic 1-forms $\varpi$.

\subsection{Metrics and complex structures}

A different point of view on Riemann surfaces, with or without punctures,  is obtained by considering connected two-dimensional orientable manifolds endowed with a Riemannian metric. This point of view is important in Physics and is widely regarded as the starting point for string perturbation theory in the Polyakov formulation.  Consider a two-dimensional oriented manifold $\Sigma$ endowed with a Riemannian metric  $\mg$. In a system of local real coordinates $\xi ^m = (\xi^1, \xi^2)$ the metric takes the form,\footnote{Throughout, the Einstein convention is used to sum over a pair of repeated upper and lower indices. }
\bea
\mg = \mg_{mn} (\xi) d\xi ^m d \xi ^n
\eea
We review the following notions:
\begin{itemize}
\item A \textit{diffeomorphism} is a smooth map $\xi \to \xi'(\xi)$  under which $\mg$ is invariant, and the components $\mg_{mn}$ transform as follows,
\bea
\mg'_{mn} (\xi') { \p \xi^{\prime \, m} \over  \p \xi^p} { \p \xi^{\prime \, n} \over  \p \xi^q} = \mg_{pq}(\xi)
\eea
Diffeomorphisms form an infinite-dimensional group ${\rm Diff}(\Sigma)$. 
\sm
\item A \textit{Weyl transformation} rescales the metric $\mg$ by an arbitrary positive function $\lambda$ on $\Sigma$, namely $\mg_{mn}(\xi)  \to \mg'_{mn}(\xi) = \lambda(\xi)  \mg_{mn}(\xi)$. Weyl transformations form an infinite-dimensional group ${\rm Weyl}(\Sigma)$. 
\sm
\item Generally, a \textit{conformal transformation} preserves all angles. All Weyl transformations are conformal transformations in this sense. A \textit{conformal coordinate transformation}, or  \textit{conformal diffeomorphism}  is  a diffeomorphism which is  conformal, which requires that $\mg_{mn}' (\xi') = \mu (\xi) \mg_{mn}(\xi)$ for some positive real function $\mu$.
\sm
\item An \textit{isometry} of $\mg$ preserves all distances. A diffeomorphism is an isometry of $\mg$ provided $\mg'_{mn}(\xi') = \mg_{mn}(\xi)$. Every conformal diffeomorphism is an isometry, but there exist conformal transformations, such as scaling by a constant, that are not isometries. 
\end{itemize}

We have the following theorem,

{\thm An orientable connected two-dimensional surface $\Sigma$ endowed with a Riemannian metric is automatically a Riemann surface.}

\sm

 To establish this result, we show that the combination of orientability and the presence of  a Riemannian metric induces a unique complex structure. The orientation may be specified by choosing an oriented volume form, 
\bea
d \mu _\mg = \sqrt{\det \mg} \, \ep _{m n} \, d\xi ^m \wedge d\xi^n \hskip 1in \ep_{mn}=-\ep_{nm} , ~ \ep _{12}=+1
\eea
A reversal of orientation corresponds to reversing the sign of $\ep$. We denote by $T_p(\Sigma)$ the tangent space to $\Sigma$ at the point $p\in \Sigma$, and by $T^*_p(\Sigma)$ the cotangent space.  A convenient basis  is given by the partial derivatives $\p_m= \p/ \p \xi^m$ for $T_p (\Sigma)$ and by the differentials $d\xi^m$ for $T^* _p (\Sigma)$. The tangent space $T_p(\Sigma)$ may be viewed as the space of  vectors $v^m \p_m$ at $p$, while the cotangent space may be viewed as the space of forms $\om_m d\xi^m$ at $p$.  An \textit{almost complex structure} $\cJ$ is a map from $T_p (\Sigma)$  to itself, or equivalently from $T^*_p(\Sigma)$ to itself, whose square is minus the identity map, 
\bea
\cJ : T_p (\Sigma) \to T_p (\Sigma) & \hskip 0.6in & v^m \to \cJ^m {}_n \, v^n \hskip 0.8in \cJ^2 = - I 
\no \\
\cJ : T_p^* (\Sigma) \to T_p^* (\Sigma) & \hskip 0.6in & \om_m \to \cJ_m{}^n \, \om_n
\eea
Having a metric and an orientation, it is straightforward to exhibit this map,
\bea
\cJ^m {}_n & = & \sqrt{\det \mg} \, \ep _{rn} \, \mg^{rm}
\no \\
\cJ_m {}^n & = & \sqrt{\det \mg} \, \ep _{mr} \, \mg^{rn}
\eea
The metric is a covariantly constant tensor with respect to the affine connection and its associated covariant derivative $\nabla _s$, so that we have $\nabla _s \mg_{mn}=0$. As a result, $\cJ$ obeys $\nabla _s \, \cJ_m {}^n=0$ and is therefore also a covariantly constant tensor acting on $T_p(\Sigma)$ and $T^*_p(\Sigma)$. Furthermore, $\cJ$ is invariant under arbitrary Weyl rescaling and therefore depends only on the conformal class of $\mg_{mn}$.  Finally, $\cJ$ is integrable in the sense that the system of equations, 
\bea
\label{B.4.holo}
\cJ_m {}^n \, \p_n f = i \, \p_m f
\eea
is integrable and thereby defines locally holomorphic functions $f$ on $\Sigma$ which endow $\Sigma$ with a \textit{complex structure}.\footnote{In dimensions higher than two, the system (\ref{B.4.holo}) has non-trivial integrability conditions, but in two dimensions there are none since the operator $\cJ-iI$ has a one-dimensional null space.}  
We may use the solutions to (\ref{B.4.holo}) to define a system of local complex coordinates $z,\bar z$ in which the metric $\mg$ is locally conformally flat,
\bea
\mg = \mg_{z \bar z} |dz|^2
\eea 
for a real positive function $\mg_{z \bar z}$. The conformally flat form of the metric is invariant under locally holomorphic diffeomorphisms $z \to z' (z)$. The complex manifold $\Sigma$ may be constructed by covering $\Sigma$ with coordinate charts $\cU_\alpha$ in each of which we have  local complex coordinates $(z_\alpha, \bar z_\alpha)$. In the intersection $\cU_\a \cap \cU_\b$ of the two coordinate charts, the local coordinates are related by transition functions $\f_{\a \b}(z)$ which are holomorphic and nowhere vanishing via the relations $z_\alpha(z) = \f_{\a \b} (z) z_\beta (z)$. Thus $\Sigma$ is  a Riemann surface.

\subsubsection{The Euler characteristic in terms of a metric}

The Riemann tensor on a connected two-dimensional Riemannian manifold has only one independent component, which may be expressed in terms of the Gaussian curvature $R_\mg$. The Euler characteristic of a compact Riemann surface $\Sigma$ with metric $\mg$ may be expressed in terms of $R_\mg$ by the following classic formula,
\bea
\label{B.4.Euler}
\chi (\Sigma) = { 1 \over 2 \pi} \int _\Sigma d \mu _\mg \, R_\mg
\eea
The normalization of the Gaussian curvature $R_\mg$ is such that the round sphere $S^2$ of radius 1 has area $4 \pi$ and $R_\mg=1$, so that (\ref{B.4.Euler}) indeed corresponds to genus zero. The Euler characteristic is a topological invariant. Indeed, the integral in (\ref{B.4.Euler}) is independent of the metric $\mg$ because $d \mu_\mg \, R_\mg$ is locally an exact $(1,1)$-form.

\subsection{Uniformization}

In this subsection, we present the uniformization theorem for simply connected Riemann surfaces, and then use its results to produce concrete constructions for arbitrary compact Riemann surfaces with punctures. 

\sm

{\thm[The uniformization theorem]

A simply connected Riemann surface is conformally isomorphic to either one of the following Riemann surfaces,
\begin{enumerate}
\itemsep=-0.05in
\item The Riemann sphere $\hat \CC$ which is the one-point compactification of $\CC$  (also denoted $S^2$);
\item The complex plane $\CC$;
\item The complex upper half-plane $\cH$ which is isomorphic to the unit disc $\cD$. 
\end{enumerate}
For an arbitrary Riemannian metric $\mg$ on $\Sigma$ in each case, there exists a Weyl transformation to a metric $\hat \mg$ such that $R_{\hat \mg}=1$ for $\hat \CC$; $R_{\hat \mg}=0$ for $\CC$; and $R_{\hat \mg}=-1$ for $\cH$ and $ \cD$. The corresponding metrics are given as follows, 
\bea
\hat \mg _{S^2} = { 4 |dz|^2 \over (1+|z|^2)^2} 
\hskip 0.5in 
\hat \mg _\CC = |dz|^2 
\hskip 0.5in 
\hat \mg _{\cH} = { |dz|^2 \over (\Im z)^2}
\hskip 0.5in 
\hat \mg _{\cD} = { 4 |dz|^2 \over (1-|z|^2)^2} 
\eea
where $z \in \hat \CC$, $z \in \CC$, $\Im(z)>0$, and $|z|<1$ respectively.}

\sm

Each one of these spaces is maximally symmetric and may be expressed as the quotient of its isometry group by the isotropy group, 
\begin{align}
\hat \CC & = SU(2)/U(1) &
\hskip 0.4 in
\CC = ISO(2)/SO(2) 
\no \\
\cH & = SL(2,\RR)/SO(2) & 
\hskip 0.4 in
\cD  = SU(1,1) / U(1) 
\end{align}
 The isometry groups $PSL(2,\RR)= SL(2,\RR)/ \ZZ_2$ and $PSU(1,1)=SU(1,1)/\ZZ_2$ are isomorphic to one another. The action on the coordinate $z$ in either case is given by,
\bea
z \to { az + b \over cz + d} 
\eea
where $a,b,c,d \in \RR$ with $ad-bc=1$ for $\cH$ and $c=b^*, d=a^*$ and $|a|^2-|b|^2=1$ for $\cD$. In both cases  the elements $I$ and $-I$ leave $z$ invariant and  are identified under the center $\ZZ_2$.

\sm

The geodesics of $\hat \CC$ are the great circles;  those of $\CC$ are straight lines; and those of $\cH$ are half-circles centered on the real axis. The geodesics of $\cD$ are the circle segments in $\cD$, centered at any $\a \in \CC$ with  $|\a|>1$, satisfying  $|z-\a|^2 = |\a|^2-1$. They intersect the unit circle $|z|=1$ orthogonally to it. The system of geodesics of $\cH$ and $\cD$ makes these spaces  into standard models for two-dimensional hyperbolic (or no-Euclidean) geometry.

\begin{figure}[htb]
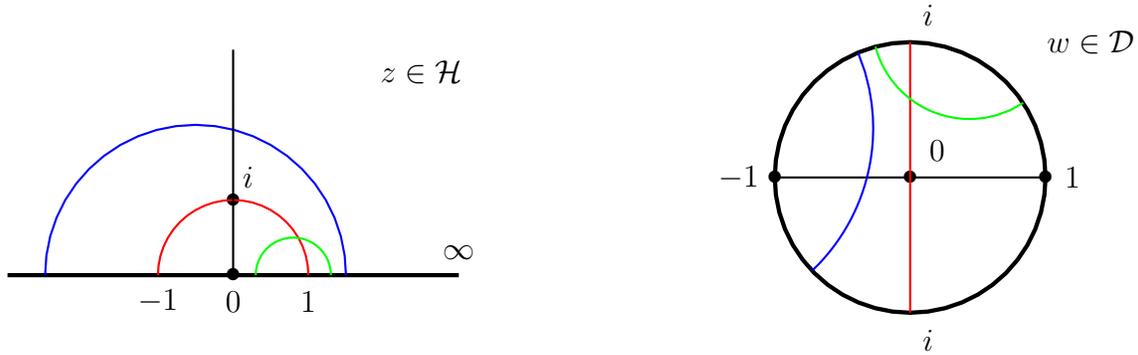

\begin{center}
\tikzpicture[scale=1]
\scope[xshift=0cm,yshift=0cm]
\draw [ultra thick] (-3,0) -- (3,0);
\draw (0.2,1.3) node{$i$};
\draw (0,1) node{$\bullet$};
\draw (3,0.3) node{$\infty$};
\draw (0,0) node{$\bullet$};
\draw (0,-0.35) node{$0$};
\draw (-1,-0.35) node{$-1$};
\draw (1,-0.35) node{$1$};
\draw  [thick, color=red, domain=0:180] plot ({cos(\x)},{sin(\x)});
\draw  [thick, color=blue, domain=0:180] plot ({-0.5+2*cos(\x)},{2*sin(\x)});
\draw  [thick, color=green, domain=0:180] plot ({0.8+0.5*cos(\x)},{0.5*sin(\x)});
\draw [thick] (0,0) -- (0,3);

\draw (2.5,2.7) node{$z \in \cH$};
\endscope
\scope[xshift=9cm,yshift=-0.5cm, scale=1.2]
\draw  [ultra thick, domain=0:180] plot ({1.5*cos(\x)},{1.5+1.5*sin(\x)});
\draw  [ultra thick, domain=180:360] plot ({1.5*cos(\x)},{1.5+1.5*sin(\x)});
\draw (0.3,1.8) node{$0$};
\draw (0,1.5) node{$\bullet$};
\draw (1.5,1.5) node{$\bullet$};
\draw (-1.9,1.5) node{$-1$};
\draw (-1.5,1.5) node{$\bullet$};
\draw (1.8,1.5) node{$1$};
\draw [thick] (-1.5,1.5) -- (1.5,1.5);
\draw [thick, color=red] (0,0) -- (0,3);
\draw (0.2,3.3) node{$i$};
\draw (0.2,-0.3) node{$i$};
\draw  [thick, color=blue, domain=0:180] plot ({1.5*(-13+8*cos(\x))/(21+16*sin(\x)-8*cos(\x))},{1.5+1.5*(-4+16*cos(\x))/(21+16*sin(\x)-8*cos(\x))});
\draw  [thick, color=green, domain=0:180] plot ({1.5*(0.55-4*cos(\x))/(9.45+5*sin(\x)+4*cos(\x))},{1.5+1.5*(1.6+cos(\x))/(1.89+sin(\x)+0.8*cos(\x))});
\draw (2,3) node{$w \in \cD$};
\endscope

\endtikzpicture
\caption{\textit{Geodesics in the upper half-plane $\cH$ are semi-circles centered on $\RR$ depicted in the left figure,  with their corresponding images under the conformal map $w=(1+iz)/(1-iz)$ from $\cH$ into the unit disc $\cD$ depicted in the right figure.}
\label{fig:B.2}}
\end{center}
\end{figure}

The uniformization theorem provides us with a powerful tool to classify and construct Riemann surfaces even when they are not simply connected.  The key to the construction is the first homotopy group $\pi_1(\Sigma)$, which is trivial for simply connected $\Sigma$ but non-trivial when $\Sigma$ is not simply-connected.

\subsubsection{Genus one}

Any compact Riemann surface $\Sigma$ of genus 1 has $\pi_1(\Sigma) = \ZZ^2$ whose generators may be chosen to be the $\mA$ and $\mB$ cycles of the first homology group. The universal simply connected covering of a compact genus one surface $\Sigma$  is the complex plane $\CC$. The fundamental group is Abelian and acts on $\CC$ by translations  via a representation $\rho : \ZZ^2 \mapsto \CC$ which defines a  lattice $\Lambda$.  To specify the representation, and the lattice $\Lambda$, it suffices to assign the periods $\rho(1,0) = \om_1$ and $\rho(0,1) = \om_2$ so that the lattice is given by $\Lambda = \om_1 \ZZ \oplus \om _2 \ZZ $. The lattice is two-dimensional (over $\RR$) provided $\om_2/\om_1= \tau \in \cH$. The action of the fundamental group, via the representation $\rho$ on $\CC$, is transitive and the  quotient   $\CC/ \Lambda$ is the original genus one compact Riemann surface $\Sigma$.  Since translations in $\CC$ are isometries of the flat Euclidean metric $|dz|^2$ on $\CC$, the genus-one surface $\CC/ \Lambda$ inherits the flat metric $|dz|^2$ subject to the identifications $z \approx z+\om_1$ and $z \approx z+ \om_2$.

\subsubsection{Higher genus $g \geq 2$}
\label{sec:B4}

{\thm Every compact Riemann surface $\Sigma$ of genus $g \geq 2$ admits a Riemannian metric $\hat \mg$ of constant negative curvature $R_{\hat \mg}=-1$.}

\sm

The curvature $R_\mg$ of an arbitrary metric $\mg$ on $\Sigma$ is related to the curvature $R_{\hat \mg}$ of a metric $\hat \mg$  related to $\mg$ by a Weyl transformation $\mg = e^{2 \sigma} \hat \mg$ by the Liouville equation, 
\bea
R_\mg  = e^{-2 \sigma} R_{\hat \mg} +  \Delta_\mg \sigma
\hskip 1in
\Delta_\mg \sigma = - { 1 \over \sqrt{\det \mg }}  \p_m ( \sqrt{\det \mg } \, \mg^{mn} \p_n \sigma )
\eea
where $\Delta _{\mg}$ is the Laplace-Beltrami operator on scalar functions for the metric $\mg$. 
To prove the theorem one shows that, for an arbitrary metric $\mg$, the Liouville equation has a solution for $R_{\hat \mg}=-1$. The Liouville equation first arose precisely  in this context.  Before presenting the construction of higher genus surfaces we introduce the concept of a Fuchsian group.

\subsection{Fuchsian groups}
\label{sec:Fuchs}

A systematic approach to the classification and explicit construction of Riemann surfaces is through the use of Fuchsian groups. A Fuchsian group is defined to be isomorphic to a discrete subgroup of $SL(2,\RR) $. Fuchsian groups are special cases of Kleinian groups which are discrete subgroups of $SL(2,\CC)$. The elements $\gamma \not = \pm I$ of a Fuchsian group $\Gamma \subset SL(2,\RR)$ belong to one of the following three types, 
\bea
\label{B.4.a}
\gamma \hbox{ \textit{elliptic} } \hskip 0.25in & & | \tr (\gamma)| < 2
\no \\
\gamma \hbox{ \textit{parabolic} } \hskip 0.08in &&  \tr (\gamma) = \pm 2
\no \\
\gamma \hbox{ \textit{hyperbolic} } &&  |\tr(\gamma)| >2
\eea
Since $\det (\gamma)=1$, $\gamma$ must be conjugate, under $SL(2,\RR)$, to one of the following matrices,
\bea
\label{B.diag}
\left ( \bma \lambda & 0 \cr 0 & \lambda ^{-1} \cr \ema \right )
\hskip 1in
\left ( \bma \pm 1  & 1 \cr 0 & \pm 1 \cr \ema \right )  
\eea
The left matrix is for elliptic type with $|\lambda|=1$  and hyperbolic type with $\lambda \in \RR$, both with $\lambda \not = \pm 1$. The right matrix is  for parabolic type.

\subsubsection{Action of $\Gamma$ on $\cH$ and $\RR \cup \{ \infty \}$}

Elements $\gamma \in SL(2,\RR)$ acts on $\cH$ by M\"obius transformation, 
\bea 
\gamma z = \gamma (z) = { az+b \over cz + d} 
\hskip 1in 
\gamma = \left ( \bma a & b \cr c & d \cr \ema \right )
\eea
Since the element $\gamma =-I$ leaves every point of $\cH$ invariant, the action is effectively by the normal subgroup $PSL(2,\RR) = SL(2,\RR)/\{\pm I \}$.  The action of $SL(2,\RR)$ on $\cH$ is transitive since an arbitrary point $z=x+iy$ is the image of the point $ i \in \cH$, namely $z=\gamma_z(i) $,  for a group element $\gamma_z$ whose entries satisfy $c=0, d=a^{-1}$ and $x=ab, y=a^2$. The isotropy group of the point $i$ is the $SO(2)$ subgroup of $SL(2,\RR)$ obtained by setting $c=-b$ and $d=a$.  The isotropy subgroup of an arbitrary point $z$ is the conjugate of $SO(2)$ by the element $\gamma_z$ given by $  \gamma _z \, SO(2) \, \gamma _z^{-1}$. As a result, $\cH$ may be represented by the coset space, 
 \bea
 \cH = SL(2,\RR) / SO(2) 
 \eea
The action of $SL(2,\RR)$ on $\cH$ extends to a transitive action on $\RR \cup \{ \infty \} $  so that every point $x \in \RR \cup \{ \infty \}$ may be expressed as $x=\gamma _x(\infty)$ for some $\gamma _x \in SL(2,\RR)$. 

\sm

An important role will be played by the fixed points of various elements $\gamma \in SL(2,\RR)$. The classification of the elements of $SL(2,\RR)$ into elliptic, parabolic, and hyperbolic given in (\ref{B.4.a}) is equivalent to the following classification in terms of their fixed points,
\bea
\label{B.4.b}
\hbox{ \textit{elliptic element} } \hskip 0.25in & & \hbox{ one fixed point } z \in \cH, \hbox{ the other at } \bar z
\no \\
\hbox{ \textit{parabolic element} } \hskip 0.08in &&  \hbox{ one fixed point, which is in } \RR \cup \{ \infty \}
\no \\
\hbox{ \textit{hyperbolic element} } &&  \hbox{ two fixed points, which are in } \RR \cup \{ \infty \}
\eea
In particular, the classifications of (\ref{B.4.a}) and (\ref{B.4.b})  apply to the elements of an arbitrary discrete subgroup $\Gamma \subset SL(2,\RR)$. One defines,
\bea
\hbox{ \textit{elliptic point of}  } \Gamma && 
\hbox{is a point } z \in \cH \hbox{ such that there exists} 
\no \\  && \hbox{ an elliptic }  \gamma \in \Gamma  \hbox{ with } \gamma (z) = z 
\no \\
\hbox{ \textit{cusp of} }  \Gamma \qquad &  & 
\hbox{is a point } x \in \RR \cup \{ \infty \}  \hbox{ such that there exists}
\no \\ && \hbox{ a parabolic } \gamma \in \Gamma \hbox{ with } \gamma (x) = x
\eea
If $z$ is an elliptic point of $\Gamma$, then $\gamma (z)$ is an elliptic point of $\Gamma$ for all $\gamma \in \Gamma$. If $x$ is a cusp of $\Gamma$, then $\gamma (x)$ is a cusp of $\Gamma$ for every $\gamma \in \Gamma$.

{\prop 
(a) If $z \in \cH$ is an elliptic point of $\Gamma$ then the set $\Gamma _z=\{ \gamma \in \Gamma \,  | \, \gamma(z)=z \}$ is a finite cyclic group. (b)  The set of elements of finite order in $\Gamma$ consist of the elliptic elements of $\Gamma$ together with $\pm I$. } 

\sm

To prove (a) we note that $\Gamma _z$ is a subgroup of $\Gamma$. Representing $z=\gamma _z (i)$ we see that  $\Gamma_z$ is the set of $\gamma \in \Gamma$ for which $\gamma _z^{-1} \gamma \gamma_z (i) = i$. The isotropy group of the point $i$ was earlier  identified as $SO(2)$ so that $\Gamma _z \subset SO(2)$. Since $\Gamma_z$ is discrete and a subgroup of the compact group $SO(2)$,   $\Gamma _z$ must be finite. Since all finite subgroups of $SO(2)$ are cyclic, $\Gamma_z$ must also be cyclic.
To prove (b), we use the fact that an element $\gamma \in \Gamma$  is conjugate under $SL(2,\CC)$  to a diagonal matrix with diagonal entries $\lambda, \lambda ^{-1} \in \CC$ as in (\ref{B.diag}). Finite order implies that there exists a positive integer $n$ such that $\lambda ^n=1$. When $n=1,2$ we have $\gamma = \pm I$ while for $n >2$, $\gamma$ must be elliptic.
Conversely, every elliptic element is of finite order by (a).

\subsection{Construction of Riemann surfaces via Fuchsian groups}

The fundamental group $\pi_1(\Sigma)$ for an arbitrary compact Riemann surface $\Sigma$ of genus $g \geq 2$  is discrete and non-Abelian. The universal (i.e. simply connected) covering space $\hat \Sigma$ of $\Sigma$ may be realized as a fiber bundle over $\Sigma$ with structure group isomorphic to $\pi_1(\Sigma)$. Since $\Sigma$ admits a metric $\hat \mg$ of constant negative curvature $R_{\hat \mg}=-1$ by the above theorem, the universal covering surface admits a metric of constant negative curvature as well. Therefore, by the uniformization theorem, we conclude that $\hat \Sigma = \cH$ or the conformaly isomorphic $ \cD$. Conversely, the fundamental group $\pi_1(\Sigma)$ must act on $\hat \Sigma$ by an isometry of the constant curvature metric. Since we have $\hat \Sigma= \cH$, whose isometry group is $PSL(2,\RR)$, the action of $\pi_1(\Sigma)$ on $\cH$ must be by a representation $\rho$,
\bea
\rho : \pi_1(\Sigma) \mapsto PSL(2,\RR)
\eea
which assigns an element of $PSL(2,\RR)$ to each equivalence class of closed curves on $\Sigma$ with a specified, though arbitrary, base point $P$. To realize these assignments concretely, we choose a base point $P \in \Sigma$, as well as a set of homology generators $\mA_I$ and $\mB_J$ for $I,J=1,\cdots, g$ passing through $P$ and intersecting one another no-where else, as shown in Figure \ref{fig:B.1}. 

\sm

\begin{figure}
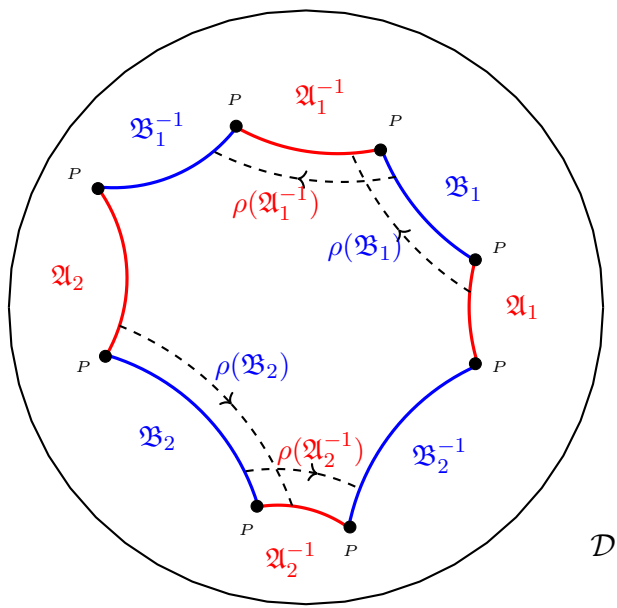

\begin{center}
\tikzpicture[scale=3.95]

\scope[xshift=0cm,yshift=0cm]
\draw  [ thick, domain=0:180] plot ({cos(\x)},{sin(\x)});
\draw  [ thick, domain=180:360] plot ({cos(\x)},{sin(\x)});
\draw  [very  thick, color=red, domain=166:198] plot ({cosh(0.6)*cos(0)+sinh(0.6)*cos(\x)}, {cosh(0.6)*sin(0)+sinh(0.6)*sin(\x)});
\draw  [very  thick, color=red, domain=241:282] plot ({cosh(0.65)*cos(85)+sinh(0.65)*cos(\x)}, {cosh(0.65)*sin(85)+sinh(0.65)*sin(\x)});
\draw  [very  thick, color=red, domain=330:396] plot ({cosh(0.5)*cos(175)+sinh(0.5)*cos(\x)}, {cosh(0.5)*sin(175)+sinh(0.5)*sin(\x)});
\draw  [very  thick, color=red, domain=54:100] plot ({cosh(0.4)*cos(265)+sinh(0.4)*cos(\x)}, {cosh(0.4)*sin(265)+sinh(0.4)*sin(\x)});
\draw  [very  thick, color=blue, domain=201:239] plot ({cosh(0.7)*cos(40)+sinh(0.7)*cos(\x)}, {cosh(0.7)*sin(40)+sinh(0.7)*sin(\x)});
\draw  [very  thick, color=blue, domain=264:323] plot ({cosh(0.5)*cos(125)+sinh(0.5)*cos(\x)}, {cosh(0.5)*sin(125)+sinh(0.5)*sin(\x)});
\draw  [very  thick, color=blue, domain=17:73] plot ({cosh(0.7)*cos(225)+sinh(0.7)*cos(\x)}, {cosh(0.7)*sin(225)+sinh(0.7)*sin(\x)});
\draw  [very  thick, color=blue, domain=114:169] plot ({cosh(0.7)*cos(315)+sinh(0.7)*cos(\x)}, {cosh(0.7)*sin(315)+sinh(0.7)*sin(\x)});
\draw  [thick, color=black, dashed, ->, domain=238:220] plot ({cosh(0.9)*cos(40)+sinh(0.9)*cos(\x)}, {cosh(0.9)*sin(40)+sinh(0.9)*sin(\x)});
\draw  [thick, color=black, dashed,  domain=220:203] plot ({cosh(0.9)*cos(40)+sinh(0.9)*cos(\x)}, {cosh(0.9)*sin(40)+sinh(0.9)*sin(\x)});
\draw (0.148,-0.74) [fill=black] circle(0.02cm) ;
\draw  [thick, color=black, dashed, ->, domain=68:42] plot ({cosh(0.9)*cos(225)+sinh(0.9)*cos(\x)}, {cosh(0.9)*sin(225)+sinh(0.9)*sin(\x)});
\draw  [thick, color=black, dashed,  domain=42:19] plot ({cosh(0.9)*cos(225)+sinh(0.9)*cos(\x)}, {cosh(0.9)*sin(225)+sinh(0.9)*sin(\x)});
\draw  [thick, color=black, dashed, ->, domain=281:261] plot ({cosh(0.85)*cos(85)+sinh(0.85)*cos(\x)}, {cosh(0.85)*sin(85)+sinh(0.85)*sin(\x)});
\draw  [thick, color=black, dashed, domain=261:243] plot ({cosh(0.85)*cos(85)+sinh(0.85)*cos(\x)}, {cosh(0.85)*sin(85)+sinh(0.85)*sin(\x)});
\draw  [thick, color=black, dashed, ->, domain=99:77] plot ({cosh(0.6)*cos(265)+sinh(0.6)*cos(\x)}, {cosh(0.6)*sin(265)+sinh(0.6)*sin(\x)});
\draw  [thick, color=black, dashed, domain=77:63] plot ({cosh(0.6)*cos(265)+sinh(0.6)*cos(\x)}, {cosh(0.6)*sin(265)+sinh(0.6)*sin(\x)});
\draw (0.57,0.16) [fill=black] circle(0.02cm) ;
\draw (0.251,0.53) [fill=black] circle(0.02cm) ;
\draw (-0.7,0.4) [fill=black] circle(0.02cm) ;
\draw (-0.235,0.61) [fill=black] circle(0.02cm) ;
\draw (-0.675,-0.165) [fill=black] circle(0.02cm) ;
\draw (-0.165,-0.67) [fill=black] circle(0.02cm) ;
\draw (0.57,-0.19) [fill=black] circle(0.02cm) ;

\draw (0.15, -0.82) node{\tiny $P$};
\draw (0.65, 0.2) node{\tiny $P$};
\draw (0.3, 0.63) node{\tiny $P$};
\draw (-0.78, 0.45) node{\tiny $P$};
\draw (-0.24, 0.7) node{\tiny $P$};
\draw (-0.75, -0.2) node{\tiny $P$};
\draw (-0.2, -0.75) node{\tiny $P$};
\draw (0.65, -0.2) node{\tiny $P$};
\draw [color=red] (0.73,0) node{\small $\mA_1$};
\draw [color=blue] (0.53,0.4) node{\small $\mB_1$};
\draw [color=red] (0.05,0.71) node{\small $\mA_1^{-1}$};
\draw [color=blue] (-0.5,0.6) node{\small $\mB_1^{-1}$};
\draw [color=red] (-0.8,0.1) node{\small $\mA_2$};
\draw [color=blue]  (-0.5,-0.44) node{\small $\mB_2$};
\draw [color=red] (-0.05,-0.85) node{\small $\mA_2^{-1}$};
\draw [color=blue]  (0.45,-0.5) node{\small $\mB_2^{-1}$};
\draw [color=red] (-0.1,0.35) node{\small $\rho(\mA_1^{-1})$};
\draw [color=red] (0.05,-0.48) node{\small $\rho(\mA_2^{-1})$};
\draw [color=blue] (0.2,0.2) node{\small $\rho(\mB_1)$};
\draw [color=blue]  (-0.18,-0.2) node{\small $\rho(\mB_2)$};
\draw (1,-0.8) node{{$\cD$}};
\endscope
\endtikzpicture
\caption{\textit{A fundamental domain for  a compact genus-two Riemann surface $\Sigma$  in the hyperbolic disc $\cD$. The surface $\Sigma$ is recovered by  identifying the curves $\mA_I $ with $ \mA_I^{-1}$  and identifying the curves $\mB_I$ with $ \mB_I^{-1}$.} \label{fig:B.3}}
\end{center}
\end{figure}

We now cut the surface $\Sigma$ along each one of the closed curves $\mA_I$ and $\mB_J$ to unfold $\Sigma$ into a polygonal simply connected domain in $\CC$, as shown schematically for genus 2 in Figure \ref{fig:B.3}. To reconstruct the surface from the fundamental domain represented in Figure \ref{fig:B.3}, we pairwise identify the curves corresponding to $\mA_I$ and $\mA_I^{-1}$ by the transformation $\rho(\mB_I)$ thereby closing the curves $\mB_I$. Similarly, we identify the curves $\mB_I$ and $\mB_I^{-1}$ by the transformation $\rho(\mA_I^{-1})$. Clearly, the composition of all commutators must give the identity, so we must have, 
\bea
\rho(\mC_1) \rho(\mC_2) \cdots \rho(\mC_g) = I \hskip 1in \rho(\mC_I) = \rho(\mA_I) \rho( \mB_I) \rho(\mA_I)^{-1} \rho(\mB_I)^{-1}
\eea
Each homotopy class of $\pi_1(\Sigma)$ contains a unique geodesic, or closed curve of minimal length in the hyperbolic metric. Choosing  the closed curves $\mA_I$ and $\mB_J$ to be geodesics, we obtain a representation of $\Sigma$ in $\cD$ by a domain $\Sigma _\cD$ whose boundary is the union of circular arcs each of which is a geodesic of $\cD$, as represented in Figure \ref{fig:B.3}. The hyperbolic  length of the closed curve $\mA_I$ is identical to the length of $\mA_I^{-1}$ and similarly the lengths of the segments $\mB_I$ and $\mB_I^{-1}$ are equal to one another.

\subsubsection{Compactification of $\Gamma \backslash \cH$}

We choose the standard topology of $\cH$ generated by open sets given by open metric discs.  Consider a Fuchsian group $\Gamma \subset SL(2,\RR)$ whose cusps form a set $\cC \subset \RR \cup \{ \infty \}$. To $\Gamma$ we associate the space,
\bea
\bar \cH = \cH \cup \cC
\eea
To specify the topology of $\bar \cH$ we add open sets that contain the cusps in $\cC$ to the topology of $\cH$. For the cusp $x= \infty$ and the cusps $x \not = \infty$,  these open sets may be chosen as follows,
\bea
x = \infty & \hskip 0.5in & \{ \infty \} \cup \big \{ z \in \cH \, \hbox{ such that } \, \Im(z) > y >0 \hbox{ for } y \in \RR^+ \big \}
\no \\
x \not= \infty & \hskip 0.5in & \{ x \} \cup \big \{ z \in \cH \, \hbox{ such that } \, |z-x-iy| < y \hbox{ for }  y \in \RR^+ \big \}
\eea
The space $\bar \cH$ equipped with this topology is clearly not compact. In fact, it is not even locally compact.\footnote{A topological space $X$ is locally compact if every point in $X$ has an open neighborhood which is contained in a compact subset of $X$.}  The group $\Gamma$ consistently acts of $\bar \cH$ which allows us  to define the quotient,
\bea
X(\Gamma) = \Gamma \backslash \bar \cH
\eea
The quotient space $\Gamma \backslash \bar \cH$ is locally compact. Finally, we state the following proposition  here without proof.

{\prop \label{prop:B6} If the space $\Gamma \backslash \bar \cH$ is compact then the number of $\Gamma$-inequivalent cusps (resp. the number of elliptic points) is finite.}

\subsection*{$\bullet$ Bibliographical notes}

Classic references on Riemann surfaces are the books by Gunning~\cite{Gunning2} and by Farkas and Kra~\cite{Kra}. Comprehensive lecture notes by Bost may be found in the collection~\cite{Bost}. A detailed treatment of modular curves and their compactification may be found in Chapters 2 and 3 of \cite{DS}, where a proof of Proposition \ref{prop:B6} is also given. 

 \newpage
 
\section{Line bundles on Riemann surfaces}
\setcounter{equation}{0}
\label{sec:LB}

In this appendix, we shall present a brief review of complex line bundles over Riemann surfaces. We begin by 
giving a general mathematical definition of a complex line bundle over a Riemann surface $\Sigma$, the topological classification of line bundles, divisors, the Riemann-Roch theorem, and various dimension formulas including the dimension of moduli space. We then discuss the same topic of line bundles from a more physics-oriented point of view as spaces of vector fields, forms, and spinors.

\subsection{Holomorphic line bundles on a Riemann surface}

A complex line bundle $L$ on a Riemann surface $\Sigma$ provides an assignment of a one-dimensional complex vector space $L_p$ to each point $p \in \Sigma$. A local description of the line bundle $L$  may be given by using a covering of $\Sigma$ with  open sets $\cU_\a$ so that $\bigcup _\a \cU_\a = \Sigma$.  Introducing local complex coordinates $z_\a, \bar z_\a$ in each open set makes $\cU_\a$  into a coordinate chart. Locally in each coordinate chart, the bundle is a direct product $\cU_\a \times \CC$ and each section $f$ of $L$ reduces to a smooth $\CC$-valued function $f_\a$ on $\cU_\a$. In the intersection $\cU_\a \cap \, \cU_\beta$ the functions $f_\a$ and $f_\b$ are related by transition functions $\f_{\a \b}$,
\bea
\label{6.A1}
f_\a = \f_{\a \b} f_\b \hskip 0.5in \hbox{ in } \quad \cU_\a \cap \, \cU_\b
\eea
The transition functions $\f_{\a \b}$ are nowhere vanishing and, for a holomorphic line bundle, must be holomorphic functions. Considering now the intersection of three coordinate charts, we obtain a condition on products of transition functions, 
\bea
\label{6A.prod}
\f_{\a \b} \, \f_{\b \g} = \f _{\a \g} \hskip 0.5in \hbox{ in } \quad \cU_\a \cap \, \cU_\b \cap \, \cU_\g
\eea
Note that, for a holomorphic line bundle $L$, the transition functions are holomorphic, but the sections $f$ and the 
local functions $f_\a$ need not be holomorphic.

\subsubsection{Topological classification}

To obtain the topological classification of line bundles over $\Sigma$, we solve the non-vanishing condition on the transition functions $\f_{\a \b}$ in terms of exponentials, $\f_{\a \b} = \exp ( 2\pi i \psi_{\a \b})$, where $\psi_{ \a \b}$ is defined by $\f_{\a \b}$ up to the addition of an arbitrary integer $n_{\a \b}$. The product relation (\ref{6A.prod}) is then equivalent to the fact that the combinations, 
\bea
c_{\a\b\g} = \psi_{\a \b} + \psi _{\b \g} - \psi _{\a\g}
\eea
are integers since their exponential equals 1 by (\ref{6A.prod}). The transformations $\psi_{\a\b} \to \psi _{\a\b}+ n_{\a \b}$ imply  transformations of $c_{\a\b\g}$ by an \textit{exact cocyle} $n_{\a\b} + n_{\b \g} - n_{\a\g}$,
\bea
\label{6A.gauge}
c_{\a \b \g} \to c_{\a\b\g} + n_{\a\b} + n_{\b \g} - n_{\a\g}
\eea
By construction in terms of $\psi_{\a\b}$, the integers $c_{\a\b\g}$ satisfy the following \textit{closed cocycle} relation on the intersection of four coordinate charts,
\bea
\label{6A.cocycle}
c_{\a \b \g} - c_{\b \g \delta} +c_{\g \delta \a} - c_{\delta \a \b} =0
\eea
The space of closed cocycles $c_{\a\b\g}$ satisfying (\ref{6A.cocycle}) modulo exact cocycles $n_{\a\b} + n_{\b \g} - n_{\a\g}$ is the second \v Cech cohomology group of $\Sigma$ with integer coefficients, denoted $H^2(\Sigma, \ZZ)$.

\sm

One may translate the \v Cech cohomology formulation of line bundles into the language of differential forms, de Rham cohomology, and gauge fields, which is more familiar to physicists. It will be convenient here to assume that $\Sigma$ is compact so that one may extract from the covering of $\Sigma$ by open sets $\cU_\a$ a covering with only a finite number of open sets, and avoid issues of convergence.  

\sm

Each line bundle $L$ on $\Sigma$ corresponds to a cohomology class of $H^2(\Sigma, \ZZ)$, labeled by the first Chern class $c_1(L)$. In turn, the first Chern class $c_1(L)$  may be represented in terms of a de Rham cohomology class $[F]$ where $F$ is the curvature 2-form of a $U(1)$ connection 1-form. In each coordinate chart $\cU_\a$ a representative $F$ of the class $[F]$ may be expressed in terms of a local connection form $A_\a$ by  $F = d A_\a$. The transition functions for the connection are,\footnote{To be precise the de Rham theorem states an isomorphism between the de Rham cohomology group $H^2_{{\rm dR}}(\Sigma)$ and the Cech cohomology group with real coefficients $H^2(\Sigma, \RR)$.}
\bea
A_\a - A_\b = d \lambda _{\a \b}
\eea
with $c_{\a\b\g} = \lambda_{\a\b} + \lambda _{\b \g} - \lambda _{\a \g}$. The first Chern class, defined by,
\bea
\label{C.Chern1}
c_1(L) = { i \over 2 \pi} \int _\Sigma F
\eea
takes integer values, so that $H^2(\Sigma, \ZZ) =\ZZ$.  For later use, we note that the first Chern class of the tensor product of two line bundles $L_1$ and $L_2$ over $\Sigma$ is given as follows,
\bea
\label{C.Chern2}
c_1 (L_1 \otimes L_2) = c_1 (L_1) + c_1(L_2)
\eea
We conclude that the topological classification of line bundles $L$ over a compact Riemann surface $\Sigma$ is in terms of a single integer $d$, referred to as the degree of $L$,  which labels the first Chern class $c_1(L)$ of $L$.

\subsubsection{Holomorphic classification: divisors}

Holomorphic line bundles in a given topological class characterized by the degree $c_1(L)=d$ are not necessarily equivalent holomorphically. The space of all holomorphic line bundles for given degree $d$ is referred to as the Picard variety,
\bea
{\rm Pic}_d (\Sigma) = \left \{ \hbox{holomorphic line bundles } L \hbox{ on } \Sigma \hbox{ with } c_1(L)=d \right \}
\eea
The special case of $d=0$ corresponds to the \textit{Jacobian variety} $J(\Sigma) = {\rm Pic}_0(\Sigma)$. 

\sm

A convenient way to characterize a holomorphic line bundle is by its {\rm divisor}, which is a complex codimension-one subset of $\Sigma$ given by a formal sum over distinct points $p_\a \in \Sigma$,
\bea
\label{6A.div}
D = \sum _{\a=1}^N n_\a p_\a
\eea
The integer  $n_\a$ denotes the order of the point $p_\a$ in the divisor $D$ and may take positive or negative values, while points for which $n_\a=0$ are usually omitted from the sum. 

\sm

To construct the line bundle $L$ associated with a divisor~$D$, we introduce coordinate charts  $\cU_\a$  with  $p_\a \in \cU_\a$ and a local complex coordinate $z_\a$ in $\cU_\a$ that vanishes at  $p_\a$ for each $\a = 1,\cdots ,N$. By choosing each $\cU_\a$ small enough we can make them mutually disjoint $\cU_\a \cap \, \cU_\b= \emptyset $ for $\a \not= \b$. We denote the open complement to all the points $p_\a$ by  $\cU_\infty = \Sigma \setminus \{ p_1, \cdots, p_N\}$ so that $\cU_\infty \bigcup _\a \cU_\a = \Sigma$. The line bundle $L$ corresponding to the divisor $D$ is then constructed by taking transition functions in the intersections $\cU_\a \cap \, \cU_\infty$ to be $z_\a ^{n_\a}$, the integers $n_\a$ corresponding to the order of the point $p_\a$ in the divisor $D$ in (\ref{6A.div}). A section $f$ of $L$, described locally by functions $f_\a$ on $\cU_\a$ and $f_\infty$ in $\cU_\infty$, satisfies $f_\a = z_\a ^{n_\a} f_\infty$. Conversely, given a line bundle $L$ and a section $f$, the corresponding divisor may be recovered uniquely from the zeros and poles of $f$. Since the open sets $\cU_\a$ are mutually disjoint, all intersections with more than two $\cU$ coordinate charts are empty. As a result, the cocycle conditions (\ref{6A.prod}) and (\ref{6A.cocycle})  are automatically satisfied. 

\sm

Adding the contributions from each transition function to the first Chern class, we find,
\bea
d = c_1(L) = \sum_{\a =1}^N n_\a
\eea
In other words, the degree $d$ of $L$ is the total number of zeros (counted with their orders $n_\a$ when $n_\a>0$) minus the total number of poles (counted with their orders $-n_\a$ when $n_\a <0$). 
A topologically trivial bundle $L$ with $d=0$ has an equal number of zeros and poles (counted with their orders) and its meromorphic sections are the meromorphic functions on~$\Sigma$. One may introduce an equivalence relation between divisors such that $D_1$ and $D_2$ are equivalent to one another if they differ by a divisor of degree zero. When this is the case, the ratio $f_2/f_1$ of any two sections $f_1$ and $f_2$ of $D_1$ and $D_2$ respectively, has degree zero.  The equivalence class with representative divisor $D$ is denoted by $[D]$.

\subsubsection{Examples}

A familiar example in Physics is provided by the Dirac magnetic monopole, which may be thought of having a constant field strength $F$ on the Riemann sphere $\hat \CC$. Denoting the North and South poles by $p_\pm$, we  introduce coordinate charts $\cU_\pm = \hat \CC \setminus \{ p_\mp \}$ and local coordinates $z_\pm$ which vanish at  $p_\pm$. Choosing $f_-=1$ and $f_+=z_+^{n_+}$ gives $\psi _{+-} = n_+ \ln (z_+)/(2 \pi i) $ and $A_+ - A_-=  n_+ dz_+ /(2 \pi i z_+)$, so that $c_1(L)=n_+$ for a Dirac monopole of charge $n_+$.

\sm

The tangent bundle $T \Sigma$ over a Riemann surface $\Sigma$ is a bundle of rank 2 over the reals. Using the complex structure of $\Sigma$ it may be decomposed into the direct sum of its holomorphic component $T\Sigma _{(1,0)}$ and its complex conjugate $T\Sigma _{(0,1)}$. The cotangent bundle $T^*\Sigma$ may be decomposed analogously, 
\bea
T \Sigma & = & T\Sigma _{(1,0)} \oplus T\Sigma _{(0,1)}
\no \\
T^* \Sigma & = & T^*\Sigma _{(1,0)} \oplus T^*\Sigma _{(0,1)}
\eea
The bundles $T\Sigma _{(1,0)}$ and $T^*\Sigma _{(1,0)}$ are holomorphic line bundles over $\Sigma$. The holomorphic cotangent bundle $T^*\Sigma _{(1,0)}$ is also referred to as the \textit{canonical bundle} over $\Sigma$ and denoted $K$, while $ T\Sigma _{(1,0)}$ is isomorphic to $K^{-1}$.

\sm

 This decomposition may be rendered explicit on the sections of the bundles.  In terms of real local coordinates $\xi^m$ on $\Sigma$ a section of $T\Sigma$ is a  vector field and may be expressed as $v^m  \p_m$ while a section of $T^*\Sigma$ is a  differential one-form that takes the form $\om_m d\xi^m$.   In local complex coordinates $z, \bar z$ they may be decomposed as follows, 
\bea
v^m \, \p_m & = & v^z \, \p_z + v^{\bar z} \, \p_{\bar z} 
\no \\
\om_m \, d\xi^m & = & \om_z \, dz + \om_{\bar z} \, d{\bar z} 
\eea
where $v^z \, \p_z $ is a section of $T\Sigma _{(1,0)}$ and $\om_z \, dz$ is a section of $T^*\Sigma _{(1,0)}$ and similarly for their complex conjugates. Their transition functions are given as follows,
\bea
(v^z)_\a (z_\a) & = & (v^z)_\b (z_\beta) \, \left (  \p z_\a / \p z_\b \right )
\no \\
(\om_z)_\a (z_\a) & = & (\om_z)_\b (z_\b) \, \left (  \p z_\a / \p z_\b \right )^{-1}
\eea
where $\p z_\a / \p z_\b$ is a holomorphic nowhere vanishing transition function in $\cU_\a \cap \, \cU_\b$. 

\sm

Finally, a spin bundle $S$ on $\Sigma$ is a holomorphic line bundle whose tensor square is isomorphic to the canonical bundle $K$, 
\bea
S \otimes S \approx K 
\hskip 1in c_1(S) = \thalf c_1(K)
\eea 
The relation between the Chern numbers follows from  (\ref{C.Chern2}). By a slight abuse of notation, a spin bundle is some times denoted by $K^\half$. For $g=0$ the spin bundle is unique. However, for $g\geq 1$, the relation to the canonical bundle  does not specify the spin bundle $S$ uniquely. The different spin bundles are referred to as \textit{spin structures}.  On a surface of genus $g$, there are $2^{2g}$ independent spin structures.

\subsection{Holomorphic sections and the Riemann-Roch theorem}

As emphasized earlier, while the transition functions of a holomorphic line bundle are holomorphic functions, the sections of a holomorphic line bundle are not generally holomorphic. The Cauchy-Riemann operator $\bar \p _L$ for a holomorphic line bundle $L$  acts on the sections of $L$ by the ordinary Cauch-Riemann operator $\bar \p = d\bar z^\a \p_{\bar z_\a} $ with respect to the local coordinate $z_\a$ in each open set $\cU_\a$. This action is covariant without the need for a connection, as may be seen by applying $\bar \p$ to local sections in the overlap of open sets $\cU_\a$ given by (\ref{6.A1}),
\bea
\bar \p f_\a = \f_{\a \b} \, \bar \p f_\b   \hskip 0.6in \hbox{in} \hskip 0.6in \cU_\a \cap \cU_\b
\eea
The subspace of all holomorphic sections of a holomorphic line bundle plays a special role and contains a lot of information on the line bundle. It may be defined as follows. To a holomorphic line bundle $L$ we associate the vector space of its holomorphic sections,
\bea
{\rm Ker} (\bar \p_L) = \{ \hbox{holomorphic sections of } L \}
\eea
These spaces are finite-dimensional and their dimensions are related by the Riemann-Roch theorem, which will be proven, in a slightly different guise, in appendix \ref{sec:C.RR}.

{\thm The Riemann-Roch theorem for a holomorphic line bundle $L$ states
\bea
\dim {\rm Ker} \left ( \bar \p_L \right )  - \dim {\rm Ker} \left ( \bar \p _{K \otimes L^{-1} } \right ) = c_1(L) + \half \chi(\Sigma)
\eea
where $K$ is the canonical bundle, $\chi(\Sigma)$ is the Euler characteristic of $\Sigma$, and $\bar \p _L$ is the Cauchy Riemann operator acting on $L$. }

\subsection{Vanishing Theorem and dimension formulas}

In this subsection, we shall state a vanishing theorem and combine its implications with the Riemann-Roch theorem to obtain general expressions for the dimensions of the spaces of homomorphic sections of line bundles. We shall specialize to the case of compact $\Sigma$ and make use of the fact that the Euler number is then given in terms of the genus $g$ of $\Sigma$ by $\chi(\Sigma) =2 - 2g$.

\sm

We begin by obtaining some immediate consequences of the Riemann-Roch theorem. Since the only holomorphic functions on $\Sigma$ are the constant functions, the corresponding bundle $L$ is the trivial bundle with $c_1(L)=0$ and  we have ${\rm Ker} (\bar \p_L)= \CC$ and  $\dim {\rm Ker} (\bar \p_L) =1$. Using the Riemann-Roch theorem for the trivial bundle $L$, we find
\bea
\dim {\rm Ker} (\bar \p_K) =g
\eea
For $g=0$ there are no holomorphic 1-forms. For $g=1$ we recover the observation, made long ago, that the space of holomorphic 1-forms is one-dimensional. For $g \geq 1$, there are $g$ linearly independent holomorphic 1-forms on $\Sigma$ already identified in appendix \ref{sec:B.1}.    Setting $L=K$, and using the fact that $K \otimes K^{-1}$ is the trivial bundle,  the Riemann-Roch theorem implies $c_1(K)=2g-2$. For $g=1$ we recover the fact, observed long ago,  that the holomorphic 1-form $dz$ is nowhere vanishing, while for $g\geq 2$ we obtain the result that every holomorphic one-form has $2g-2$ zeros (counted with multiplicities). 

\sm

More generally, representing the divisor class $[L]$ of an arbitrary line bundle $L$ by a tensor power of the canonical bundle  $[K^n]$ for $n \in \ZZ/2$, where it is understood that $K^\half =S$ is a spin bundle with specified spin structure, we have the following vanishing theorem:

{\thm The following holds}
\bea
\label{6A.van}
\dim {\rm Ker} (\bar \p_L) =0 \hskip 0.4in 
\left \{ \bma g=0 & \hbox{ and } & n >0 \cr g\geq 2 & \hbox{ and } & n <0  \cr \ema \right .
\eea

The entry for $g=0$ states the absence of holomorphic forms or spinors of degree greater than 0 on the sphere. The entry for $g \geq 2$ states that there are no holomorphic vector fields on a higher genus Riemann surface. The proof of this theorem will be giving in appendix \ref{sec:C.VT}, in a slightly different guise. 

\sm

Combining the results of the vanishing theorem with the results obtained from the Riemann-Roch theorem for arbitrary divisor class  $[L]=[K^n]$ and using the relations $c_1(L)= c_1(K^n) = n c_1(K) $ for  $n \in \ZZ/2$ we obtain for  $n \not =  \half$, 
\bea
\label{6A.dims1}
\dim {\rm Ker} (\bar \p _L) = \left \{ \bma 
(2n-1)(g-1) & \hbox{ for } &  g \geq 2 & \hbox{and} & n \geq { 3 \over 2} & \cr  
g &  &  g \geq 0 &  & n=1 & \cr
1 &  &  g \geq 0 &  & n=0 & \cr
1 &  &  g =1 &  & n \not= \half & \cr
1-2n && g=0 && n \leq -\half \cr
\ema \right .
\eea
For $n$ a half-odd-integer, these dimensions are independent of the spin structure of~$L$.  But for $n =  \half$ the dimension of ${\rm Ker} (\bar \p _L) $ depends on the spin structure of $L$. Spin structures are partitioned into even and odd depending on whether $\dim {\rm Ker} (\bar \p _L) $ is even or odd,   
\bea
\label{6A.dims2}
\dim {\rm Ker} (\bar \p _L) = \left \{ \bma 
1 ~ (\mod 2) &  & g \geq 1 &  & n = \thalf &\hbox{(odd spin structures)} \cr
&&&& \cr
0 ~ (\mod 2) & & g \geq 1 &  & n = \thalf & \hbox{(even spin structures)} \cr
\ema \right .
\eea
The dimensions in (\ref{6A.dims2}) equal $1$ and $0$ respectively throughout moduli space for $g=1,2$ and at generic points in moduli space for arbitrary $g \geq 3$.  For $g \geq 3$, the dimensions may increase by a multiple of 2 at certain sub-varieties of moduli space, including at the locus of hyper-elliptic Riemann surfaces.

\subsection{Tensors and spinors on $\Sigma$}
\label{sec:C4}

In Physics, we will deal with scalar, vector, tensor, and spinor fields on a Riemann surface~$\Sigma$ viewed as a two real dimensional surface equipped with a Riemannian metric,
\bea
\mg = \mg_{mn} (\xi) \, d \xi^m d \xi^n
\eea
in a system of real coordinates $\xi^m$.  We shall now analyze complex line bundles, covariant derivatives and Laplace-Beltrami operators on line bundles in a tensorial formulation, and use it to prove the Riemann-Roch and vanishing theorems.

\sm

An arbitrary tensor field $\mt_{m_1 \dots  m_n}$ of rank $n$ may be decomposed into a direct sum of symmetric  traceless tensors and scalars. Indeed, any anti-symmetric pair of indices may be contracted with the tensor $\sqrt{\mg} \, \ep_{mn}$ to produce a scalar, while the metric $\mg_{mn}$ may be used to eliminate all traces. The resulting symmetric traceless tensor may be further decomposed into the eigenspaces of the complex structure $\cJ$.  The convention $v^z \p_z = v^z (dz)^{-1}$ may be used to identify a form of negative weight with a vector field of opposite weight. Henceforth,  we shall suppress the $z$ and $\bar z$  indices on these reduced tensors, so that $\mt= t_{z, \cdots z}$ and $\bar \mt = t_{\bar z \cdots, \bar z}$. Thus, all tensor fields decompose into a direct sum of the following spaces,  
\bea
T_{(n,0)}^* = \Big \{ \mt(z) : \Sigma \to \CC ~ \hbox{such that} ~ \mt'(z') (dz')^n = \mt(z) (dz)^n \Big \}
\eea 
The space  $T_{(1,0)}^*$ may be identified with the space of sections of the canonical bundle $K$, while the spaces  $T^*_{(n,0)}$ for $n >0$ may be identified with the space of sections of the $n$-th tensor power $K^n$ of the canonical bundle $K$.  Finally, we include spinors by allowing $n$ to take half-integer values which further requires specifying a spin structure.

\subsubsection{Covariant derivatives and Laplace-Beltrami operators}

On each space of sections $T_{(n,0)}^*$ we define an $L^2(\Sigma)$ inner product as follows,
\bea
\< \mt_1 |  \mt_2 \> = \int _\Sigma d\mu_\mg \, (\mg_{z \bar z})^{-n} \, \overline{ \mt_1 (z)} \, \mt_2(z)
\eea
The Cauchy-Riemann operator $\bar \p = (d\bar z) \pbz$ acts covariantly on the spaces $T_{(n,0)}^*$, without the need for a Christoffel connection, 
\bea
\bar \p : T_{(n,0)}^* \to T^*_{(n,1)} ~~ \hbox{with} ~~ \mt(z) (dz)^n \to \big ( \pbz \mt(z) \big ) (dz)^n d\bar z
\eea
It is often more convenient to use a covariant derivative which maps a space into a space of the same type. This may be achieved by the Cauchy-Riemann operator acting on $T^*_{(n,0)}$, 
\bea
\nabla ^z_{(n)} : T_{(n,0)}^* \to T_{(n-1,0)}^* \qquad \hbox{with} \qquad \mt(z) (dz)^n \to \nabla ^z_{(n)} \mt (z) = \big ( g^{z\bar z} \pbz \mt(z) \big )  (dz)^{n-1} 
\eea
To define the adjoint operator $- \nabla _z ^{(n-1)} $ of $\nabla ^z_{(n)}$ with respect to the $L^2(\Sigma)$ inner product  does require a Christoffel connection and is given by, 
\bea
\nabla _z ^{(n)} : T_{(n,0)}^* \to T_{(n+1,0)}^* ~~ \hbox{with} ~~ \mt(z) (dz)^n \to \big ( \nabla _z \mt(z) \big ) (dz)^{n+1} 
\eea
The adjoint operators, and the associated Laplace-Beltrami operators,  are as follows,
\bea
\left ( \nabla ^z _{(n)} \right )^\dagger = - \nabla _z ^{(n-1)} 
& \hskip 0.8in & 
\Delta ^+_{(n)} = - 2\nabla ^z _{(n+1)}  \nabla _z ^{(n)}
\no \\
\left ( \nabla _z ^{(n)} \right )^\dagger = - \nabla ^z _{(n+1)} 
&&
\Delta ^-_{(n)} =  - 2\nabla _z ^{(n-1)} \nabla ^z _{(n)} 
\eea
The Laplace-Beltrami  operators both map  $T_{(n,0)} \to T_{(n,0)}$. The non-zero eigenvalues and their corresponding eigenfunctions are related. To see this, we observe that,
\bea
\Delta ^+_{(n)} \psi = \lambda \psi & \quad \Longrightarrow \quad & \Delta ^-_{(n+1)} \left ( \nabla ^{(n)} _z \psi \right ) = \lambda \left ( \nabla ^{(n)} _z \psi \right ) 
\no \\
\Delta ^-_{(n)} \f = \mu \f & \quad \Longrightarrow \quad & \Delta ^+_{(n-1)} \left ( \nabla _{(n)} ^z \f \right ) 
= \mu \left ( \nabla _{(n)} ^z \f \right ) 
\eea
When $\lambda, \mu \not=0$, these maps are invertible. 
As a result, the spectra of non-vanishing eigenvalues of $\Delta ^+_{(n)}$ and $\Delta ^-_{(n+1)}$ coincide and the one-to-one map between their eigenfunctions is provided by the above relations. No such correspondence follows for the kernels of these operators. Since the Laplace-Beltrami operators are positive, their kernels are related to the kernels of the Cauchy-Riemann operators as follows,
\bea
{\rm Ker} \, \Delta ^+_{(n)} & = & {\rm Ker} \, \nabla _z ^{(n)}
\no \\
{\rm Ker} \, \Delta ^-_{(n)} & = & {\rm Ker} \, \nabla ^z _{(n)} 
\eea
In the subsequent sections, we shall relate the dimensions of these kernels to one another by the Riemann-Roch and vanishing theorems.

\subsection{Proof of the Riemann-Roch theorem}
\label{sec:C.RR}

The kernel of $\nabla _{(n)}^z$ is the space of holomorphic differentials of rank $n$ for $n \geq 1$, and holomorphic vector fields of rank $-n$ for $n\leq -1$. In this formulation, the Riemann-Roch theorem for a compact Riemann surface $\Sigma$ may be stated as follows,
\bea
\dim {\rm Ker} \, \nabla ^z _{(n+1)} - \dim {\rm Ker} \, \nabla _z ^{(n)}  = - \left ( n + \thalf \right ) \chi(\Sigma)=(2n+1)(h-1)
\eea
where $\chi (\Sigma)= 2 - 2h$ is the Euler number of $\Sigma$. The Riemann-Roch theorem may equivalently be recast in terms of Laplace-Beltrami operators, 
\bea
\label{C.RR1}
\dim {\rm Ker} \, \Delta _{(n+1)} ^- - \dim {\rm Ker} \, \Delta _{(n)}^+    = (2 n + 1 ) (h-1)
\eea
To prove this result, we use the fact established earlier  that the non-zero eigenvalues of $\Delta _{(n)}^+$ and $\Delta _{(n+1}^-$ are identical to obtain the following relation,
\bea
\label{C.dims}
\dim {\rm Ker} \, \Delta _{(n+1)} ^- - \dim {\rm Ker} \, \Delta _{(n)}^+    
= \Tr \left ( e^{ - s \Delta _{(n+1)}^- } \right )  - \Tr \left ( e^{ - s \Delta _{(n)}^+ } \right ) 
\eea
valid for all $0 < s \in \RR$. In particular, the right side may be evaluated in the limit $s \to 0$ where the short-time asymptotics of the heat-kernel may be used to derive an explicit formula. The short-time expansion, or Bott-Seeley expansion,  is given as follows, 
\bea
\Tr \left ( f \, e^{-s \Delta _{(n)}^\pm } \right ) =  { 1 \over 4 \pi s} \int _\Sigma d\mu _\mg f 
+{ 1 \pm 3 n \over 12 \pi} \int _\Sigma d \mu_\mg R_\mg f + \cO(s)
\eea
Setting $f=1$, substituting the expressions into (\ref{C.dims}), and using the integral expression for the Euler number, we 
establish (\ref{C.RR1}), which proves the Riemann-Roch theorem.

\subsection{Proof of the vanishing theorem}
\label{sec:C.VT}

In terms of the differential operators $\nabla ^z _{(n)}$ and $\nabla ^z _{(n)}$ the vanishing theorem read as follows,
\bea
\label{B.vanish}
 \dim {\rm Ker} \, \nabla ^z _{(n)} =0  & \hskip 1in & g=0 \hbox{ and } n >0
 \no \\
\dim {\rm Ker} \, \nabla ^z _{(n)} =0  & \hskip 1in & g\geq 2  \hbox{ and } n <0
\eea
The proof of these theorems uses the following equations from differential geometry,
\bea
{} [ \nabla _m, \nabla _n] \, \mt_p & = & - \mt_q \, R^q {}_{pmn}
\no \\
{} [ \nabla _m, \nabla _n] \, \mt^p & = & + \mt^q \, R^p{}_{qmn}
\eea
where $R^q{}_{pmn}$ is the Riemann tensor, and $t_q$ is a rank one tensor, i.e. a one-form. Adapted to the case of two dimensions, and working in complex coordinates,  the formulas reduce to,
\bea
{} [ \nabla _z , \nabla_{\bar z} ] \, \mt_z =  - \mg_{z \bar z} R_\mg \, \mt_z
\eea
where $R_\mg$ is the Gaussian curvature (normalized to be one for the round sphere of unit radius). Formulas for  tensors and spinors $\mt_{z \cdots z}= \mt \in T_{(n,0)}^*$ of arbitrary rank $n \in \ZZ/2$ may be derived by taking the tensor product and the square root, and we find, 
\bea
{} [ \nabla _z , \nabla_{\bar z} ] \, \mt  =  - n \, \mg_{z \bar z} \, R_\mg \, \mt
\eea
We are now ready to prove the vanishing theorem. We use the following rearrangement formula, obtained by integrating by parts (no boundary terms arise since $\Sigma$ has no boundary), 
\bea
\label{B.sq}
\int _\Sigma d \mu_\mg \, (\mg_{z \bar z})^{n-1} |\nabla _{\bar z} \, \mt |^2
= \int _\Sigma d \mu_\mg  \Bigg ( (\mg_{z \bar z})^{n-1} \, | \nabla _{\bar z} \bar \mt |^2
+ n (\mg_{z \bar z})^n R_\mg \, |\mt |^2  \Bigg ) 
\eea

$\bullet$ For the case $g=0$, we may choose a metric of constant positive curvature $R_\mg=1$ on the sphere. Since we have $n >0$ the right side of (\ref{B.sq})  is positive definite in $\mt$ so that the left side cannot vanish unless $\mt=0$. As a result there are no holomorphic differential forms of any positive weight $n >0$ on the sphere which proves the result on the first line of (\ref{B.vanish}). 

$\bullet$ For the case $g\geq 2$, we may choose a metric of constant negative curvature $R_\mg=-1$. Since we have  $n<0$ the right side of  (\ref{B.sq}) is again positive definite so that the left side cannot vanish unless $\mt=0$. As a result there are no holomorphic vector fields of any negative weight $n<0$ on a higher genus Riemann surface. In particular, this implies that there are no conformal Killing vectors or conformal Killing spinors on a Riemann surface of genus $g \geq 2$, and thus no continuous symmetries. 

$\bullet$ For the case  $g=1$, and a flat metric $R_\mg=0$ on the torus, the relation (\ref{B.sq}) imposes no restrictions.

\subsection{The dimension of moduli space}

The moduli space $\cM_g$ of  compact Riemann surfaces of arbitrary genus $g$ is the space of inequivalent complex structures or, equivalently, the space of inequivalent conformal structures. Complex structures and conformal structures may be parametrized  with the help of the Weyl-invariant tensor $\cJ_m {}^n = \sqrt{\mg} \, \ep_{mp}\, \mg ^{pn}$.  Two tensors that are related by a diffeomorphsim of $\Sigma$ are equivalent to one another, so that the moduli space is given by,
\bea
\cM_g = \{ \cJ \} / {\rm Diff}(\Sigma)
\eea
To compute the dimension of $\cM_g$, we evaluate the dimension of its tangent space at a given complex structure with local complex coordinates $z, \bar z$ in terms of which the metric is given by $\mg =  2\mg_{z \bar z} |dz|^2$, and the complex structure tensor $\cJ$ has the following components, 
\bea
\cJ_z{}^{\bar z} = - \cJ_{\bar z} {}^z = 0 
\hskip 1in
\cJ_z{}^z = - \cJ_{\bar z} {}^{\bar z} = i 
\eea 
Next, we compute the variation of the complex structure tensor, as the metric $\mg$ is varied, \textit{leaving the system of complex coordinates $z, \bar z$ unchanged. } The variation of the metric is used to first order,
\bea
\delta \mg = 2 \delta \mg_{z\bar z} \, |dz|^2 + \delta \mg_{zz} \, dz^2 + \delta \mg_{\bar z \bar z} \, d\bar z^2
\eea
The effect of the infinitesimal Weyl transformation $\delta \mg_{z \bar z}$ cancels out of the variation of $\cJ$ since the complex structure is invariant under Weyl transformations. Thus, we obtain the following variations, to first order in $\delta \mg_{zz}$ and $\delta \mg_{\bar z \bar z}$,  
\begin{align}
\delta \cJ_z{}^z & =0 & 
\delta \cJ_z{}^{\bar z} & =   + i \, \mg_{z \bar z} \, \delta \mg^{\bar z \bar z}  ~~ \in ~ T^*_{(1,-1)} 
\no \\
\delta  \cJ_{\bar z} {}^{\bar z} & = 0 &
\delta  \cJ_{\bar z} {}^z & =   -i \, \mg_{\bar z z} \, \delta \mg^{z z} ~~ \in ~ T^*_{(-1,1)} 
\end{align} 
Moduli deformations, however, are variations of complex structure modulo diffeomorphisms. The action of an infinitesimal  diffeomorphism $v^m \p_m$ on the metric is given as follows,
\bea
\delta _v \mg_{mn} = \nabla _m v_n + \nabla _n v_m
\eea
Under this variation, the complex structure transforms as follows,
\bea
\delta _v \, \cJ_z{}^{\bar z} & = & - i \, \nabla _z^{(1)}   v^{\bar z} 
\no \\
\delta _v \, \cJ_{\bar z} {}^z & = & + i \, \nabla _{\bar z}^{(-1)}    v^z 
\eea
Vectors in the tangent space $T_\mg \cM_g$  to moduli space $\cM_g$ at a given metric $\mg$,  may be identified with complex structure deformations that are in the orthogonal complement to the ranges of these operators, evaluated at the metric $\mg$,
\bea
T_\mg \cM_g = \Big ( {\rm Range } ~ \nabla _z ^{(1)} \Big ) ^\dagger 
\oplus \Big ( {\rm Range } ~ \nabla _{\bar z} ^{(-1)} \Big ) ^\dagger
\eea 
But the orthogonal complement to the range of any operator is the kernel of the adjoint of that operator. 
Since we have $(\nabla _z ^{(1)} )^\dagger = - \nabla ^z _{(2)} $, we find, 
\bea
T_\mg \cM_g = {\rm Ker} \nabla ^z _{(2)} \oplus {\rm Ker} \nabla ^{\bar z} _{(-2)}
\eea
First of all, we note that the tangent space to $\cM_g$ is split according to holomorphic and anti-holomorphic directions, so that $\cM_g$ is a complex manifold (actually, more precisely a complex orbifold). The dimension of moduli space may be read off from the table of dimensions given in the preceding subsection, 
\bea
\label{eq:modspacedim}
\dim_\CC \cM_g = \left \{ \begin{matrix} 0 & g=0 \cr 1 & g=1 \cr 3g-3 & g \geq 2 \cr \end{matrix} \right .
\eea
The combination $\delta \cJ_{\bar z} {}^z$ and its complex conjugate are often referred to as \textit{Beltrami differentials}. The elements of the dual space ${\rm Ker} \nabla ^z _{(2)}$ are the \textit{holomorphic quadratic differentials}.

\subsection*{$\bullet$ Bibliographical notes}

A summary of the ingredients of holomorphic line bundles on compact Riemann surfaces, accessible to physicists, was given in the review paper on string perturbation theory~\cite{RMP} and in the lecture notes~\cite{Bost}. A  fundamental and useful reference  is the book by Fay~\cite{Fay}. Helpful discussions of line bundles characterized by divisors may be found in the book by Farkas and Kra \cite{Kra}, while the deformation theory of complex structures for Riemann surfaces is treated quite explicitly in the book by Schiffer and Spencer \cite{Schiffer}.  More general treatments of vector bundles on complex and K\"ahler manifolds may be found in the books by Gunning \cite{gun2} and Kodaira \cite{Kodaira}. The Bott-Seeley expansion in the context of two-dimensional Riemannian manifolds is discussed and proven in \cite{Alvarez:1982zi} and \cite{RMP}.

\newpage

\section{Higher genus $\tet$-functions and meromorphic forms}
\setcounter{equation}{0}
\label{sec:Theta}

In this final appendix, we shall review the modular geometry of the Siegel half-space at higher rank, Riemann $\tet$-functions of higher rank, the embedding of higher-genus Riemann surfaces into the Jacobian variety via the Abel map,  and use these ingredients to construct the prime form, the Szeg\"o kernel, and other meromorphic differential forms on higher-genus Riemann surfaces.

\subsection{The Siegel half space}
\label{sec:A1}

The rank $g$ Siegel half space $\cH_g$  may be defined  as the space of complex $g \times g$ symmetric matrices with positive definite imaginary part, 
\bea
\label{A1}
\cH_g = \Big \{ \Omega \in \CC^{g \times g}, ~ \Omega^t = \Omega, ~ Y= \Im (\Omega)>0 \Big \}
\eea
Its dimension is  $\dim \cH_g = \half g (g+1)$. Alternatively, $\cH_g$ is given as the coset space
\bea
\label{A2}
\cH_g = Sp(2g, \RR) / U(g) 
\eea
The group $Sp(2g, \RR)$ acts on $\Omega$ by
\bea
\label{A3}
\Omega \to \Omega ' = (A\Omega +B) (C \Omega +D)^{-1}
\eea 
where the $g\times g$ real matrices $A,B,C,D$ are given in terms of $M \in Sp(2g,\RR)$ by, 
\bea
\label{A4}
M = \left ( \begin{matrix} A & B \cr C & D \cr \end{matrix} \right ) 
\hskip 0.8in 
M^t \mJ M = \mJ 
\hskip 0.8in 
\mJ = \left ( \begin{matrix} 0 & -I \cr I & 0 \cr \end{matrix} \right ) 
\eea
The unitary subgroup $U(g)$  is generated by setting $C=-B$ and $ D=A$ subject to the conditions $A^tA+B^tB=I$ and $A^tB-B^tA=0$. Since the group $U(g)$ contains a $U(1)$ factor, $\cH_g$ is a K\"ahler manifold, with the following $Sp(2g,\RR)$-invariant K\"ahler metric,
\bea
\label{A5}
ds^2_g = \sum_{I,J,K,L=1}^g (Y^{-1})^{IJ} \, d \Omega _{IK} \, (Y^{-1} )^{KL} \, d \bar \Omega _{JL}
\eea
The associated Laplace-Beltrami operator  $\Delta$ on scalar functions of $\cH_g$ is given by,
\bea
\label{A6}
\Delta = \sum _{I,J,K,L=1}^g 4 \, Y_{IJ} Y_{KL} \, \p^{IK} \, \bar \p^{JL}
\eea
where the derivatives with respect to the components of $\Omega_{IJ}$ are defined as follows,
\bea
\p^{II} = { \p \over  \p\Omega _{II}} 
\hskip 1in
\p^{IJ} = \half { \p \over  \p \Omega _{IJ}} \qquad \hbox{ for }  J \not= I
\eea 
For genus one, $\cH_1=\cH$ coincides with the Poincar\'e upper half-plane, $ds_1^2$ with the Poincar\'e metric on $\cH_1$, and $\Delta$ with the Laplace-Beltrami  operator on scalars functions on $\cH_1$.

\subsection{The Riemann theta function}

The \textit{Riemann $\tet$-functions} are complex-valued functions  on $\cH_g \times \CC^g$ that generalize the Jacobi $\tet$-functions on $\cH \times \CC$ to arbitrary rank $g$. Just as for  Jacobi $\tet$-functions, Riemann $\tet$-functions  may be considered for arbitrary complex characteristics $\delta = [\delta' |\delta '']$ with $\delta ', \delta '' \in \CC^{2g}$. In terms of local coordinates $\Omega$ on  $\cH_g$ and $\zeta = (\zeta _1, \cdots, \zeta _g)^t $ on $ \CC^g$, the Riemann $\tet$-function for characteristic $\delta$ may be defined by the following absolutely convergent series,  
\bea
\tet [\delta] (\zeta| \Omega) 
\equiv  
\sum _{n \in \ZZ^g } 
\exp \biggl (i \pi (n + \delta ') ^t \Omega (n+ \delta ') + 2\pi i (n+\delta ') ^t  (\zeta + \delta '') \biggl ) \, .
\eea
Upon shifting by $m,  n \in \ZZ^g$ we have the periodicity relations,
\bea
\tet [\delta] (\zeta + m + \Omega n| \Omega ) & = &
e^{ -i \pi n^t \Omega n - 2 \pi i n^t (\zeta + \delta ') + 2 \pi i m^t \delta '' } \, \tet [\delta] (\zeta | \Omega)
\no \\
\tet [\delta' +n | \delta'' + m ] (\zeta | \Omega ) & = &
e^{ 2 \pi i m ^t \delta'} \,  \tet [\delta' | \delta ''] (\zeta | \Omega)
\eea
The $\tet$-function without characteristics is often denoted by $\tet (\zeta| \Omega) = \tet [0] (\zeta| \Omega)$.

\sm

We shall focus on \textit{half-integer characteristics} $\delta$ with values $\delta', \delta '' \in (\ZZ/2\ZZ) ^g$ corresponding to a spin structure. The parity $4\delta ' \cdot \delta ''$ (mod 2) of a $\tet$-function with half-integer characteristics  depends on $\delta$ and, at a point $(\zeta | \Omega)$ where $\tet [\delta ](\zeta |\Omega) \not=0$,  is defined by,
\bea
\tet [\delta ] (- \zeta | \Omega ) = (-1) ^{4 \delta ' \cdot \delta ''} \tet [\delta ](\zeta | \Omega)
\eea
The spin structure $\delta$ is referred to as an even or odd spin structure according to whether the integer $4\delta ' \cdot \delta ''$ is even or odd.

\subsubsection{Modular transformations}

The discrete (or \textit{arithmetic}) subgroup $Sp(2g,\ZZ)$ of $Sp(2g, \RR)$ is the group of modular transformations,  whose action on the Siegel upper half-space $\cH_g$ is given by (\ref{A3}), but now with $M \in Sp(2g, \ZZ)$ with the entries of the matrices $A,B,C,D$ being integers.  Under a modular transformation $M \in Sp(2g,\ZZ)$, a half-integer characteristic $\delta = [\delta ' | \, \delta '']$ transforms into another half-integer  characteristic  $\ti \delta = [\ti \delta ' | \, \ti \delta '']$ given as follows,
\bea
\left ( \begin{matrix}  \tilde \delta ' \cr \tilde \delta '' \end{matrix} \right )
=
\left ( \begin{matrix}D & -C \cr -B & A \cr \end{matrix} \right )
\left ( \begin{matrix} \delta ' \cr \delta '' \cr \end{matrix} \right )
+ \half \, {\rm diag} \left ( \begin{matrix} CD^t \cr AB^t \cr \end{matrix} \right ) 
\hskip .7 in
M= \left ( \begin{matrix}A & B \cr C & D \cr \end{matrix} \right )
\eea
Under joint modular transformations of $\Omega$ and $\delta$, the $\tet$-function transforms as follows, 
\bea
\tet [\tilde \delta ] \left (\{ (C\Omega +D)^{-1} \}^t  \zeta | \tilde \Omega \right ) =
\ep (\delta, M) \, \det (C\Omega + D)^{\half} \, \tet [\delta ](\zeta | \Omega)
\eea
where $\ep(\delta, M) ^8=1$ and the precise dependence of $\ep ( \delta, M)$ on its arguments  is given in \cite{Fay,Igusa1}.

\subsubsection{Riemann relations on $\tet$-functions}

For each spin structure $\lambda$, there exists a  Riemann relation which may be expressed as a quadrilinear sum over all spin structures $\kappa$,
\be
\label{D.RR}
\sum _\kappa \<\kappa | \lambda \>
\tet [\kappa ](\zeta _1 ) \tet [\kappa ](\zeta _2)
\tet [\kappa ](\zeta _3 ) \tet [\kappa ](\zeta _4)
= 4\, 
\tet [\lambda ] (\zeta _1 ') \tet [\lambda ] (\zeta _2 ')
\tet [\lambda ] (\zeta _3 ') \tet [\lambda ] (\zeta _4 ')
\ee
The signature symbol for the pairing of two spin structures is defined by,
\bea
\< \kappa | \lambda \> = e^{4 \pi i ( \kappa' \cdot  \lambda'' - \kappa '' \cdot \lambda ')}
\eea 
and the relation between the vectors $\zeta$ and $\zeta '$ is given in terms of a matrix $\Lambda$,
\be
\left ( \begin{matrix}
\zeta _1 ' \cr  \zeta _2 ' \cr \zeta _3 ' \cr \zeta _4 ' \cr \end{matrix}  \right )
= \Lambda 
\left ( \begin{matrix}
\zeta _1  \cr  \zeta _2  \cr \zeta _3  \cr \zeta _4  \cr \end{matrix} \right )
\qquad \qquad
\Lambda =
\half \left (\begin{matrix}
 1 &  1 &  1 &  1 \cr 
 1 &  1 & -1 & -1 \cr 
 1 & -1 &  1 & -1 \cr 
 1 & -1 & -1 &  1 \cr \end{matrix} \right )
\ee 
which satisfies $\Lambda ^2 = I$. In the special case where $\zeta = \zeta '=0$, only even spin structures  $\kappa =\delta $ contribute to the sum and we recover a Riemann identity for each odd spin structure $\lambda = \nu$, 
\be
\sum _\delta \<\nu |\delta \> \tet [\delta ]^4 (0,\Omega )  =0
\ee
More generally, if $\zeta_1+\zeta_2+\zeta_3+\zeta_4=0$ then $\zeta_1'=0$,  the right side of (\ref{D.RR}) vanishes for any odd spin structure $\lambda$, and the Riemann relations become, 
\be
\sum _\kappa \<\kappa | \lambda \>
\tet [\kappa ](\zeta _1 ) \tet [\kappa ](\zeta _2)
\tet [\kappa ](\zeta _3 ) \tet [\kappa ](\zeta _4)
= 0
\ee
For genus one, there is only a single odd spin structure and the Riemann relation reduces to the famous Jacobi identity $\tet_2^4 - \tet _3^4 +\tet_4^4 =0$.

\subsection{The Jacobian, Abel map, and Riemann vanishing theorem}

We recall from appendix \ref{sec:RS} that a canonical basis may be chosen for the holomorphic $(1,0)$-forms $\om_I$ with  $I=1,\cdots, g$ on a compact Riemann surface of genus $g$ by normalizing the integrals of $\om_I$  on the $\mA$-cycles of a canonical basis $(\mA, \mB)$ of generators for the homology group $H_1 (\Sigma, \ZZ)$. The integrals on the $\mB$-cycles then give the period matrix $\Omega$ of the Riemann surface $\Sigma$ in the homology basis $(\mA, \mB)$, 
\bea
\oint _{\mA_I} \om _J = \delta _{IJ} 
\hskip 1in
\oint _{\mB_I} \om _J = \Omega _{IJ} 
\eea
Under a symplectic change of canonical homology basis, the holomorphic $(1,0)$-forms $\om$ and period matrix $\Omega$ transform as given in (\ref{B.3.4}).  In view of the Riemann relations, we have $\Omega ^t = \Omega$ and $\Im (\Omega) >0$, so that the period matrix of a compact Riemann surface takes values in the Siegel half space $\cH_g$. The set of $\mA$ and $\mB$ periods defines a lattice $\Lambda = \ZZ^g + \Omega \ZZ^g$ in $\CC^g$, whose quotient is an Abelian variety referred to as the the Jacobian variety, 
\bea
J(\Sigma) = \CC^g / \{ \ZZ^g + \Omega \ZZ^g\}
\eea
In other words, the Jacobian is a $g$-dimensional complex torus. For genus $g=1$, this torus gives an equivalent representation of the Riemann surface itself. For higher genus $g \geq 2$, this simple correspondence no longer holds, but is replaced by a more subtle identification produced by the Abel map.

\sm

Given a base point $z_0 \in \Sigma$, the Abel map sends a divisor $D$ of $n$ points $z_i \in \Sigma$ with weights $q_i \in \ZZ$ for $i=1,\cdots,n$, formally denoted by  $D=q_1 z_1 + \cdots  q_n z_n$, into $\CC^g$ by,
\bea
\label{abelmap}
D= q_1 z_1 + \cdots + q_n z_n 
\to 
\sum _{i=1} ^n q_i \int _{z_0} ^{z_i} (\omega _1, \cdots , \omega _g)
\eea
where the $g$-tuple $(\omega _1, \cdots , \omega _g)$ stands for the vector of holomorphic $(1,0)$-forms $\om_I$.
The Abel map into $\CC^g$ is multiple valued because each integral $\int _{z_0} ^{z_i} \om_I$ is multiple-valued in $\CC$, but the Abel map is single valued as a map into the Jacobian $J(\Sigma)$. 

\sm

{\thm The Riemann vanishing theorem states that $\tet(\zeta|\Omega)=0$ if and only if there exist $g-1$ points $p_1, \cdots, p_{g-1} \in \Sigma$ such that,
\bea
\label{zetaDel}
\zeta = p_1+\cdots p_{g-1} - \Delta (z_0)
\hskip 1in 
\zeta _I =  \sum_{i=1}^{g-1} \int _{z_0} ^{z_i} \om_I - \Delta _I(z_0)
\eea 
where $\Delta _I (z_0)$ are the components of the Riemann vector for base-point $z_0$, given by, 
\bea
\Delta _I (z_0) = - \half - \half \Omega_{II} + \sum_{J\not= I} \oint _{\mA_J} \om_J(z) \int ^z _{z_0} \om_I
\eea
The combination $\zeta$ is independent of the base-point. }

The proof of the Riemann vanishing theorem may be found in any book on $\tet$-functions, or in the physics literature in \cite{RMP}. For the special case of genus $g=1$, the set of points $p_i$ is empty, and the Riemann vector reduces to $\Delta = - \half - \half \tau$ which merely states that $\tet ( \thalf + \thalf \tau |\tau)=0$, a fact we established long ago.

\sm

An important application of the Riemann vanishing theorem is to the existence of the following holomorphic $(1,0)$-forms, defined for any odd spin structure $\nu$ by 
\bea
\label{omnu}
\om_\nu(z) = \sum _I \om _I(z) \p^I \tet [\nu] (0|\Omega )
\eea
Its $2(h-1)$ zeros are all double zeros at points $p_i$ such that $\zeta_I=\nu_I$ in (\ref{zetaDel}).  Therefore, $\om_\nu(z)$ admits a holomorphic square root $h_\nu(z)$. Since the square of $h_\nu$ is  $\om_\nu(z)$, which is a section of the canonical bundle, $h_\nu$ is a section of a spin bundle with spin structure $\nu$. For each odd spin structure $\nu$, the holomorphic $(\half, 0)$-form $h_\nu$ is unique up to a sign.

\subsection{The prime form}

To construct the  \textit{prime form} we consider an arbitrary  odd spin structure $\nu$. Its associated holomorphic section $h_\nu(z)$ of the spin bundle with spin structure $\nu$ has $g-1$ zeros. The prime form is a $(-\thalf,0)$-form in both variables $z$ and $w$, defined by,
\bea
E(z,w|\Omega) = {\tet [\nu ] (z-w| \Omega) \over h_\nu (z) h_\nu (w)}
\eea
The argument $z-w$ of the $\tet$-function stands for the Abel map of (\ref{abelmap}) with divisor  $D=z-w$. The form $E(z,w|\Omega)$ defined this way is  independent of $\nu$. It is holomorphic in $z$ and $w$ and has a unique simple zero at $z=w$. It is single valued when $z$ is moved around $\mA_I$ cycles, but has non-trivial monodromy when $z\to z+\mB_I$ is moved around $\mB_I$ cycles,
\bea
E(z+\mA_I,w|\Omega) & = &E(z,w|\Omega) 
\no \\
E(z+\mB_I,w |\Omega) & = & - \exp \biggl ( -i \pi \Omega _{II} + 2 \pi i \int ^z _w \!
\omega _I \biggr ) E(z,w|\Omega) 
\eea
Therefore, it is properly defined on the simply connected covering space of $\Sigma \times \Sigma$, or it may be defined in a fundamental domain for $\Sigma$ in the complex plane.

\subsection{Holomorphic differentials}

Combining the Riemann-Roch and vanishing theorems, the dimension of the space of holomorphic $(n,0)$-forms is denoted by $\Upsilon(n) = \dim {\rm Ker} \, \pbz ^{(n)}$ and is given by, 
\be
\Upsilon (n) 
=
\left \{ 
\begin{matrix}
0 & n < 0\, , \ {\rm and} \ n=1/2 \ {\rm even \ spin \ structure} \cr  
1 & n=0\, ,   \ {\rm and} \ n=1/2 \ {\rm odd \ spin \ structure}  \cr 
g & n=1 \cr  (2n-1) (g-1) & n \geq 3/2, \ {\rm and} \, g\geq 2  \end{matrix}
\right . 
\ee
The dimensions listed for $n=1/2$ are for generic moduli and are valid for exceptional moduli mod 2. A basis  of holomorphic differentials is denoted by $\phi _a ^{(n)}$, $a=1, \cdots , \Upsilon (n)$. They are holomorphic sections of the line bundles $K^n$, the $n$-th power of the canonical bundle $K$, for which the number of zeros is given by, 
\be
\# \ {\rm zeros}  \ \phi ^{(n)} _a = c_1(K^n) = 2n(g-1) 
\ee 
For a given $a$, there are $\Upsilon(n)-1$ zeros fixed to be $z_b $ with $b \not=a$, and another $g$ zeros due to the Riemann vanishing theorem, giving a total of $2n(g-1)$, in agreement with the value of the first Chern class. 
For $n=0$, they are constants, for $n=1/2$ and $\nu $ odd they are denoted by $h_\nu (z)$, while for $n=1$ they are the holomorphic Abelian differentials  $\omega _I$, for $I=1,\cdots ,g$.

\sm

Given any set of $\Upsilon (n)$ points $z_1, \cdots , z_{\Upsilon (n)}$ on the surface, we may choose a basis $\hat \phi ^{(n)} _a$ for the holomorphic $n$-differentials normalized at the points $z_b$ by,
\be
\label{diffnorm}
\hat \phi _a ^{(n)} (z_b) = \delta _a ^b
\ee
The holomorphic differentials with this normalization may be exhibited explicitly in terms of the prime form $E(z,w)$, the $h/2$ differential $\sigma (z)$ and $\vartheta$-functions. For $n \geq 3/2$, we have, 
\be
\label{diffexp}
\hat \phi _a ^{(n)} (z)
= 
{\vartheta [\delta ](z-z_a + \sum z_b -(2n-1)\Delta) \over 
 \vartheta [\delta ] ( \sum z_b -(2n-1)\Delta)}
{\prod _{b\not= a} E(z,z_b) \over \prod _{b\not= a} E(z_a,z_b)}
\biggl ( {\sigma (z) \over \sigma (z_a)} \biggr ) ^{2n-1}
\ee
Here, $\sigma (z)$ is a holomorphic $(\tfrac{h}{2}, 0)$-form without zeros or poles defined, up to a constant, by the following ratio,
\bea 
{\sigma (z) \over \sigma (w)} & = & 
{\vartheta (z-\sum p_i +\Delta) \over \vartheta (w-\sum p_i +\Delta)}
\prod _{i=1} ^h {E(w,p_i) \over E(z,p_i)}
\eea
where $p_i$, $i=1,\cdots , h$ are arbitrary points on the surface. Note that $\sigma(z)$ is single valued around $\mA_I$ cycles but multivalued around $\mB_I$ cycles in the following way
\bea
\sigma (z+\mB_I) &=& \sigma (z) \exp \bigl \{ -i\pi (h-1) \Omega _{II} + 2 \pi i
\Delta _{Iz} \bigr \}
\eea
Besides the $\Upsilon(n)-1$ zeros $z_b$, $b\not=a$, the differential $\hat \phi _a ^{(n)}(z)$ has $h$ additional zeros. The differential  $\hat \phi _a ^{(n)}(z)$ is an $(n,0)$-form in $z$, a $(-n,0)$-form in $z_a$ and a $(0,0)$-form $z_b$ with $b\not=a$. For $n=1$, we have, 
\be
\hat \phi ^{(1)} _a (z)
= 
{\tet  (z-z_a  + \sum z_b -w_0 - \Delta ) \over 
 \tet (  \sum z_b -w_0 - \Delta ) \ E(z,w)} \
\prod _{b\not= a} {E(z,z_b) \over E(z_a,z_b)} 
{ E(z_a,w_0) \over E(z, w_0)}
\ {\sigma (z) \over \sigma (z_a)} \, .
\ee

\subsection{Meromorphic differentials}

Meromorphic differentials play the role of Green functions for the Cauchy-Riemann operators acting on holomorphic line bundles. The  Green function $G_{n}(z,w)= G_n (z,w;z_1,\cdots, z_{\Upsilon(n)})$ for the operator $\p _{\bar z} ^{(n)}$ is a meromorphic $(n,0)$ form in $z$ and $(1-n,0)$-form in $w$. When $n\geq 3/2$ for general spin structure $\delta$  and $n=1/2$ for even spin structure $\delta$, the Green function $G_n$ may be defined by the following relations,
\bea
\p _{\bar z} ^{(n)} G _n (z,w)  & = & + 2 \pi \delta (z,w)
\\
\p _{\bar w} ^{(1-n)} G_n(z,w) & = & - 2 \pi \delta (z,w)
+ 2\pi \sum _{a=1} ^{\Upsilon (n)} \hat \phi _a ^{(n)} (z) \delta (w,z_a)
\eea
The properly normalized holomorphic $n$-differentials $\hat \phi _a ^{(n)}$ are defined in (\ref{diffnorm}) and (\ref{diffexp}). Setting $z=z_a$, we have $\p _{\bar w} ^{(1-n)} G_n(z_a,w)=0$, so that $G_n(z_a,w)=0$.
Explicit expressions for the Green's function are 
\be
G_n (z,w) 
= 
{\vartheta [\delta ]\big (z-w + \sum z_b -(2n-1)\Delta \big ) \over 
 \vartheta [\delta ]\big ( \sum z_b -(2n-1)\Delta \big ) \ E(z,w)} \
{\prod _{a} E(z,z_a) \over \prod _{a} E(w,z_a)} \
\biggl ( {\sigma (z) \over \sigma (w)} \biggr ) ^{2n-1}
\ee
For $n=1/2$, this reduces to the standard form of the Szeg\" o kernel, usually denoted by
\be
S_\delta (z,w) = 
{\vartheta [\delta ]\big (z-w  \big ) \over 
 \vartheta [\delta ]\big ( 0 \big ) \ E(z,w)}
\ee
For $n=1$, the Green's function $G_1(z,w)=G_1(z,w;z_1,\cdots ,z_h, w_0)$
is the Abelian differential of the third kind, satisfying
\bea
\p _{\bar z} ^{(1)} G _1 (z,w)  & = & + 2 \pi \delta (z,w) - 2\pi
\delta (z,w_0)
\\
\p _{\bar w} ^{(0)} G_1 (z,w) & = & - 2 \pi \delta (z,w)
+ 2\pi \sum _{a=1} ^h \hat \phi _a ^{(1)} (z) \ \delta (w,z_a)
\eea
and explicitly given by the following expression
\be
G_1(z,w) 
= 
{\vartheta \big (z-w-w_0 + \sum z_b - \Delta \big ) \over 
 \vartheta \big ( -w_0 + \sum z_b - \Delta \big ) \ E(z,w)} \
{\prod _{a} E(z,z_a) E(w,w_0) \over \prod _{a} E(w,z_a) E(z, w_0)} \
 {\sigma (z) \over \sigma (w)} \, .
\ee
The combination $\p_z \p_w \ln E(z,w)$ is a meromorphic differential (Abelian of the second kind) with a single double pole at
$z=w$. Its integrals around homology cycles are given by
\bea
\label{prf}
\oint _{\mA_I} \! dz \p_z \p_w \ln E(z,w) & = & 0
\nonumber \\
\oint _{\mB_I} \! dz \p_z \p_w \ln E(z,w) & = & 2 \pi i \omega _I(w)
\eea
and will be of use throughout.

\subsection{The  $b c$ system}
\label{app.bc}

A useful unification of the formulas for holomorphic and meromorphic forms is provided by the conformal field theory correlators of anti-commuting fields $b$ and $c$ of weights $(n,0)$ and $(1-n,0)$.  This quantum field theory was already introduced in section \ref{sec:5.bc} for a Riemann surface of genus 1, namely a torus. Here,  we shall  generalize the correlators to arbitrary genus $g \geq 2$. Since the roles of $b$ and $c$ are swapped by letting $n \to 1-n$, we shall  restrict to considering the cases $n \geq \thalf$ without loss of generality.

\sm

As in the case of the torus, the fields $b,c$ are locally holomorphic, 
\bea
\pbz b = \pbz c=0
\eea
However, as  $b$ and $c$  fields approach one another poles may develop. The poles may be represented schematically by the singular contribution to the operator product expansion, given by a simple pole with unit residue,
\bea
b(z) \, c(w) \sim { 1 \over z-w}
\eea
All poles must arise from these operator coincidences.  In view of the anti-commuting nature of the fields, the operator product of two like fields produces a simple zero,
\bea
b(z_1) b(z_2) & \sim & (z_1-z_2) \, \p b(z_2)  \ b(z_2)
\no \\
c(w_1) c(w_2) & \sim & (w_1-w_2) \, \p c(w_2) \, c(w_2)
\eea
In the case of the torus we used Fourier analysis to solve for the fields and their correlators. Since higher-genus Riemann surfaces have no continuous isometries, such as translation symmetry, Fourier analysis is no longer applicable. 

\sm

Now consider the correlator of an arbitrary number of the $b$ and $c$ fields of weight $(n,0)$ and $(1-n,0)$ respectively,
\bea
\< b(z_1) \cdots b(z_N) \, c(w_1) \cdots c(w_M)\>
\eea
Viewed as a function of $z_1$, the correlator has simple zeros at $z_2, \cdots, z_N$ and simple poles at $w_1, \cdots, w_M$.  But since all poles arise from operator coincidences, and no other poles can appear, the number of poles in $z_1$ is exactly $M$, and thus the $n$-form must have $M+2n(g-1)$ zeros. Now $N-1$ of these zeros are specified by the positions of the points $z_2, \cdots, z_N$, leaving an extra $g$ zeros due to the Riemann vanishing theorem, for a total of $N-1+g$ zeros. Equating the number of zeros computed in these two different ways gives,
\bea
\label{D.const}
N=M+(2n-1)(g-1)
\eea 
The correlator must vanish unless the number of $b$ and $c$ fields satisfies this relation.

\subsubsection{The case $n = \thalf$ with even spin structure}

In the case $n = \half$ with even spin structure $\delta$ and generic moduli,  neither the $b$ nor the $c$ field has zero modes. 
By using Wick contractions, the correlator is given by,
\bea
\< b(z_1) \cdots b(z_M) \, c(w_M) \cdots c(w_1)\> = \det  S_\delta (z_i, w_j)
\eea
where $S_\delta$ is the Szeg\"o kernel for even spin structure $\delta$, the determinant is taken of the $M \times M$ matrix $S(z_i, w_j)$, and the numbering of the points $w_j$ has been chosen  in descending order so that the sign multiplying the determinant on the right side is positive.  An alternative formula is obtained by matching poles and zeros in each variable, and we obtain, 
\bea
\< b(z_1) \cdots b(z_M) \, c(w_M) \cdots c(w_1) \> =
{ \tet [\delta] (D|\Omega) \over \tet [\delta] (0|\Omega) } { \prod _{i<j} E(z_i,z_j) E(w_i,w_j) \over \prod _{i,j} E(z_i, w_j)}
\eea
where the divisor is $D=(z_1-w_1)+\cdots + (z_M - w_M)$. This formula is often referred to as the \textit{bosonized correlator}  as it may be obtained by representing the spin-$\thalf$ fermion fields $b,c$ in terms of the chiral half of a complex scalar field. 

\subsubsection{The case $n = \thalf$ with odd spin structure}

In the case $n=\half$ with  odd spin structure $\nu$ and generic moduli, there is a single zero mode $h_\nu$ of the Dirac equation. Clearly, the two-point correlator $\<b (z) c(w)\>$  must be saturated by the zero modes and be proportional to $h_\nu(z)h_\nu(w)$.  For higher point functions, we need a Green function. The Green function may be defined by an equation that is orthogonal to the Dirac zero mode. Such an equation is not unique, but different definitions will produce the same correlator. Preserving meromorphicity, we choose the following definition, 
\bea
\pbz S_\nu(z,w) = 2 \pi \delta(z,w) - 2 \pi \delta (z,w_0) { h_\nu (w) \over h _\nu (w_0)}
\eea
which is orthogonal to $h_\nu(z)$ by construction. The general correlator is then given by,
\bea
\< b(z_1) \cdots b(z_M) \, c(w_M) \cdots c(w_1) \>
= \sum _{i,j=1}^M (-)^{i+j} \< b(z_i) c(w_j) \> \, \det S_\nu (z_k, z_l) \Big |_{{k \not= i, \atop  l \not= j}} 
\eea
The constant of proportionality in $\<b (z) c(w)\> \sim h_\nu(z)h_\nu(w)$ is given by the functional determinant of the Dirac operator on $n=\half$ sections with odd spin structure $\nu$.

\subsubsection{The case $n = 1$}

In any non-vanishing correlator for $n=1$, the $b$ field has $g$ zero modes, while the $c$ field has one zero mode, namely the constants.  We shall normalize the correlator so that the simplest non-zero amplitude is given by,
\bea
\cG_0(z_1, \cdots, z_g; w) = \< b(z_1) \cdots b(z_g) c(w) \> = \det \om_I (z_i)
\eea
where the determinant is taken of the $g \times g$ matrix with entries $\om_I(z_i)$ for $i, I=1,\cdots, g$. The correlator with an arbitrary number of $b,c$ fields subject to the condition (\ref{D.const}), 
\bea
\cG_M(z_1, \cdots w_{M+1} ) = \< b(z_1) \cdots b(z_{g+M})\,  c(w_1) \cdots c(w_{M+1}) \>
\eea
with this normalization is given by
 \bea
\cG_M (z_1,  \cdots, w_{M+1}) =  { \tet \left ( D \right )  \over Z^3 }  \, 
{ \prod _{i<j} E(z_i,z_j) \prod _{a<b} E(w_a,w_b) \prod _i \sigma (z_i)  \over \prod _{i,a} E(z_i,w_a) \prod _a \sigma (w_a)  } 
\label{defcgn}
\quad
\eea
where $i,j =1, \cdots, g+M$ and $a,b=1,\cdots, M+1$ and $D$ is defined by
\bea
D = \sum _{i=1}^{g+M}  z_i - \sum _{a=1}^{M+1} w_a - \Delta
\eea 
The factor $Z$ is the chiral boson partition function  determined by setting $M=0$,  
\bea
\det \om_I(p_i) = { \tet \left ( p_1 + \cdots p_g -q -\Delta \right )  \over Z^3 }  \, 
{ \prod _{i<j} E(p_i,p_j) \prod _i \sigma (p_i)  \over \prod _{i} E(p_i,q) \sigma (q)  } 
\eea
for arbitrary points $q$ and $p_i$ with $i=1,\cdots, g$. 
With this normalization the correlator transforms under modular transformations by,
\bea
\tilde \cG_M(z_1, \cdots, w_{M+1}) = \det (C \Omega +D)^{-1} \cG_M(z_1, \cdots, w_{M+1})
\label{trflawG}
\eea

\subsubsection{The case $n \geq \tfrac{3}{2}$}

For $n \geq \tfrac{3}{2}$, the $b$ field has $\Upsilon = (2n-1)(g-1)$ zero modes, while the $c$ field has no zero modes. 
We have the following bosonization formula, 
\bea
\< b(z_1) \cdots b(z_{M+\Upsilon}) \, c(w_1) \cdots c(w_M) \> =
{ \tet [\delta] (\zeta |\Omega) \over Z} { \prod _{a<b} E(z_a,z_b) \prod _{i<j} E(w_i,w_j) \prod _a \sigma (z_a) \over \prod _{i,a} E(z_a, w_i) \prod _i \sigma (w_i)} 
\no \\
\eea
where $a,b=1,\cdots , M+ \Upsilon$, while $i,j=1, \cdots , M$ and $\zeta$ is given by,
\bea
\zeta = \sum _{a=1}^{M + \Upsilon} z_a - \sum _{i=1}^M w_i - (2n-1) \Delta
\eea
In the special case where $M=0$, we recover a purely holomorphic correlator, which is entirely saturated by the zero modes of the field $b$, 
\bea
\< b(z_1) \cdots b(z_\Upsilon)  \> =
{ \tet [\delta] (\zeta |\Omega) \over Z}  \prod _{a<b} E(z_a,z_b) \prod _a \sigma (z_a) 
\eea

\subsection*{$\bullet$ Bibliographical notes}

Classic introductions to Siegel modular forms and higher rank modular geometry may be found in the books by Siegel~\cite{Siegel2,Siegel3}, Igusa~\cite{Igusa1}, and Klingen~\cite{Klingen}. Useful lecture notes on higher rank modular forms are those by van der Geer~\cite{vdGeer}. The connection between higher rank $\tet$-functions and Riemann surfaces is spelled out in the classic book by Fay~\cite{Fay}. The connection between higher rank $\tet$-functions and non-linear equations may be found in the survey article by Dubrovin \cite{Dubrovin}. The solution of the $bc$ quantum system in terms of $\tet$-functions, using the work of \cite{Fay}, was given in \cite{Verlinde:1986kw}.

\newpage

\bibliography{bib}
\bibliographystyle{JHEP}

\end{document}